\documentclass[12pt,a4paper,twoside]{book}

\usepackage{thesis}

\usepackage{graphicx, epstopdf}
\usepackage{amsmath}

\usepackage[utf8]{inputenc}
\usepackage[sorting=none]{biblatex} 
\usepackage[colorlinks]{hyperref} 
\usepackage[toc,acronym,nonumberlist]{glossaries} 
\usepackage[labelfont=bf]{caption} 
\usepackage{textcomp, gensymb}  
\usepackage{pifont,mdframed} 
\usepackage{physics}  
\usepackage{subcaption} 
\usepackage[font=small,labelfont=bf]{caption}
\usepackage{minted} 
\usepackage{xcolor} 
\usepackage{amssymb} 
\usepackage{pdflscape} 
\usepackage{dirtytalk} 
\usepackage{setspace} 
\usepackage{chngcntr} 
\usepackage{caption} 
\usepackage{makecell, booktabs}  

\usepackage{CJKutf8}

\hypersetup{
    linktocpage,
    colorlinks = true,
    linkcolor = blue,
    citecolor = red
}

\definecolor{LightGray}{gray}{0.9}

\makeatletter
\newcommand*{\glsplainhyperlink}[2]{%
    \begingroup%
      \hypersetup{hidelinks}%
      \hyperlink{#1}{#2}%
    \endgroup%
}
\let\@glslink\glsplainhyperlink
\makeatother

\makeglossaries
\newacronym{wcd}{WCD}{water Cherenkov detector}
\newacronym{ssd}{SSD}{scintillator surface detector}
\newacronym{eas}{EAS}{extensive air shower}
\newacronym{mip}{MIP}{minimum ionising particle}
\newacronym{vem}{VEM}{vertical equivalent muon}
\newacronym{ldf}{LDF}{lateral distribution function}
\newacronym{sdp}{SDP}{shower detector plane}
\newacronym{sd}{SD}{surface detector}
\newacronym{fd}{FD}{fluorescence detector}
\newacronym{oa}{OA}{opening angle}
\newacronym{snrem}{SNR}{supernova remnant}
\newacronym{snr}{SNR}{signal-to-noise ratio}
\newacronym{cmb}{CMB}{cosmic microwave background}
\newacronym{gzk}{GZK}{Greisen–Zatsepin–Kuz’min}
\newacronym{dsa}{DSA}{diffusive shock acceleration}
\newacronym{em}{EM}{electromagnetic}
\newacronym{uv}{UV}{ultraviolet}
\newacronym{agasa}{AGASA}{Akeno Giant Air Shower Array}
\newacronym{fast}{FAST}{Fluorescence detector Array of Single-pixel Telescopes}
\newacronym{tdr}{TDR}{top-down reconstruction}
\newacronym{fadc}{FADC}{flash analog to digital converter}
\newacronym{heat}{HEAT}{High Elevation Auger Telescopes}
\newacronym{mc}{MC}{Monte Carlo}
\newacronym{ism}{ISM}{interstellar medium}
\newacronym{agn}{AGN}{active galactic nuclei}
\newacronym{pmt}{PMT}{photomultiplier tube}
\newacronym{wls}{WLS}{wavelength-shifting}
\newacronym{uhecr}{UHECR}{ultra-high energy cosmic ray}
\newacronym{auger}{Auger}{the Pierre Auger Observatory}
\newacronym{ta}{TA}{the Telescope Array experiment}
\newacronym{nkg}{NKG}{Nishimura-Kamata-Greisen}
\newacronym{corsika}{CORSIKA}{COsmic Ray SImulations for KAscade}
\newacronym{icrc}{ICRC}{International Cosmic Ray Conference}
\newacronym{grb}{GRB}{gamma-ray burst}
\newacronym{gh}{GH}{Gaisser-Hillas}
\newacronym{ccd}{CCD}{charge-coupled device}
\newacronym{fov}{FOV}{field of view}
\newacronym{gcos}{GCOS}{Global Cosmic Ray Observatory}
\newacronym{brm}{BRM}{Black Rock Mesa}
\newacronym{ll}{LL}{Los Leones}
\newacronym{psf}{PSF}{point-spread-function}
\newacronym{yap}{YAP}{yttrium-aluminium-perovskite}
\newacronym{clf}{CLF}{central laser facility}
\newacronym{nn}{NN}{neural network}
\newacronym{dnn}{DNN}{deep neural network}
\newacronym{lstm}{LSTM}{Long-Short Term Memory}
\newacronym{tsfel}{TSFEL}{Time-Series Feature Extraction Library}
\newacronym{pe}{p.e.}{photoelectrons}
\newacronym{fir}{FIR}{finite impulse response}

\title{Development of the Reconstruction Procedure of the Fluorescence detector Array of Single-pixel Telescopes for measuring Ultra-High Energy Cosmic Rays}

\author{Fraser Bradfield}
\date{\today}

\begin{document}
\newcommand{\Xmax}{$X_{\textrm{max}}$}
\newcommand{\gcm}{\,g\,cm$^{-2}$}
\newcommand{\REF}{\textcolor{red}{[ref?]}}

\def\Offline{\mbox{$\overline{\textrm%
{Off}}$\hspace{.05em}\protect\raisebox{.4ex}%
{$\protect\underline{\textrm{line}}$}}\space}

\renewcommand{\thepage}{\roman{page}}
\renewcommand{\bibname}{References}

\pagestyle{empty} 
\input titlepage

\cleardoublepage
\pagestyle{headings}
\markboth{Contents}{Contents}
\tableofcontents

\pagestyle{empty} 
\cleardoublepage
\chapter*{Abstract}
The \gls{fast} is an experiment aiming to deploy an array of simplified, autonomous fluorescence telescopes over an area of $\sim60,000$\,km$^{2}$ to observe \glspl{uhecr}. The unprecedented size of such an array will enable measurements of cosmic rays with energies above 10$^{20}$\,eV with large statistics, providing new insights into UHECR sources. To achieve the low-cost per telescope required for an array of this scale, each FAST telescope is equipped with just four \glspl{pmt}. As such, traditional bottom-up methods for reconstructing the properties of observed cosmic-ray-induced extensive air showers are not applicable. Instead, a top-down approach is used where simulations are directly compared to data and the best match chosen via a maximum likelihood estimation.
This method is known as the \say{\gls{tdr}} and requires an accurate \say{first guess} of the shower parameters to be successful.
In this work, improvements to the current \gls{tdr} and first guess estimation methods are made. 
By ensuring a smooth relationship between the shower parameters and output traces the reconstruction efficiency of the \gls{tdr} using a full FAST array is improved from $\sim30\%$ to $\sim95\%$ at 10$^{20}$\,eV. 
The overall precision of the \gls{tdr} is also improved by implementing a cut which excludes events where the position of the shower maximum (\Xmax{}) is outside the field of view of triggered telescopes.
Two methods for estimating a first guess of the shower parameters are then developed.
The first utilises machine learning techniques, expanding upon previous work, and the second a library of templates. 
Both methods are shown to be effective for predicting the shower geometry for events which trigger $>2$ \glspl{pmt} or are observed from two or more locations (stereo observation). 
In particular, the combined performance of the machine learning approach and \gls{tdr} is shown to achieve resolutions in the arrival direction, \Xmax{} and energy of $\sim{}2\degree$, $\sim{}30$\gcm{} and $\sim{}7\%$ respectively for events observed in stereo.
Finally, the improved reconstruction is applied to data from the current FAST prototype installations. 
The FAST results are shown to agree reasonably well with those from other experiments. Degeneracies in the reconstructed shower energies and core positions resulting from different first guesses highlight the importance of stereo observation. Finally, the first measurements of the UHECR energy spectrum and composition by FAST are presented. 

\newpage

\addcontentsline{toc}{chapter}{Abstract}

\cleardoublepage
\chapter*{Declaration}
I certify that this work contains no material which has been accepted for the award of any other degree or diploma in my name, in any university or other tertiary institution and, to the best of my knowledge and belief, contains no material previously published or written by another person, except where due reference has been made in the text. In addition, I certify that no part of this work will, in the future, be used in a submission in my name, for any other degree or diploma in any university or other tertiary institution without the prior approval of Osaka Metropolitan University and where applicable, any partner institution responsible for the joint award of this degree.

\vspace{10mm}

\textbf{Signed:}\hspace{40mm}\textbf{Date:} 2025/09/01
\addcontentsline{toc}{chapter}{Declaration}

\cleardoublepage
\chapter*{Acknowledgements}
Reflecting on the last three years, I feel deeply grateful to have had such a wonderful experience in Japan whilst pursuing my PhD. I attribute this to the incredibly kind, intelligent and hardworking people I have met along the way. To all those not mentioned here who have made this journey so special, thank you.

\vspace{5mm}

First and foremost, I would like to thank my supervisor, Associate Professor Toshihiro Fujii, for his support and guidance throughout my PhD. Your enthusiasm has been infectious and I have learnt much from our time working together. To Professor Yoshiki Tsunesada, thank you for your advice and fruitful discussions regarding my research, and for always correcting my Japanese.

\vspace{5mm}

To all members of the cosmic ray research lab at OMU, past and present, thank you my making my time in Japan so memorable. I will look back fondly on our travels together across Japan, our drinking parties, and everyday life in the lab. In particular I would like to thank my fellow FAST members, Haruka Tachibana and Shunsuke Sakurai, for being thoroughly enjoyable to work with, and Takuro Kobayashi, for his early friendship and assistance in my transition to living in Japan. 

\vspace{5mm}

To all those who encouraged me to pursue a PhD overseas, thank you. It is no exaggeration to say I would not be here without you. Special mention is given to the Yonezawa family, whose kindness and support both in Australia and Japan has been immeasurable, and which I hope to repay some day.

\vspace{5mm}

To my friends and family back in Australia, I have missed you all, but have always felt your love and encouragement from afar. Thank you.

\vspace{5mm}

Lastly, to my girlfriend Miki. I cannot imagine these last three years without you. Being with you has made Japan feel like home. I hope we can share many more laughs together in the future.

\begin{CJK}{UTF8}{min}
\chapter*{謝辞}

この3年間を振り返ると、日本での博士課程で素晴らしい経験を積むことができたことに深く感謝しています。これは、これまで出会った親切で賢明な方々のおかげです。ここに名前を挙げていない皆様にも、この旅を特別なものにしてくださったことに、心から感謝いたします。

\vspace{5mm}

まず第一に、博士課程を通して支えとご指導をいただいた指導教員の藤井俊博准教授に感謝申し上げます。先生の熱意は私にも伝染し、共に研究した時間から多くのことを学びました。常定芳基教授には、研究に関するアドバイスと有意義な議論、そして常に私の日本語を訂正していただき、ありがとうございました。

\vspace{5mm}

宇宙線物理学研究室の過去および現在のメンバーの皆様には、日本での時間を本当に思い出深いものにしていただき、ありがとうございました。共に日本中を旅して過ごしたこと、飲み会、そして研究室での日々の生活をこれからも懐かしく想い出すでしょう。特に、FASTの仲間である橘春花さんと櫻井駿介さんには、一緒に仕事をするのが本当に楽しかったです。そして、小林拓郎さんには、来日から最初の友達であることと日本生活に慣れるまでサポートしてくださったことに感謝したいと思います。

\vspace{5mm}

海外で博士号を取得するよう励ましてくださった皆様にも感謝いたします。皆様がいなければ、今の私はここにいないと言っても過言ではありません。特に、オーストラリアと日本の両方で計り知れないほどの親切なサポートをしてくださった米澤一家には感謝の意を表します。いつか恩返ししたいと思っています。

\vspace{5mm}

オーストラリアの友人や家族の皆さん、皆さんがいなくて寂しい思いをしましたが、遠くからでもいつも皆さんの愛情と励ましを感じていました。ありがとうございました。

\vspace{5mm}

最後に、恋人のミキさん。あなたのいないこの3年間は想像もできません。
あなたと一緒にいると、日本がまるで故郷のように感じられます。これからも、たくさん笑い合えることを願っています。
\end{CJK}

\addcontentsline{toc}{chapter}{Acknowledgements}


\makeatletter
\def\cleardoublepage{%
  \clearpage
  \if@twoside
    \ifodd\c@page
    \else
      \hbox{}
      \thispagestyle{empty}
      \newpage
      \if@twocolumn \hbox{}\newpage \fi
    \fi
  \fi
}
\makeatother

\mainmatter  
\pagestyle{headings}
\chapter*{Introduction}

Despite having been studied for more than a century, the origin and nature of the most energetic particles in the universe, so-called \say{ultra-high energy cosmic rays (UHECRs)}, remains unclear. 
Primarily protons and other atomic nuclei, these particles originate in extreme astrophysical environments, reaching energies more than a million times larger than achievable with current man-made particle accelerators on Earth. 
By measuring the cascades of secondary particles produced when UHECRs interact with the Earth's atmosphere known as \say{\glspl{eas}}, the properties of the original cosmic rays can be reconstructed.
Cosmic rays with energies above 10$^{20}$\,eV are of particular interest since the effect of extragalactic and galactic magnetic fields on their trajectories is minimal. 
This means their arrival directions as measured at Earth should point back to their sources, making them a promising candidate to reveal the origins of UHECRs. 
However, detecting a significant number of these particles is challenging due to their exceedingly low flux at Earth, less than 1 particle/km$^2$/century. 

\vspace{5mm}

The Fluorescence detector Array of Single-pixel Telescopes (FAST) is a next-generation cosmic ray experiment aiming to address this challenge. 
FAST intends to deploy simplified, low-cost, autonomous fluorescence telescopes, which measure the fluorescence light emitted from EASs, over an area of $\sim60,000$\,km$^2$ - an order of magnitude larger than the current largest cosmic ray experiments. 
With this increased exposure, FAST will be able to measure an unprecedented number of cosmic rays with energies above 10$^{20}$\,eV using the fluorescence technique. 
To meet cost constraints, the FAST telescope design consists of just four photomultiplier tubes (PMTs).
This sits in contrast to the several hundred used by current fluorescence telescopes. 
The reduction in resolution necessitates an alternative method for reconstructing the properties of cosmic rays. 

\vspace{5mm}

The current solution utilises a top-down approach whereby the PMT traces from data are directly compared with those from simulations using a likelihood function. 
The set of shower parameters which give the maximum likelihood, found using an optimiser, are chosen as the reconstructed parameters. This procedure is known as the \say{top-down reconstruction (TDR)}. 
The performance of the TDR is heavily influenced by the initial parameters or \say{first guess} passed to the optimiser. 
At the time of writing, no robust method for producing a first guess has been developed. 
Although previous attempts using machine learning have shown promising results, they have only been applicable to a future, full-sized FAST array. 
For the current and near-future FAST prototype installations, the combined performance of the first-guess estimation and TDR is essentially unknown. 
Understanding the capabilities of these installations and verifying them through the analysis of events observed in coincidence with other experiments is crucial to the success of the FAST project. 

\vspace{5mm}

With these points in mind, the overarching goal of this thesis is to develop the FAST reconstruction procedure such that it can be reliably applied to the current and near-future FAST prototype installations. The structure of the thesis is as follows;

\vspace{5mm}

\noindent\textbf{Chapter 1:}
Introduces cosmic rays and the primary observables used to study them, referencing the latest results in the field. A brief summary of the history of cosmic ray research is also given.

\vspace{5mm}

\noindent\textbf{Chapter 2:}
Details the characteristics of EASs and the various methods by which they are detected/reconstructed. Examples of past, present and future cosmic ray observatories are provided.

\vspace{5mm}

\noindent\textbf{Chapter 3:}
Presents a detailed overview of the FAST experiment including hardware specifications, calibration, optical performance measurements and the software framework used for simulations/reconstruction.

\vspace{5mm}

\noindent\textbf{Chapter 4:}
Fixes the primary issue with the current TDR procedure, namely the decrease in reconstruction efficiency with 
energy. The degeneracy between the fitted depth of shower maximum and energy for certain showers is also addressed.

\vspace{5mm}

\noindent\textbf{Chapter 5:}
Investigates the performance of both previously developed and newly developed machine learning models for obtaining a first guess of the shower parameters with the current and near-future FAST telescope layouts. The combined performance of the machine learning first guess and TDR is analysed.

\vspace{5mm}

\noindent\textbf{Chapter 6:}
Introduces an alternative first-guess estimation method using templates of PMT traces from a single telescope. The method is designed to compensate for the weaknesses of the machine learning approach.

\vspace{5mm}

\noindent\textbf{Chapter 7:}
Applies the improved TDR and first guess methods developed to data from the current FAST prototypes. First measurements of the cosmic ray energy spectrum and composition with FAST are made. 

\vspace{5mm}

\noindent\textbf{Chapter 8:}
Concluding remarks

\chapter{Cosmic Rays}
\label{ch:CR}

Cosmic rays are highly energetic particles, predominantly protons and other atomic nuclei, originating in outer space. 
They are accelerated in extreme astrophysical environments and span energies from approximately $10^{9}\sim10^{20}$\,eV. 
Cosmic rays with energies greater than $10^{18.5}$\,eV are known as ultra-high energy cosmic rays (UHECRs) and arrive at Earth with a flux of $\sim1$\,particle/km$^{2}$/year. 
The origin of UHECRs remains an open question in modern astroparticle physics and underpins much of the current research performed by cosmic-ray, gamma-ray and neutrino observatories.

\vspace{5mm}

To study the origins and acceleration mechanisms of cosmic rays, scientists measure the energy spectrum, mass composition and anisotropy of arrival directions of cosmic rays which arrive at Earth. 
This chapter introduces each of these components, highlighting recent results in the field. 
Before this, a brief overview of the history of cosmic ray research is given.

\section{A Brief History of Cosmic Ray Research}
In 1785 Charles Augustin de Coulomb made perhaps the first recorded observation of the ionising property of cosmic rays. 
He found that a metallic conductor when placed in air would spontaneously discharge \cite{de1785troisieme}. 
This phenomenon would be observed by several other physicists over the next century including Michael Faraday in 1835 \cite{FaradayDiscovery}, Carlo Matteucci in 1850 \cite{wilson1901ionisation} and William Crookes in 1879 \cite{crookes1879electrical}, the latter two of whom both showed that the rate of discharge decreased at lower atmospheric pressures.  
In 1900, German physicists Elster and Geitel quantitatively measured the rate of spontaneous discharge, concluding that the discharge was due to the presence of free ions in the atmosphere \cite{elster1900electricity}. 
The result would be confirmed in 1901 by Wilson \cite{wilson1901ionisation}. 
However the source of these ions remained unknown.

\vspace{5mm}

At the beginning of the 20th century, scientists began to systematically measure the mysterious ionisation in a variety of different environments to uncover its source. 
These included measurements in the lab with metallic shielding,
in tunnels and salt mines, 
above the Atlantic Ocean,
and at the top of the Eiffel tower \cite{carlson2011nationalism, de2014atmospheric, falkenburg2012ultra}.
A particularly notable contribution came from Italian physicist Domenico Pacini, who found a significant decrease in the ionisation rate below water. He concluded that, contrary to consensus at the time, the ionisation was primarily coming from the atmosphere and not radioactive materials in the Earth \cite{pacini1912radiazione}. 

\vspace{5mm}

\begin{figure}
    \centering
    \includegraphics[width=0.9\linewidth]{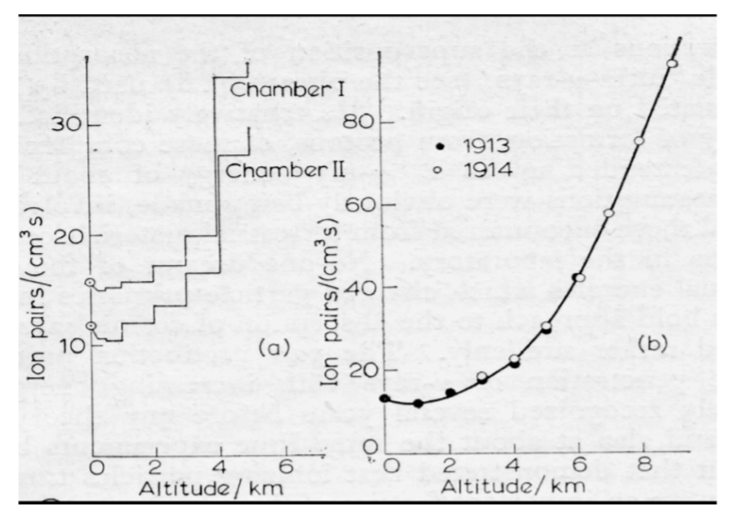}
    \caption{Measurements made by Hess in August 1912 (left) and Kolhörster in 1913/14 (right) of the ionisation rate as a function of altitude. From \cite{percarlson}.}
    \label{fig:HessResults}
\end{figure}

The final breakthrough came from the measurements of Austrian born physicist Victor Hess. Over the course of ten balloon flights between 1911 and 1913, Hess measured the ionisation rate as a function of altitude. Measurements from his flight in August 1912 \cite{hess1912beobachtungen} and their later confirmation by Kolhörster \cite{kolhorster1913measurements} are shown in Figure \ref{fig:HessResults}. The results show an increase in the ionisation rate above $\sim2$\,km. This led Hess to conclude that the ionisation was of extraterrestrial origin, winning him the 1936 Nobel Prize in Physics.

\vspace{5mm}

Studies by Compton and others over the ensuing decades revealed that the newly discovered \say{cosmic rays} were primarily charged particles, and in fact mostly protons \cite{carlson2011nationalism}. In 1939,
Pierre Auger and colleagues would measure near simultaneous pulses in separated detectors, concluding that the signals were attributable to a common high energy cosmic ray of energy $\sim10^{15}$\,eV initiating cascades or \say{showers} of secondary particles in the upper atmosphere \cite{auger1939extensive}. These showers are now known as extensive air showers (EASs) and are the primary method of studying UHECRs today. The details of EASs and their observation are discussed in Chapter \ref{ch:EAS}.

\section{Cosmic Ray Energy Spectrum}
\begin{figure}[t]
    \centering
    \includegraphics[width=0.9\textwidth]{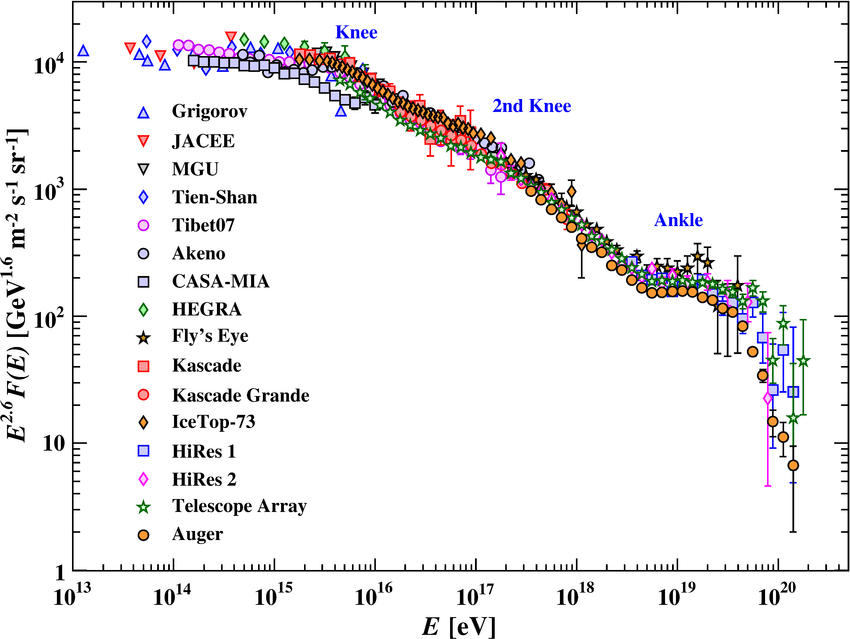}
    \caption{The all particle energy spectrum of cosmic rays. Measurements spanning seven orders of magnitude from several different experiments are shown (see legend). From \cite{tanabashi2018review}.}
    \label{fig:energyspectrum}
\end{figure}
\begin{figure}[t]
    \centering
    \includegraphics[width=0.7\linewidth]{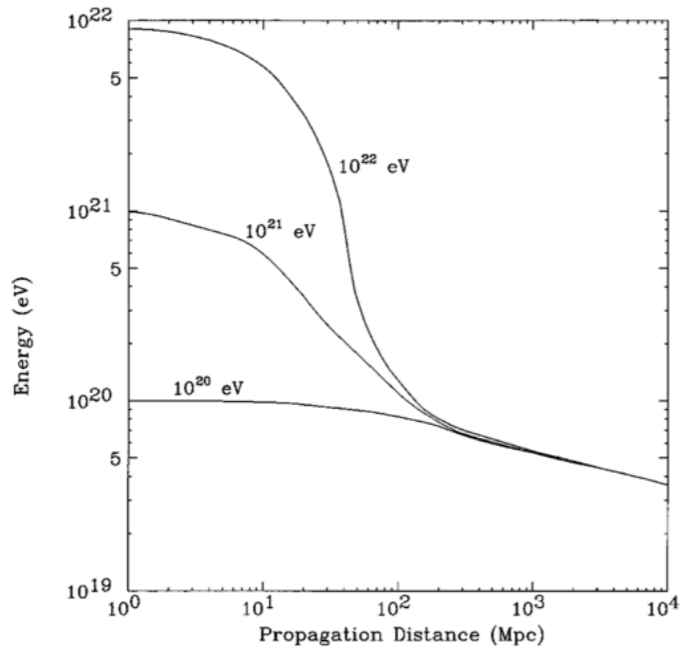}
    \caption{Energy loss of cosmic ray protons during propagation as a result of interactions with CMB photons. From \cite{cronin1999cosmic}.}
    \label{fig:GZK}
\end{figure}

The cosmic ray energy spectrum describes the flux of cosmic rays arriving at Earth as a function of energy. The spectrum broadly follows a steeply falling power law, which can be expressed as
\begin{equation}
    \frac{\mathrm{d}N}{\mathrm{d}E}\propto{}E^{-\gamma}
\end{equation}
where $N$ is the number of particles, $E$ is the particle energy and $\gamma$ is the spectral index.  
A value of $\gamma=3$ models the spectrum over the entire energy range (GeV to 100's of EeV) reasonably well. 
Slight variations in the index across the spectrum are thought to indicate transitions between different source populations and/or acceleration mechanisms for the different energy cosmic rays. 
To easily visualise the changes in $\gamma$, the flux of cosmic rays is typically multiplied by some power of $E$ when plotted. 
This is demonstrated in Figure \ref{fig:energyspectrum} which shows the all particle energy spectrum, i.e. the energy spectrum of all cosmic rays irrespective of their masses. 
Here, the flux has been multiplied by $E^{2.6}$, highlighting various key features of the spectrum. 
These features are discussed below.

\vspace{5mm}

At around $3\times10^{15}$\,eV the spectral index $\gamma$ changes from $\sim2.6$ to $\sim3$. 
This point in the spectrum is known as the \say{knee} and has been shown by the KASCADE experiment to be due to a \say{cutoff} (sharp reduction in flux) in the spectra of protons and lighter elements \cite{kampert2004cosmic}. 
The cutoff may be due to lighter elements reaching their maximum attainable energy via acceleration from \glspl{snrem} or galactic magnetic fields no longer being able to confine these nuclei to the galactic disk \cite{blumer2009cosmic}.  

\vspace{5mm}

The spectrum steepens further at $\sim10^{17}$\,eV to $\gamma\approx-3.3$. 
This feature is referred to as the \say{second knee}. 
One interpretation of the second knee is that iron primaries are reaching their cutoff energies from SNR acceleration at $E\approx26E_k$ where $E_k$ is the energy of the first knee \cite{blumer2009cosmic}. 
A contribution from low energy extragalactic protons may also exist \cite{abu2018knee}.

\vspace{5mm}

At approximately $10^{18.5}$\,eV the spectral index transitions back to $\gamma\approx2.6$, marking the location of the so-called \say{ankle}. 
The ankle is traditionally considered to be the region in which the flux of extra-galactic cosmic rays begin to dominate the spectrum. 
One piece of evidence for this comes from anisotropy measurements. 
If cosmic rays around/above the ankle were coming from the galaxy then they would trace an almost straight line back to their sources due to their large gyro radius (see Section \ref{sec:candidatesources}). 
This would give rise to considerable small-scale anisotropies in cosmic ray arrival directions.
As no such anisotropy is observed, the majority of these particles are thought to be of extragalactic origin \cite{pierre2017observation, abbasi2017search}.

\vspace{5mm}

Finally, above $\sim5\times10^{19}$\,eV, there is a sharp cutoff in the spectrum where $\gamma$ increases to $\sim5$.
This cutoff is generally thought to be a result of the highest energy cosmic rays losing energy through interactions with \gls{cmb} photons. 
Specifically, these cosmic rays undergo
\begin{equation*}
    p+\gamma_{\textrm{CMB}}\rightarrow{}n+\pi^+,
\end{equation*}
\begin{equation*}
    p+\gamma_{\textrm{CMB}}\rightarrow{}p+\pi^0.
\end{equation*}
The higher in energy the cosmic ray, the faster it loses energy. 
This is shown in Figure \ref{fig:GZK}. From the figure, it can be seen that for a cosmic ray above $10^{20}$\,eV to be observed at Earth, the source must be within $\sim100$\,Mpc. 
This theory was first put forward independently by Greisen, and by Zatsepin and Kuzmin, in 1966 and is known as the GZK effect
\cite{zatsepin1966upper, greisen1966end}. 
Another possible reason for the cutoff is simply that cosmic ray sources are are unable to accelerate particles to higher energies.

\subsection{Comparisons at the Highest Energies}
\begin{figure}[t]
    \centering
    \includegraphics[width=0.8\linewidth]{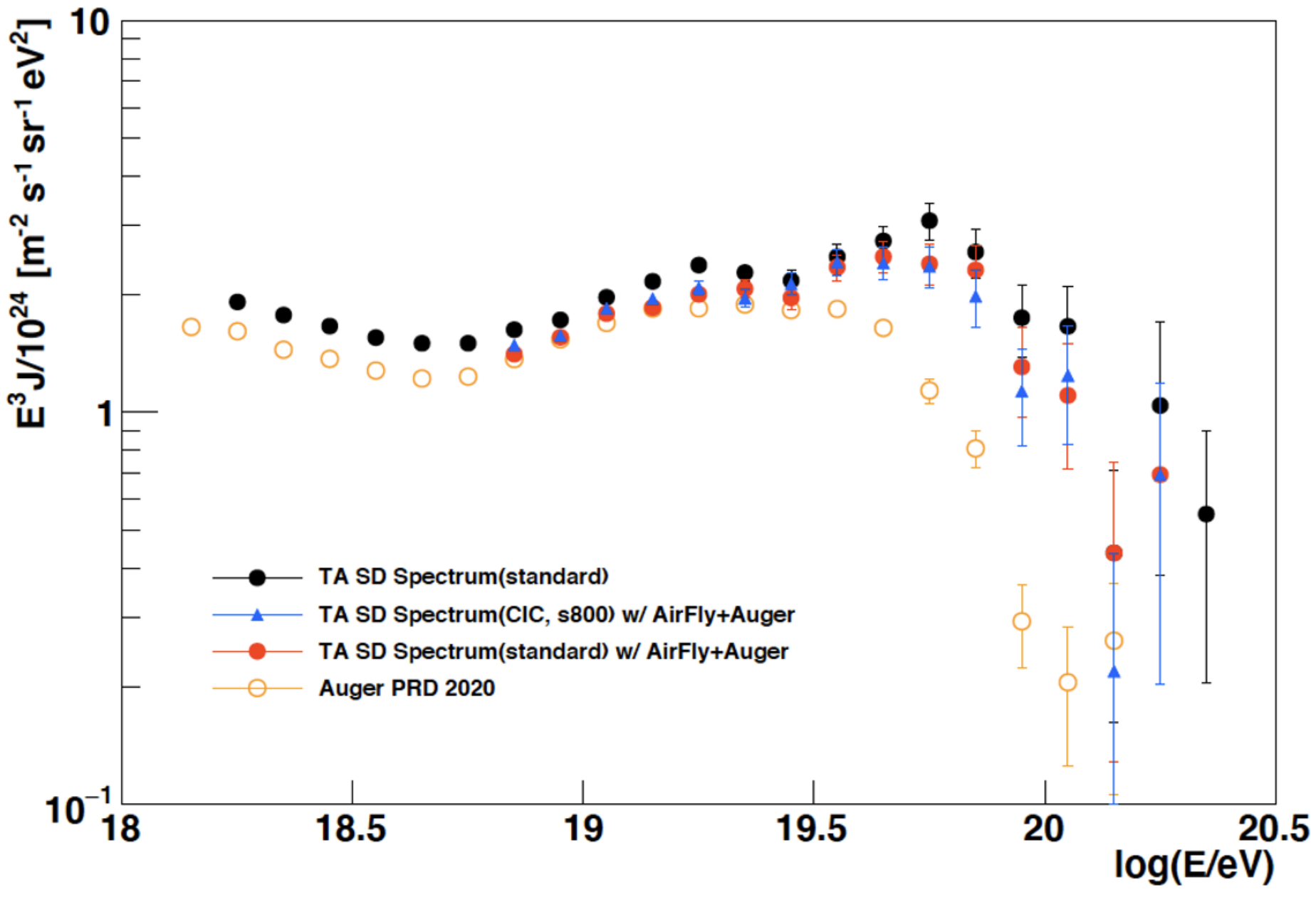}
    \caption{Measurements of the cosmic-ray energy spectrum above 10$^{18}$\,eV by Auger and TA as reported at the \gls{icrc} 2023. When identical fluorescence yield models and event selection criteria are used, the systematic offset below $10^{19.5}$ is removed. From \cite{kim2023highlights}.}
    \label{fig:higherenergyES}
\end{figure}

Figure \ref{fig:higherenergyES} presents another view of the cosmic-ray energy spectrum, this time focusing on energies above $10^{18}$\,eV. 
The results shown here come from the two largest cosmic ray observatories currently in operation, \gls{auger} in Mendoza, Argentina, and \gls{ta} in Utah, United States \cite{pierre2015pierre, abu2012surface, tokuno2012new}.
These observatories focus on measuring the highest energy cosmic rays and predominantly observe the southern/northern hemispheres respectively. 
Working groups consisting of members from both collaborations have been set up to share and closely compare measurements of the energy spectrum, mass composition and anisotropy. 
In the figure, the cosmic ray energy spectrum as measured by TA is shown with black points and the spectrum measured by Auger with yellow points. 
The orange points show the TA spectrum recalculated using the same fluorescence yield model (model describing the amount of fluorescence light produced by excited nitrogen molecules in air) as Auger. 
The blue points further apply the same shower selection process as used by Auger to the TA analysis.   
The results show excellent agreement up to $10^{19.5}$\,eV. 
Beyond this energy TA observes a higher flux than Auger. 
It remains to be seen whether this is a result of a source population only visible to TA or detector/analysis effects still yet unaccounted for \cite{kim2023highlights}.

\section{Acceleration Mechanisms}

Whilst several different possible mechanisms of UHECR acceleration are discussed in the literature (e.g. unipolar inductors, magnetic reconnection, re-acceleration in sheared jets etc.) \cite{kotera2011astrophysics, aloisio2017acceleration}, only the traditional Fermi acceleration and diffusive shock acceleration theories are discussed in detail here. 
In 1949 Enrico Fermi proposed that cosmic rays are accelerated through elastic collisions with magnetised interstellar gas clouds \cite{fermi1949origin}. 
This theory is often referred to as 2$^{\textrm{nd}}$ order Fermi acceleration, since the average fractional change in cosmic-ray energy predicted over many collisions is proportional to $\beta^2$ where $\beta=V/c$ is the speed of each gas cloud as a fraction of the speed of light. 
Fermi's ideas were expanded upon in the 1970's by considering the same acceleration mechanism taking place in the presence of supernova shocks \cite{blandford1978particle}. 
This theory is known as diffusive shock acceleration and it predicts an average fractional energy change proportional to $\beta$. 
For this reason the theory is also referred to as 1$^{\textrm{st}}$ order Fermi acceleration. Details of each mechanism are given below.

\subsection{Fermi Acceleration}
\begin{figure}[t]
    \centering
    \includegraphics[width=0.7\linewidth]{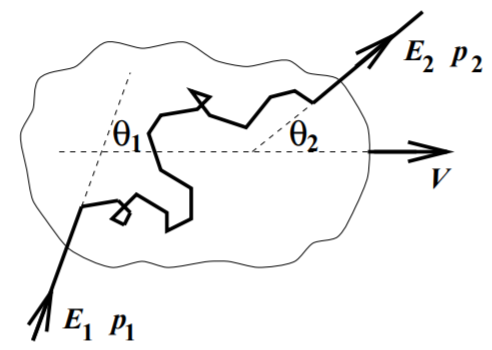}
    \caption{Schematic of Fermi's original idea of cosmic ray acceleration - cosmic rays scattering off of interstellar gas clouds. From \cite{protheroe1998acceleration}.}
    \label{fig:FermiAccel}
\end{figure}

In Fermi's original theory cosmic rays are accelerated through consecutive scatterings with gas clouds in the interstellar medium. 
Each cloud has a random velocity on the order of 15\,km/s and possesses a magnetic field due to the cloud being partially ionised. 
A single collision is envisioned as a cosmic ray with energy $E_1$ and momentum $p_1$ incident on a cloud travelling with velocity $V$. 
The angle of the incident cosmic ray with respect to the cloud velocity is labelled $\theta_1$. 
The cosmic ray is then randomly scattered by the cloud's magnetic field, exiting the cloud at an angle $\theta_2$ with energy and momentum $E_2$ and $p_2$ respectively. 
Figure \ref{fig:FermiAccel} shows a schematic diagram of the process. 
The energy of the incoming cosmic ray in the cloud frame (primed) is
\begin{equation}
    E_1'=\gamma_\textrm{cloud}E_1(1-\beta\cos{\theta_1}).
    \label{eqn:E1}
\end{equation}
After scattering, the energy of the cosmic ray in the lab frame is given by the reverse Lorentz transformation
\begin{equation}
    E_2=\gamma_\textrm{cloud}E_2'(1+\beta\cos{\theta_2'}).
    \label{eqn:E2}
\end{equation}
As the scattering is elastic $E_2'=E_1'$. Substituting Equation \ref{eqn:E1} into the right-hand side of Equation \ref{eqn:E2} and rearranging gives
\begin{equation}
    \frac{\Delta E}{E_1}=\frac{1-\beta\cos\theta_1+\beta\cos\theta_2'-\beta^2\cos\theta_1\cos\theta_2'}{1-\beta^2}-1.
    \label{eqn:fracEnergy}
\end{equation}
Equation \ref{eqn:fracEnergy} is the fractional energy change for a single collision. 
The average change can be obtained by calculating the average values of $\cos\theta_1$ and $\cos\theta_2'$. 
Since the scattering inside the cloud is random,  
\begin{equation}
    \expval{\cos\theta_2'}=0.
\end{equation}
The average value of $\cos\theta_1$ depends on the relative velocities between the cosmic rays and clouds. 
It can be shown that
\begin{equation}
    \expval{\cos\theta_1}=-\beta/3.
\end{equation} 
Thus the average fractional change in energy is given by
\begin{equation}
     \expval{\frac{\Delta E}{E_1}} = \frac{1+\frac{1}{3}\beta^2}{1-\beta^2}-1 \approx \frac{4}{3}\beta^2
     \label{eqn:fracEnergyApprox}
\end{equation}
since $\beta\ll1$. 
Physically speaking, the process exhibits only a slight positive average energy gain because the velocity of the clouds compared to the cosmic rays is very small. 
This results in head-on collisions (energy gain) being only slightly more frequent than tail-on collisions (energy loss) \cite{protheroe1998acceleration}.

\subsection{Diffusive Shock Acceleration}
\begin{figure}[t]
    \centering
    \includegraphics[width=0.8\linewidth]{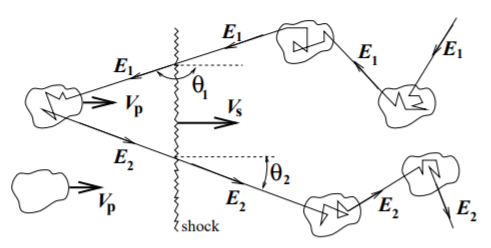}
    \caption{Schematic diagram of diffusive shock acceleration, where cosmic rays travel back and forth across a shock front, each time being accelerated by head on collisions with magnetised gas clouds. From \cite{protheroe1998acceleration}.}
    \label{fig:DSA}
\end{figure}

In diffusive shock acceleration magnetised clouds are present either side of an astrophysical shock (e.g. supernova shock). 
Cosmic rays are then accelerated back and forth across the shock front by consecutive head-on collisions with the clouds as shown in Figure \ref{fig:DSA}. 
As there are only head on collisions this process is much more efficient for gaining energy. 
In diffusive shock acceleration the fundamental mechanism of scattering off gas clouds remains the same. 
Thus Equation \ref{eqn:fracEnergy} can be applied. 
The only changes are the average values of $\cos\theta_1$ and $\cos\theta_2'$. 
These are now
\begin{equation}
    \expval{\cos\theta_1} = -2/3 \quad \textrm{and} \quad \expval{\cos\theta_2'} = 2/3.
\end{equation}
Hence for diffusive shock acceleration
\begin{equation}
     \expval{\frac{\Delta E}{E_1}} = \frac{1+\frac{4}{3}\beta+\frac{4}{9}\beta^2}{1-\beta^2}-1 \approx \frac{4}{3}\beta
\end{equation}
where the first order dependence on $\beta$ signifies the greater efficiency of the mechanism.

\section{Mass Composition}
\label{sec:MassComposition}
\begin{figure}[t]
    \centering
    \includegraphics[width=0.8\textwidth]{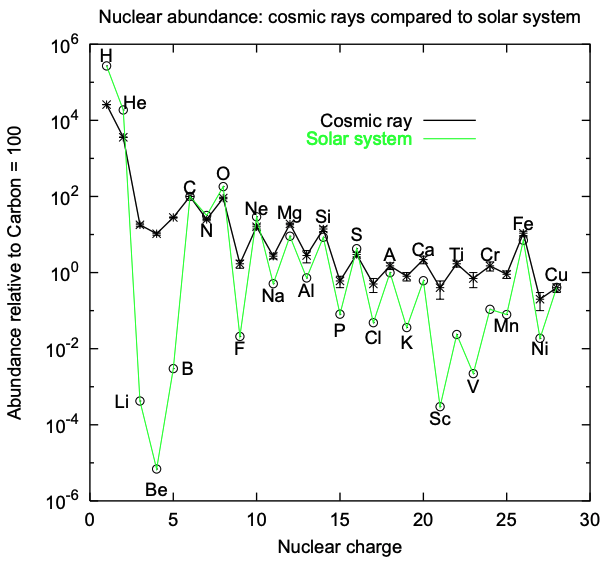}
    \caption{Relative abundances of ordinary nuclei and cosmic rays in the solar system. From \cite{gaisser2006high}.}
    \label{fig:crsoloarsystem}
\end{figure}

The term \say{mass composition} refers to what nuclei make up observed cosmic rays. 
Since cosmic rays are primarily charged particles and are thus influenced by magnetic fields during their acceleration/propagation, knowledge of their mass composition can assist in deducing information about potential sources. 
At low energies, the flux of cosmic rays is sufficiently high for satellite based experiments, such as the Alpha Magnetic Spectrometer on the International Space Station \cite{aguilar2021alpha} or the Calorimetric Electron Telescope \cite{torii2007calet}, to measure the composition of cosmic rays directly. 
Figure \ref{fig:crsoloarsystem} shows the relative abundances of cosmic rays (from satellite data) and regular nuclei in the solar system. The excess in cosmic rays of elements Li, Be, B and Sc, Ti, V, Cr, Mn arises from spallation of heavier cosmic rays with the interstellar medium \cite{gaisser2006high, gaisser2016cosmic}.

\vspace{5mm}

\begin{figure}[t]
    \centering
    \includegraphics[width=1\linewidth]{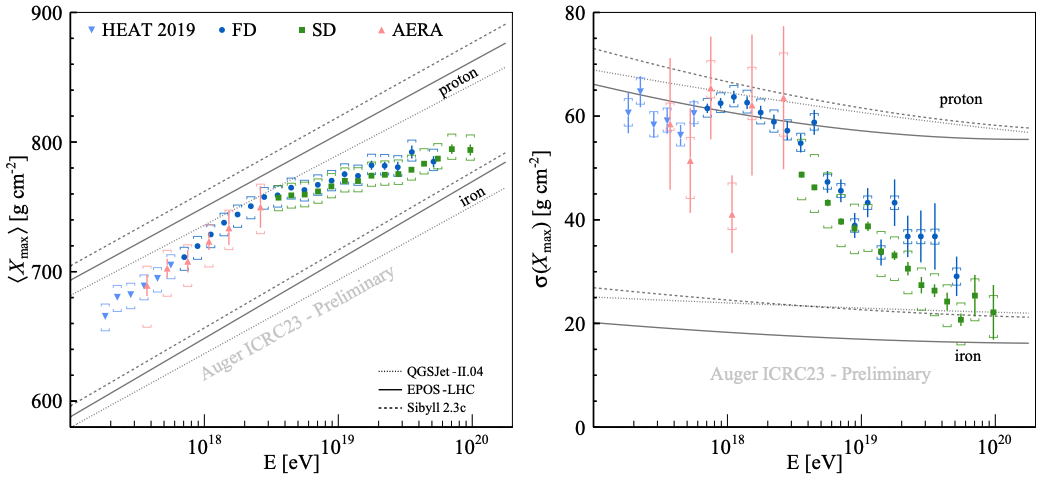}
    \caption{\Xmax{} moments (\textit{left:} mean, \textit{right:} standard deviation) measured by different detectors at Auger as presented at ICRC 2023. From \cite{mayotte2023measurement}.}
    \label{fig:AugerXmaxICRC2023}
\end{figure}

At higher energies the flux of cosmic rays is too low for direct detection to be viable. 
Instead, observables from the aforementioned EASs are analysed to estimate the properties of the original cosmic rays. 
The most common observable used to estimate mass composition is the distribution of a quantity known as \Xmax{} (see Section \ref{sec:EASintro}). Figure \ref{fig:AugerXmaxICRC2023} shows the means and standard deviations of the \Xmax{} distributions measured by Auger as a function of energy. 
The different sets of points represent different detection methods. 
These include fluorescence detection (HEAT 2019/FD), radio detection (AERA), and detection using ground based surface detectors (SD). 
The SD result is particularly noteworthy as it extends the average \Xmax{} measurements to the highest energies by applying machine learning techniques \cite{aab2021deep}. 
Discussion on the various detection methods can be found in Section \ref{sec:detectionMethods}. The measured data is plotted alongside expected values from simulations using different hadronic models with either a purely proton or purely iron composition. This allows one to quickly infer whether the measured composition is lighter (towards proton), heavier (towards iron) or somewhere in-between at any particular energy.
The analysis can be extended by assuming distributions of \Xmax{} for different primary groups and then fitting the fractions each group contributes to the data in each energy bin. 
The primary groups most commonly used are H, He, CNO and Fe. 

\vspace{5mm}

\begin{figure}[]
    \centering
    \includegraphics[width=1\linewidth]{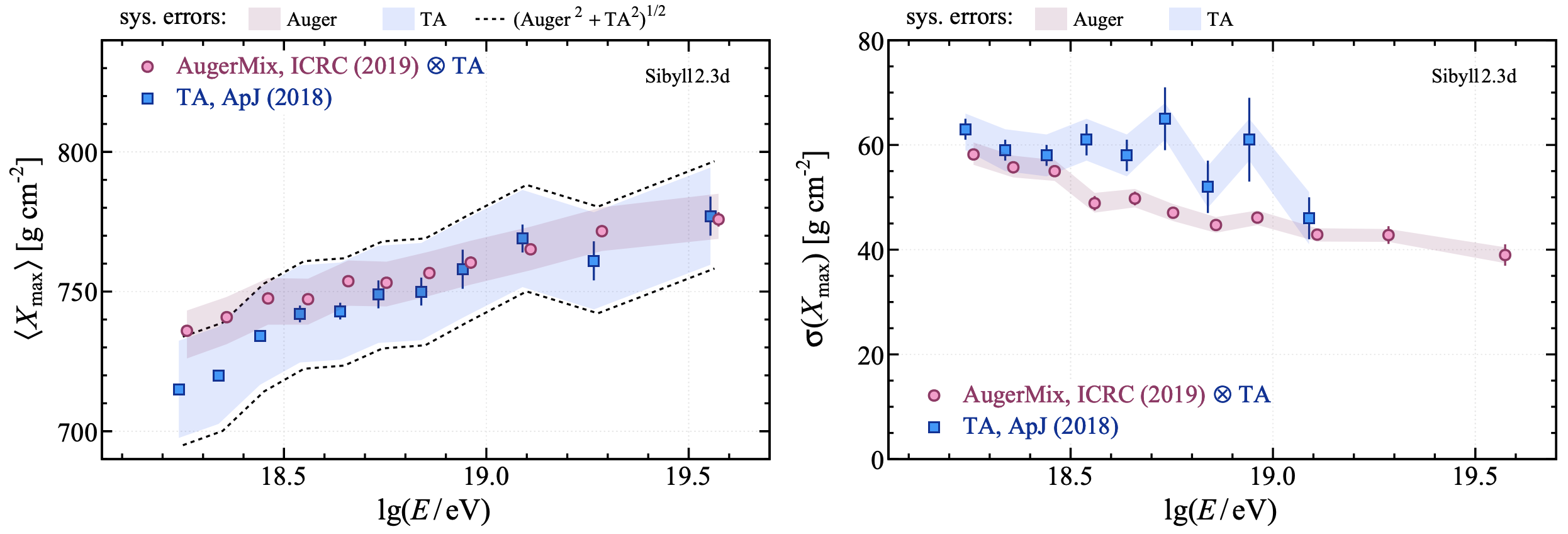}
    \caption{Mean (left) and standard deviation (right) measurements of the \Xmax{} distributions reported by TA (blue) compared to Auger data transferred into the TA detector, i.e. ``Auger Mix'' (red). Results reported at ICRC 2023. From \cite{yushkov2023depth}.}
    \label{fig:TAXmax}
\end{figure}

The \Xmax{} moment measurements from TA are shown in comparison to the Auger results in Figure \ref{fig:TAXmax}. 
The two are mostly consistent with some discrepancy appearing in the width of the \Xmax{} distributions between $10^{18.5}$ and $10^{19}$\,eV. 
This comparison is performed by taking the mass composition (fractions for each primary group) estimated by Auger in each energy bin and sending them through the TA detector response \cite{yushkov2023depth}. 
Together, these results show that the average composition of cosmic rays appears lighter at around $3\times10^{18}$\,eV and tends towards a heavier composition at the highest energies.

\section{Candidate Sources}
\label{sec:candidatesources}
Although firm conclusions as to what astrophysical objects accelerate UHECRs remain elusive, there are several constraints one can place on potential sources based on observations/modelling assumptions. 
For example, if one assumes that diffusive shock acceleration is the mechanism by which UHECRs are accelerated then astrophysical shocks must be present at the acceleration site. 
A more general condition is that a candidate source must be large enough to confine a UHECR during acceleration. 
Known as the \say{Hillas criterion}, this condition requires that the characteristic size of a source $R$ must be larger than the gyro-radius $r_g$ of the particle it is accelerating. 
Mathematically,
\begin{equation}
    R>r_g=\frac{p}{ZB}\approx{}\frac{E}{cZB}
\end{equation}
where $p$ and $Z$ are the particles momentum and charge respectively, and $B$ is the magnetic field strength. In the last approximation $p\approx{}E/c$ is used since the particles are relativistic.
\begin{figure}[t]
    \centering
    \includegraphics[width=0.9\textwidth]{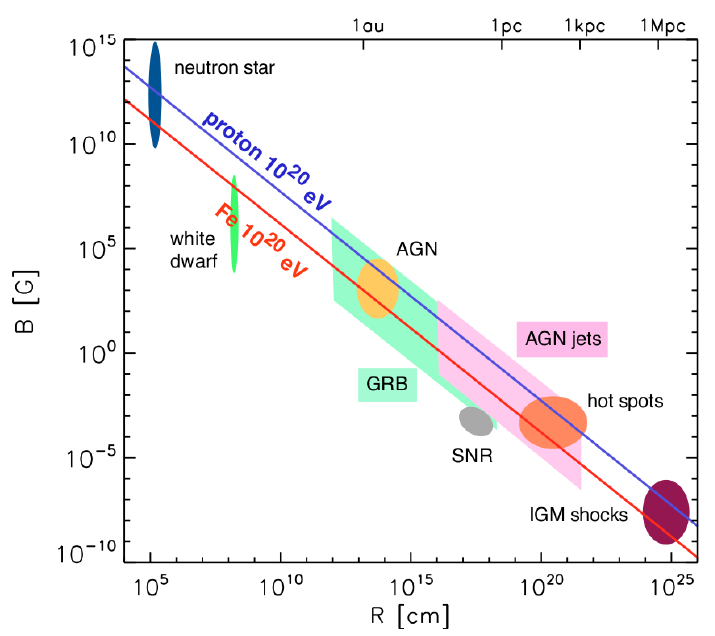}
    \caption{An updated version of the traditional Hillas plot, showing the size and magnetic field strength of various astrophysical objects. The red (blue) lines show the requirements for accelerating a proton (iron) nucleus to $10^{20}$\,eV. From \cite{kotera2011astrophysics}.}
    \label{fig:hillasplot}
\end{figure}
The above equation can be rearranged to give the maximum attainable energy for a cosmic ray accelerated at a source with given $R$ and $B$. 
This gives rise to the standard \say{Hillas plot}, which shows the $R$ and $B$ values for different possible sources, together with the values required to accelerate a proton/iron cosmic ray to a particular energy. 
An example is shown in Figure \ref{fig:hillasplot}. Some sources which satisfy the Hillas criterion for a 10$^{20}$\,eV proton include \cite{kotera2011astrophysics, aloisio2017acceleration}
\begin{itemize}
    \item \textbf{\Gls{agn} and their jets:} AGN are compact regions in the centres of galaxies which emit powerful jets and winds. They are powered by supermassive black holes accreting matter from a surrounding accretion disk. Depending on the specific properties of the AGN, cosmic rays may be accelerated in/around the jets through shear or shock acceleration processes, or in the magnetosphere of the black hole via electrostatic acceleration.
    \item \textbf{Neutron stars:} Neutron stars are the incredibly dense cores of collapsed stars which have undergone a supernova explosion. The strong magnetic fields ($\sim10^{15}$\,G) around fast-spinning, magnetised neutron stars, also known as magnetars, could generate the required electric fields to accelerate UHECRs. 
    \item \textbf{\Glspl{grb}:} GRBs are extremely bright, short-lived flashes of gamma-rays arising from either supernova explosions or the merger of compact objects (neutron stars/black holes). The shocks generated in these explosions are possible acceleration sites for UHECRs.
\end{itemize} 
Further discussion on different candidate sources can be found in \cite{torres2004astrophysical}.

\vspace{5mm}

Other conditions such as the number density of sources, total energy injection, universal maximum rigidity and extremely high energy events such as the Amaterasu particle \cite{telescope2023extremely} must also be considered when proposing potential UHECR sources. 
A recent theory put forward by Farrar \cite{farrar2024binary} states that binary neutron star mergers have the potential to meet each of these criterion, and that UHECRs are produced as the end result of an r-process. 
At present, the theory appears to reasonably satisfy all observational constraints. 
Additional multi-messenger observations and measurements at the highest energies will be necessary to further test the theory.

\section{Anisotropy}

\begin{figure}[t]
    \centering
    \includegraphics[width=0.9\linewidth]{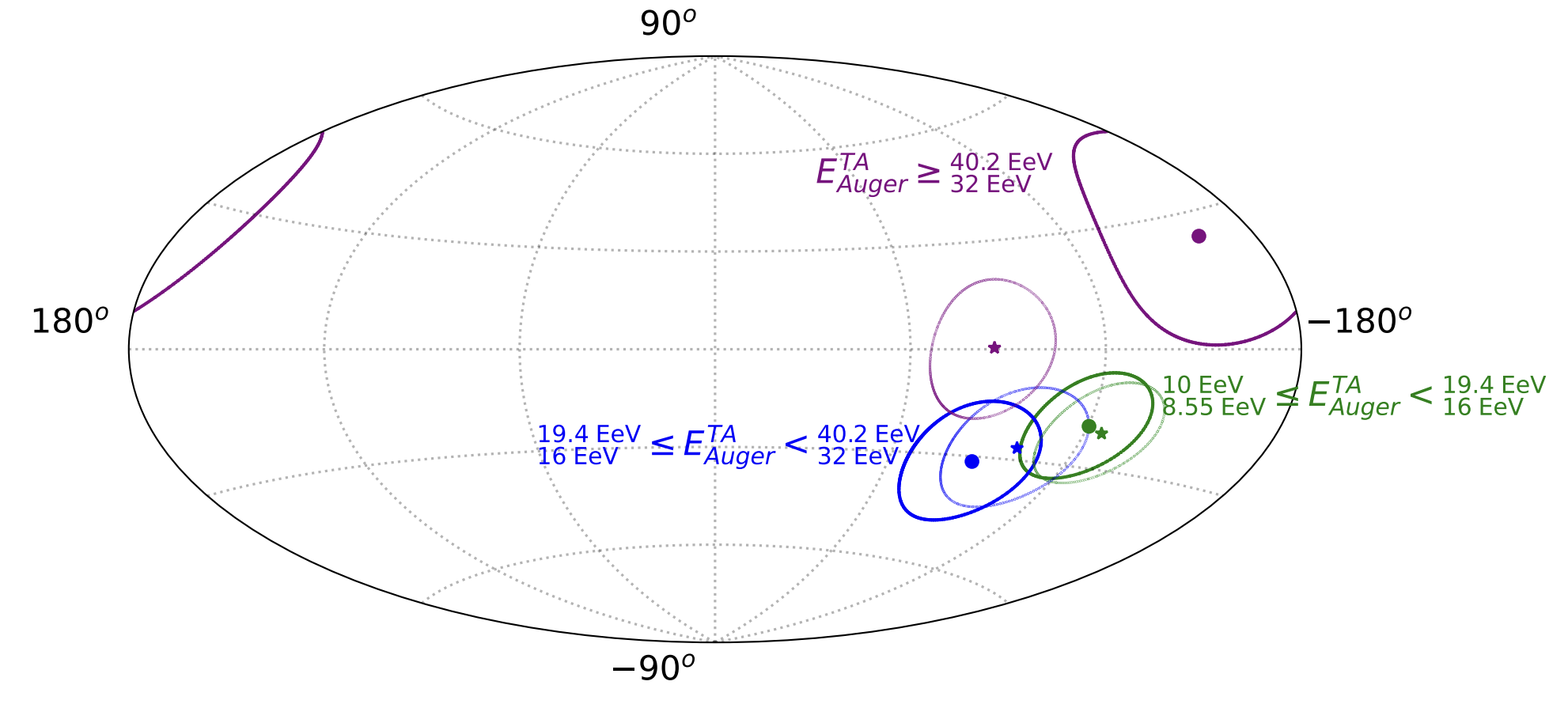}
    \caption{Evolution of the direction of the dipole with energy using the combined Auger and TA datasets. Results shown in galactic coordinates. The directions  for the lowest two energy bins are consistent with the Auger only result (thin circles). Reported in ICRC 2023. From \cite{caccianiga2023update}.}
    \label{fig:AugerTAWorkingGroupDipoleICRC2023.png}
\end{figure}

\begin{figure}[]
    \centering
    \includegraphics[width=0.9\linewidth]{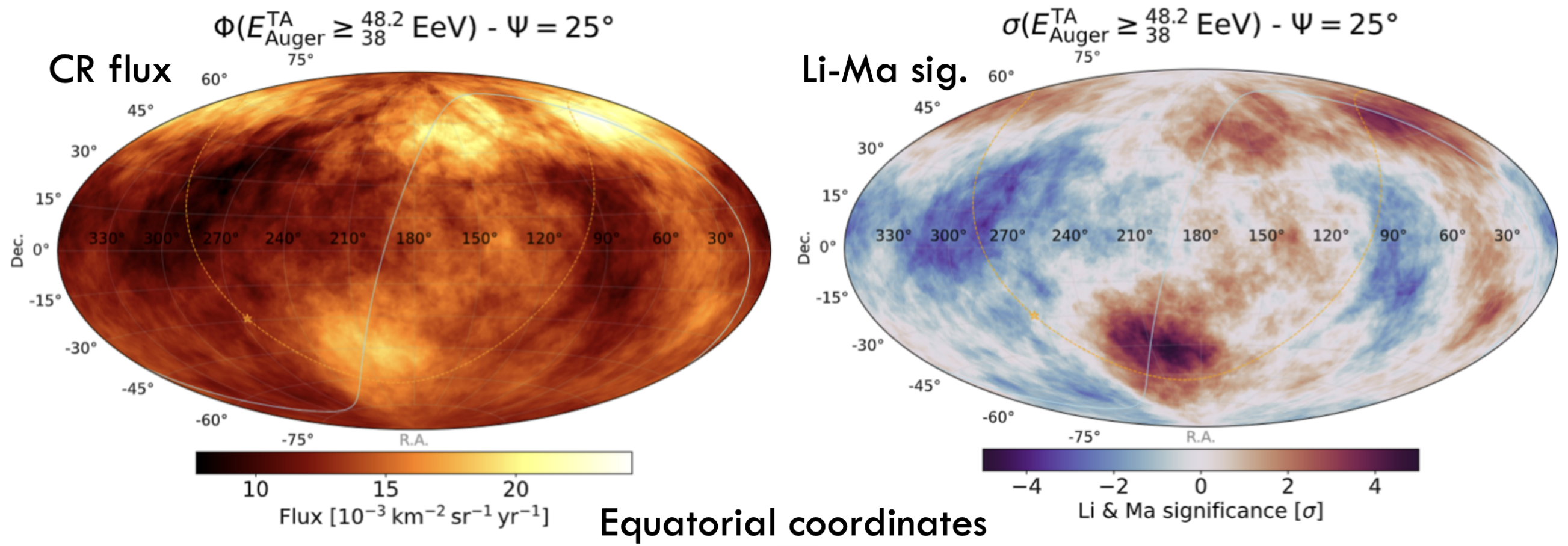}
    \caption{ICRC 2023 results of the flux map (left) and Li-Ma significance map (right) of cosmic rays measured by Auger and TA with top-hat smoothing of radius $\psi=25\degree$. From \cite{kim2023highlights}.}
    \label{fig:hotspots}
\end{figure}

Deviations from isotropy in the arrival directions of cosmic rays is the most direct method of determining their origins. This is made difficult however by intervening galactic and extragalactic magnetic fields which deflect cosmic rays from their original trajectory by an amount proportional to their rigidity $E/Z$. Thus only by focusing on the highest energy cosmic rays can \say{cosmic-ray astronomy}, where detected events are traced back along their arrival direction to a particular source, be performed.

\vspace{5mm}

Anisotropy studies at the highest energies are typically performed at large and intermediate angular scales. At large scales the main structure of interest is a dipole in cosmic ray arrival directions.
Using both Auger and TA datasets to cover the whole sky, the Auger-TA anisotropy working group has measured this dipole in different energy bins as shown in Figure \ref{fig:AugerTAWorkingGroupDipoleICRC2023.png}. The position and amplitude of the dipole vary slightly with energy but in all cases the direction points away from the galactic centre (exception being the highest energy bin where the dipole is not statistically significant). For events with energies $E^{\textrm{TA}}_{\textrm{Auger}}>^{10}_{8.55}$\,EeV, the dipole has an amplitude of $6.51\%\pm0.93\%\pm0.65\%$ (first uncertainty is statistical, second is due to energy calibration) and direction in equatorial coordinates of $\left(\alpha,\delta\right)=\left(97.1\degree\pm9.4\degree\pm0.1\degree, -35.7\degree\pm8.7\degree\pm7.8\degree\right)$ \cite{caccianiga2023update}.

\vspace{5mm}

As for intermediate scale anisotropies, Auger reports an excess in the direction of the Centaurus A region with significance 4.0$\sigma$ \cite{golup2024update}.
TA reports two excesses, one in the direction of the Ursa Major constellation above $E=57$\,EeV and one in the direction of the Perseus-Pisces supercluster above $E=10^{19.4}$\,eV. Both excesses have estimated significances of $3.2\sigma$ \cite{kim2023updates}. These excesses can be seen in Figure \ref{fig:hotspots} which shows the flux map and Li-Ma significance map for cosmic rays observed by Auger and TA with energies $E^{\textrm{TA}}_{\textrm{Auger}}>^
{48.2}_{38}$\,EeV. Additionally, by comparing the combined data set with the expected distributions from a catalogue of starburst galaxies, a departure from isotropy with a significance of 4.6$\sigma$ has been measured \cite{caccianiga2023update}. Assuming these hot-spots truly are regions of increased cosmic ray production, further measurements from both Auger and TA, combined with those from new experiments with increased exposures, will be necessary to bring the statistical significance above the desired 5$\sigma$ level. 

\chapter{Extensive Air Showers and Detection Methods}
\label{ch:EAS}

At energies larger than approximately 10$^{15}$\,eV, the flux of cosmic rays arriving at earth reduces to $\sim1$\,particle/m$^2$/yr. This makes direct detection methods via satellites or balloon borne experiments, which typically utilise detecting areas of only a few square metres, impractical for measuring a large number of events. Instead, cosmic rays at these energies are observed indirectly by measuring so-called \say{extensive air showers} (EASs). The properties of the original or \say{primary} cosmic ray which induce an EAS, such as the energy and arrival direction, are inferred from these measurements. This chapter gives an overview of extensive air showers and the typical methods by which they are detected/measured.

\section{Extensive Air Showers}
\label{sec:EASintro}
A cosmic ray which enters Earth's atmosphere will interact with an atmospheric nucleus. This interaction sparks a cascade of particle production which may spawn upwards of 10$^{11}$ particles depending on the original cosmic ray's energy \cite{heck1998corsika}. Since these particles originate from the interaction of a \say{primary} cosmic ray, they are often referred to as \say{secondary particles}. It is this cascade of secondary particles that is known as an extensive air shower. 

\vspace{5mm}

\begin{figure}[t]
    \centering
    \includegraphics[width=1\linewidth]{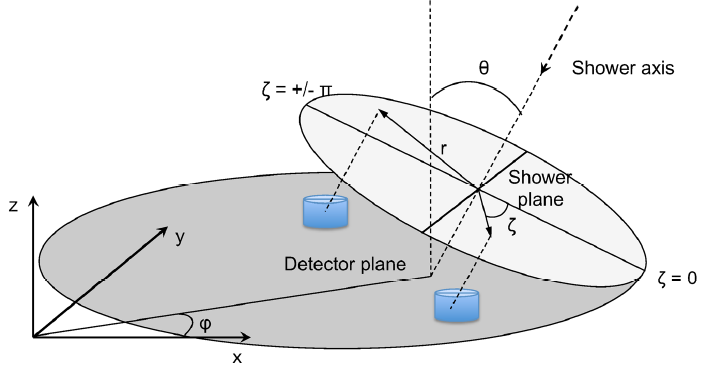}
    \caption{Diagram of the geometry of an EAS. From \cite{aab2016azimuthal}.}
    \label{fig:showerGeom}
\end{figure}

The particles in an EAS are primarily concentrated around the \say{shower axis}, defined as the path the primary cosmic ray would have travelled if it had not interacted. The point where the shower axis intersects the ground is known as the \say{shower core} and the angle of the shower axis with respect to the vertical is the shower zenith angle, commonly labelled $\theta$. The azimuthal direction of the shower is labelled $\phi$ and is typically measured with respect to one of the four cardinal directions. The plane perpendicular to the shower axis is the \say{shower plane}. The secondary particle density at ground level is typically measured as a function of the distance $r$ between the shower axis and (projected) detector positions in the shower plane. Angles in the shower plane are labelled $\zeta$ with $\zeta=0\degree$ defined to be beneath the shower axis. Figure \ref{fig:showerGeom} shows the basic geometry of an EAS.

\vspace{5mm}

The longitudinal development (or simply development) of a shower refers to the number of particles generated as the shower moves through the atmosphere. The number of particles generated is dependent on the density of the atmosphere, hence the development is typically measured as a function of the amount of matter traversed by the shower. This quantity is referred to as the atmospheric slant depth (or just \say{slant depth}) and is given by the column density 
\begin{equation}
\label{eqn:slantDepth}
    X(d) = \int_0^d \rho(l)\,\mathrm{d}l 
\end{equation}
where $d$ is the distance along the shower trajectory as measured from the top of the atmosphere and $\rho$ is the density of air. The slant depth is most commonly measured in \gcm{} units. The slant depth at which the shower reaches its maximum size, measured as either the maximum number of secondary particles or the maximum energy deposited by the secondary particles per unit slant depth, is known as \Xmax{} and is an important parameter for mass composition studies (see Section \ref{sec:MassComposition}). The process of estimating EAS parameters from data is called \say{reconstruction} and varies between experiments. The core location (shower core), arrival direction ($\theta$ and $\phi$), \Xmax{} (or some other mass sensitive variable) and primary cosmic ray energy are the six parameters typically reconstructed, allowing for measurements of the cosmic ray energy spectrum, mass composition and anisotropy. 

\vspace{5mm}

\begin{figure}[t]
    \centering
    \includegraphics[width=1\linewidth]{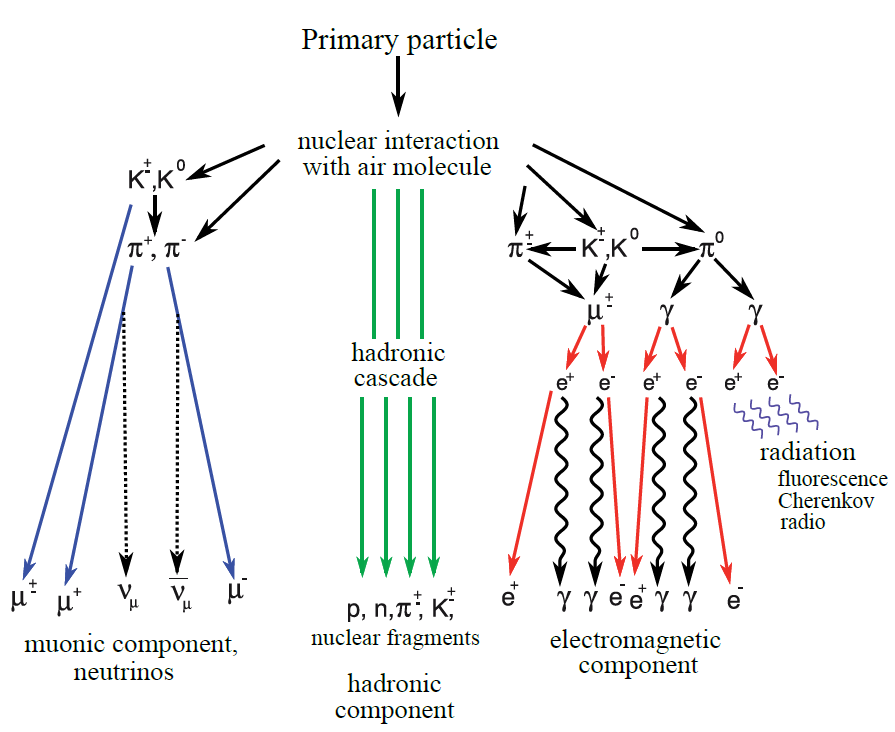}
    \caption{Schematic of the different components of an EAS and how they arise from the initial interaction between a primary cosmic ray and air molecule. From \cite{haungs2015kcdc}.}
    \label{fig:EASModel}
\end{figure}

The particles produced in an extensive air shower can be separated into three main components as illustrated in Figure \ref{fig:EASModel}. The hadronic component consists primarily of charged/neutral pions and kaons. The decay of charged pions creates muons and neutrinos, forming the muonic component, whilst the decay of neutral pions into photons initiates the \gls{em} cascades which make up the electromagnetic component. The basic theory behind EM and hadronic showers is discussed below.

\subsection{Electromagnetic Cascades}
The theory of EM showers presented here was first developed by Heitler \cite{heitler1984quantum}. Although the model is simple compared to more recent efforts (e.g. \cite{selivanov2024final}), it correctly predicts the fundamental features of EM shower development.

\vspace{5mm}

In Heitler's model, electrons, positrons and photons repeatedly \say{split} into two new particles after travelling a depth $d=\lambda_\textrm{r}\ln2$ where $\lambda_\textrm{r}$ is the radiation length in the medium. For dry air $\lambda_\textrm{r}= 37$\,g\,cm$^{-2}$. The \say{splitting} of $e^{\pm}$ is the emission of a photon via Bremsstrahlung, whereas photons split via pair production. Both interactions occur in the presence of an air nucleus and can be written as
\begin{equation}
    N+e^{\pm}\xrightarrow{}N+e^{\pm}+\gamma\quad\mathrm{and}\quad N+\gamma\xrightarrow{}N+e^{+}+e^{-}
\end{equation}
After a splitting the outgoing particles are assumed to each possess half the energy of the parent particle. Particles continue to split until they reach a critical energy $E_c^e$ where the average collisional energy losses become greater than radiative losses. In air $E_c^e\approx85$\,MeV. The maximum number of particles during the shower development $N_\textrm{max}$ will be reached when all particles have energy $E_c^e$. Hence, for a shower initiated by a single $e^{\pm}$ or $\gamma$ with energy $E_0$, we have 
\begin{equation}
\label{eqn:HeitlerE0}
    E_0=E_c^e N_\textrm{max}.
\end{equation}
The atmospheric depth at which $N_\textrm{max}$ is reached, \Xmax{}, is given by the number of splittings required to reach the critical energy $n_\textrm{c}$ multiplied by $d$. Since $N_\textrm{max}=2^{n_\textrm{c}}$, $n_\textrm{c}=\ln\left[E_0/E_c^e\right]/\ln2$ and thus 
\begin{equation}
\label{eqn:XmaxEM}
X_\textrm{max}^{\gamma}=n_\textrm{c}\lambda_\textrm{r}\ln2=\lambda_\textrm{r}\ln\left[E_0/E_c^e\right].
\end{equation}
The $\gamma$ superscript indicates that this equation is only valid for purely electromagnetic showers. Differentiating Equation \ref{eqn:XmaxEM} with respect to $\log_{10}(E_0)$ gives the elongation rate
\begin{equation}
    \Lambda=\frac{\textrm{d}X_\textrm{max}}{\textrm{d}\log_{10}E_0}=2.3\lambda_\textrm{r}\approx85\textrm{\,g\,cm}^{-2}.
\end{equation}
The Heitler model's predictions of a proportional relationship between primary cosmic-ray energy and maximum shower size (Equation \ref{eqn:HeitlerE0}), as well as the elongation rate of 85\,g\,cm$^{-2}$ have been verified by simulations and experiments \cite{matthews2005heitler}.

\subsection{Hadronic Cascades}
A basic understanding of hadronic showers can be obtained by generalising the Heitler model as done by Matthews \cite{matthews2005heitler}. 
Similar to electromagnetic cascades, Matthews assumes that hadronic particles interact after traversing a fixed depth $\lambda_\textrm{I}\ln2$ where $\lambda_\textrm{I}\approx120$\,g\,cm$^{-2}$ is the interaction length for pions in air. At each interaction, two-thirds of the particle's energy is converted to charged pions and the remaining third to neutral pions. The neutral pions decay to photons, initiating EM showers, whilst the charged pions continue to travel through the atmosphere and interact. The charged pions stop interacting when they reach a critical energy of $E_c^\pi$, estimated as the energy at which the decay length of a charged pion becomes less than the distance to the next interaction point.

\vspace{5mm}

For a primary hadron with energy $E_0$, the number of charged pions produced after $n$ interactions is $N_\pi=\left(N_\textrm{ch}\right)^n$ where $N_\textrm{ch}$ is the multiplicity of charged pions produced in the interaction. In his model Matthews takes $N_\textrm{ch}=10$. The energy per charged pion after $n$ interactions is thus
\begin{equation}
    E_\pi=\frac{E_0}{\left(\frac{3}{2}N_\textrm{ch}\right)^n}.
\end{equation}
After $n_\textrm{c}$ interactions $E_\pi$ will fall below the critical energy $E_c^\pi$. Strictly speaking, the critical energy varies with $E_0$, since whether or not the \textit{distance} to the next interaction point is greater than the decay length after $n$ interactions depends on $E_\pi$, a function of $E_0$. 
For simplicity however, Matthews takes $E_c^\pi=20$\,GeV. Upon reaching the critical energy, all charged pions decay to muons, giving $N_\mu=N_{\pi,\textrm{max}}$. As in Equation \ref{eqn:HeitlerE0}, the primary energy $E_0$ can be expressed as a sum of the maximum shower size multiplied by the critical energy for each component,
\begin{equation}
    E_0=E_c^e N_\textrm{max}+E_c^\pi N_\mu\approx0.85\textrm{\,GeV}\left(N_e+24N_\mu\right).
\end{equation}
Since physical detectors primarily detect the electrons/positrons from the EM component, $N_\textrm{max}$ has been replaced with $N_e$, the total number of $e^\pm$, in the last approximation. In Matthews' model, $N_e=N_\textrm{max}/10$. To calculate the depth of shower maximum, Matthews considers the first generation of photon showers generated from $\pi^0$ decays. As the first interaction generates $(1/2)N_\textrm{ch}$ neutral pions, which then decay to $N_\textrm{ch}$ photons, each photon will have an energy of $E_0/(3N_\textrm{ch})$. Applying Equation \ref{eqn:XmaxEM} to a shower with energy $E_0/3N_\textrm{ch}$ and including an offset to account for the depth of the first interaction, Matthews gives the following formula for the \Xmax{} of proton generated hadronic showers,
\begin{equation}
\label{eqn:HadronicXmax}
    X_\textrm{max}^p=\left(470+58\log_{10}[E_0/\textrm{PeV}]\right)\textrm{\,g\,cm}^{-2}.
\end{equation}
Although Equation \ref{eqn:HadronicXmax} is a systematic underestimation of \Xmax{}, due to only considering the first generation of photon initiated EM cascades, the predicted elongation rate of 58\,g\,cm$^{-2}$ agrees closely with simulations \cite{matthews2005heitler}.

\section{Detection Methods}
\label{sec:detectionMethods}
The three most common methods for detecting extensive air showers are surface detector arrays, fluorescence detection and radio detection. The following section gives a basic overview of each of these methods. See the attached references for further details.

\subsection{Surface Detector Arrays}
\label{sec:surfaceDetectorArrays}
One method for detecting EASs is by directly detecting the secondary particles of a shower at ground level. This is typically achieved by arranging particle detectors into an 
array on the ground. Such an array is referred to as a \say{\gls{sd} array}. Note that \say{SD} can be used to refer to the entire set of detectors constituting an array, i.e. \say{the TA/Auger SD} or to individual detectors. The number of detectors and area over which they are installed varies depending on what region of the cosmic ray energy spectrum is being studied. This is due to the steeply falling flux and increasing \say{size} of the showers (extent over which the secondary particles are spread) with energy. At $\sim10^{16}$\,eV, a few tens of detectors with spacing $\sim100$\,m is sufficient, whereas cosmic rays with energies $\geq10^{18.5}$\,eV require hundreds of detectors spread over several hundred/thousands of square kilometres \cite{pierre2015pierre, abu2012surface,iwasaki2023performance}. 

\vspace{5mm}

The particle detectors used are typically some form of \gls{wcd} or scintillation detector. WCDs rely on the detection of Cherenkov radiation, light emitted by charged particles travelling through a medium faster than the speed of light in that medium. Charged secondary particles which penetrate the detector and are travelling faster than $c/n_\textrm{water}\approx0.75c$ will produce Cherenokv light in the water. This light is then collected by photomultiplier tubes (PMTs) and converted into an electrical signal for analysis. WCDs have a known bias towards the detection of muons. This is because typical muons in an EAS are sufficiently energetic to pass through the entire volume of water whilst producing Cherenkov radiation. Electrons meanwhile usually only penetrate a short distance into the water before losing all their energy.
Photons from an extensive air shower which penetrate the detector can also be detected provided they undergo pair production and produce sufficiently relativistic electrons \cite{allekotte2008surface}. 

\begin{figure}
    \centering
    \includegraphics[width=0.43\linewidth]{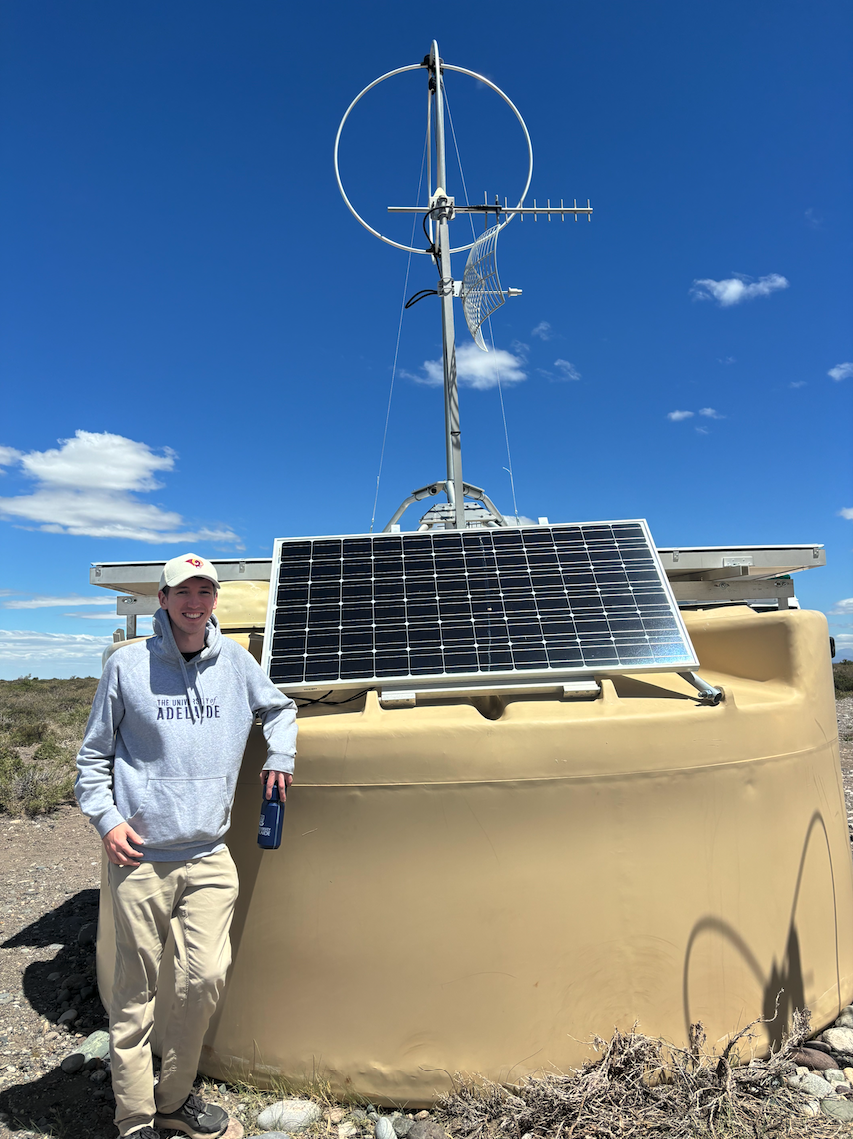}
    \includegraphics[width=0.56\linewidth]{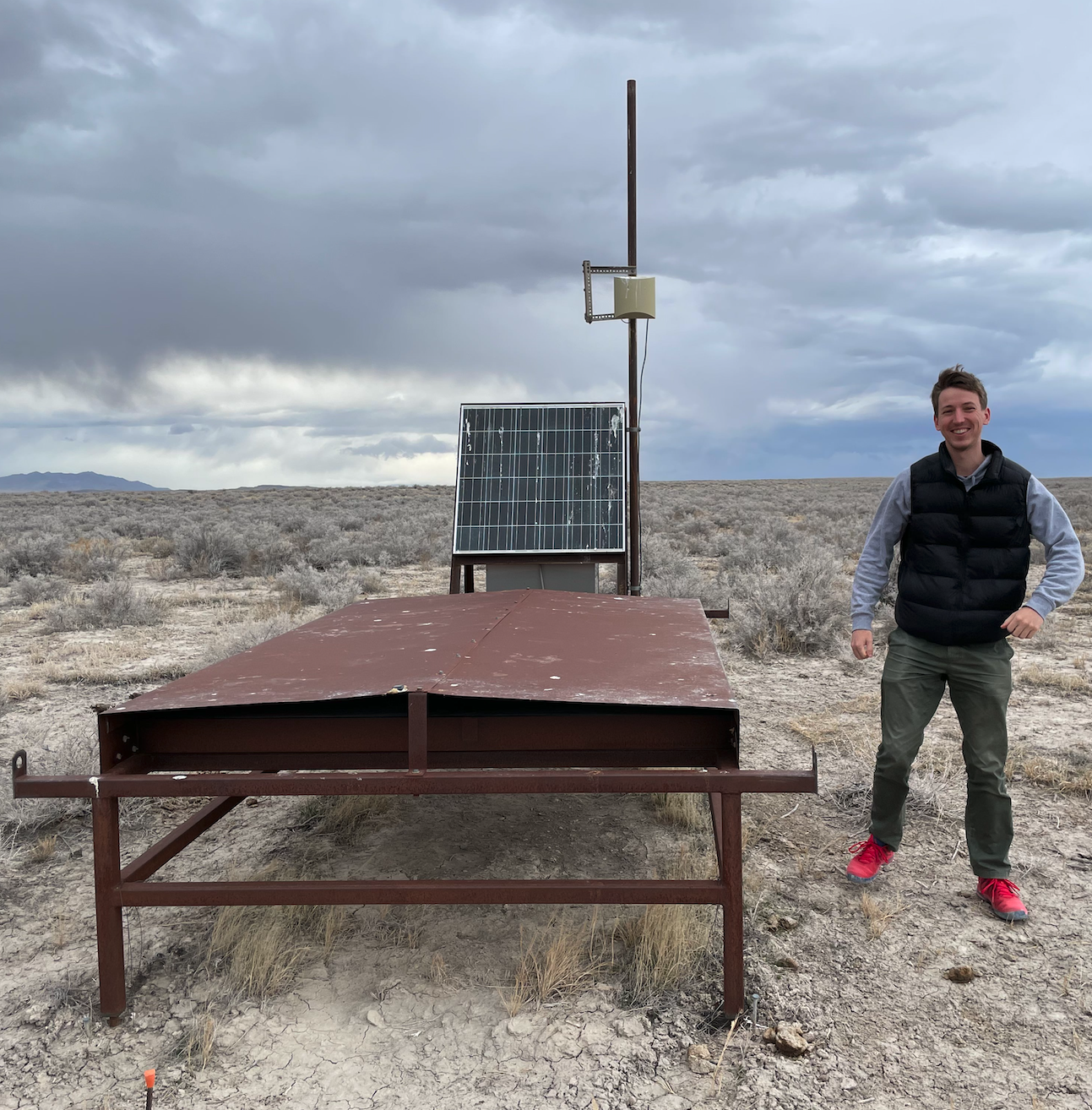}
    \caption{\textit{Left:} Photo of the WCD used by Auger. Visible on top of the tank are the solar panel used for generating power and the scintillator and radio detectors installed as part of the recent AugerPrime upgrade \cite{castellina2019augerprime}. \textit{Right:} The scintillation detector used by TA. The scintillators and PMTs are housed underneath a 1.2\,mm thick iron roof which shields the detector from the surrounding desert environment \cite{abu2012surface}.}
    \label{fig:surfaceDetectorsFraser}
\end{figure}

\vspace{5mm}

Scintillation detectors rely on the property of scintillation, luminescence stimulated by ionising radiation, to detect secondary particles. When charged secondary particles penetrate the detector and excite the scintillating material, light in the ultra-violet/visible range of the electromagnetic spectrum is produced. This light is then collected and channelled to PMTs, often through wavelength shifting fibres which adjust the wavelength of the emitted light to increase each PMT's collection efficiency. Scintillation detectors have an equal response to the muonic and electromagnetic components of air showers. Photons are also detectable if they pair-produce within the scintillating medium \cite{abu2012surface, castellina2019augerprime}. Photos of the WCD used by the Pierre Auger Observatory and the scintillator detector used by the Telescope Array experiment are shown in Figure \ref{fig:surfaceDetectorsFraser}. The Author is pictured beside each detector. 

\vspace{5mm}

\begin{figure}
    \centering
    \includegraphics[width=1\linewidth]{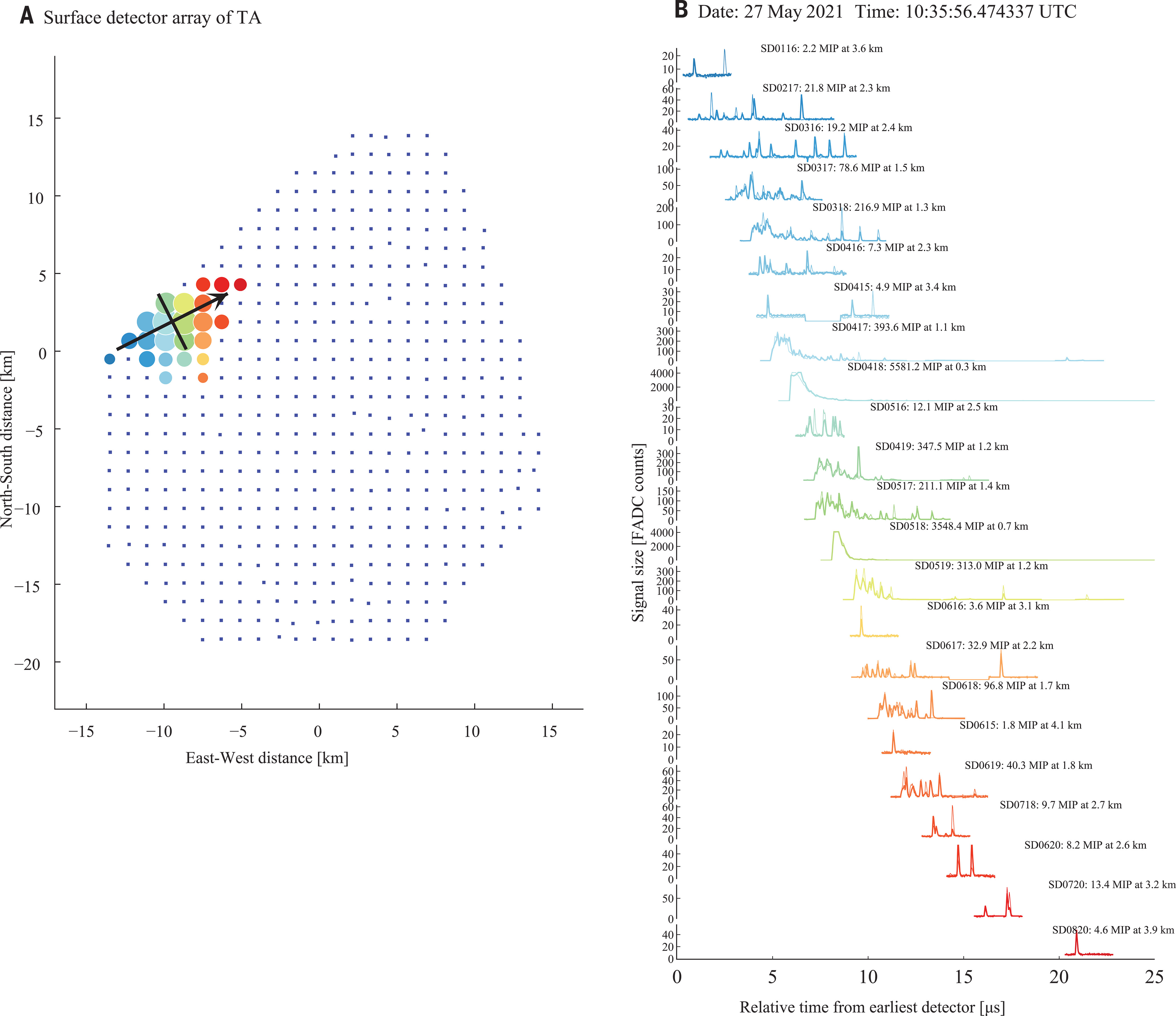}
    \caption{\textit{Left:} Map of the TA surface detector showing which detectors were triggered by the Amaterasu particle event. The colours indicate the relative trigger timing of each detector. The size of each circle represents the recorded signal. \textit{Right:} Waveforms from each triggered detector shown from earliest (top) to latest (bottom). The same colour code as the left plot is used. From \cite{telescope2023extremely}.}
    \label{fig:amaterasu}
\end{figure}

\begin{figure}
    \centering
    \includegraphics[width=0.8\linewidth]{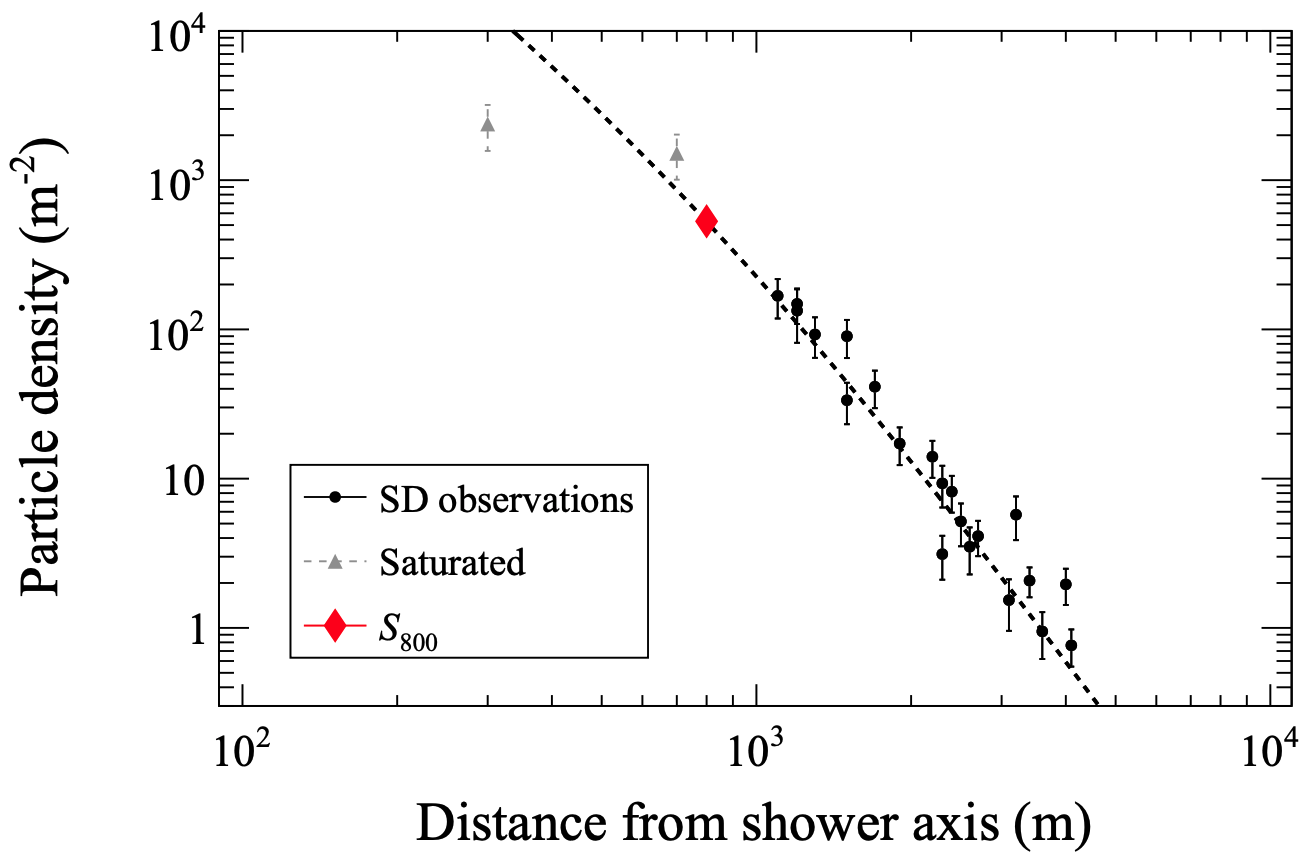}
    \caption{The LDF fit to the signals recorded in the Amaterasu particle event. Gray triangles represent detectors with saturated signals. The value of $S_{800}$ is used to estimate the primary energy. From \cite{telescope2023extremely}.}
    \label{fig:amaterasuLDF}
\end{figure}

The fundamental method of extracting the primary cosmic ray parameters from surface-detector-array data is the same regardless of the type of detector used. First, a series of checks are performed to determine which detectors recorded signal. This is usually achieved by looking for spatial and temporal coincidences of detectors which observe signal above a given threshold. The timing information and particle densities of each selected detector are then used to calculate the shower geometry, i.e. arrival direction and core location. This defines the shower axis. The particle density is then analysed as a function of the distance $r$ from the shower axis. Typically some empirical function, often referred to as a \say{\gls{ldf}}, is fit to this data and interpolated to give the expected particle density at a fixed distance. This value is then used in combination with the zenith angle of the shower to estimate the energy of the primary cosmic ray. Figure \ref{fig:amaterasu} shows an example of an event detected by the TA surface detector. This particular event is the second most energetic cosmic ray ever detected, known as the \say{Amaterasu} particle \cite{telescope2023extremely}. Figure \ref{fig:amaterasuLDF} shows the LDF fit to the signals from each triggered detector in the event. The value labelled $S_{800}$, in conjunction with the reconstructed zenith angle $\theta=38.6\pm0.4\deg$, was used to estimate the primary energy as $244\pm29^{+51}_{-76}$\,EeV. Details of the reconstruction procedures for the Auger and TA surface detectors can be found in \cite{aab2020reconstruction} and \cite{abbasi2023energy} respectively.

\vspace{5mm}

SD arrays have an up-time of nearly 100\% and are relatively cheap to deploy at scale. 
By themselves they traditionally struggle to accurately determine the mass composition of measured cosmic rays, whether by inferring \Xmax{} or using other observables such as the muon number. 
They also rely heavily on simulations to relate the measured particle densities to the primary energy \cite{abdul2024testing}. These relations come from extrapolations of accelerator data at much lower interaction energies than present in an EAS and vary depending on the hadronic interaction model used.
A hybrid setup (see Section \ref{sec:hybridDetection}) where FD measurements are used to calibrate the energy scale of the SD array, reduces this hadronic model dependence. 

\vspace{5mm}

The \say{resolution} of an SD array, or any other detection method/setup for that matter, refers to the degree of precision with which the shower parameters can be reconstructed. This is normally defined as the event-to-event statistical uncertainty. For SD arrays the resolution naturally depends on the spacing of detectors. The standard Auger and TA SDs, with spacings of 1500\,m and 1200\,m respectively, have resolutions of around 1$\degree$ in arrival direction, $\sim100$\,m in core location and $\sim10$\% in energy \cite{aab2020reconstruction, abbasi2023energy}. 

\subsection{Fluorescence Detection}
\label{sec:FluorescenceDetection}

\begin{figure}[t]
    \centering
    \includegraphics[width=0.95\linewidth]{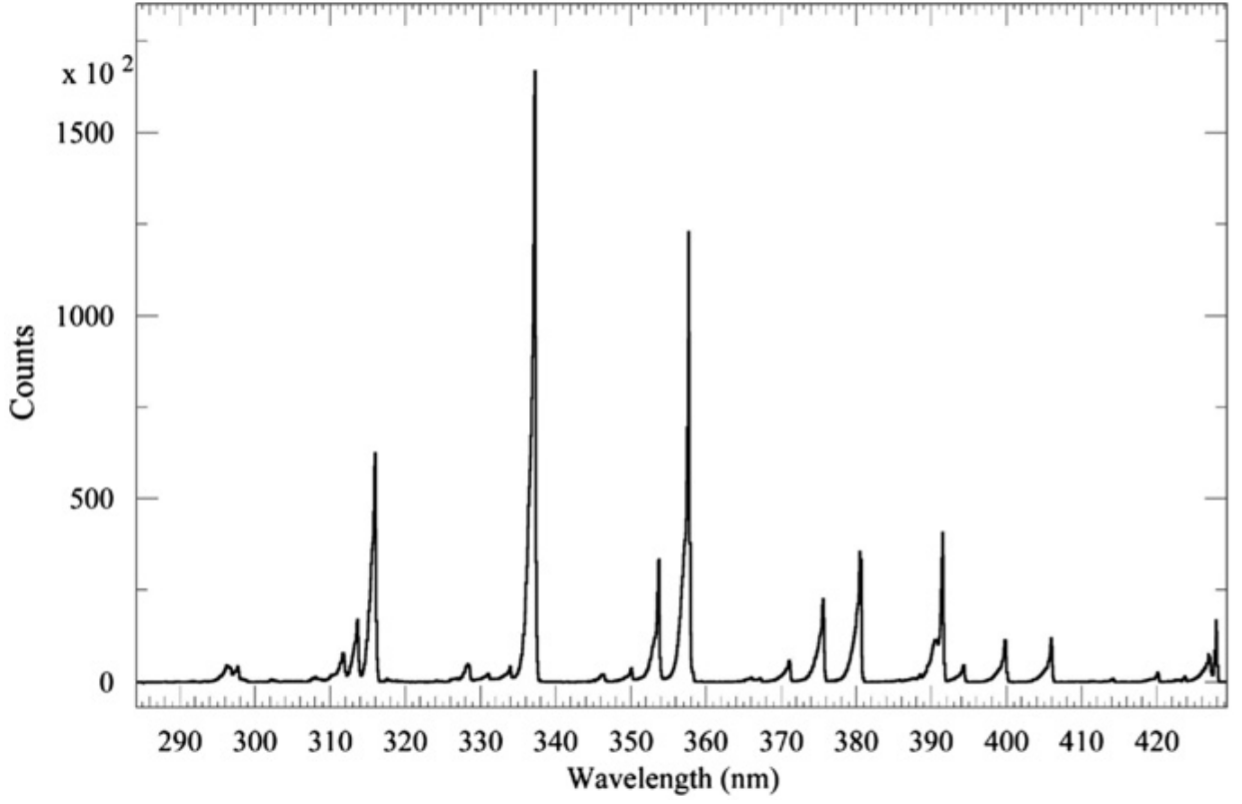}
    \caption{Fluorescence spectrum of nitrogen in dry air. Measurements taken at 800 hPa and 293 K. From \cite{abraham2010fluorescence}.}
    \label{fig:fluspec}
\end{figure}

During the development of an EAS the atmosphere acts like a calorimeter whereby the charged particles in the air shower deposit energy into the atmosphere by exciting atmospheric nitrogen/oxygen molecules. When these molecules relax back to the ground state they emit fluorescence light isotropically. The majority of light is emitted in the \gls{uv}/visible region with wavelengths between 300 - 420\,nm. The spectrum of emitted wavelengths from nitrogen in dry air is shown in Figure \ref{fig:fluspec}. The total number of fluorescence photons produced is proportional to the energy deposited by the secondary particles.
Thus by measuring the amount of fluorescence light the energy of the primary cosmic ray can be determined. 

\vspace{5mm}

\begin{figure}[t]
    \centering
    \includegraphics[width=0.8\linewidth]{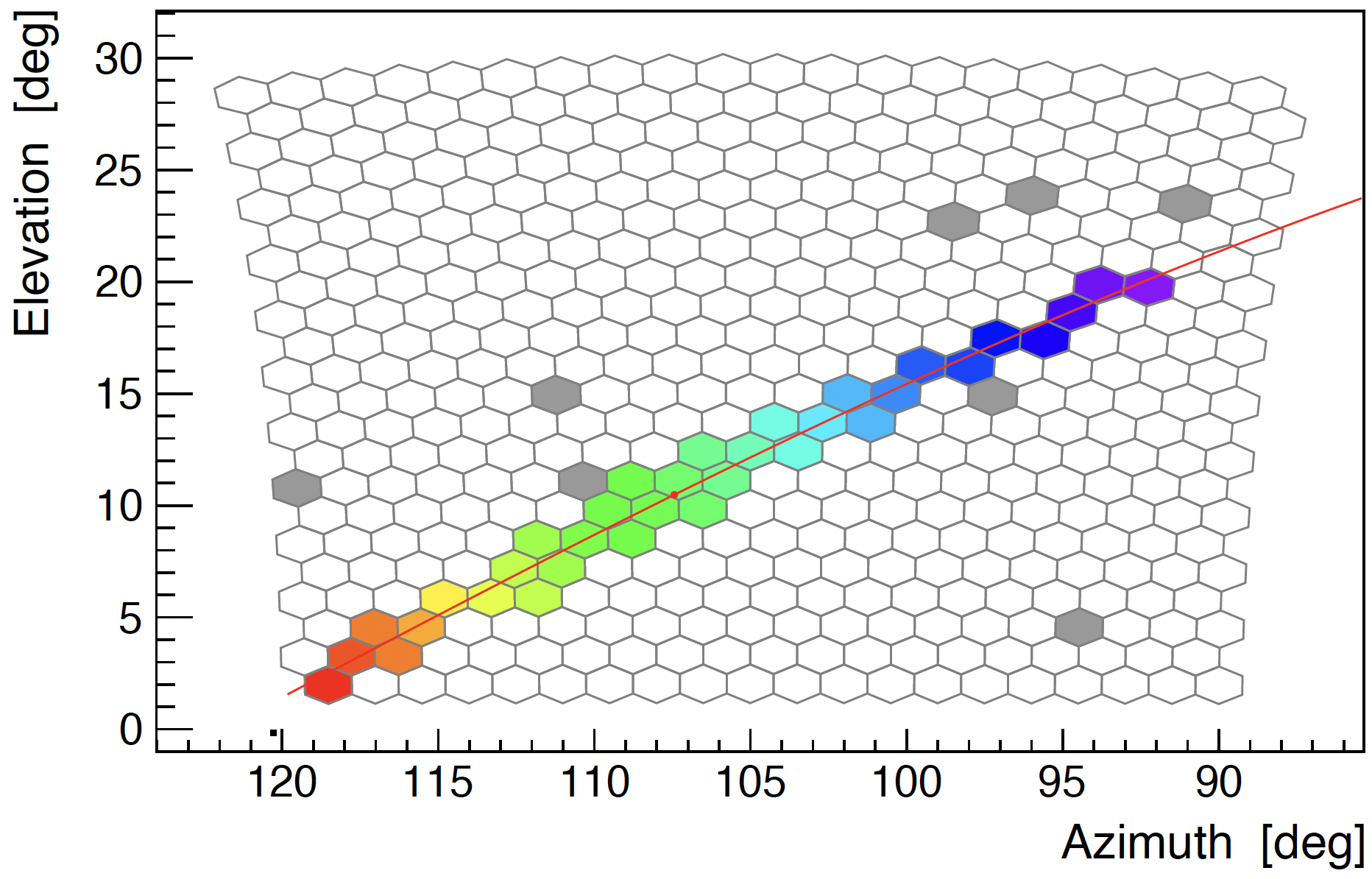}
    \caption{Example of triggered pixels across an FD camera. This example comes from the Auger fluorescence telescopes. The colours represent the timing of the pixel pulses, with purple being the earliest and red the latest. The red line is the best-fit SDP projected onto the camera. The grey pixels are triggered pixels which have been removed from the reconstruction due to being too far away in angle or in time from the best-fit SDP. From \cite{justin2020extending}.}
    \label{fig:augerCameraExample}
\end{figure}

Measurements of fluorescence light from EASs are performed using specially designed fluorescence telescopes. Similar to SD arrays, the term \say{\gls{fd}} can refer to either a collection of fluorescence telescopes operating together to observe EASs or to individual telescopes. Although the design of fluorescence telescopes vary, the general detection mechanism is the same. First, the light arriving at the telescope is focused (by way of a mirror or lens) onto a grid of PMTs. The optical setup ensures that photons that reach a given PMT must have entered the telescope from a specific direction in the sky. As a result, each PMT corresponds to a particular viewing direction, and the entire telescope can be regarded as a \say{camera} that maps different regions of the sky onto different PMTs. For this reason each PMT is also referred to as a \say{pixel}. Before arriving at the PMTs the light is passed through a UV filter to remove unwanted background light. This filter may be placed at the telescope aperture, as is the case for the Auger fluorescence telescopes, or just before each PMT's surface as done by TA. 
As the shower develops and the location of the bulk of the shower particles approaches ground, the fluorescence light arriving at the camera will pass over several pixels forming a \say{track}. This line of pixels and their viewing directions defines the plane in which both the shower axis and telescope lie, the \gls{sdp}. An example of this is shown in Figure \ref{fig:augerCameraExample}. The relative timing of the pulses in each pixel is then used to determine the direction of the shower axis inside the SDP. Note that this method of determining the shower geometry is for monocular reconstruction, that is using only information from fluorescence telescopes from one location. It relies on having a relatively large number of triggered pixels each with a small field of view. If telescopes from two or more locations view the same shower it is known as stereo observation. In this case the intersection of the shower planes determined at each location defines the shower axis. This method removes the need for a timing fit to the triggered pixels, which can be degenerate for shorter tracks, and improves the reconstruction of the geometrical parameters.

\begin{figure}[h]
    \centering
    \includegraphics[width=0.8\linewidth]{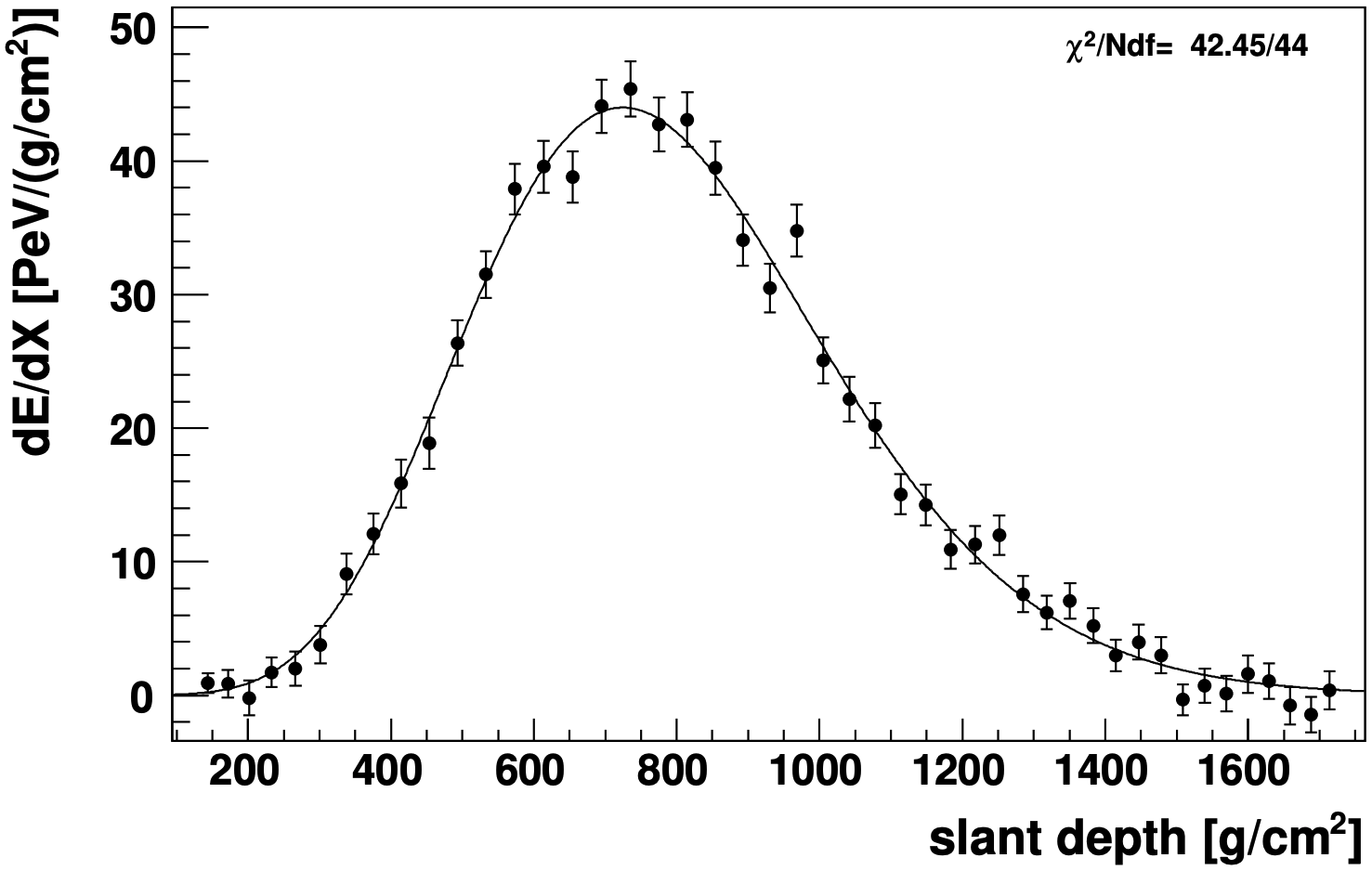}
    \caption{Example from Auger of a fitted GH function to d$E$/d$X$ measurements as a function of slant depth. From \cite{abraham2010fluorescence}.}
    \label{fig:gaisserHillasExample}
\end{figure}

\vspace{5mm}

With the geometry of the shower known, the longitudinal profile of the shower can be reconstructed. This profile describes the energy deposited into the atmosphere per unit slant depth as a function of slant depth and is often modelled using the \gls{gh} function
\begin{equation}
\label{eqn:GH}
    \frac{\mathrm{d}E}{\mathrm{d}X}(X)=\left(\frac{\mathrm{d}E}{\mathrm{d}X}\right)_{\textrm{max}}\left(\frac{X-X_0}{X_{\textrm{max}}-X_0}\right)^{\frac{X_{\textrm{max}}-X_0}{\lambda}}\exp\left(\frac{X_{\textrm{max}}-X}{\lambda}\right).
\end{equation}
Here, $X$ is the slant depth, $X_0$ and $\lambda$ are fitted shape parameters, and \Xmax{} is the slant depth at which the energy deposit per unit slant depth is a maximum, i.e. $(\mathrm{d}E/\mathrm{d}X)_\textrm{max}$. Exactly how this profile is used to reconstruct \Xmax{} and primary energy differs across experiments. For Auger the shower geometry is used to calculate the corresponding slant depth $X$ for each pixel directly. The associated $(\mathrm{d}E/\mathrm{d}X)\Delta{}X$ values, where $\Delta{}X$ is the slant depth viewed by each pixel, are calculated based on the signal in each pixel. A direct fit of Equation \ref{eqn:GH} is then performed giving \Xmax{}. Integrating the fitted function gives the total calorimetric energy of the shower \cite{abraham2010fluorescence}. An example of this is shown in Figure \ref{fig:gaisserHillasExample}. On the other hand, TA uses an inverse Monte Carlo method. The ratios between integrated PMT signals from data are compared to those from simulations using various values of \Xmax{} in Equation \ref{eqn:GH}. The \Xmax{} value giving the best match is chosen. The calorimetric energy is estimated by the ratio between the observed total signal and expected signal, assuming the expected signal sums to unity \cite{abbasi2016energy}. 
An additional energy contribution known as the invisible energy must be added to the calorimetric energy to obtain the total energy of the shower. This energy comes from neutrinos which are not observed by the FD and, depending on the calorimetric energy definition, from muons which reach the surface without depositing all their energy into the atmosphere \cite{abbasi2023energy, arqueros2008air}. The fraction of Cherenkov light which is captured by the telescopes
must also be carefully accounted for to obtain an accurate reconstruction.
Further details on the Auger and TA FD reconstructions can be found in \cite{abraham2010fluorescence} and \cite{abbasi2016energy}. 

\vspace{5mm}

The primary strengths of the fluorescence detection technique are the near calorimetric measurement of the shower energy and direct observation of \Xmax{}. However, fluorescence telescopes can only operate on clear, moonless nights, bringing the duty-cycle down to $\sim15\%$ \cite{pierre2015pierre}. Cuts requiring \Xmax{} to be in the field of view of the telescopes to ensure a reliable reconstruction further reduce statistics. Another hurdle for fluorescence observation is the continuous monitoring of the atmosphere required to accurately predict the fraction of light reaching the telescope and thus primary cosmic-ray energy. In particular, not accounting for the attenuation of fluorescence light by atmospheric aerosols can induce biases in the energy reconstruction by 8 - 25\% \cite{pierre2013techniques}. For FD events observed in conjunction with a companion SD, allowing for improved geometrical reconstruction, resolutions in the shower parameters of 0.6$\degree$ in arrival direction, $50$\,m in core location, $\leq10\%$ in energy and 20 - 30\gcm{} in \Xmax{} can be achieved \cite{abraham2010fluorescence, huege2016radio}.

\subsection{Radio Detection}
\begin{figure}
    \centering
    \includegraphics[width=1\linewidth]{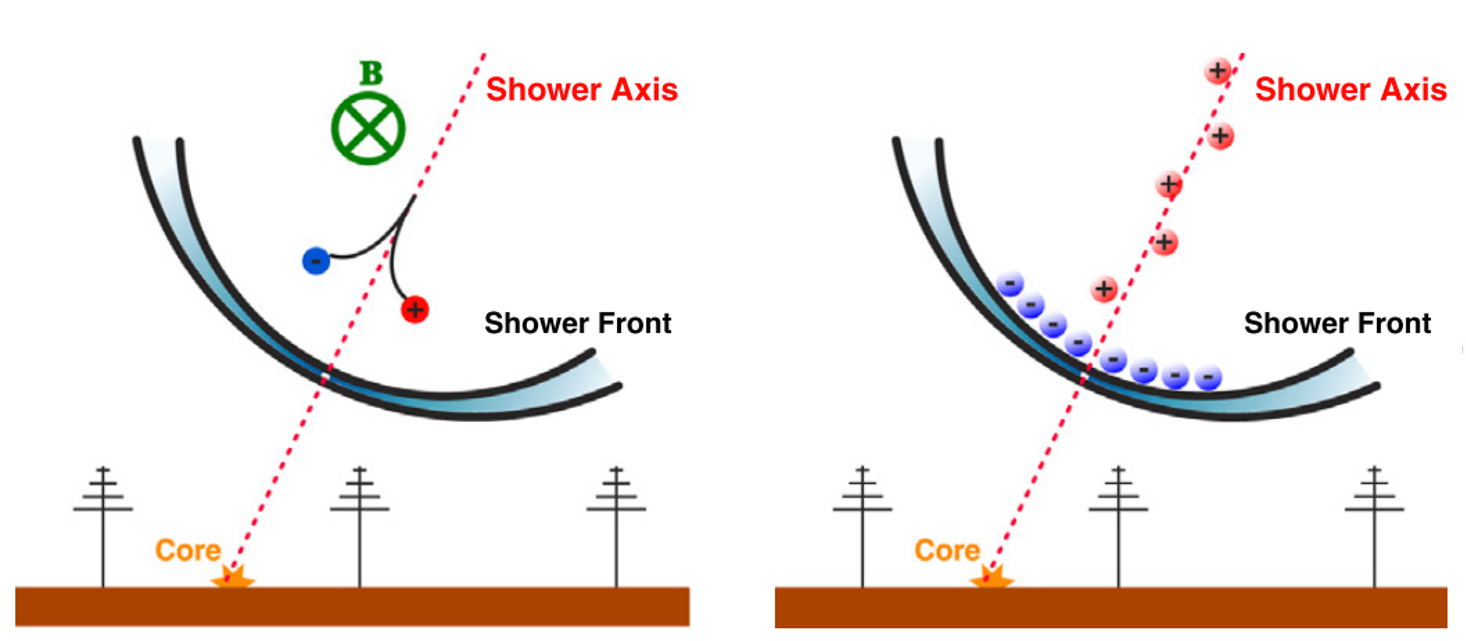}
    \caption{Illustrations of the two primary mechanisms of radio emission from extensive air showers. \textit{Left:} Geomagnetic emission coming from time-varying transverse currents. \textit{Right:} Negative charge excess from accumulated electrons. Adapted from \cite{huege2016radio}.}
    \label{fig:radioEmissionMechanisms}
\end{figure}

The pulsed radio emission from extensive air showers was first identified in the 1960s and arises from the acceleration of secondary electrons/positrons. The primary emission mechanism is the acceleration due to the geomagnetic field. In particular, the electrons and positrons \say{drift} in opposite directions as governed by the Lorentz force
\begin{equation}
    \va{F} = q\va*{v}\times\va{B}
\end{equation}
where $q$ is the particle charge, $\va*{v}$ the particle velocity and $\va{B}$ the magnetic field. The drift direction is perpendicular to the shower axis for particles travelling parallel with the shower axis. Thus these currents are known as \say{transverse currents}. The mechanism is illustrated in the left panel of Figure \ref{fig:radioEmissionMechanisms}. The time variation of these transverse currents, which arise as the total number of particles in the shower varies, generates the radio emission \cite{huege2016radio}.

\vspace{5mm}

The other primary mechanism is a charge anisotropy built up over the course of the shower development. Specifically, some positrons will annihilate with the electrons of air atoms producing photons. These photons may then ionise other air atoms, from which the stray electrons will be swept up in the particle front. This generates a negative charge excess of roughly $20\%$. Once again it is the time variation of this charge excess as the shower develops which produces the radio signal \cite{huege2016radio, schroder2017radio}. This effect is also known as the Askaryan effect and is illustrated in the right panel of Figure \ref{fig:radioEmissionMechanisms}.

\vspace{5mm}

For radio data, the arrival direction and core location can be estimated using the arrival times and amplitudes respectively of the radio signal in triggered detectors. An LDF can then be fit to the amplitudes and, similar to SD array reconstructions, the shower energy determined using the estimated radio amplitude at a predetermined distance. 
This estimate is a calorimetric measurement of the energy contained within the electromagnetic component of the shower. This is because the amplitude of the radio signal is proportional to the number of electrons/positrons in the cascade, which is itself proportional to the energy of the shower. Furthermore, the radio emission from air showers is mostly coherent, meaning the intensity of the radio signal scales quadratically with the shower energy.
\Xmax{} can be estimated based on the shape of the radio footprint at ground, specifically the slope of the radio LDF, which varies based on the geometrical distance to the shower maximum \cite{huege2016radio}. The radio experiment LOFAR has \say{combined} the above methods by developing a non-rotationally symmetric LDF with six free parameters. These parameters can be used to estimate each of the above shower parameters \cite{nelles2015radio}.

\vspace{5mm}

Downsides to the radio detection method include difficulty in self-triggering due to anthropogenic noise, a zenith and azimuth dependent detection threshold and the relatively dense detector spacing required. Conversely, radio detection is very economical, with individual detectors being $\sim$1000 USD, and has a duty-cycle of almost 100\%, with only thunderstorms causing down time. Other advantages include negligible absorption of the radio emission by the atmosphere, meaning consistent atmospheric monitoring is not necessary, and the calorimetric measurement of the shower energy. The resolutions of current radio detectors are on the order of $0.5\degree$ in arrival direction, $\sim15\%$ in energy and $\sim50$\gcm{} in \Xmax{} \cite{huege2018radio}.

\subsection{Secondary Particle Detection with CCDs}
\begin{figure}[t]
    \centering
    \includegraphics[width=1\linewidth]{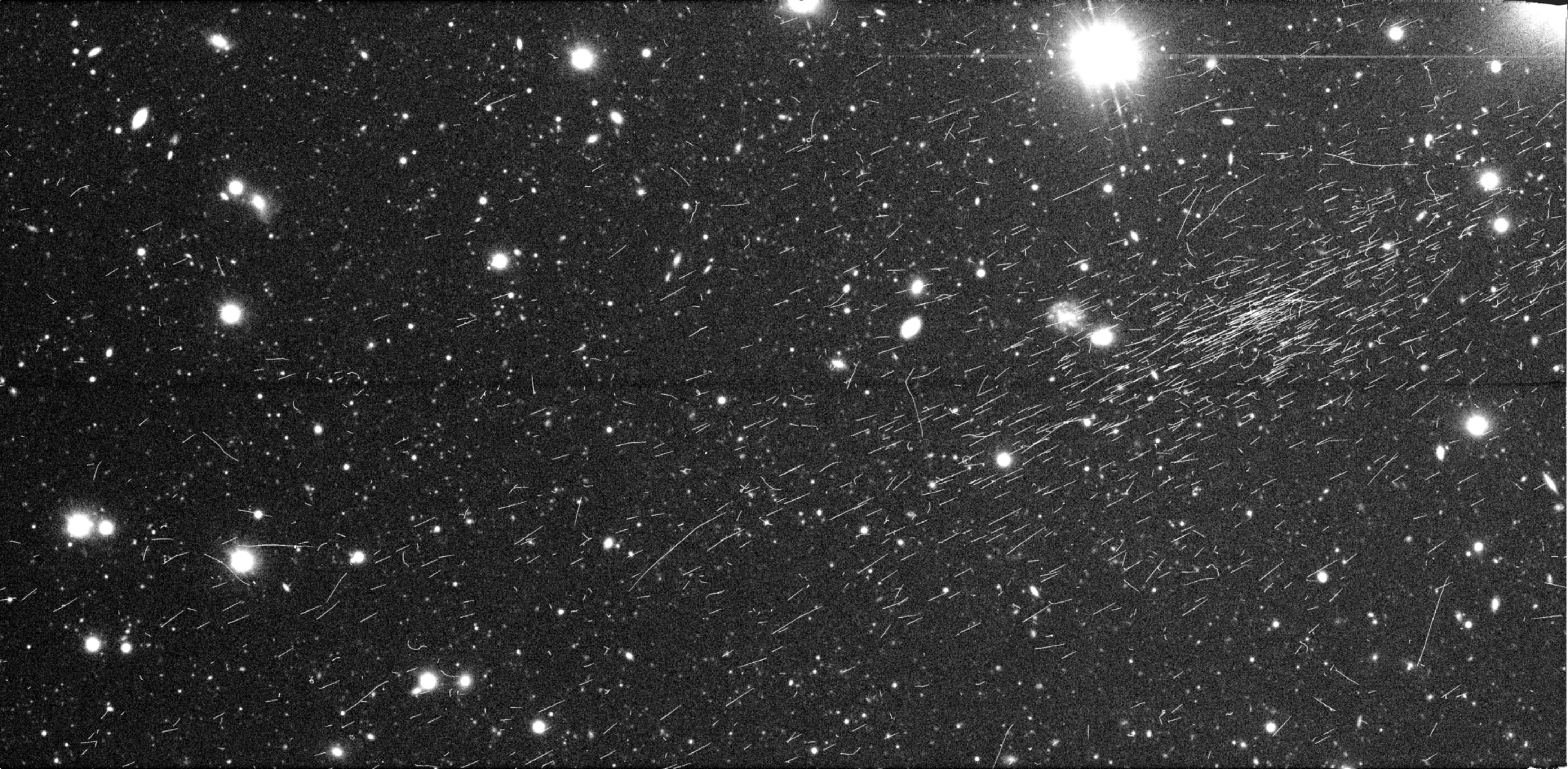}
    \caption{Example of a cosmic-ray extensive air shower recorded by a CCD of Subaru Hyper Suprime-Cam. The high density of aligned tracks, indicative of an EAS, is clearly visible in the upper right portion of the image. From \cite{kawanomoto2023observing}.}
    \label{fig:SubaruCCDshower}
\end{figure}

Although not directly related to the observation of UHECRs, a recent paper, of which the Author was a co-author \cite{kawanomoto2023observing}, highlights a new method for detecting low energy EASs - the direct detection of secondary particles using \glspl{ccd}. The serendipitous observations were made with the Subaru Hyper Suprime-Cam (Subaru HSC). Subaru HSC is an optical/infrared telescope located atop Maunakea in Hawaii which typically observes stars, galaxies and other astronomical objects. During standard analysis, several images taken by the telescope were found to contain a greater than usual number of background \say{tracks}; imprints left by charged particles penetrating the CCD depletion layer. Such tracks are generally treated as unwanted noise and are removed. However for these images the tracks were found to be of similar directions and lengths, indicating a common source for the particles, i.e. a cosmic ray induced EAS. An example of one such image is shown in Figure \ref{fig:SubaruCCDshower}. Although unlikely to become a stand-alone method of EAS observation, pairing similar CCD setups with low-cost ground arrays could allow for detailed studies of secondary particles and their interactions.

\subsection{Hybrid Detection}
\label{sec:hybridDetection}
Using two or more methods for observing the same EAS is known as hybrid observation. Hybrid observation is the most effective way of observing air showers, since the strengths/weaknesses of each detection method are typically complementary. Both Auger and TA employ hybrid observation, using the calorimetric measurements of the shower energy taken with their respective FD to calibrate the energy measurements from their respective SD. Event-by-event reconstructions are also improved, for example when using the timing information from an SD to improve FD monocular reconstruction. The recent upgrade to the Pierre Auger Observatory, AugerPrime \cite{castellina2019augerprime}, and the plans for future experiments such as GCOS (Section \ref{sec:GCOS}), show a push for future cosmic-ray observatories to utilise all the detection methods at their disposal to extract as much information as possible from each individual cosmic ray event.

\section{Cosmic Ray Experiments}
This section provides some examples of past, current and planned UHECR experiments. 

\label{sec:CRexperiments}

\subsection{Volcano Ranch}
\begin{figure}
    \centering
    \includegraphics[width=0.5\linewidth]{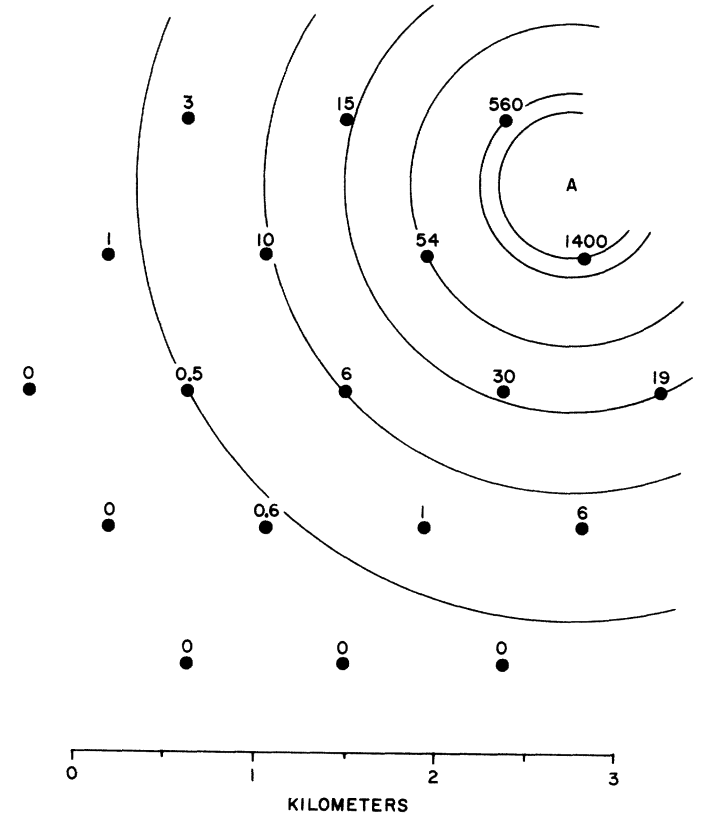}
    \caption{Detector locations and measured signals for the 10$^{20}$\,eV event detected by the Volcano Ranch experiment. From \cite{linsley1963evidence}.}
    \label{fig:volcanoRanch}
\end{figure}

Headed by American physicist John Linsley, the Volcano Ranch experiment was located near Albuquerque, New Mexico and operated between 1959 and 1978. The experiment consisted of an array of 19 3.3\,m$^2$ plastic scintillation detectors spread over an area of 8.1\,km$^2$ \cite{linsleyFermiLab}. In 1962 Volcano Ranch became the first experiment to measure a cosmic ray above 10$^{20}$\,eV in energy. The particle densities measured by each detector and estimated core position of the event are shown in Figure \ref{fig:volcanoRanch} \cite{linsley1963evidence}. The observations made at Volcano Ranch were the first to suggest that some cosmic rays might originate outside the galaxy \cite{linsley1963primary}. 

\subsection{Akeno Giant Air Shower Array}
The \gls{agasa} was located 100\,km west of Tokyo in Akeno, Japan. Operated between 1990 and 2004, the array consisted of 111 scintillation detectors each with an area of 2.2\,m$^{2}$ spread over $\sim100$\,km$^{2}$. Muon detectors were also installed at 27 stations for the purpose of measuring the mass composition  \cite{chiba1992akeno,shinozaki2006agasa}. AGASA was the first experiment with an aperture large enough to probe the GZK region, measuring about 1000 events above an energy of 10$^{19}$\,eV. Initial results surprisingly found no evidence of the predicted cutoff, however a re-analysis of the data reduced the overall energy scale and brought the results in line with other experiments. The mass composition measured with the muon detectors was consistent with the heavy to light transition between $2\times10^{17}$ and $2\times10^{19}$\,eV observed by Fly's Eye/HiRes (see following section). No anisotropy was observed in the directions of the highest energy events, excluding galactic source models \cite{shinozaki2006agasa,sokolsky2007highest}.

\subsection{Fly's Eye and HiRes}
\label{sec:FlysEyeAndHisRes}
The Fly's Eye experiment and it's subsequent upgrade HiRes were the first large scale experiments to use the fluorescence detection technique for measuring EASs. Operated between 1982 and 1992, the original Fly's eye was located in Dugway, Utah and consisted of two stations, Fly's Eye I and Fly's Eye II, separated by 3.3\,km. At each site, spherical mirrors with $12\sim14$ PMTs located at their focal plane were positioned so as to image the night sky. The \gls{fov} of each PMT was roughly $5\degree\times5\degree$. Each mirror unit was housed in a motorised, corrugated steel pipe which could be tilted down during the day to avoid exposure to the elements. A total of 67 units were located at Fly's Eye I and 36 units at Fly's Eye II \cite{baltrusaitis1985utah}. 

\vspace{5mm}

\begin{figure}
    \centering
    \includegraphics[width=0.8\linewidth]{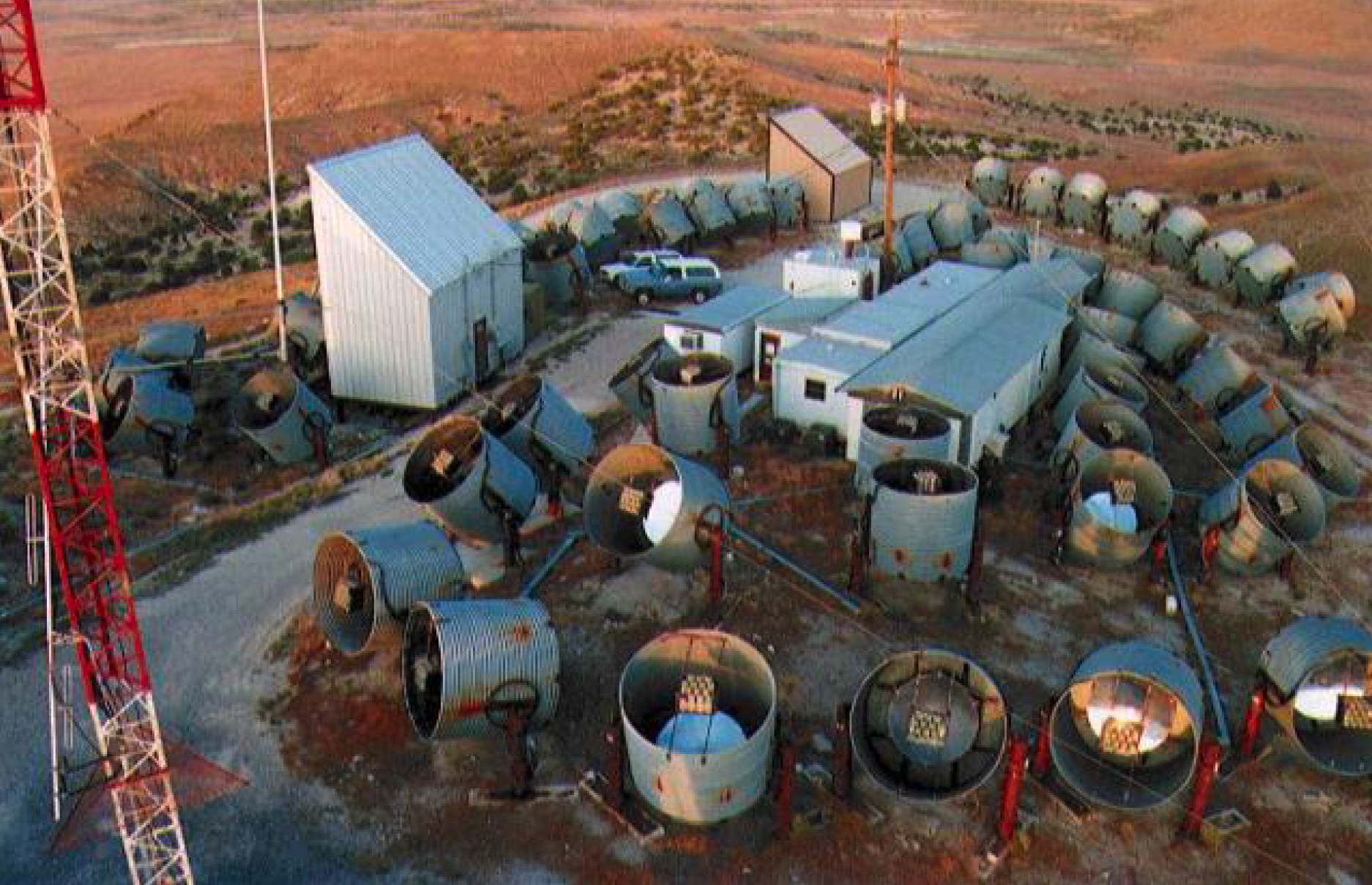}
    \caption{A photo of Fly's Eye I. From \cite{saffi2020analysis}.}
    \label{fig:flysEye}
\end{figure}

Analysis of events observed in stereo using both Fly's Eye I and II showed a clear break in the cosmic ray spectrum around 3\,EeV (the ankle). Mass composition results indicated a predominately heavy composition around 0.1\,EeV, becoming lighter with energy up to $\sim10$\,EeV. Fly's eye also detected the highest energy cosmic ray event ever recorded with an energy of 320\,EeV \cite{bird1995results}. An aerial shot of Fly's Eye I is shown in Figure \ref{fig:flysEye}.

\vspace{5mm}

The High Resolution Fly's Eye experiment or HiRes was a significant upgrade of the original Fly's Eye. Once again two stations (HiResI and HiResII) were built, this time 12.6\,km apart. HiResI (II) operated between 1997 (1999) and 2006. The increased exposure and resolution afforded by the larger scale of the experiment, combined with new electronics and a reduced FOV for each pixel of $\sim1\degree\times1\degree$, allowed HiRes to confirm the presence of the GZK cutoff at the $5\sigma$ level. Composition measurements with improved \Xmax{} resolution confirmed the light to heavy transition seen by Fly's Eye \cite{sokolsky2007highest}.

\subsection{The Pierre Auger Observatory}
The Pierre Auger Observatory is the largest cosmic ray experiment currently in operation. Completed in 2008, the observatory is located in Malargüe, Argentina and utilises both fluorescence telescopes and ground based particle detectors to measure EASs. The base ground array consists of 1660 WCDs spaced by 1500\,m. The detectors are spread over $\sim3000$\,km$^2$. Two \say{infill arrays} with detector spacings of 750\,m and 433\,m are located inside the base array. These are used to extend the energy threshold down to 10$^{17}$\,eV. A recent upgrade to the observatory known as \say{Auger Prime} has seen the addition of both scintillator detectors and radio antennas on top of every WCD. The upgrade seeks to enhance the ability of the ground array to determine the primary mass composition by extracting additional/complementary information from each individual shower \cite{pierre2015pierre, castellina2019augerprime}. 

\vspace{5mm}

\begin{figure}
    \centering
    \includegraphics[width=0.49\linewidth]{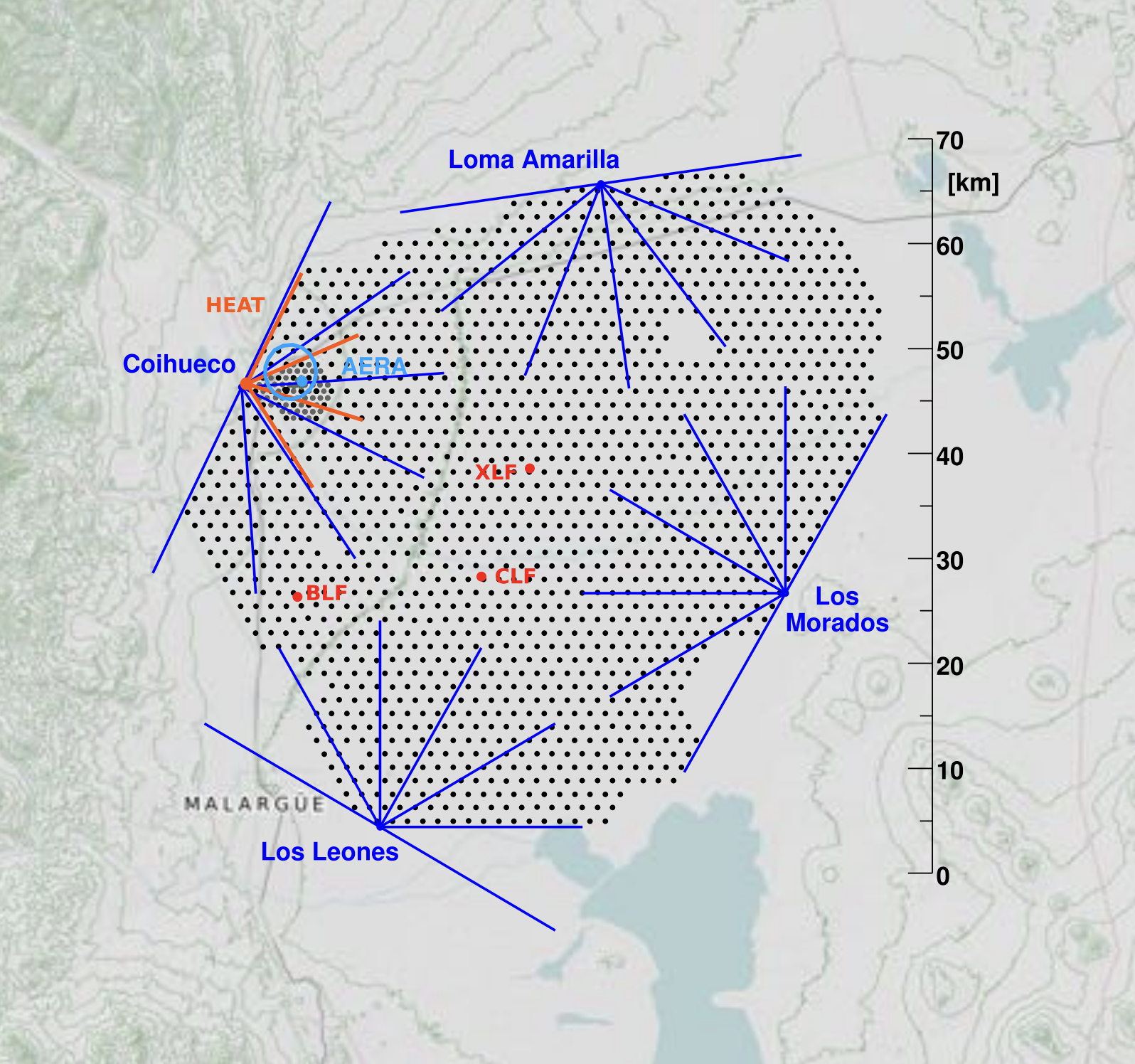}
    \includegraphics[width=0.49\linewidth]{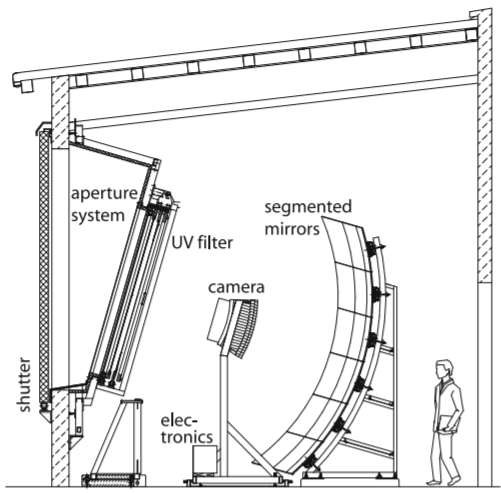}
    \caption{\textit{Left:} Map of the Pierre Auger Observatory. Each black dot shows the location of a WCD station. The FOV of each fluorescence telescope is shown by blue lines (red for HEAT). From \cite{aab2016pierre}. \textit{Right:} Schematic of the fluorescence telescopes used by Auger. From \cite{abraham2010fluorescence}.}
    \label{fig:paoStuff}
\end{figure}

The ground array is overlooked by 27 fluorescence telescopes spread across four sites. Each telescope covers a $30\degree\times30\degree$ FOV and consists of a UV filter, 10\,m$^2$ spherical mirror and 440 pixel camera. Three of these telescopes, known as the High Elevation Angle Telescopes or HEAT, have the ability to be tilted up to $29\degree$ in elevation for the purpose of measuring lower energy showers which develop higher up in the atmosphere (smaller \Xmax{}) \cite{pierre2015pierre}. Figure \ref{fig:paoStuff} shows the layout of the experiment and the design of the fluorescence telescopes. Results regarding the cosmic ray energy spectrum, anisotropy and mass composition from Auger can be found in Chapter \ref{ch:CR}.

\subsection{The Telescope Array Experiment}
The Telescope Array experiment is currently the largest cosmic ray observatory in the northern hemisphere. Like Auger, TA uses a hybrid approach for the detection of EASs, with full hybrid operation since 2008. The TA ground array consists of 507 plastic scintillator detectors. These detectors are spaced by 1200\,m and spread over an area of $\sim700$ km$^{2}$. Each scintillator consists of two 12\,mm layers of scintillating material with 1\,mm of stainless steel in between. 
The addition of the \say{TALE} and \say{TALE-infill} arrays, which possess graded detector spacings from 600\,m down to 100\,m, allow for the observation of cosmic rays with energies down to $10^{15.5}$\,eV \cite{abu2012surface, iwasaki2023performance, thomson2011telescope}. 

\vspace{5mm}

\begin{figure}
    \centering
    \includegraphics[width=0.45\linewidth]{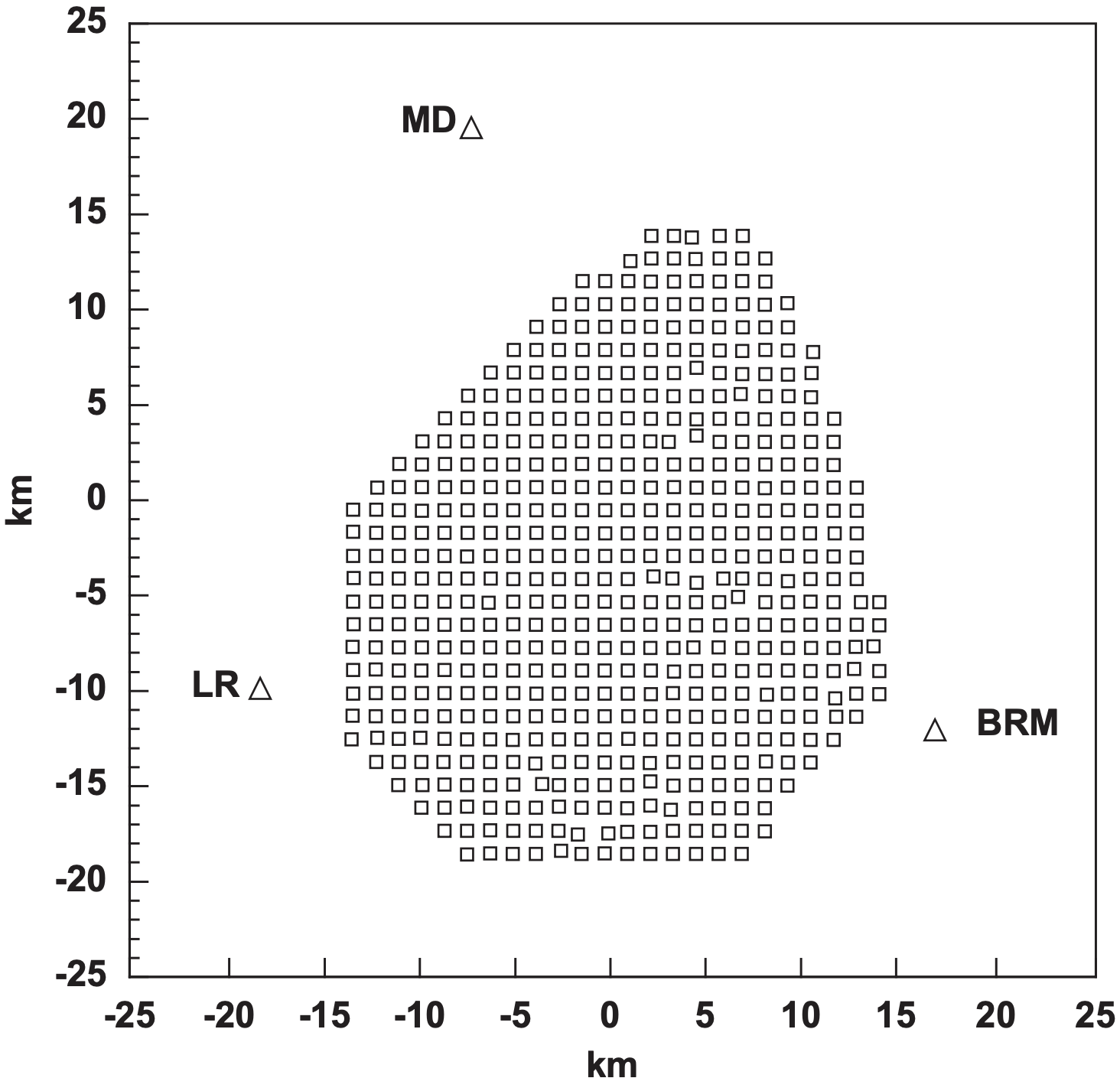}
    \includegraphics[width=0.54\linewidth]{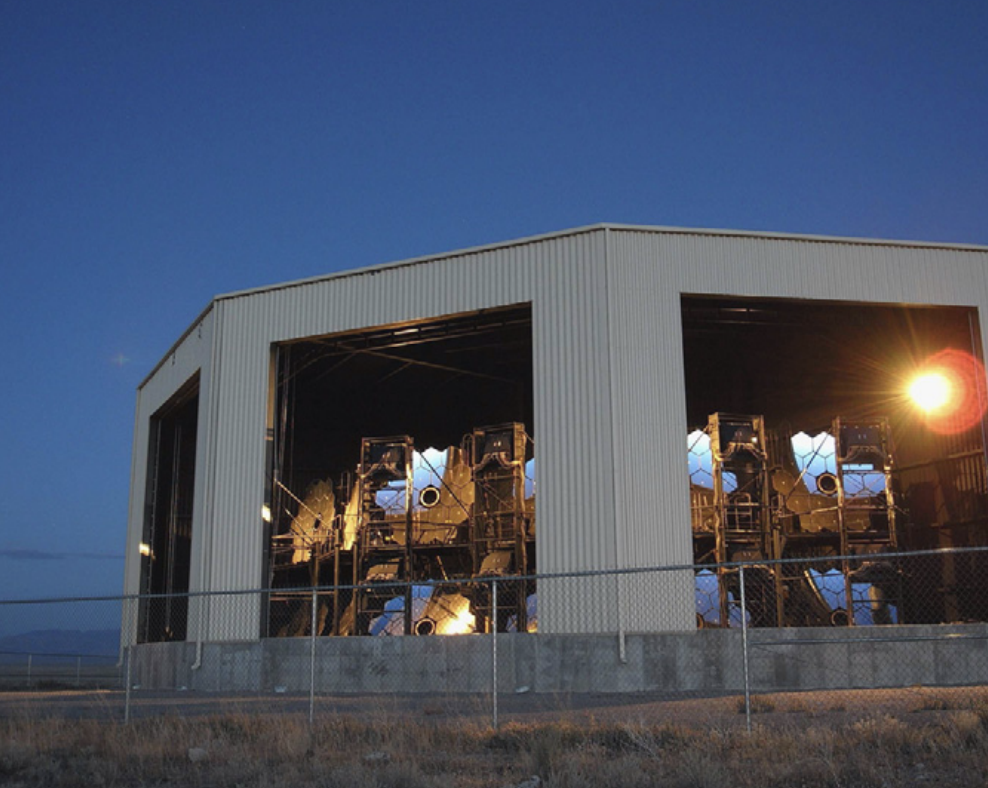}
    \caption{\textit{Left:} Map of the Telescope Array experiment. The squares show the location of each scintillator detector and the triangles show the FD locations. From \cite{abu2012surface}. \textit{Right:} Photo of the TA FD at Black Rock Mesa. From \cite{tokuno2012new}.}
    \label{fig:taStuff}
\end{figure}

Looking over the ground array are 38 fluorescence telescopes spread over three sites. Each telescope has a FOV of $15\degree$ in elevation and $18\degree$ in azimuth and consists of a 6.8\,m$^{2}$ spherical mirror focusing light onto a 256 pixel camera. The FOV of each individual PMT is approximately $1\degree$ \cite{tokuno2012new}.
Figure \ref{fig:taStuff} shows a map of TA and a photo of the fluorescence telescopes at the \gls{brm} site.
TA is currently undergoing an expansion known as TA$\times$4. Upon completion, the upgrade will see the detection area of TA increased to $\sim2800$\,km$^{2}$, comparable to that of Auger \cite{kido2018ta}. Recent results from TA can be found in Chapter \ref{ch:CR}.

\subsection{The Global Cosmic Ray Observatory}
\label{sec:GCOS}
The \gls{gcos} is a planned future cosmic ray observatory. The effective area of GCOS is intended to be an order of magnitude larger than current experiments. Sites in both the northern and southern hemisphere are envisioned so as to provide full sky coverage. Although the type of detectors to be used are still under consideration, the success of the Auger and TA experiments over the last two decades supports the hybrid approach to EAS observation.
Indeed, one option for GCOS is to simply expand the size of the existing arrays. The benefits of such an approach are that the technology is already proven and existing infrastructure can be utilised \cite{coleman2023ultra, ahlers2025ideas}. However, a ten fold increase in size using current designs is likely financially impractical. As such cost-effective alternatives are being considered. For fluorescence detection, detector designs include FAST and the Cosmic Ray Air
Fluorescence Fresnel lens Telescope (CRAFFT) \cite{yamamotocosmic}. A layered/nested water Cherenkov detector design, which has shown promising capabilities in mass discrimination, is being considered for the ground array, with the potential addition of radio antennas on top of each station \cite{ahlers2025ideas}.

\vspace{5mm}

Importantly, any proposed design will need to meet strict observational requirements such that the main science objective, uncovering the origin of the highest energy cosmic rays, can be satisfied. These include an angular resolution $<1\degree$, an energy resolution of $\leq10\%$ and ability to identify the mass of the primary particle. For \Xmax{} observations this translates to a resolution $<30$\,g\,cm$^{-2}$. 
The beginning of construction for GCOS is targeted for the 2030s \cite{ahlers2025ideas}.

\chapter{The Fluorescence detector Array of Single-pixel Telescopes}
\label{ch:FAST}
The last two decades have seen remarkable progress in our understanding of UHECRs, in large part due to the efforts and collaboration of Auger and TA. 
However, the origins of UHECRs and the physical mechanisms by which they are produced still remain unestablished. Further insights will require greater statistics above $\sim10^{19.5}$\,eV, where deflections to particle trajectories from intervening galactic and extra-galactic magnetic fields are minimal, allowing for cosmic-ray astronomy.
To achieve this future observatories will need detecting areas larger than those of current experiments by at least one order of magnitude. The Fluorescence detector Array of Single-pixel Telescopes (FAST) is one such next-generation experiment. FAST aims to utilise low-cost, easily deployable, autonomous fluorescence telescopes to observe EASs over an area of $\sim60,000$\,km$^2$. Currently, there are four prototype FAST telescopes in operation, three at \acrshort{ta} and one at \acrshort{auger}. The following chapter provides an overview of the history, development, current status and future prospects of FAST.

\section{Conception and Early Tests}
\label{sec:FASTEarlyTests}
\begin{figure}
    \centering
    \includegraphics[width=\textwidth]{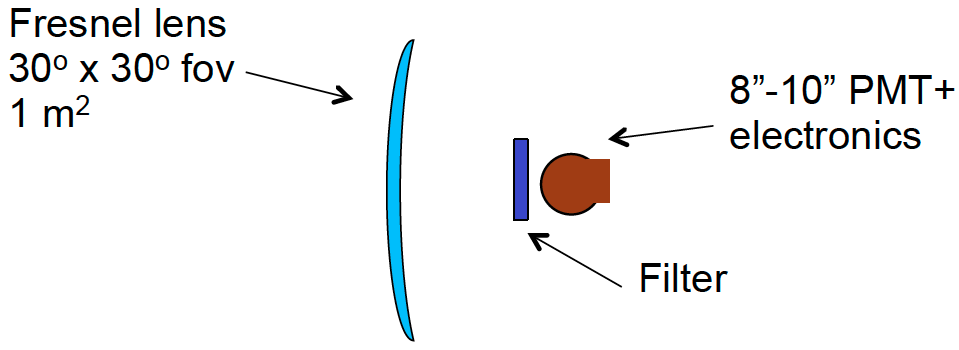}
    \caption{An early schematic design for a simplified, low-cost fluorescence detection system. From \cite{privitera2012}.}
    \label{fig:FASTEarlyDesignConcept}
\end{figure}

The initial conception of FAST can be traced back to presentations in late 2011 and early 2012 given by Prof. Paolo Privitera. In his presentation at the 2012 International Symposium on Future Directions in UHECR physics \cite{privitera2012}, Privitera put forward an idea for a then future experiment focusing on measuring cosmic rays above $10^{19.5}$\,eV with unprecedented statistics and with the ability to measure mass composition (e.g. \Xmax{}). He proposed a large ground array of simplified fluorescence detectors, each consisting of a simple 1\,m$^2$, $30\degree\times30\degree$ FOV Fresnel lens, followed by a filter and \textit{single} PMT (hence the name \say{FAST}). A basic schematic of this early design is shown in Figure \ref{fig:FASTEarlyDesignConcept}. A set of 12 detectors arranged to give full 360$\degree$ coverage in azimuth would form a single ``station". Note the term \say{Eye} will also be used when referring to a collection of FAST telescopes at a single location, regardless of the number.
With this design, $\sim170$ stations spaced by 20\,km could cover an area of $\sim60,000$\,km$^2$ for a fraction of what it would cost to scale up the existing \acrshort{auger} or \acrshort{ta} FDs to an equivalent area.

\vspace{5mm} 

The concept was tested in 2014 when a single 200\,mm PMT (R5912-03, Hamamatsu) was installed in the JEM-EUSO prototype telescope at TA (EUSO-TA telescope) \cite{takizawa2013ta, bisconti2016}. The EUSO-TA telescope consists of two planar 1\,m$^2$ Fresnel lenses giving a light collecting area of $\sim1$\,m$^2$ and a circular FOV of radius $\sim7\degree$. This made for a suitable test environment. Over the 83\,hr of operation time, a total of 16 shower candidates were detected in coincidence with TA, along with laser shots which were used to calibrate the timing between the two experiments. Later comparisons between simulations and the measured laser shots/showers showed good agreement, validating the basic design \cite{fujii2016detection}. 

\vspace{5mm}

In addition to on-site tests, preliminary investigations into the expected performance for a FAST array were also performed around this time. By using a modified version of the Auger \Offline{} software (see Section \ref{sec:Offline}), simulations of the response of a triangular arrangement of three FAST stations to proton and iron showers were analysed. 
These showers were generated using the CORSIKA simulation software \cite{heck1998corsika}. The reference telescope design consisted of a 1\,m$^2$ light collecting area, mirror, filter and \textit{four} PMTs. Compared to the original concept, the Fresnel lens approach was dropped in favour of a mirror due to technological constraints \cite{mandat2017prototype}. The additional PMTs were added to achieve a lower energy threshold and improved efficiency/resolution at lower energies, enabling better comparison between FAST and current generation FDs. These benefits, together with the improved accuracy and precision expected in the reconstruction, were deemed to outweigh the additional expense.

\vspace{5mm}

\begin{figure}[t!]
    \centering
    \includegraphics[width=\textwidth]{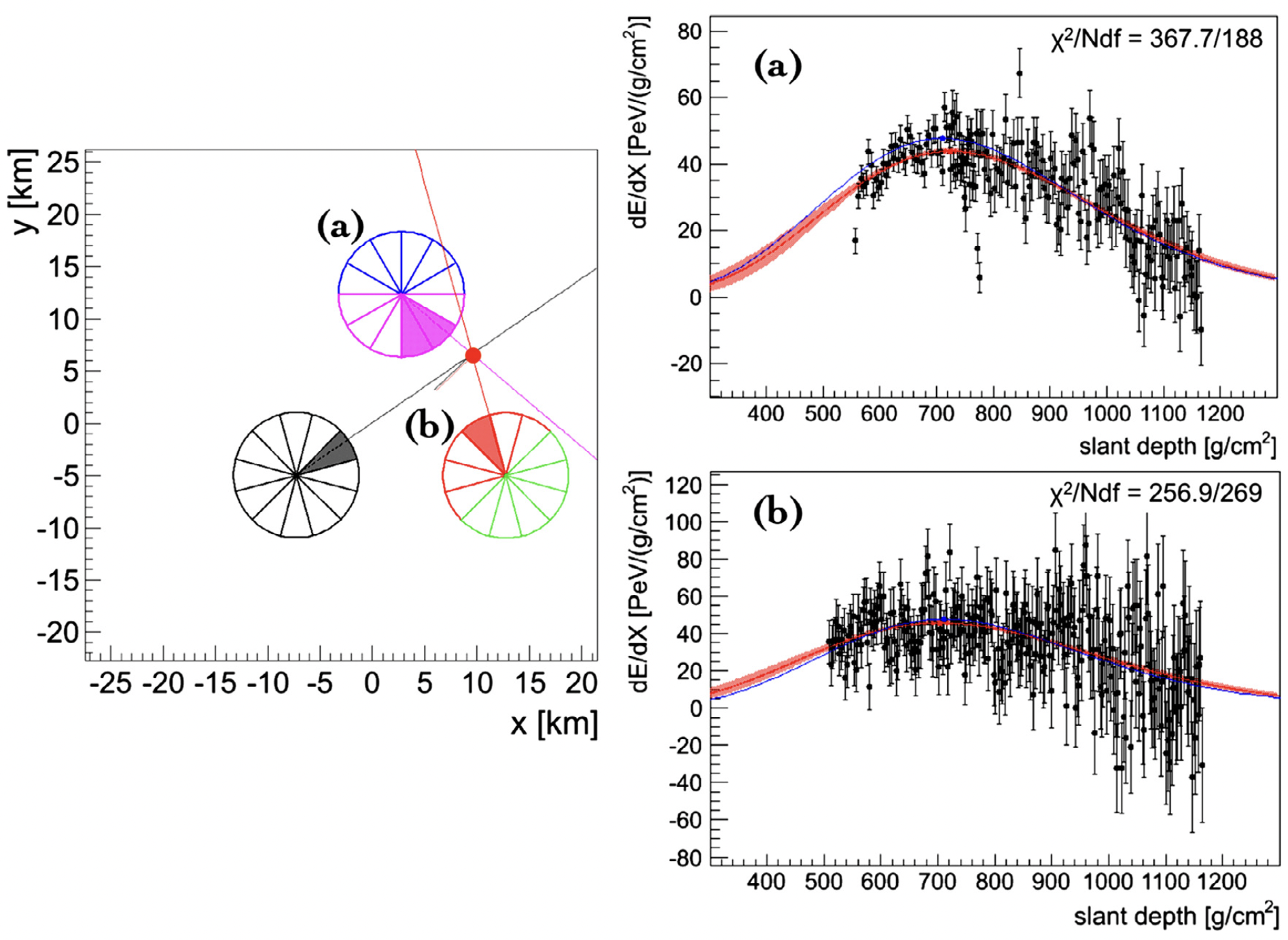}
    \caption{Early FAST reconstruction of a 10$^{19.5}$\,eV simulated shower. \textit{Left:} Layout of the FAST stations in the simulation (separation=20\,km). The red dot marks the core location of the shower. \textit{Right:} Reconstructed energy deposit profiles for two stations. The blue line shows the simulated profile. The red line shows the reconstructed profile, with the shaded region indicating the uncertainty in the fit. From \cite{fujii2016detection}.}
    \label{fig:FASTRecExample2015}
\end{figure}

\begin{figure}
    \centering
    \includegraphics[width=\textwidth]{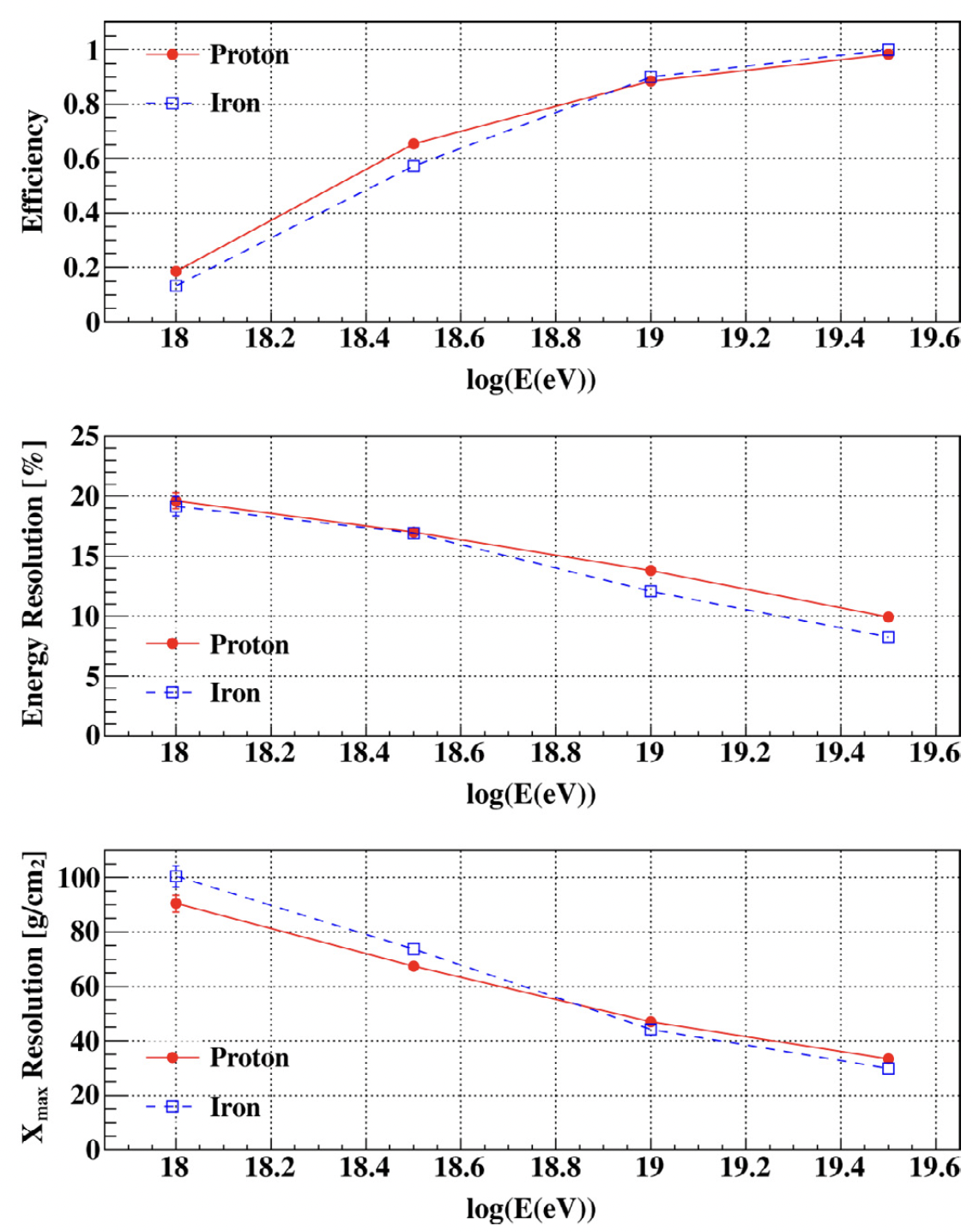}
    \caption{From top to bottom: The reconstruction efficiency, energy resolution and \Xmax{} resolution as a function of energy from early FAST simulations using 
    a modified version of the Auger \Offline{} software framework. Results are shown for proton (red) and iron (blue) showers.}
    \label{fig:FASTRecResults2015}
\end{figure}

The simulated showers were reconstructed via the Auger method of calculating the energy deposited as a function of slant depth and fitting a Gaisser-Hillas profile (see Section \ref{sec:FluorescenceDetection}). The shower geometry was taken from the simulated values smeared by $1\degree$ in arrival direction and 100\,m in core position, emulating the typical resolution of a surface detector (see Section \ref{sec:CheckingRecon} for an explanation of smearing).
An example of a simulated $10^{19.5}$\,eV event and its reconstruction is shown in Figure \ref{fig:FASTRecExample2015}. Each data point (black) represents an individual time bin of the combined sum of all PMT traces for that telescope/set of telescopes (as opposed to individual pixels). The reconstruction efficiency, energy resolution and \Xmax{} resolution derived from these simulations is shown as a function of energy in Figure \ref{fig:FASTRecResults2015}. At the target energy of $10^{19.5}$\,eV the reconstruction efficiency was almost 100\% for all the triggered events, whilst the energy and \Xmax{} resolutions were $\sim10\%$ and $\sim34$\,\gcm{} respectively. Further details on these simulations and event selection criteria can be found in \cite{fujii2016detection}. Overall, these early tests demonstrated the viability of the FAST concept and encouraged the development of the first full-scale prototype telescopes.

\section{First Prototype Telescopes}
\label{sec:FASTFirstGenPrototypes}
\begin{figure}[t!]
    \centering
    \begin{subfigure}[b]{0.59\textwidth}
        \includegraphics[width=\textwidth]{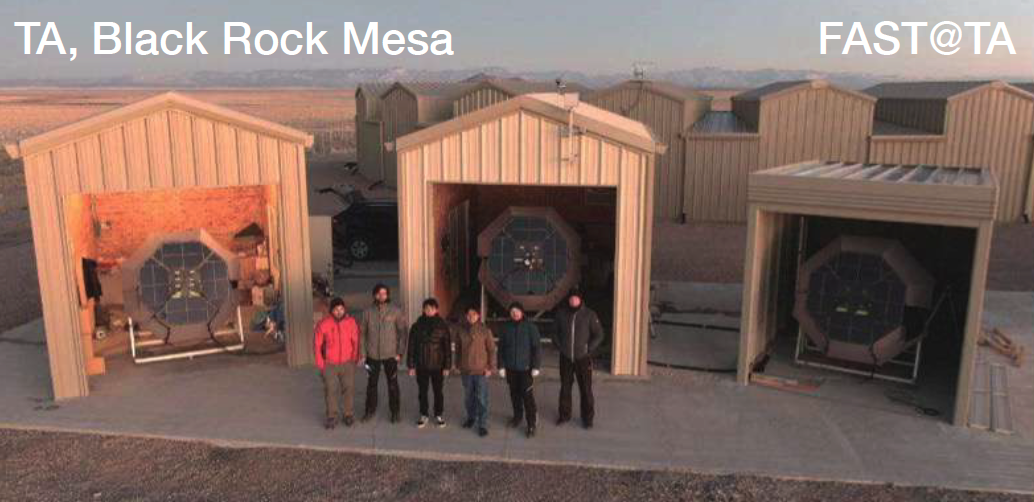}
        \caption{}
        \label{fig:FAST-TAFrontOn}
    \end{subfigure}
    \begin{subfigure}[b]{0.4\textwidth}
        \includegraphics[width=\textwidth]{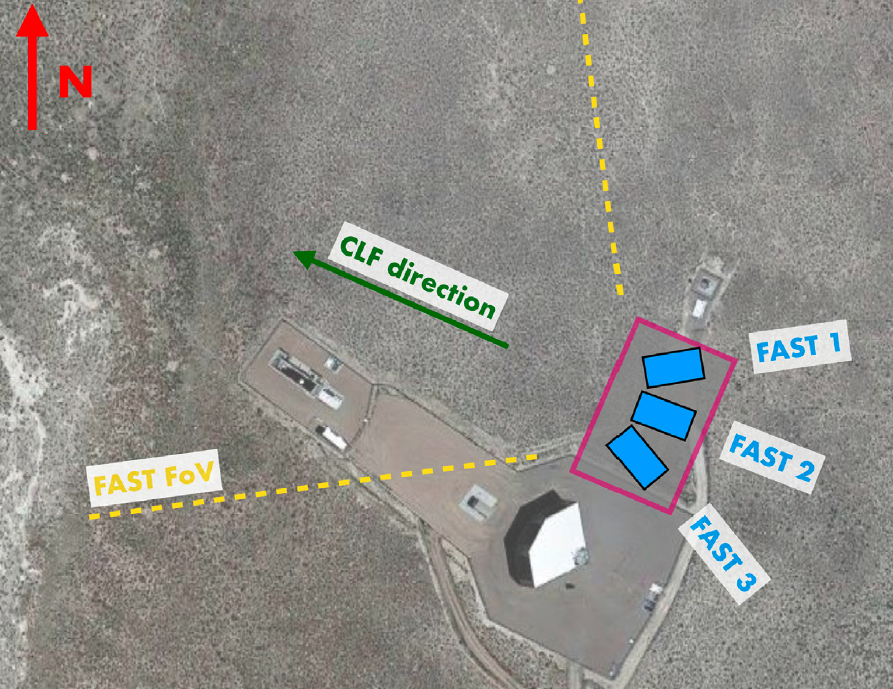}
        \caption{}
        \label{fig:FAST-TAOverhead}
    \end{subfigure}
    \begin{subfigure}[b]{0.59\textwidth}
        \includegraphics[width=\textwidth]{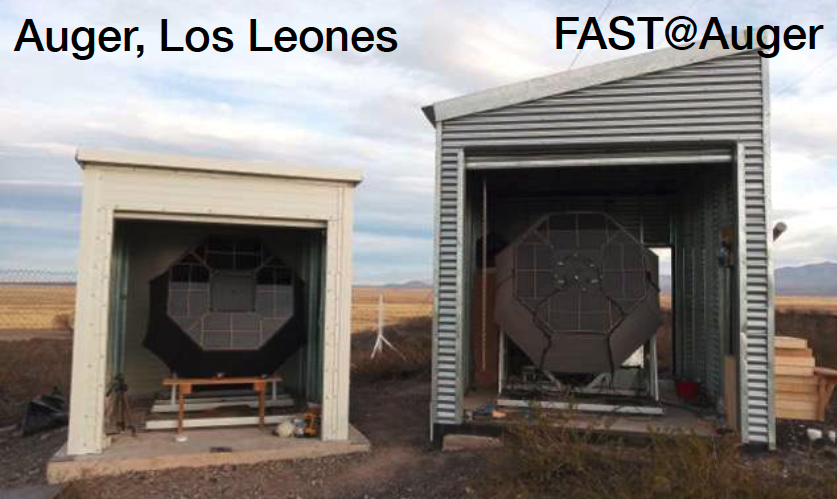}
        \caption{}
        \label{fig:FAST-AugerFrontOn}
    \end{subfigure}
    \begin{subfigure}[b]{0.4\textwidth}
        \includegraphics[width=\textwidth]{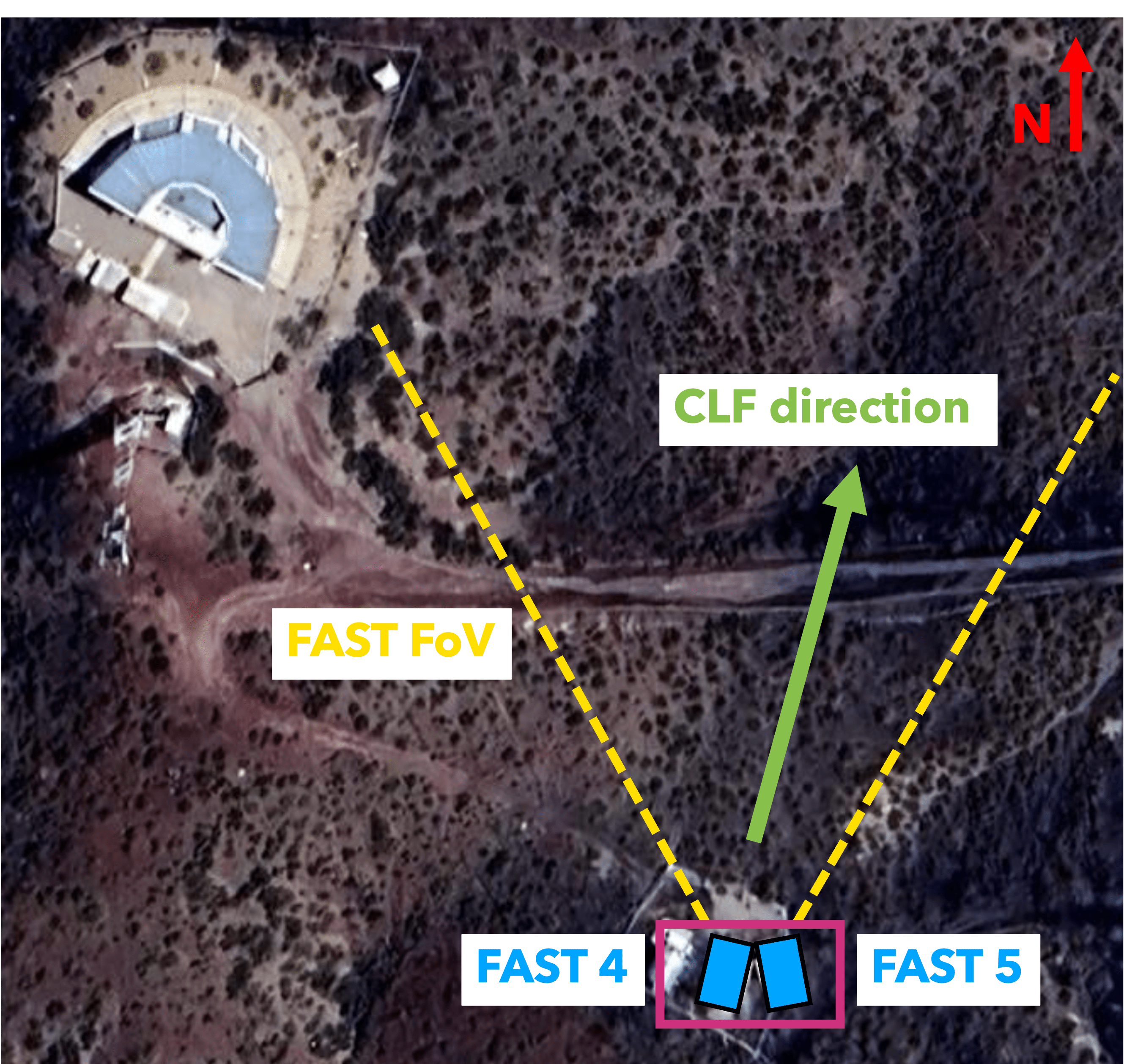}
        \caption{}
        \label{fig:FAST-AugerOverhead}
    \end{subfigure}
    \caption{\textbf{(a)} FAST prototype telescopes at BRM (FAST@TA). From \cite{fujii2023recent}. \textbf{(b)} Overhead view of the FAST@TA telescopes, showing the collective field of view and telescope labelling. From \cite{malacari2020first}. \textbf{(c)} FAST prototype telescopes at LL (FAST@Auger).  From \cite{fujii2023recent}. \textbf{(d)} Same as (b) but for FAST@Auger.}
    \label{fig:FAST-TA&PAOPrototypes}
\end{figure}

In October 2016, the first full-scale FAST prototype telescope was installed at the BRM site of TA. Two additional telescopes followed in September 2017 and October 2018. Combined, the three telescopes cover $30\degree$ in elevation and 90$\degree$ in azimuth, all of which is contained in the FOV of the TA fluorescence telescopes at BRM. In April 2019 and June 2022, prototypes of the same design were deployed at the \gls{ll} site of Auger, overlooking the FOV of two of the Auger telescopes. Figure \ref{fig:FAST-TA&PAOPrototypes} shows front on and overhead views of each set of telescopes, along with the naming/telescope labelling conventions used. The three prototypes at TA are labelled ``FAST 1/2/3" and are collectively referred to as ``FAST@TA". ``FAST 4/5" refer to the two prototypes at Auger, together forming ``FAST@Auger". Due to missing the required electronics FAST 5 has yet to take data, and so ``FAST@Auger" will refer to just FAST 4. These naming conventions will be used throughout this thesis. 

\subsection{Telescope Design}
\begin{figure}[t!]
    \centering
    \includegraphics[width=0.45\textwidth]{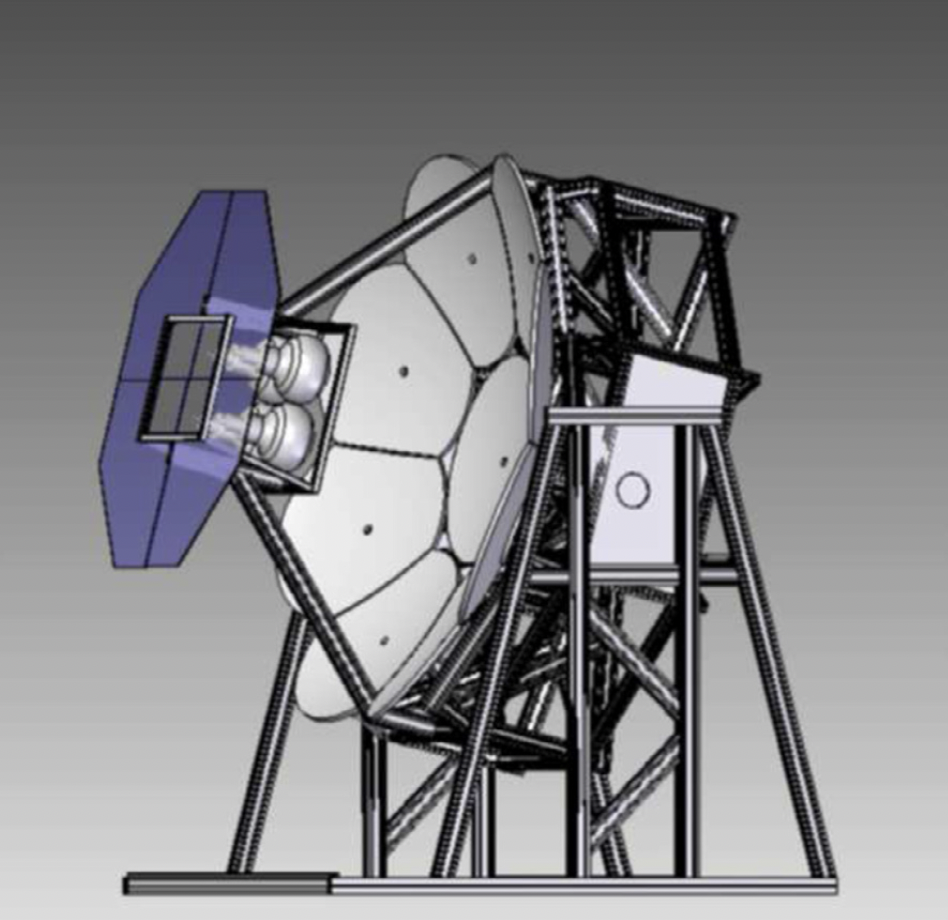}
    \includegraphics[width=0.45\textwidth]{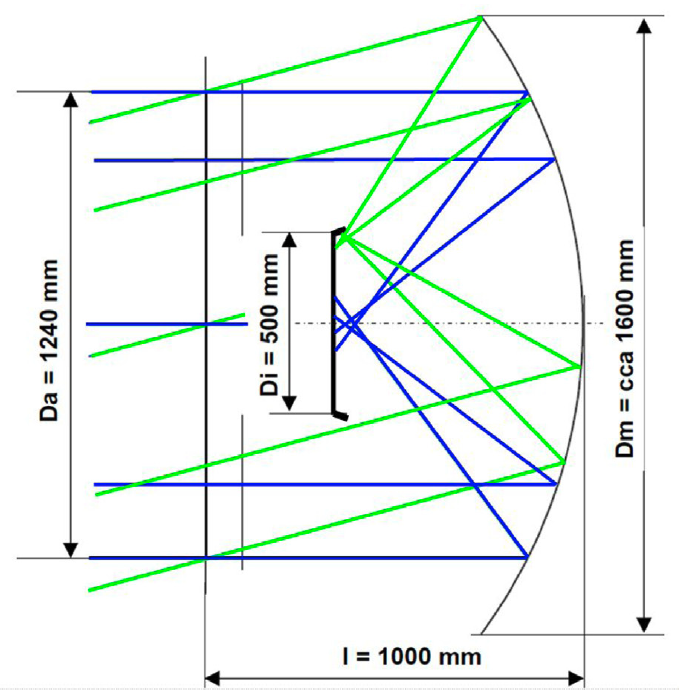}
    \caption{\textit{Left:} Schematic design of the first full-scale FAST prototype with a 9-segment spherical mirror and four photomultiplier tubes. \textit{Right:} Dimensions of the prototype mirror/aperture/camera box system. $D_a$, $D_m$ and $D_i$ represent the diameters of the aperture, mirror and camera box respectively, with $l$ the mirror-aperture distance. From \cite{malacari2020first}.}
    \label{fig:FAST9MirrorPrototype}
\end{figure}

The first full-scale FAST prototype telescopes follow a lensless Schmidt type optical design, utilizing a 1.6\,m diameter spherical mirror focusing light onto a camera box containing four 200\,mm PMTs arranged in a $2\times2$ grid. An octagonal aperture of height 1.24\,m is located 1\,m from the centre of the mirror. After accounting for the camera shadow, the telescope has an effective area of 1\,m$^2$ and a $30\degree\times30\degree$ FOV, meeting the original design goals \cite{malacari2020first}. Figure \ref{fig:FAST9MirrorPrototype} shows a schematic of the telescope and a side on view with dimensions annotated. As shown in the schematic, the spherical mirror is made up of nine segments, one central circular mirror and eight side mirrors or ``petals". Whilst a single mirror would have been ideal, the segmented design was chosen due to being technologically simpler and less expensive to produce \cite{mandat2017prototype}. The mirrors are produced at the Joint Laboratory of Optics of Palacky University and the Institute of Physics of the Academy of Sciences of the Czech Republic using a borosilicate glass substrate. Vacuum coated layers of Al and SiO$_2$ make up the reflective surface. The reflectivity of the mirror over the fluorescence wavelength range is shown in Figure \ref{fig:FASTPrototypeSpectralEfficiency} together with the transmittance of the UV filter and overall spectral efficiency. 
The filter blocks photons with wavelengths $>400$\,nm. It also serves to protect against dust and aerosols \cite{malacari2020first}.

\vspace{5mm}

The support structure of the telescope, which can be seen in Figure \ref{fig:FAST9MirrorPrototype}, is made of commercially available, light weight aluminium profiles. 
The mirror is mounted on a stand which can be adjusted to discrete elevation angles ($0\degree$, $15\degree$, $30\degree$ and $45\degree$ above the horizon) and contains mounts for each of the nine mirrors, each with two degrees of freedom. This allows each segment to be accurately aligned. The camera box and filter are mounted on a support structure connected to the perimeter of the mirror. Four flat side mirrors are attached to the edges of the camera box to reflect light that would be lost back into the PMTs. A cover is then wrapped around the entire support structure/mirror and attached to the edges of the filter.
This prevents light from outside the telescope FOV entering the optical system whilst also protecting it from the surrounding environment \cite{malacari2020first}. The telescopes are housed in individual huts designed to withstand the harsh desert conditions and protect the telescope during daylight hours. Each hut is equipped with a shutter which can be opened and closed remotely.

\subsection{Optical Performance}
\label{sec:OpticalPerformance}
A standard measure for the quality of an optical system is the \gls{psf}. This function describes how light from a point source is blurred after passing through an optical system. For the finely grained cameras used by Auger and TA, the size or angular spread of the PSF is typically constrained to be roughly the same as the FOV of a single pixel $\approx1-1.5\degree$ \cite{tokuno2012new,abraham2010fluorescence}. This prevents light from a ``single-point" in the shower being spread over multiple pixels, which would negatively impact the reconstruction resolution. 
For FAST however, the coarse $2\times2$ layout and 200\,mm diameter of the PMTs allows for more leniency in the size and shape of the PSF. 

\vspace{5mm}

Figure \ref{fig:FASTPrototypePSF} shows a comparison of the PSF of the FAST optical system as calculated from ray tracing simulations (top) and as measured in-situ at FAST 2 (bottom). Note the scale of the axes matches the dimensions of the PMTs. The ray tracing simulation was performed with the Zemax software package\footnote{https://www.ansys.com/products/optics/ansys-zemax-opticstudio} using collimated beams incident on the aperture at on-axis, $7\degree$ and $11\degree$ angles. In line with the prototype design, the image plane in the simulations has been moved 25\,mm closer to the mirror (relative to the focal surface). This is a design choice made to ensure that light focused towards the dead-space between all four PMTs from on-axis beams is not completely lost. A Tyvec diffusing material has been applied to the camera box surface in this space to further reduce lost signal. The on-site measurements were performed by imaging a point-like light source placed $\sim150$\,m from the telescope onto a flat screen in front of the camera box. The overall size and shape of the PSF agrees well between simulations and measurements, verifying the optical system's performance. The \say{star} shape of the PSF is caused by the octagonal aperture, whilst the additional structure seen in the measured PSF is the result of a chain link fence located between the telescope and light source \cite{malacari2020first}.

\begin{figure}[t]
    \centering
    \begin{subfigure}[b]{0.55\textwidth}
        \includegraphics[width=\textwidth]{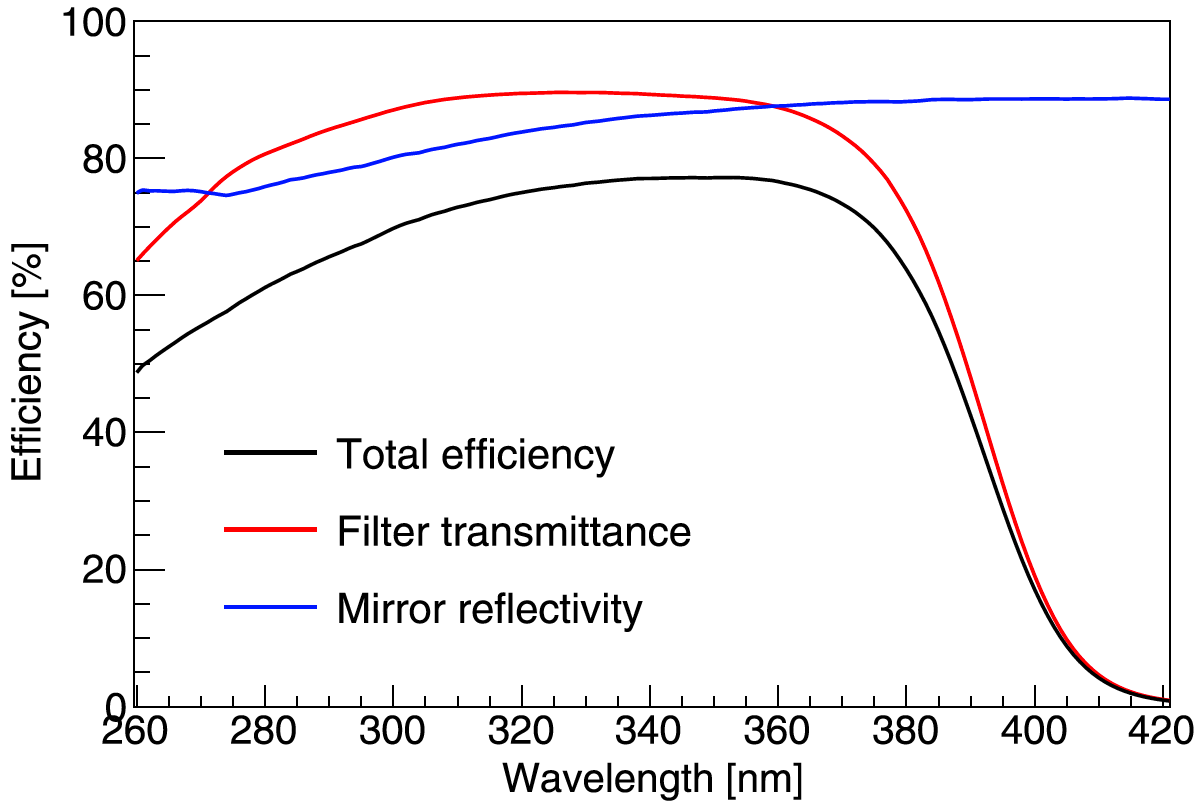}
        \caption{}
        \label{fig:FASTPrototypeSpectralEfficiency}
    \end{subfigure}
    \begin{subfigure}[b]{0.44\textwidth}
        \includegraphics[width=\textwidth]{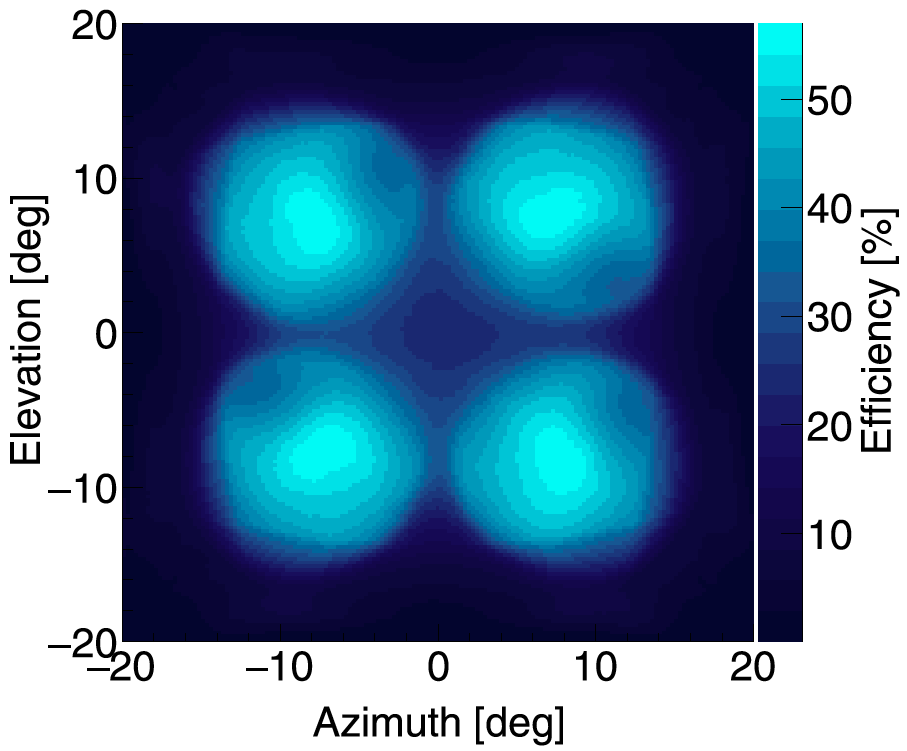}
        \caption{}
        \label{fig:FASTPrototypeRayTrace}
    \end{subfigure}
    \caption{\textbf{(a)} Spectral efficiency of the mirror, filter and combined optical system for the first generation FAST prototypes. \textbf{(b)} Wavelength-independent directional efficiency of the prototype telescope. From \cite{malacari2020first}.}
\end{figure}

\begin{figure}[t!]
    \centering
    \includegraphics[width=\textwidth]{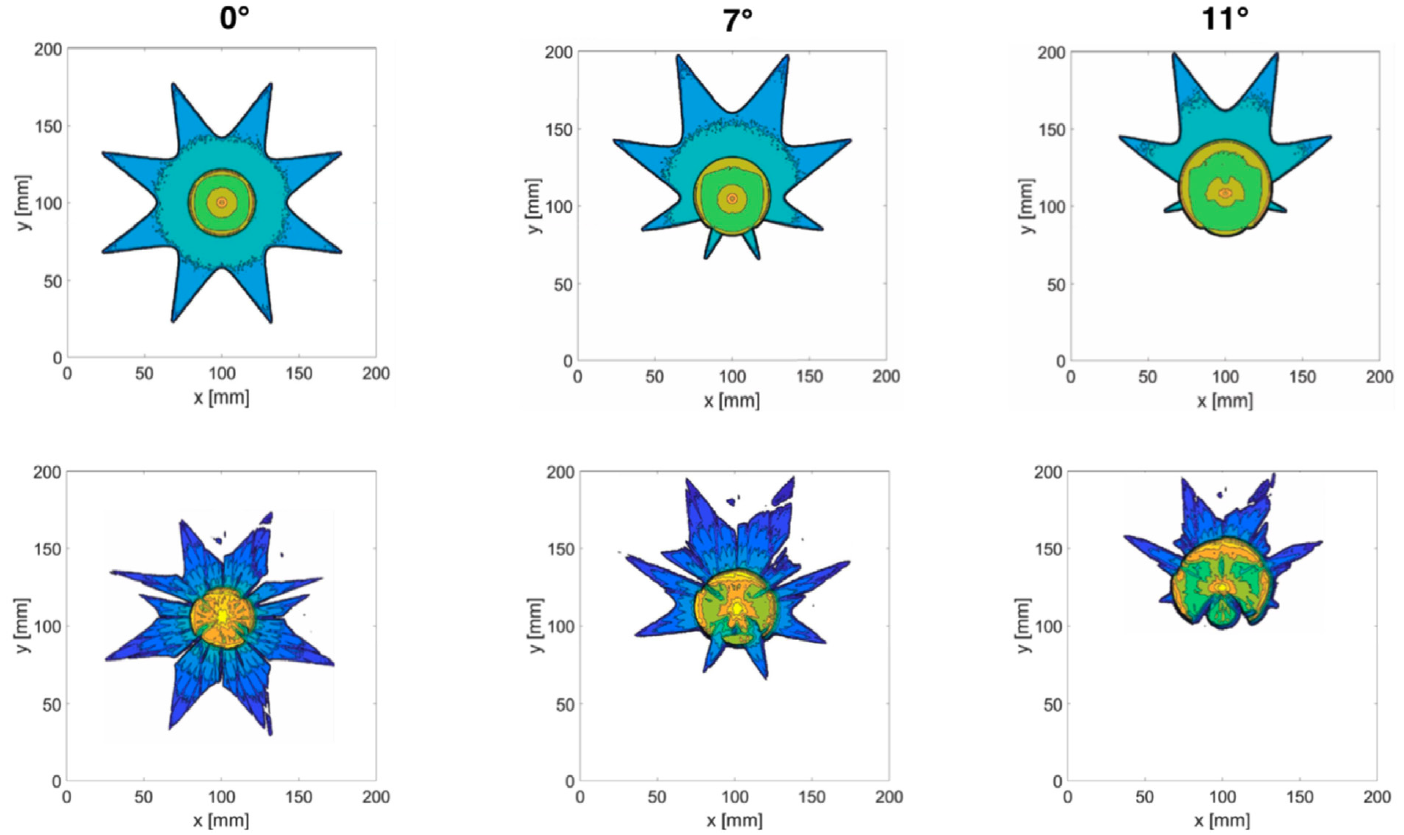}
    \caption{PSF of the FAST telescopes. \textit{Top:} Simulation results using dedicated ray tracing software. \textit{Bottom:} In-situ measurements performed at FAST 2. From \cite{malacari2020first}.}
    \label{fig:FASTPrototypePSF}
\end{figure}

\vspace{5mm}

The directional efficiency of the prototype design, i.e. how the optical system responds to light incident on the telescope from different angles, has also been measured, in this case with a full, wavelength-independent ray tracing simulation of the telescope. The results are shown in Figure \ref{fig:FASTPrototypeRayTrace}. For a given elevation and azimuthal angle, measured with respect to the mirrors optical axis, a fixed number of parallel light rays were set incident on the telescope aperture and traced through the optical system. The total number of photons collected by the PMTs divided by the original number of incident photons gave the collection efficiency for that particular combination of elevation and azimuth. The simulation includes all the elements of the telescope previously introduced, including the segmented spherical mirror, telescope support structure, filter, camera box with side mirrors and diffusing material, and the four PMTs \cite{malacari2020first}. Fresnel losses which occur at the air-glass interface of the PMT surfaces and the spatially dependent collection efficiency of the PMTs, measured at Chiba University \cite{abbasi2010139}, are also accounted for. The spatial dependence in particular gives rise to the \say{cold spots} (regions of lower efficiency) on the edges of each circular region seen in Figure \ref{fig:FASTPrototypeRayTrace}. Further information on the spatial dependence including updated in-situ measurements can be found in Section \ref{sec:recBias}.

\subsection{Electronics and Data Acquisition}

\begin{figure}
    \centering
    \includegraphics[width=0.75\textwidth]{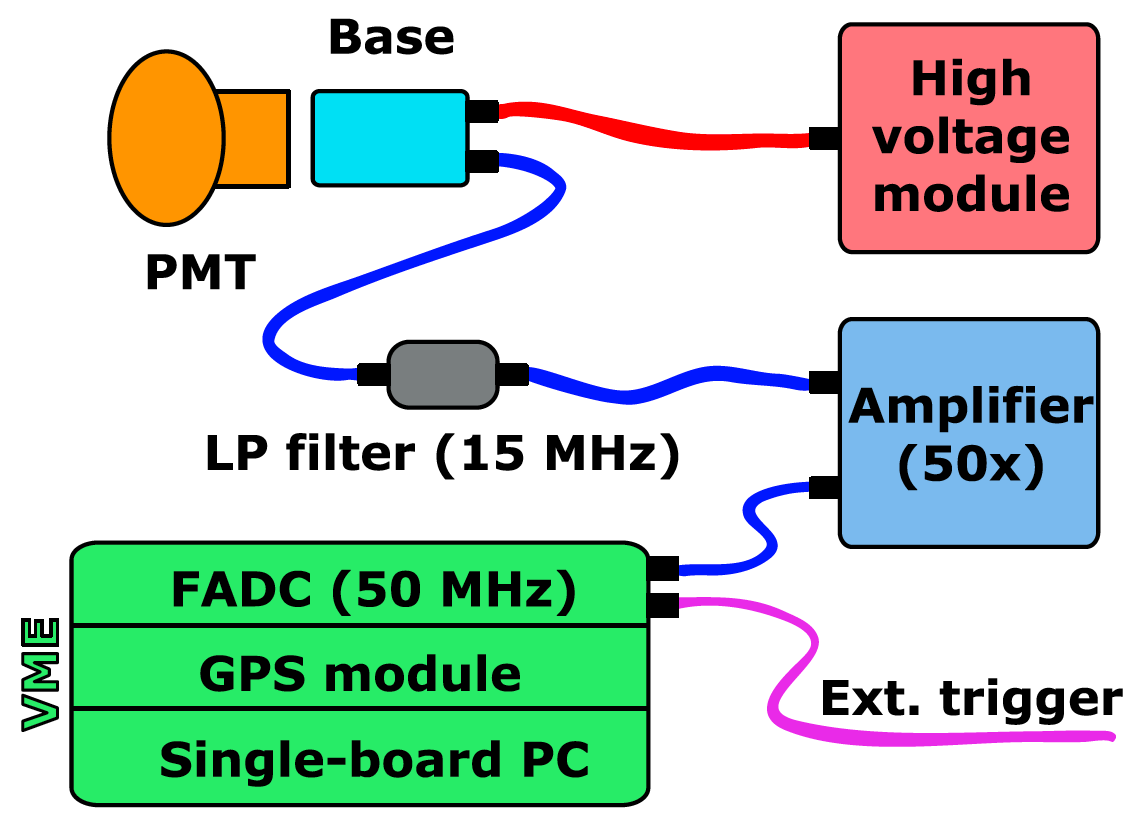}
    \caption{Schematic of the electronics chain for a single FAST PMT. From \cite{malacari2020first}.}
    \label{fig:FASTPrototypeElectronics}
\end{figure}

To convert the fluorescence light from air showers to a digital signal which can be analysed, the first generation FAST prototypes utilise the electronics chain shown in Figure \ref{fig:FASTPrototypeElectronics}. The telescope camera contains four R5912-03 Hamamatsu PMTs, each connected to an E7694-01 Hamamatsu base supplied by a high voltage module. 
Low frequency noise is removed from the PMT signals by a 15\,MHz low-pass filter. The signals are then amplified by a factor of 50 before being digitised by a 50\,MHz FADC (Flash Analogue to Digital Converter). The FADC is housed in a VME crate alongside a GPS module for recording time-stamps and a single-board PC which runs the data-acquisition (DAQ) software. Details on the model components can be found in \cite{malacari2020first}. The FAST@TA and FAST@Auger installations derive their power and internet connections from the existing infrastructure at TA and Auger respectively. 

\vspace{5mm}

During standard operation, the FAST telescopes receive external triggers (signals to record data) from the FD of their associated companion experiment. For FAST@TA, this means receiving a signal from the BRM FD station (sent via a BNC cable connected between the two buildings) whenever 5 adjacent pixels detect signals above a certain threshold within a 12.8\,\textmu{}s window \cite{malacari2020first}. For FAST@Auger, an external trigger is formed when bay 4 of the Los Leones FD passes the Auger second level trigger condition, which similarly searches for adjacent pixels passing a dynamically set threshold \cite{abraham2010fluorescence}. Both FAST@TA and FAST@Auger have the capacity to trigger independently, however the current trigger algorithm, which simply checks for the number of FADC counts to be above a set threshold, is not sophisticated enough to efficiently select air-shower-candidate events. 

\subsection{Air Shower Measurements}
\begin{figure}
    \centering
    \includegraphics[width=\textwidth]{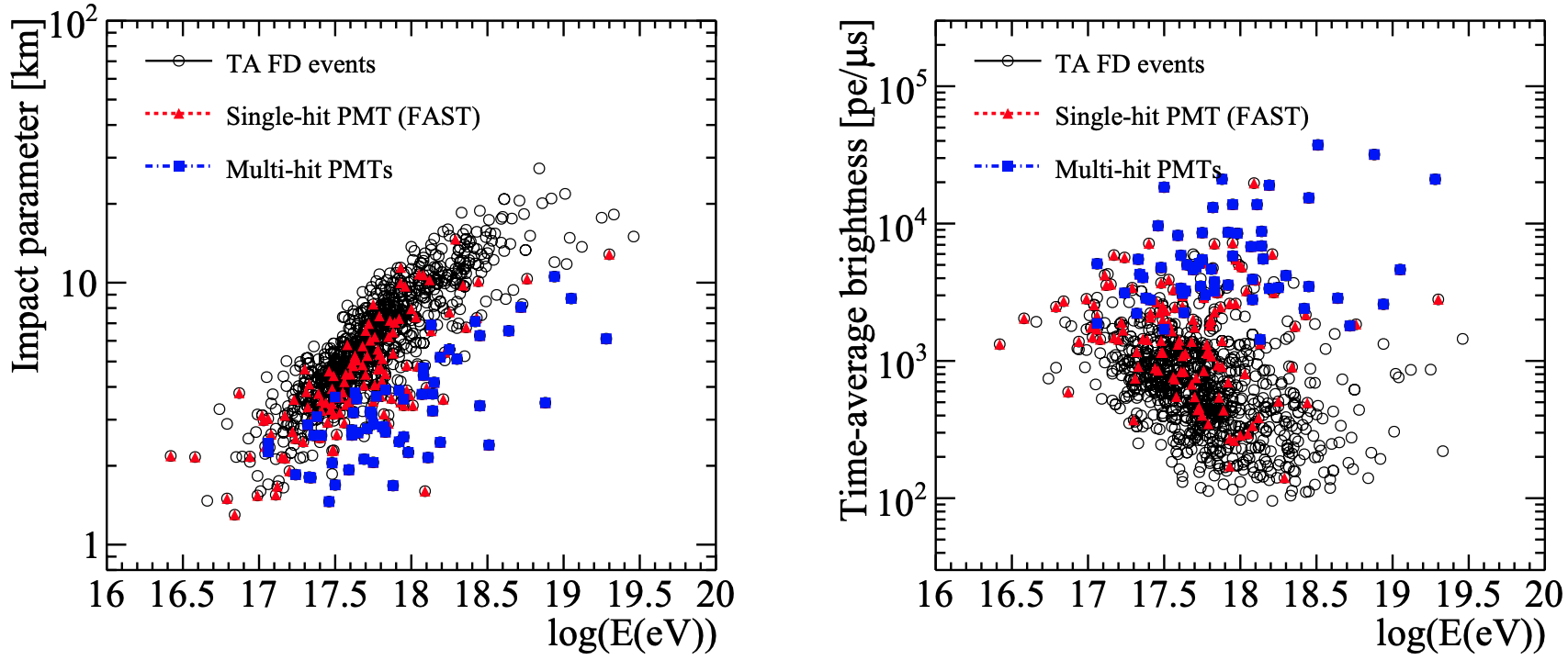}
    \caption{\textit{Left:} Impact parameter as a function of energy for the single-hit PMT (red) and multi-hit PMT (blue) events found in the signal search from \cite{fujii2021latest}. The open circles show the candidate TA events. \textit{Right:} Same as the left plot but for the time-average brightness. From \cite{fujii2021latest}.}
    \label{fig:FAST-TAEventSearch}
\end{figure}

\begin{figure}
    \centering
    \includegraphics[width=\textwidth]{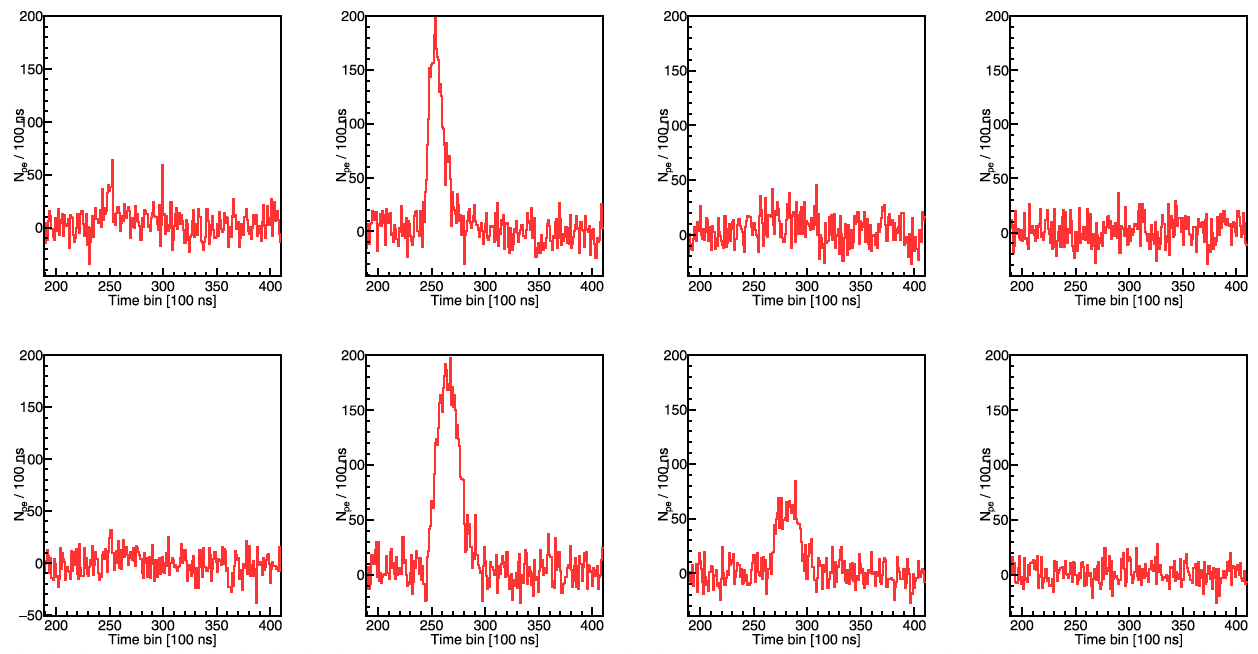}
    \caption{PMT traces of the 19\,EeV event recorded by FAST 1 and 2 on May 15$^{\textrm{th}}$ 2018. From \cite{justin2020extending}.}
    \label{fig:FAST-TAExampleEvent}
\end{figure}

Since 2016, the first generation of FAST prototypes have collected roughly 4000\,hrs of data; $\sim750$\,hr from FAST@TA and $\sim3250$\,hr from FAST@Auger. Despite its later installation, FAST@Auger has more than four times the number of observation hours due to a dedicated shift schedule and being operational during the COVID-19 pandemic. In \cite{fujii2021latest}, a search for air-shower events in the FAST@TA data set constituting approximately 224\,hrs of data found 179 significant events, with 59 events having more than one PMT with significant signal. The impact parameter $R_\textrm{p}$ (distance of closest approach of the shower) and time-average brightness of these events, as measured by TA, are shown as functions of energy in Figure \ref{fig:FAST-TAEventSearch}. For a given energy, there is expected to be a maximum distance up to which FAST can detect showers. By inspection of the $R_\textrm{p}$/energy plot, it appears that FAST may be able to detect showers up to $\sim20$\,km at 30\,EeV. The PMT traces for the highest energy shower amongst the detected events are shown in Figure \ref{fig:FAST-TAExampleEvent}. The $y$-axis shows the number of \gls{pe} per 100\,ns. The event had a TA reconstructed energy of $19$\,EeV, \Xmax{}$\approx850$\,g\,cm$^{-2}$ and $\theta\approx55\degree$. These signal search findings are summarised here to give context to the new results presented in Chapter \ref{ch:REAL}.

\vspace{5mm}

In total, the cost for the first-generation prototypes came to $\sim25$\,k\,USD/telescope. Although the newer and cheaper second-generation of prototypes will soon be deployed, all FAST data collected thus far has been with the first-generation prototypes. Furthermore, these prototypes are expected to continue operation for several years to come. Therefore, an accurate understanding of their capabilities is critical for reconstruction/analysis and evaluating the performance of FAST thus far. 


\section{Second Generation Prototypes}
\label{sec:FASTFieldTelescope}
\begin{figure}[th]
    \centering
    \includegraphics[width=0.32\textwidth]{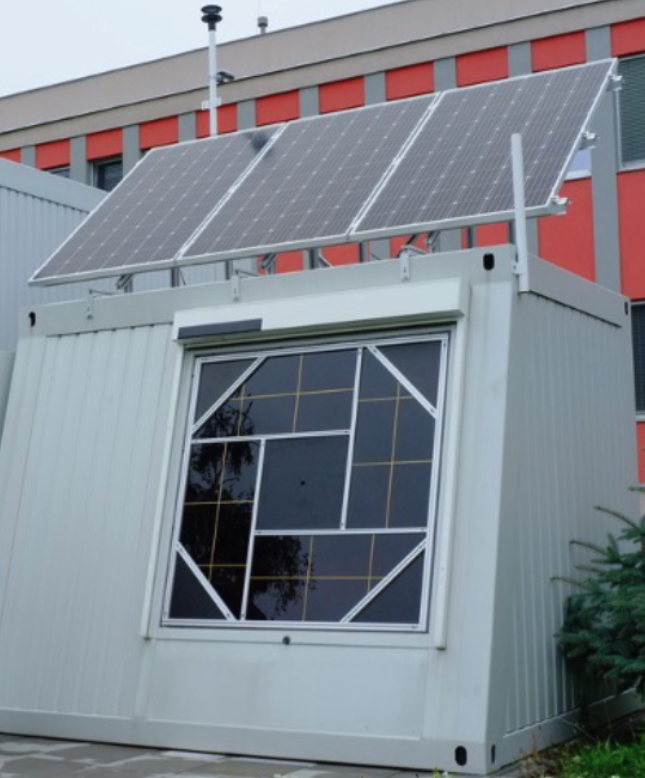}
    \includegraphics[width=0.32\textwidth]{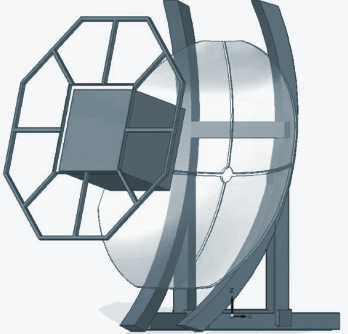}
    \includegraphics[width=0.32\textwidth]{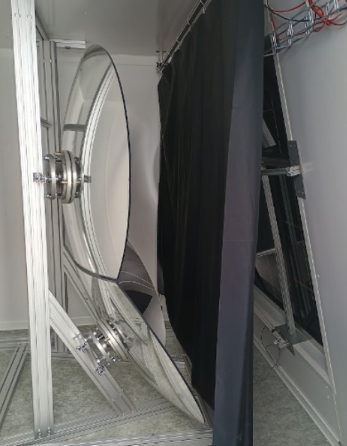}
    \caption{\textit{Left:} Outside view of the FAST-Field telescope enclosure. The housing for the rolling shutter is visible above the filter. \textit{Middle:} Schematic of the new telescope design with just four mirror segments. \textit{Right:} Photo of the telescope inside the enclosure with fail-safe curtain. From \cite{Hamal:2024pC}.}
    \label{fig:FASTGen2}
\end{figure}

Two second-generation FAST prototypes are scheduled to be installed at Auger in late 2025. These prototypes are cheaper, easier to manufacture and more autonomous than the first generation telescopes, all of which are crucial for the large-scale deployment of FAST. As they are designed to deployed \say{in the field}, without the need for existing infrastructure, these telescopes have been labelled \say{FAST-Field telescopes}. Figure \ref{fig:FASTGen2} shows photos of the updated design. Key differences from the first full-scale prototypes are the enclosure, telescope design and electronics.

\vspace{5mm} 

The enclosures for the second generation telescopes are far more compact then the huts used at FAST@TA, with two enclosures able to fit inside a standard size shipping container. They are autonomous in that the required power for the electronics systems is provided by a set of solar panels and battery. Sensors for wind, rain and atmospheric monitoring are installed on the roof of the enclosure. The telescope aperture and UV filter have been integrated into the side of the enclosure. This means the telescope and electronic components are never exposed to the outside conditions. A rolling shutter is installed above the filter which is lowered for protection during daylight hours. The enclosure is also water/moisture proof, thermally insulated, and contains a safety curtain located between the mirror and camera box. This acts as a fail safe in the event of a power failure where the shutter may not be able to be closed. 

\vspace{5mm}

As for the telescope itself, the new design consists of a mirror with just four segments, each produced through the \say{hot slumping} technique. The polishing and grinding processes previously applied to the mirrors are now skipped as the considerable time and labour costs they add to the manufacturing process were found to only marginally improve the optical systems response. The mirrors are now supported by a much simpler support structure. Finally, the new electronics are designed to support an entire station of FAST telescopes and consume less power. 

\section{Calibration and Atmospheric Monitoring}

Accurate reconstruction of an EAS's properties requires each component of the fluorescence detection setup to be properly calibrated and maintained. Degradation of the optical components, gain variance of the PMTs and changes in the atmosphere and night-sky background (NSB) can all impact the amount of signal detected.

\subsection{PMT Calibration}

To accurately estimate the number of photons incident on the detector, and hence the properties of detected air showers, each PMT must be correctly calibrated. The absolute calibration of the FAST PMTs is performed at the University of Chicago. This involves measuring the detection efficiency, differential linearity and gain as a function of high-voltage for each PMT \cite{malacari2020first}. Once deployed, changes in the gain of a subset of the PMTs due to variations in temperature are monitored using a \gls{yap} pulser. This pulser emits UV photons with a peak wavelength of 370\,nm and a 20\,ns FWHM pulse width at a rate of $\sim50$\,Hz. During data taking and at the beginning and end of each run, measurements of the signal received from the YAP pulser are taken using a high-threshold trigger. By integrating the YAP signal and comparing the results with temperature measurements, the temperature dependence of the PMT gain can be determined. Using measurements taken at FAST@TA over a one year period, 
where the telescopes were exposed to the surrounding environmental conditions during observation, the gain of the PMTs was found to change by roughly $-0.411\pm0.001\%/\degree$C. All temperature and YAP pulser measurements are stored in an SQL database for later use in analysis \cite{malacari2020first}.

\subsection{Mirror and Filter Cleaning}
\label{sec:cleaning}
Exposure to the environment, particularly the build up of dust and other particulates, can affect the quality of the optical components of the FAST telescope. The telescope mirror and camera box (PMT surfaces and side mirrors) largely avoid this problem, thanks to the protection afforded by the cover of the first-generation prototype design and enclosure of the second-generation prototype design. However in both cases the UV filter is exposed to the surrounding desert environment during data-taking and is tilted at an angle of 15$\degree$, making it liable to collect contaminants on its surface. For a future, large-scale FAST array, regular cleaning of the telescope components will be impractical, so it is important to understand how the components degrade over time and how this effects the feasibility of such an array. Measurements of the UV filter transmittance and mirror reflectivity for FAST 1 (two years after deployment) and FAST 2 (one year after deployment) were made in October 2018. The reduction in mirror reflectivity was found to be negligible whilst the UV filter transmittance had decreased by 5.5\% and 8.5\% after one and two years respectively \cite{malacari2020first}. Further measurements over a long-term period will be necessary to properly account for the reduction in transmittance in the reconstruction.

\subsection{Atmospheric Monitoring}
The fluorescence detection technique relies on using the atmosphere as a giant calorimeter to measure extensive air showers. This makes changes in the atmospheric conditions one of the largest sources of systematic uncertainty in FD measurements. Understanding and accounting for these changes, which occur over seasonal, nightly and even hourly timescales, is critical for the accurate reconstruction of the shower energy and \Xmax{}. The primary observables of interest are the degree of cloud cover, background light and the molecular and aerosol profiles of the atmosphere.

\vspace{5mm}

\begin{figure}[t]
    \centering
    \includegraphics[width=0.49\textwidth]{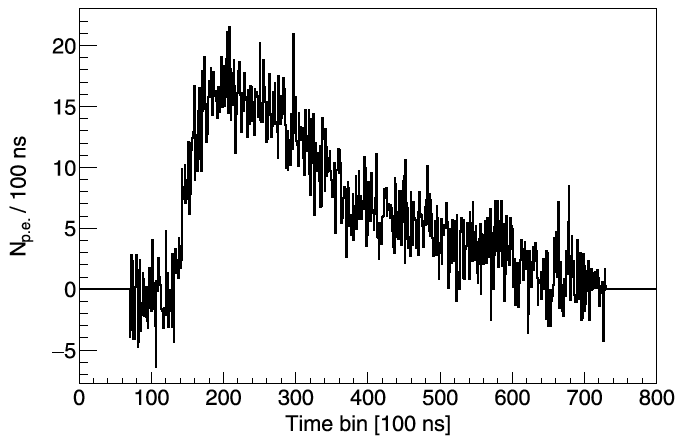}
    \includegraphics[width=0.49\textwidth]{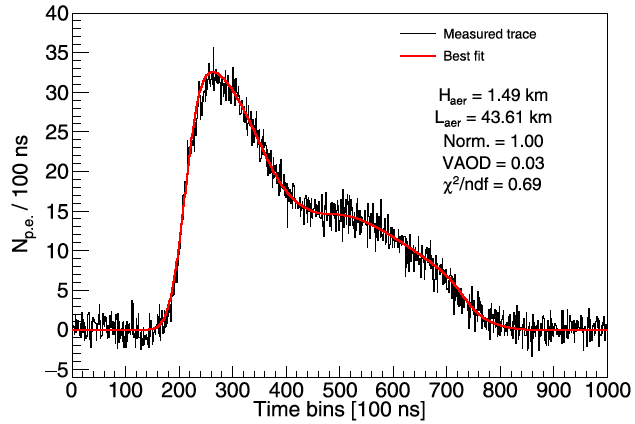}
    \caption{\textit{Left:} Measurement of the TA CLF signal by FAST 2 averaged over 200 shots. \textit{Right:} Estimation of the aerosol atmosphere properties by comparing (simulated) measurements to simulations of CLF laser shots. Details in the text. From \cite{malacari2020first}.}
    \label{fig:FASTAtmosStuff}
\end{figure}

To measure these properties, FAST employs several atmospheric monitoring tools. The FD buildings at TA and Auger possess weather stations used to check for high wind speeds and rain. Both observatories are also equipped with a \gls{clf} which fires a UV laser of known wavelength and energy at regular intervals vertically into the atmosphere \cite{fick2006central, takahashi2011central}. The TA CLF is located 21\,km from FAST@TA, whilst the Auger CLF is located 26\,km away from FAST@Auger. FAST has been given access to the data from these monitoring tools. Measurements of these laser pulses can be used for calibration and to monitor the cloud cover/aerosol properties of the atmosphere. An example of a CLF trace (averaged over 200 shots) measured by FAST 2 is shown on the left in Figure \ref{fig:FASTAtmosStuff}.

\begin{figure}[]
    \centering
    \includegraphics[width=0.49\textwidth]{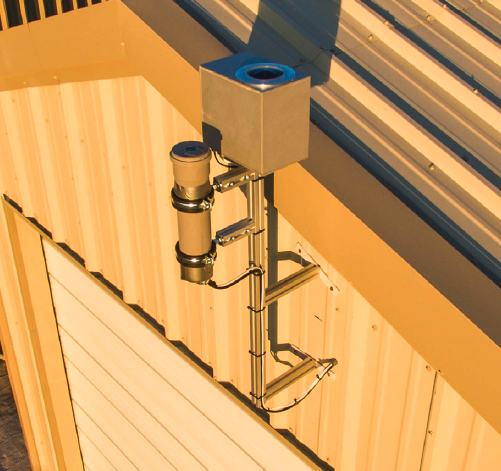}
    \includegraphics[width=0.49\textwidth]{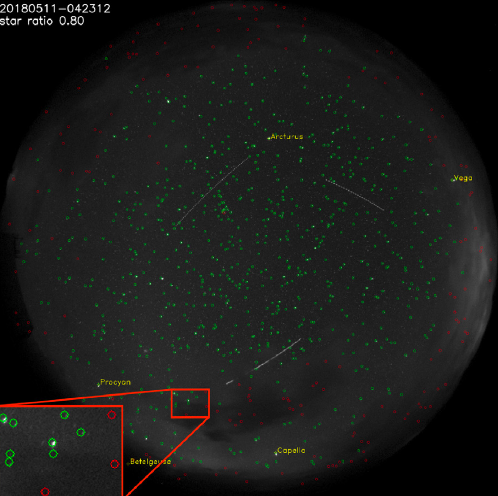}
    \caption{\textit{Left:} The FASCam and SQM attached to the top of the FAST 2 hut. \textit{Right:} Example image from FASCam. From \cite{malacari2020first}.}
    \label{fig:FASCam}
\end{figure}

\vspace{5mm}

In principle, comparing such CLF measurements to simulations can allow for the aerosol content of the atmosphere to be estimated. The right plot in Figure \ref{fig:FASTAtmosStuff} shows an example of the technique using purely simulated data. The expected signal at FAST 2 (including background noise) from a TA CLF laser shot passing through an atmosphere with a horizontal attenuation length of 40\,km and scale height of 1.5\,km is shown in black. Taking the horizontal attenuation length and scale height as free parameters, simulated laser shots free of background noise are then compared to the data with a $\chi^2$ fit. The best fit result is overlaid in red, with parameters given on the right. Efforts to apply this technique to measured data are ongoing. Further details on the simulation of laser shots can be found in \cite{malacari2020first}. 

\vspace{5mm}

Two additional, purpose-built sensors are used by FAST to monitor the atmosphere, the FAST All-Sky Camera (FASCam) and the Sky-Quality Monitor (SQM). FASCam estimates the cloud coverage by comparing the positions of known stars in the Tycho-2 catalogue to those able to be seen on any given night. A clear night is defined as an observable star fraction $\geq$80\%. The SQM measures the night-sky brightness in magnitudes per square arc second. Figure \ref{fig:FASCam} shows the FASCam and SQM attached to the hut of FAST 2 and an example image from FASCam. All calibration and atmospheric monitoring data is stored in a database which can be queried during reconstruction \cite{chytka2020automated}.

\section{Simulating EASs with FAST}

An accurate simulation of the FAST telescope's response to extensive air showers is integral not only to expected performance studies and analysis based calculations (detector exposure etc.) but also to the FAST reconstruction process (see Section \ref{sec:FASTreconstruction}). The simulation software currently used by FAST, \say{FAST-sim}, is based on an old version of the \Offline{} software framework \cite{argiro2007} used by the Pierre Auger Collaboration. The basic functioning of \Offline{} and FAST-sim are detailed in this section.

\subsection{The \texorpdfstring{\Offline}{}Software Framework}
\label{sec:Offline}
\Offline{} is a software framework developed by the Pierre Auger Collaboration for the simulation and reconstruction of extensive air showers. Written in C++, the framework is object orientated in its design and consists of three primary components.
\begin{itemize}
    \item \textbf{Processing modules:} A collection of self-contained processing-steps or algorithms. By combining modules in various sequences, a wide variety of tasks can be accomplished. This design allows users to easily share and compare different algorithms. The sequencing and configuration of the modules is specified using an XML-based language.
    \item \textbf{Event data:} A set of classes containing all raw, calibrated, reconstructed and simulated data of an event. Information is shared between modules by reading and writing to the event data.
    \item \textbf{Detector description:} Data relating to the detector configuration, performance and atmospheric conditions at any point in time. This data is read only.
\end{itemize}
Supporting the above components are classes and utilities for error-logging, handling of units, geometrical transformations etc. \cite{argiro2007}.

\subsection{FAST-sim}
\label{sec:FASTSimulation}
In 2020, the FAST collaboration adopted an old version of \Offline{} to use for simulation and reconstruction purposes with FAST. The modular nature of the framework allowed purpose built FAST modules to be easily written utilizing the \Offline{} infrastructure and classes. These modules were used to build the simulation program for FAST, \say{FAST-sim}. The sequence file (i.e. the set of instructions for what modules to run and in which order) for simulating a single shower with FAST-sim is shown in Figure \ref{fig:FASTSimModuleSeq}.

\begin{figure}[h]
\begin{minted}[
frame=lines,
framesep=2mm,
baselinestretch=1.2,
bgcolor=LightGray,
fontsize=\footnotesize,
linenos
]
{xml}
<!-- A sequence for FAST simulation --> 
<sequenceFile>
  <enableTiming/>
  <moduleControl>
    <loop numTimes="1" pushEventToStack="yes">
      <module> FASTProfileSimulatorCG </module>
      <module> FASTEventGeneratorCG </module>
      <module> ShowerLightSimulatorKG </module>
      <module> FASTSimulator </module>
      <module> FASTEventFileExporter </module>
    </loop>
   </moduleControl>
</sequenceFile>
\end{minted}
\caption{Sequence file showing the order of modules to run to simulate a single shower with the FAST-sim program.}
\label{fig:FASTSimModuleSeq}
\end{figure}

\vspace{5mm}

As with \Offline{}, XML-based configuration files are used to specify the parameters of the simulation. For FAST-sim, this primarily involves specifying the parameters of the simulated shower (i.e. \Xmax{}, energy, zenith, azimuth and core position) and the layout of the FAST telescopes, although many other parameters can also be adjusted. A rough description of each module is provided below. 

\subsubsection{FASTProfileSimulator}
Creates a Gaisser Hillas profile of the form in Equation \ref{eqn:GH}. First, the number of particles in the shower is calculated in bins of atmospheric depth using the user specified \Xmax{}, energy and zenith angle. These values are then multiplied by the average energy deposit per particle at each depth, giving the total energy deposit as a function of atmospheric depth. Note that there is no treatment of the invisible energy in these calculations, i.e. the user specified energy is treated as the total calorimetric energy of the shower. The values of $X_0$ and $\lambda$ are fixed to their average values of $-121$\gcm{} and 61\gcm{} respectively as FAST is not expected to be sensitive enough to reliably reconstruct them. The event timing and azimuth angle of the shower are also configured in this module. 

\subsubsection{FASTEventGenerator}
Defines the coordinate system to be used. The core location and direction of the shower are set based on these coordinates. 

\subsubsection{ShowerLightSimulatorKG}
Calculates the light produced along the shower axis. This light includes fluorescence photons calculated using the AIRFLY fluorescence model and Cherenkov photons calculated from the number of shower electrons above the Cherenkov threshold in air.
Taken directly from \Offline{}.

\subsubsection{FASTSimulator}
Calculates the number of photons (wavelength dependent) reaching each simulated telescope. This includes direct fluorescence photons, direct Cherenkov photons, and scattered Cherenkov photons. Typically, parametric models of the molecular and aerosol atmospheres are used. However, molecular atmospheric profiles of the US Standard Atmosphere and monthly average atmospheres at Malargüe are also available. Photons reaching the telescope aperture are propagated through the FAST optics, accounting for the transmission of the filter, spectral response of the mirror, directional efficiency, and PMT quantum efficiency. 
The signal in each PMT is then calculated in 100\,ns time bins to generate the simulated traces.

\subsubsection{FASTEventFileExporter}
Background noise is added to the simulated traces by sampling from a Gaussian with mean 0 and user specified variance. The results of the simulation are written to an output file in the \verb|FASTEventFile| format. 

\vspace{5mm}

\begin{figure}[t]
    \centering
    \includegraphics[width=1\textwidth]{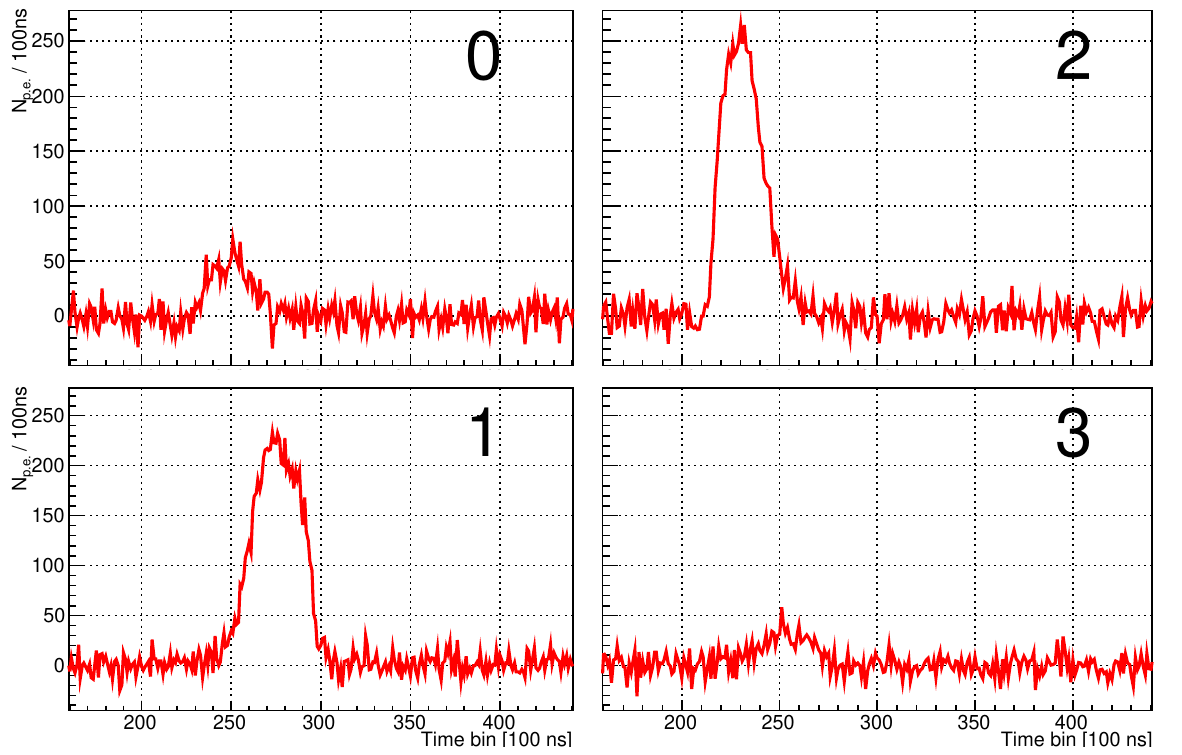}
    \caption{Example of simulated PMT traces for a single telescope with the FAST-sim software. The PMTs are laid out in the sky view orientation. Simulated shower parameters in the text.}
    \label{fig:FASTSimExampleEvent}
\end{figure}

An example of the traces from a simulated event is shown in Figure \ref{fig:FASTSimExampleEvent}. This event was simulated with $E=10^{19}$\,eV, $X_\textrm{max}=900$\gcm{}, $\theta=40\degree$, $\phi=40\degree$, core location of $(x,y)=(-1000,5000)$\,m, and observed by a single FAST telescope located at (0,0) with $\textrm{elevation}=15\degree$ and azimuthal direction $90\degree$. Note the FAST azimuth convention is to measure anti-clockwise from east with $\phi\in(-180\degree,180\degree)$. The numbers in the top right corner of each panel indicate the numbering of the PMTs for a single telescope. The traces are presented in \say{sky-view orientation}, i.e. the position of each PMT trace reflects the region of the telescope FOV seen by that pixel. Thus PMTs 0 and 2 view the upper left and upper right regions of the telescope's FOV, and PMTs 1 and 3 view the lower left and lower right regions. This will be the default layout when showing PMT traces from a telescope.

\section{Reconstruction Methods}
\label{sec:FASTreconstruction}
In Section \ref{sec:FASTEarlyTests} it was demonstrated that if the geometry of the shower is known then FAST can potentially utilise the familiar method of fitting a Gaisser Hillas profile to determine the energy and \Xmax{} of a given shower. However, FAST acting independently cannot achieve a precise geometrical reconstruction with standard shower-track-fitting techniques due to the coarse $2\times2$ PMT layout. To circumvent this, previous FAST Collaboration member Justin Albury, in his PhD thesis \cite{justin2020extending}, developed a top-down approach whereby the PMT traces from data are directly compared with those from simulations using a likelihood function. The reconstructed parameters are those which maximise the likelihood function. This procedure is called the \say{top-down reconstruction (TDR)} and is computationally expensive as it requires running many simulations. Furthermore, Albury showed that the initial parameters or \say{first guess} passed to the optimiser significantly impact the final result. To reduce the required time and provide a reasonable first guess for the TDR, Albury, and later Fujii \cite{fujii2021latest}, investigated machine learning techniques. An overview of both the TDR and machine learning approaches is given here. Further details are provided in later chapters where relevant. Readers are recommend to Albury's thesis for a thorough discussion of the initial implementation and testing of the TDR. 

\subsection{Top-Down Reconstruction}
\label{sec:topdownrec}
The inspiration for applying a top-down approach to event reconstruction with FAST comes from work done by Shussler for Auger \cite{schussler2005top}. Shussler found that this approach yielded improved resolutions for the shower parameters when compared to standard methods, however the computationally intense nature of the procedure meant it was only considered for rare, very-high energy showers. For FAST, this improved resolution is necessary when reconstructing the shower parameters without having the shower geometry provided by an external source e.g. companion SD.

\subsubsection{Likelihood function}
The top-down reconstruction is a maximum likelihood based estimation of the shower parameters. This approach attempts to find the most probable values of the shower parameters $\vec{a}$=($E$, \Xmax{}, $\theta$, $\phi$, core $x$, core $y$) by maximising the likelihood function

\begin{equation}
\label{eqn:TDLikelihood}
    \mathcal{L}(\vec{x}|\vec{a})=\prod_i^NP(x_i|\vec{a}).
\end{equation}
Equation \ref{eqn:TDLikelihood} says that the likelihood $\mathcal{L}$ of observing the measurements $\vec{x}$ given the shower parameters $\vec{a}$ is equal to the product of the probabilities for observing each $x_i$ given $\vec{a}$, where $x_i$ are the signals recorded in each bin of the PMT traces. The parameters $\vec{a}$ which maximise $\mathcal{L}$ for a given set of measurements $\vec{x}$ are the reconstructed shower parameters. In practice, the natural logarithm of Equation \ref{eqn:TDLikelihood} is taken and the negative log-likelihood
\begin{equation}
\label{eqn:TDLogLikelihood}
    -2\ln\mathcal{L}(\vec{x}|\vec{a})=-2\sum_i^N\ln\left(P(x_i|\vec{a})\right)
\end{equation}
minimised instead. This is done for convenience and consistency with $\chi^2$ fitting. Instead of comparing measured data to a pre-simulated library of events and finding the minimum value of $-2\ln\mathcal{L}(\vec{x}|\vec{a})$, the simulation is built into the minimisation procedure. This means that after each evaluation of the likelihood function, a new simulation is performed with an updated set of parameters, allowing for a more efficient search of the parameter space and for time-dependent effects (atmospheric conditions, detector status etc.) to be accounted for.

\subsubsection{Signal probability density function}
The probability for observing a signal $x_i$ in a given time bin depends on the expected signal and background fluctuations in that bin. The background fluctuations in a single time bin follow a Poisson distribution and arise primarily from the night-sky background, with a smaller component coming from noise generated by the electronics. Since the number of background photons is large, the Poisson distribution can be well-approximated by a Gaussian distribution of the same mean and variance. Therefore, the probability of observing a signal $x_i$ in a single time bin given an expected signal $\mu_i$ and variance $\sigma^2$ is
\begin{equation}
\label{eqn:BinProb}
    P(x_i|\mu_i,\sigma)=\frac{1}{\sigma\sqrt{2\pi}}\exp\left[-\frac{(x_i-\mu_i)^2}{2\sigma^2}\right].
\end{equation}
Here, the values of $\mu_i$ for each time bin come from the PMT traces obtained by simulating a shower with parameters $\vec{a}$ and with no background noise. The variance $\sigma^2$ is given by  
\begin{equation}
    \sigma^2=\sigma_{\textrm{bckg}}^2+\mu_i(1+V_g)
\end{equation}
where $\sigma_{\textrm{bckg}}^2$ is the background variance of the signal and $V_g$ is the gain variance of the PMT \cite{justin2020extending}.

\subsubsection{Event likelihood function}
By calculating Equation \ref{eqn:BinProb} for every time bin in every pixel in an event, the \say{event likelihood function} can be defined. For a shower simulated with parameters $\vec{a}$, the event log-likelihood is the sum of the probabilities over all pixels and time bins
\begin{equation}
\label{eqn:eventLikelihood}
    \ln\mathcal{L}(\vec{x}|\vec{a})=\sum_k^{N_{\textrm{pix}}}\sum_i^{N_{\textrm{bins}}}\ln\left(P_k(x_i|\vec{a})\right).
\end{equation}
$P_k(x_i|\vec{a})$ is the probability of measuring $x_i$ photoelectrons in the $i^{\textrm{th}}$ time bin of pixel $k$. Once again in practice the negative log-likelihood is minimised. Minimisation is performed using the MIGRAD method of the Minuit2 minimizer in ROOT \cite{brun1997root}. The uncertainty on a given parameter is the value by which that parameter must change in order to give an increase of $x_c$ to the likelihood function. The value of $x_c$ depends on the number of parameters reconstructed simultaneously. Table \ref{tab:uncertainties} shows these values for the number of reconstructed parameters ranging from $1\sim6$. All uncertainties for the TDR which will be shown are calculated based on this method.

\begin{table}[h]
    \centering
    \begin{tabular}{c c c c c c c}
    \hline
        $n$ & | 1 & 2 & 3 & 4 & 5 & 6 \\
        \hline
        $x_c$ & | 1 & 2.28 & 3.53 & 4.72 & 5.89 & 7.04 \\
    \hline
    \end{tabular}
    \caption{Values $x_c$ by which the negative log-likelihood must be increased by to give the uncertainties on the $n$ reconstructed parameters.}
    \label{tab:uncertainties}
\end{table}

\begin{figure}[t]
    \centering
    \includegraphics[width=1\textwidth]{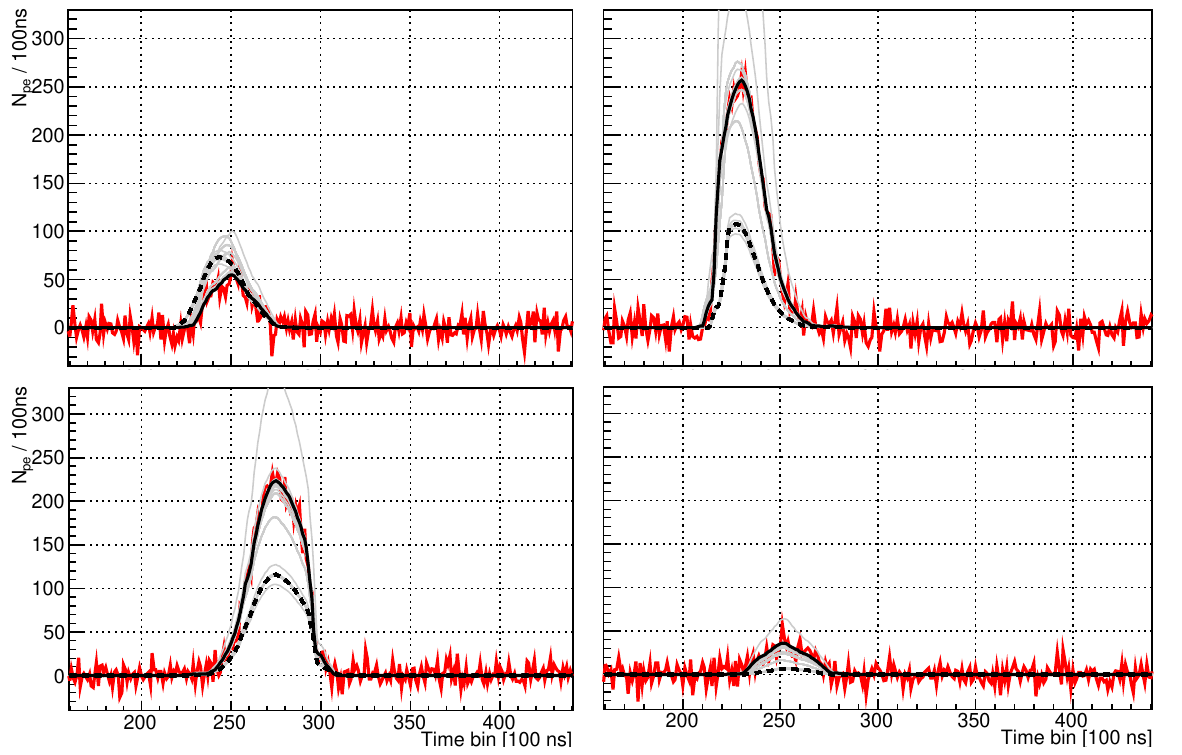}
    \caption{Example of the top-down reconstruction applied to the example event in Figure \ref{fig:FASTSimExampleEvent}. For demonstration purposes, only the energy, \Xmax{} and zenith angle parameters have been fitted, with the other geometrical parameters kept fixed to their true values. The dashed black lines shows the first guess, corresponding to $E=6\times10^{18}$\,eV, $X_\textrm{max}=750$\gcm{} and $\theta=30\degree$. The grey lines show the trialled traces. The solid black lines shows the best fitted traces, corresponding to $E=(1.0\pm0.1)\times10^{19}$, $X_\textrm{max}=902\pm12$\gcm{} and $\theta=40.01\pm0.60\degree$.
    }
    \label{fig:TopDownExample}
\end{figure}

\begin{figure}[]
\begin{minted}[
frame=lines,
framesep=2mm,
baselinestretch=1.2,
bgcolor=LightGray,
fontsize=\footnotesize,
linenos
]
{xml}
<!-- A sequence for FAST shower reconstruction -->
<sequenceFile >
<enableTiming/>
<moduleControl >
<loop numTimes="unbounded" pushEventToStack="yes">
<module> FASTEventFileReaderUA </module>
<module> FASTTopDownReconstructorUA </module>
<module> FASTEventFileExporterUA </module>
</loop >
</moduleControl >
</sequenceFile >
\end{minted}
\caption{Sequence file to perform the top-down reconstruction.}
\label{fig:FASTRecModuleSeq}
\end{figure}

In general, when performing the top-down reconstruction the absolute time-offset of the traces must also be determined. For a given set of shower parameters, this is done by shifting the simulated traces in 1 bin steps, each time recalculating Equation \ref{eqn:eventLikelihood}, and then returning the smallest value of $-2\ln\mathcal{L}$ found back to the minimizer.  
The offset calculated for the shower parameters which give the minimum value of $-2\ln\mathcal{L}$ is the final absolute time offset. An example of the top-down reconstruction applied to a simulated event is shown in Figure \ref{fig:TopDownExample}. The sequence file for the top-down reconstruction is shown in Figure \ref{fig:FASTRecModuleSeq}. Like FAST-sim, the modules shown here have been built utilising the \Offline{} framework. Briefly, \verb|FASTEventFileReader| reads in the ROOT file containing the data to be reconstructed, \verb|FASTTopDownReconstructor| handles the main processing steps, including the minimisation and simulations, and \verb|FASTEventFileExporter| is the same module as used in FAST-sim. Each iteration of the minimisation involves a call to the first four modules shown in Figure \ref{fig:FASTSimModuleSeq} using the shower parameters requested by Minuit.

\subsection{Neural Network First Guess}
\begin{figure}
    \centering
    \includegraphics[width=0.8\linewidth]{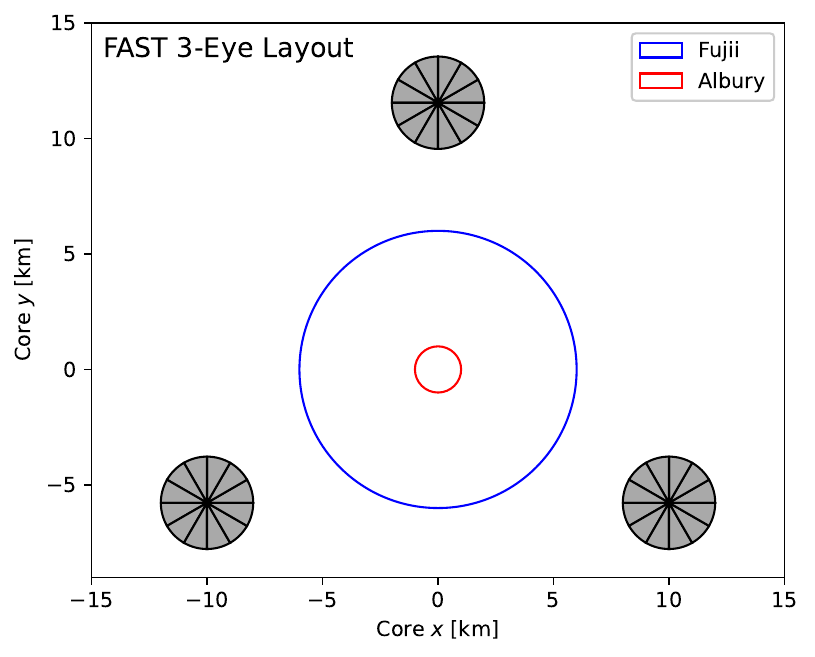}
    \caption{FAST 3-Eye layout used by Albury and Fujii for training NNs to predict the shower parameters. The core locations for showers falling in the central core region in Albury's study and the core locations used in Fujii's study are shown by the interior of the red and blue circles respectively.}
    \label{fig:NNcore}
\end{figure}

In the TDR, the first iteration of the minimisation procedure requires an initial guess of the shower parameters. Albury investigated using machine learning techniques to provide this \say{first guess}, and found success utilising a simple feed-forward, fully connected, deep \gls{nn} \cite{justin2020extending}. Albury trained this NN on simulated showers incident on an array of three FAST stations (12 telescopes/station, 360$\degree$ coverage) arranged in an equilateral triangle with spacing 20\,km. This layout is shown in Figure \ref{fig:NNcore} and will be referred to as the \say{FAST 3-Eye layout}. Albury trained models to estimate the shower parameters for showers in three separate, circular core regions, each with radius 1\,km. The central region is shown in Figure \ref{fig:NNcore} in red. It was this region in which Albury calculated the efficiency of the full NN + TDR procedure, details of which can be found in Section \ref{sec:receffprob}. Fujii \cite{fujii2021latest} extended Albury's initial study by training the same network on a wider range of core positions, shown by the blue circle in Figure \ref{fig:NNcore}. 
The network inputs for these models were the centroid time, integrated signal, and height of the pulse from each PMT containing significant signal. The network outputs were the standard six shower parameters ($E$, \Xmax{}, $\theta$, $\phi$, core $x$, core $y$). Fujii's results showed that at 40\,EeV resolutions in energy and \Xmax{} of $\sim8\%$ and 30\,g\,cm$^{-2}$, and angular and core resolutions of $\sim4\degree$ and $\sim450$\,m could be obtained. Additional details on the NNs tested by Albury and Fujii can be found in Chapter \ref{ch:ML}.

\vspace{5mm}

The combined performance of the neural-network first guess $+$ TDR was investigated by Albury for the FAST 3-Eye layout. The approach was shown to work well above 10$^{19}$\,eV, with resolutions in the reconstructed parameters comparable to that of stereo detection by Auger/TA. Additionally, biases in the NN first guess, arising from model predictions on data containing different atmospheric conditions than used in training, were able to be effectively removed by the TDR \cite{justin2020extending}. 

\begin{figure}
    \centering
    \includegraphics[width=\textwidth]{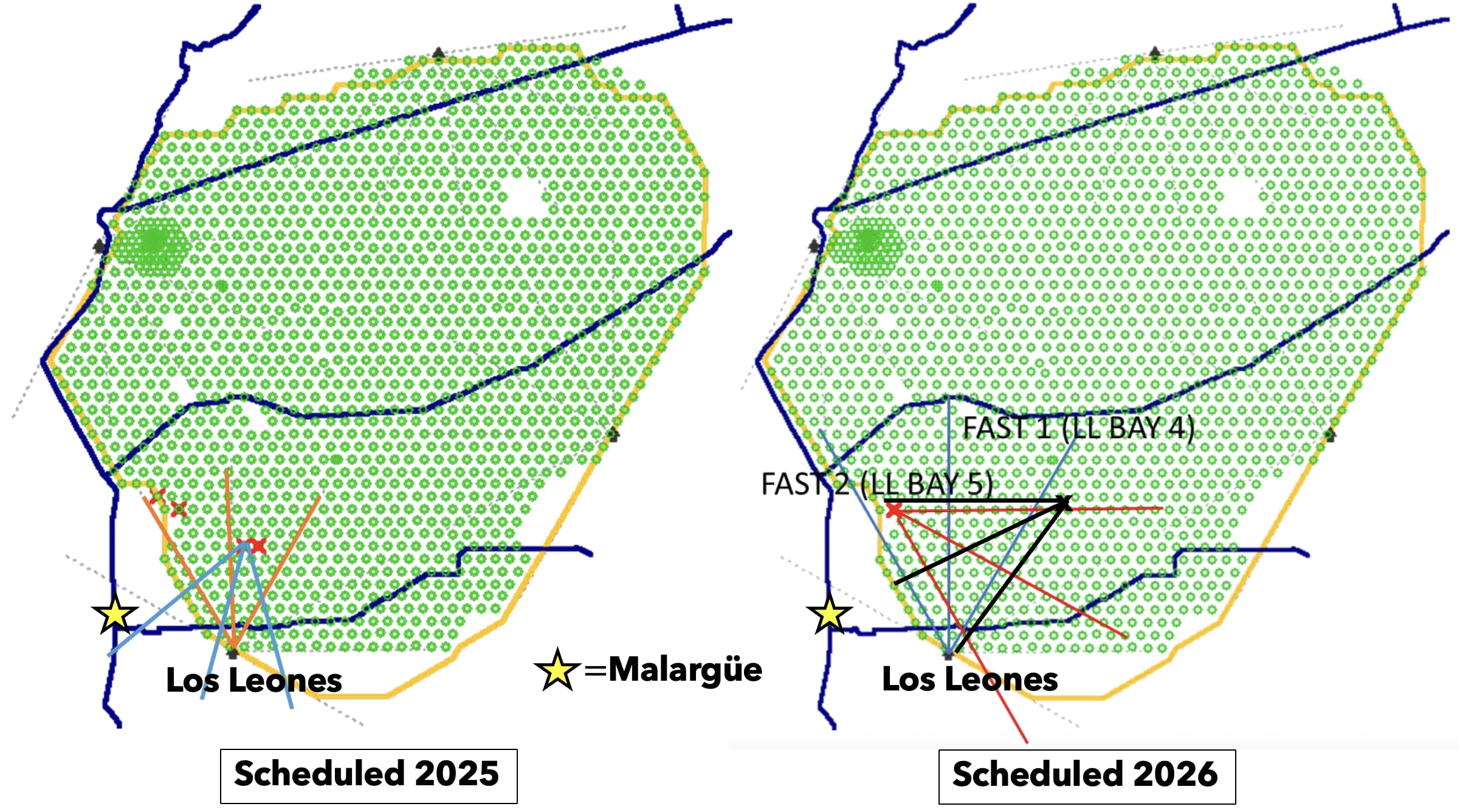}
    \caption{The stage one (left) and stage two (right) layouts for FAST mini-array, shown with respect to the Auger SD.}
    \label{fig:FASTMiniPlans}
\end{figure}

\section{Future Developments} 
\label{sec:FASTFuture}
The next milestone for FAST will be the completion of the so-called \say{FAST mini-array}. The first stage of the mini-array, scheduled for late 2025, will see two FAST-Field telescopes installed 12\,km from the LL site facing the original FAST@Auger. The current FAST5 prototype will also have new electronics installed. This will give four active FAST telescopes at Auger and will serve as a test for the stereo observation capabilities of FAST. The second stage of the mini-array will install two more FAST-Field telescopes at Auger. The FAST-Field telescopes installed as part of the first stage will be moved and the two additional telescopes placed so as to form a triangle with FAST@Auger. This stage is planned for late 2026. Figure \ref{fig:FASTMiniPlans} shows (roughly) the first and second stage layouts overlayed onto the Auger SD.

\chapter{Top-Down Reconstruction Improvements}
\label{ch:TDR}

Previous work on developing the FAST reconstruction chain has primarily focused on understanding and evaluating the performance of the FAST 3-Eye layout (see Figure \ref{fig:NNcore}) using simulations. In particular, the current machine-learning-based method for obtaining a first guess of the shower parameters has only been tested with this configuration. Whilst these results have been promising and are likely applicable to a future, sufficiently large FAST array, they provide limited insight into the expected performance of both the current prototype setups (FAST@TA/FAST@Auger) and the soon-to-be installed FAST mini-array. In addition, both the TDR and first guess estimation have yet to be thoroughly tested on real data. In line with these observations, the primary goals of this thesis are as follows,

\begin{itemize}
    \item Improve and extend the current reconstruction procedure such that data from the FAST@Auger, FAST@TA and FAST mini-array installations can be reliably reconstructed. This will involve
    \begin{itemize}
        \item Resolving issues with the current reconstruction procedure
        \item Developing a method for obtaining a first guess of the shower parameters with each installation.
    \end{itemize}
    \item Characterise the performance of each layout with the entire reconstruction chain
    \item Apply the updated reconstruction procedure to real data and check the results against expectations from simulations/other experiments. 
\end{itemize}
Verifying the performance of the FAST telescopes and software at these smaller scales will form an important stepping stone to the widespread deployment of FAST. This chapter focuses on improvements to the TDR, serving as an extension to the work of Albury \cite{justin2020extending}.

\section{Reconstruction Efficiency Problem}
\label{sec:receffprob}
\begin{figure}[t]
    \centering
    \includegraphics[width=\textwidth]{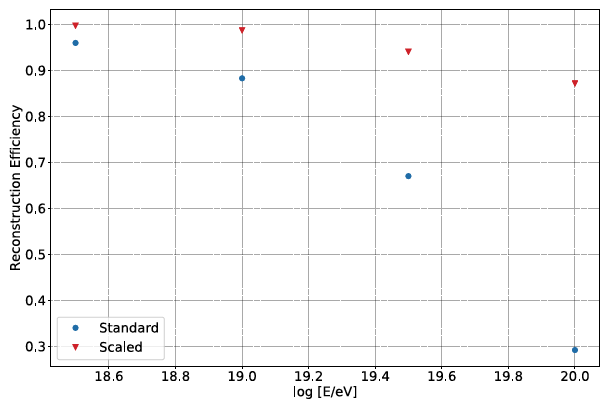}
    \caption{Albury's result for the reconstruction efficiency of the top-down reconstruction as a function of energy. Details in the text. From \cite{justin2020extending}.}
    \label{fig:JustinRecEfficiency}
\end{figure}

The most significant issue with the current FAST reconstruction process is the decrease in reconstruction efficiency as a function of energy, first identified by Albury in his PhD thesis. When evaluating the performance of the entire reconstruction chain for the FAST 3-Eye layout, Albury found a decrease in the reconstruction efficiency of the TDR with increasing energy. His results are shown in Figure \ref{fig:JustinRecEfficiency}. Here, the reconstruction efficiency is defined as
\begin{equation}
    \epsilon_i=\frac{N^{\textrm{rec}}_i}{N^{\textrm{trig}}_i}
\end{equation}
where $N^{\textrm{rec}}_i$ and $N^{\textrm{trig}}_i$ are the number of successfully reconstructed and number of triggered showers in energy bin $i$, respectively. The TDR is considered successful if the minimiser returns a \say{valid minimum} as determined by the minimiser itself (Minuit). Although the precise trigger condition used in this plot is not clearly stated by Albury (i.e. whether only one or multiple stations were required to trigger\footnote{A triggered station being one in which at least one PMT had an SNR above a given threshold.}), Figure 8.17 in \cite{justin2020extending} shows that the trigger efficiency for either one, two or three triggered stations was essentially 100\% above 10$^{19}$\,eV for the simulation setup used. This setup is shown in Figure \ref{fig:NNcore} with core positions simulated inside the red circle. Thus the decrease in efficiency can be attributed purely to more failed reconstructions at higher energies. The blue points in Figure \ref{fig:JustinRecEfficiency} show the reconstruction efficiency when using the standard TDR method described in Section \ref{sec:topdownrec}. In this case, for showers with an energy of 10$^{20}$\,eV, the minimiser \textit{fails} to find a valid minimum on $\sim70$\% of occasions. This is a significant issue for FAST which aims to observe the highest energy cosmic rays. 

\vspace{5mm}
  
Whilst not explored in detail, Albury attributed the decreasing efficiency to unexpected \say{jumps} or discontinuities in the likelihood function arising from small inaccuracies in the FAST simulation. The presence of such discontinuities presents difficulties for standard minimisers, which typically rely on accurate estimates of the derivative of the likelihood function. Albury reasoned that higher energy showers were more susceptible to the problem as signal from these showers is generally greater and so the impact from any inaccuracies in the simulation would be amplified. 

\vspace{5mm}

Albury provided a temporary solution to the efficiency problem by scaling the likelihood function by a constant factor of $10^{-5}$. For a fixed tolerance, i.e. level below which the minimiser is satisfied that it has found the minimum, this has the effect of causing the minimiser to sample the parameter space in larger steps. This leads to 
a comparatively smoother likelihood function and hence easier convergence to a minimum. The reconstruction efficiency using this scaled likelihood is shown by the red points in Figure \ref{fig:JustinRecEfficiency}. Although the solution significantly increased the reconstruction efficiency at higher energies, it came at the cost of decreased precision in the reconstructed shower parameters. Furthermore, the constant multiplicative factor of $10^{-5}$ is only applicable to the specific number of telescopes used in the FAST 3-Eye layout, 12 telescopes $\times$ 3 stations = 36 telescopes. Since the likelihood is calculated as a sum over each bin in each pixel, applying the same factor to a single telescope with just four pixels would not be suitable given the same tolerance. One could potentially adjust the tolerance level depending on the number of telescopes, however this does not fix the underlying issue of discontinuities in the likelihood function.

\begin{figure}[]
    \centering
    \includegraphics[width=\textwidth]{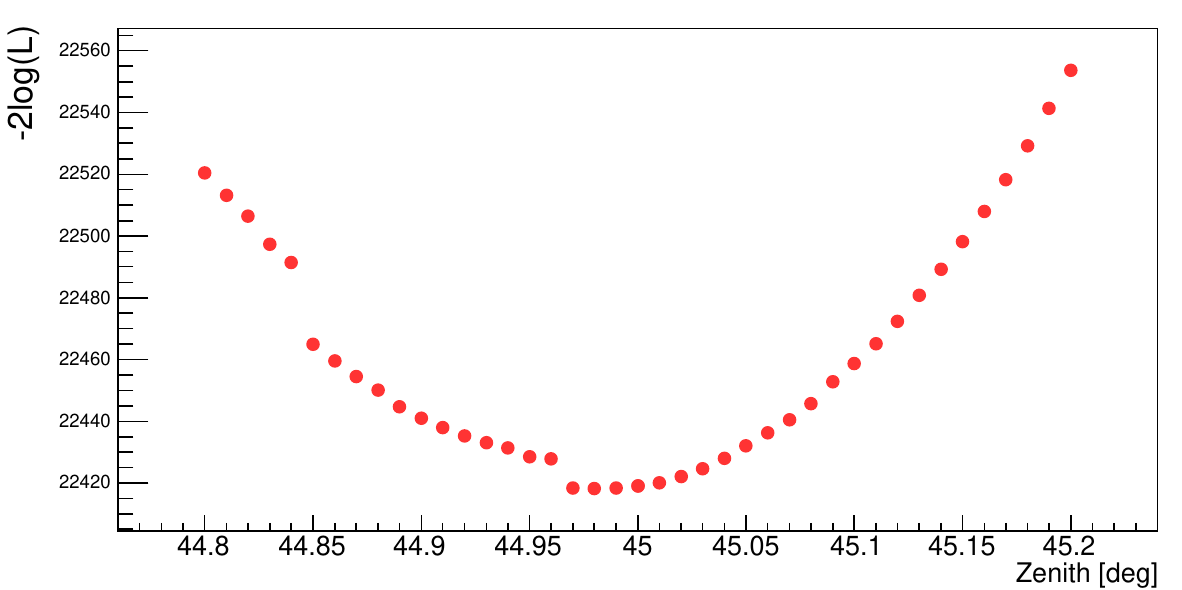}
    \caption{Scan of the likelihood function over zenith angle for the example event described in Section \ref{sec:UnderstandingEfficiency} using the original simulation. The jumps between 44.84 - 44.85$\degree$ and 44.96 - 44.97$\degree$ come from discontinuities in the binning of the shower axis (see Section \ref{sec:disconSources}).}
    \label{fig:RecEffExampZenLikeScan}
\end{figure}

\vspace{5mm}

In this section the sources of the discontinuities are investigated directly. The sources are indeed found to be attributable to several inaccuracies in the FAST simulation. By fixing these inaccuracies the reconstruction efficiency of the TDR is shown to dramatically improve.

\subsection{Understanding the Reduction in Efficiency}
\label{sec:UnderstandingEfficiency}
To understand the cause of the discontinuities, consider the contribution to the likelihood function coming from a single bin $i$ of a PMT trace. This contribution is calculated using the natural logarithm of Equation \ref{eqn:BinProb} where $x_i$ is the signal from data in bin $i$ and $\mu_i$ is the expected signal (no noise) in bin $i$ given simulated parameters $\vec{a}$=($E$, \Xmax{}, $\theta$, $\phi$, core $x$, core $y$). As the parameters $\vec{a}$ are smoothly varied, the expected signal $\mu_i$ should also change smoothly. If $\mu_i$ were to abruptly change between two sets of simulated parameters that are very close, then a discontinuity in the likelihood function would occur.  

\vspace{5mm}

\begin{figure}[t]
    \centering
    \includegraphics[width=\textwidth]{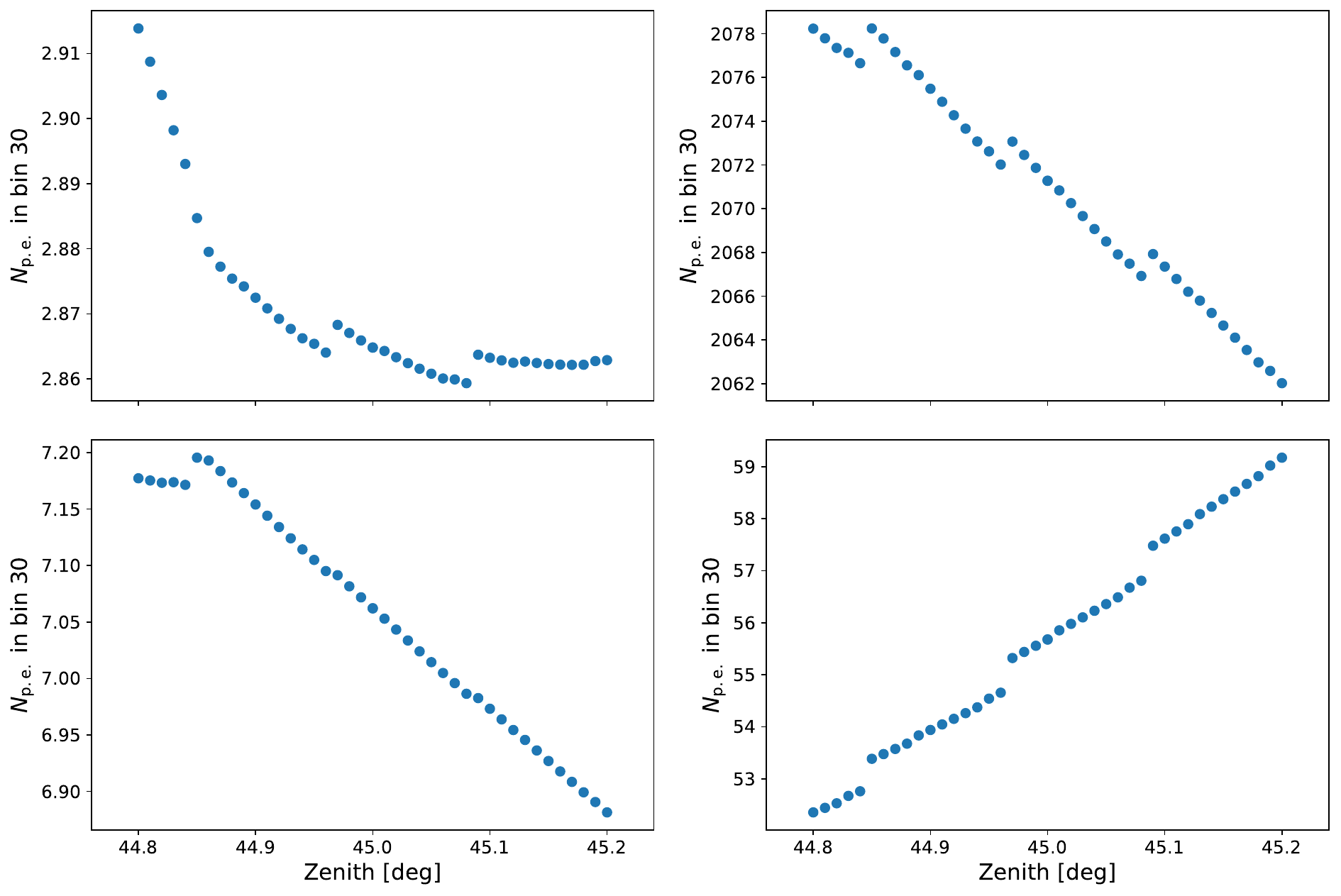}
    \caption{Number of photoelectrons in the 30$^\textrm{th}$ bin (arbitrarily chosen) of each of the PMT traces as a function of zenith for the example event. The location of the discontinuities in the signal match the location of the discontinuities in the likelihood function (Figure \ref{fig:RecEffExampZenLikeScan}).}
    \label{fig:oneBinSig}
\end{figure}

As an example consider a simulated shower with $E=10^{20}$\,eV, $X_\textrm{max}=800$\gcm{}, $\theta=45\degree$, $\phi=40\degree$ and core location of ($-1000$\,m, 6000\,m) incident on a single FAST telescope at (0\,m, 0\,m) pointing along the $y$-axis. Performing a scan of the likelihood function in zenith angle (keeping all other parameters fixed) over a small range around the simulated value gives the graph shown in Figure \ref{fig:RecEffExampZenLikeScan}. Two features are immediately noticable. First, the minimum of the function is slightly less than the simulated value of $45\degree$. This is due to the added background noise and is perfectly acceptable. The second feature is the sharp jumps in the likelihood function between 44.84 - 44.85$\degree$ and 44.96 - 44.97$\degree$. Such discontinuities are not expected at this scale and can result in failed minimisations. Plotting the simulated signal $\mu_i$ (i.e. number of p.e.) in a single arbitrary bin from each PMT as a function of zenith angle gives Figure \ref{fig:oneBinSig}. The discontinuities in the signal values occur between the same zenith angles as the jumps in the likelihood function. This analysis also reveals a third discontinuity between 45.09 - 45.1$\degree$ which is not immediately obvious just by visual inspection of the likelihood function.

\vspace{5mm}

This demonstration shows that, fundamentally, it is not the energy of the showers but rather the size of the signals upon which the minimisation success rate depends. To see this, first make the reasonable assumption (not strictly proven) that for a given tolerance level, the larger the discontinuities in the likelihood function are the more likely it is for a minimisation to fail. As the discontinuities in the likelihood function come from discontinuities in the signal of individual bins, it follows that the larger these signal discontinuities are the more likely a failed minimisation. As for why Albury's results show a decrease in efficiency with energy, the \textit{specific} simulation setup used by Albury (Figure \ref{fig:NNcore}) would have yielded, on average, larger signals and thus larger signal discontinuities for higher energy showers. This is not necessarily representative of data from a single FAST eye (e.g. FAST@TA), as the increased volume over which to observe showers with increasing distance from the telescope and the greater amount of light emitted from higher energy showers means that higher energy showers are generally observed much further away than lower energy showers. Hence, the average signal as a function of energy for data is unlikely to vary as drastically as for simulations confined to a limited range of core positions. That being said, for showers incident on the \textit{interior} of a full-sized FAST array, there will be some energy above which all showers trigger at least one station. Above that energy, it would be the case that higher energy showers have greater average signals and thus fail the (current) reconstruction more often. 

\vspace{5mm}

Of the six shower parameters which can be adjusted in the simulation, only changes in the geometrical parameters ($\theta$, $\phi$, core $x$, core $y$) were found to cause discontinuities in the PMT traces. This can be understood as both \Xmax{} and energy only enter into the simulation as parameters of an analytic GH function. This means, for a fixed geometry, any changes in \Xmax{} and/or energy only result in smooth changes to the total number of photons in the shower and where they are \say{produced} along the shower axis. These changes are reflected smoothly in the output traces.

\vspace{5mm}

\begin{figure}[t]
    \centering
    \includegraphics[width=0.8\textwidth]{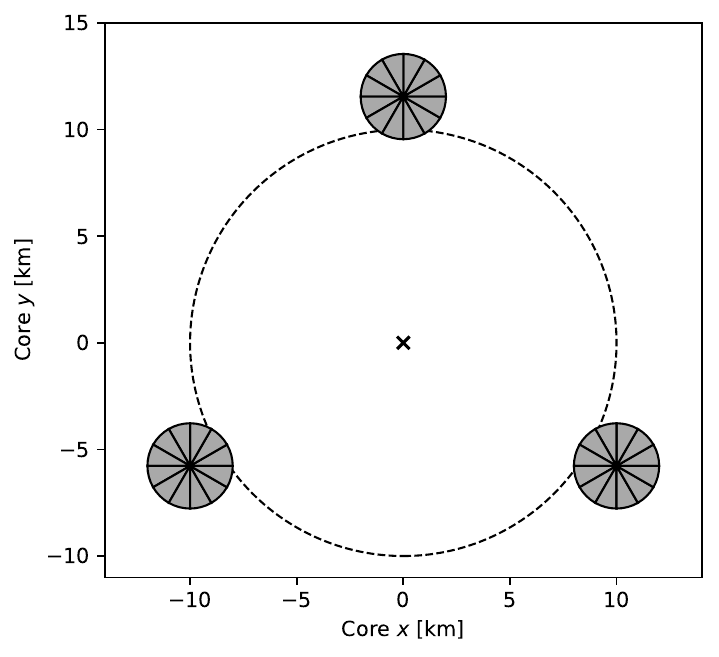}
    \caption{Diagram showing the FAST 3-Eye layout and core positions (interior of the dashed circle) of the simulated showers.}
    \label{fig:3eyeSims}
\end{figure}

The causes of the discontinuities in the FAST simulation are discussed in the following sub-sections. For reference, a set of 1000 simulated showers incident on the FAST 3-Eye layout with a fixed energy of 30\,EeV was prepared. The simulation parameters and telescope/core position layout are shown in Table \ref{tab:simparsExamp} and Figure \ref{fig:3eyeSims} respectively. The zenith angles of these showers was then reconstructed using the current TDR and a first guess of $\theta$ equal to the true value. All other parameters were fixed to their true values. Of the 1000 reconstructions, 274 failed with the minimiser returning an invalid minimum. Additionally, 232 events had an uncertainty in the reconstructed zenith $\theta_\textrm{err}>100\degree$. For this particular simulation setup, where deviation from $\theta_\textrm{true}$ and the size of $\theta_\textrm{err}$ should only be due to added noise, such uncertainties are unrealistic and are indicative of discontinuities in the simulation. The same data set will be reconstructed after fixing each of the discontinuity sources described below to check the effect of each improvement.

\begin{table}[]
    \centering
    \begin{tabular}{|c|c|}
    \hline\hline
        \Xmax & EPOS-LHC (500 - 1200\gcm{}) \\
        \hline
         Energy & Fixed (30\,EeV) \\
         \hline
         $\theta$ & $\sin\theta\cos\theta$ (0 - 80$\degree$) \\
         \hline
         $\phi$ & Uniform (0 - 360$\degree$) \\
         \hline
         Core & Uniform (in circle centred at (0,0), $r=10$\,km) \\
         \hline\hline
    \end{tabular}
    \caption{The distributions sampled from for each shower parameter to generate the test data set for evaluating the TDR after each discontinuity fix. The range of each distribution is shown in brackets. Excluding the fixed energy, each of the distributions sampled from were continuous.}
    \label{tab:simparsExamp}
\end{table}

\subsection{Discontinuity Sources and Fixes}
\label{sec:disconSources}
Three sources of discontinuities were found in the FAST simulation. Each source and its corresponding fix are explained below.


\subsubsection{Shower axis binning}
To calculate the amount of light reaching the telescope from different points along the shower axis, the FAST simulation divides the section of the shower axis in the FOV of the telescope into several bins. The number of bins is calculated by dividing the time between where the shower enters/exits the telescope FOV, $\Delta{}t$, by the \textit{initial} time length of one bin, typically set to 10\,ns. In the original simulation both the number of bins and $\Delta{}t$ are calculated as integers ($\Delta{}t$ measured in ns). Moreover, instead of using the initial value of 10\,ns, the \textit{final length} of each time bin is calculated by dividing $\Delta{}t$ by the (integer) number of bins. Regardless of whether these were deliberate choices or simple oversights, there are two main problems with the above approach. 

\vspace{5mm}

\begin{figure}[]
    \centering
    \includegraphics[width=\textwidth]{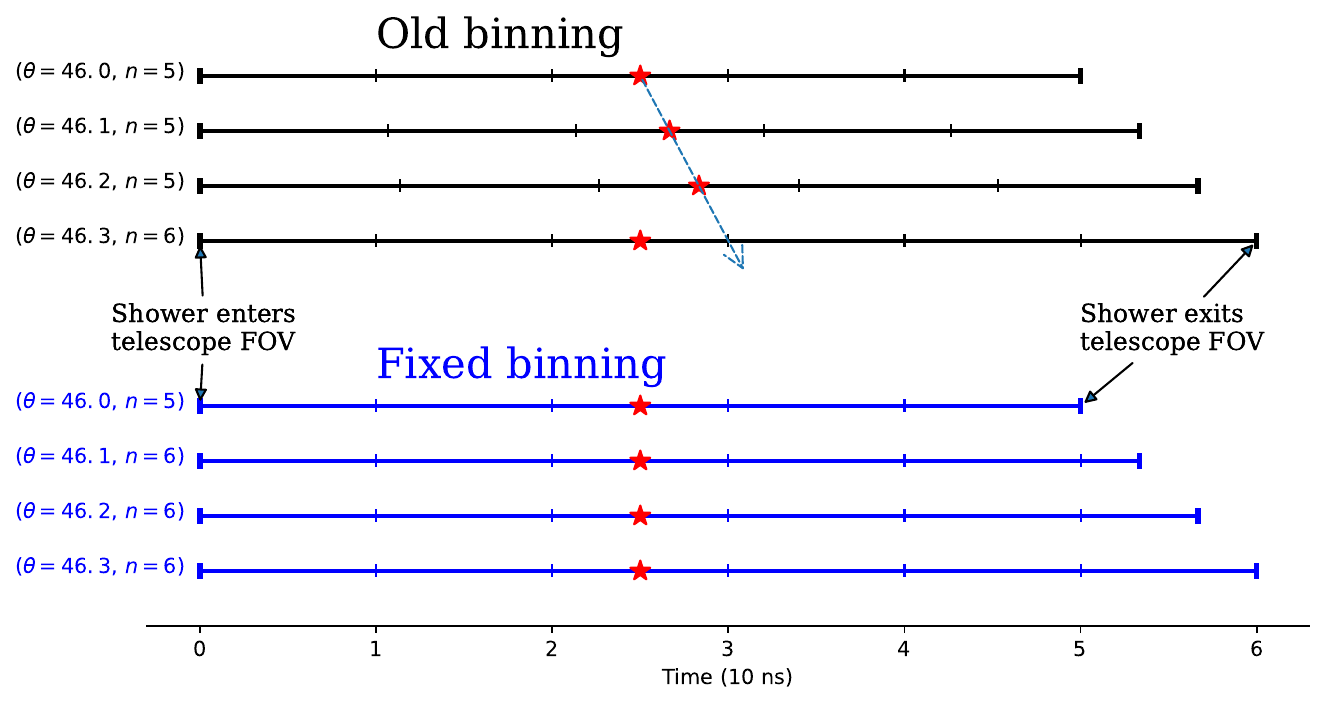}
    \caption{Diagram showing the old (top) and new (bottom) methods for binning the shower axis in the FAST simulation. In the old method, discontinuities in the size and position of individual bins would arise when the total number of bins increased by one. The discontinuity in position is shown by the offset between where the dashed arrow intersects the bottom shower axis (expected bin centre if the bin locations were changing smoothly) and the position of the red star on this axis (actual bin centre using old binning method). By fixing the time length of each bin to be exactly 10\,ns, with a variable-length final bin, the discontinuities in the bin position and size disappear. 
    }
    \label{fig:binningProblem}
\end{figure}

\begin{figure}[]
    \centering
    \includegraphics[width=\textwidth]{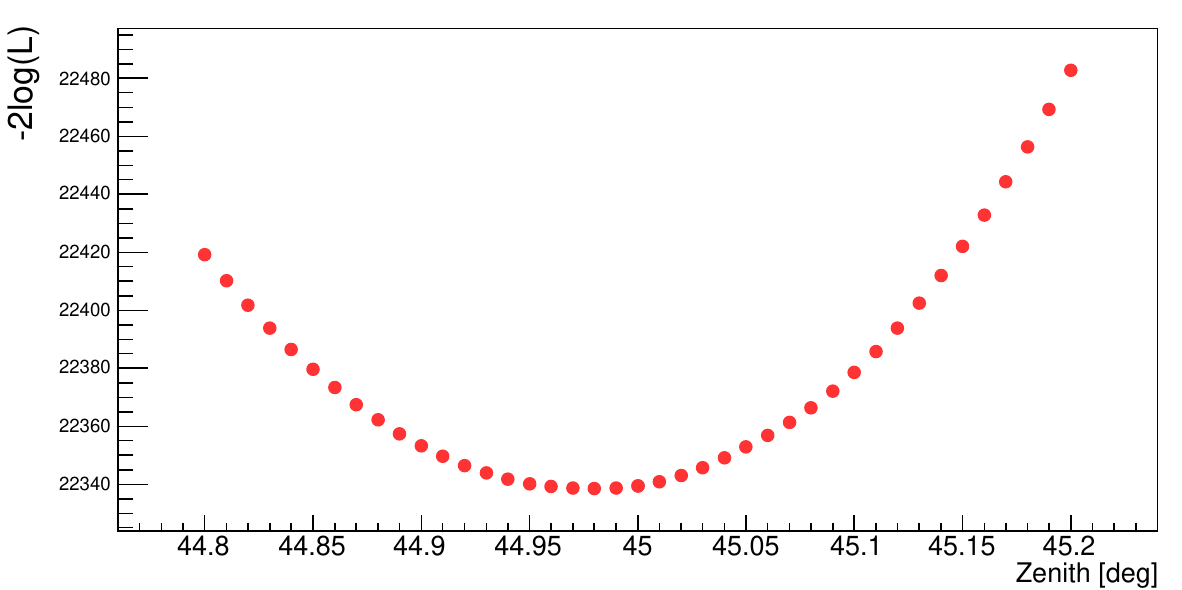}
    \caption{Scan of the likelihood function for the example event in Section \ref{sec:UnderstandingEfficiency} using the corrected binning method. Compared to Figure \ref{fig:RecEffExampZenLikeScan} there are no obvious discontinuities.}
    \label{fig:correctedBinningLike}
\end{figure}

\begin{figure}[]
    \centering
    \includegraphics[width=\textwidth]{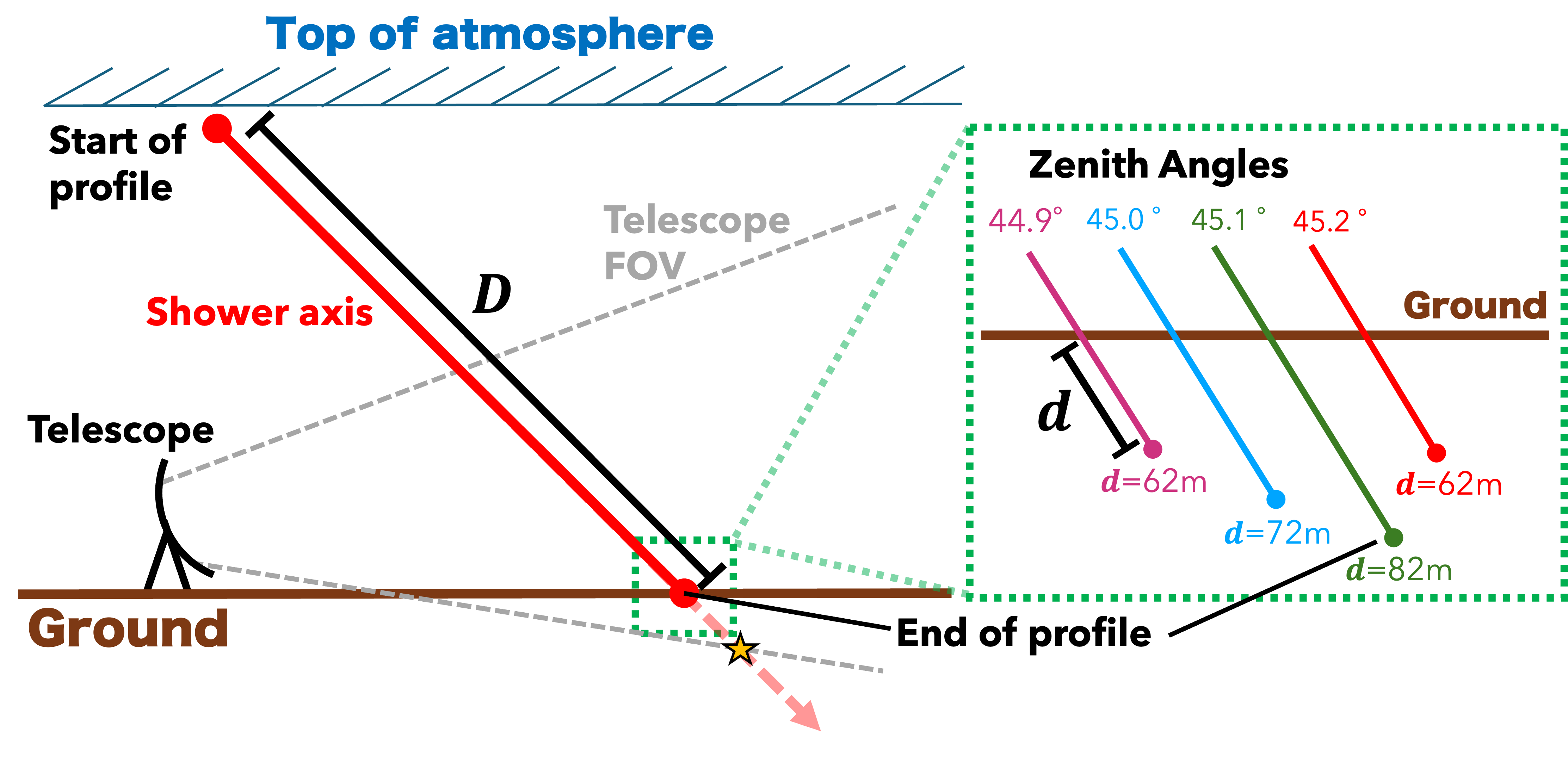}
    \caption{Diagram showing the instances in which $D$ is used for calculating the shower axes binning. The orange star shows the location of the intersection between the shower axes and telescope FOV cone. As this intersection occurs beneath the ground, past the point where both the fluorescence and Cherenkov profiles end, $D$ is used. The zoomed-in panel shows how, with the original simulation, small changes in the shower zenith angle cause non-smooth changes in the profile end point and thus $D$. The example values of $\theta$ and $d$ have been taken from Figure \ref{fig:distStopCore}.}
    \label{fig:endOfProfileDiagram}
\end{figure}

\begin{figure}[]
    \centering
    \includegraphics[width=\textwidth]{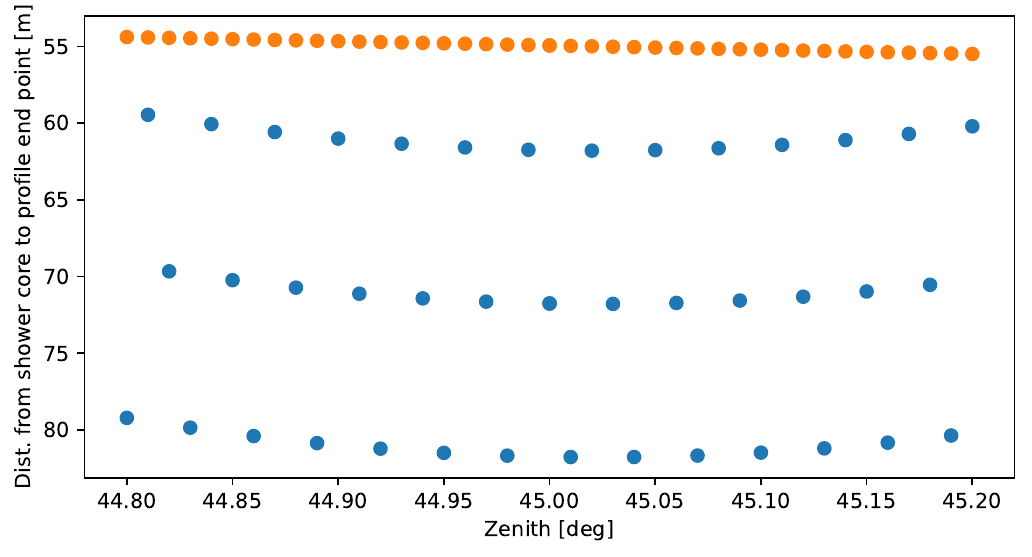}
    \caption{Distance from the shower core to the end point of the shower profile as a function of zenith angle for the example event in Section \ref{sec:UnderstandingEfficiency}. The blue (orange) points show the values before (after) the correction to the profile end point calculation.}
    \label{fig:distStopCore}
\end{figure}

Firstly, $\Delta{}t$ represents a physical quantity with essentially arbitrary precision. As such it should be calculated as a floating point number, not an integer. This is fixed by changing the data type of the appropriate variable in the simulation code. Secondly, setting the \textit{final length} of each bin in the above way will (in general) cause discontinuities in the signal from individual bins as the shower geometry changes. This is because the signal from an individual bin depends on the length of the bin and the location of the bin centre, both of which experience jumps when the number of bins increases by one. This effect is illustrated in the top half of Figure \ref{fig:binningProblem}. The black segmented lines represent the binned shower axis for the different zenith angles shown on the left. The number of bins $n$ (calculated using the above approach) is also shown. For ease of visualisation, the scenario is idealised such that the shower axis enters the telescope FOV at a fixed time and exits at successively later times. 
The red stars highlight the centres of the third bin of each line. For the top \say{shower axes}, each with five bins, the location of the bin centre and the length of each bin change evenly. However in the fourth shower axis, where the number of bins increases to six (60\,ns/10\,ns), both quantities \say{jump} to smaller values.

\vspace{5mm}

The solution to this problem is simply fixing the length of every bin, with the exception of the last bin, to be exactly 10\,ns. The length of the final bin is set to the time difference between the end of the second to last bin and where the shower exits the telescope FOV,
\begin{equation}
    \textrm{Length of last bin = 10\,ns$\times$($\Delta{}t$ mod 10\,ns)}
\end{equation} 
where \say{mod} represents the modulo operation. The corrected binning method is shown in the bottom half of Figure \ref{fig:binningProblem}. Ignoring the final bin, the lengths of each bin and the position of the bin centres now remain constant with changes in the zenith angle. This better matches the expected behaviour. Performing the same scan over the likelihood function for the example event in Figure \ref{fig:RecEffExampZenLikeScan} with the corrected binning gives Figure \ref{fig:correctedBinningLike}. The likelihood function now changes smoothly with no discontinuities, as desired. 

\vspace{5mm}

A second mistake in the shower axis binning was also identified. Specifically, the distance between the beginning and end of the shower profile, $D$, was found to fluctuate non-smoothly with small changes in the shower geometry. $D$ is used in place of the distance to where the shower axis exits the telescope FOV when this intersection occurs past the end of the profile. 
When used, fluctuations in $D$ lead to discontinuities in the shower axis binning and in turn signal in individual bins. In the original simulation, $D$ is calculated as the difference between the start and end points of the fluorescence/Cherenkov light profiles. 
The end points of these profiles were found to vary non-smoothly due to the use of integer values when determining the profile binning\footnote{Note that this calculation is handled by an Auger module. Although this may mean the same \say{bug} is present in the Auger simulation, it likely has minimal impact on their reconstruction performance due to the bottom-up approach implemented.}. Figure \ref{fig:endOfProfileDiagram} illustrates the problem. The left hand side shows the scenario in which $D$ is used whilst the right hand side shows a zoomed in view of the region around the end point of the profile. In the zoomed in view, multiple shower axes with slightly different zenith angles are plotted. The value $d$ represents the distance between the shower core and the profile end point as measured along the shower axis. Note it is typical for these profiles to extend slightly below ground as depicted here. The non-monotonicity of $d$ e.g. the \say{jump} back to a smaller $d$ value going from 45.1$\degree$ to 45.2$\degree$, causes discontinuities.

\vspace{5mm}

Using floating point values to calculate the profile binning instead, the end points and hence $D$ were made to change smoothly as a function of the shower geometry. Figure \ref{fig:distStopCore} shows the distance from the shower core to the end point of the fluorescence profile (i.e. $d$) of the example event described in Section \ref{sec:UnderstandingEfficiency} as a function of zenith angle. The blue (orange) points show the uncorrected (corrected) values. Note that because $D$ is not used in the shower axis binning for this particular event, the discontinuities in the uncorrected values do not impact the likelihood function in Figure \ref{fig:RecEffExampZenLikeScan}. Reconstructing the test data set of 1000 showers with the above corrections to the shower axis binning, the number of failed reconstructions reduced from 274 to 109, whilst the number of events with $\theta_\textrm{err}>100\degree$ reduced from 232 to 20. 

\begin{figure}[t]
    \centering
    \includegraphics[width=0.85\textwidth]{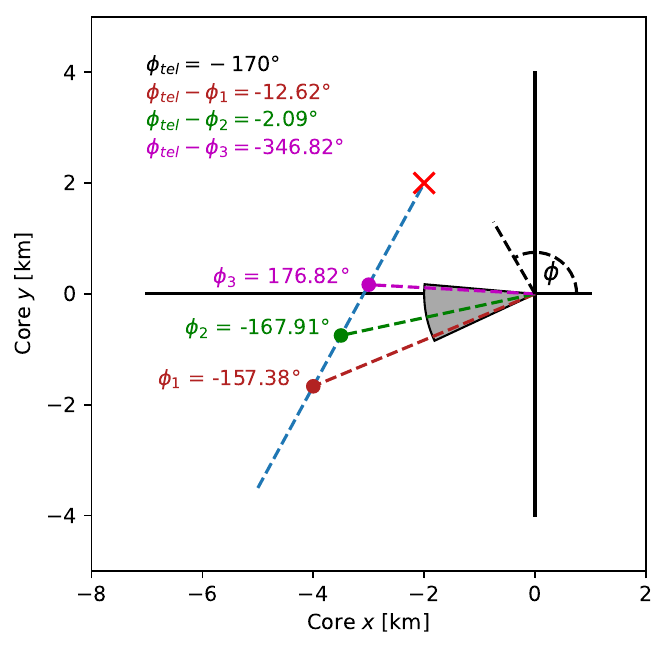}
    \caption{Diagram showing how, in the original simulation, not accounting for the discontinuity at $\phi=\pm180\degree$ lead to erroneous results in the azimuthal difference between a telescopes pointing direction and points along the shower axis.}
    \label{fig:azimuthMiss}
\end{figure}

\subsubsection{Azimuthal range}
To account for the directional efficiency of the FAST telescope, the difference between the azimuth of the telescope pointing direction, $\phi_\textrm{tel}$, and the azimuthal positions of points along the shower axis must be calculated. An error in this calculation was found to cause discontinuities in the original FAST simulation. Figure \ref{fig:azimuthMiss} provides a visual explanation. In the Figure, a telescope located at the origin with $\phi_\textrm{tel}=-170\degree$ is observing a shower (dashed blue line, red $\times$ is the shower core) with an axis which crosses the $y$-axis. In this setup the azimuth of points along the shower axis has a discontinuity at $\phi=\pm180\degree$ where the values change from negative to positive. This is shown by three example points $\phi_1$, $\phi_2$ and $\phi_3$. If a simple subtraction between $\phi_\textrm{tel}$ and azimuths of points along the shower axis is performed, then this difference will also have a discontinuity. This is shown in the top left of the figure using the three example points. In particular, the difference $\phi_\textrm{tel}-\phi_3$ lies outside the nominal azimuth range of $(-180\degree,180\degree)$. In the original simulation differences $<-180\degree$  were \textbf{not} redefined to be in this range, resulting in several discontinuities and in turn failed reconstructions. Fixing this issue further reduced the number of failed fits from 109 to 82 and the number of fits with $\theta_\textrm{err}>100\degree$ from 20 to 8.

\subsubsection{Signal re-binning}
The final cause of discontinuities identified in the FAST simulation was the re-binning of the PMT traces which occurs at the end of the simulation. Before re-binning, each bin of the PMT traces corresponds to a single 10\,ns bin along the shower axis. The standard re-binning uses a factor of 10, meaning the signal from groups of 10 consecutive bins are added to produce 100\,ns bins in the final trace. Two issues were found in this procedure. The first was once again a mishandling of variable data types which occasionally led to the final bin of the re-binned trace being set to zero. The second issue relates to the application of the directional efficiency map (Section \ref{sec:OpticalPerformance}) and requires more explanation. 

\vspace{5mm}

\begin{figure}[]
    \centering
    \includegraphics[width=0.9\textwidth]{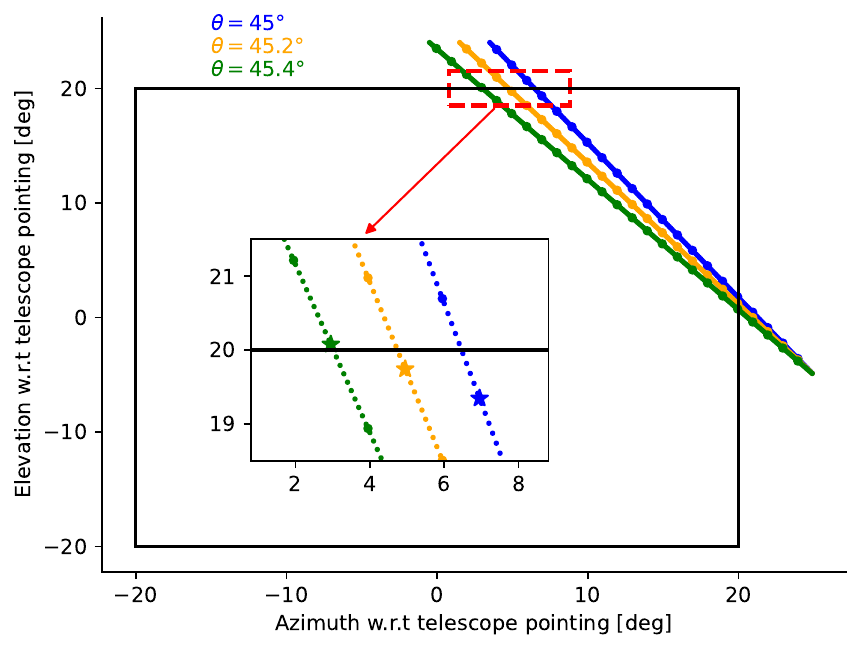}
    \caption{Diagram illustrating how changes in the shower geometry can cause the centre of a bin on the edge of the telescope FOV (stars in the inset) to shift from inside to outside the FOV. Once outside the FOV the directional efficiency drops to 0, causing a signal discontinuity in this bin.}
    \label{fig:rebinningPMTMap}
\end{figure}

Consider a single set of ten 10\,ns bins being summed to give one re-binned 100\,ns bin. When the re-binning is performed the relative elevation and azimuth values of each 10\,ns bin are averaged to give the elevation and azimuth position for the re-binned bin. This elevation/azimuth pair corresponds to a directional efficiency between 0 and 1, given by interpolating a directional efficiency \say{map} (\mbox{2-D} histogram) such as Figure \ref{fig:FASTPrototypeRayTrace}. The efficiency value is used to scale the signal in the 100\,ns bin. The problem arises in the transition between points lying inside and outside the range of each map. The standard range of the directional efficiency maps used by FAST is between $(-20\degree,20\degree)$ for both elevation and azimuth. For elevation/azimuth pairs outside this range the directional efficiency is set to 0. Small changes in the shower geometry can cause the elevation/azimuth values of a bin inside the map range to shift outside the map range. If the edges of the maps were to tend to 0 sufficiently smoothly then this would not be an issue, since the signal from a bin near the edge of the map would decrease to arbitrarily small values as it approached the boundary. 
In reality however, the maps used in the FAST simulation have sections where the edge values are on the order of 0.02 - 0.03. This may not seem large, but if the \Xmax{} of a shower coincides with one of these edge locations, then a bin moving from inside to outside the map range could experience a drastic drop in signal, leading to a discontinuity. Figure \ref{fig:rebinningPMTMap} shows a schematic of the problem. The black box represents the map edge, with coloured lines representing example tracks for showers with slightly different zenith angles.
The inset shows a zoomed in view of the area inside the red rectangle. This reveals that the coloured \say{lines} are actually many individual points. Each small circle represents the centre of a single 10\,ns bin. The large circles, also visible in the zoomed out view, represent the centres of the 100\,ns bins. The stars in the inset show the 4$^\textrm{th}$ 100\,ns bin of each shower track. As the zenith angle is adjusted the position of the star shifts until it falls just outside the map range, at which point the signal from this bin would be set to 0, causing a discontinuity.

\vspace{5mm}

\begin{figure}[]
    \centering
    \includegraphics[width=0.9\textwidth]{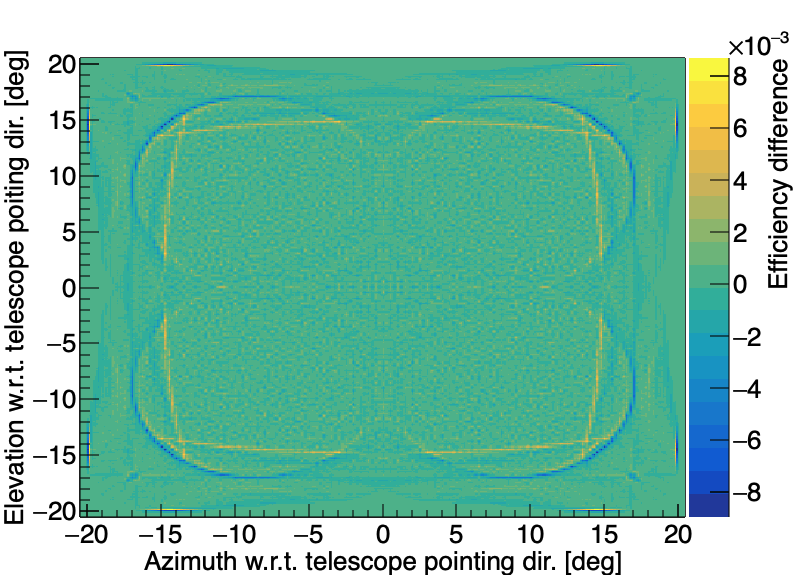}
    \caption{Difference in efficiency between the smoothed and non-smoothed ideal directional efficiency map. The difference does not exceed more than $\pm1\%$ over the whole map.}
    \label{fig:smoothMapDiff}
\end{figure}

The ideal solution to this problem would be to produce maps which extend far enough in elevation and azimuth such that the values on the edges are \say{sufficiently} close to 0. Whilst this is beyond the scope of the current work, the effect can be roughly emulated by smoothing the current maps. Specifically, bins containing 0 were added to the current 2-D histograms between ($-20.5\degree$,$-20\degree$) and ($20\degree$,$20.5\degree$) in both elevation and azimuth. The entire histogram was then smoothed using the \verb|Smooth| function of the \verb|TH2| class in ROOT \cite{brun1997root}. This has the effect of slightly extending the map edges and reducing the values of bins on the edge of the histogram by a factor of $5\sim10$. 
Figure \ref{fig:smoothMapDiff} shows the difference between the smoothed version and original version of the \say{ideal} directional efficiency map that was used to create the test data set. Note this map is different from Figure \ref{fig:FASTPrototypeRayTrace} and can be found in Appendix \ref{fig:idealDirEffMap}. The maximum difference between the smoothed/non-smoothed versions across the entire map is less than $\pm1\%$. A proper extension of the directional efficiency maps using actual measurements should be considered in future work.

\vspace{5mm}

In addition to creating a smoothed directional efficiency map, the calculation of the signal in bins on the edge of the map was adjusted. First, based on the shower geometry, the first (last) set of ten 10\,ns bins to enter (exit) the map range is determined. The signal in the corresponding 100\,ns bin/s is then calculated by summing the scaled contributions (using the \textit{smoothed} ideal directional efficiency map) of each 10\,ns bin. This further decreases the impact of any remaining discontinuity between the interior and exterior of the smoothed maps, since each individual bin contains less overall signal and thus jumps to zero are less impactful on the final likelihood. This \say{fine binning calculation} is only applied to the 100\,ns bins which cross the map edges as doing so for all 100\,ns bins
would increase the necessary simulation time significantly. Implementing the new re-binning method and smoothed maps into the FAST simulation reduced the remaining number of failed reconstructions and number of events with $\theta_\textrm{err}>100$ in the test data set from 82 to 79 and 8 to 2 respectively.

\subsection{Minimiser Strategy}
At this stage, nearly all events which originally had $\theta_\textrm{err}>100\degree$ have been fixed. However there are still a large number of showers failing the reconstruction. Inspecting the sampled likelihood functions of these showers revealed no obvious discontinuities, so the output from the minimiser was investigated directly. This showed that the cause of failure for these events was an estimated distance to the minimum larger than required for successful minimisation (this threshold is determined by the tolerance level and number of fit parameters). Moreover all but one failed reconstruction had a reconstructed zenith identical to the simulated value. This suggested that the minimiser was not exploring the parameter space thoroughly enough to accurately predict the minimum due to the first guess (i.e. the true values) already being close to the best possible fit. To fix this the \verb|strategy| parameter of the minimiser was adjusted. Roughly speaking, this parameter controls the thoroughness of the parameter search. For the TDR, where function calls involve an entire shower simulation, the strategy setting which attempts to minimise the function in the least number of calls (\verb|strategy=0|) is typically used. By changing the \verb|strategy| setting such that the minimiser checked the parameter space more carefully (\verb|strategy=1|), the number of failed reconstructions was reduced to a single shower. The number of reconstructions with $\theta_\textrm{err}>100\degree$ also decreased by one. With this result in mind, the TDR was altered for all subsequent studies in the following way,
\begin{itemize}
    \item Initially, attempt to reconstruct the shower parameters using  \verb|strategy=0|
    \item If this fails, attempt with \verb|strategy=1|
\end{itemize}
In practice, it is extremely unlikely that the first guess passed to the TDR will be accurate enough to cause the minimiser to fail in this way, particularly if multiple parameters are being reconstructed simultaneously. However, in the event that the minimiser fails for some other reason, re-performing the fit with a slightly modified search methodology may improve the success rate.

\vspace{5mm}

\begin{figure}[]
    \centering
    \includegraphics[width=0.9\linewidth]{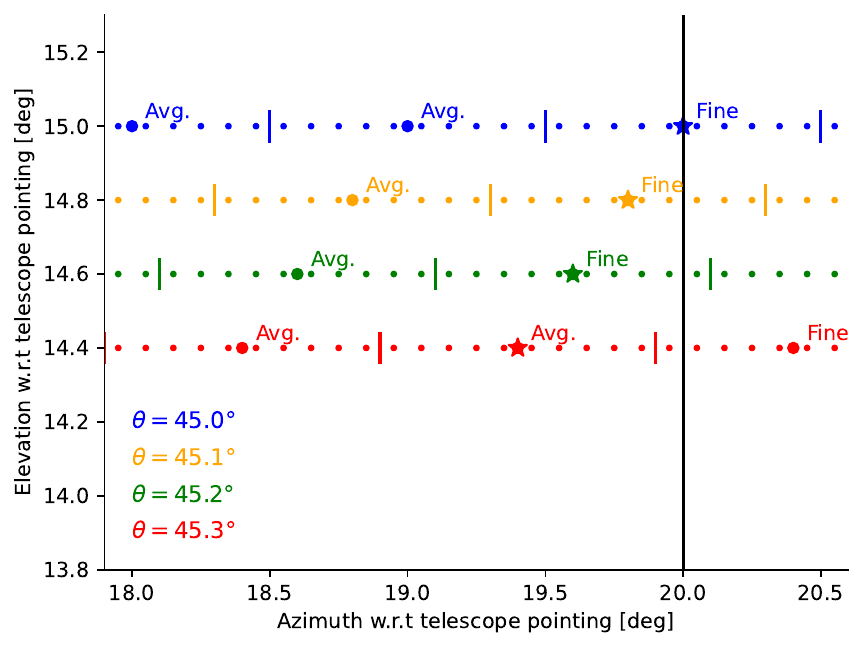}
    \caption{Illustration of the issue with the final remaining failed reconstruction. Like Figure \ref{fig:rebinningPMTMap}, the small dots represent the centres of each 10\,ns bin and the large dots/stars the centres of the re-binned 100\,ns bins. The ``Avg". or ``Fine" labels next to each 100\,ns bin indicate the method for calculating the signal (details in the text). The starred bin begins on the boundary with half of it's 10\,ns bins inside and half outside the map range. As the zenith angle changes the starred bin shifts left until all 10\,ns bins are within the interior of the map, at which point the calculation method changes. This causes a small discontinuity in the signal of the starred bin. Note the differences in elevation/azimuth between the different zenith angle paths have been exaggerated for ease of visualisation.}
    \label{fig:remainingShowerProblem}
\end{figure}

Table \ref{tab:recImprovements} shows a summary of how each of the above changes reduced the number of failed reconstructions and reconstructions with $\theta_\textrm{err}>100\degree$. The one remaining reconstruction failure is a result of the imperfect solution to the signal re-binning. The problem is illustrated in Figure \ref{fig:remainingShowerProblem}. With small changes in the shower geometry, the range of a single 100\,ns bin on the map edge may shift from including a small section beyond the map range to being completely inside the map range. When this occurs, the fine binning calculation (\say{Fine} in the figure) for the signal is no longer used for this bin. Instead, the standard azimuth/elevation pair averaging and scaling method (\say{Avg.} in the figure) is applied. This can result in a discontinuity due to the small differences in the calculation methods. In the absence of additional measurements, a more robust solution, perhaps where the two calculation methods are smoothly connected, could be considered in future work.

\begin{table}[t]
    \centering
    \begin{tabular}{c|c|c}
        \textbf{State} & \textbf{Failed reconstructions} & \textbf{$\theta_\textrm{err}>100\degree$}\\
        \hline\hline
        Original simulation &  274 & 232\\ 
        + Shower axis binning (sim.) & 109 & 20\\ 
        + Azimuthal difference (sim.) & 82 & 8 \\ 
        + Signal re-binning (sim.) & 79 & 2 \\
        + Strategy Improvement (rec.) & 1 & 1
        
    \end{tabular}
    \caption{Table showing how the number of failed reconstructions and reconstructions with $\theta_\textrm{err}>100\degree$ of the 1000 events in the test data set decreased with improvements to the simulation/reconstruction procedure. The sim./rec. labels indicate whether the fix related to the FAST simulation or to the overall reconstruction procedure.}
    \label{tab:recImprovements}
\end{table}

\subsection{Checking the Improved Reconstruction}
\label{sec:CheckingRecon}
This section checks whether the improvements in reconstruction efficiency seen in the previous section translate to the reconstruction of all six shower parameters at once. Additional showers with the same parameters and setup as shown in Table \ref{tab:simparsExamp} and Figure \ref{fig:3eyeSims} were simulated but now at energies of $10^{18.5}$\,eV, $10^{19}$\,eV and $10^{20}$\,eV. One thousand showers at each energy were simulated. This data set (including the 1,000 $10^{19.5}$\,eV showers) will be referred to as the \say{Large Core Region}. The larger range of core positions, compared to Albury's original test (see Figure \ref{fig:NNcore}), makes for a more robust test of the efficiency as it better emulates the range of data which will be collected by a future full-sized FAST array. However, a \say{direct} comparison to Albury's results is also desirable. To this end, a further 100 showers at each of the four energies were simulated but with core positions inside the same central core region used by Albury (circle of radius 1\,km centred at (0,0)). This data set will be labelled the \say{Small Core Region}. Note that the $40\sim50$ showers falling in the same 1\,km circle in the Large Core Region data set will also be used as part of the Small Core Region data set for greater statistics. 
The first guess values were set to the true values randomly smeared by 40\,g\,cm$^{-2}$ in \Xmax{}, 10\% in energy, 3$\degree$ in arrival direction and 300\,m in core location respectively. This roughly simulates the resolution in the machine learning first guess used by Albury in his study. The TDR with the above improvements was then run on each data set and the reconstruction efficiency calculated as a function of energy. Before presenting the results, a precise explanation of the aforementioned smearing is given. 

\subsubsection{Parameter smearing}
To explain smearing, it is helpful to introduce some additional terminology. Consider first the shower geometry. Let $\theta_0$, $\phi_0$, $x_0$ and $y_0$ represent the true values of the shower zenith, azimuth and core $x/y$ parameters respectively. Now let $\theta_1$, $\phi_1$, $x_1$ and $y_1$ represent the estimated or reconstructed values of the parameters. The angle between the true and estimated vectors for the arrival direction is known as the \say{opening angle} $\alpha$ and is given by
\begin{equation}
    \alpha=\arccos\left(\sin(\theta_0)\sin(\theta_1)\cos(\phi_1-\phi_2) + \cos(\theta_0)\cos(\theta_1)\right).
\end{equation}
The \say{core distance} $d$ between the true and estimated core location is given by 
\begin{equation}
    d=\sqrt{(x_1-x_0)^2+(y_1-y_0)^2}.
\end{equation}
Assuming the differences between the true and estimated values of each parameter are independently normally distributed with mean 0 and variances $\sigma_\textrm{ang}^2$ (for $\theta$ and $\phi$) and $\sigma_\textrm{core}^2$ (for core $x/y$), then both $\alpha$ and $d$ follow a Rayleigh distribution
\begin{equation}
\label{eqn:rayleighDist}
    f(z)=\frac{z}{\sigma^2}\exp\left(-\frac{z^2}{2\sigma^2}\right).
\end{equation}
Here $z$ is either $\alpha$ or $d$ and $\sigma$ is the appropriate value of $\sigma_\textrm{ang}$ or $\sigma_\textrm{core}$. The angular resolution $\psi$ and core resolution $\delta$ are defined as the 68\% confidence interval of Equation \ref{eqn:rayleighDist}, i.e. the values below which 68\% of opening angles and 68\% of core distances lie respectively. Thus for a given angular (or core) resolution $\psi$
\begin{equation}
    \int_0^\psi\frac{z}{\sigma^2}\exp\left(-\frac{z^2}{2\sigma^2}\right)\,\textrm{d}z=1-\exp\left(-\frac{\psi^2}{2\sigma^2}\right)=0.68.
\end{equation}
Solving for $\sigma$ gives $\sigma=\psi/1.51$. Thus to \say{smear} the shower geometry by some resolutions $\psi$ and $\delta$ is then simply a matter of sampling Equation \ref{eqn:rayleighDist} using values of sigma equal to $\psi/1.51$ and $\delta/1.5$ to obtain a randomly varied opening angle and core distance. These values are then used together with the true values $\theta_0$, $\phi_0$, $x_0$ and $y_0$ to generate a new set of geometrical parameters $\theta_1$, $\phi_1$, $x_1$ and $y_1$. This emulates the uncertainty in some reconstruction procedure (first guess, SD reconstruction etc.). Smeared values of \Xmax{} and energy are obtained by sampling from Gaussians with means equal to the true values and standard deviations equal to the desired resolutions. For the use case presented here, the smeared parameters are used as the starting guess for the TDR. 

\vspace{5mm}

Note that when \textit{evaluating} the angular or core resolution of a reconstruction procedure, the underlying opening angle/core distance distributions may not \textit{exactly} follow Equation \ref{eqn:rayleighDist}. This is because the variances $\sigma_\theta^2$ and $\sigma_\phi^2$, and $\sigma_x^2$ and $\sigma_y^2$, are not always equal. However the resolutions $\psi$ and $\delta$ can still be calculated directly from the 68\% regions of the distributions. The terms opening angle, core distance and smearing will be used throughout the remainder of this thesis. 

\begin{figure}[t]
    \centering
    \includegraphics[width=0.92\textwidth]{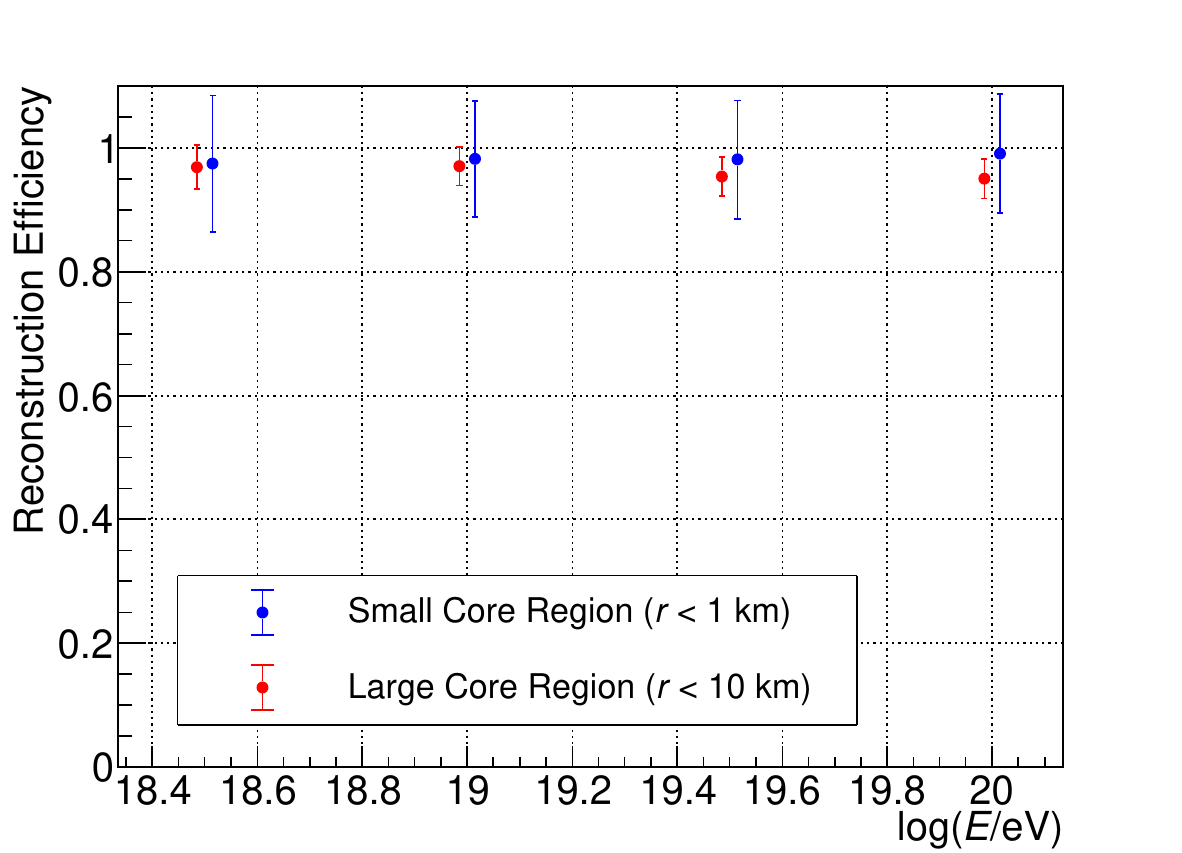}
    \caption{Reconstruction efficiency as a function of energy for the improved TDR. The simulated showers had energies of $10^{18.5}$, $10^{19}$, $10^{19.5}$ and $10^{20}$\,eV. The blue and red points, corresponding to the efficiency for the Small Core Region and Large Core Region data sets respectively, are plotted at $x$ values slightly above/below these energies for visibility.}
    \label{fig:newReconEff}
\end{figure}

\begin{figure}[t]
    \centering
    \includegraphics[width=0.49\linewidth]{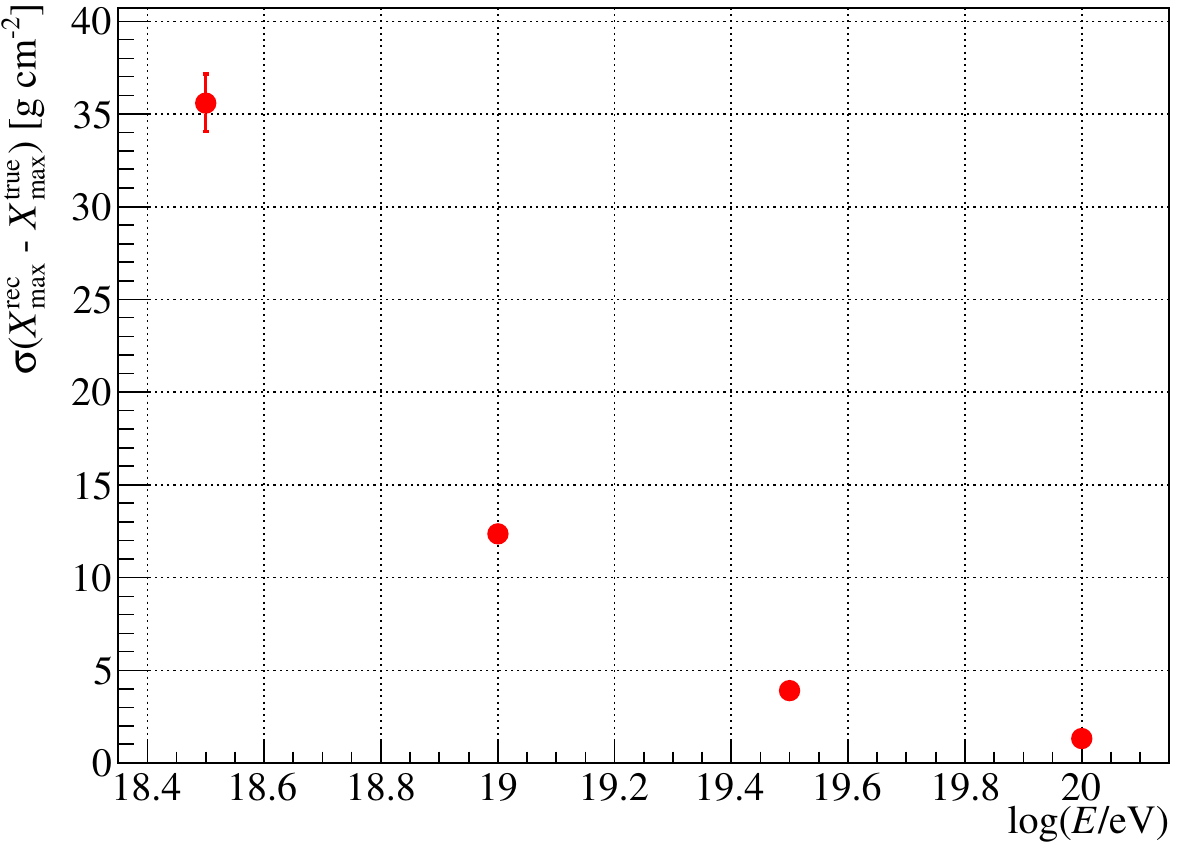}
    \includegraphics[width=0.49\linewidth]{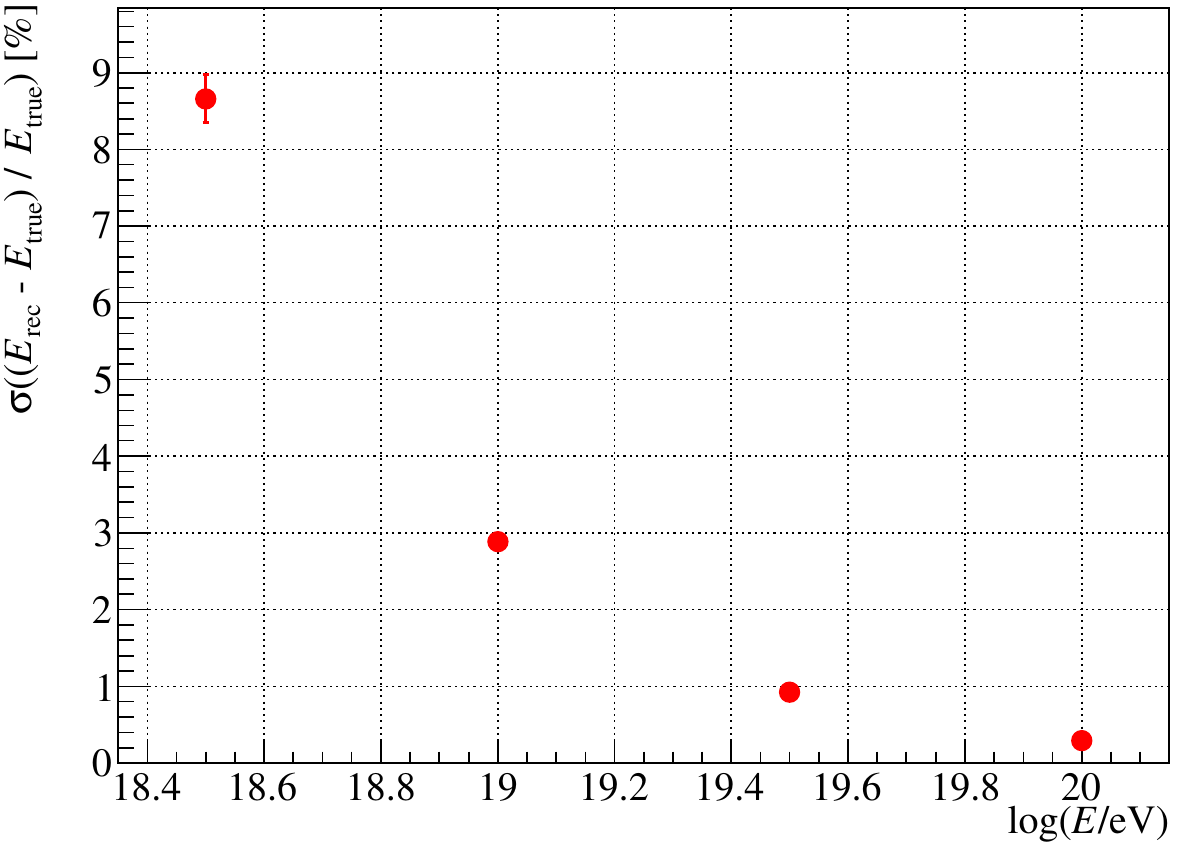}
    \includegraphics[width=0.49\linewidth]{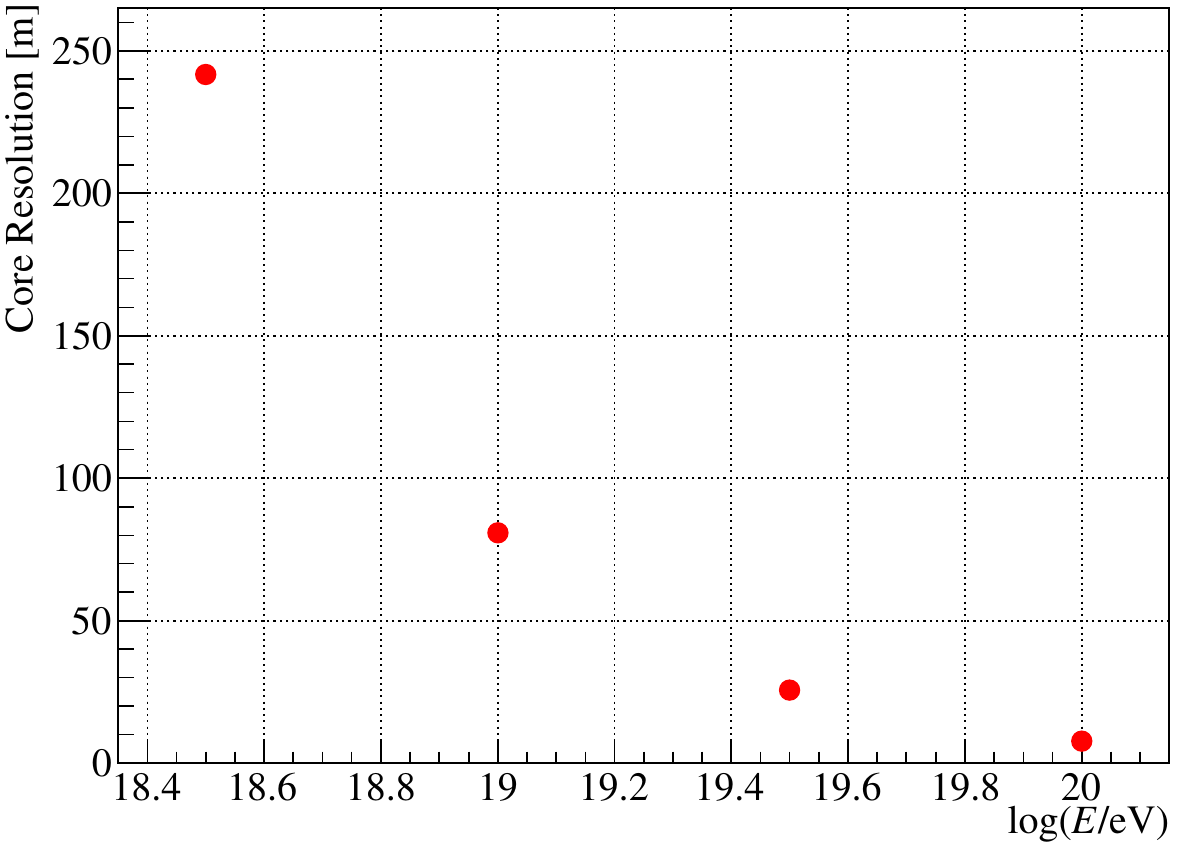}
    \includegraphics[width=0.49\linewidth]{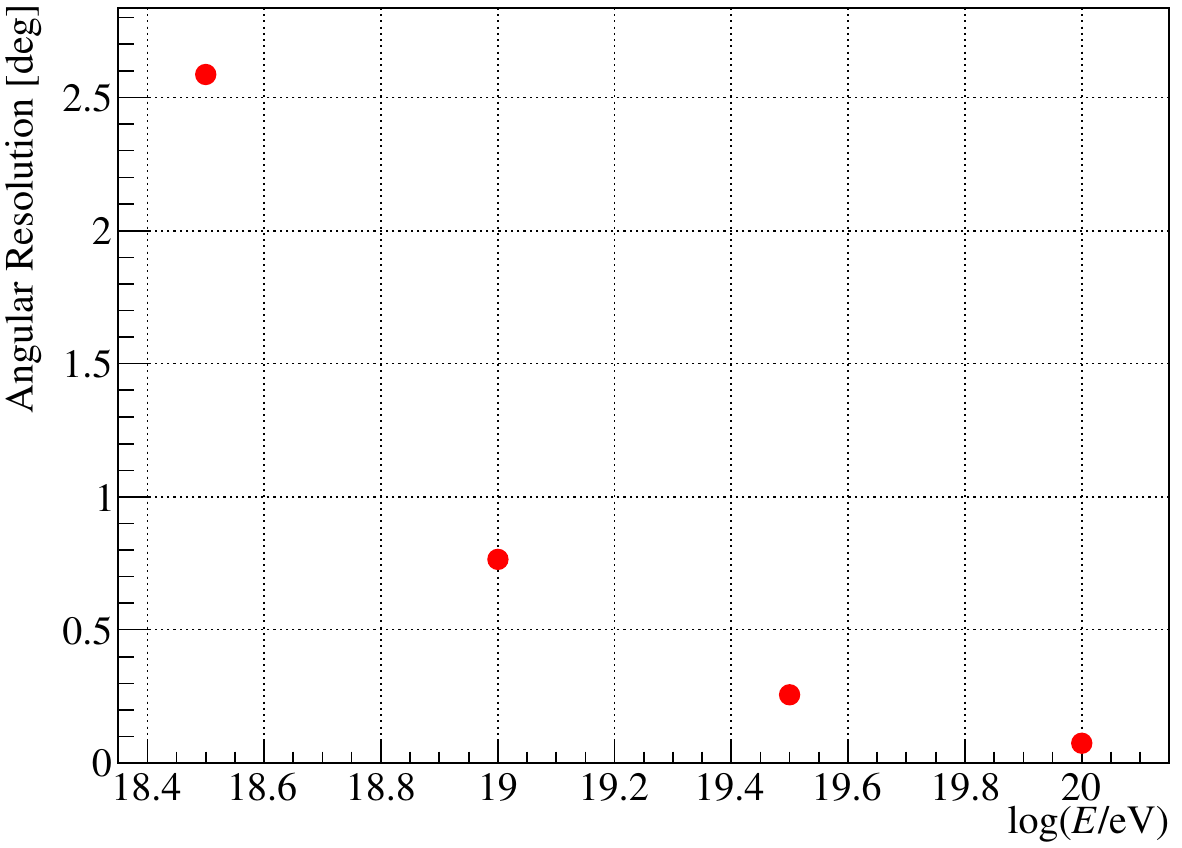}
    \caption{Resolutions in \Xmax{} (top left), energy (top right), core position (bottom left) and arrival direction (bottom right) as function of energy for the TDR applied to simulations of the FAST 3-Eye layout (data set 1). See the text for details on the first guess conditions.}
    \label{fig:newReconRes}
\end{figure}

\subsubsection{Results}
Figure \ref{fig:newReconEff} shows the reconstruction efficiency of the updated TDR at the four simulated energies. The trigger condition was that at least one PMT had a signal-to-noise ratio (SNR) $>6$ (see Section \ref{sec:snrcalc}). The red (blue) points show the efficiency for showers in the Large Core Region and Small Core Region data sets respectively. Directly comparing the results from the Small Core Region to Albury's, there is a dramatic improvement, with the efficiency being above 95\% across the whole energy range. As for the Large Core Region, the efficiency is $\sim97\%$ at both 10$^{18.5}$ and 10$^{19}$\,eV and $\sim95$\% at 10$^{19.5}$\, and 10$^{20}$\,eV, showing that the improvements hold for a wider range of showers.
Furthermore, there is no longer any reliance on an artificial scaling of the likelihood function. Of course, successful minimisation by itself will likely not be a sufficient condition for the success of the entire reconstruction chain. The success of the first guess method employed, together with various cuts on the quality of the data and reconstruction, will need to be taken into account for a more accurate description of the reconstruction efficiency. It should also be noted that the TDR performed here \textit{did not} fit the absolute time offset of the traces. Although doing so is not strictly necessary for simulations (as the starting time is fixed), it should be factored in when estimating the TDR's efficiency on data. This is left for future work. 
Nevertheless, the improvements demonstrated here are an important step towards the reliable application of the FAST reconstruction process to data.

\vspace{5mm}

Focusing on the results from the Large Core Region, of the 10$^{20}$\,eV showers which failed the minimisation, roughly 90\% failed due to reaching the maximum allowable number of simulations, nominally set to 500. Increasing this limit could allow more showers to be successfully reconstructed at the cost of additional computation time. Another alternative may be to fit the parameters over several iterations. For example, the first iteration could fix the \Xmax{}/energy values and fit only the arrival direction/core position. The second iteration could then fit all parameters using the newly estimated geometry. Investigating the performance of different minimisers/minimisation techniques may also be worthwhile, particularly those which are less affected by discontinuities in the objective function. In short, further investigation into the possible fail states of the TDR, particularly when fitting multiple parameters simultaneously, and different possible fitting routines should be considered for future work.

\vspace{5mm}

To conclude this section, the \Xmax{}, energy, core and angular resolutions of the reconstruction results from the efficiency test using the Large Core Region data set as a function of energy are shown in Figure \ref{fig:newReconRes}. The Gaussian fits to the event-by-event difference histograms for \Xmax{} and energy, and the cumulative distributions used to determine the core and angular resolutions are shown in Appendix \ref{apx:improvedReconFits}. At 10$^{19}$\,eV, the resolutions are already extremely good (\Xmax{}\,$\sim$\,13\gcm{}, $E$\,$\sim$\,3\%, core\,$\sim$\,80\,m, angular\,$\sim$\,0.8$\degree$) easily satisfying the GCOS requirements (Section \ref{sec:GCOS}). These resolutions only improve at higher energies, to a degree where they seem \say{too good to be true}. Indeed these results are unrealistic in the sense that they do not reflect what the true resolutions of a future FAST 3-Eye layout would be. The two primary reasons for this are listed below.
\begin{itemize}
    \item \textbf{First guess accuracy:} Unlike the constant values used for smearing here, the accuracy of a realistic first guess estimate would vary both with energy and the shower geometry (e.g. the parameters of a shower falling in the centre of the array, seen by multiple telescopes, would be better estimated than a shower falling on the edge, seen by only one/two telescopes). Furthermore, the smearing values used are overly optimistic, with the geometrical resolutions in particular being better than obtained by Fujii \cite{fujii2021latest}.
    \item \textbf{Idealised simulations:} The only source of randomness in these simulations (and indeed for all simulations in this thesis) was the Gaussian background noise added to the PMT traces. No shower-to-shower fluctuations, atmospheric/calibration uncertainties, or signal-baseline fluctuations were considered. Moreover there is no simulation of the effects of the electronics' response (e.g. signal saturation) or lateral width of the shower profile in the current FAST simulation. 
\end{itemize}
Despite these points however, it is not unreasonable to expect that a full sized FAST array could achieve resolutions on par with or possibly even surpassing those of Auger/TA for events observed in stereo. This is because FAST uses the timing and signal in each individual bin as information. Thus for large signals in multiple PMTs across 2/3 locations, there are in principle many more data points constraining the shower parameters than are typically used by the Auger/TA FD reconstructions ($\approx$number of triggered pixels in each camera). This is of course provided the FAST simulation is made to precisely and accurately reproduce data.

\section{$X_{\textrm{max}}$ and Energy Correlation}
\label{sec:xmaxEnergyCorrelatoin}
\begin{figure}[t]
    \centering
    \includegraphics[width=0.8\textwidth]{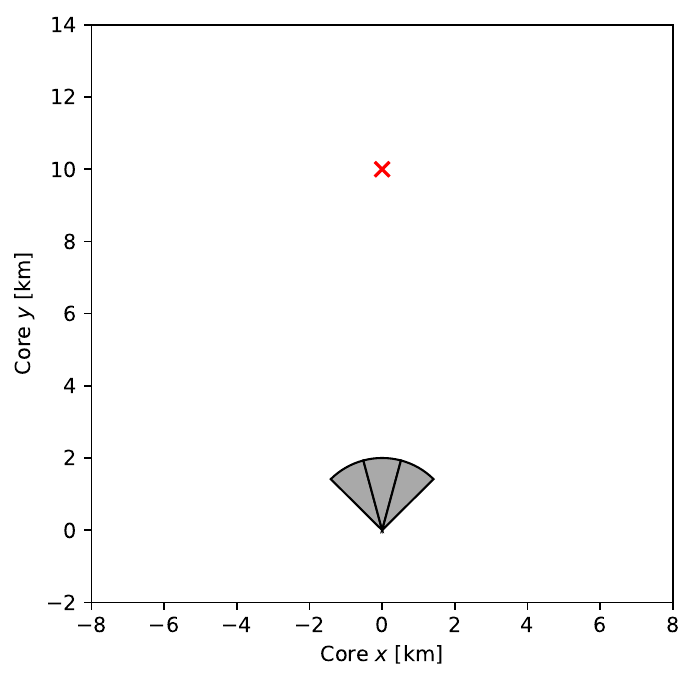}
    \caption{Telescope layout and core position (red $\times$) for the simulated showers in Section \ref{sec:xmaxEnergyCorrelatoin}.}
    \label{fig:XmaxEnergyCorrSimSetup}
\end{figure}

The second section of this chapter focuses on another issue with the TDR, namely a correlation between the reconstructed values of energy and \Xmax{}. This problem was identified when inspecting Albury's initial results regarding the performance of FAST in \say{hybrid mode} \cite{justin2020extending}. In hybrid mode, the shower geometry is provided by an independent SD and FAST only reconstructs \Xmax{} and energy. To investigate the issue the conditions of Albury's original study were reproduced. Albury used a set of 1000 simulations for his study, where the showers were set incident on the FAST@TA layout with a fixed core at (0\,km, 10\,km). All other simulation parameters were identical to the setup in Table \ref{tab:simparsExamp}. Figure \ref{fig:XmaxEnergyCorrSimSetup} shows the telescope layout and core position used. During the minimisation procedure, the geometrical parameters were kept fixed to their true values, leaving only the energy and \Xmax{} parameters to vary. The first guess values for energy and \Xmax{} were set to 30\,EeV and 850\,g\,cm$^{-2}$ respectively.

\begin{figure}[t]
    \centering
    \includegraphics[width=0.49\linewidth]{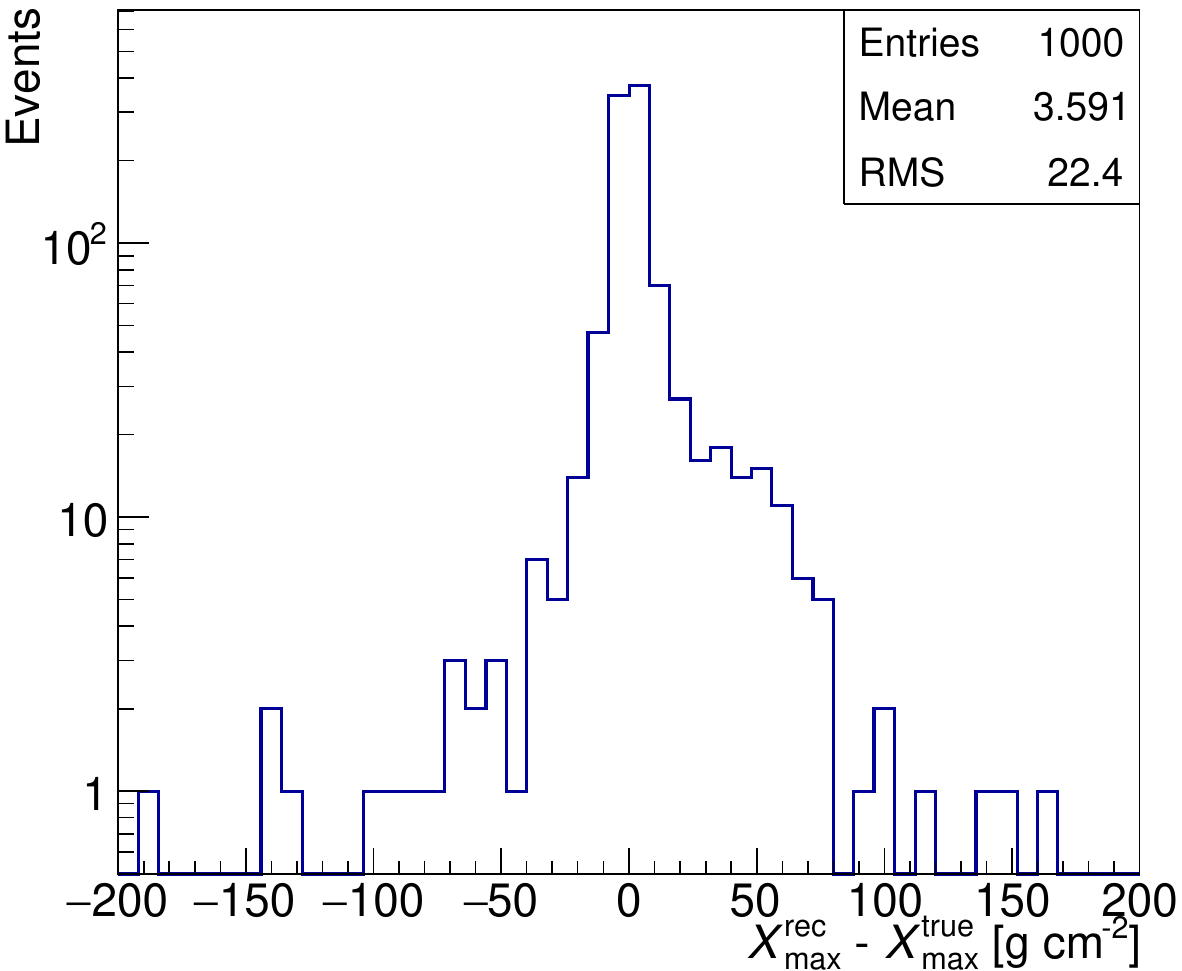}
    \includegraphics[width=0.49\linewidth]{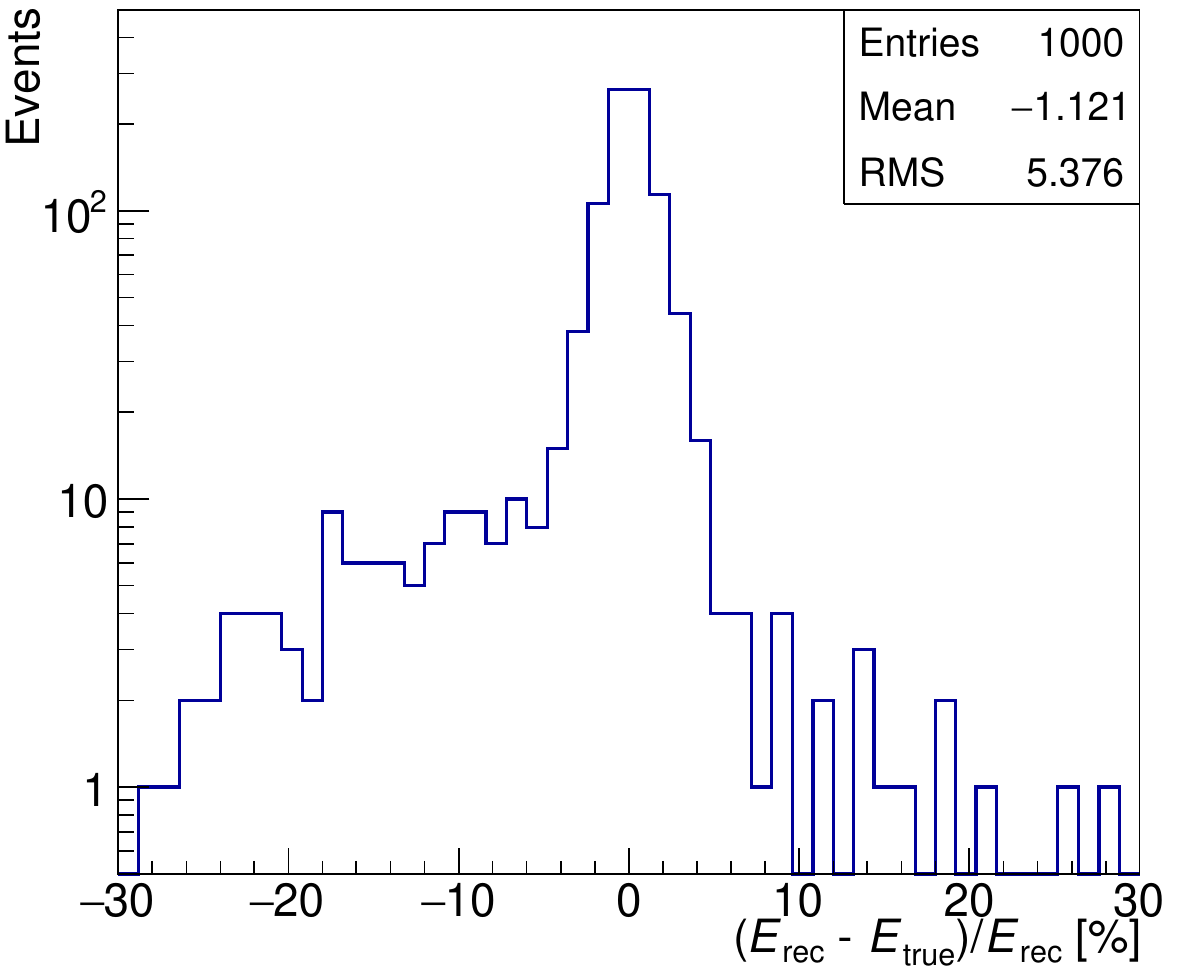}
    \caption{Results of reconstructing \Xmax{} and energy simultaneously for 1000 simulated events incident on the FAST@TA setup. The geometry is fixed to the true values.}
    \label{fig:hybridModeXmaxEnergy}
\end{figure}

\vspace{5mm}

\begin{figure}[]
    \centering
    \includegraphics[width=0.9\linewidth]{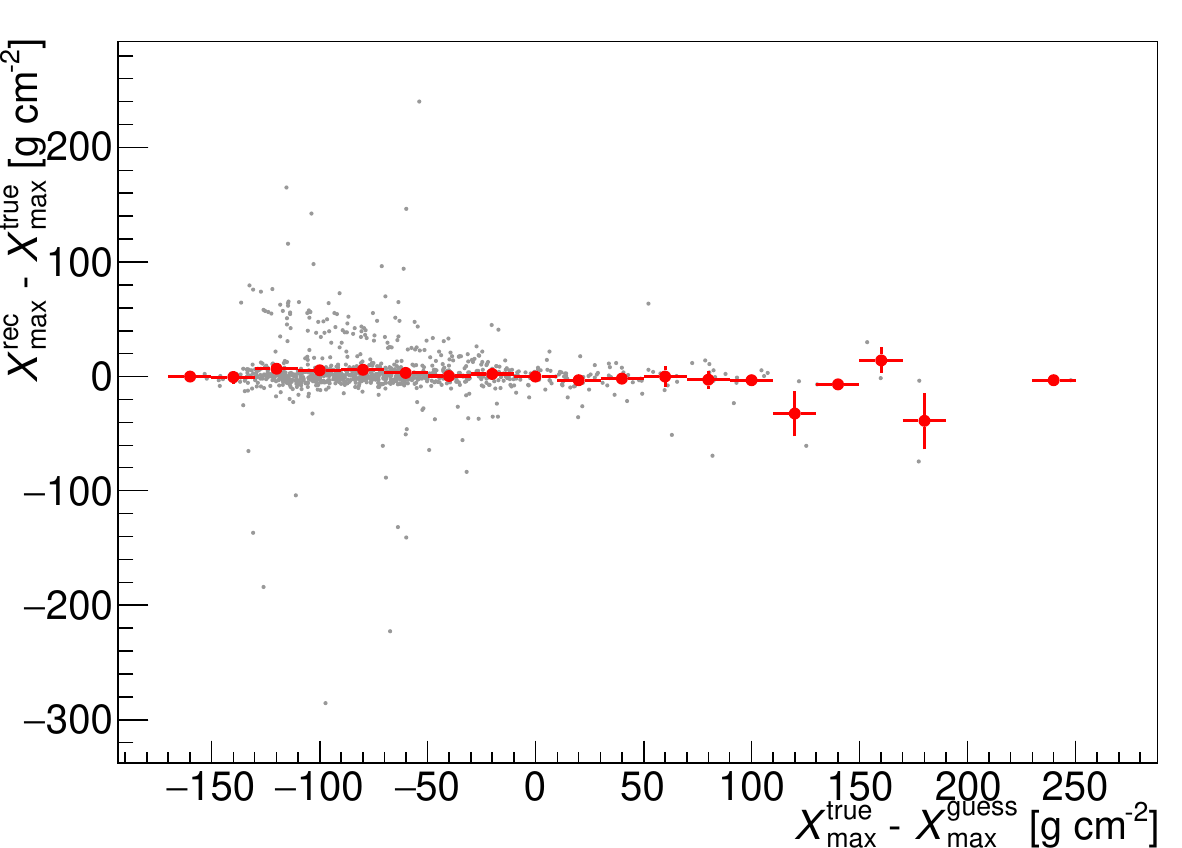}\\
    \includegraphics[width=0.9\linewidth]{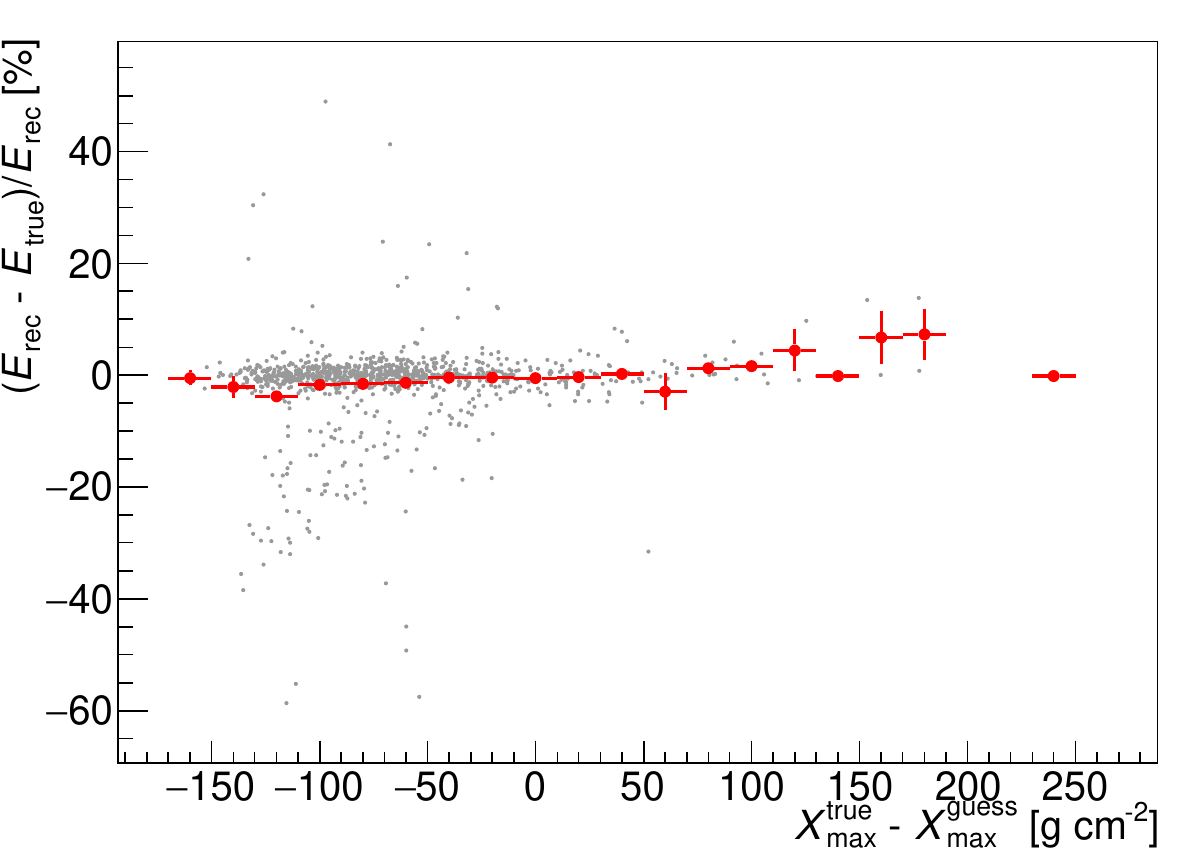}
    \caption{Difference between the reconstructed and true values for \Xmax{} (top) and energy (bottom) as a function of the difference between the true \Xmax{} and $X_\textrm{max}^\textrm{guess}=850$\gcm{}.}
    \label{fig:XErec_extension}
\end{figure}

After producing simulations with the same setup and reconstructing the showers as described, distributions of $\Delta$\Xmax{}$=X_\textrm{max}^\textrm{rec}-X_\textrm{max}^\textrm{true}$ and $\Delta{}E=(E_\mathrm{rec}-E_\mathrm{true})/E_\mathrm{true}$ were calculated. In future chapters, where the differences in reconstructed and true energies may be larger, $\Delta{}E=\ln(E_\mathrm{rec}/E_\mathrm{true})$ will be used. Future chapters will also make use of this $\Delta$ notation for other parameters - the convention will be $\Delta{x}=x_\textrm{rec} - x_\textrm{true}$. The $\Delta$\Xmax{} and $\Delta{}E$ distributions are shown in Figure \ref{fig:hybridModeXmaxEnergy} and are consistent with Albury's results. Both distributions have a central peak around 0 followed by a positive/negative tail respectively. One explanation as to the source of these tails is that, for showers with a true \Xmax{} far away from the first guess of 850\,g\,cm$^{-2}$, the TDR settled in a local minimum closer to the first guess. This hypothesis was tested by plotting $\Delta$\Xmax{} and $\Delta{}E$ against the difference between the true \Xmax{} and first guess \Xmax{}=850\gcm{}. The results are shown in Figure \ref{fig:XErec_extension}. The grey points represent individual events while the red points display the average of the grey points in bins of \Xmax{}. From these plots the following observations are made;
\begin{figure}[]
    \centering
    \includegraphics[width=0.92\linewidth]{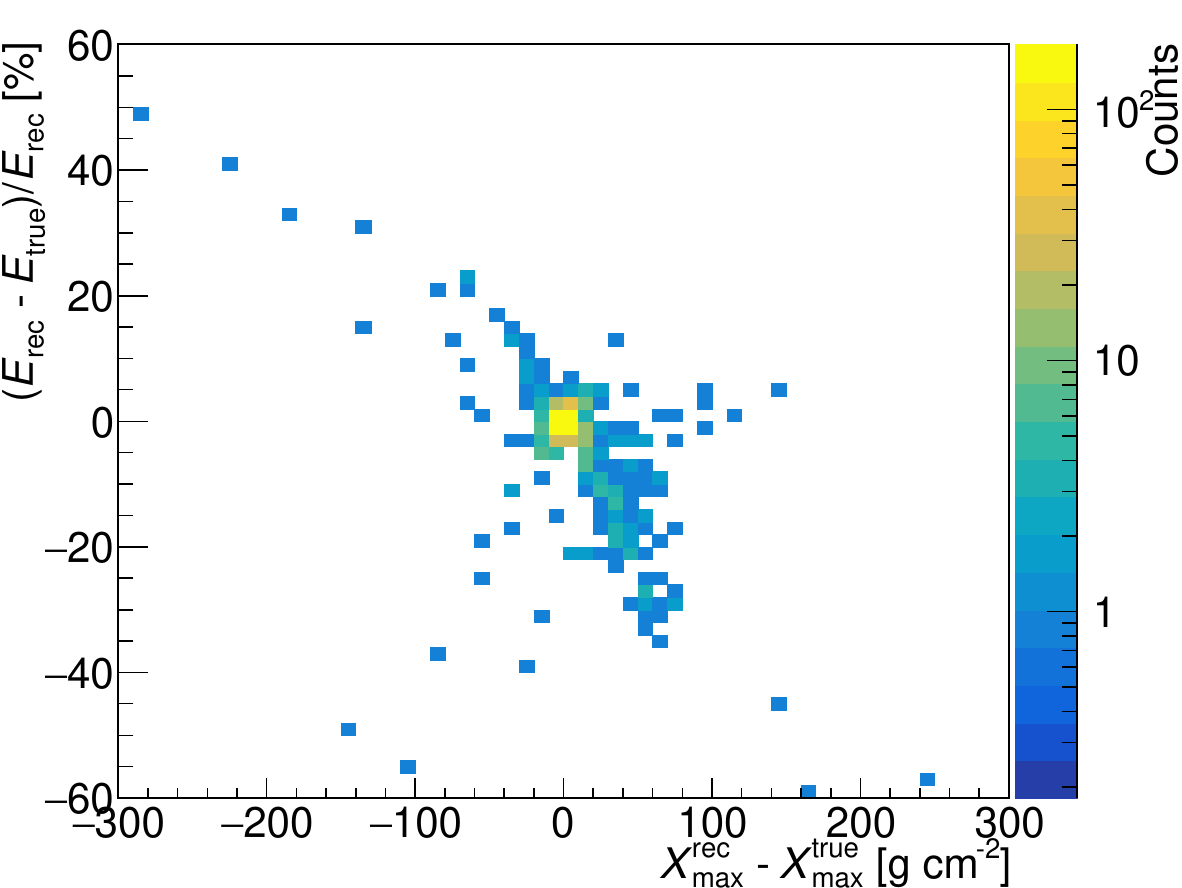}
    \includegraphics[width=0.88\linewidth]{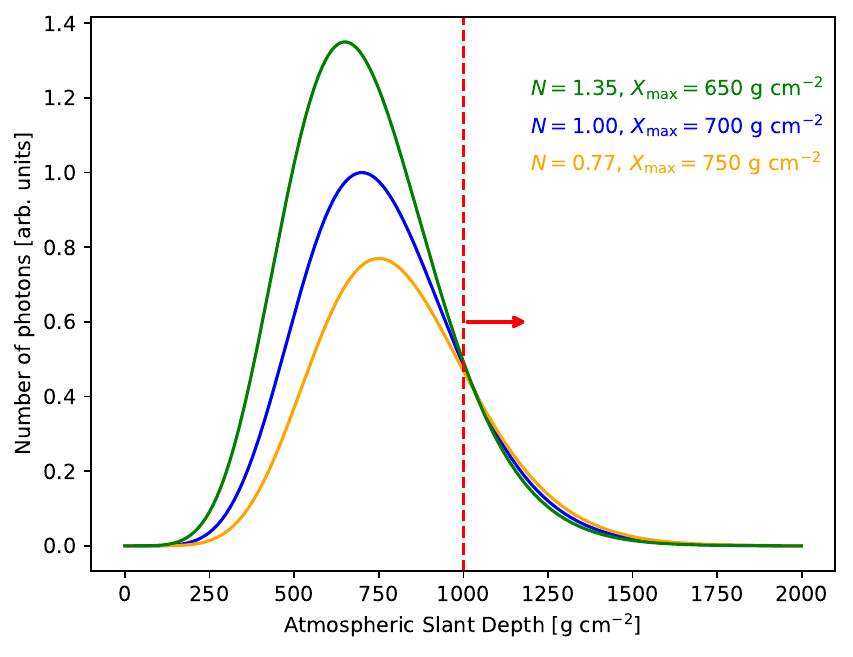}
    \caption{\textit{Top:} Correlation between the \Xmax{} difference and energy difference. \textit{Bottom:} Schematic diagram demonstrating how a degeneracy can arise when fitting \Xmax{} and energy. Right of the dashed red line represents the telescope FOV. The three Gaisser-Hillas profiles shown (normalisation/\Xmax{} in top right) appear very similar to the telescope.}
    \label{fig:diffXEcomp}
\end{figure}
\begin{figure}[]
    \centering
    \includegraphics[width=0.9\linewidth]{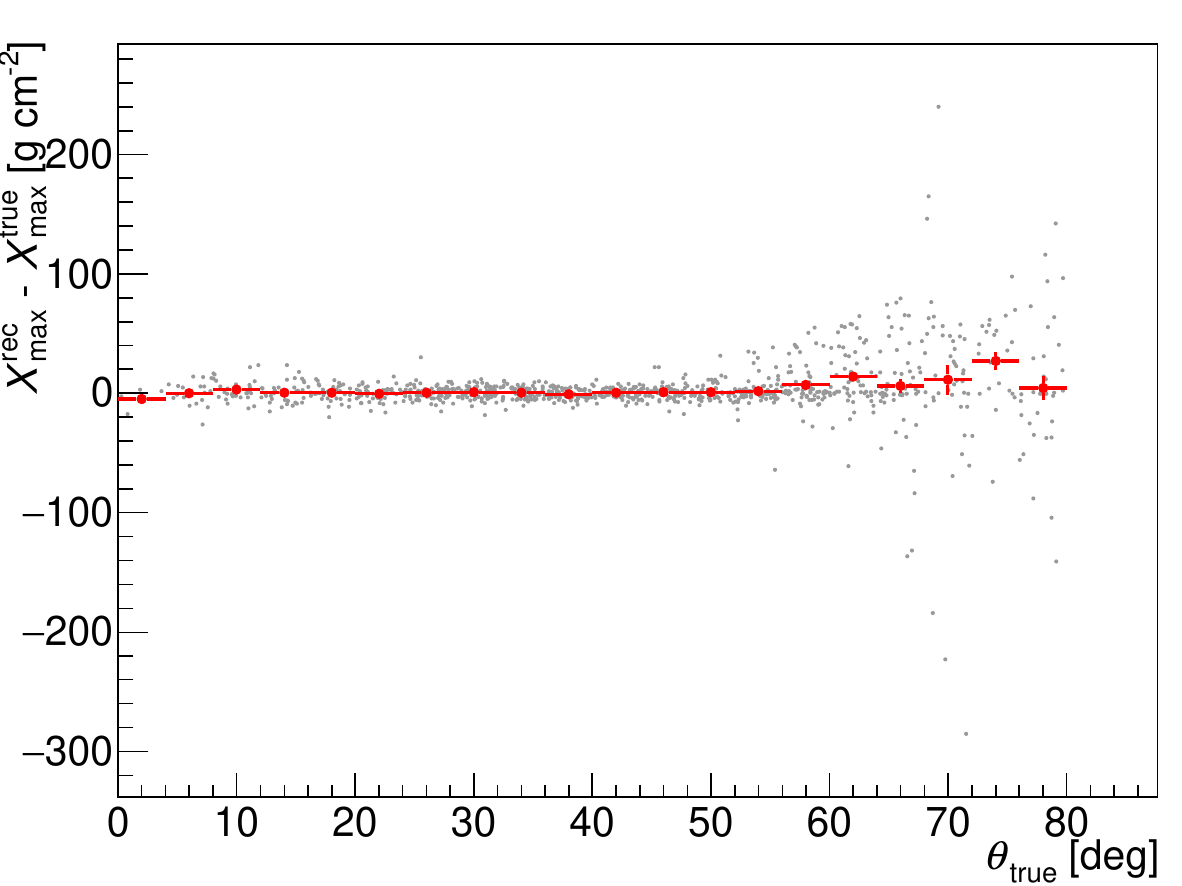}
    \includegraphics[width=0.9\linewidth]{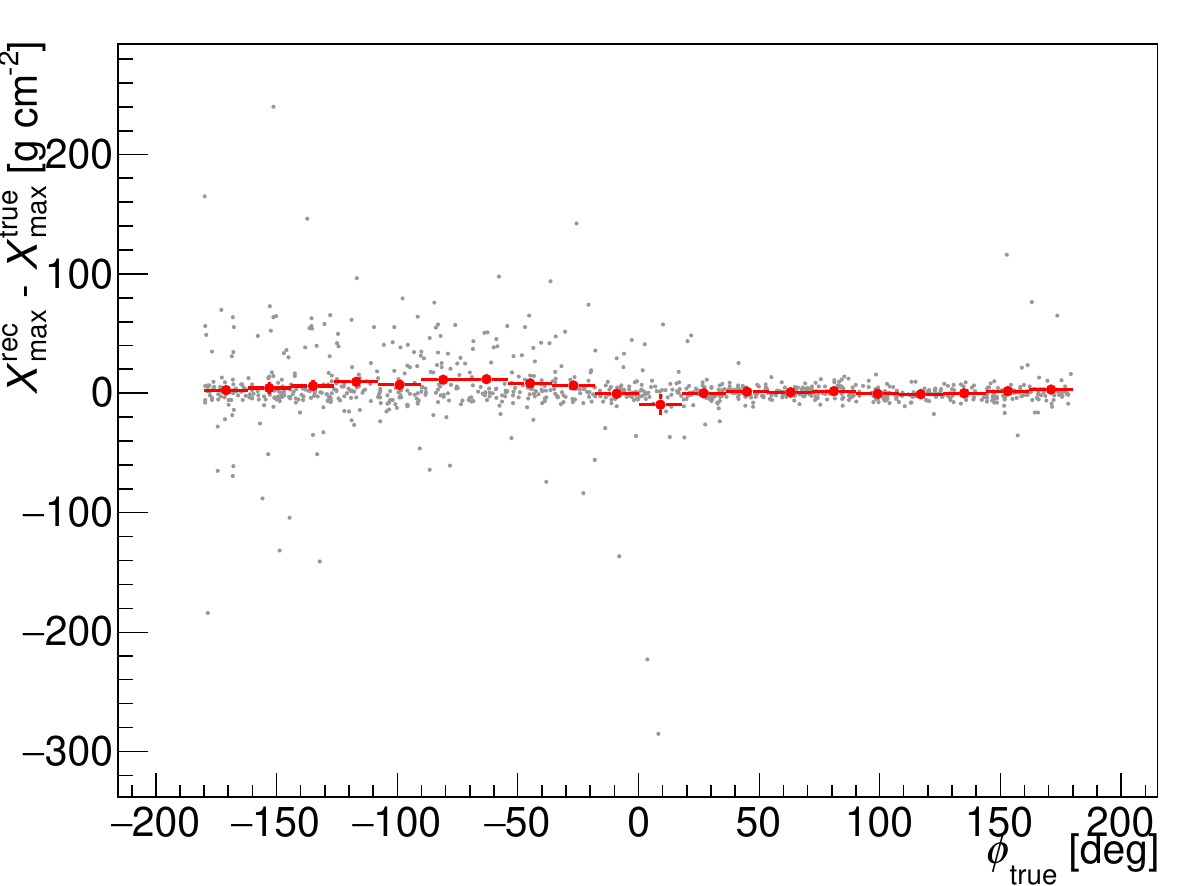}
    \caption{Difference in reconstructed and true \Xmax{} as a function of the shower zenith (top) and azimuth (bottom).}
    \label{fig:XEgeomdep}
\end{figure}
\begin{itemize}
    \item The tails of the histograms in Figure \ref{fig:hybridModeXmaxEnergy} are represented as diagonal streaks of grey points tending from high/low to 0 for $\Delta$\Xmax{}/$\Delta{}E$ as $X_\textrm{max}^\textrm{true}-X_\textrm{max}^\textrm{guess}$ approaches 0.
    \item The majority of events with $X_\textrm{max}^\textrm{true}-X_\textrm{max}^\textrm{guess}<-50\,$g\,cm$^{-2}$ are still reconstructed well. Thus the \say{failures} cannot be entirely attributed to a poor first guess. It is likely the particular geometry, specifically the arrival direction (since the core is fixed), of those events which caused the minimiser to settle in a local minimum.
\end{itemize}
The tails going in opposite directions in similar regions hints at a correlation between the reconstructed \Xmax{} and energy. To check this is the case, $\Delta$\Xmax{} vs. $\Delta{}E$ was plotted as a 2-D histogram. The result is shown in the top plot of Figure \ref{fig:diffXEcomp}. As expected there is a dense circle of points around (0,0) where the majority of events were reconstructed well, and a smaller sample of events showing a negative correlation.
To understand why this correlation arises, consider a single FAST telescope. In the FAST simulation, \Xmax{} and energy define the Gaisser-Hillas profile to be used when calculating the amount of light reaching the telescope. If this telescope observes a section of the Gaisser-Hillas where different combinations of \Xmax{} and energy give similar looking profiles then a degeneracy in the reconstructed \Xmax{} and energy values will arise. This is demonstrated in the bottom plot of Figure \ref{fig:diffXEcomp}. Here, the right side of the red dashed line indicates the telescope FOV. One can see that, from the telescope's perspective, the three profiles look very similar and may be indistinguishable depending on the signal to noise ratio. To properly distinguish the three profiles would require observing the showers over a larger atmospheric depth range, ideally including \Xmax{}.

\vspace{5mm}

Figure \ref{fig:XEgeomdep} shows $\Delta$\Xmax{} as a function of zenith and azimuth. This shows that the poor reconstructions primarily occur at large zenith angles $\gtrsim50\degree$ and for negative azimuth angles, which in this setup corresponds to showers coming from behind the telescope. 
This algins with expectations since inclined showers coming from behind the telescope will only be inside the telescope FOV for a small depth range. Moreover, the more inclined a shower the deeper the depth range over which the shower is observed. This leads to greater degeneracy between \Xmax{} and energy as the difference between profiles grows smaller at larger slant depths. This may explain why the variation in $\Delta$\Xmax{} appears to increase with larger values of $\theta$.

\vspace{5mm}

\begin{figure}[t]
    \centering
    \includegraphics[width=0.49\linewidth]{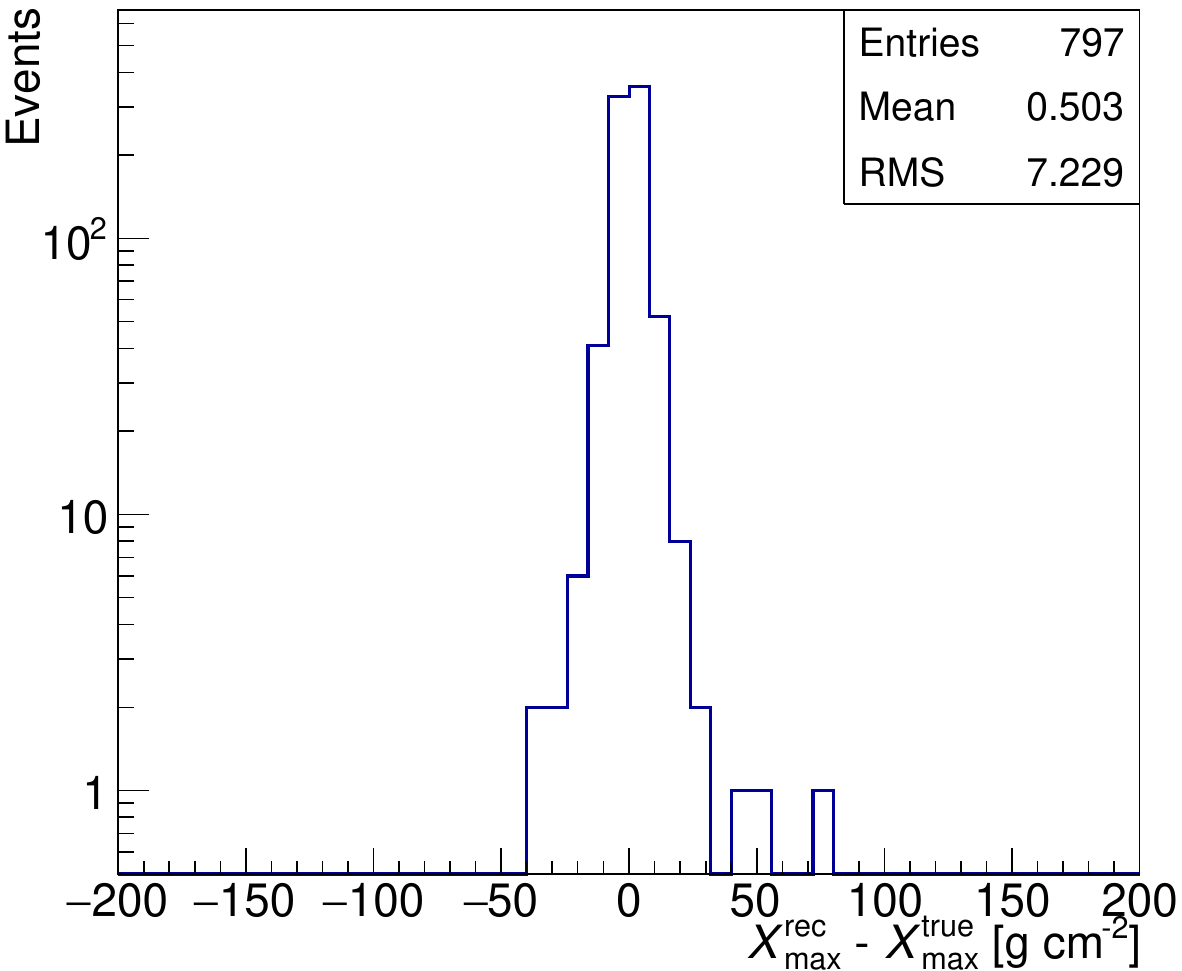}
    \includegraphics[width=0.49\linewidth]{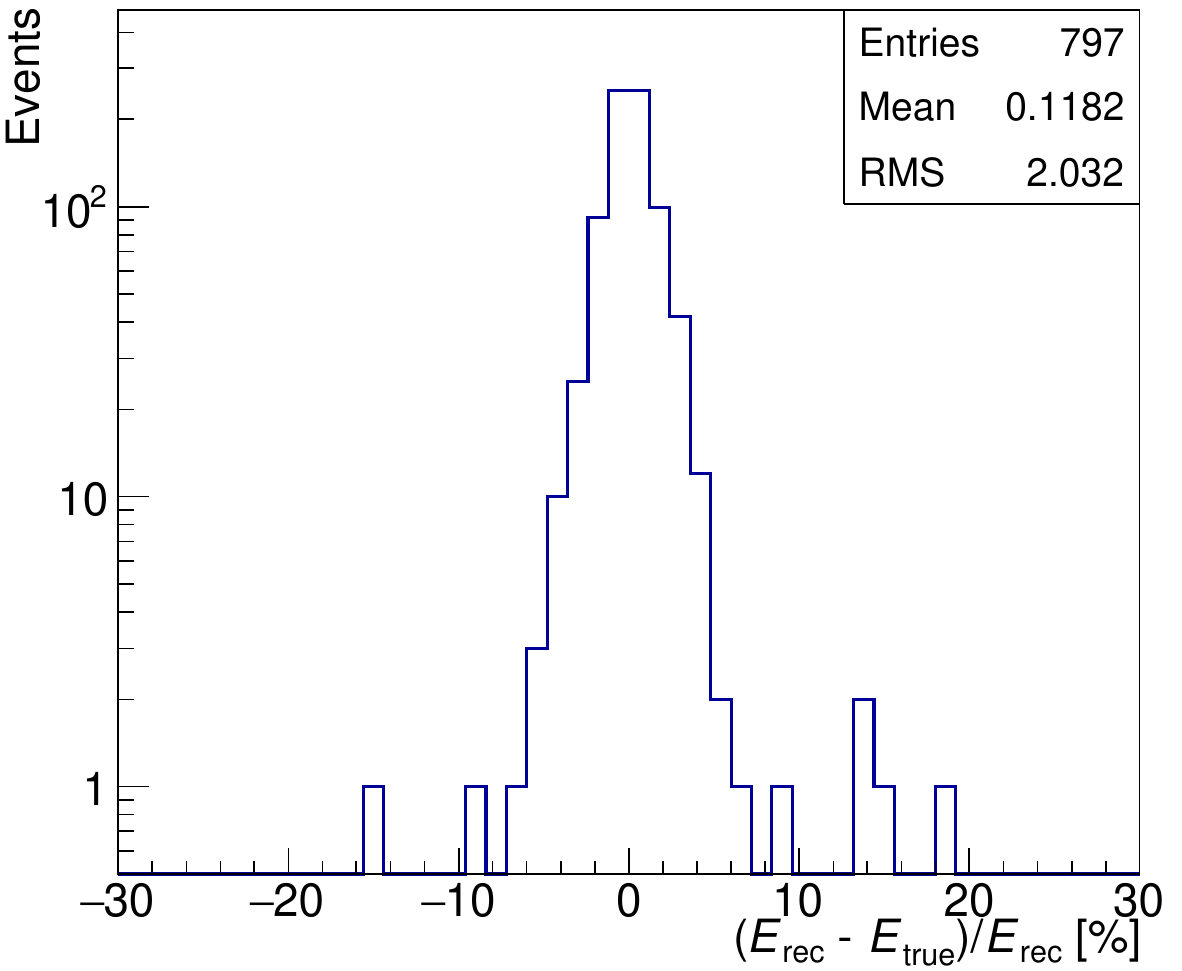}
    \caption{Same as Figure \ref{fig:hybridModeXmaxEnergy} but now including a cut on whether \Xmax{} lies in the FOV of the triggered telescopes. The tails of the distributions are no longer present.}
    \label{fig:hybridModeXmaxEnergyCut}
\end{figure}

To improve the overall precision of the reconstruction results, all events which did not have a reconstructed \Xmax{} in the FOV of at least one triggered telescope were removed. For this study a triggered telescope was a telescope with at least one PMT with a SNR $>6$. The resulting $\Delta$\Xmax{} and $\Delta{}E$ histograms are shown in Figure \ref{fig:hybridModeXmaxEnergyCut}. There are no longer tails in the distributions and the previous small positive/negative biases in $\Delta$\Xmax{}/$\Delta{}E$ are no longer present.
These results demonstrate that FAST, similar to current generation telescopes, can only reliably reconstruct \Xmax{} and energy when \Xmax{} is in the telescope FOV. Thus, unless stated otherwise, all analyses which make use of the TDR from here on will only include showers with \Xmax{} inside the FOV of at least one triggered telescope.
Since the TDR fits all six shower parameters simultaneously, this cut must be performed after the fitting. Specifically, once the TDR reaches a minimum, the fitted geometry and \Xmax{}, combined with the telescope locations and pointing directions, will be used to determine if the found minimum has an \Xmax{} in the FOV of at least one of the triggered telescopes. If not, that shower will be removed from the analysis.

\section{Summary}
In this chapter two problems with the original TDR procedure have been addressed. The first of these, a decrease in the reconstruction efficiency with energy, was solved by amending several sections of the FAST simulation which were causing discontinuities in the likelihood function. For a top-down approach which compares simulated traces to data bin-by-bin, it was found that the shower axis binning and signal re-binning methods must be carefully handled so as to ensure smooth changes in individual bin signals, thus giving a smooth likelihood function. The changes made to these methods to ensure smoothness were physically motivated (e.g. requirement for the edges of the directional efficiency map to tend smoothly to zero) and should be adopted by similar experiments aiming to reconstruct the properties of air showers via direct waveform fitting. The updated TDR, without any artificial likelihood scaling, has an improved efficiency at all energies compared to previous work, most notably at 10$^{20}$\,eV where the efficiency has increased from 30\% to 95\%. The second problem was a correlation between \Xmax{} and energy when fitted simultaneously. This problem was solved by removing showers from the analysis where \Xmax{} was not in the FOV of at least one triggered telescope, similar to the FOV cuts performed by Auger and TA on their FD data. This increased the precision of the TDR and removed slight biases in the $\Delta$\Xmax{} and $\Delta{}E$ distributions. Future work should focus on improving the signal re-binning solution implemented here, ideally by extending the range of the directional efficiency maps with real measurements, and on further analysis of the failure modes and possible degeneracies of the TDR. 

\chapter{First Guess Estimation I - Neural Networks}
\label{ch:ML}

The previous chapter focused on improving the underlying minimisation process by which FAST reconstructs the shower parameters, the TDR. Crucial to the success of this minimisation is the first guess of the shower parameters passed to the minimiser. If the first guess is not sufficiently close to the true parameters, then the minimiser may fall into a local minimum, or worse fail to reconstruct the shower parameters entirely. Accurately determining the shower geometry is the most important part of the first guess. This is because the shower geometry largely governs which telescopes/which PMTs view the shower. 
Previous efforts into estimating a first guess by Albury \cite{justin2020extending} and later Fujii \cite{fujii2021latest} have focused on applying machine learning techniques to simulated data from the FAST 3-Eye layout. Fujii's result showed that for showers seen by all three eyes with $E\gtrsim40$\,EeV, a simple feed-forward, \gls{dnn} can estimate the shower parameters with resolutions in arrival direction of $4.2\degree$, core location of 465\,m, \Xmax{} of 30\,g\,cm$^{-2}$ and energy of 8\%. Although these results are promising,
they are not applicable to the current prototype installations or to the soon-to-be deployed FAST mini-array. Developing a robust first guess method which can be applied to data from these installations is critical for validating the entire FAST reconstruction chain. The following two chapters investigate two different approaches for this purpose. This chapter focuses on using neural networks, expanding upon previous studies.

\section{Machine Learning Approach - Overview}
\label{sec:machineLearning}
Considering the reasonable success of the machine learning approach implemented in previous work with the FAST 3-Eye layout, it is natural to question whether a similar methodology could be applied to the FAST@Auger, FAST@TA and FAST mini-array layouts. Moreover, is there room to improve upon the basic model structure/architecture that was used in past studies? The following chapter attempts to address these questions by first replicating the basic feed-forward, DNN architecture previously used with FAST (with some minor adjustments) and then testing it on the above layouts. Variations of the model with additional complexity are then explored. The basic introduction to machine learning and neural networks given below provides most of the necessary background for understanding what follows.

\section{Machine Learning and Neural Networks}
Machine learning is a branch of artificial intelligence focused on having computers/machines \say{learn} to perform certain tasks based on exposure to (typically) large amounts of data. Broadly speaking, this learning process involves having the machine learning model make a decision/prediction based on some data, evaluating the accuracy/performance of this decision, and then updating the internal parameters of the model such that it's performance improves on the given task. There are several different types of machine learning algorithms, each with its own use case. Here, artificial neural networks in the context of regression problems will be introduced, since this is the architecture that has been used previously with FAST.

\vspace{5mm}

An artificial neural network is a type of learning algorithm modelled after the human brain. Like the brain, where information is passed between neurons via synapses, information in an artificial neural network is passed between layers of nodes via weights. Each node stores some information, for example a real continuous variable, and this information is passed to nodes in subsequent layers via the \textit{learned} weights. A typical network consists of an input layer (data from the user), a series of intermediate or hidden layers where the inputs are transformed, and finally an output layer (prediction).

\vspace{5mm}

\begin{figure}
    \centering
    \includegraphics[width=1\linewidth]{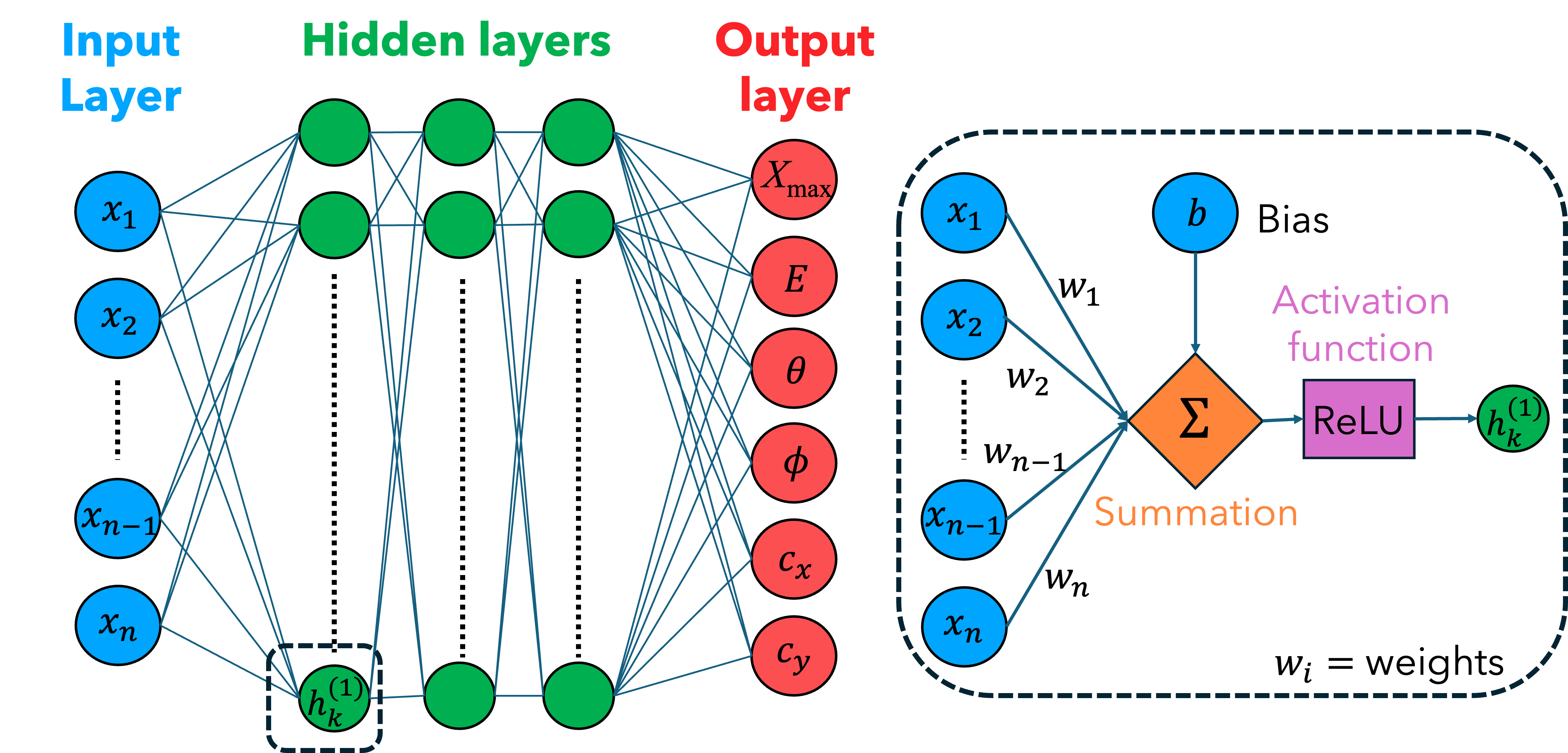}
    \caption{Diagram showing the basic structure of a DNN using the previous implementations tested with FAST as a reference. For these implementations the output parameters were the standard six shower parameters ($c_{x/y}$ here refer to core $x$/core $y$) and the activation function used for the hidden layers was ReLU. The box on the right shows a zoomed in view of how the value of a node in one of the hidden layers, in this case node $h_k^{(1)}$, is calculated based on the information from nodes in the previous layer.
    }
    \label{fig:NNdiagram}
\end{figure}

To model the often complex, non-linear relationships between input and output variables, activation functions are applied between node layers. The transformation of data $\vec{x}$ between a layer $i$ and the subsequent layer given an activation function $f(x)$ can be written as
\begin{equation}
    \vec{x_{i+1}}=f(W\vdot{}\vec{x_i}+\vec{b})
\end{equation}
where $W$ is a matrix of learned weights and $\vec{b}$ is a learned bias term. Choices of activation function vary depending on the application, however a popular choice for regression problems (like estimating the continuous shower parameters) is the rectified linear unit or ReLU, defined as 
\begin{equation}
    \textrm{ReLU} = \begin{cases} 
      x & x> 0 \\
      0 & x\leq0. 
   \end{cases}
\end{equation}
Training the network involves having the network predict the outputs and then comparing the prediction to the true values using a \say{loss function}. The loss function used in this work is the standard mean squared error (MSE), defined as 
\begin{equation}
    \textrm{MSE} = \frac{1}{n}\sum^n_{i=1}\left(y_i^{\textrm{pred}}-y_i^{\textrm{true}}\right)^2
\end{equation}
for output parameters $\{y_1,...y_n\}$. The weights of the neural network are then updated, typically via a process called back-propagation, in order to minimise the loss function. The task of an artificial neural network then is to find the optimal weights which minimise the loss function. An illustration of a basic neural network is shown in Figure \ref{fig:NNdiagram}, using previous implementations with FAST as a reference. Further information on machine learning and neural networks can be found in \cite{heaton2018ian}.

\section{SNR Threshold Calculation}
\label{sec:snrcalc}
Before making an estimate of the shower parameters, it is important to identify which PMTs contain \say{useful} signal from the observed event. 
The term \say{useful} here loosely refers to signals which, when incorporated into some first guess method, improve the accuracy of the overall estimate. For example, including the characteristics/features of a PMT trace which is dominated by background noise is likely to only add additional uncertainty to the result. Such a trace would not be considered \say{useful} in this context. That being said, it is also important not to overlook small but valuable signals which could help constrain the shower parameters. Since a single FAST telescope consists only of four pixels, extracting as much information as possible from each individual trace is vital for obtaining the best possible first guess.

\vspace{5mm}

The first step in deciding whether or not a pixel should be included in some first guess procedure is to determine whether the pulse from said pixel surpasses an appropriate \gls{snr} or \say{significance} threshold. Pixels which meet this threshold will be referred to as having passed the \say{threshold trigger}. Note that in both simulations and data (depending on the algorithm used), pixels passing the threshold trigger may not contain signal from an extensive air shower. This could be due to random fluctuations in the signal or, in the case of data, from artificial light sources such as aeroplanes or laser light from the central laser facilities at Auger and TA. Pixels which are deemed to contain signal from an extensive air shower will be said to have passed the \say{event level trigger} and will be used in the first guess methods explored in this work. See Section \ref{sec:additionalTriggering} for further discussion on the event level trigger.

\subsection{Cross Check with Previous Work}
In Albury's original attempt to use machine learning techniques for the first guess estimation, he utilised the same algorithm as used by Auger to determine the SNR of each PMT pulse. The algorithm is as follows; for a single PMT trace, the bins $k_{\textrm{start}}$ and $k_{\textrm{stop}}$ which maximise the ratio
\begin{equation}
\label{eqn:newsnr}
    \textrm{SNR} = \frac{S}{N} = \frac{\sum_{i=k_{\textrm{start}}}^{k_{\textrm{stop}}-1}S_i}{\sigma_{\textrm{nsb}} \sqrt{k_{\textrm{stop}}-k_{\textrm{start}}}}
\end{equation}
are determined. Here, $S_i$ is the signal in the $i^{\textrm{th}}$ bin and $\sigma_{\textrm{nsb}}$ is the standard deviation of the background noise estimated using a region of the trace deemed to contain no signal. Although in data this value varies from night-to-night, the average value, based on the analysis of FAST@TA coincidence data (see Section \ref{sec:coincAnal}), is $\sim11$\,p.e./100\,ns. This roughly matches the nominal value $\sigma_{\textrm{nsb}}=10$\,p.e./100\,ns used in simulations when adding artificial noise. Lastly, the minimum and maximum number of bins allowed between $k_{\textrm{start}}$ and $k_{\textrm{stop}}$ are set to 3 and 29 respectively. Note that in Albury's thesis this equation was written incorrectly with the signals $S_i$ being summed up to the $k_\textrm{stop}$ bin rather than the $k_\textrm{stop}-1$ bin. 

\begin{figure}[]
    \centering
    \includegraphics[width=0.9\textwidth]{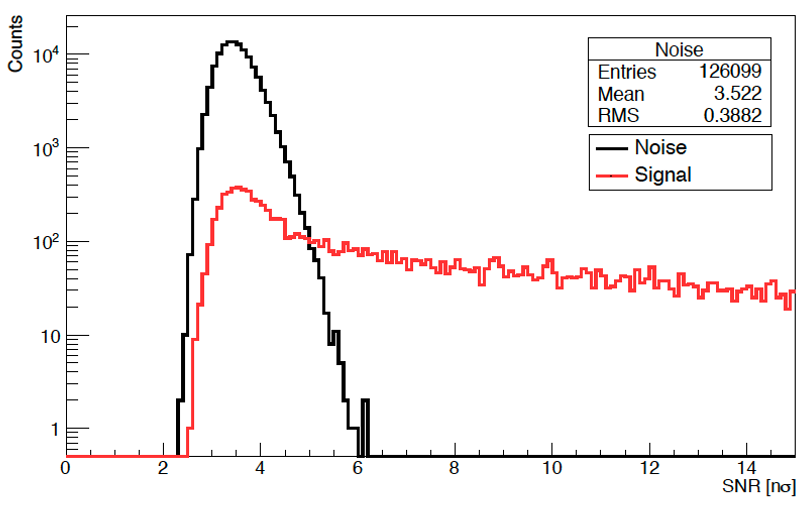}
    \caption{Histograms showing the SNR for traces with (red) and without (black) signal from simulated events as calculated by Albury \cite{justin2020extending}.}
    \label{fig:JustinSNRThreshResult}
\end{figure}
\begin{figure}[]
    \centering
    \includegraphics[width=0.9\textwidth]{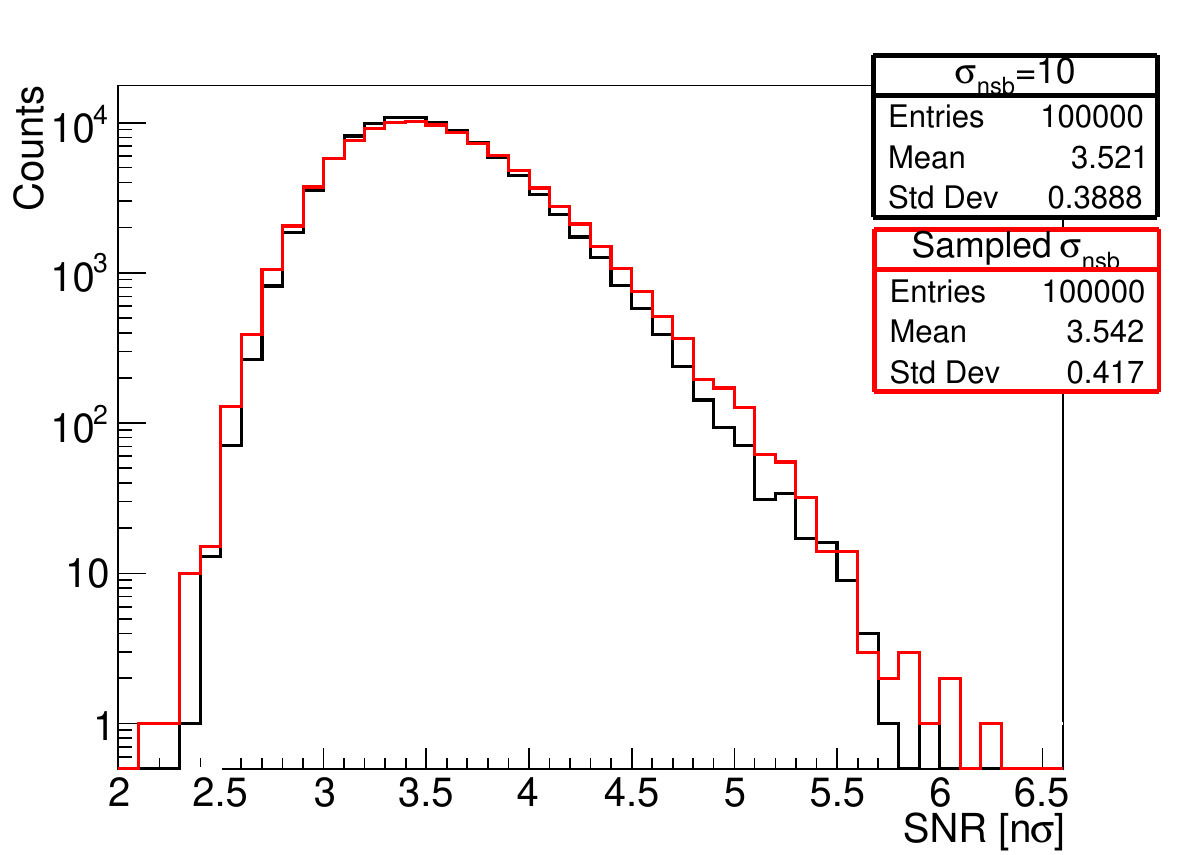}
    \caption{Distributions of SNR as calculated using Equation \ref{eqn:newsnr} applied to Gaussian noise generated from $\mathcal{N}(\mu=0,\sigma_\textrm{nsb}=10)$. The results when using a value of $\sigma_\textrm{nsb}$ estimated from the first 200 samples from of the trace and when using a fixed value of $\sigma_\textrm{nsb}=10$ are shown in red and black respectively.}
    \label{fig:jusnoise}
\end{figure}

\vspace{5mm}

To determine the appropriate SNR threshold above which a PMT is considered to have useful/significant signal, Albury used a simulated dataset and compared the distribution of SNRs for traces with no signal (i.e. pure Gaussian noise with $\sigma_{\textrm{nsb}}=10$\,p.e./100\,ns) vs. traces with signal\footnote{Exactly how these two cases were defined is not explained.}. The corresponding plot is shown in Figure \ref{fig:JustinSNRThreshResult}. From this plot and analysing the rate of false positives for different thresholds, Albury settled on an SNR cutoff of 5 for significant pulses. When attempting to reproduce this result as a cross-check, a slight inconsistency in the noise response between Albury's result and the above algorithm was identified. To reproduce the noise response, 100,000 traces of length 1,000 with points sampled from $\mathcal{N}(\mu=0,\sigma_\textrm{nsb}=10)$ were generated. Equation \ref{eqn:newsnr} was then used to determine the maximum SNR of each trace using the first 200 bins to estimate $\sigma_{\textrm{nsb}}$ (should be $\sim10$). The result is shown by the red histogram in Figure \ref{fig:jusnoise}. The RMS of the distribution is noticeably different from Albury's result, implying some difference in the calculation method. Performing the same analysis but using a fixed value of $\sigma_{\textrm{nsb}}=10$ in Equation \ref{eqn:newsnr} gives the black histogram in Figure \ref{fig:jusnoise}. With this change, both the mean and RMS of the distribution agree with Albury's result within statistical uncertainty. For the purposes of applying this algorithm to search for signal in data, $\sigma_{\textrm{nsb}}$ should be estimated from the trace and not assumed to be a particular value, as the background level can vary significantly across different nights (see Figure \ref{fig:fundamentalPlots}). Therefore, from this point onwards when determining the SNR of a FAST pixel pulse, Equation \ref{eqn:newsnr} will be used with $\sigma_{\textrm{nsb}}$ estimated from the trace as described above.

\begin{figure}[t!]
    \centering
    \includegraphics[width=0.49\textwidth, trim={0.8cm 1cm 1.2cm 0.5cm}]{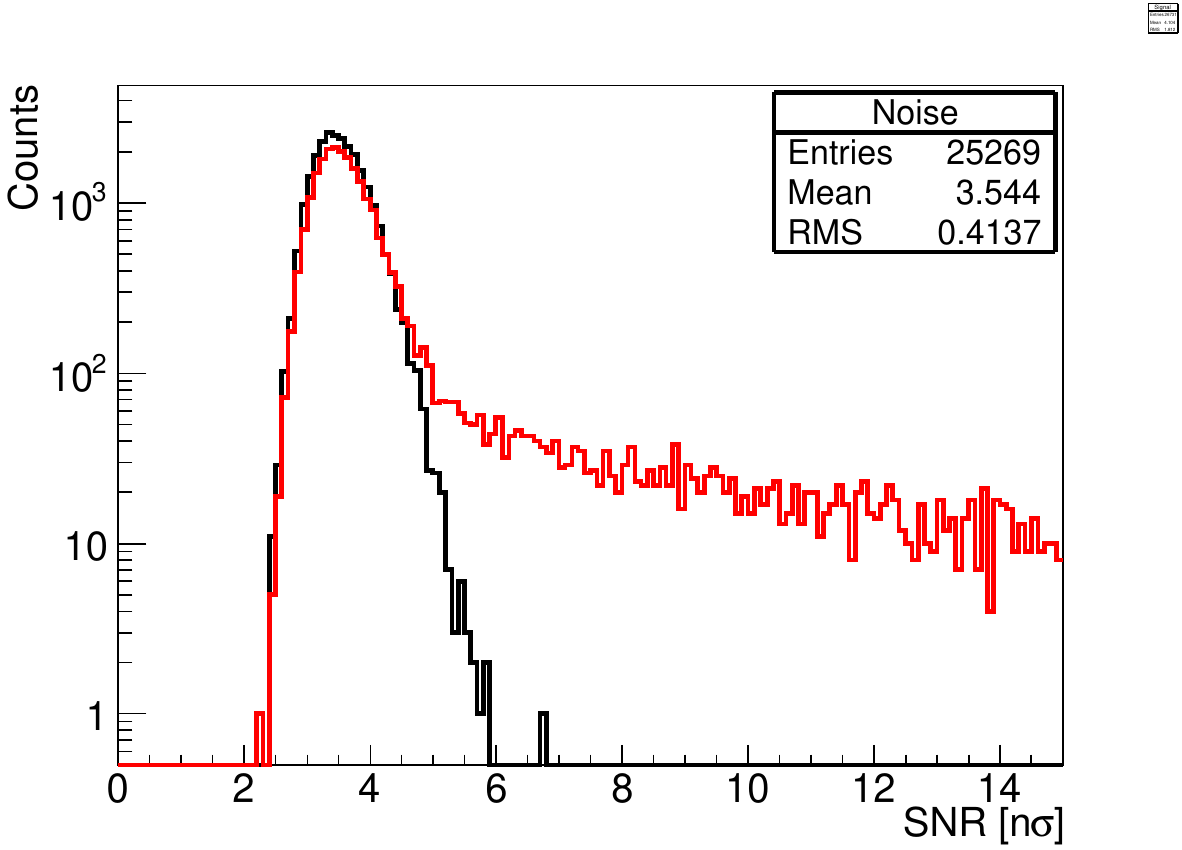}
    \includegraphics[width=0.49\textwidth, trim={0.8cm 1cm 1.2cm 0.5cm}]{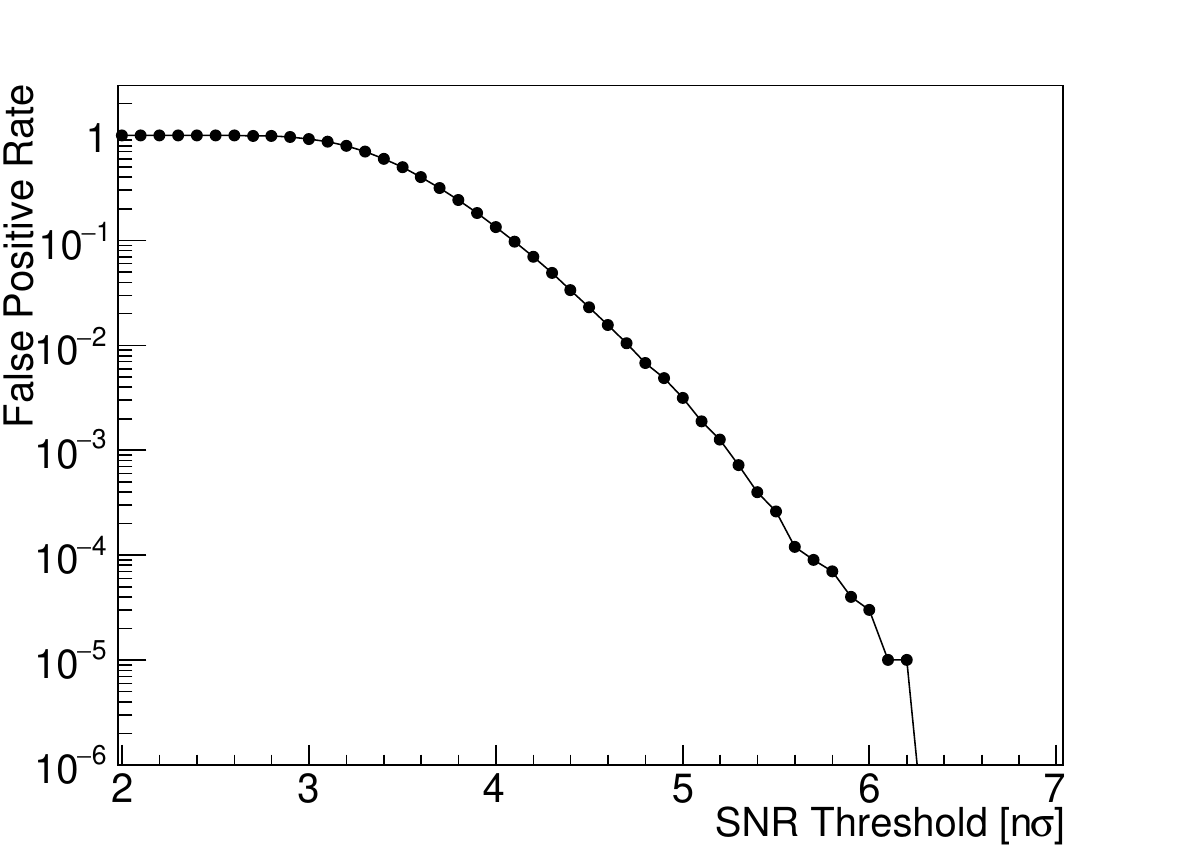}
    \caption{\textit{Left:} Maximum SNR values calculated using Equation \ref{eqn:newsnr} for simulated traces 
    with (red) and without (black) signal. \textit{Right:} False positive rate as a function of the SNR threshold. The rate of false-positives above an SNR threshold of 6 is negligible.}
    \label{fig:threshdetermine}
\end{figure}

\subsection{Threshold Determination}
A similar analysis to Albury's is now performed to confirm the appropriate SNR threshold. PMT traces from the simulated dataset in Section \ref{sec:MLdataset} are used. Traces with all 0 values (before adding artificial noise) are labelled as containing no signal. All other traces are treated as containing signal regardless of how small.Histograms of the SNR for pixels with (red) and without (black) signal are shown in the left plot of Figure \ref{fig:threshdetermine}. The long tail of the signal distribution and alignment of the signal and noise distribution peaks are consistent with Albury's result. The difference in the relative scales of the distributions is due to the dataset used here having a higher percentage of PMTs viewing the showers.  The right plot of Figure \ref{fig:threshdetermine} shows the false positive rate as a function of the SNR threshold calculated using the red histogram from Figure \ref{fig:jusnoise}. The false positive rate is defined as the number of noise traces falsely classified as signal for a given threshold divided by the total number of noise pulses. For a maximum SNR threshold of 5, the false positive rate is $\sim10^{-3}$. For a threshold of 6 it drops to $\sim10^{-5}$. Although there are potentially many valuable signals with SNR values between 5 and 6, as can be seen in the left plot of Figure \ref{fig:threshdetermine}, in this work a conservative approach is taken and the SNR threshold set to 6. This is done to reduce the chance that pixels not viewing the shower randomly pass the threshold trigger 
and subsequently negatively impact the first guess. For example, for the FAST mini-array layout, with a total of 24 pixels, a false positive rate of $10^{-3}$ equates to approximately 1 in $40\sim50$ events having a false threshold trigger. Moreover, a higher threshold will result in higher quality signals, in theory leading to a more precise reconstruction of the shower parameters.  
Although setting a higher SNR threshold reduces the sensitivity of FAST to lower energy/more distant showers, the trade-off is considered worthwhile for more reliability in the first guess. A more sophisticated signal search algorithm and/or additional trigger conditions could be used to help remove random threshold triggers and lower the SNR threshold. This is partly explored in Section \ref{sec:additionalTriggering}. For the following chapter however, the threshold trigger and event level trigger will be considered equivalent i.e. all pixels in an event with an SNR $>6$ will be used when making the first guess.

\section{Data Sets}
\label{sec:MLdataset}
 
The ultimate goal of training any machine learning model is to apply said model to new, unseen data. It is therefore crucial that the range of shower parameters aiming to be reconstructed is sufficiently represented in the training process. Although the target energy range of FAST is primarily $>10^{19}$\,eV, the majority of events observed by the FAST prototypes thus far have energies between $10^{17}$ and $10^{18.5}$\,eV (see Section \ref{sec:coincAnal}). Thus, in order to apply these models to data, it is necessary to train on a wide range of energies. The target energy range of $17.5<\log(E\mathrm{/eV})<20$ is chosen for this work. This covers more than $70\%$ of the coincidence events observed by the FAST prototypes thus far. 

\vspace{5mm}

\begin{figure}[]
    \centering
    \includegraphics[width=0.9\linewidth]{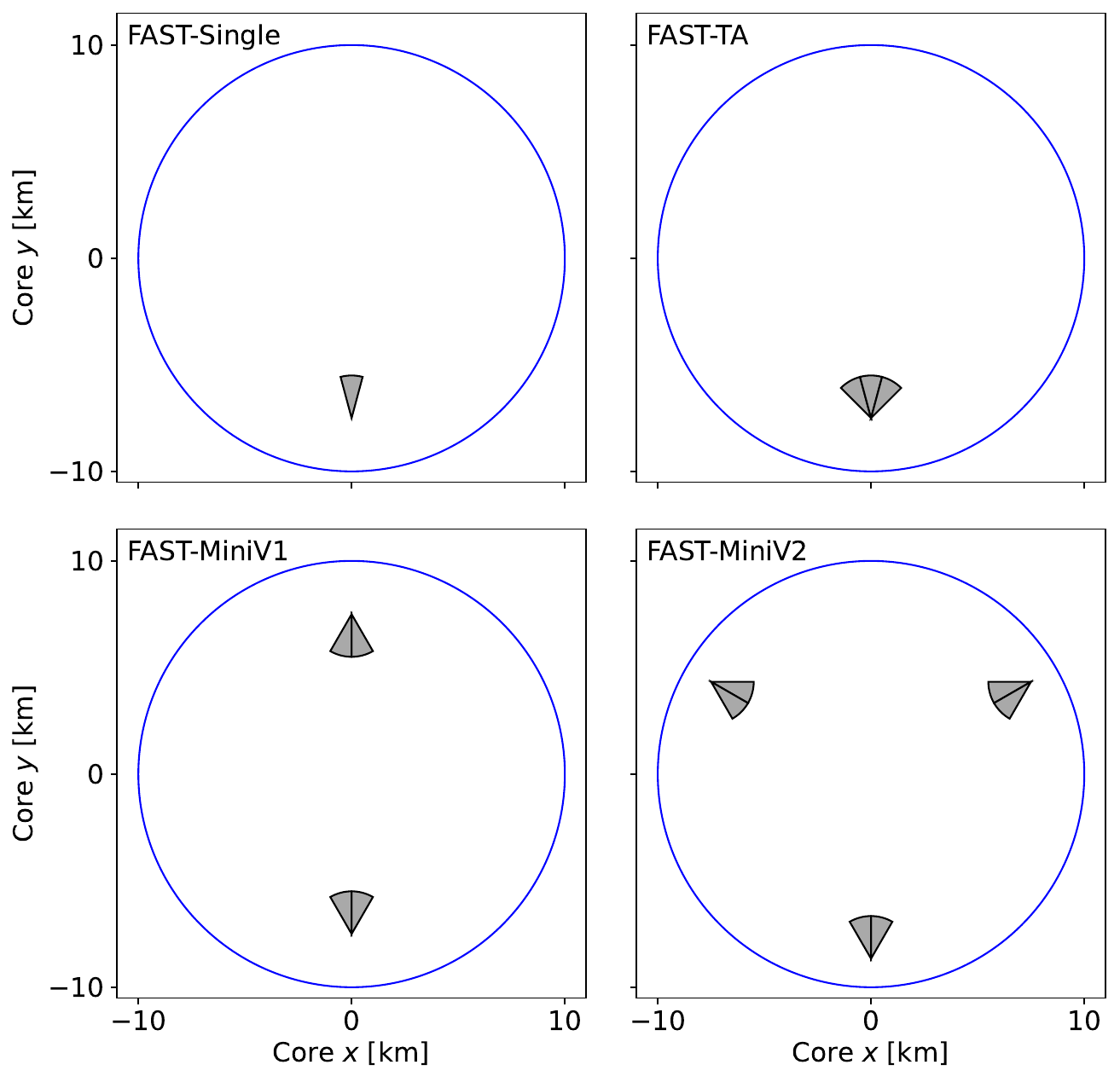}
    \caption{Telescope positions for each of the four layouts studied; FAST-Single (top left), FAST-TA (top-right), FAST-MiniV1 (bottom left) and FAST-MiniV2 (bottom right). Core positions of the simulated showers lie inside the blue circles. }
    \label{fig:mlCorePos}
\end{figure}

A total of 10$^6$ showers were simulated, incident on four different telescope layouts. The simulated parameter distributions are shown in Table \ref{tab:mlDataPars} and the layouts/core positions in Figure \ref{fig:mlCorePos}. Note the minimum simulated energy is set to 10$^{17.3}$\,eV so that the minimum target energy does not lie on the boundary of the simulated parameter space. This is done to increase the reliability of the model's estimates around the minimum target energy. 
As the highest energy shower observed by the FAST prototypes thus far has an energy of $\sim10^{19.5}$\,eV, a maximum simulated energy of $10^{20}$\,eV is sufficiently high. As for \Xmax{}, a uniform distribution was chosen to try and force the model to better learn the relevant patterns in the data over the full range of \Xmax{} values. Although this exposes the network to some \say{unrealistic showers} e.g. a 10$^{20}$\,eV event with $X_\textrm{max}=500$\gcm{}, the goal was to avoid a scenario where the model simply learns to guess the mean of some strongly peaked \Xmax{} distribution. The influence of this choice on the final results should be checked in future work by sampling from parameterisations of realistic \Xmax{} distributions.  

\begin{table}[]
    \centering
    \begin{tabular}{|c|c|}
    \hline\hline
        \Xmax{} & Uniform (500 - 1200\,g\,cm$^{-2}$) \\
        \hline
         Energy & Uniform in log($E/\textrm{eV}$) (17.3 - 20) \\
         \hline
         $\theta$ & $\sin\theta\cos\theta$ (0 - 80$\degree$) \\
         \hline
         $\phi$ & Uniform (0 - 360$\degree$) \\
         \hline
         Core & Uniform (in circle centred at (0,0), $r=10$\,km) \\
         \hline\hline
    \end{tabular}
    \caption{The distribution sampled from and range (in brackets) of each shower parameter used to generate the simulations. Each of the distributions sampled from were continuous.
    }
    \label{tab:mlDataPars}
\end{table}

\vspace{5mm}

Regarding Figure \ref{fig:mlCorePos}, the four layouts are referred to as FAST-Single, FAST-TA, FAST-MiniV1 and FAST-MiniV2. The FAST-Single and FAST-TA layouts are representative of the current prototype layouts at Auger and TA respectively, with each telescope located at (0\,km, $-7.5$\,km). The FAST-MiniV1 and FAST-MiniV2 layouts correspond to the two scheduled stages of the FAST mini-array deployment (see Section \ref{sec:FASTFuture}), albeit with slightly different spacings between the telescopes. Here, both the FAST-MiniV1 and FAST-MiniV2 telescopes are separated by 15\,km. The FAST-MiniV1 telescopes are positioned at (0\,km, $\pm7.5$\,km), whilst the FAST-MiniV2 telescopes are positioned at (0\,km, $-8.66$\,km) and ($\pm7.5$\,km, 4.33\,km). The response of all four telescope layouts is simulated simultaneously for each shower i.e. each shower is viewed by 13 telescopes (the centre telescope of FAST-TA is used as FAST-Single). This significantly lowers the required computational time.
The standard level of background noise $\sigma_{\textrm{nsb}}=10$\,p.e./100\,ns is added to the simulated traces. Likewise, the standard parametric models for the molecular and aerosol atmospheres are used, and the ideal directional efficiency map is applied to each telescope. The total number of showers with at least one PMT passing the threshold trigger (SNR $>6$) for each layout is; FAST-Single: $\sim2.0\times10^5$, FAST-TA: $\sim4.1\times10^{5}$, FAST-MiniV1: $\sim4.7\times10^{5}$ and FAST-MiniV2: $\sim5.5\times10^{5}$.

\section{Output Parameter Normalisation}
The goal of the machine learning models trained here is to predict the six shower parameters ($E$, \Xmax{}, $\theta$, $\phi$, core $x$, core $y$). These parameters will also be referred to as output or target parameters. To ensure that each output parameter is weighted equally when calculating the loss, each parameter is scaled to the range (-1, 1). The scaling is achieved via the following transformation,
\begin{equation}
    x_\textrm{new}=\frac{x_\textrm{old}-x_\textrm{mid}}{x_\textrm{diff}}
\end{equation}
where $x_\textrm{mid}=\left(x_\textrm{max}+x_\textrm{min}\right)/2$ and $x_\textrm{diff}=\left(x_\textrm{max}-x_\textrm{min}\right)/2$. The reverse transformation is applied after the model predicts the shower parameters (using $x_\textrm{max}$ and $x_\textrm{min}$ from the training data). 

\vspace{5mm}

\begin{figure}
    \centering
    \includegraphics[width=1\linewidth]{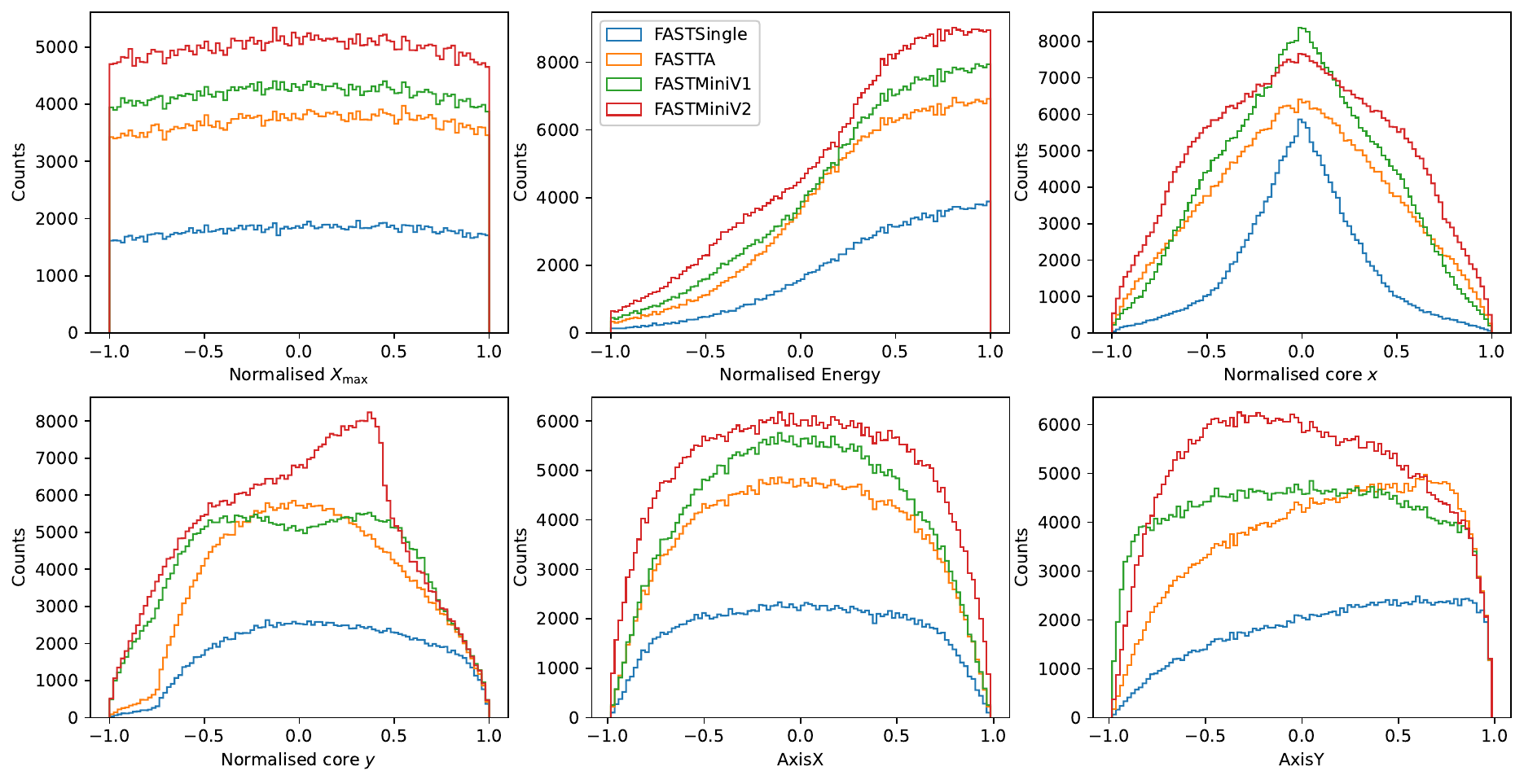}
    \caption{Normalised output parameters for each layout.}
    \label{fig:outputNormPlot}
\end{figure}

For the shower geometry, the models do not associate $\phi=\pm180\degree$ as the same value. Thus $\theta$ and $\phi$ are converted to Cartesian coordinates and the models are trained to predict the X/Y components of the shower axis unit vector instead. These values will be labelled \say{AxisX} and \say{AxisY}. Note the $z$ component or \say{AxisZ} is not needed since $z=\sqrt{1-x^2-y^2}$. When reconstructing the shower parameters using the trained model, the AxisX and AxisY outputs are transformed back to $\theta$ and $\phi$. The distributions of the target parameters for each layout are shown in Figure \ref{fig:outputNormPlot}. The \Xmax{} distributions are mostly flat and the AxisX distributions are symmetric as expected. The energy distribution is heavily biased towards higher energies since these showers remain visible to the FAST telescopes over a larger distance. The shape of the core $x$, core $y$ and AxisY distributions meanwhile reflect the geometry of each layout. To give one example, only a small number of showers falling behind the FAST-Single or FAST-TA telescopes will actually be visible to the telescopes. Even fewer of such showers will leave a significant signal. This phenomenon manifests as the short tails on the left side of the core $y$ histograms for the FAST-Single and FAST-TA layouts. The output distributions shown here will be the same for each machine learning model tested.

\section{Basic DNN}
\label{sec:basicDNN}
The first model tested has the same basic architecture as those used by Albury and Fujii. This model is a simple feed-forward, DNN consisting of a single input layer, several hidden layers, and an output layer. The model will be referred to as the \say{Basic DNN}. The term feed-forward describes how the flow of information in the network is only in one direction, from the input, through the hidden layers, to the output. The network is \say{deep} in that the number of hidden layers is greater than the minimum threshold for what is considered \say{deep learning}, typically 2 \cite{schmidhuber2015deep}. Each layer is \say{dense} or \say{fully connected}, meaning every node in a layer is connected to all nodes in the previous layer. In this model the ReLU activation function is used for the hidden layers. The inputs to the network are three features from each PMT. These features are the integrated signal $S$, centroid time $\bar t$ and pulse height $h$ of each PMT pulse. PMTs which do not pass the threshold trigger have each of their features set to 0. These PMTs still provide useful information to the network which can be used to help constrain the shower parameters. For a PMT $i$ with a SNR $>6$ between bins $k_\textrm{start}$ and $k_\textrm{stop}-1$, $S_i$, $\bar t_i$ and $h_i$ are calculated as
\begin{equation}
\label{eqn:pmtSig}
    S_i=\sum_{j=k_\textrm{start}}^{k_\textrm{stop}-1}s_j,
\end{equation}
\begin{equation}
\label{eqn:centroidTime}
    \bar t_i=\frac{\sum_{j=k_\textrm{start}}^{k_\textrm{stop}-1}s_jt_j}{\sum_{j=k_\textrm{start}}^{k_\textrm{stop}-1}s_j},
\end{equation}
and
\begin{equation}
    h_i=\max\left(\{s_{k_\textrm{start}}, s_{k_\textrm{start}+1}, ...,s_{k_\textrm{stop}-1}\}\right)
\end{equation}
where $s_j$ and $t_j$ are the signal and time of the $j^\textrm{th}$ bin respectively. Each of these values is then normalised before being passed to the network. A slight change between the implementation here and previous implementations is that when normalising the signals and heights, the logarithm base 10 of each value is taken,
\begin{equation}
    \hat S_i=\log_{10}(S_i) \quad\mathrm{and}\quad \hat h_i=\log_{10}(h_i).
\end{equation}
Each value is then divided by the average \textit{total} logarithmic signal $\hat S_0$ or height $\hat h_0$ (respectively) over the entire training data set, i.e.
\begin{equation}
    \hat{\hat{S_i}} = \frac{\hat{S_i}}{\hat{S_0}} \quad\mathrm{and}\quad \hat{\hat{h_i}} = \frac{\hat{h_i}}{\hat{h_0}}
\end{equation}
where 
\begin{equation}
    \hat{S_0}=\left<\sum{}\hat{S_i}\right> \quad \textrm{and} \quad \hat{h_0}=\left<\sum{}\hat{h_i}\right>
\end{equation}
are the averages over all events. Remember that the $\hat{S_i}$ and $\hat{h_i}$ in the sums are only for pixels which passed the threshold trigger. Taking the logarithm of the signals and heights in this way improved the stability and performance of the model during training. The centroid time normalisation is identical to previous implementations. Since only the time differences between PMTs are physically meaningful, the earliest centroid time in each event is subtracted from the other centroid times in the same event,
\begin{equation}
    \hat{\bar{t_i}}=\bar{t_i}-\bar{t_0},
\end{equation}
where $\bar{t_0}$ is the earliest centroid time. Each $\hat{\bar{t_i}}$ is then divided by the standard deviation of all centroid times over the entire training data set $\sigma_t$ giving
\begin{equation}
    \hat{\hat{\bar{t_i}}}=\frac{\hat{\bar{t_i}}}{\sigma_t}.
\end{equation}
The values $\hat{\hat{S_i}}$, $\hat{\hat{h_i}}$ and $\hat{\hat{\bar{t_i}}}$ from each PMT are the inputs to the neural network. Histograms of the input parameter distributions for each layout can be found in Appendix \ref{apx:basicDNNplots}. For small, feed-forward networks like this, the structure of the network can be written concisely by simply listing the number of nodes in each layer with a \say{/} after each value. The preliminary structure for this model was $(3\times\textrm{Number of PMTs})$/128/64/6 for each layout.

\subsection{Training Setup}
For building and training the model, Python with Keras \cite{chollet2015keras} on top of Tensorflow was used. The common features of the \textit{initial} training setups for the Basic DNN model and alternative models tested in Section \ref{sec:alternativeModels} are listed below.
\begin{itemize}
    \item \textbf{Optimiser} - Adam (Adaptive Moment Estimation) with a learning rate of 0.001. The optimiser is the algorithm used to search the high dimensional parameter space of network weights to find those which minimise the loss function. The learning rate controls the \say{step size} of this search. 
    \item \textbf{Train:Test Split} - 90\%:10\%. Ten percent of the entire data set is set aside for testing and does not enter into the training process at all. The remaining 90\% is used to train the model. When training the model a single time, the training data is split into a \say{training set} and \say{validation set}. The model's parameters are updated by training on the training set, with the goal of minimising the \say{training loss} i.e. the loss function applied to the predictions of the model on the training set. The validation set is used to monitor the model's learning. Information from the validation set is not used when updating the network parameters. At the end of each epoch (see below) the \say{validation loss} i.e. the loss function applied to the predictions of the model on the validation set, is calculated, providing an estimate of the models performance on unseen data.
    \item \textbf{Epochs} - 100, with a patience parameter of 10. One epoch is a single pass over the entire training dataset. The patience parameter refers to how many epochs to wait without improvement in the validation loss i.e. for a value of 10, training will be stopped if there are 10 consecutive epochs without improvement in the validation loss. The model weights which correspond to the lowest validation loss are saved.
\end{itemize} 
The batch size for the initial Basic DNN training was set to 64. The batch size is the number of samples to analyse before updating the network weights. This value will vary in later models.

\subsection{Initial Tests}
\begin{figure}
    \centering
    \includegraphics[width=0.49\linewidth]{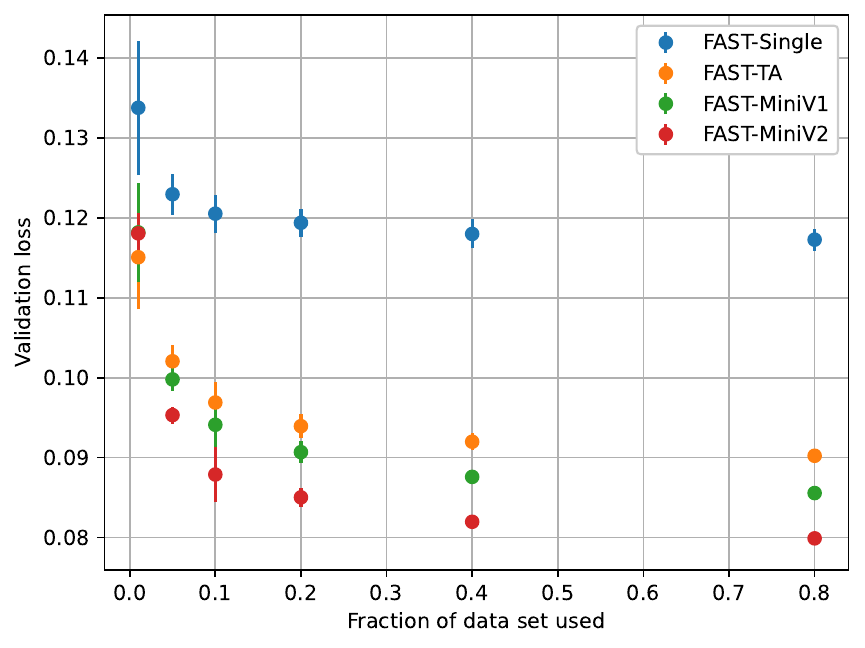}
    \includegraphics[width=0.49\linewidth]{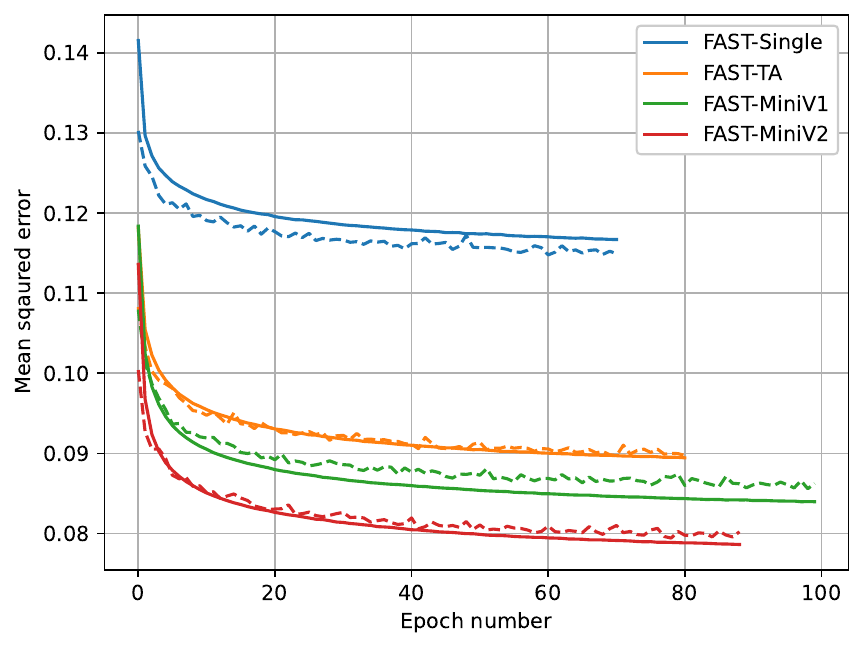}
    \caption{\textit{Left:} Validation loss for each layout using varying fractions of the full dataset. \textit{Right:} Training loss (solid lines) and validation loss (dashed lines) for each layout when using 80\% of the full dataset.}
    \label{fig:basicDNNtraining}
\end{figure}

The left plot of Figure \ref{fig:basicDNNtraining} shows the results of training the Basic DNN model on each layout using successively larger samples of the training dataset. This is done to assess how the model improves with additional data and whether or not additional data would meaningfully increase the model's performance. The $x$-axis shows the fraction of the dataset used and the $y$-axis the \textit{mean} validation loss. Dataset fractions of 0.01, 0.05, 0.1, 0.2, 0.4 and 0.8 were tested. For a single data point, the data corresponding to that point (e.g. 10\% of the training data) was split into 10 equal parts or \say{folds}. The model was then trained 10 times, each time using a different fold as the validation set. The mean ($y$ value) and standard deviation (error bar) of the 10 validation losses was then calculated. This technique is known as $k$-folds cross validation (here $k=10$) and is typically used when optimising model hyper-parameters (see Section \ref{sec:basicDNNtuning}) to prevent over-fitting\footnote{Over-fitting occurs when the model learns a mapping of inputs to outputs which work well only for the training set and do not generalise to other data.}. The method is used here to verify the stability of the model performance. The right plot of Figure \ref{fig:basicDNNtraining} shows the training curves for the best run (out of the 10 folds) of each layout when using 80\% of the dataset. The training loss is shown by the solid lines and the validation loss by the dashed lines.
Below are some basic observations; 
\begin{itemize}
    \item For each layout, the validation loss decreases with additional data, as is the expected behaviour. The validation loss appears to start plateauing between $\sim20-40\%$ for each layout. This indicates that additional data is unlikely to dramatically improve performance.
    \item The layout with the best performance is FAST-MiniV2, followed by FAST-MiniV1, FAST-TA and FAST-Single. This also matches expectations as, with more PMTs, there is more information available to estimate the shower parameters. Additional inputs also mean a larger number of weights with which the models can learn the relationship between inputs and outputs (for the fixed hidden layer structure used here). 
    \item The size of the error bars tends to decrease with additional data, becoming almost negligible beyond $\sim20\%$. This verifies the stability of the model performance.
    \item The shapes of the training curves are very similar, with the majority of the learning occurring in the first $\sim40$ epochs, after which only gradual improvements are seen.
\end{itemize}
These simple checks are important to verify that the models are behaving as expected and are not overfitting. Before applying these models to the test datasets, a series of additional checks and optimisations are performed to try and obtain better performance i.e. a lower validation loss.

\subsection{Model Tuning}
\label{sec:basicDNNtuning}
The parameters which govern the overall model architecture and training process, such as the learning rate, chosen optimiser or number of hidden layers in a model, are called hyper-parameters. Optimising these parameters is known as hyper-parameter optimisation and is a crucial step in obtaining the best possible performance from any machine learning model. For the Basic DNN model described here, the most relevant hyper parameters are the number of hidden layers $L$, the number of nodes in each hidden layer $(N_{1},N_{2},...,N_{L})$ and the learning rate $l_r$. To check how the model's performance changes with these hyper parameters, four different hidden layer structures and five different learning rates were tested with each layout. The learning rates tested were 0.01, 0.003, 0.001, 0.0003 and 0.0001, and the hidden layer structures tested were (128/64), (180/100/50), (256/128/64/32) and (512/256/128/64/32). To save on computational time, these tests were performed using just 20\% of each training dataset. Similar to above, the mean validation loss for each set of hyper-parameters was calculated using $k$-folds cross validation with $k=10$. However, the validation loss was only calculated for five out of the ten folds, again to save time. The number of epochs in training was also set to 50. These choices were made based on the stability and learning behaviour of the model demonstrated in Figure \ref{fig:basicDNNtraining}. 

\vspace{5mm}

\begin{figure}[t]
    \centering
    \includegraphics[width=\linewidth]{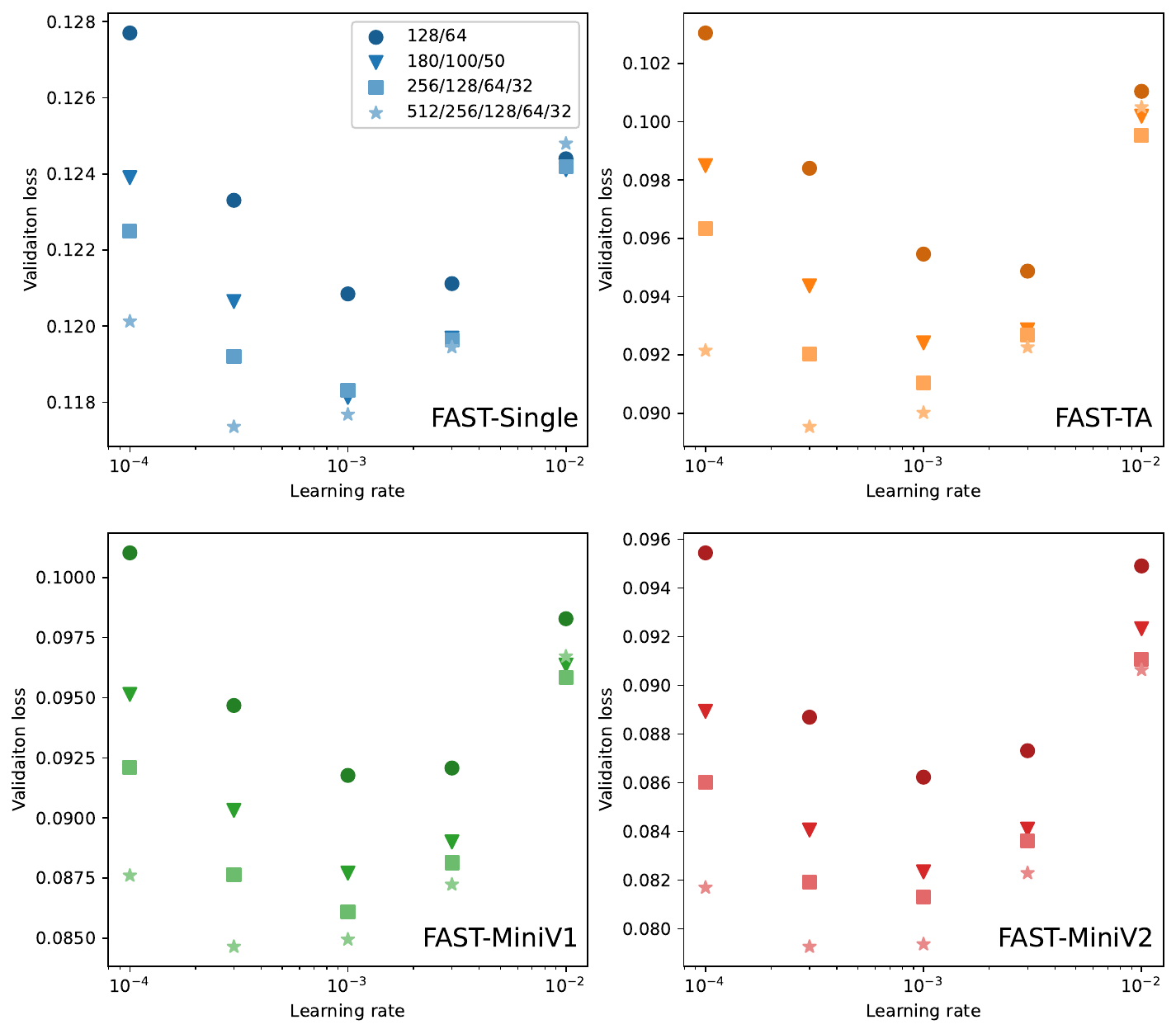}
    \caption{The validation loss when using different hidden-layer configurations (see the legend) as a function of learning rate. The results are shown for each layout using the same colour code as applied in Figures \ref{fig:outputNormPlot} and \ref{fig:basicDNNtraining}.
    }
    \label{fig:basicHyperParams}
\end{figure}

The validation losses for each hidden layer structure and layout are shown as a function of learning rate in Figure \ref{fig:basicHyperParams}. The behaviour for each layout is very similar, with the best performance being obtained with the five-layer structure. Again this is not particularly surprising considering the additional layers equate to roughly 10 times the number of weights. The learning rate is clearly the most important hyper-parameter (in these tests), with the nominal value of 0.001 used in the initial model testing generally being the best choice. For the five layer structure a slightly lower learning rate appears optimal. Overall, compared to the initial training setup, a 5 - 10\% improvement in the validation loss for each layout is achieved when using the best set of hyper-parameters out of those tested. Tests using batch sizes of 32 and 128 in training with the optimal five-hidden-layer structure were also performed. No significant difference was observed in the validation loss.

\subsection{Basic DNN Performance}
\label{sec:basicDNNperformance}
\begin{figure}[]
    \centering
    \includegraphics[width=0.9\linewidth]{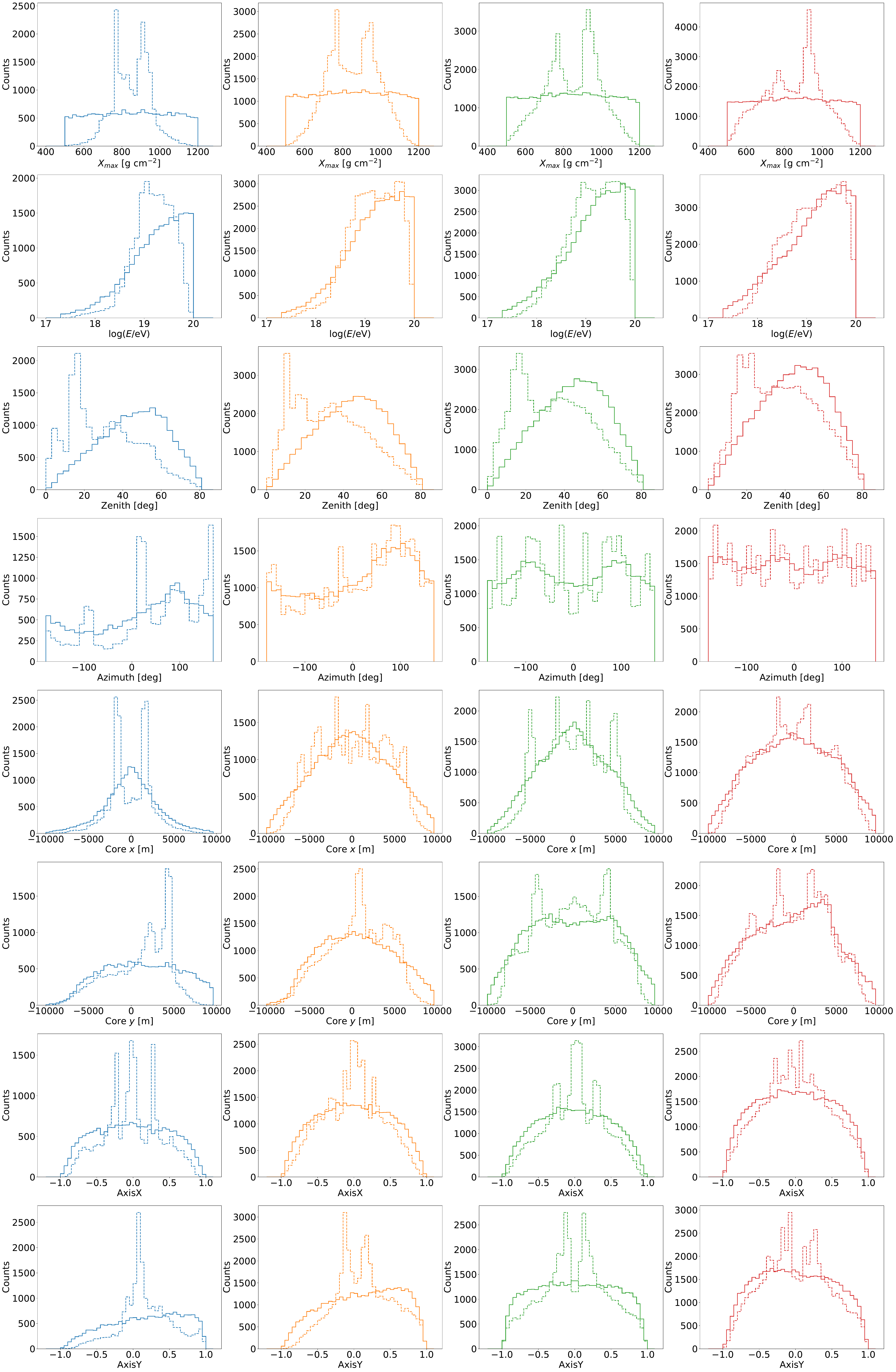}
    \caption{True (solid lines) and reconstructed (dashed lines) distributions for each output parameter. From left to right, the layouts are FAST-Single, FAST-TA, FAST-MiniV1 and FAST-MiniV2.}
    \label{fig:basicDNNhists}
\end{figure}

\begin{figure}[]
    \centering
    \includegraphics[width=1\linewidth]{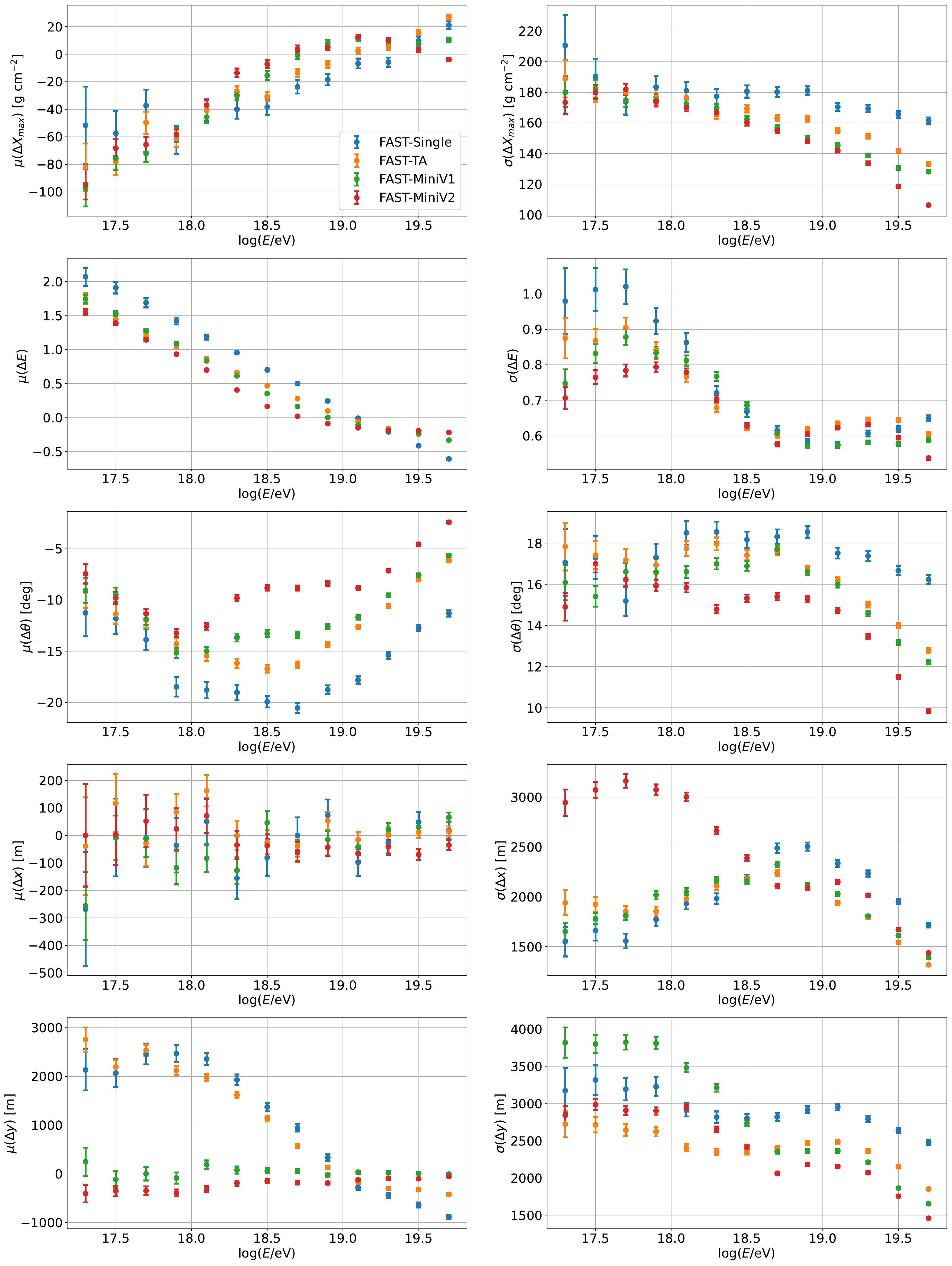}
    \caption{The biases (left) and resolutions (right) in each reconstructed parameter as a function of energy. The non-monotonic nature of the resolution plots is thought to be due to how the observable distribution of each parameter changes as a function of energy. Note the different $z$-axis scales between different slices.}
    \label{fig:basicDNNresolutions}
\end{figure}

Based on the above results, the final models for each layout were trained using the five layer structure with a learning rate of 0.0003. The entire training data set was used, running for the standard 100 epochs with a patience parameter of 10. The trained models were then used to estimate the shower parameters of the test data set. Note the same threshold trigger and data processing steps used during training were applied to the test data set. Figure \ref{fig:basicDNNhists} shows the true and reconstructed distributions for each output parameter and each layout over all energies. 
Both the AxisX/AxisY and $\theta$/$\phi$ combinations are shown for completeness. Figure \ref{fig:basicDNNresolutions} shows the biases and resolutions of each reconstructed output parameter (with the exception of $\phi$) as a function of energy.

\vspace{5mm}

In Figure \ref{fig:basicDNNhists}, the reconstructed output parameter distributions exhibit an overall tendency towards the centre of the true distributions. This is highlighted by the sharp peaks either around or on either side of the central values in many of the distributions. The effect 
is indicative of the model not having fully learned the precise relationship between all possible inputs and outputs. For regions of the parameter space where this is the case, for example where only one or two pixels trigger with low SNRs, it is likely that the model simply chooses the central values as, statistically, this has the best chance of minimising the loss. The effect is particularly egregious for FAST-Single, though is present in each layout. This behaviour is connected to the bias seen in zenith angle. This is because a guess near the centre of the distribution for both AxisX and AxisY corresponds to the $x$-axis and $y$-axis components of the shower axis \textit{unit} vector being $\sim0$. This results in a large $z$ component and thus small zenith angle. The bias appears to decrease slightly with additional telescopes, behaviour which seems consistent when looking at the AxisX and AxisY reconstructed distributions for each layout.

\vspace{5mm}

Regarding Figure \ref{fig:basicDNNresolutions}, each point represents either the mean (bias) or standard deviation (resolution) of the difference between the reconstructed and true values for one of the output parameters for a particular energy bin. Here $\Delta{}E=\ln(E_\textrm{rec}/E_\textrm{true})$ and $\Delta{}x/\Delta{}y$ refer to the differences in core $x/y$ respectively. The error bars correspond to the standard error in the mean $=\sigma/N$ and the standard error in the standard deviation $\approx\sigma/\sqrt{2(N-1)}$ \cite{rao1973linear}. The biases are shown by the plots on the left hand side, and the resolutions on the right hand side. For each output parameter, the layout with the least bias is FAST-MiniV2, followed by FAST-MiniV1, FAST-TA and finally FAST-Single. This matches the order of the validation losses. The same order mostly holds for the resolution plots with a few exceptions. In general, the parameter resolutions are expected to improve with energy since there are more showers from which to learn the relationship between inputs and outputs. Additionally, the average SNR of the signals is larger at higher energies, in theory leading to a more accurate reconstruction. 

\begin{figure}[]
    \centering
    \includegraphics[width=1\linewidth]{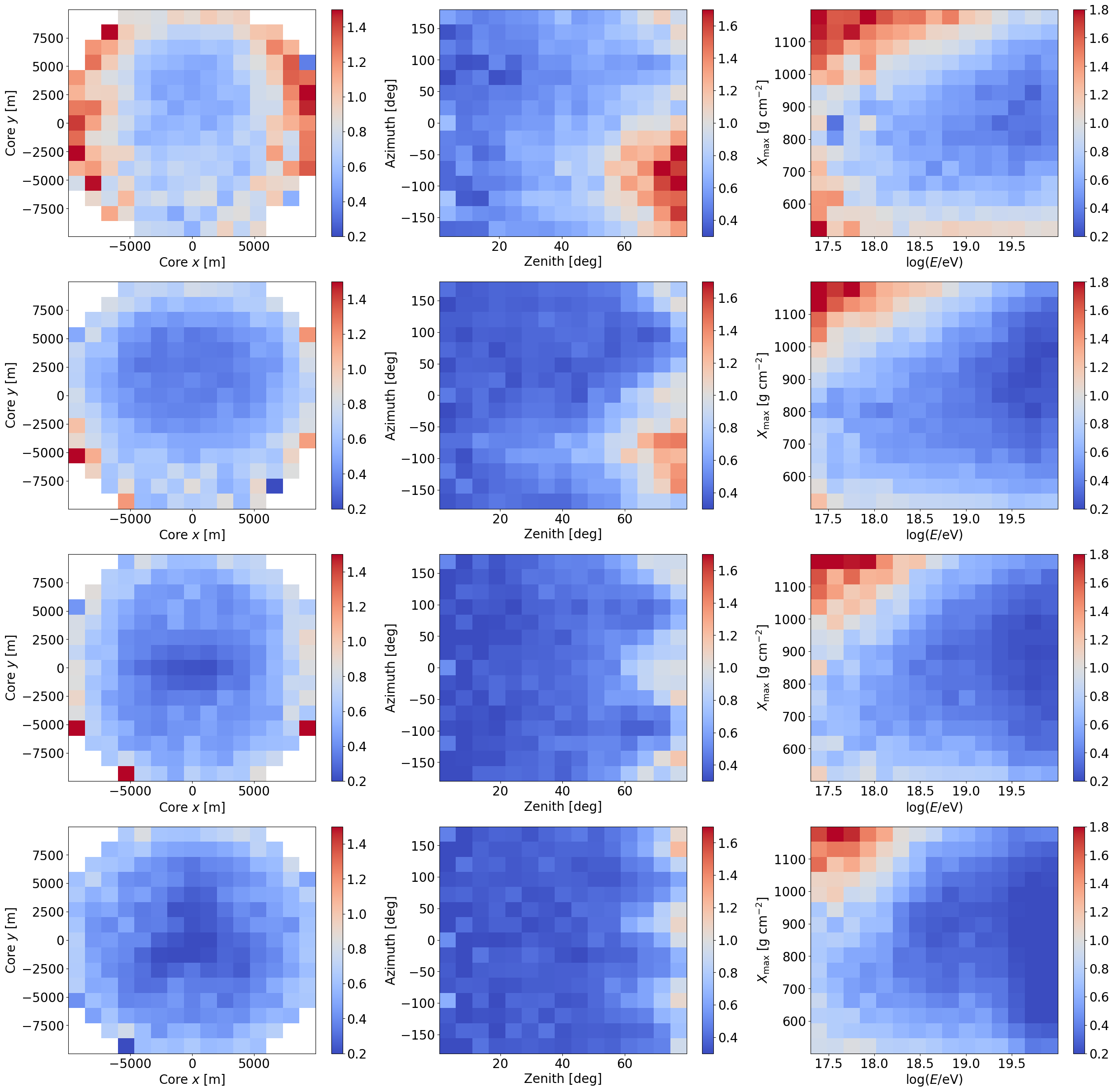}
    \caption{Average MSE when evaluating the Basic DNN model on the test data set shown in slices of core position (left), arrival direction (centre), and \Xmax{}/energy (right). From top to bottom the results are shown for FAST-Single, FAST-TA, FAST-MiniV1 and FAST-MiniV2.}
    \label{fig:basicDNNmaps}
\end{figure}

\begin{figure}[t!]
    \centering
    \includegraphics[width=1\linewidth]{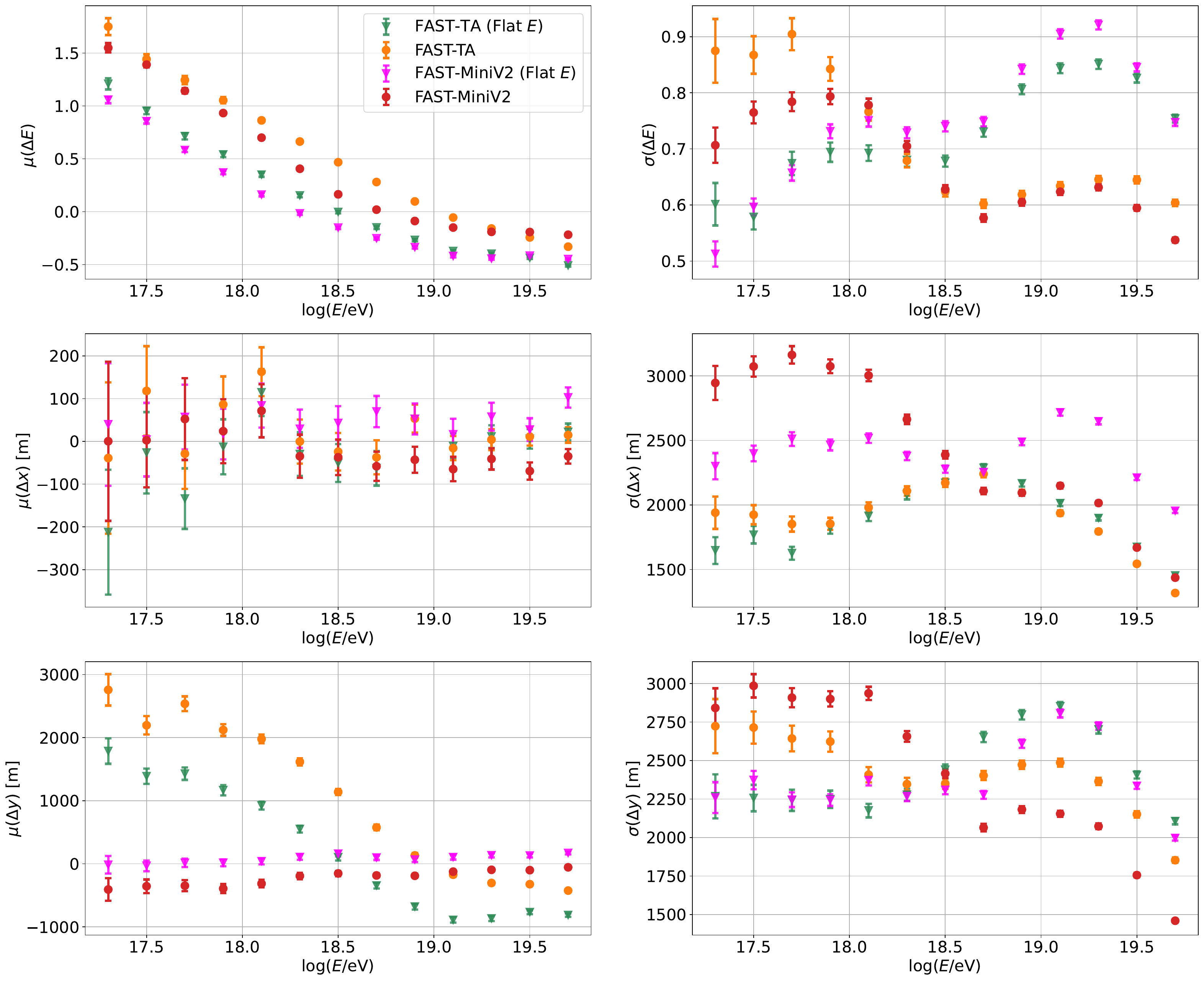}
    \caption{The standard bias and resolution in the reconstructed energy (top panels) and core position (bottom two panels) for the FAST-TA and FAST-MiniV2 layouts compared to those when using a flat distribution of energies during training.}
    \label{fig:basicDNNflatEresolutions}
\end{figure}

\vspace{5mm}

For \Xmax{} and energy, the respective biases show a roughly linearly trend from negative/positive values to 0, flattening at around 10$^{19}$\,eV. This behaviour appears similar to the correlation observed in Section \ref{sec:xmaxEnergyCorrelatoin}. The resolution in \Xmax{} gradually decreases from $\sim180$\,g\,cm$^{-2}$ at $E=10^{17.5}$ to between 110 - 170\,g\,cm$^{-2}$ depending on the layout. The energy resolution for each layout is less straightforward, showing a small increase up until $\sim10^{18}$\,eV, then a decrease up until $\sim10^{18.7}$\,eV, after which the resolution is a roughly constant at around 0.6. Similar \say{breaks} in the shape of the resolution graphs at broadly similar energies are also present in the zenith, core $x$ and core $y$ parameters. The cause of this behaviour is not fully understood but is thought to be related to how the parameters of the showers observed by each layout change as a function of energy. 

\vspace{5mm}

Take for example the worsening in the core $x$ resolution seen by FAST-Single, FAST-TA and FAST-MiniV1. These layouts all have telescopes positioned and pointing along (or near to) the $y$-axis. This means at low energies, where showers are only observable up until a few kilometres, the range of core $x$ values for showers observed by these installations is narrow. As the energy of the showers and hence distance up to which showers can be observed increases, this range also increases. With an increased range of possible core $x$ values, poor core $x$ estimates e.g. those which simply guess the centre of the distribution, will cause the resolution to worsen. This trend continues until the energy of the showers is such that the entire range of possible core $x$ values, defined by the dataset, stops increasing with energy. Beyond this energy the resolution improves in accordance with the expectation outlined above. The resolution for FAST-MiniV2 in this case begins very poor because the positions of the telescopes are such that, even for low energy showers, the range of possible core $x$ values is large. 

\vspace{5mm}

Another contributing factor is the difficulty for the FAST telescopes to reconstruct the component of the core position which aligns with the telescope pointing direction. This is clearly shown in the core $y$ bias plot for FAST-Single and FAST-TA. Both of these layouts only view showers from a single location and primarily point along the $y$-axis. This gives rise to a degeneracy where lower energy showers with small core $y$ values (close to the telescope/s) are mistaken for higher energy showers with large core $y$ values (far from the telescopes/s). The degeneracy is shown explicitly in Appendix \ref{apx:basicDNNplots}. This phenomenon may be another reason why the core $x$ resolution for FAST-MiniV2 is poor at lower energies. An exploration of the precise factors determining the various behaviour seen in the bias and resolution plots will not be performed here, but is recommended for future work should similar structure be observed.

\vspace{5mm}

Figure \ref{fig:basicDNNmaps} shows the average MSE in slices of core position, arrival direction and \Xmax{}/energy for each layout. These maps are useful to quickly see where each model struggles to reconstruct the shower parameters accurately. For example, low energy showers with large \Xmax{} values are reconstructed poorly by all models. Considering that such showers are essentially not observed in data this is not a significant issue. At large zenith angles showers which come from behind the telescopes for FAST-Single, FAST-TA and FAST-MiniV2 are reconstructed comparatively poorly. For FAST-MiniV1, the fact the telescopes are facing each other means that such showers are in fact reconstructed better than those which cross the FOV horizontally. As for core positions, showers on the boundary are reconstructed worse by all layouts than those falling in the centre. This is because showers falling in the centre are more likely to be seen by multiple telescopes and thus there is more information to accurately reconstruct them.

\vspace{5mm}

To check whether the cause of the strong energy dependence seen in the above results was related to the models being trained on a very uneven distribution of energies (see Figure \ref{fig:outputNormPlot}), the models were retrained with the same training setup but now using an almost flat output energy distribution. Specifically, the first 1000 events in each energy bin of the energy histograms in Figure \ref{fig:outputNormPlot} were selected for training. The new models were then used to reconstruct the same test data set. Figure \ref{fig:basicDNNflatEresolutions} shows the biases and resolutions in the reconstructed energy and core parameters using the new and old models for the FAST-TA and FAST-MiniV2 layouts. Since the total amount of training data is different only the \say{shape} of the results is analysed. For the parameters shown, the models trained on the flat energy distribution perform comparatively better at lower energies but worse at higher energies. This applies both to the biases and resolutions in the parameters. The other output parameters showed no significant change. Although better estimates at lower energies are perhaps more favourable for applying these models to the current coincidence data, the trade off of reduced performance at higher energies, the target of FAST, is not desirable long term. Further investigation into how changing the distributions of target parameters affects the final biases and resolutions could be considered for future work.

\subsection{Discussion}
This section has attempted to extract the best possible performance using the Basic DNN network structure. A hyper-parameter search was performed and small improvements in the data processing over the original method were made. The model learning behaviour, suitable data-set size, batch size, and impact of changing the target energy distribution have been investigated. Overall, the resolutions \textit{as presented here} are rather poor, with the best performance by FAST-MiniV2 only achieving resolutions of $\sim110$\,g\,cm$^2$ in \Xmax{}, $\sim50\%$ in energy, 10$\degree$ in zenith, and a few thousand metres in core resolution. Applying stricter cuts on the input data to the networks and/or reconstructed data, such as the number of triggered PMTs or number of different Eyes observing the shower, would improve these results. Training only on showers which trigger a large number of PMTs/are observed in stereo may also improve results, with the trade off being the inability to apply the model to lower energy showers and/or showers which trigger only a few PMTs i.e. the majority of the coincidence data. The above suggestions are investigated in Section \ref{sec:TSFELDNNperformance}. At this stage, rather than further fine tuning or implementing cuts with the Basic DNN model, it is expected that altering the model architecture and/or adding additional inputs to characterise the PMT traces will be more efficient in obtaining greater accuracy in the first guess. This is tested in the following section.

\section{Alternative Models}
\label{sec:alternativeModels}
There are two main options for improving upon the Basic DNN. The first is to use the same feed-forward, deep neural network architecture but with additional features for each PMT trace. Additional inputs characterising the traces should in principle allow the shower parameters to be better estimated. The second option is to alter the underlying model architecture. This could involve using different types of layers such as recurrent or convolutional layers, or using an entirely different learning paradigm e.g. Random Forests. Both approaches are tested below. Note however that thoroughly testing every possible extension/viable learning method is beyond the scope of this work. As such only two alternative models are compared.

\subsection{Model 1: TSFEL DNN}
This model is a straightforward extension to the Basic DNN above. The name comes from the Python library used to extract the additional input features, \say{\gls{tsfel}\footnote{https://tsfel.readthedocs.io/en/latest/index.html}}. Whilst it may be possible to \say{handcraft} features which better characterise the FAST traces, leaving the task of feature extraction to a predefined library saves considerable time and effort. This library allows for the calculation of four different types of features from a given time series - spectral, fractal, temporal and statistical. It is expected that the features of the FAST traces which are most important for estimating the shower parameters are the temporal and statistical features. Hence only these are extracted. This gives a total of 45 additional features for each PMT. The features were calculated using a fixed 600\,bin length segment of the traces selected in the following way. First, the closest bin to the centroid time of the earliest triggered PMT was determined. Then, with respect to this bin, the previous 150 and subsequent 450 bins from all PMT traces were extracted (using 100\,ns bin traces). The values of 150 and 450 were chosen based on analysing traces containing no noise. See Appendix \ref{apx:binCheck} for details. Selecting the trace segments to analyse in this way ensures that the features are calculated independent of the absolute time offset of the traces. 

\vspace{5mm}

Of the 45 extracted features, 14 were found not to vary between traces and so were removed. All but a few of the remaining 31 features had distributions which were either entirely negative or positive. Therefore the following transformation was applied to each feature,
\begin{equation}
    \hat x = \begin{cases} 
      \log_{10}(|x|) & x\neq0 \\
      0 & x=0 \\ 
   \end{cases}
\end{equation}
with the exception of the \say{Median features}. This ensured that the values of each feature were reasonably small with the majority being between (-10, 10). At this stage, the original three features used in the Basic DNN model along with the width of the max SNR region (given by $k_\textrm{stop}-k_\textrm{start}$) were added to the list of 31 features from TSFEL. This gave a total of 35 possible input features for each PMT. To remove redundant information, the correlation between each feature associated with a single PMT was checked. If two features had a correlation of more than 0.95 then one was removed. After this process a total of 11 features from each PMT remained. These were used as the inputs to the TSFEL DNN. For the full list of input features
and their respective distributions see Appendix \ref{apx:TSFELfeatList}.

\vspace{5mm}

As with the Basic DNN, pixels which did not pass the threshold trigger had their input features set to 0. All other settings (training setup, model structure, activation layers, loss function etc.) were initially set identical to the Basic DNN. The only difference was the number of input features. For a \say{fair} comparison with the Basic DNN, a basic hyper-parameter search was performed using 20\% of the dataset, averaging the validation loss across 5 folds. The same four layer configurations as trialled with the Basic DNN and the learning rates 0.0003, 0.001 and 0.003 were tested. Once again the 5 layer, 0.0003 learning rate model was found to perform best. This is the model used for comparison in Section \ref{sec:modelComparison}. Future work could investigate whether using trace segments centred on each pixel pulse with fixed/variable widths improves performance.

\subsection{Model 2: LSTM Network}
This model utilises a different architecture known as \say{\gls{lstm}} \cite{hochreiter1997long}. An LSTM network is an enhanced version of a Recurrent Neural Network able to learn long term dependencies in sequential data. For FAST, an LSTM network can be used to learn the salient features of PMT traces. These features can then be combined, as done above with the feed-forward neural network architectures, to predict the shower parameters. Put simply, instead of the user defining and calculating the features to be extracted and learned from, an LSTM \textit{learns} what features are useful. This architecture is used in a wide variety of applications and was used by Auger for extracting information from WCD signals to predict \Xmax{} \cite{aab2021deep}. A simple description of how an LSTM transforms an input time series to an output is given in Appendix \ref{apx:LSTMexplanation}. 

\vspace{5mm}

For this model, the PMT traces are input directly to the network. The input trace segments are the same as those used for calculating the TSFEL features above. However, to avoid processing unnecessary noise in the LSTM layer and to reduce the required memory/computation time, the 600\,bin traces were re-binned by a factor of 4. The first 10 and final 40 bins of the rebinned trace were then removed. This left one hundred 400\,ns bins for each PMT trace. Examples of the input traces to the network can be found in Appendix \ref{apx:LSTMInputs}. Each of the input traces are then passed to the same LSTM layer (i.e. same set of weights). This is done so that the same features are extracted from each trace. The output dimension of the LSTM layer is set to 64, meaning each PMT has 64 features extracted. These features are concatenated before being passed to a \say{dropout layer} with a rate of 30\%. This layer randomly sets 30\% of features in the previous layer to 0, helping to prevent over-fitting and reducing the reliance on specific nodes. The output from the dropout layer is then connected to a dense layer of 64 nodes with the ReLU activation function, before finally being connected to the output layer. The batch size is set to 256 to reduce training time.

\vspace{5mm}

Once again PMTs which did not pass the threshold trigger had their trace values set to 0. After being processed by the LSTM, these zeroed traces will in general give some non-zero output vector. Since the zeroed traces are considered to contain no useful information, these output vectors were also set to 0, requiring a slight modification of the standard LSTM layer provided in Keras. The same loss function, number of epochs (with patience) and train:test split as used for the Basic DNN were applied to this model. As LSTM networks take considerably longer to train, no hyper-parameter optimisation was performed here. Future work could investigate such optimisations, for example varying the output dimension of the LSTM, adding additional layers of processing after the LSTM layer, or training using shorter input traces.

\subsection{Model Comparison}
\label{sec:modelComparison}
The TSFEL DNN and LSTM network models introduced above were trained on the full dataset of each layout. The validation losses are summarised in Table \ref{tab:modelComp} with the Basic DNN results shown for reference. Note that due to time constraints each value is from only a single training run. Both alternative models perform better than the Basic DNN with a roughly 15\% improvement in the validation loss across all layouts. With the exception of FAST-Single, where there is an approximately 10\% difference in the validation loss, the TSFEL DNN and LSTM network perform roughly the same. The LSTM model does appear marginally better however a more robust comparison, where each network is trained multiple times and the results averaged using $k$-folds cross validation, would be required to confidently make this claim. 

\vspace{5mm}

Assuming the performance is comparable between the TSFEL DNN and LSTM network, the choice of which model to proceed with largely comes down to each models speed, interpretability and robustness when applied to data from the prototype FAST telescopes. For interpretability, the TSFEL DNN is superior in the sense that the features with which the model predicts the shower parameters are well understood. The TSFEL DNN is also significantly quicker during inference than the LSTM. This is a minor consideration though as both models can produce estimates of the shower parameters for several thousand events in less than a few seconds. Lastly, the accuracy of both models is expected to decline when applied to data due to features of the data, such as fluctuating signal baselines and single bin peaks, not being accounted for in simulations. It is not clear however which approach would be superior. As such, for the remainder of this chapter and indeed thesis, the TSFEL DNN will be taken as the machine learning model to use for first guess purposes. 

\begin{table}[]
    \centering
    \begin{tabular}{||c|c|c|c|c||}
        \hline
         & \textbf{FAST-Single} & \textbf{FAST-TA} & \textbf{FAST-MiniV1} & \textbf{FAST-MiniV2} \\
         \hline
         \hline
         \textbf{Basic} & 0.1144 & 0.0827 & 0.0765 & 0.0685 \\
         \hline
         \textbf{TSFEL} & 0.1006 & 0.0689 & 0.0649 & 0.0570 \\
         \hline
         \textbf{LSTM} & 0.0906 & 0.0672 & 0.0632 & 0.0565 \\
         \hline
    \end{tabular}
    \caption{Table comparing the validation losses for the two alternative models tested with the original Basic DNN for each layout.}
    \label{tab:modelComp}
\end{table}

\section{TSFEL DNN Performance}
\label{sec:TSFELDNNperformance}

\begin{figure}
    \centering
    \includegraphics[width=1\linewidth]{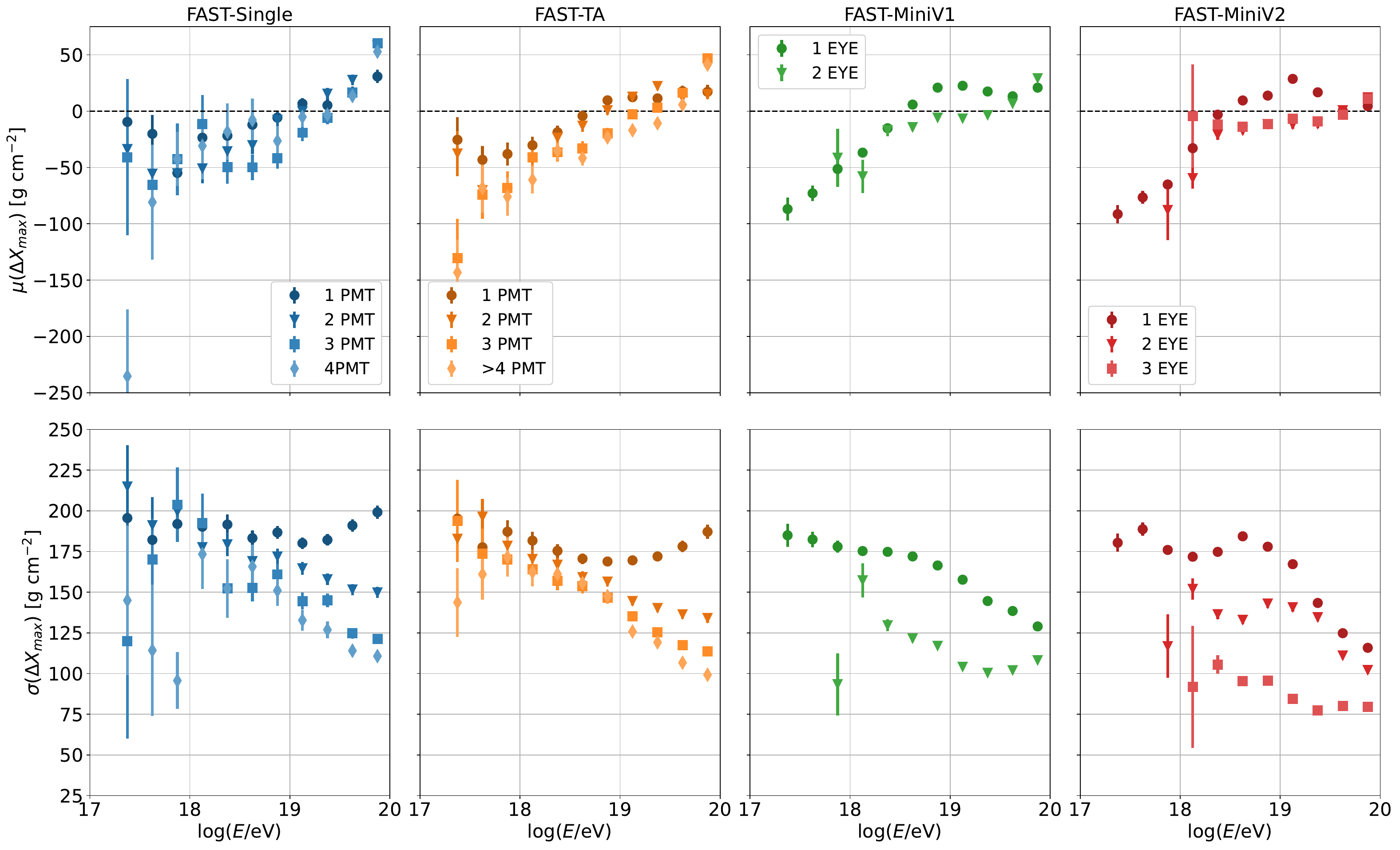}
    \includegraphics[width=1\linewidth]{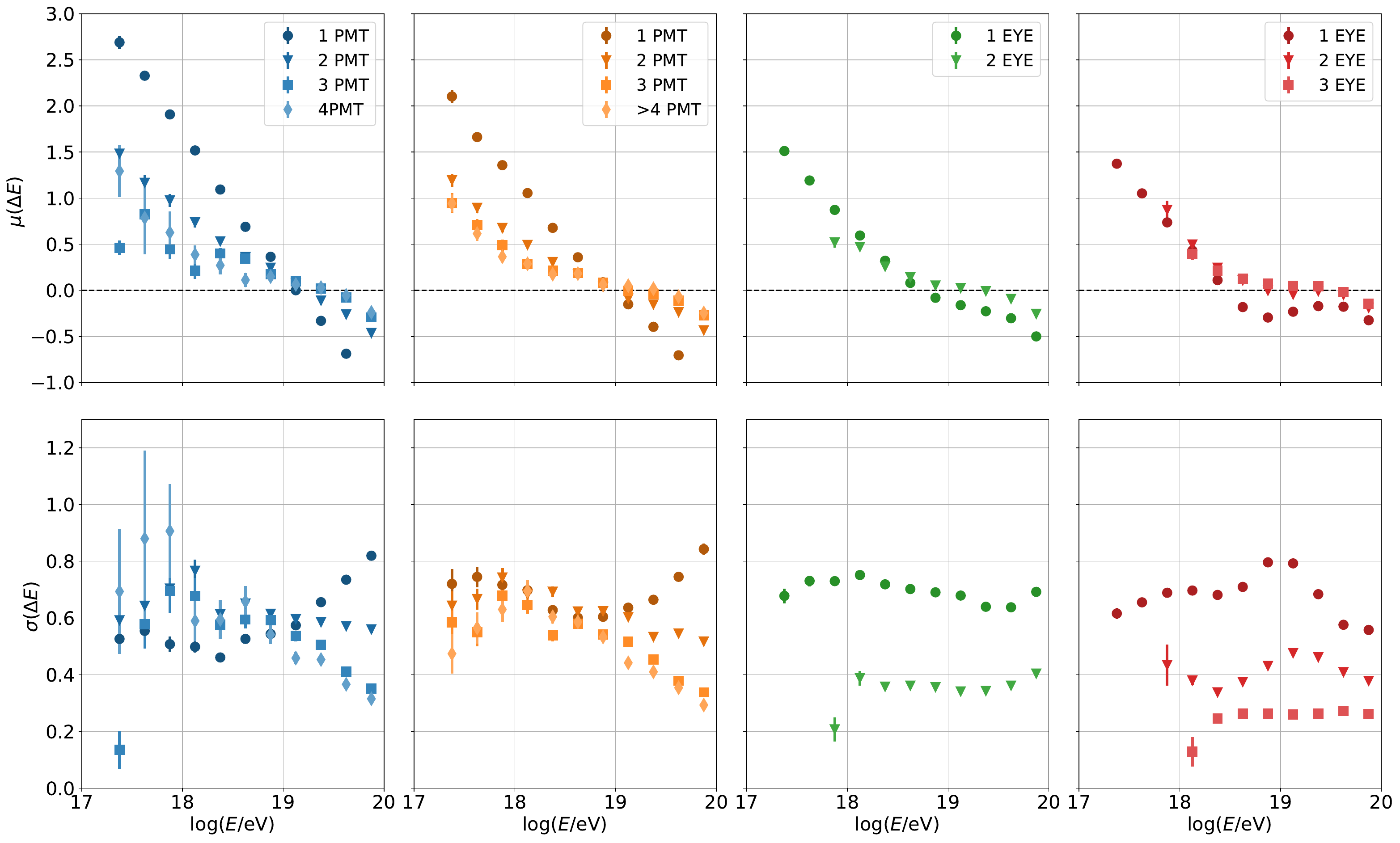}
    \caption{Biases and resolutions for \Xmax{} (top panels) and energy (bottom panels) as a function of energy. Results are from applying the TSFEL DNN model to the test data set of each layout. The performance is shown for different numbers of PMTs (FAST-Single/FAST-TA) or different numbers of Eyes (FAST-MiniArrayV1/2) observing the events.}
    \label{fig:TSFELex}
\end{figure}

\begin{figure}
    \centering
    \includegraphics[width=1\linewidth]{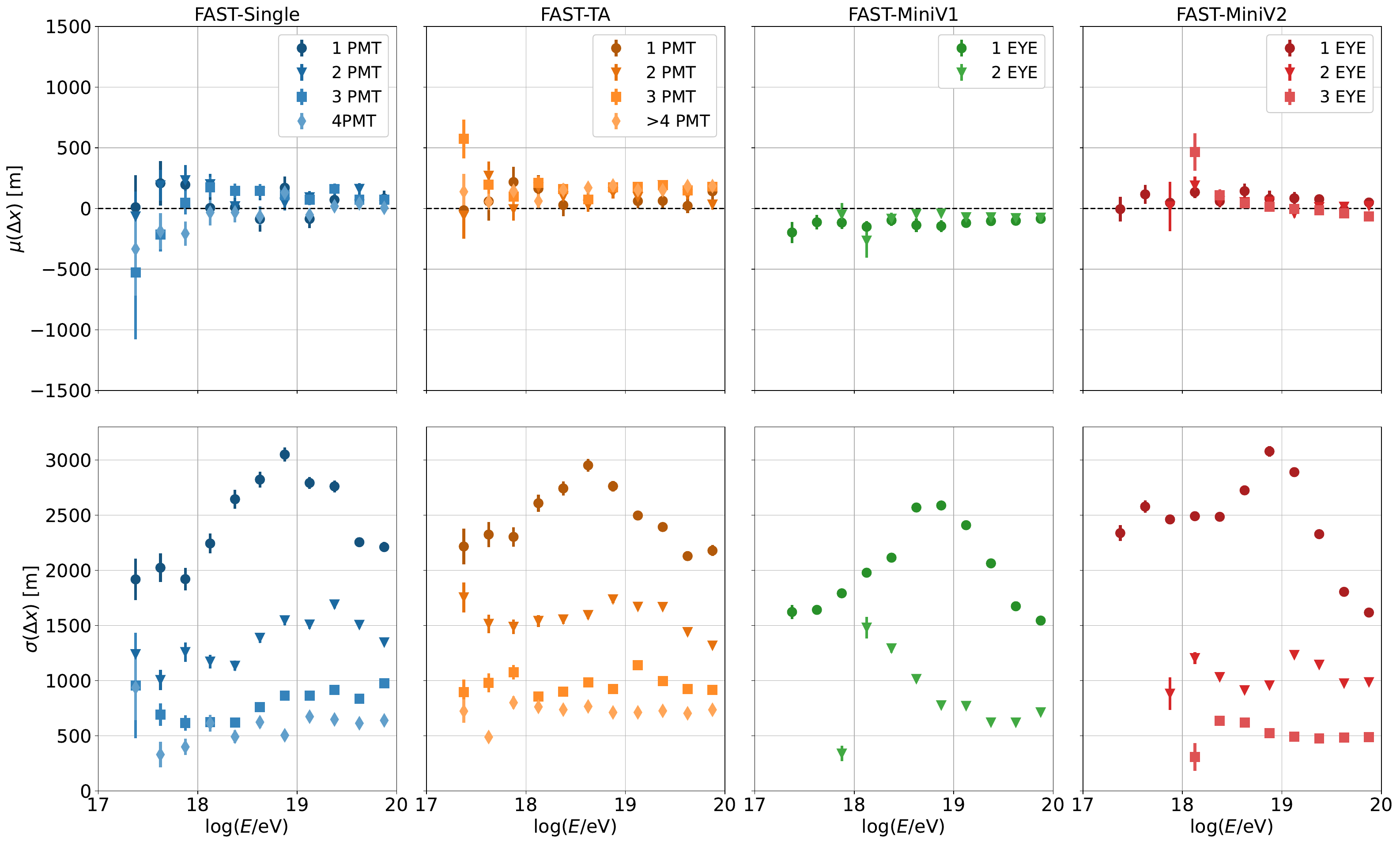}
    \includegraphics[width=1\linewidth]{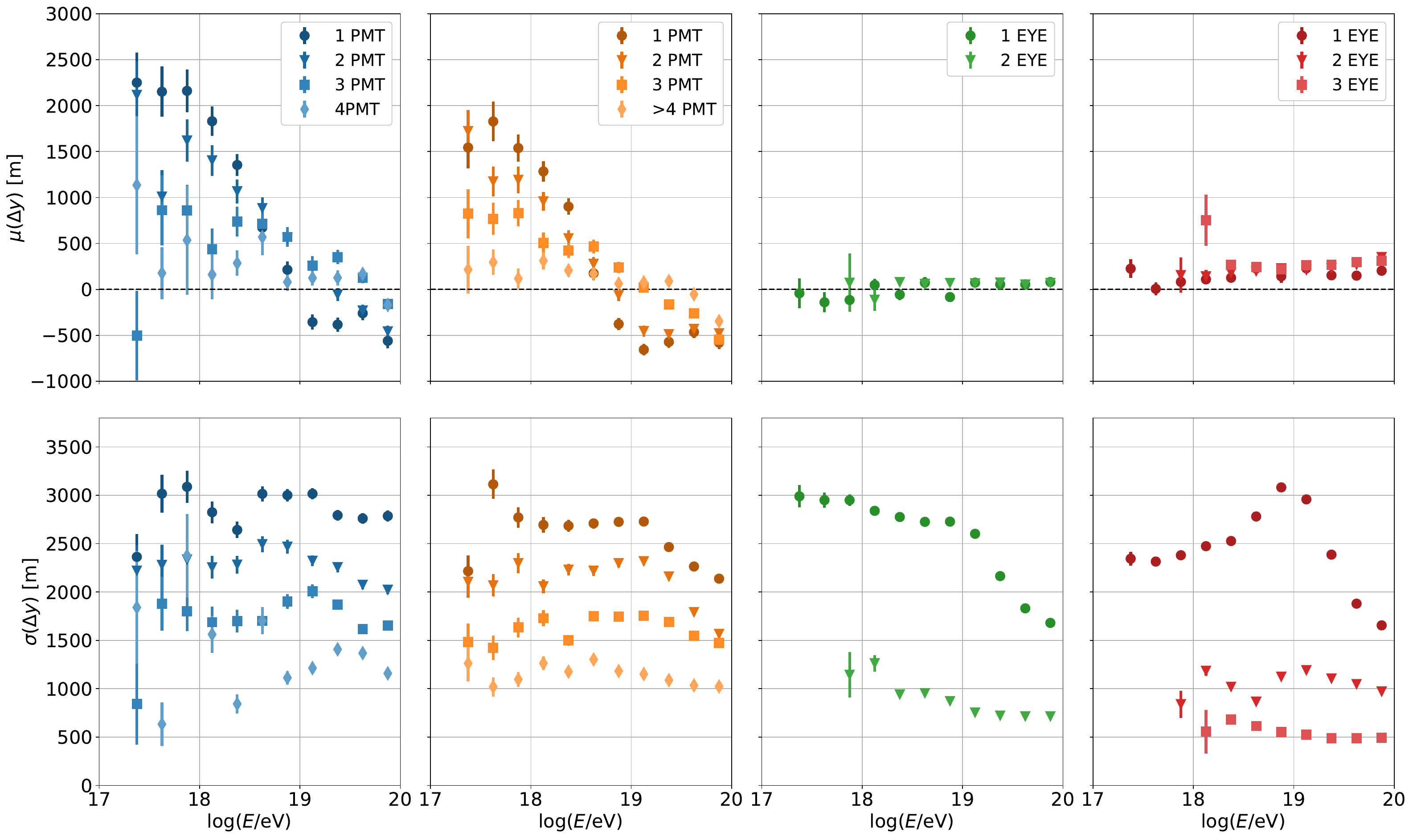}
    \caption{Same as Figure \ref{fig:TSFELex} but for the core $x$ (top panels) and core $y$ (bottom panels) parameters.}
    \label{fig:TSFELcore}
\end{figure}

\begin{figure}
    \centering
    \includegraphics[width=1\linewidth]{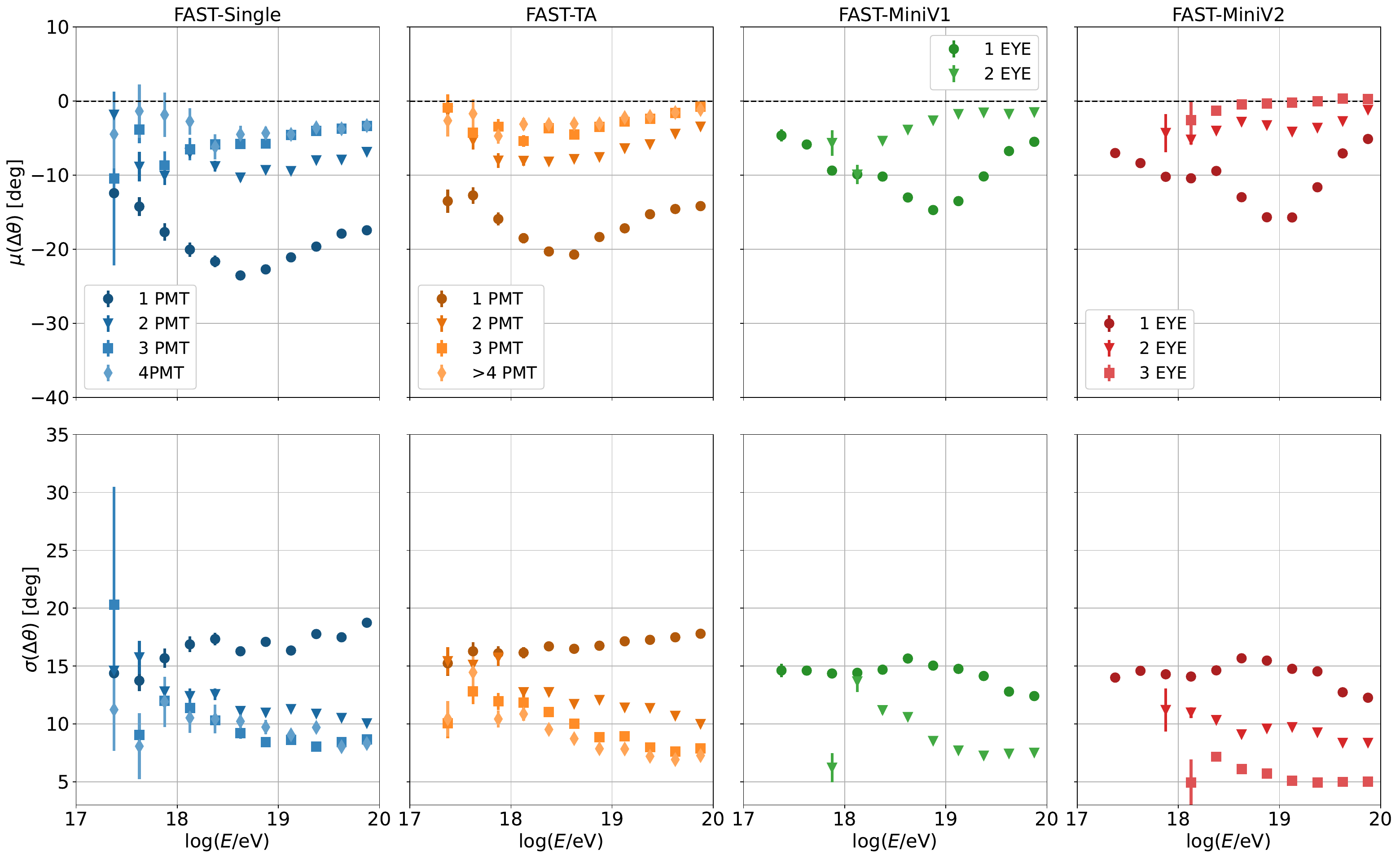}
    \includegraphics[width=1\linewidth]{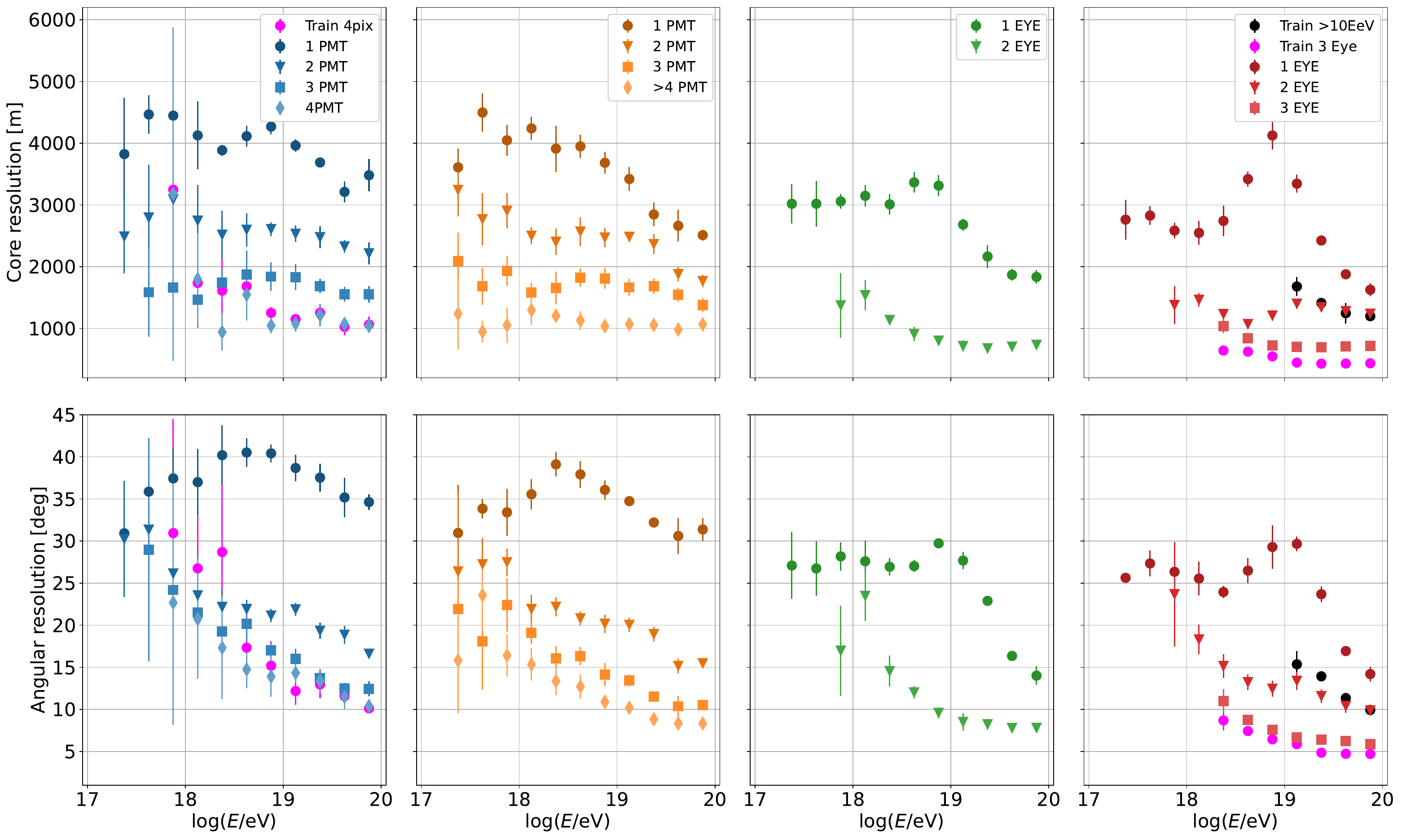}
    \caption{(Top panels) Same as Figure \ref{fig:TSFELex} but for zenith angle. (Bottom panels) Core resolution and angular resolution as a function of energy for the TSFEL DNN applied on the test data of each layout.}
    \label{fig:TSFELzRes}
\end{figure}

\begin{figure}
    \centering
    \includegraphics[width=1\linewidth]{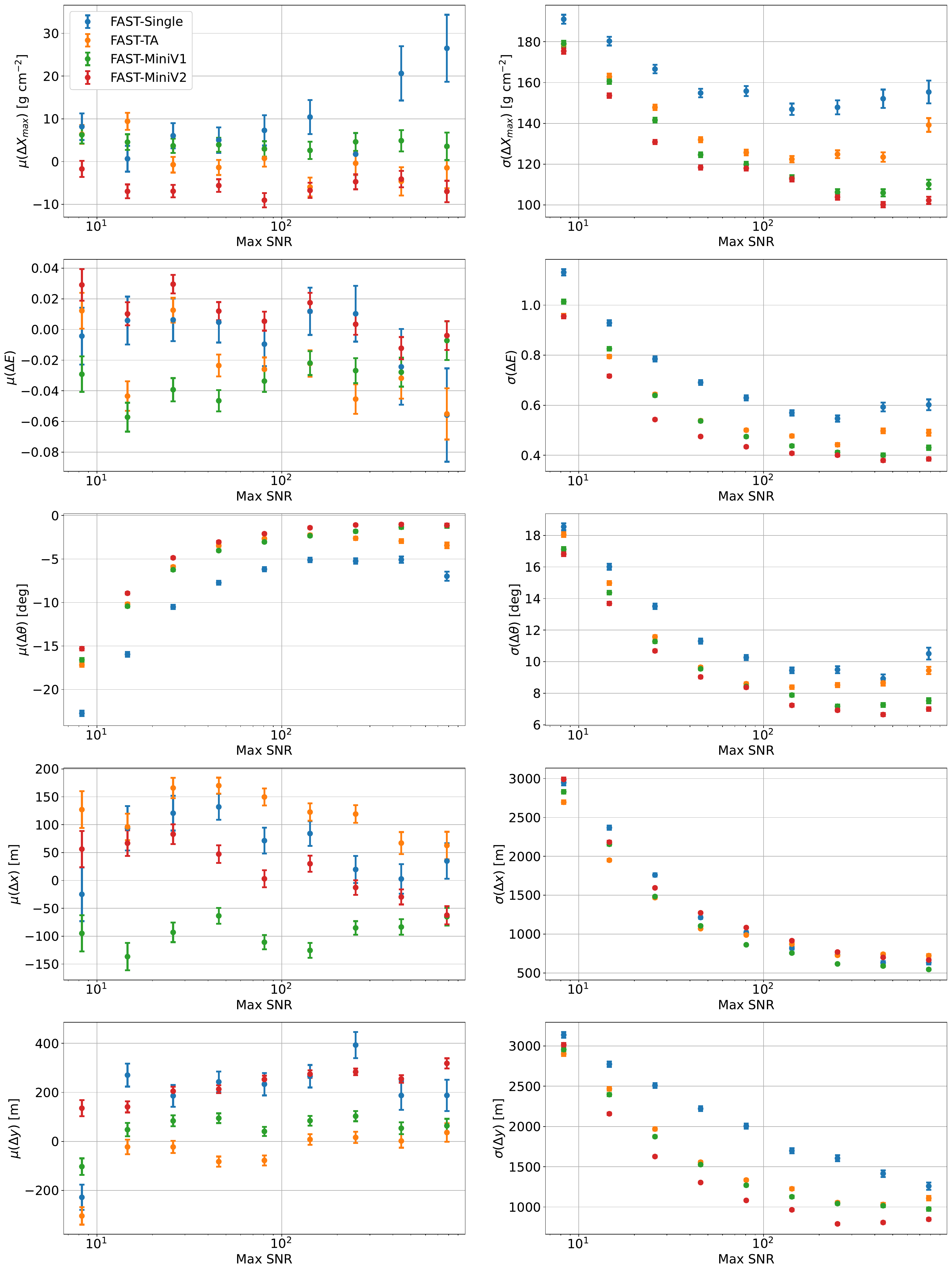}
    \caption{The biases and resolutions in the output parameters as a function of maximum SNR. The shape of the resolution plots matches the initial expectations outlined in Section \ref{sec:basicDNNperformance}.}
    \label{fig:tsfelSNR}
\end{figure}

The overall performance of the TSFEL DNN model on the test data set is shown across Figures \ref{fig:TSFELex}, \ref{fig:TSFELcore} and \ref{fig:TSFELzRes}. These plots show the biases and resolutions in each reconstructed parameter for each layout as a function of energy. The bottom two rows of panels in Figure \ref{fig:TSFELzRes} show the core and angular resolutions. The performance of each layout is categorised by either the number of triggered PMTs (for FAST-Single and FAST-TA) or the number of eyes viewing the shower (FAST-MiniArrayV1/2). Note that, unlike traditional bottom-up methods for estimating the shower parameters, where uncertainties in the geometrical parameters directly influence the reconstructed \Xmax{}/energy, the feed-forward neural network utilised here \say{reconstructs} each parameter simultaneously, meaning the parameter resolutions are somewhat independent of each other.

\vspace{5mm}

Overall, the shapes of each graph are similar to the behaviour observed with the Basic DNN. For FAST-Single, the biases and resolutions in each output parameter improve with more triggered PMTs up to the maximum of four, where the \Xmax{} and energy resolutions reach $\sim125$\gcm{} and 40\% respectively. For showers with energy $>10^{19}$\,eV the core resolution is $\sim1000$\,m and the angular resolution is $<15\degree$. Training on data containing only showers with all four pixels triggered does not impact the geometrical resolutions as shown by the magenta points in the core/angular resolution graphs. For showers which trigger only 1 PMT the network clearly does not have enough information to estimate the shower parameters any sort of accuracy, with core/angular resolutions of $\sim4000$\,m and $\sim35\degree$. Although the results with two/three triggered PMTs show marked improvement, it is thought that resolutions of $\lesssim1000$\,m and $\lesssim10\degree$ will be required for the TDR to meaningfully improve upon the first guess.
For FAST-TA the results are very similar with only minor improvements in the geometrical resolutions. Above 10$^{19}$\,eV the angular resolution reaches $10\degree$ for showers triggering $\geq4$ PMTs. For the same showers the core resolution is quite stable around 1000 - 1250\,m across the whole energy range. 

\vspace{5mm}

For FAST-MiniV1 and V2, the resolutions for a single Eye largely match the overall results for FAST-Single and FAST-TA. This seems to indicate that the additional information of telescopes at other locations \textit{not observing} any signal does not majorly improve the overall precision. This is likely because, for a particular set of telescopes, the phase space of showers which do not leave a signal in those telescopes is large, and thus the shower parameters are not significantly constrained when another set of telescopes far away does observe the shower. However the almost zero bias seen in reconstructed core $y$ values does indicate that the additional Eye/s help to remove some degeneracy. A sensitivity study investigating how much information is given by telescopes not observing the shower could be performed in future work. 

\vspace{5mm}

For showers observed in stereo the results are clearly improved with generally less overall bias and better resolutions across the whole energy range. Above $10^{19}$\,eV, \Xmax{} and energy are estimated within approximately 100\gcm{}/40\% and 75\gcm{}/30\% for 2-Eye stereo observation with FAST-MiniV1 and 3-Eye stereo observation with FAST-MiniV2 respectively. The geometrical resolutions for the same energy range and setups are roughly 750\,m/$8\degree$ and 750\,m/$6.5\degree$. Stereo observation with both eyes from the FAST-MiniV1 layout performs slightly better than stereo observation with two eyes using FAST-MiniV2. This is thought to be due to the angle at which the showers are viewed. Telescopes which face each other may better account for the degeneracy in the component of the core aligned with their pointing direction. A clear limit on the lowest energy detectable via stereo is seen in both the FAST-Mini plots, roughly $10^{18}$\,eV for two eyes and $10^{18.4}$\,eV for three eyes. The magenta points in the FAST-MiniV2 core and angular resolutions plots show the results when the model is trained using only events observed in stereo with 3-Eyes (\say{3-Eye stereo}). This test is done to check whether focusing only on high quality events will improve the reconstruction of said events. There is a small but noticeable improvement, however such a model cannot be applied to non 3-Eye stereo events. This is rather inconvenient and would require other models to be trained to analyse other data from the same layout. Moreover the geometrical resolutions are likely already sufficient for successful reconstruction with the TDR. This being said, there may be potential for future work to train separate models for different classes of events given a fixed layout i.e. FAST-3-Eye layout.

\vspace{5mm}

Lastly, Figure \ref{fig:tsfelSNR} shows the biases and resolutions of each parameter over all events for each layout as a function of the maximum SNR in an event. Unlike energy, the parameters of showers observed by the telescopes should not change significantly as a function of maximum SNR. Thus the shape of the resolution graphs are expected to decrease with SNR before flattening, as was predicted for energy in Section \ref{sec:basicDNNperformance}. This is indeed what is observed, with limited improvement in the resolutions above a maximum SNR of $\sim100$.

\section{Discussion}
Overall, there are two main takeaways from the above results. The first is that estimating the shower parameters with just one or two pixels in a single eye appears very challenging. In the worst case (one triggered pixel) a total of just 11 values are being used to estimate the 6 shower parameters (ignoring the additional 0s from non-triggered PMTs). In these cases, it is likely that significant degeneracies arise between the network inputs and possible output shower parameters, leading to poor resolutions. Examples of these degeneracies are shown explicitly in the next chapter. If a method of accurately estimating the shower parameters with only one/two pixels does exist, then it will likely only be viable for large signals and possibly require additional signal processing (i.e. filters to reduce the amount of noise) and/or a different/more complex model architecture, for example the highly successful transformer architecture being used to produce large language models \cite{vaswani2017attention}. 
Investigating the best-case scenario where the PMT traces contain no background noise may provide some insight into the limitations of the FAST telescope design. Such tests are left for future studies. For FAST telescopes operating in hybrid mode however, where the shower geometry is provided by a companion SD, one/two pixels may be sufficient to obtain a reasonable first guess of \Xmax{} and energy. A brute force search, although more time consuming, is also likely to be feasible in this case, thanks to the sufficiently small parameter space.

\vspace{5mm}

\begin{figure}
    \centering
    \includegraphics[width=\linewidth]{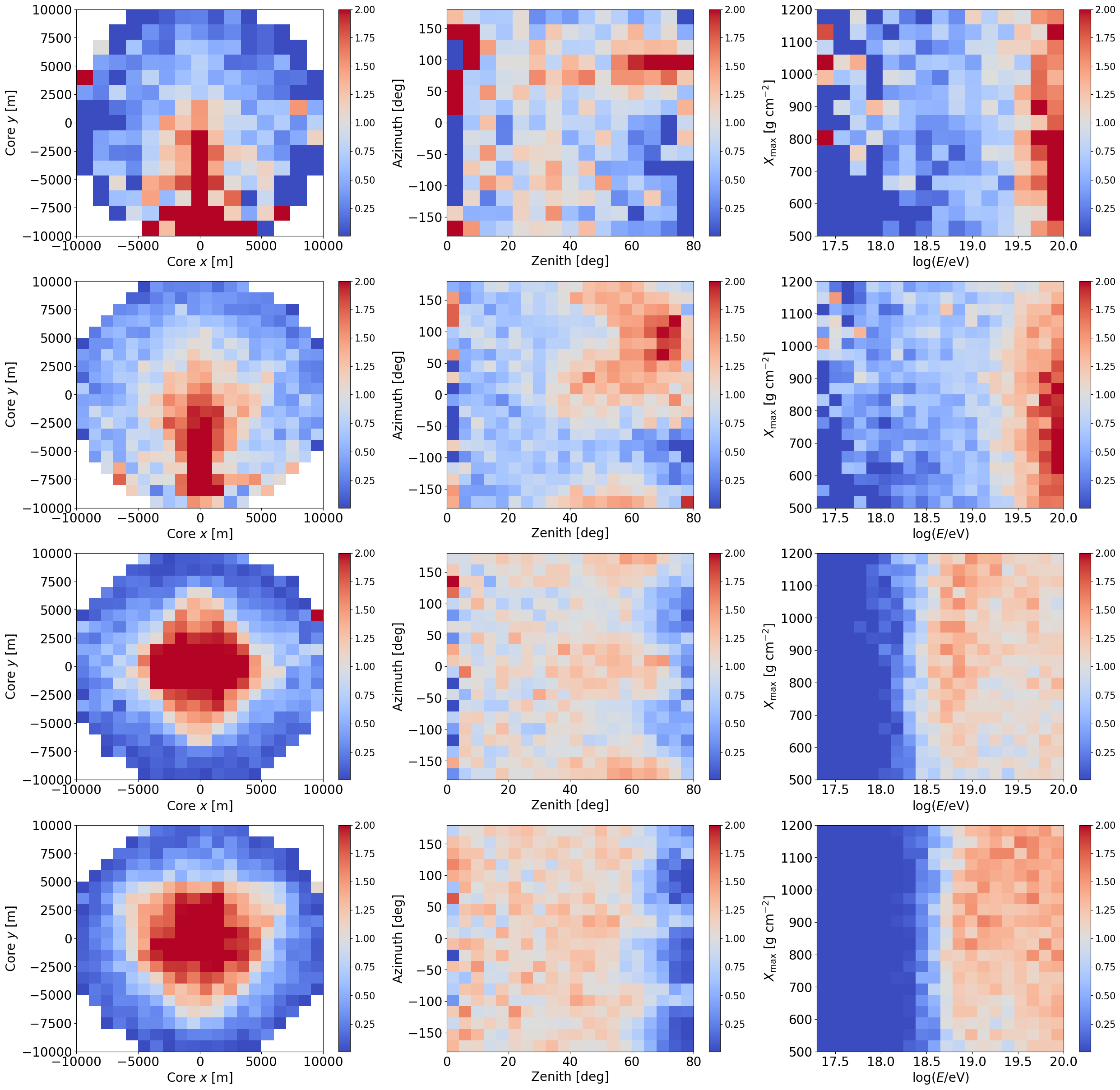}
    \caption{Ratios between the normalised number of events expected to be observed with FAST-Single (FAST-TA, FAST-MiniV1, FAST-MiniV2) given that 4 pixels ($\geq4$ pixels, 2 Eyes, 3 Eyes) triggered and the normalised number of events expected to be observed with at least one triggered pixel, split into core position, arrival direction and \Xmax{}/energy slices. From top to bottom the results are for FAST-Single, FAST-TA, FAST-MiniV1 and FAST-MiniV2.}
    \label{fig:fourPixelDiff}
\end{figure}

The second takeaway is that showers which trigger $\gtrsim4$ pixels in a single eye or are observed in stereo can be reconstructed with reasonable accuracy. However showers which meet these criteria necessarily posses a particular geometry and/or set of shower parameters. For example, only sufficiently energetic showers can be observed in stereo. Applying the first guess method only to these showers would introduce a bias in the showers that are reconstructed. This is illustrated in Figure \ref{fig:fourPixelDiff} which shows ratios between the normalised number of events expected to be observed with 4 pixels ($\geq4$ pixels, 2 Eyes, 3 Eyes) and the normalised number of events expected to be observed with at least one triggered PMT for FAST-Single (FAST-TA, FAST-MiniV1, FAST-MiniV2). Similar to Figure \ref{fig:basicDNNmaps}, the parameter space is shown using slices of core position, arrival direction and \Xmax{}/energy. For FAST-Single (FAST-TA), requiring observation with 4 pixels ($\geq4$ pixels) would result in higher energy showers and showers falling near/pointing directly towards the telescope/s being (relatively) $1.5\sim2$ times more abundant in the dataset. For FAST-MiniV1/2, requiring 2-Eye/3-Eye stereo observation would actually reduce the relative number of showers observed which point towards one of the telescope sets. The conditions would also increase the relative number of higher energy showers observed and the relative number of showers with core positions in the interior of the array. As for possible improvements to the stereo results, the filtering/alternate architecture suggestions made above also apply here. Further hyper-parameter optimisation and tuning of the current model could also be performed, ideally using a dedicated library such as Optuna \cite{optuna_2019}. Additionally, including the locations/pointing directions of triggered telescopes as inputs to the model may improve the first guess. Even if the current resolutions are sufficient, a better first guess will always be desirable as the subsequent minimisation using the TDR will require less time. It is recommended that future work focus on improving the FAST-MiniV2 results, as this is the layout which will form the fundamental shape of the FAST prototype arrays moving forward. 

\section{Shortcomings} 
Despite the positive results obtained here, there are a few noticable shortcomings of the models trained in this chapter. First, the models were only able to be trained on a fixed detector layout. Once trained, they could not be directly applied to other telescope configurations. This meant different datasets had to be prepared for each layout and the training performed separately, requiring additional computational time and resources compared to training a single model. Moreover, considering that the large scale deployment of FAST would necessarily take place over several years, with telescope layouts potentially changing and the spacings between different Eyes unlikely to be exactly as implemented during simulations/training, a model which can generalise to different layouts would be ideal. Whilst creating a layout independent model is not impossible, it is expected to require significantly more data and computing power. In addition, the final model may prove to be inferior to a similar model trained only on one specific layout, similar to how training only on stereo showers with FAST-MiniV2 was found to improve the geometrical resolutions over the original setup. 

\vspace{5mm}

The second issue is that no event-by-event uncertainty in the first guess estimate is provided with the current models. Although such an estimate may not strictly be necessary, given that the TDR does provide one, it would be useful to assess how confident the model is in it's guess. This could be achieved by having the models predict the uncertainty on each parameter or by applying Bayesian techniques. However doing so would once again necessitate far more computation and/or a more complex model architecture. The next chapter attempts to address both of these concerns using a different first guess approach.

\section{Combined Machine Learning and Top-Down Reconstruction Performance}
\label{sec:combinedMLcheck}
To end this chapter, the combined performance of the TSFEL DNN first guess estimate followed by the TDR is investigated. A sample of 10,000 showers was simulated using the same telescope layouts and parameter distributions as in Section \ref{sec:MLdataset}, with the exception of energy which was sampled from a $E^{-2}$ distribution. This was done to ensure sufficient statistics at low energies once the SNR threshold cut had been applied. For each of the four layouts, the TSFEL DNN for that layout was applied to showers passing the threshold trigger, giving first guesses of the shower parameters. These guesses were then used as the initial parameters for the TDR. 

\vspace{5mm}

The quality cuts applied to the final reconstruction results were; successful minimisation, \Xmax{} in the FOV of at least one triggered telescope, relative errors in the \Xmax{} and energy reconstructions ($X_\mathrm{max}^\textrm{err}/$\Xmax{} \& $E_\textrm{err}/E$) $<0.5$ and absolute errors in the core $x$ and core $y$ values $<1000$\,m. The differences between the reconstructed and true parameters were then analysed. Events were separated based on the number of triggered PMTs/number of triggered Eyes, as done in Figure \ref{fig:TSFELex}. The results for FAST-MiniArrayV2 are shown in Figures \ref{fig:xmaxDiffMini2ml} - \ref{fig:angDiffMini2ml} below. The results for the other layouts are shown in Appendix \ref{sec:additionalMLTDRplots}. For $\Delta{}$\Xmax{} and $\Delta{}E$
Gaussian functions have been fit to better estimate the true means and standard deviations of the underlying distributions. Investigating the source of the several outliers seen in these distributions is planned for future work. Figure \ref{fig:tsfelWithTDRsummary} summarises the results for each layout.

\vspace{5mm}

The overall findings mirror those from Section \ref{sec:TSFELDNNperformance}. For just one or two triggered PMTs in a single FAST Eye, there does not appear to be enough information to estimate the shower parameters with the desired accuracy - this being the requirements for a future GCOS candidate i.e. resolutions in \Xmax{}, energy and arrival direction of $\sim30$\gcm{}, 10\% and $1\degree$ respectively. For three or more PMTs however the results significantly improve for each layout. The \Xmax{} and energy resolutions are now roughly 40 - 60\gcm{} and 10\% respectively. The angular and core resolutions also improve to 500 - 800\,m and $5$ - 8$\degree$ depending on the layout. It is also interesting to see that the results for FAST-TA appear slightly better than those for FAST-MiniV1/2 \textit{for a single eye}. This is likely due to the three telescopes for FAST-TA vs. the two telescopes at each eye for the FAST-Mini layouts. 

\vspace{5mm}

For stereo observation with two eyes, the core and angular resolutions improve further, down to a few hundred metres and $\sim2.5\degree$ respectively. The resolutions from FAST-MiniV1 are slightly better than those from FAST-MiniV2, again similar to the machine learning results. This further verifies that the best reconstruction of showers observed from two locations is when the telescopes are facing each other. Finally, for 3-fold stereo observation, the  angular resolution improves to better than $1\degree$, reaching the target resolution. It should be stressed here that, like Figure \ref{fig:newReconRes}, these results are for perfectly ideal simulations and do not account for several real-world effects/features present in data (see the discussion in Section \ref{sec:CheckingRecon}). Still, the results are encouraging and show that, for events with either many triggered pixels or stereo observation, the FAST telescopes do have the potential to accurately reconstruct the shower parameters. This is, once again, provided that the FAST simulations on which both the first guess method and TDR are based can be made to accurately replicate the time-dependent observation conditions of the telescope and surrounding environment.

\begin{figure}
    \centering
    \includegraphics[width=1\linewidth, trim={0 0 1cm 0}]{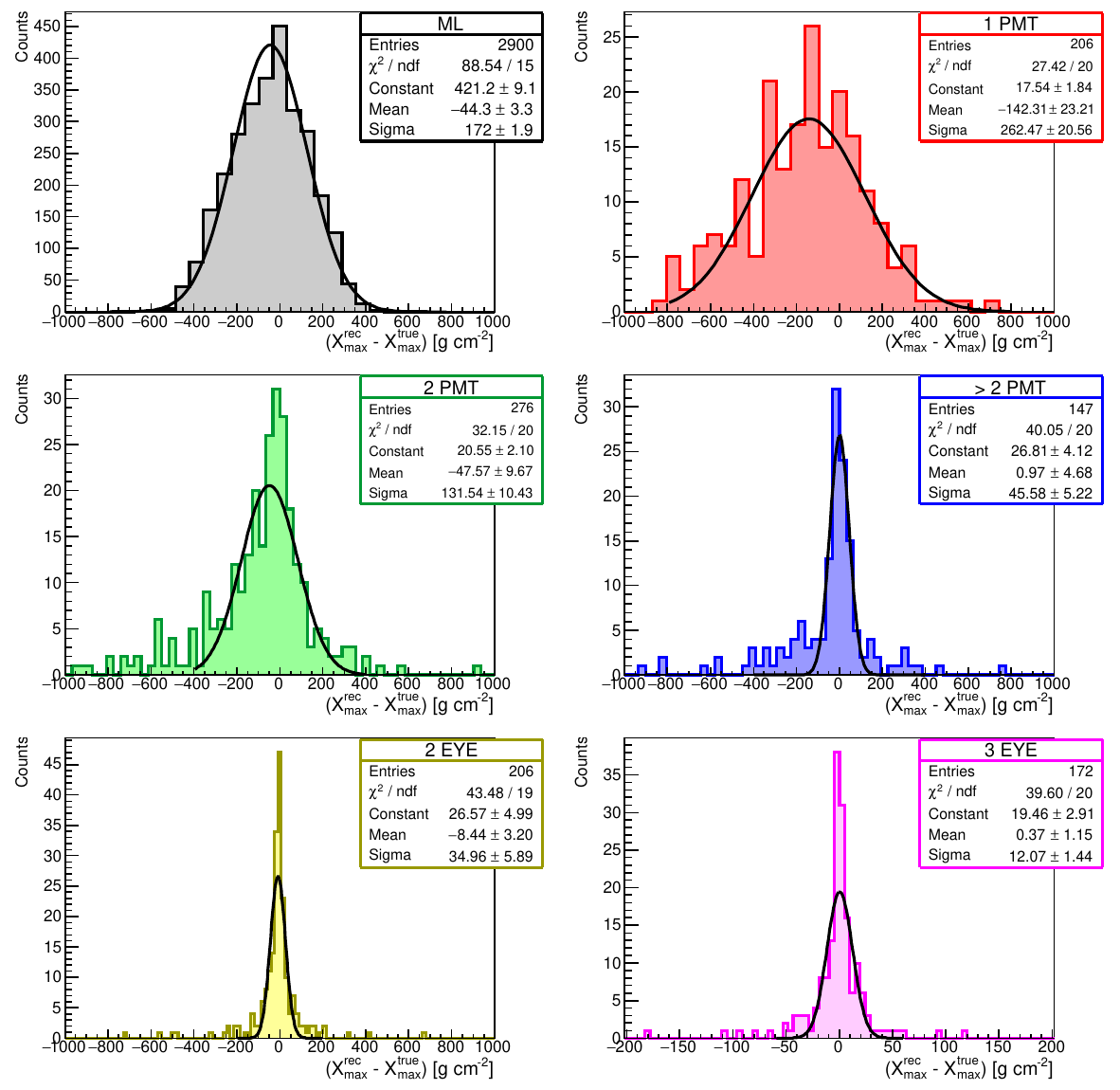}
    \caption{Difference between the reconstructed and true \Xmax{} for events triggering different numbers of PMTs/eyes. The red, green and blue histograms show the results for 1, 2 and $>2$ triggered PMTs (1 Eye). The yellow and pink histograms are for stereo observation with 2 and 3 triggered Eyes respectively (total number of PMTs triggered$\geq2$). The 3-Eye histogram has been zoomed in to better view the central part of the distribution. This causes one event at $\Delta$\Xmax{}$\sim-700$\gcm{} to not be shown.}
    \label{fig:xmaxDiffMini2ml}
\end{figure}

\begin{figure}
    \centering
    \includegraphics[width=1\linewidth, trim={0 0 1cm 0}]{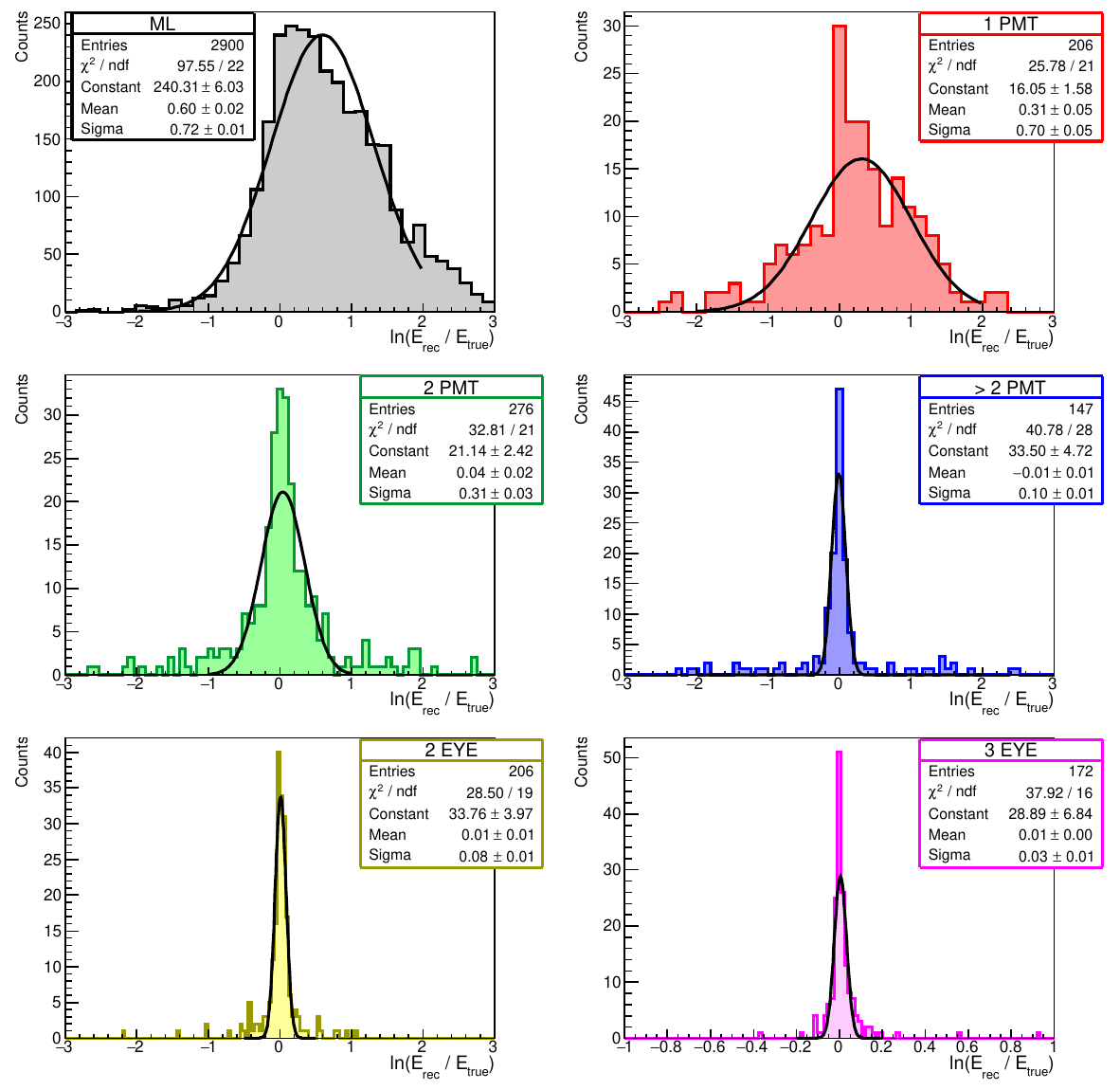}
    \caption{Same as Figure \ref{fig:xmaxDiffMini2ml} but for energy. The 3-Eye histogram has been zoomed in to better view the central part of the distribution. This causes one event at $\Delta{}E\sim1.2$ to not be shown.}
    \label{fig:energyDiffMini2ml}
\end{figure}

\begin{figure}
    \centering
    \includegraphics[width=1\linewidth, trim={0 0 1cm 0}]{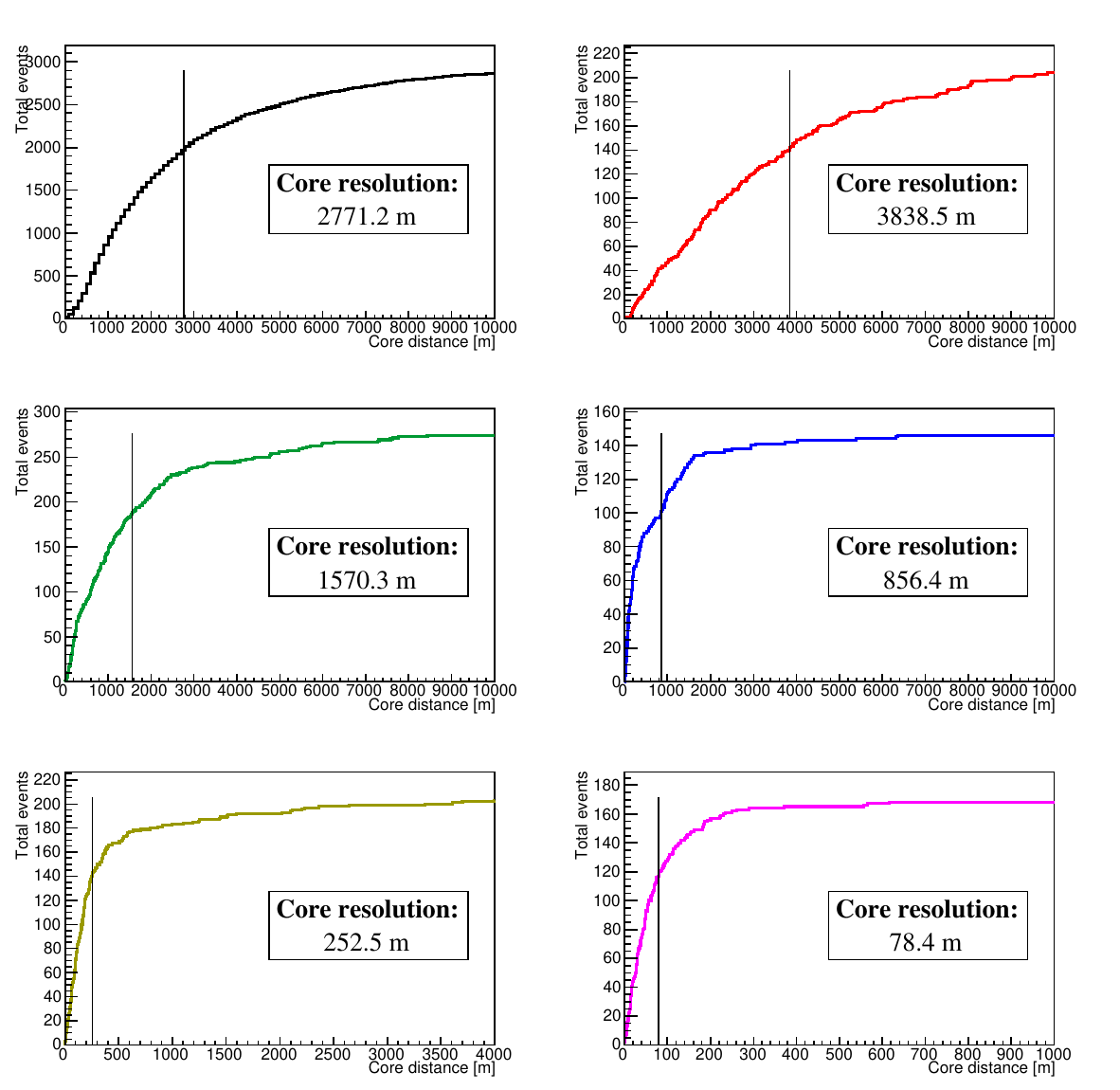}
    \caption{Cumulative distributions of the difference in core position for each of the different event types (same layout as Figure \ref{fig:xmaxDiffMini2ml}). The point where the cumulative distributions exceed 68\% of the total number of events is indicated by the black line. The numerical values are given in the boxes. The 2-Eye and 3-Eye plots have been zoomed in to better view the detail at smaller core distances.}
    \label{fig:coreDiffMini2ml}
\end{figure}

\begin{figure}
    \centering
    \includegraphics[width=1\linewidth, trim={0 0 1cm 0}]{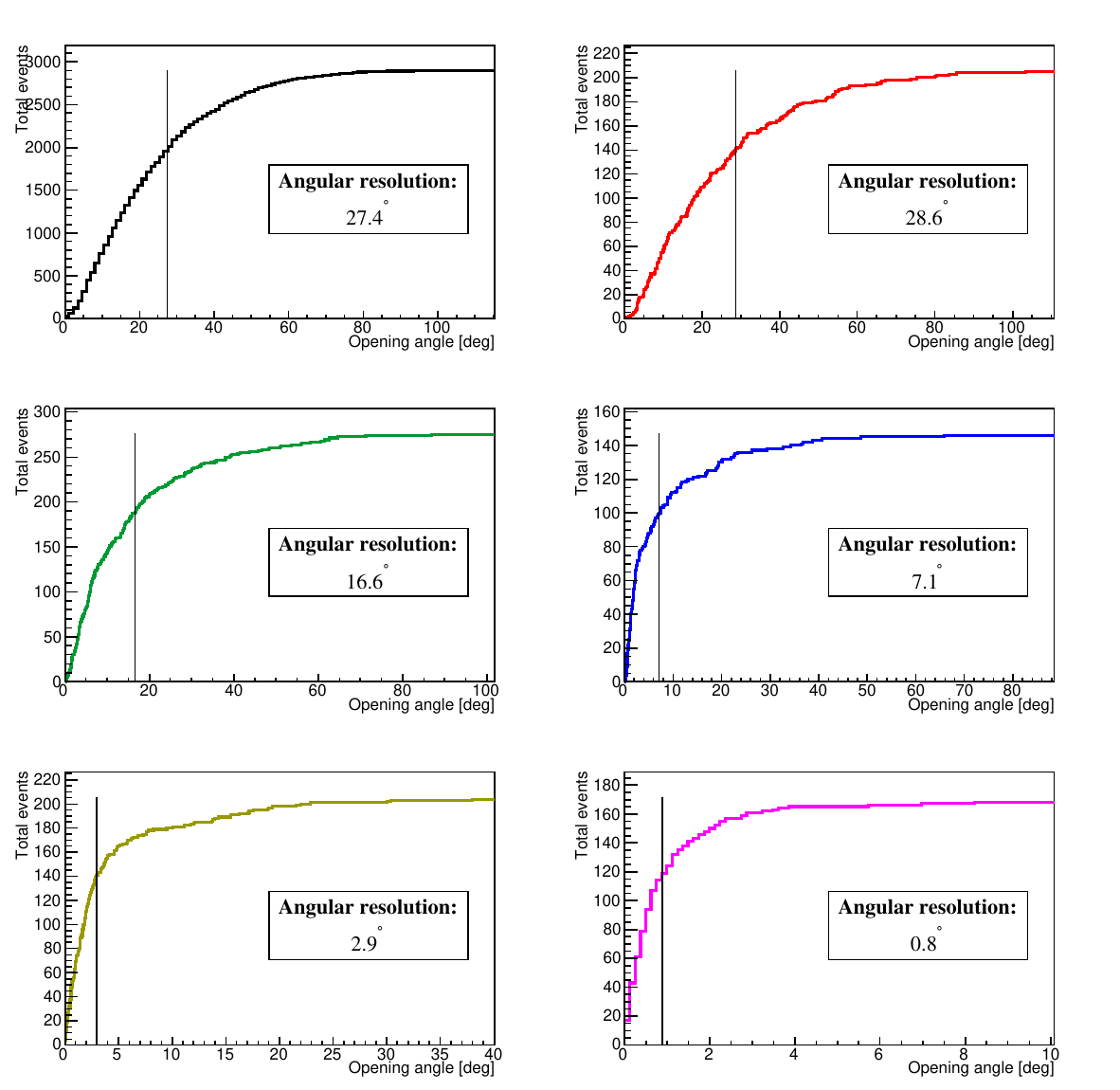}
    \caption{Same as Figure \ref{fig:coreDiffMini2ml} but for the angular resolution - the histograms show the cumulative distribution of the opening angle. Similar to Figure \ref{fig:coreDiffMini2ml}, the 2-Eye and 3-Eye plots have been zoomed in to better view the detail at smaller opening angles.}
    \label{fig:angDiffMini2ml}
\end{figure}

\begin{figure}
    \centering
    \includegraphics[width=0.99\linewidth]{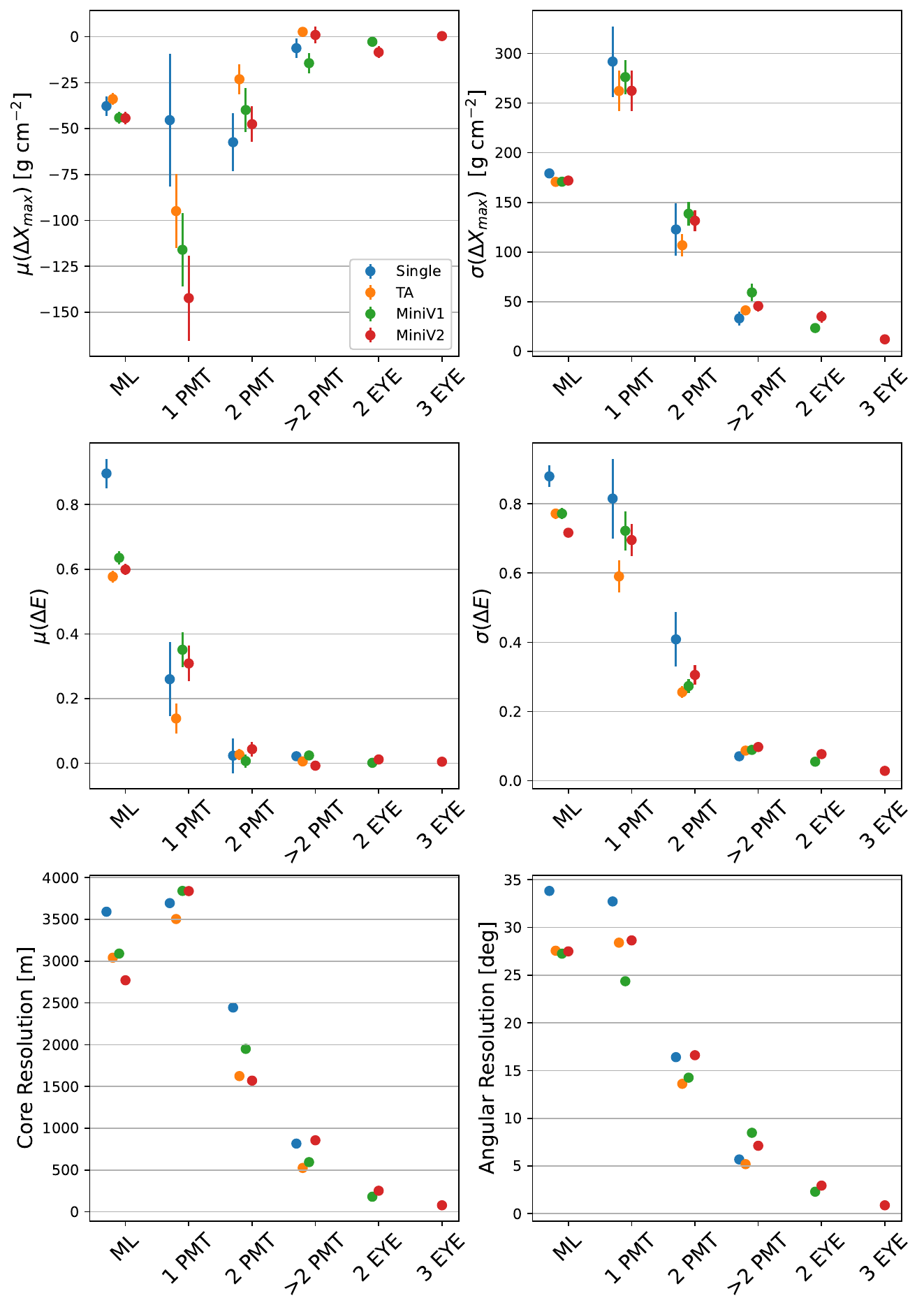}
    \caption{The reconstructed \Xmax{} (top panels) and energy (middle panels) biases and resolutions using the TSFEL DNN + TDR for each of the four layouts. The results are shown for events with different numbers of triggered PMTs/triggered Eyes. The TSFEL DNN estimates which encompass all event types are shown as the leftmost points.}
    \label{fig:tsfelWithTDRsummary}
\end{figure}

\section{Summary}
This chapter has presented the first investigation into the potential for using neural networks to predict the shower parameters for the current and near future FAST-prototype layouts. Of the three different models tested, the TSFEL DNN - an extension of the architecture previously used with FAST - was found to be the best overall choice. This model was a feed-forward DNN which utilised 11 different features from each PMT trace as inputs. The model was found to perform best when reconstructing showers observed in stereo or triggering $\gtrsim4$ pixels in a single Eye. The best resolutions were obtained with the FAST-MiniV2 model above a true energy of $10^{19}$\,eV. The \Xmax{}, energy, core and angular resolutions were 75\gcm{}, 30\%, 750\,m and $6.5\degree$ respectively. Future work could investigate generalising the model such it can be applied to different layouts and/or improving the reconstruction accuracy with additional trace processing. The combined performance of the TSFEL DNN + TDR was also tested, with results showing resolutions in \Xmax{} and energy of $\sim40$ - 60\gcm{} and $\sim10$\% for all layouts when $>2$ PMTs were triggered in a single eye. In particular, 3-Eye stereo observation with the FAST-MiniV2 layout was, in perfectly ideal conditions, shown to be able to achieve the desired parameter resolutions for a future GCOS style detector, namely $<30$\gcm{} in \Xmax{}, $<10\%$ in energy and $\sim1\degree$ in arrival direction. Including additional sources of fluctuations/uncertainty in the simulations which are present in data, such as saturated signals and varying atmospheric conditions, and/or testing the TDR's ability to recover any induced biases because of neglecting such effects during model training is recommended for future work.

\chapter{First Guess Estimation II - Template Method}
\label{ch:TEMP}
Following on from Chapter \ref{ch:ML}, this chapter presents an alternative method for obtaining a first guess of the shower parameters. Rather than attempting to learn the relationship between PMT trace features and shower parameters, a library of template traces corresponding to many different combinations of the possible shower parameters is created. A first guess estimate can be made by simply comparing a set of data traces to each template in the library and choosing the best match. This approach will be referred to as the \say{Template Method}. The motivation for the Template Method comes from the two main shortcomings identified in the previous chapter when using machine learning models, namely the need to train different models for different telescope layouts and the lack of an uncertainty estimate on the first guess parameters. The chapter begins by introducing two additional constraints on the triggering of individual PMTs, before explaining the methodology and analysing the performance of the method.

\section{Additional Trigger Conditions} 
\label{sec:additionalTriggering}
In this chapter, two additional trigger conditions must be fulfilled in conjunction with the threshold trigger (SNR $>6$) for a pixel to be considered as having passed the event level trigger. The conditions relate to the absolute timing of said threshold trigger/s and whether or not the triggered pixel forms part of a group of triggered pixels.

\subsection{Absolute Timing}
\begin{figure}
    \centering
    \includegraphics[width=0.9\linewidth]{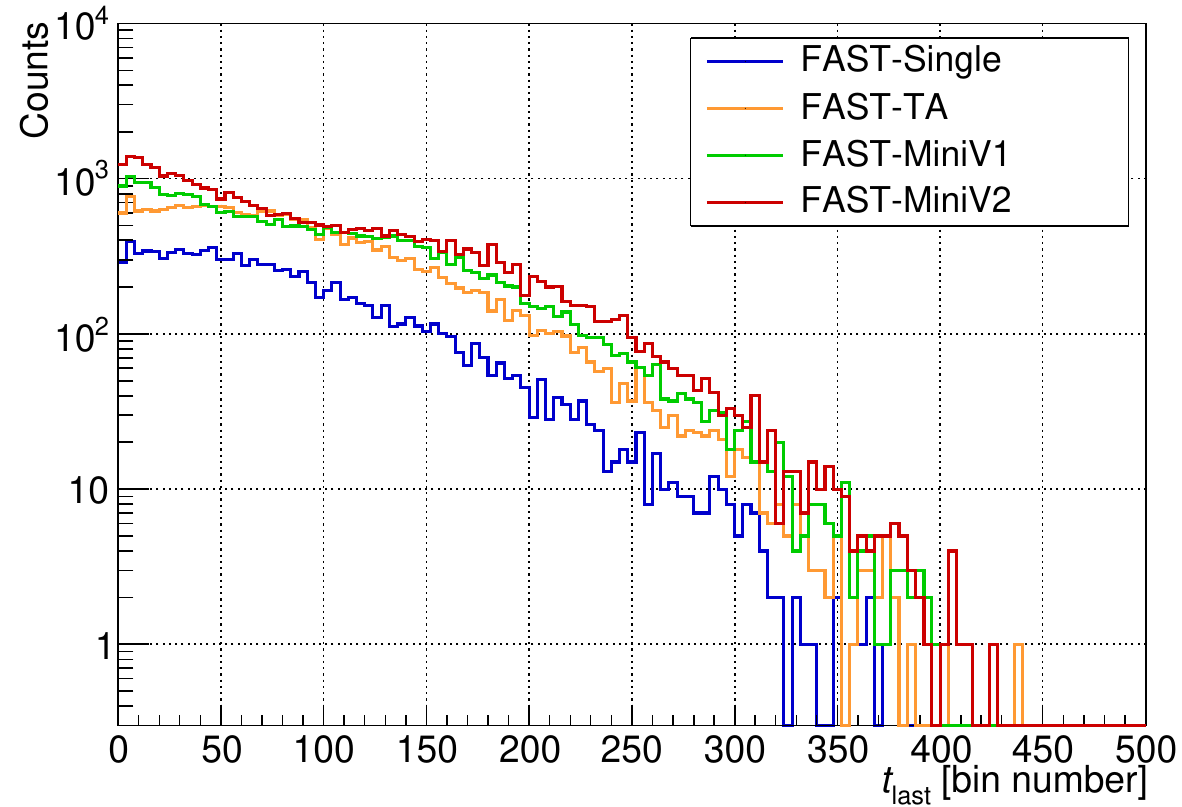}
    \caption{Distributions of $t_\textrm{last}$ calculated from subsets of the datasets introduced in Chapter \ref{ch:ML} for the four different telescope configurations; FAST-Single, FAST-TA, FAST-MiniV1 and FAST-MiniV2 (see Figure \ref{fig:mlCorePos}).}
    \label{fig:absoluteTimeChecking}
\end{figure}

The absolute timing condition checks whether the centroid time of each pixel, as calculated using Equation \ref{eqn:centroidTime}, lies within an expected range determined by the telescope layout and distribution of showers expected to be observed. For the events which will be analysed in this chapter, which are a sample of the simulations generated in Section \ref{sec:MLdataset}, the appropriate range is between bins 0 - 500 (for 100\,ns bin traces). This was determined by analysing the timing of the peak photo-electron counts in each PMT for these events. Specifically, for a single event, the bin containing the maximum number of photo-electrons for each PMT trace was found. Recall that these traces contain no background noise and so the maximum is used as a proxy for the centroid time. Of these bins, the bin corresponding to the latest time, $t_\textrm{last}$, was taken. This procedure was performed for each event and the distribution of $t_\textrm{last}$ plotted. Figure \ref{fig:absoluteTimeChecking} shows the results when applied to the four different layouts. The values of $t_\textrm{last}$ range from 0 to $\sim400$ irrespective of the layout. Thus in simulations, for these layouts, it can be expected that the centroid time of a pixel which contains true signal (i.e. not random noise) should lie between bins 0 - 500. 

\vspace{5mm}

For data, it may seem that cutting out threshold triggers based on their absolute time could lead to ignoring important signals as, unlike in simulations, the time at which the shower enters the atmosphere and hence the absolute time offset of the signal is not known. However in a data trace the rough position of the signal \textit{is} generally known. This is because when a trigger is met, a set number of samples before the trigger location are saved. Thus threshold triggers found either far before the expected trigger location or (for the above examples) more than 500 bins after, can be safely assumed to be noise. The 0 - 500 range is valid for the simulations used in this chapter, however in general the limits of $t_\textrm{last}$ will depend on the telescope configuration and distribution of showers being observed. A more robust condition might be to limit the acceptable centroid times in a single FAST-Eye to be within 500 bins of the earliest centroid time in that eye (as shown by the FAST-TA histogram). This could be implemented in future work.

\subsection{Pixel Grouping}
Pixels which have a maximum SNR above the threshold and possess centroid times within the expected range are then checked for grouping. The logic applied here is similar to how Auger and TA form FD triggers by looking for connected groups of triggered pixels, though at a coarser scale. The logic is applied to each FAST eye separately. A step by step explanation of the algorithm is given below. Figure \ref{fig:groupingLogic} illustrates each of the listed cases using the pixel layout for FAST@TA.
\begin{enumerate}
    \item For each triggered pixel in a single FAST eye, check which (if any) neighbouring pixels are triggered. Neighbouring pixels are those which view a portion of the sky adjacent to the FOV of the triggered pixel. This includes pixels from the same telescope and two pixels from either the left or right neighbouring telescope. Triggered pixels with no neighbouring triggered pixels will be referred to as \say{isolated pixels}. Pixels with 1 or more triggered neighbours will be said to form a single \say{pixel group}.
    \item If the FAST eye contains a single isolated pixel, then this pixel is considered to be part of the event (case \textbf{(a)}). The large diameter of the FAST PMTs means many events, particular low energy events, will only trigger a single PMT in a FAST eye, so isolated pixels should be kept. If there is more than one isolated pixel, only the pixel with the largest SNR is kept (case \textbf{(b)}).
    \item If there is only one pixel group in a FAST eye, then this group is considered part of the event (case \textbf{(c)}). Any isolated pixels are disregarded (case \textbf{(d)}). 
    \item In the event of two or more pixel groups, the pixel group with the larger number of triggered pixels is chosen (case \textbf{(e)}). For two groups of identical size the group with the largest maximum SNR is chosen (case \textbf{(f)}).
\end{enumerate}

\begin{figure}[t]
    \centering
    \includegraphics[width=1\linewidth]{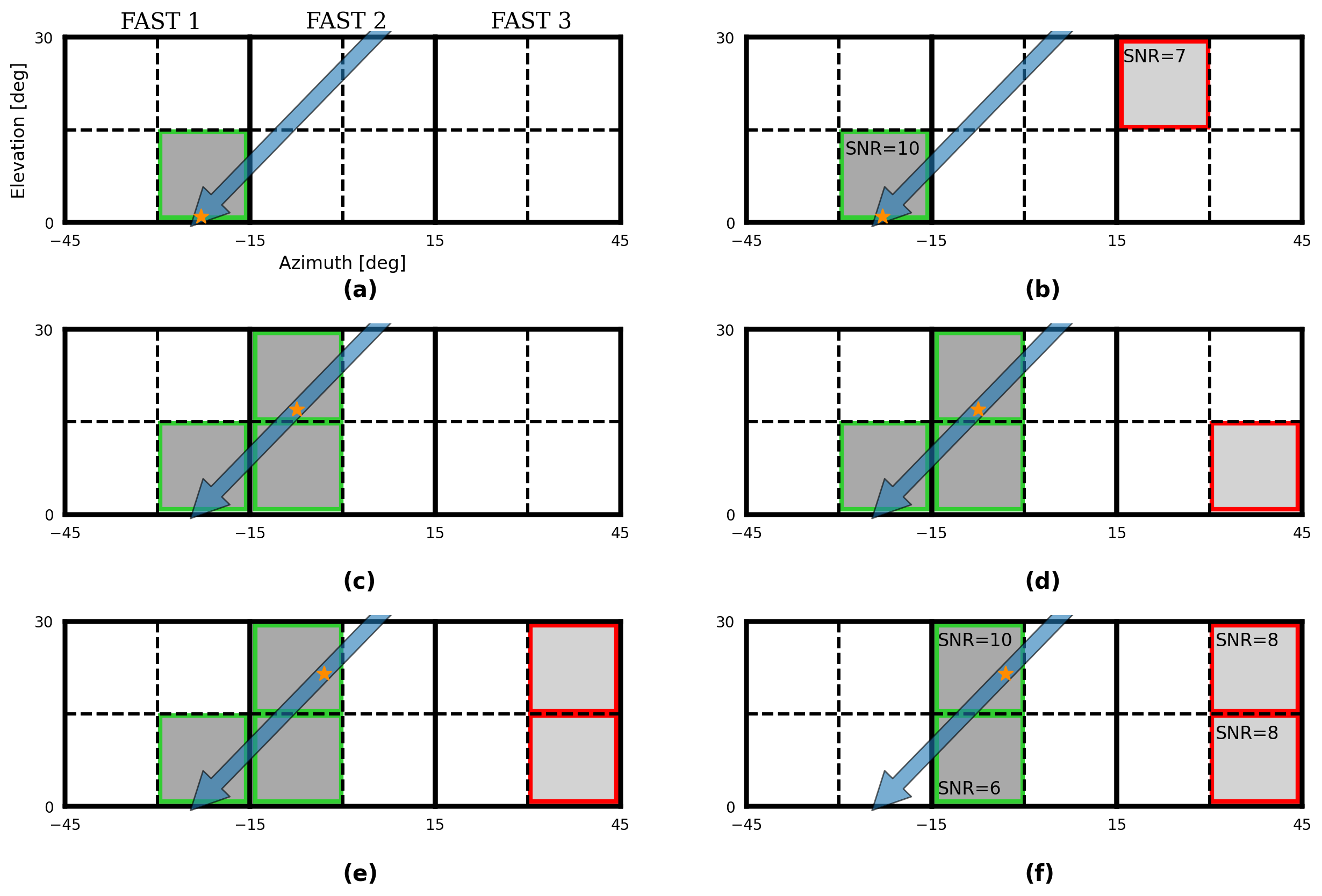}
    \caption{Diagrams showing examples of the pixel grouping logic. Pixels outlined in green (red) have passed (not passed) the grouping check. The blue arrows and orange stars represent examples of how the shower axis and \Xmax{} location respectively \textit{might} appear in the FOV of the telescopes for these examples.}
    \label{fig:groupingLogic}
\end{figure}

It may seem redundant to apply these additional constraints given the false positive rate of the threshold trigger (SNR $>$6) is $\sim10^{-5}$. Indeed when applied to simulations, where added noise is sampled from a simple Gaussian with a mean of zero, these conditions are essentially unnecessary. However, both the magnitude and baseline of the background noise from data taken by the current prototypes can vary noticeably more than in simulations. Appendix \ref{apx:baseline} shows an example of a data trace with a fluctuating, non-zero baseline. Such traces cause many more false triggers than in simulations when simply applying Equation \ref{eqn:newsnr}, since this equation assumes a noise baseline of zero. The extra conditions outlined above, though imperfect, help to remove these false triggers and thus allow for the Template Method (and all other first guess methods) to be more reliably applied to data. The conditions are introduced here so that the following performance analysis reflects the performance on data as closely as possible. Of course, a modified SNR formula which incorporates the possibility of a fluctuating baseline could be composed. Such a function is used in Section \ref{sec:coincEventSearch} when searching for coincidences between Auger/TA data and the FAST prototypes. However even with an improved SNR formulation, pixels which have an SNR above the threshold but \textit{do not} pass both the above criteria (or a refined version of them) should not be considered as containing signal from air showers.

\section{Basic Methodology}
As with any template matching technique, the fundamental idea of the Template Method is very simple. First, generate many examples of the different signals expected to be observed by a \textit{single} FAST telescope using simulations. In this case a single template corresponds to a set of four PMT traces. Then, for some measured data (assuming a single telescope), compare the traces with each template via some metric. The shower parameters corresponding to the template which \say{best match} the data traces are chosen as the first guess values. Equation \ref{eqn:eventLikelihood} is used to quantitatively compare each template with data, i.e. the same negative log-likelihood function used by the TDR. Thus the template which gives the minimum log-likelihood when compared to the data is considered to be the best match. Furthermore, the uncertainty in the first guess parameters can, in principle, be estimated. This would be done by evaluating the one sigma contours of the likelihood function using the set of values obtained by calculating each template's likelihood. Altogether, the Template Method operates similarly to the TDR. The difference is that, in order to account for time-dependent effects, the TDR simulates the \say{templates} from scratch one after the other with a minimiser choosing each set of shower parameters to test. In the case of multiple telescopes observing the same shower, the template matching described above can be performed for each telescope separately. Provided that a method of combining the results from each telescope can be devised (see Section \ref{sec:multipleTels}) this means that the Template Method can be applied to any configuration of telescopes.

\section{Template Library}
\label{sec:tempDataset}
The library of templates was simulated as follows. A single FAST telescope was placed at (0,0) pointing along the $y$-axis, with showers simulated on a grid of core positions as shown in Figure \ref{fig:templateCores}. For the ideal directional efficiency map used in these simulations (see Appendix \ref{fig:idealDirEffMap}), the response of the FAST telescope is symmetric. As such only showers with core $x$ values $\leq0$ were simulated. This is indicated by the red line and grayed out points on the right hand side of the plot. At each core location showers were simulated with steps in \Xmax{} of 100\,g\,cm$^{-2}$, zenith angle of $5\degree$ and azimuth angle of $10\degree$. The precise values for each parameter are summarised in Table \ref{tab:templateData}. Note that the energy is fixed as, in the FAST simulation, changing the energy simply scales the simulated traces. Thus during inference the template traces can be scaled to find the best matching energy. The templates were simulated with no added noise and using the standard parametric atmosphere as done for the machine learning dataset.

\begin{table}[h!]
    \centering
    \begin{tabular}{|c|c|}
    \hline\hline
        \textbf{Parameter} & \textbf{Values} \\
        \hline
        \Xmax & 500, 600, 700, ..., 1200\,g\,cm$^{-2}$ \\
        \hline
        Energy & Fixed (10$^{19}$\,eV) \\
        \hline
        $\theta$ & 0, 5, 10, ..., 80$\degree$ \\
        \hline
        $\phi$ & $-170$, $-160$, $-150$, ..., 180$\degree$ \\
        \hline
        Core $x$ & $-15$, $-14$, $-13$, ..., 15\,km \\
        \hline
        Core $y$ & $-5$, $-4$, $-3$, ..., 25\,km \\
    \hline\hline
    \end{tabular}
    \caption{The values of each shower parameter used to create the template data set.}
    \label{tab:templateData}
\end{table}

\begin{figure}[t]
    \centering
    \includegraphics[width=0.8\linewidth]{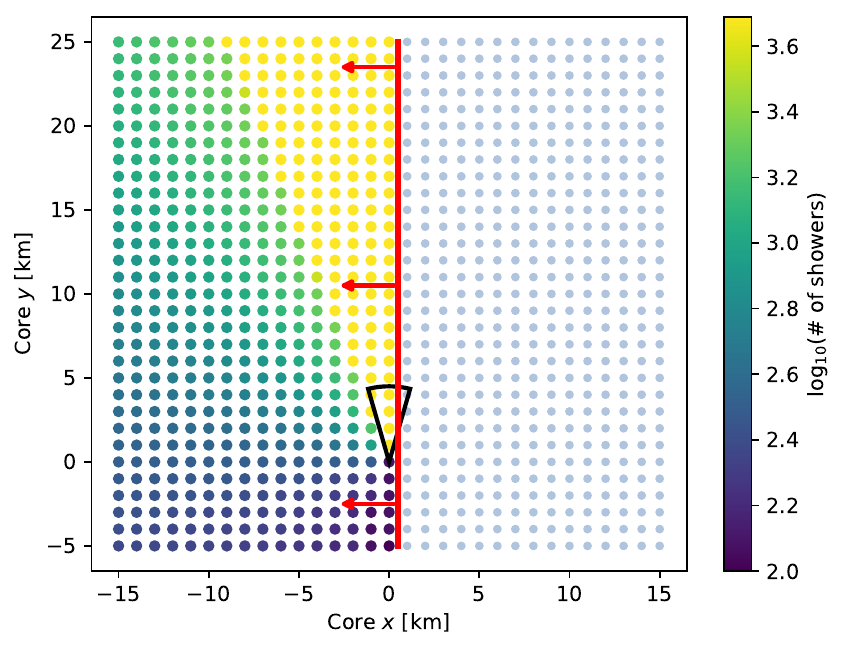}
    \caption{Core locations of the template showers. Only showers with core locations left of the red line were simulated. The colour of each point represents the number of simulated showers at that core location after removing showers outside the FOV of the telescope. The telescope is shown in black.}
    \label{fig:templateCores}
\end{figure}
 
This setup gives 8 different values for \Xmax{}, 17 for $\theta$, 36 for $\phi$ and 480 different core positions, equating to approximately 2.35\,million showers. Many of these showers do not lie in the FOV of the telescope (e.g. a shower landing at $(-15\,\textrm{km},-5\,\textrm{km})$ with an azimuth of $-90\degree$). Removing these showers leaves $\sim980,000$ templates. Including showers with core $x>0$ gives a total of $\sim1.7$\,million possible templates with which to compare. The spacing of the shower parameters for this library was decided based on a trade-off between resolution and required simulation time. In the ideal case, the Template Method described below would always choose the template with shower parameters \say{closest} to those of the data event. Note that closest here refers to the Euclidean distance in the shower parameter space. This terminology will be used throughout the rest of the chapter. For this scenario, the upper limit on the difference between the true and reconstructed \Xmax{} and core position would be
\begin{equation}
     \left(\Delta X_\textrm{max}\right)_\textrm{max}\leq50\textrm{\,g\,cm}^{-2} 
\end{equation}
\begin{equation}
     \left(\Delta \textrm{Core}\right)_\textrm{max}\leq\sqrt{500^2+500^2}\approx700\,\textrm{m}.
\end{equation}
For estimates of $\theta$ and $\phi$ within $2.5\degree$ and $5\degree$ respectively (half the size of the parameter spacing), the average opening angle $\bar\alpha$ can be estimated by
\begin{equation}
    \bar\alpha \approx \frac{1}{5\times10\times80}\int_{-2.5}^{2.5}\mathrm{d}\gamma \int_{-5}^{5}\mathrm{d}\delta \int_0^{80}\mathrm{d}\theta \quad g(\theta,\gamma,\delta)
\end{equation}
where
\begin{equation}
    g(\theta,\gamma,\delta) = \arccos\left( \sin(\theta)\sin(\theta+\gamma)\cos(\delta)+\cos(\theta)\cos(\theta+\gamma) \right)
\end{equation}
Numerical integration gives $\bar\alpha\approx$2.14$\degree$. Although the Template Method will of course not always return the closest event, these ideal \say{resolutions} were considered sufficient for a first guess. Moreover, reducing the step size by half in one or more of the parameters would extend the required simulation time from several days to potentially several weeks.

\section{Template Comparison Method}
\label{sec:tempCompMethod}
When comparing a set of template traces $\mathcal{T}$ to a set of data traces $\mathcal{D}$, the goal is to find the time-shifted and scaled version of $\mathcal{T}$ which gives the minimum negative log-likelihood with respect to $\mathcal{D}$. Doing this for every template ensures a fair comparison between the likelihoods. Let $t_\textrm{off}$ be the optimal time-offset between $\mathcal{D}$ and $\mathcal{T}$ and $A$ be the \say{energy scale factor}, i.e. the optimal value by which $\mathcal{T}$ should be scaled to match $\mathcal{D}$. Ideally, these values would be found by checking all possible time-offsets and performing some fitting procedure for $A$ at each offset. However, as will be shown in the next section, performing a fit for $A$ at even just a few offset values greatly increases the required computation time per template. The method outlined below describes a quicker alternative. Note that whilst this method may not always select the optimal values for $t_\textrm{off}$ and $A$, the considerable increase in speed makes it more appropriate as part of a first guess estimation.

\vspace{5mm}

An initial time-offset $t_\textrm{init}$ is determined by taking the difference between the centroid time of the first triggered PMT in $\mathcal{D}$ and the centroid time from the same PMT in $\mathcal{T}$ and rounding the difference to the nearest bin. This roughly aligns the peak of the first triggered PMT in data with the peak of the corresponding PMT in the template. The template traces are then shifted in one bin steps by $t_\textrm{test}\in(t_\textrm{init}-5,t_\textrm{init}+5)$.  For each $t_\textrm{test}$ the corresponding scaling factor $A_\textrm{test}$ is estimated as follows. Let $S_i^\textrm{data}$ be the signal of a PMT $i$ passing the event level trigger. This signal is calculated using Equation \ref{eqn:pmtSig} and the start and stop bins, $k_{\textrm{start},i}$ and $k_{\textrm{stop},i}$, which maximise the SNR. The signal in the corresponding \textit{shifted} template PMT trace is
\begin{equation}
    S_i^\textrm{temp} = \sum_{j=k_{\textrm{start},i}}^{k_{\textrm{stop},i}-1}s_j^\textrm{temp}
\end{equation}
where $s_j^\textrm{temp}$ is the signal in the $j^\textrm{th}$ bin of the template PMT trace. After calculating $S_i^\textrm{data}$ and $S_i^\textrm{temp}$ for each PMT passing the event level trigger, $A_\textrm{test}$ is then given by
\begin{equation}
\label{eqn:energyScaleFactor}
    A_\textrm{test}=\frac{\sum_iS_i^\textrm{data}}{\sum_iS_i^\textrm{temp}}
\end{equation}
which corresponds to a first guess energy of
\begin{equation}
\label{eqn:templateEnergy}
    E_\textrm{test}=10^{19}\times A_\textrm{test}.
\end{equation}
Note that $A_\textrm{test}$ is restricted to the range $(0.001,100)$, meaning the minimum and maximum possible values of $E_\textrm{test}$ are $10^{16}$ and $10^{21}$\,eV respectively. The (negative log) likelihood of measuring $\mathcal{D}$ given the observed shower has parameters $\vec{a}$ identical to those of the template $\mathcal{T}$ but now with an energy $E_\textrm{test}$ is calculated via 
\begin{equation}
\label{eqn:scaleFactorLikelihood}
    -2\ln\mathcal{L}\left(\vec{x}|\vec{a},A_\textrm{test}\right) = -2\sum_k^{N_\textrm{pix}}\sum_i^{N_\textrm{bin}}\ln\left(P_k\left(x_i|\vec{a},A_\textrm{test}\right)\right)
\end{equation}
where 
\begin{equation}
    P\left(x_i|\vec{a},A_\textrm{test}\right)=\frac{1}{\sqrt{2\pi\sigma}}\exp\left[-\frac{\left(x_i-A_\textrm{test}\mu_i\right)^2}{2\sigma^2}\right]
\end{equation}
with
\begin{equation}
    \sigma^2=\sigma_{\textrm{bckg}}^2+A_\textrm{test}\mu_i(1+V_g).
\end{equation}
These equations are akin to Equations \ref{eqn:eventLikelihood}, \ref{eqn:BinProb} and \ref{eqn:pmtSig} but re-written to include the scale factor $A_\textrm{test}$. Note that $\mu_i$ denotes the template signal in the $i^\textrm{th}$ bin (where the template shower energy is 10$^{19}$\,eV). The final values of $t_\textrm{off}$ and $A$ for $\mathcal{T}$ are the values $t_\textrm{test}$ and $A_\textrm{test}$ which give the minimum value of $-2\ln\mathcal{L}$ over all values of $t_\textrm{test}$. 

\vspace{5mm}

\begin{figure}
    \centering
    \includegraphics[width=0.9\linewidth]{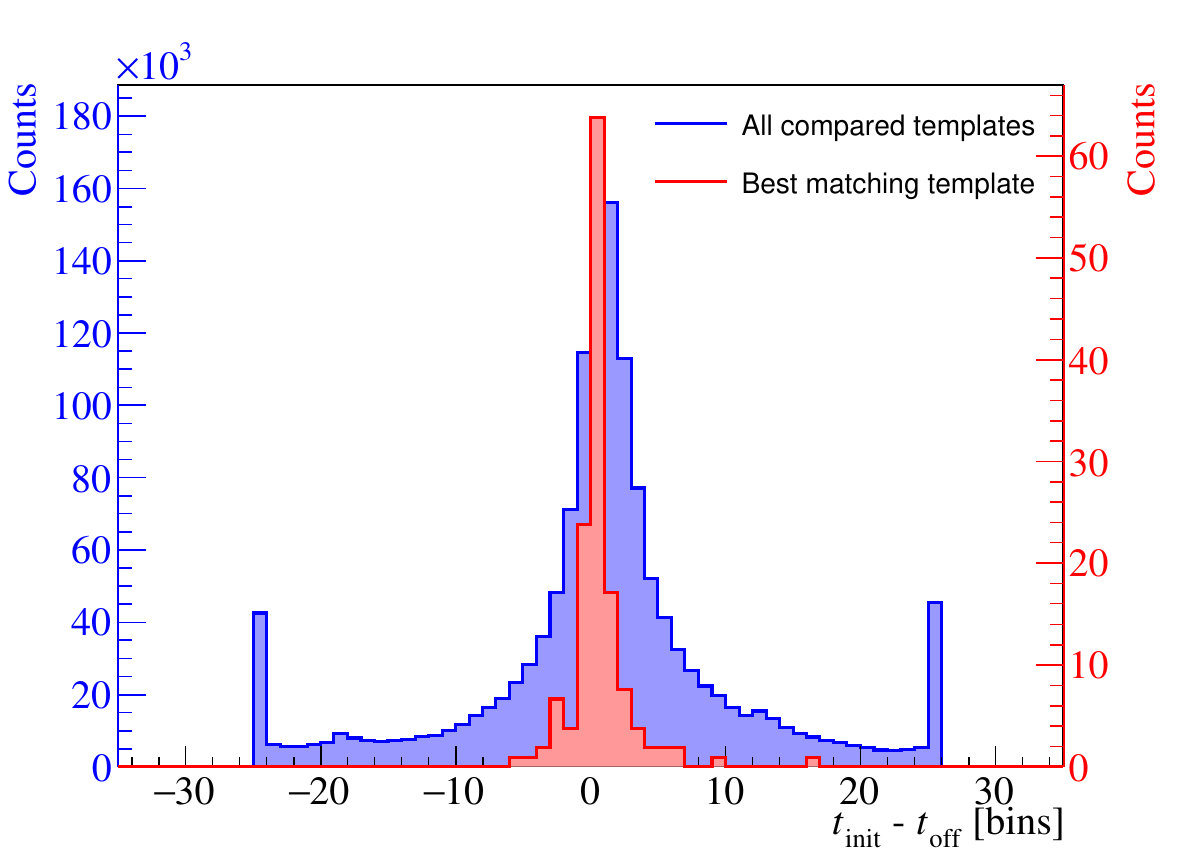}
    \caption{Difference between $t_\textrm{init}$ and $t_\textrm{off}$ for all templates compared (blue) and the best fitting templates for each event (red). The majority of the best matching templates have a time offset within $\pm5$ bins of $t_\textrm{init}$. One bin in this case is 200\,ns.}
    \label{fig:templateTimeOffsetCheck}
\end{figure}

In the above procedure, multiple $t_\textrm{test}$ values are tested to accommodate variations in the data due to noise, different template trace shapes and the slightly different calculation methods for the template and data centroid times (see the following section). The choice of $\pm5$ bins was made based on analysing a small sample of $\sim150$ events. These events were samples from the test data set for FAST-Single used in the previous chapter. The events all contained two or more PMTs passing the event level trigger. For each event the number of templates with which to compare was narrowed down using the criteria outlined in Section \ref{sec:templateSelection} $\left(T_\textrm{max}=5,S_\textrm{max}=0.1\right)$. Figure \ref{fig:templateTimeOffsetCheck} shows the difference between the initial offset $t_\textrm{init}$ and the best fit time offset $t_\textrm{off}$ for each of the templates compared. Here $t_\textrm{off}$ values between $t_\textrm{init}\pm25$\,bins were tested. The blue histogram shows the difference for every template compared across all events, whilst the red histogram shows the difference only for the best fitting template (one per event). All but two events had a best fitting template with $t_\textrm{off}$ greater than $\pm5$ bins away from $t_\textrm{init}$. Moreover, the bulk of all compared templates had a best fit offset within $\pm5$ bins of $t_\textrm{init}$, with templates outside of this range generally having a worse likelihood. Unlike the approach outlined above, this analysis fitted the energy scale factor $A$ for each template. Although not usually recommended due to the increased computation time, this was done to achieve the best possible accuracy. 

\begin{figure}
    \centering
    \includegraphics[width=0.9\linewidth, trim={0 0 1cm 0}]{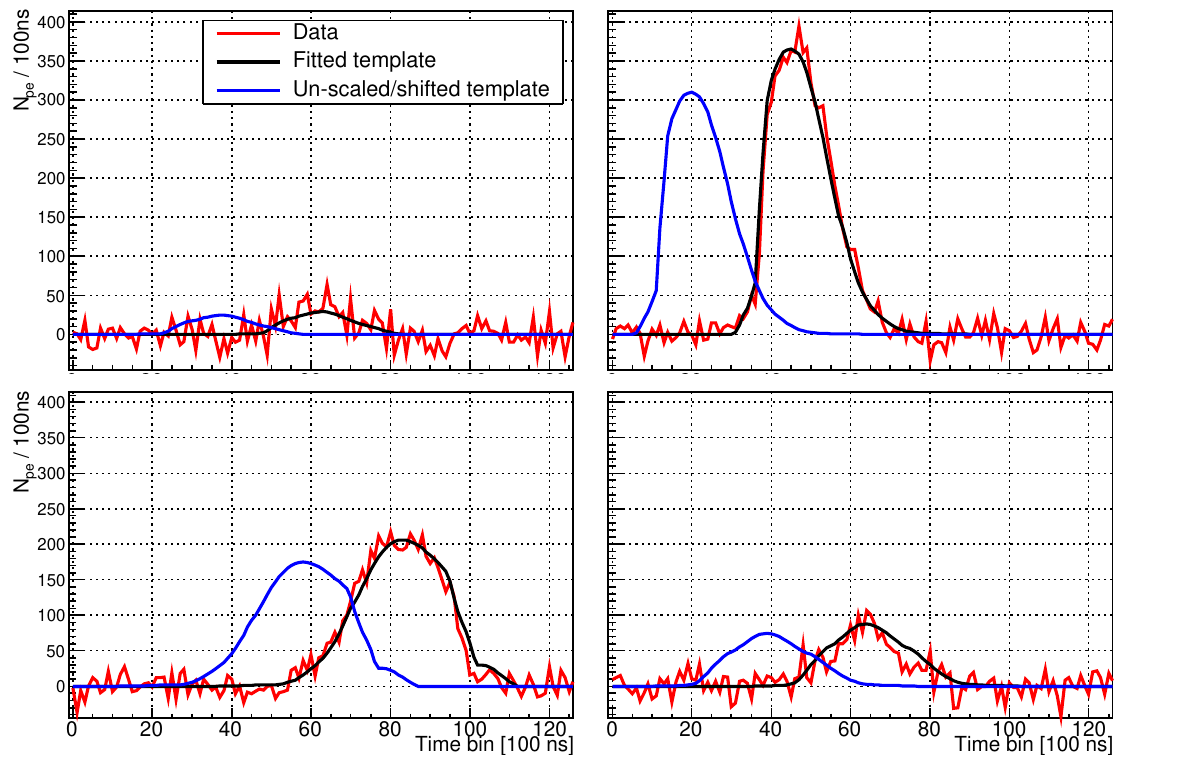}
    \caption{Example of how a set of template traces (blue) are shifted and scaled (black) to fit a set of data traces (red).}
    \label{fig:templateComparisonExample}
\end{figure}

\vspace{5mm}

Finally, note that the length of both the data traces and template traces are deliberately shortened to speed up the comparison process. This is done by keeping only the 500 bin region of the trace most likely to contain signal and re-binning it by a factor of two. Thus each comparison is made with traces of length 250 bins (each bin is 200\,ns). For simulated data and the templates, the relevant region is the same as used for the absolute timing cut, i.e. the first 500 bins. For data from the FAST prototypes, the relevant region is between bin 150 and 650 (assuming 100\,ns bin traces). Since the majority of time required for the Template Method comes from the comparison of the traces, this reduces the required time by roughly a factor of four. An example of the template comparison procedure is shown in Figure \ref{fig:templateComparisonExample}. The red traces represent the (simulated) data to fit to, the blue traces are the initial template traces and the black traces are the time-shifted and scaled template traces.

\section{Template Selection}
\label{sec:templateSelection}
\begin{figure}
    \centering
    \includegraphics[width=0.8\linewidth]{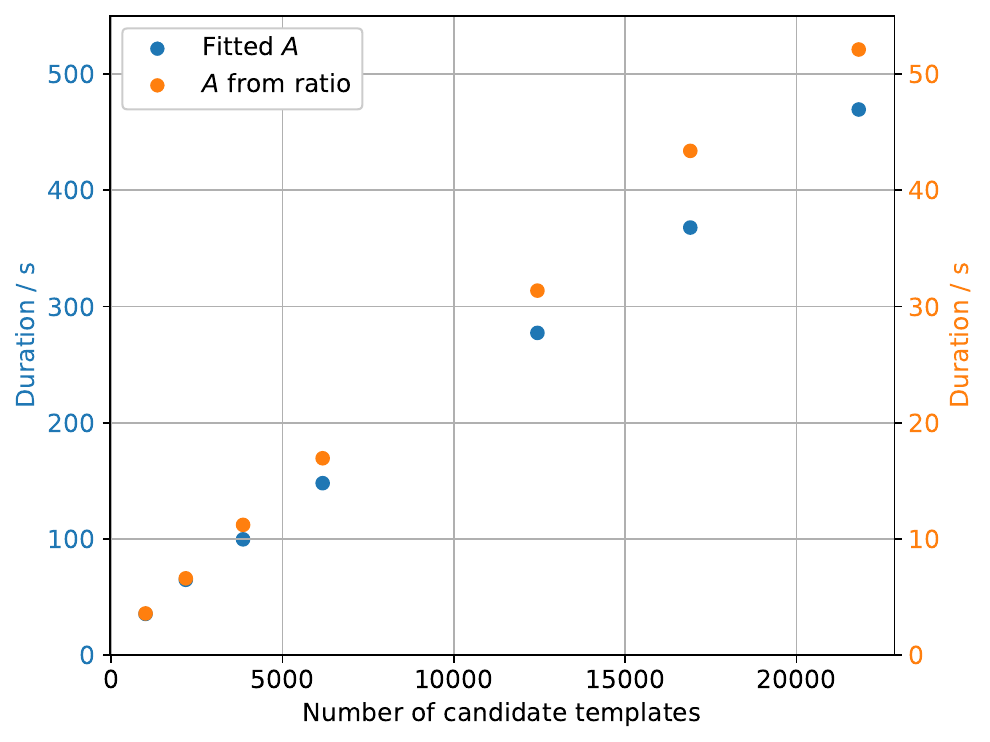}
    \caption{Required time to compare different numbers of templates using the methods outlined in Section \ref{sec:tempCompMethod}.}
    \label{fig:templateMethodDuration}
\end{figure}

Comparing one set of data traces to every template using the procedure outlined in Section \ref{sec:tempCompMethod} would be very time consuming. Figure \ref{fig:templateMethodDuration} shows the number of seconds required to run the method over varying numbers of templates (using $t_\textrm{test}\in(t_\textrm{init}-5,t_\textrm{init}+5)$). The tests were run using a single core on an Apple M1 chip. The blue points show the results when fitting the scale factor $A$, whilst the orange points show the results for estimating $A$ using Equation \ref{eqn:scaleFactorLikelihood}. The relationship in both cases is essentially linear as expected. When fitting $A$ the required time to evaluate the best possible likelihood for $\sim$10,000 templates exceeds $\sim4$\,min. For reference comparing every template in the library would take $>10$\,hr. This duration is a result of requiring many evaluations of the likelihood function for each $t_\textrm{test}$. The required time could be shortened by reducing the number of $t_\textrm{test}$ values tested or further re-binning of the traces, however these options come at the cost of losing discriminating power between similar templates. On the other hand, when estimating $A$ from the signal ratio in Equation \ref{eqn:scaleFactorLikelihood}, and thus only calculating $-2\ln\mathcal{L}$ once per template, the necessary duration is roughly a factor of 10 less. In other words, estimating $A$ from the signal ratio can allow 10 times the number of templates to be evaluated in the same amount of time. This allows for a more thorough evaluation of the likelihood space, which becomes critical when combining measurements from multiple telescopes (see Section \ref{sec:multipleTels}). In addition, with the TDR already requiring up to $\sim10$\,mins per event of computational time, a first-guess method with similar or greater computational requirements is not ideal. 

\vspace{5mm}

Overall then, estimating $A$ using the signal ratio as opposed to fitting, despite losing some accuracy, appears the more suitable choice for the Template Method. Even so, comparing every template in the library would still require more than an hour of computation. Fortunately, not every template has to be compared to the data traces to find a good fit. The relevant search space can be greatly narrowed down by only testing templates with similar features to the data. Since the energy of the templates is fixed, only energy independent features are able to be used for this purpose. The most obvious energy independent features are the relative timing and signal ratio between triggered PMTs. The following subsections detail how these features are used to select \say{candidate templates} for telescopes with either multiple triggered PMTs or a single triggered PMT.

\subsection{Telescopes with Multiple Triggered PMTs}
For multiple triggered PMTs in a single telescope, the centroid times and signals are calculated as described in Equations \ref{eqn:pmtSig} and \ref{eqn:centroidTime}. The signal ratios between triggered PMTs are obtained by dividing each signal by the total signal (i.e. the sum of all triggered PMT signals). The centroid times and signal ratios are then ordered from earliest to latest and smallest to largest respectively, and the \textit{differences} between subsequent values calculated. These differences will be labelled $\Delta T_{i,j}^\textrm{data}$ and $\Delta S_{k,l}^\textrm{data}$. 
The $i,j,k,l\in\{1,2,3,4\}$ subscripts identify which PMTs the difference is between, with (for example) the $i^\textrm{th}$ PMT's value being subtracted from the $j^\textrm{th}$ PMT's value. Thus $\Delta T_{1,3}^\textrm{data}$ is the centroid time of the third PMT minus the centroid time of the first PMT. Different subscripts are used for $\Delta T$ and $\Delta S$ as the triggering order of the PMTs may be different to the order of signal ratios\footnote{In practice there is likely no difference when just using the triggering order, but this is the implementation tested here.}. 

\begin{figure}[t]
    \centering
    \includegraphics[width=1\linewidth]{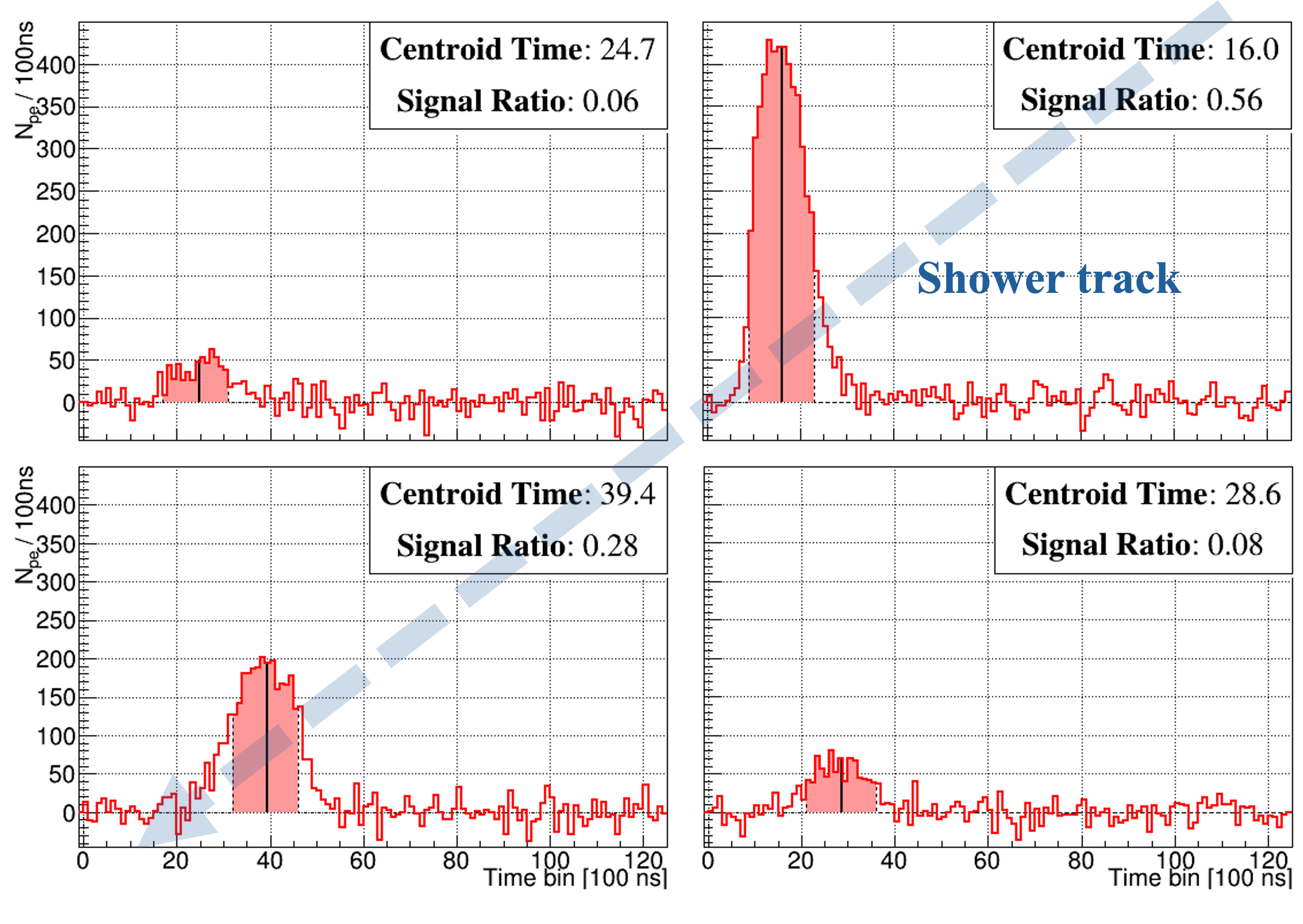}
    \caption{An example of how the number of templates is narrowed down using the centroid time and signal ratio \textit{differences} between PMTs. Each PMT trace is shown in red. The filled area shows the maximum SNR region and the black line represents the centroid time. From earliest to latest, the centroid times are (in bins of 100\,ns) \{16, 24.7 28.6, 39.4\}, giving differences of $\Delta T_{2,0}^\textrm{data}=8.7$, $\Delta T_{0,3}^\textrm{data}=3.9$ and $\Delta T_{3,1}^\textrm{data}=10.8$. From smallest to largest, the signal ratios are \{0.06, 0.08, 0.28, 0.56\}, giving differences of $\Delta S_{0,3}^\textrm{data}=0.02$, $\Delta S_{3,1}^\textrm{data}=0.2$ and $\Delta S_{1,2}^\textrm{data}=0.28$. Each $\Delta T_{i,j}^\textrm{data}$ and $\Delta S_{k,l}^\textrm{data}$ is then compared to those from every template in the library. Templates with all values of $\Delta T_{i,j}^\textrm{temp}$ and $\Delta S_{k,l}^\textrm{temp}$ within $T_\textrm{max}$ and $S_\textrm{max}$ of their data counterparts are selected as candidates.}
    \label{fig:templateSelectionExample}
\end{figure}

Each $\Delta T_{i,j}^\textrm{data}$ and $\Delta S_{k,l}^\text{data}$ are then compared to the corresponding values of each template in the library, $\Delta T_{i,j}^\textrm{temp}$ and $\Delta S_{k,l}^\textrm{temp}$. Note that for the template traces the centroid times and signals are calculated using the entire trace, i.e. $k_\textrm{stop}=1000$ (for a 1000\,bin trace) and $k_\textrm{start}=0$, as the traces contain no noise. A template is deemed to be a candidate if 
\begin{equation}
\label{eqn:templateConditions}
    |\Delta T_{i,j}^\textrm{data}-\Delta T_{i,j}^\textrm{temp}|<T_\textrm{max} \quad \textrm{and} \quad 
    |\Delta S_{k,l}^\textrm{data}-\Delta S_{k,l}^\textrm{temp}|<S_\textrm{max}
\end{equation}
for each pair of subsequently triggered PMTs $i,j$ and each pair of PMTs with successively larger signal ratios $k,l$. Figure \ref{fig:templateSelectionExample} shows an example of the procedure.

\vspace{5mm}

\begin{figure}
    \centering
    \includegraphics[width=0.9\linewidth]{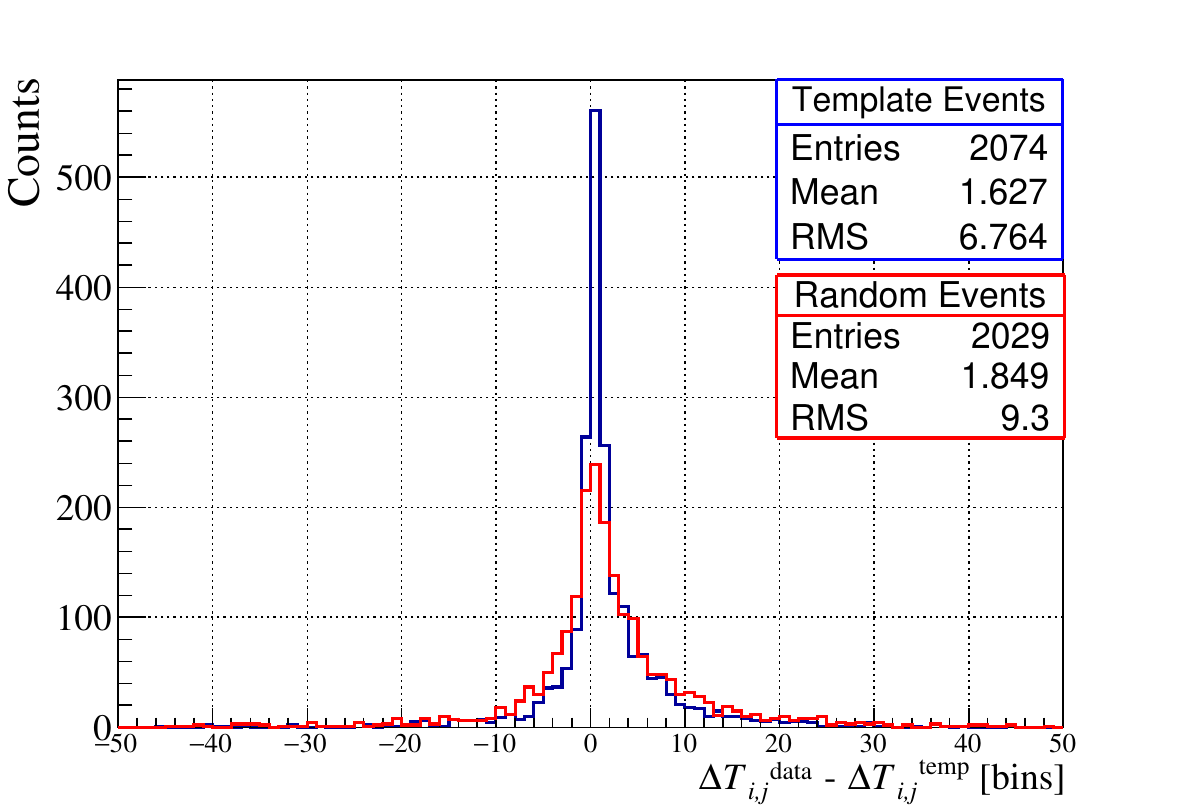}
    \includegraphics[width=0.9\linewidth]{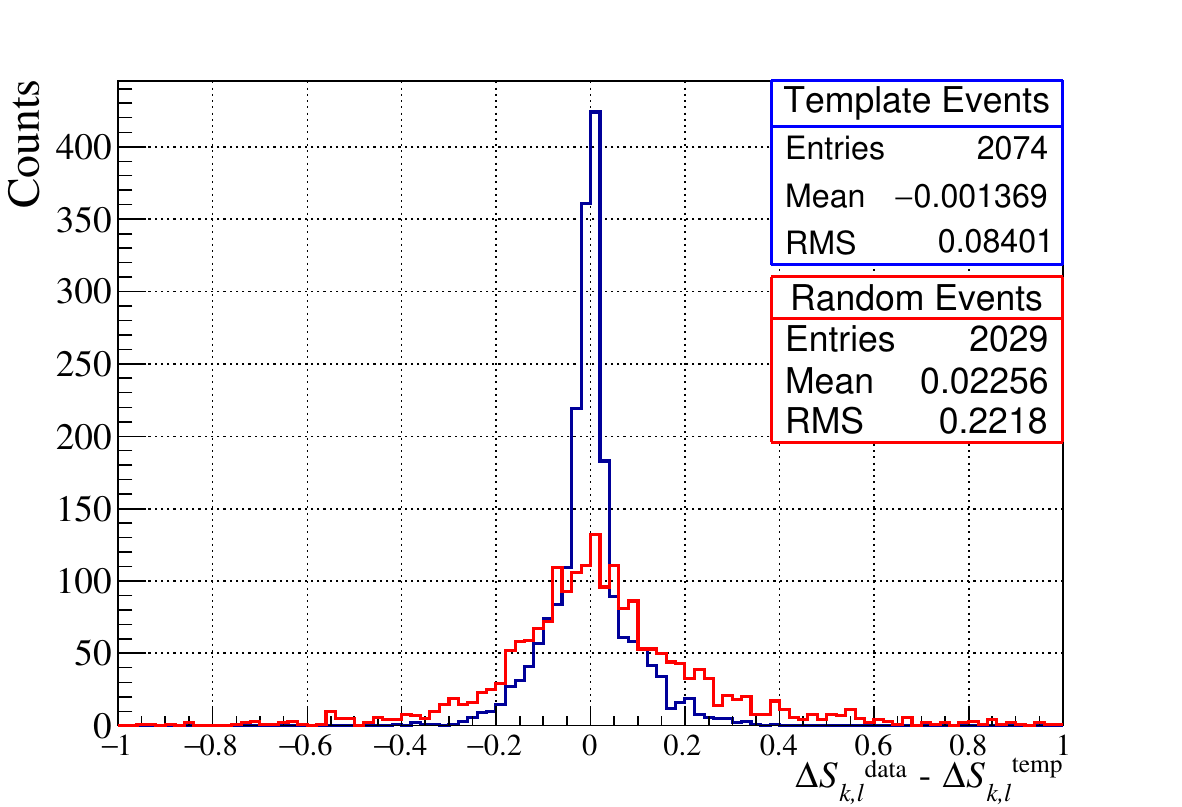}
    \caption{\textit{Top:} (Blue) The difference in relative timing for subsequently triggered PMTs between template events with and without noise. (Red) Same as blue but between random events and the ``closest" template in the library (Euclidean distance in the shower parameter space). \textit{Bottom:} Same as the top plot but for the signal ratios between PMTs. The differences are taken between PMTs with successively larger signals.}
    \label{fig:templateSelectionSpacings}
\end{figure}

To gauge the appropriate threshold values $T_\textrm{max}$ and $S_\textrm{max}$, a random sample of templates were selected and Gaussian noise added to each of the traces using the standard $\sigma_\textrm{nsb}=10$\,p.e./100\,ns. Using only those PMTs with traces passing the threshold trigger, the values $\Delta T_{i,j}^\textrm{data}$ and $\Delta S_{k,l}^\textrm{data}$ were calculated. The difference between the original templates $\Delta T_{i,j}^\textrm{temp}$ and $\Delta S_{k,l}^\textrm{temp}$ values were then determined and the results plotted as the blue histograms in Figure \ref{fig:templateSelectionSpacings}. In the majority of cases the difference in relative timing changes by less than 10\,bins and the difference in signal ratios by less than 20\%. Thus if a shower with parameters equivalent to one of the templates was simulated and a first guess estimated with this method, values of $T_\textrm{max}=10\,\textrm{bins}\times$100\,ns/bin = 1\,ms (the centroid times are calculated on the 100\,ns traces) and $S_\textrm{max}=0.2$ would (almost) ensure that the shower was compared to the template with the same parameters. The differences tend to zero the greater the average SNR of the subsequent PMTs. This is shown in Appendix \ref{fig:templateTimeSigSNR}. 

\vspace{5mm}

If the same procedure is performed for events with random shower parameters, this time comparing $\Delta T_{i,j}^\textrm{data}$ and $\Delta S_{k,l}^\textrm{data}$ to the $\Delta T_{i,j}^\textrm{temp}$ and $\Delta S_{k,l}^\textrm{temp}$ of the closest template in the library, the red histograms in Figure \ref{fig:templateSelectionSpacings} are obtained. These histograms are wider, with RMS values roughly $1.5$ and $2.5$ times larger for the relative time differences and signal ratio differences respectively. From these graphs, one might conclude that values of $T_\textrm{max}\approx2$\,ms and $S_\textrm{max}\approx0.40$ are required for the Template Method to be \say{successful}, i.e. choose the template with the closest parameters. However, it is important to recall that while the goal of the Template Method is to provide an accurate estimate of the shower parameters for use by the TDR, it does so by finding the set of template traces which give the minimum negative log-likelihood when compared to the data. This set of traces may not always be the closest template. Even if the best matching template has a set of shower parameters radically different from the true shower parameters (which of course are typically unknown), it is these values that should be passed to the minimiser as a first guess. Therefore, the values of $T_\textrm{max}$ and $S_\textrm{max}$ should be set not to encompass the closest template, but to select templates which will give the best match, i.e. minimum negative log-likelihood. For this purpose the thresholds estimated using the template events $T_\textrm{max}=1$\,ms and $S_\textrm{max}=0.2$ are appropriate.  

\vspace{5mm}

\begin{figure}[t]
    \centering
    \includegraphics[width=0.9\linewidth]{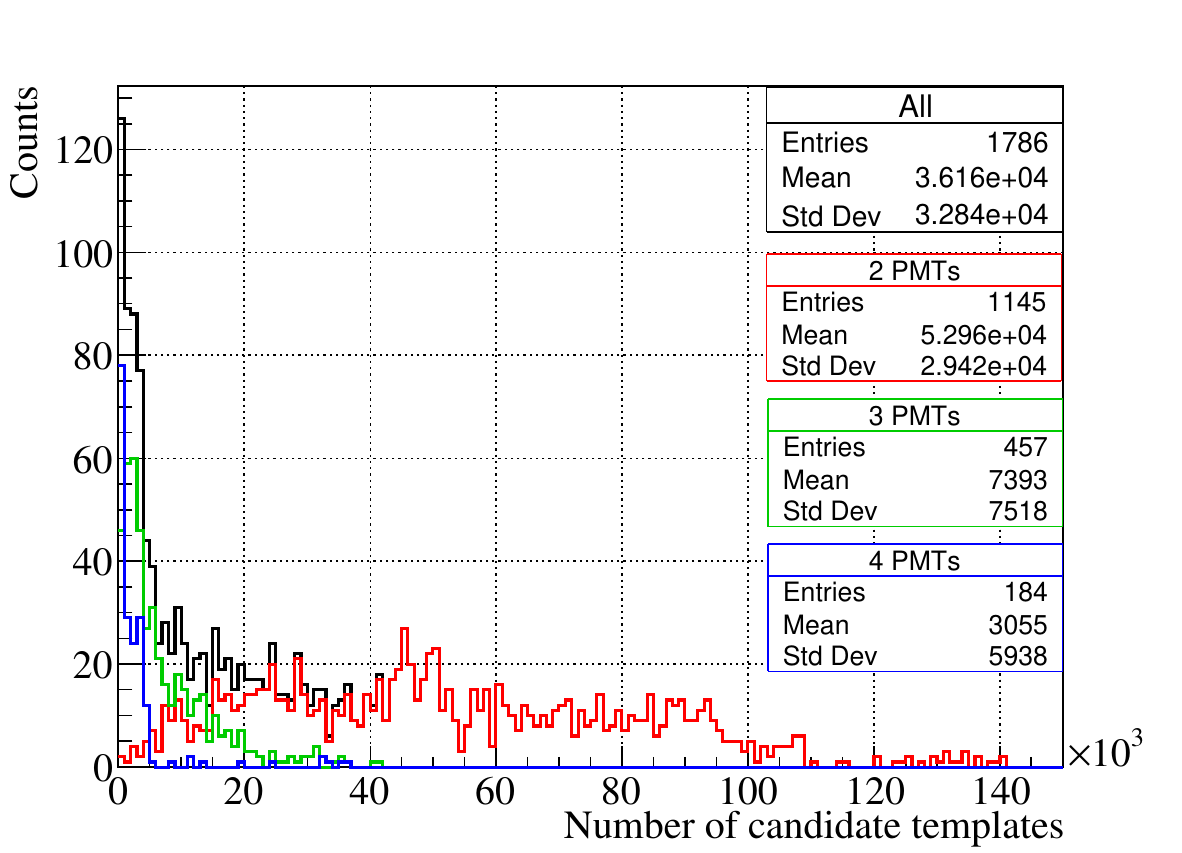}
    \caption{The number of candidate templates selected using threshold values of $T_\textrm{max}=1$\,ms and $S_\textrm{max}=0.2$ for two (red), three (green) and four (blue) triggered pixels. The black histogram shows the combined total.}
    \label{fig:templateCandidateNumber}
\end{figure}

With these thresholds, the number of candidate templates for telescopes with multiple triggered PMTs typically ranges from a few thousand to $\sim100,000$. Figure \ref{fig:templateCandidateNumber} shows histograms of the number of candidate templates selected for different numbers of triggered pixels. In general, the more PMTs which trigger, the fewer the number of candidate templates. To understand why, consider the case of two triggered PMTs. In this scenario a template only has to satisfy the pair of inequalities in Equation \ref{eqn:templateConditions} once (for the single pair of PMTs) to be considered a candidate. However for four triggered PMTs, the same inequalities must be met for all three pairs of PMTs for both the timing and signal ratios. This is generally a much stricter requirement, and thus the number of candidate templates is less. With the average number of candidate templates being approximately 40,000, the average length of time required to run the Template Method on a single telescope is roughly $2$\,minutes (based on Figure \ref{fig:templateMethodDuration}). This is considered acceptable for this \say{first version} of the Template Method.
Finally, if the maximum number of photo-electrons in a single bin across all PMT traces in a template is less than $P_\textrm{min}=0.01\times$10\,p.e./100\,ns, then the template is not included in the list of possible candidates. This pulse height threshold is chosen based on the upper limit of $A$ (100 times more energetic than the simulated template) and by assuming that signal from an air shower will require a maximum pulse height of at least 10\,p.e./100\,ns to be detected by FAST.

\subsection{Telescopes with a Single Triggered PMT}
\begin{figure}[t]
    \centering
    \includegraphics[width=0.8\linewidth]{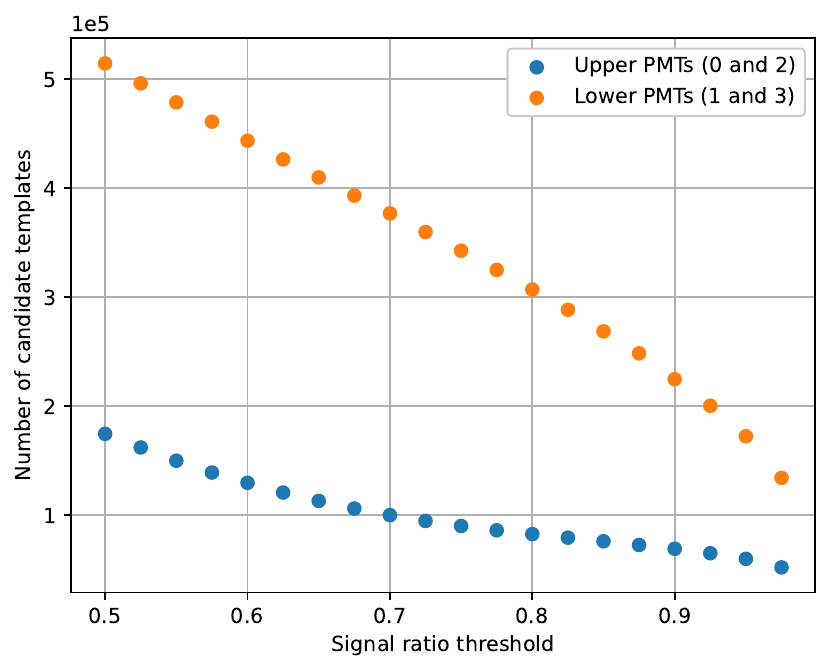}
    \caption{Number of candidate templates for a single triggered PMT as a function of the applied signal ratio threshold. The blue (orange) points show the results for the upper (lower) PMTs.}
    \label{fig:singlePixelCandidateCheck}
\end{figure}

If only a single PMT in a telescope triggers there is no relative timing difference or signal ratio difference with which to narrow down the number of candidate templates. One option is to take all templates which have a signal ratio greater than some threshold in the corresponding PMT as candidates. Figure \ref{fig:singlePixelCandidateCheck} shows the number of potential candidates for PMTs observing the upper portion of the sky (blue) and the lower portion of the sky (orange) as a function of the applied signal ratio threshold. The number of candidate templates selected using a signal ratio threshold of 0.5 is roughly 175,000 for the upper PMTs, and 515,000 for the lower PMTs. With the current comparison method, this would equate to $\sim10$\,minutes and $\sim25$\,minutes of computational time respectively for a \textit{single} telescope. This is simply too long and would not be suitable for application to a large FAST array.

\vspace{5mm}

\begin{figure}
    \centering
    \includegraphics[width=1\linewidth,trim={0cm 0cm 1cm 0cm}]{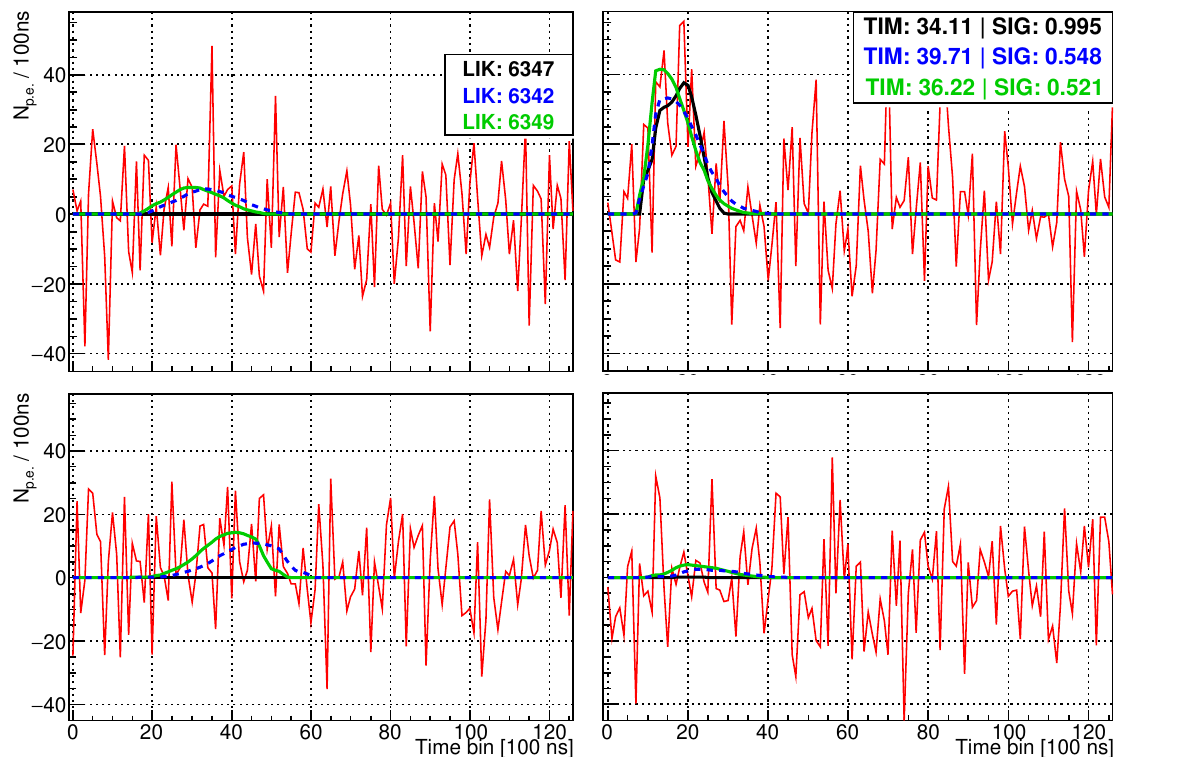}
    \caption{Example of how limiting the candidate templates for a single triggered pixel by applying a high signal ratio threshold removes viable templates. The original signal from the shower is shown in green. The same signal with added background noise (data) is shown in red. The template closest to the shower parameters (shifted and scaled) is shown by the dashed blue lines. The black lines shown the best fit template to the data when only considering showers with a signal ratio in PMT 2 above 0.9. The signal ratios for PMT 2 from each of the raw traces are shown in PMT 2's panel. The negative log-likelihood values for each set of raw traces compared to the data are shown in the top-left panel. The original traces and traces from the template closest to the true shower parameters show small signals in PMTs 0 and 1. Although not visible above the background noise, these signals are non-negligible relative to the raw signals in PMT 2, giving signal ratios in PMT 2 around 0.5. By only considering templates with a signal ratio above 0.9 the blue template would be ignored and the black template selected as the best fit instead - despite having a worse overall likelihood.}
    \label{fig:pix1problem}
\end{figure}

\begin{figure}[t]
    \centering
    \includegraphics[width=1\linewidth]{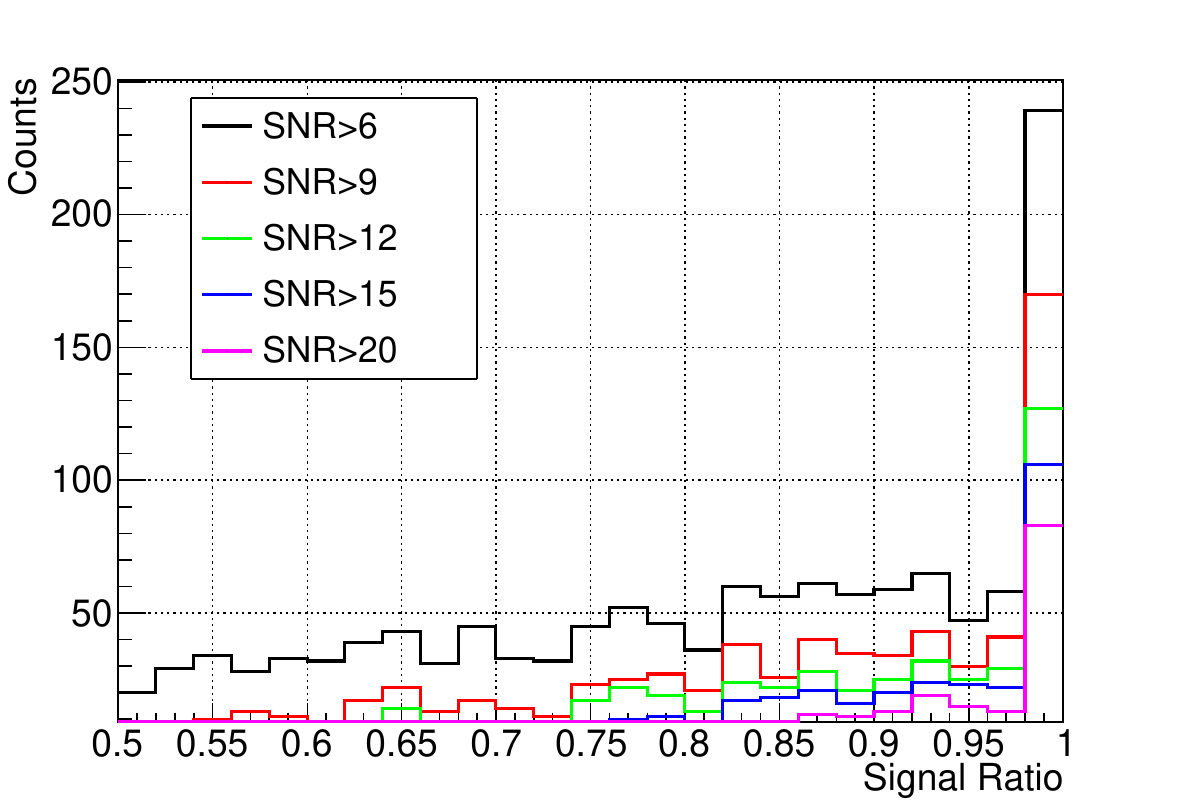}
    \caption{Distributions of signal ratios from templates with a single triggered PMT after adding background noise. The signal ratios are from the triggered PMTs original signal ratio as part of the template. The distributions have been separated into different (measured) SNR thresholds. Thus, for showers with a single triggered PMT with a SNR $>20$ (for example), only templates with a signal ratio in that PMT $\gtrsim$ 0.85 need to be checked.}
    \label{fig:singlePixelThreshold}
\end{figure}

To resolve this, the number of candidates could be limited to say $50,000$ showers by increasing the thresholds to 0.9 for the upper PMTs and $>0.975$ for the lower PMTs. However such a high cutoff removes many viable templates. An example of this is shown in Figure \ref{fig:pix1problem}. Removing templates in this way becomes an issue when trying to combine the results of a telescope with one triggered PMT with other telescopes (see Section \ref{sec:multipleTels}), as the likelihood space may not be sufficiently sampled for accurate interpolation. To ensure that the majority of viable templates are considered, an appropriate threshold based on the SNR of the signal could be estimated. Figure \ref{fig:singlePixelThreshold} shows an example of this. Here, a sample of templates has been taken and the standard $\sigma=10$\,p.e./100\,ns noise added to the traces. Events with a single triggered PMT were then selected and distributions of the original signal ratio for these PMTs plotted. In the Figure, these distributions have been separated into successively higher SNR cutoffs. From this plot one can deduce how low to set the signal ratio threshold for a particular SNR. For instance, the appropriate cutoff for pixels with an SNR $>12$ may be $\approx0.75$. Once again though, the number of candidate templates selected via these \say{appropriate} thresholds would be too great to analyse within the desired 2 - 3\,minutes/telescope time-frame, particularly for PMTs viewing the lower portion of the sky. 

\vspace{5mm}

\begin{figure}[]
    \centering
    \includegraphics[width=0.64\linewidth]{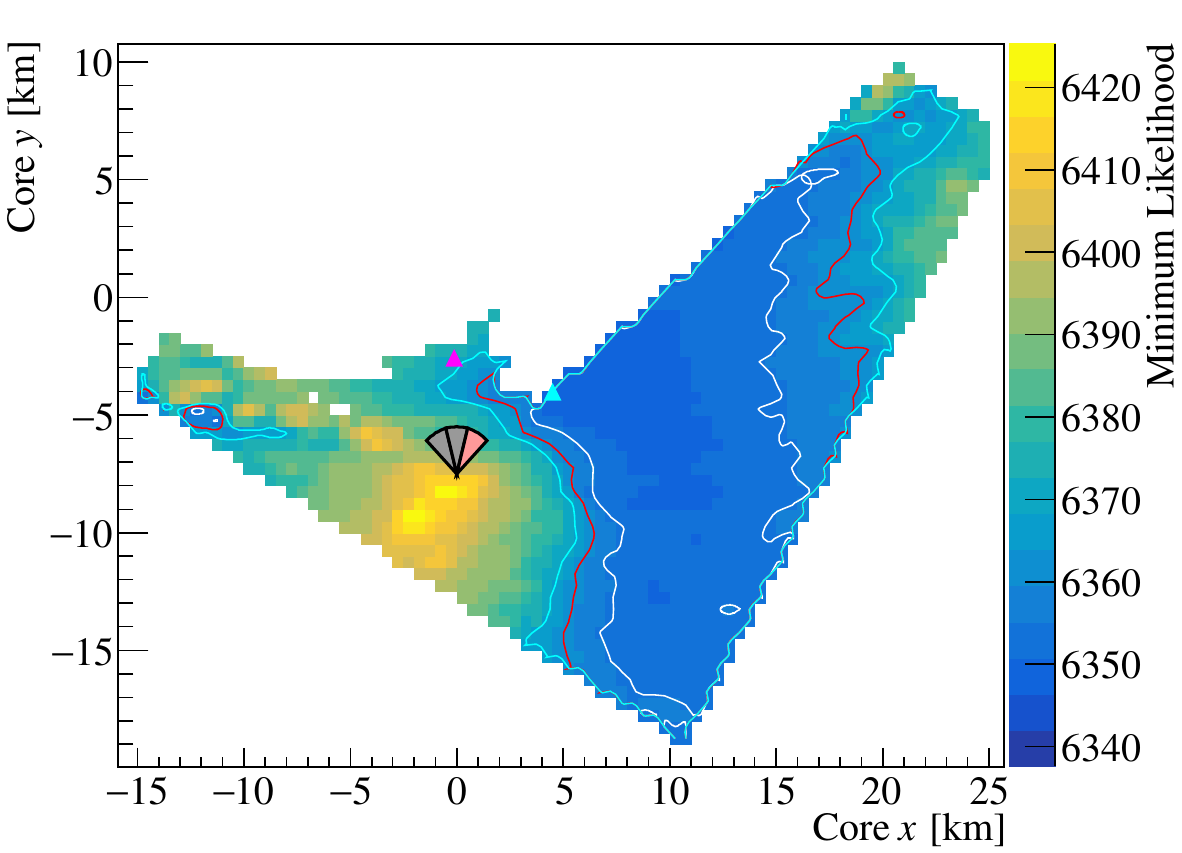}
    \includegraphics[width=0.64\linewidth]{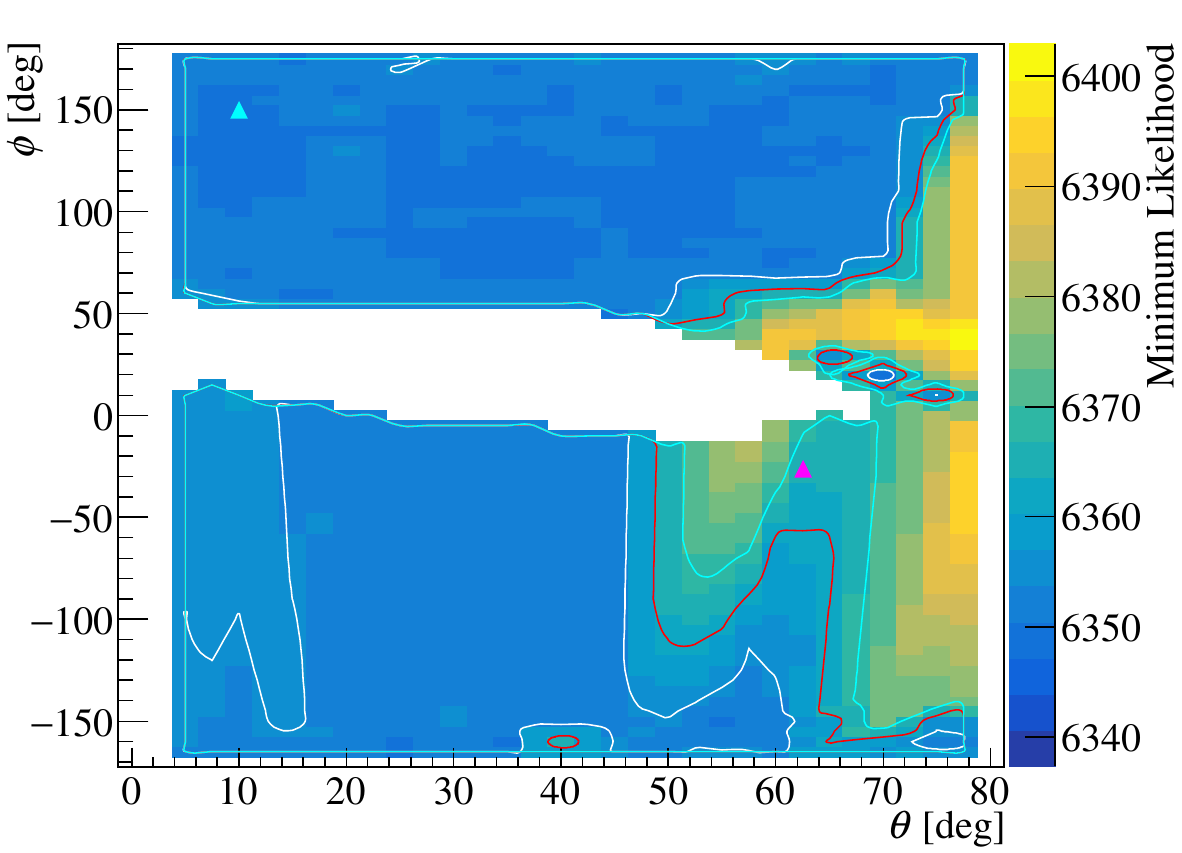}
    \includegraphics[width=0.64\linewidth]{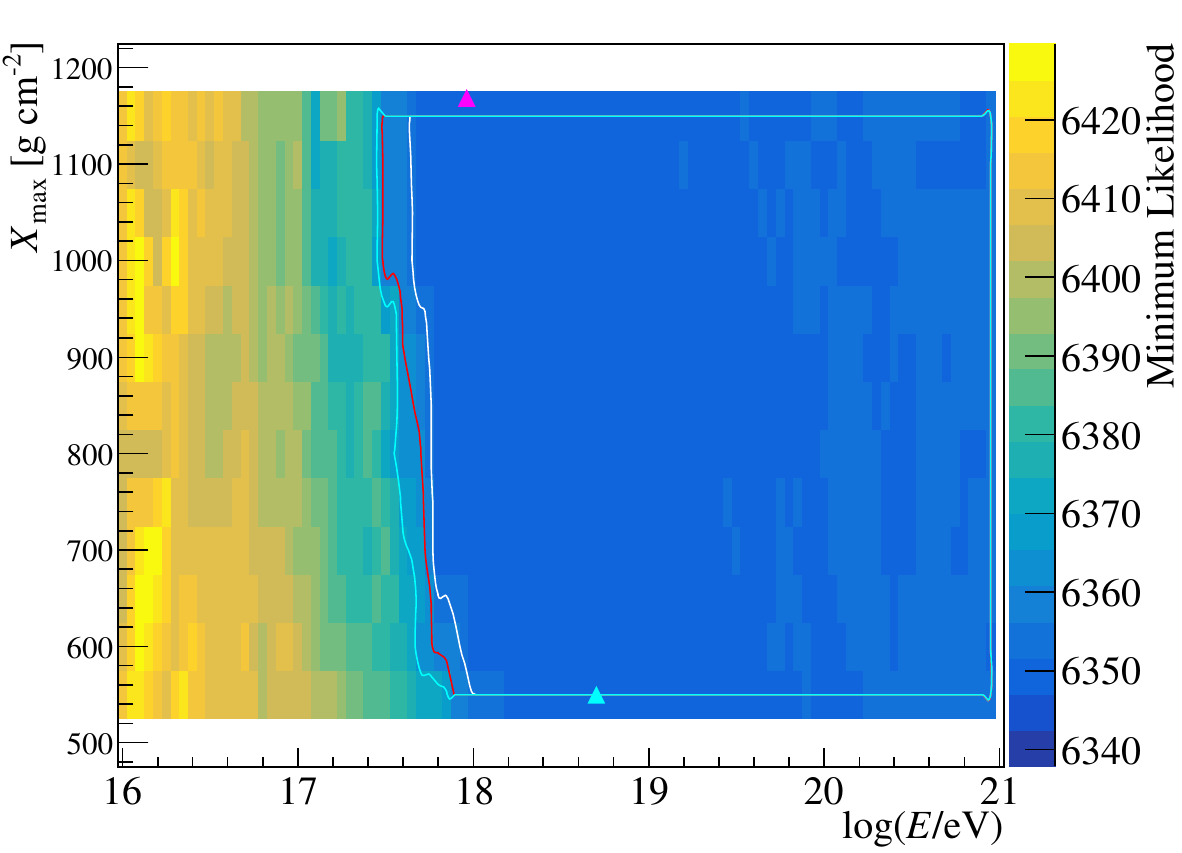}
    \caption{2-D histograms showing the minimum negative log-likelihood in slices of core location (top), arrival direction (middle) and \Xmax{}/energy (bottom) for the example event in Figure \ref{fig:pix1problem} (considering only candidate templates with a signal ratio $>0.9$ in PMT 2). The white/red/cyan contour lines show the 1/2/3$\sigma$ uncertainty regions. The cyan (magenta) triangles show the locations of the reconstructed (true) shower parameters.}
    \label{fig:singlePixelDegen}
\end{figure}

Ultimately, it was decided to sacrifice some accuracy to achieve the desired speed by adjusting the method described in Section \ref{sec:tempCompMethod} when analysing telescopes with a single triggered PMT. The adjustment was to only test a single value of $t_\textrm{off}=t_\textrm{init}$. With this change, and the signal ratio threshold set to 0.5, comparing the $\sim175,000$ templates for the upper PMTs and $\sim515,000$ showers for the lower PMTs takes roughly 1.5\,minutes and 3\,minutes respectively. Note however that telescopes with a single triggered pixel will only be included in the reconstruction of events where more than one telescope has observed the shower. This decision was made for two reasons. Firstly, unlike the machine learning method, the Template Method does not provide a straightforward way to use the knowledge that some telescopes \textit{did not} see the shower in the first guess estimate. Thus even for a large number of telescopes, for example the FAST 3-Eye layout, if only a single telescope sees the shower with one triggered PMT, the performance will be functionally equivalent to that of a single telescope. Additionally, the machine learning results in Section \ref{sec:TSFELDNNperformance} indicated that one pixel does not provide sufficient information to estimate the shower parameters. This is supported by the degeneracy observed in the typical core, arrival direction and \Xmax{}/energy slices of the likelihood function used to determine the first guess (see Section \ref{sec:templateParameterDetermination}) for events with a single triggered PMT. An example of said degeneracy can be seen in Figure \ref{fig:singlePixelDegen}. The fact that the one/two/three sigma contours cover the majority of the parameter space shows that the first guess estimate is highly uncertain, i.e. many different combinations of shower parameters give similar likelihood values.

\vspace{5mm}

Future work could attempt to reduce the required computational time by further cutting unnecessary signal from the end of the traces (perhaps on an event by event basis), additional re-binning for sufficiently wide traces and/or potentially altering the comparison method such that only the traces of triggered PMTs are compared when calculating the likelihood. Each of these adjustments would need to be done carefully so as not to lose the discriminating power between different templates. Another possibility is to use other energy independent features to narrow down the search space, for example those used in the machine learning section such as kurtosis and skewness.

\section{Parameter Estimation}
\label{sec:templateParameterDetermination}

The procedure for estimating the first guess parameters of an event differ depending on whether said event was observed by a single telescope or multiple telescopes. Both procedures are outlined below. Before estimating the final parameters, the following steps are performed for each telescope with signal. First, each candidate template is compared to the data traces. For each candidate the minimum negative log-likelihood (hereafter just \say{likelihood}) is found by scanning over $t_\textrm{test}$ and estimating the energy scaling via Equation \ref{eqn:energyScaleFactor}. These values are then saved. Out of these values, the template parameters corresponding to the minimum likelihood, $(-2\ln{}\mathcal{L})_\textrm{min}$, are labelled $\vec{a}_\textrm{min}$. The distribution of the saved likelihood values in the six-dimensional parameter space (\Xmax{}, $E$, $\theta$, $\phi$, core $x$, core $y$) is then visualised using three 2-D \say{slices}. Specifically, 2-D histograms for the parameters corresponding to the core location (core $x$, core $y$), arrival direction ($\theta$, $\phi$) and longitudinal profile (\Xmax{}, $E$) are constructed. For each histogram, the content of each bin is set to the minimum over all likelihood values for templates with parameters which correspond to that bin. In other words, each map is the minimum likelihood surface for its set of parameters. These histograms will be referred to as the core location, arrival direction and profile \textit{likelihood maps} respectively. An example of these maps has already been shown in Figure \ref{fig:singlePixelDegen}.

\subsection{Single Telescope}
\begin{figure}[]
    \centering
    \includegraphics[width=0.95\linewidth,trim={0cm 0cm 1cm 0cm}]{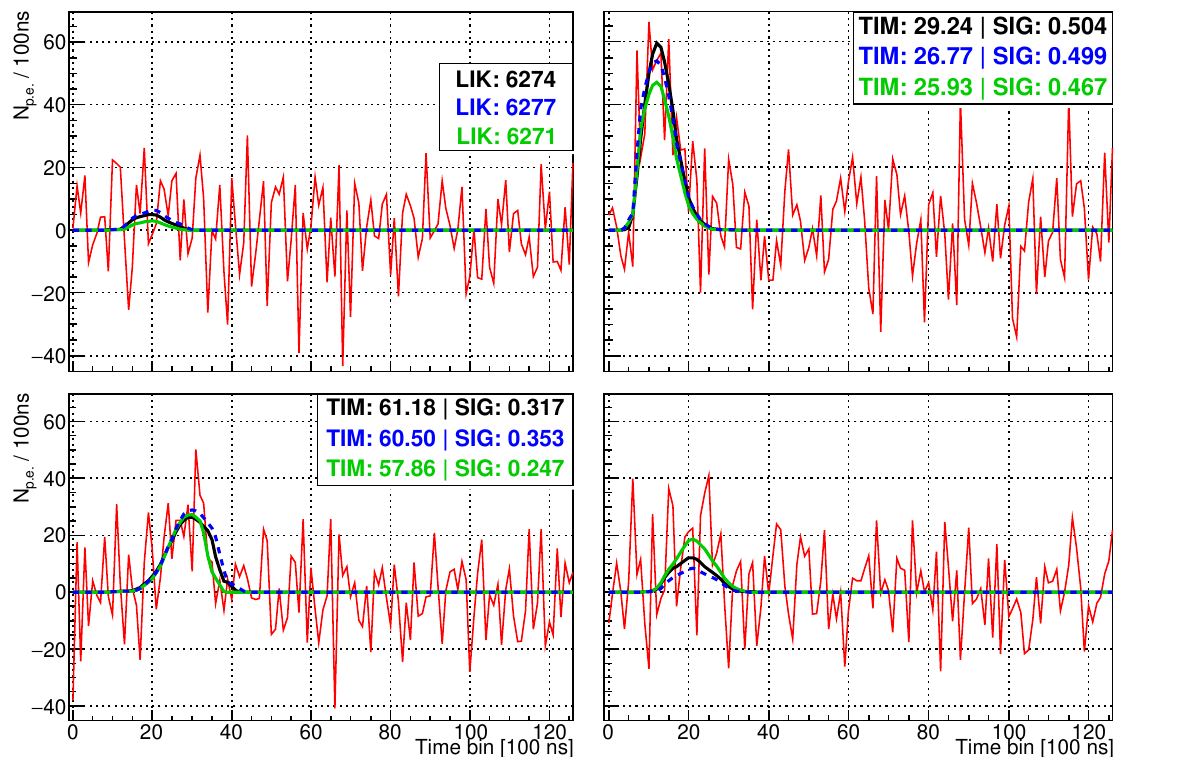}
    \includegraphics[width=0.49\linewidth]{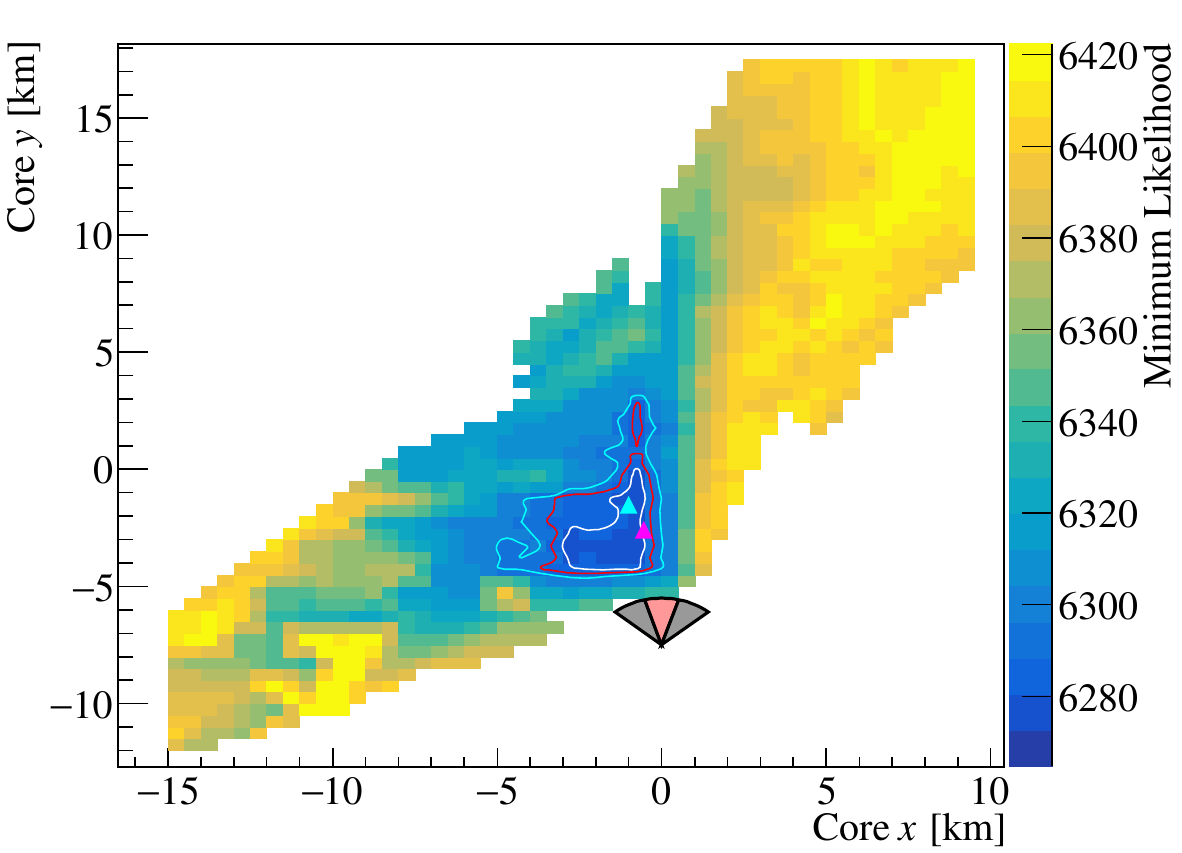}
    \includegraphics[width=0.49\linewidth]{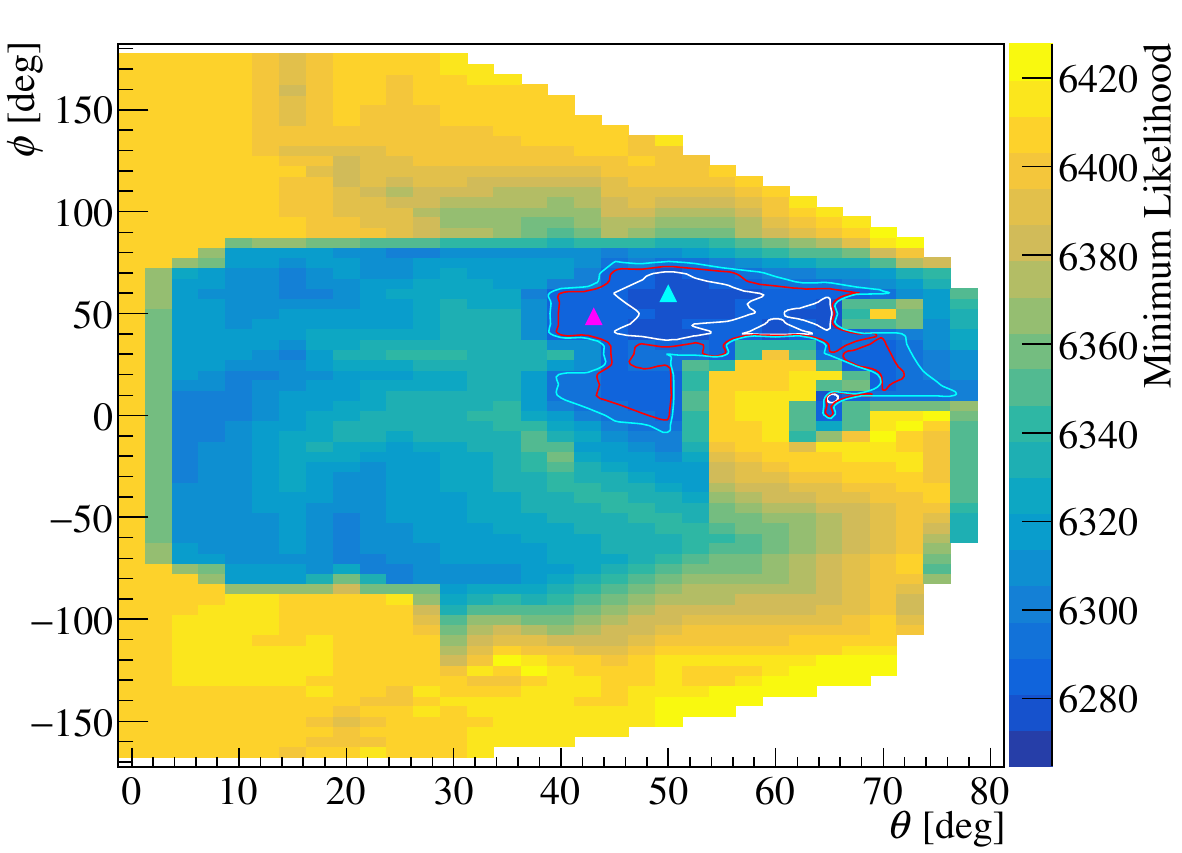}
    \includegraphics[width=0.49\linewidth]{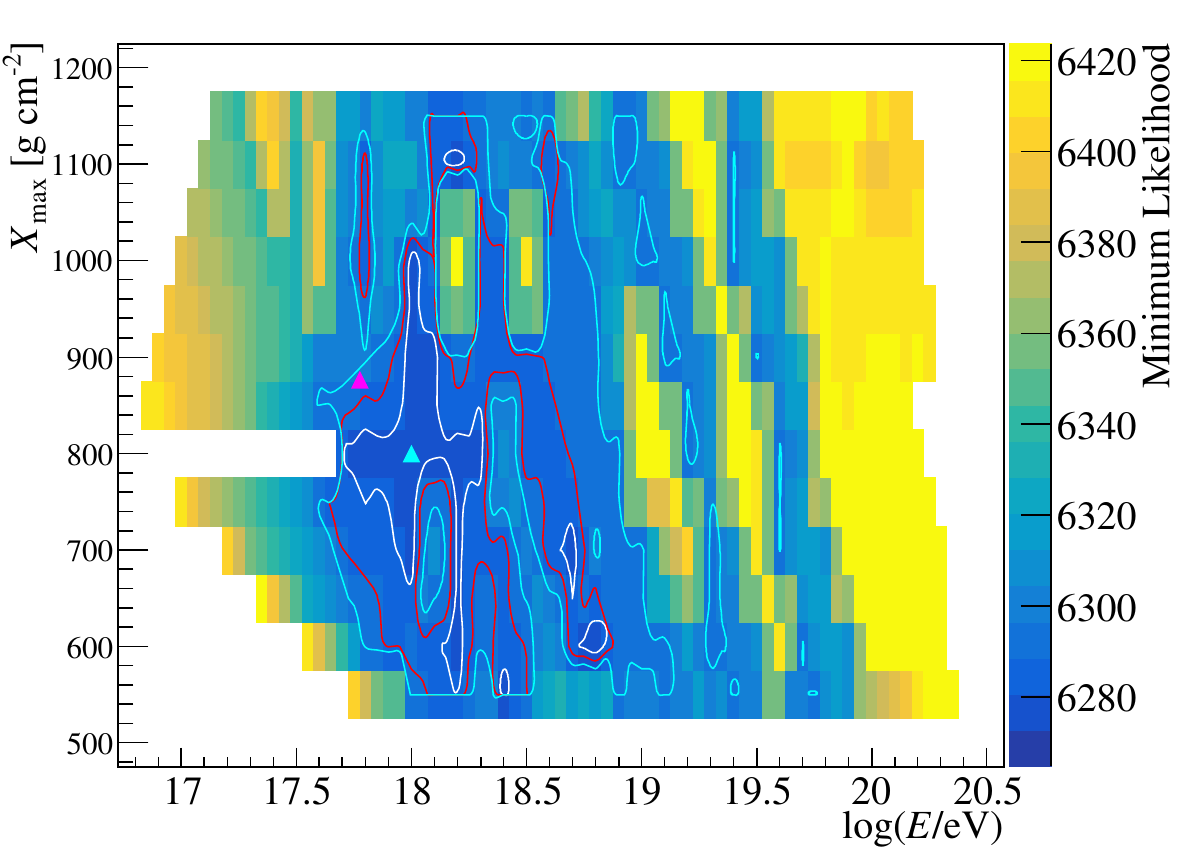}
    \caption{Example of the Template Method applied on a simulated shower observed by a single telescope with two triggered pixels. The traces in the top part of the figure follow the same layout as in Figure \ref{fig:pix1problem}. The bottom plots show the core location (top left), arrival direction (top right) and profile (bottom) likelihood maps.}
    \label{fig:pix2Example}
\end{figure}

\begin{figure}[]
    \centering
    \includegraphics[width=0.95\linewidth,trim={0cm 0cm 1cm 0cm}]{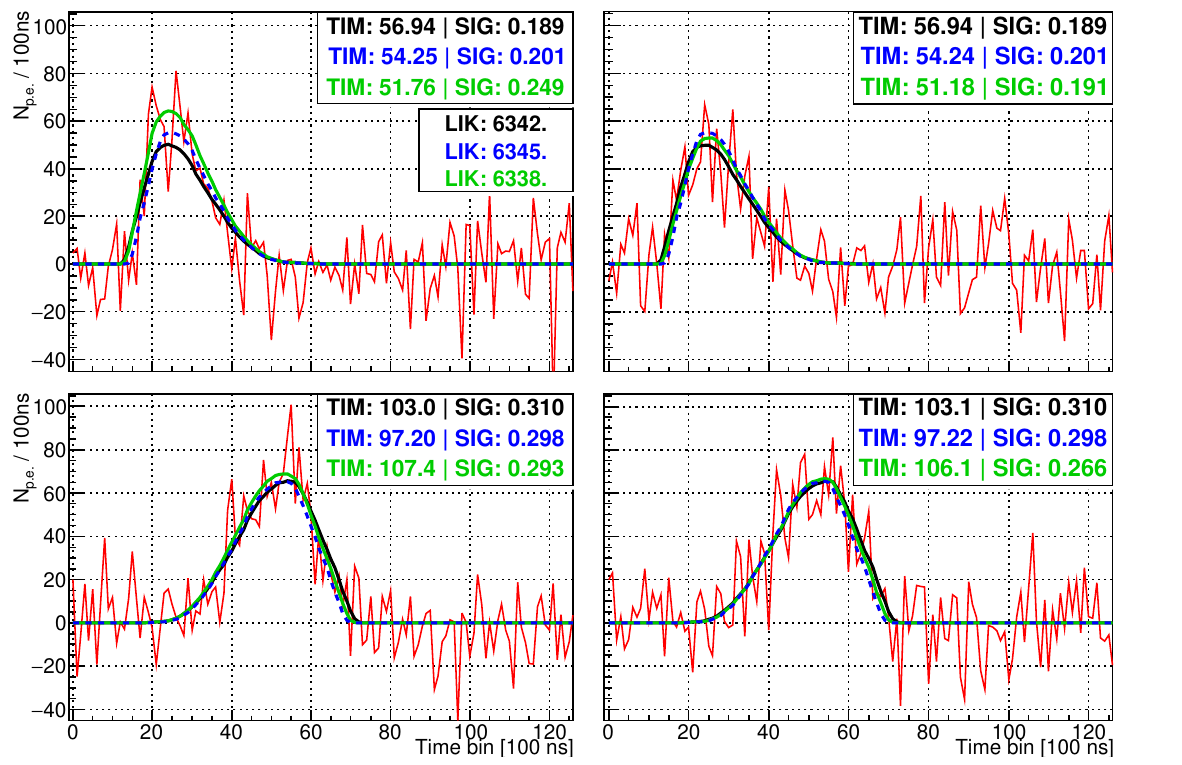}
    \includegraphics[width=0.49\linewidth]{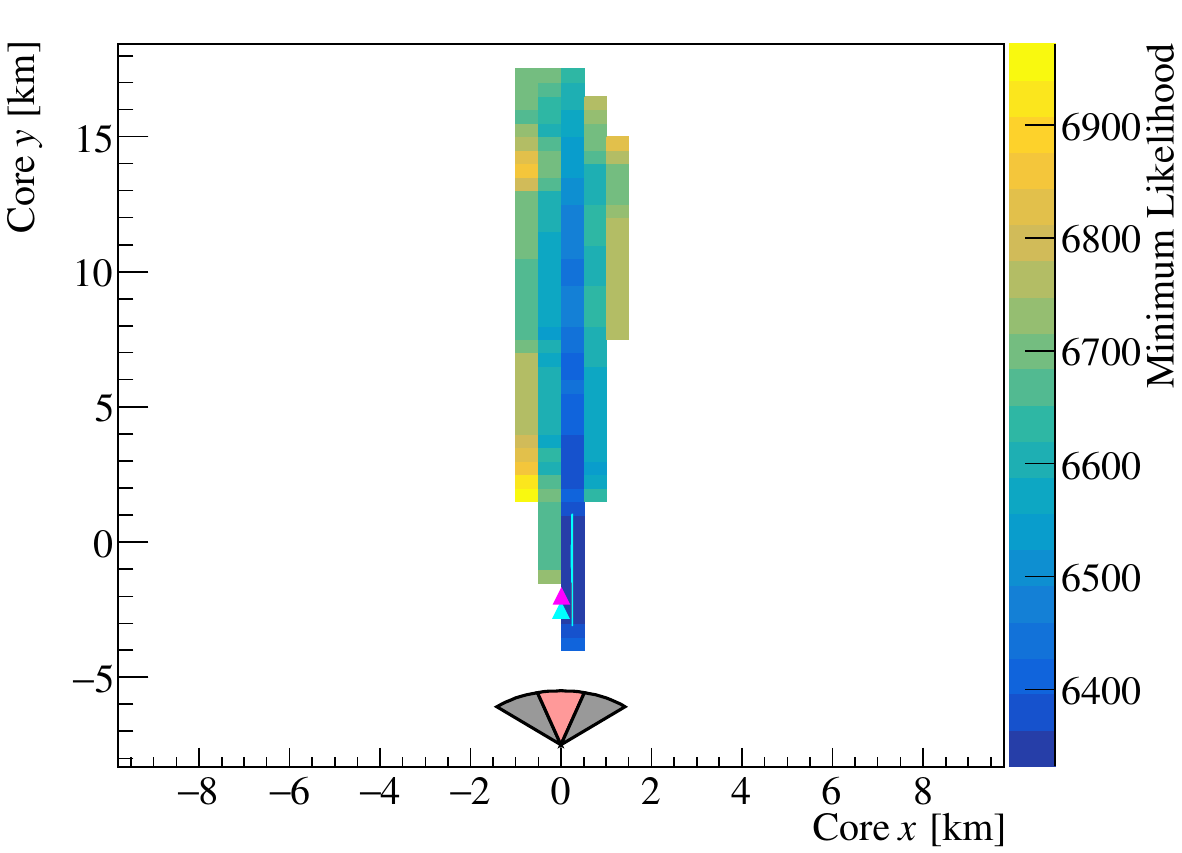}
    \includegraphics[width=0.49\linewidth]{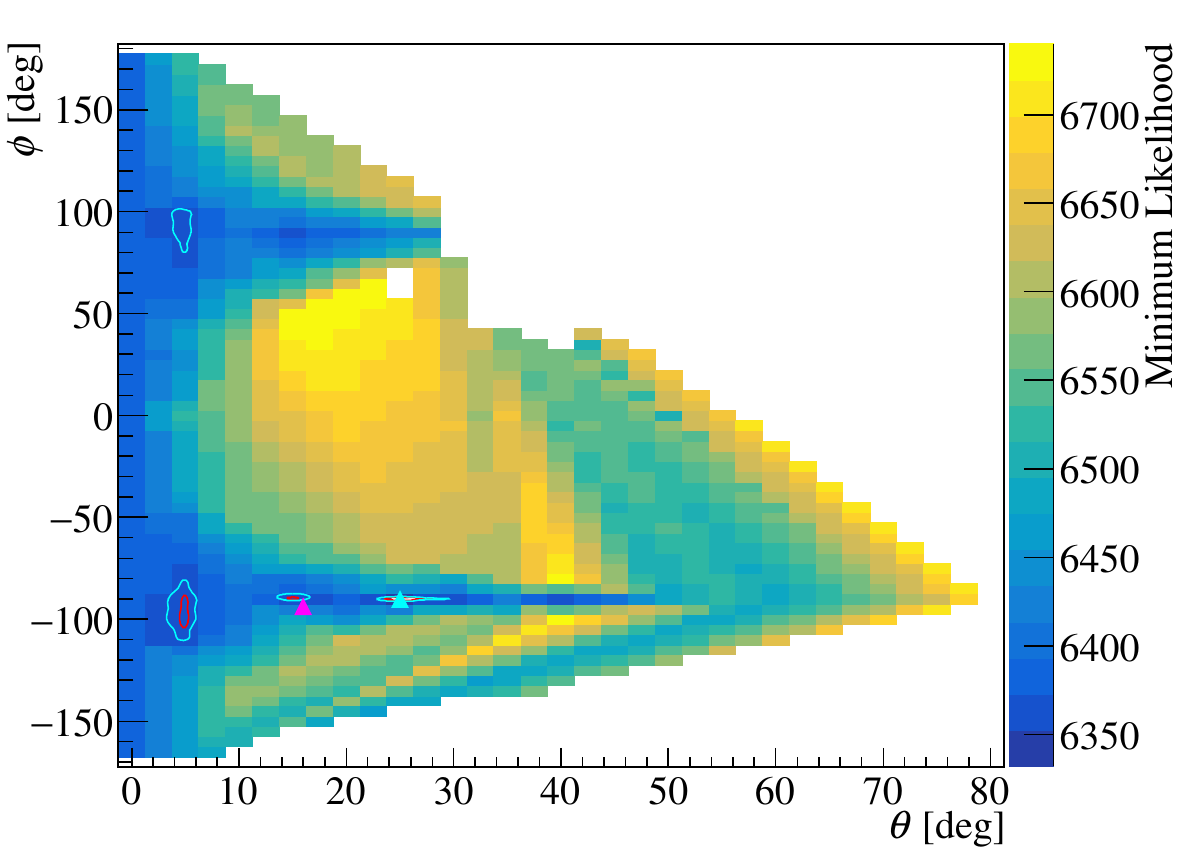}
    \includegraphics[width=0.49\linewidth]{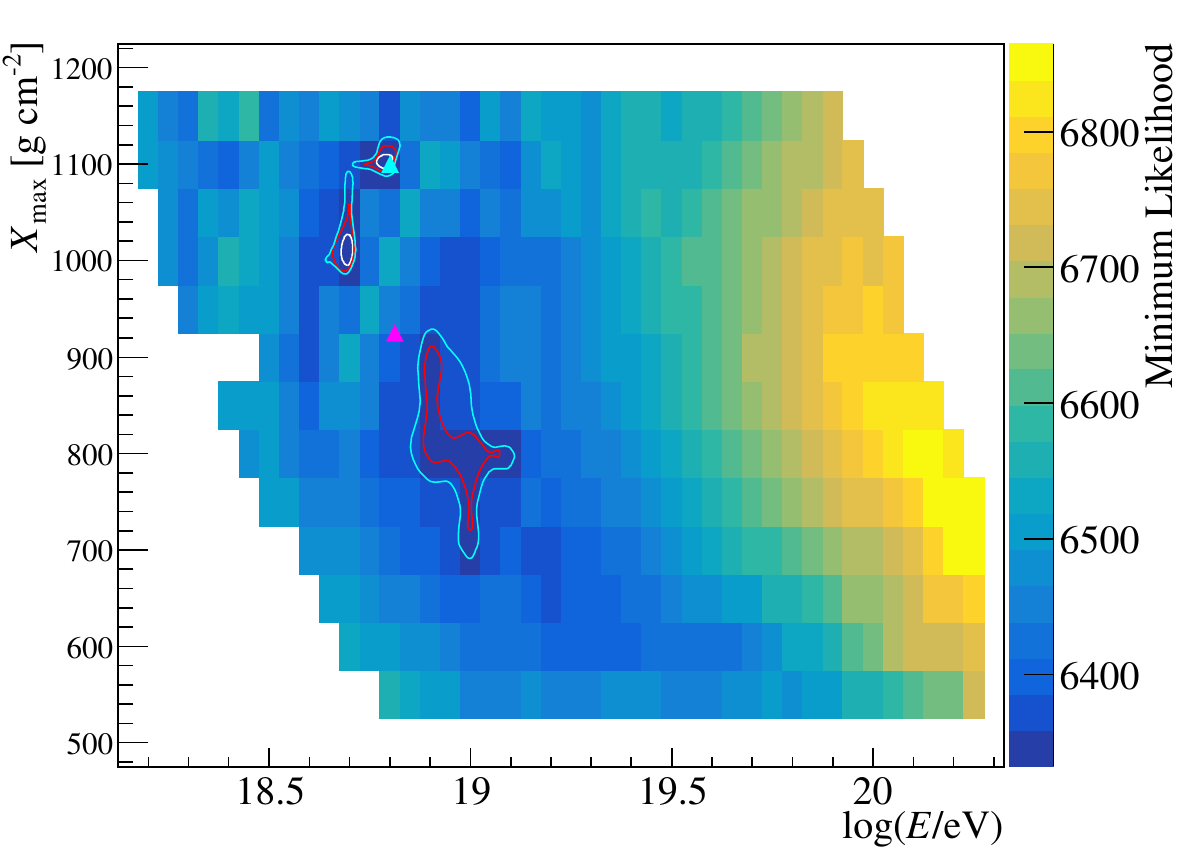}
    \caption{Another example of the Template Method, this time applied to a single telescope with four triggered pixels. The plots are laid out the same as in Figure \ref{fig:pix2Example}.}
    \label{fig:pix4Example}
\end{figure}

For an event observed by a single telescope, the first guess parameters are simply given by $\vec{a}_\textrm{min}$. Figures \ref{fig:pix2Example} and \ref{fig:pix4Example} show examples of the core location, arrival direction and profile likelihood maps for simulated showers observed by a single telescope with two and four triggered pixels respectively. The locations of the true parameters are shown in the likelihood maps by magenta triangles and the best fit parameters by cyan triangles. Another possible option for estimating the parameters is to smooth each likelihood map and then take the locations of the minimum of each map to be the first guess. This type of approach is applied when combining the results from multiple telescopes. For a single telescope, this method was found to improve the estimate in some cases but also worsen it in others. For events where the minimum is located in a sharp trough of the likelihood space, smoothing each map would occasionally eliminate the trough, causing the minimum to be at a different, less accurate location.

\subsection{Multiple Telescopes}
\label{sec:multipleTels}
\begin{figure}
    \centering
    \includegraphics[width=0.68\linewidth]{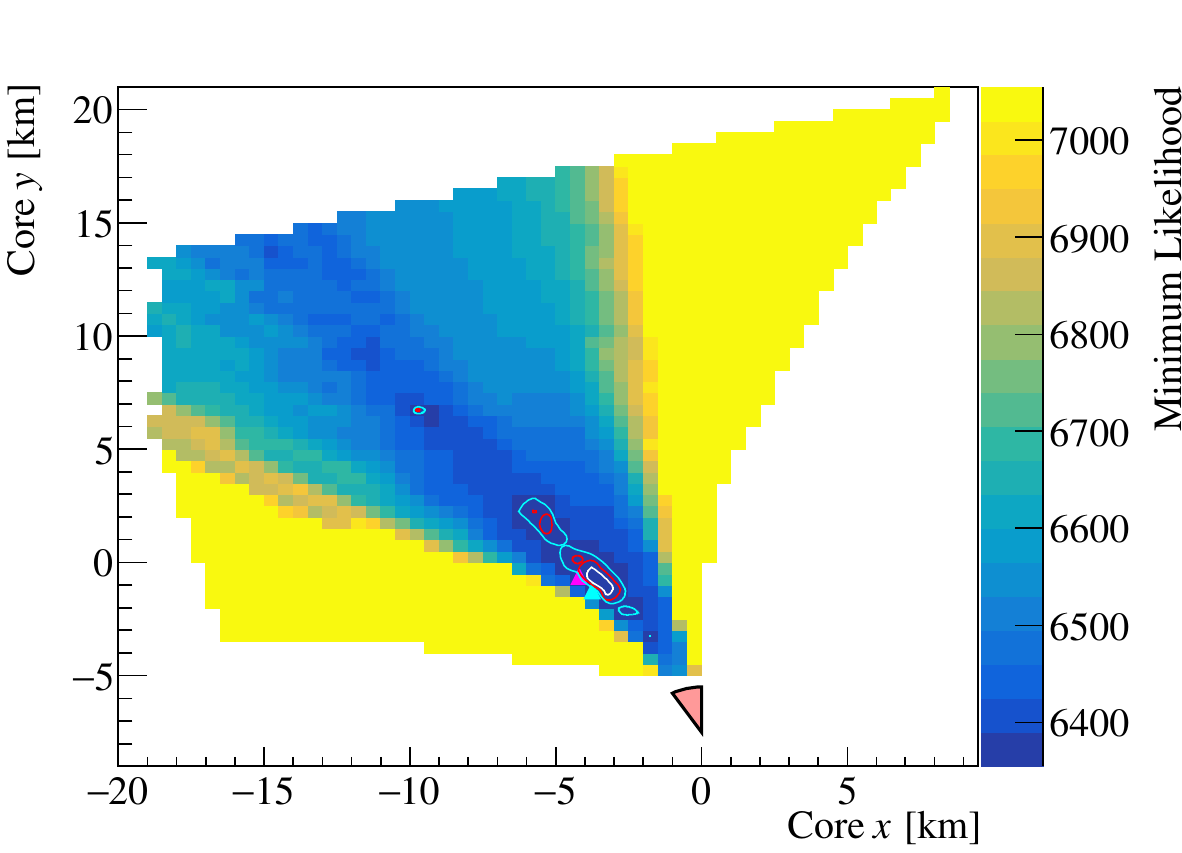}
    \includegraphics[width=0.68\linewidth]{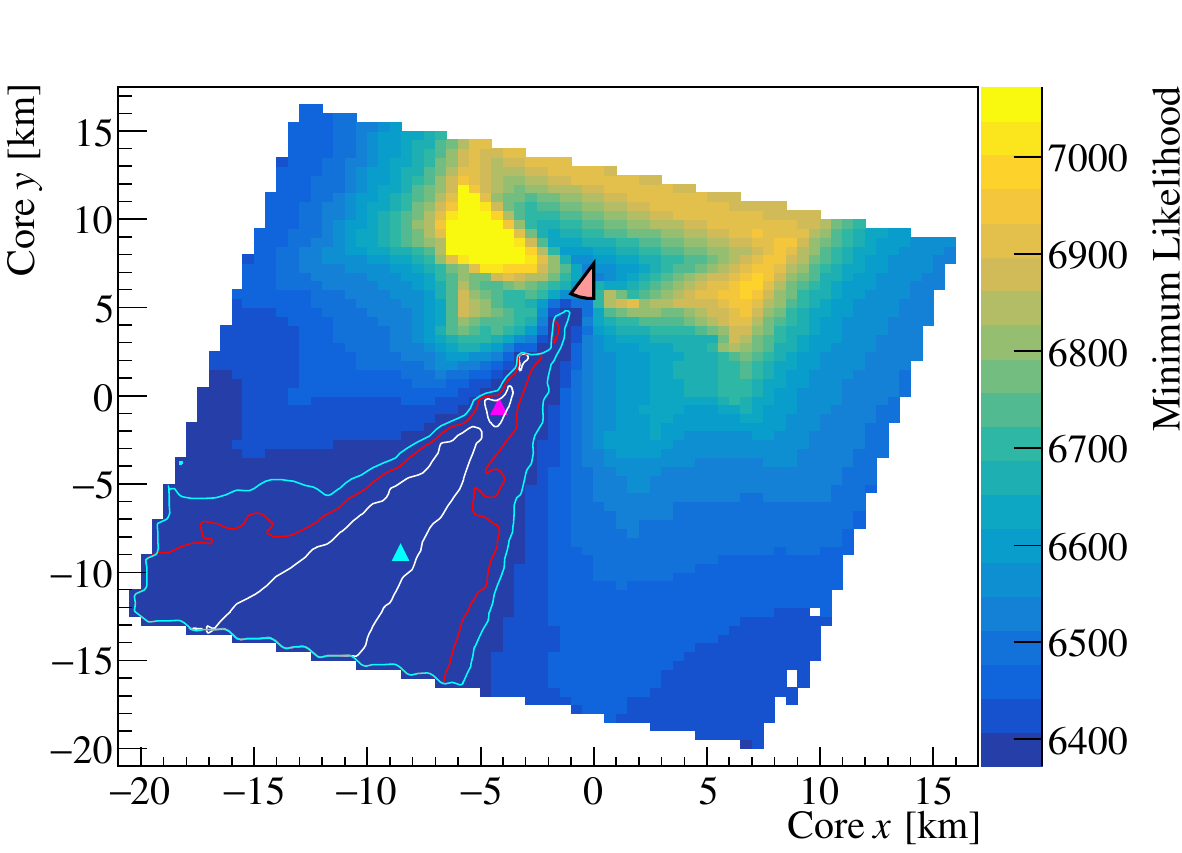}
    \includegraphics[width=0.68\linewidth]{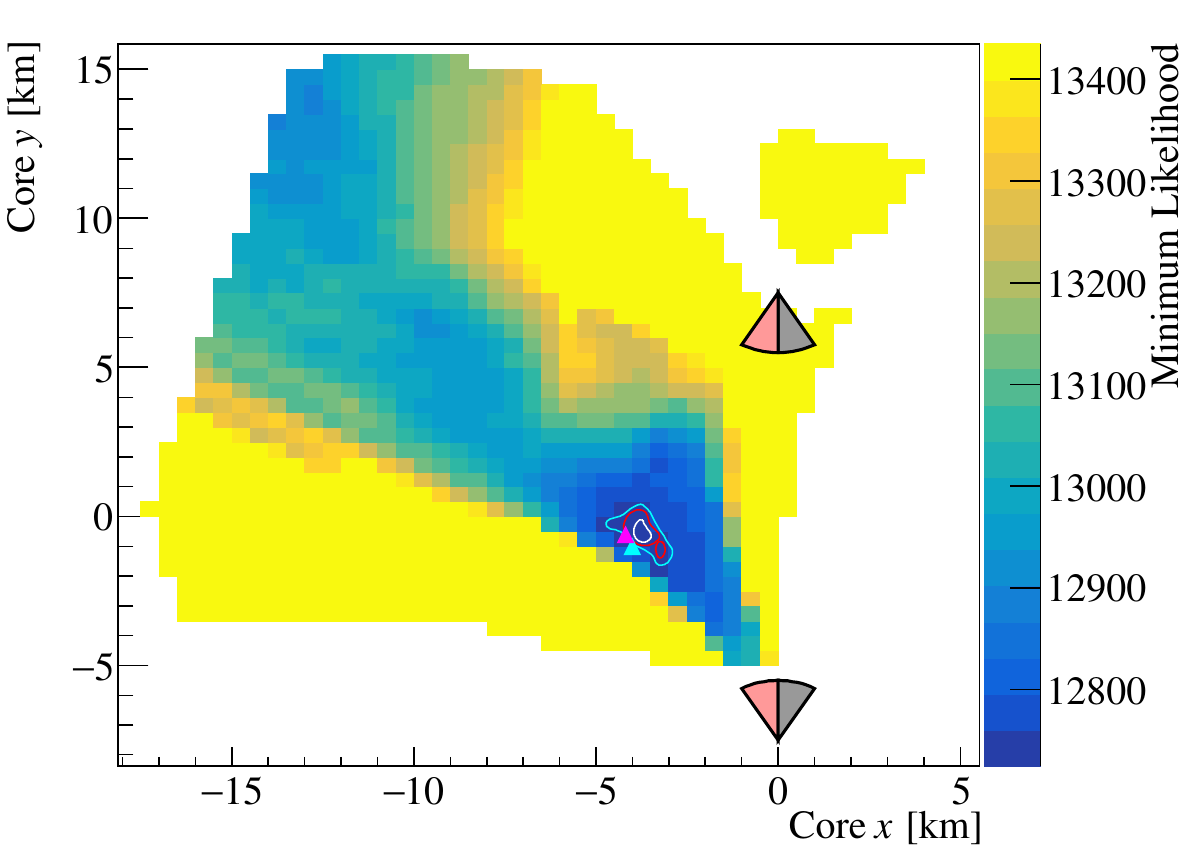}
    \caption{Example of combining the interpolated core location likelihood maps from two telescopes observing a simulated shower from different locations.}
    \label{fig:multiTelExamp}
\end{figure}

For multiple telescopes the core location, arrival direction and profile likelihood maps are calculated in each telescope's frame of reference. The core location and arrival direction maps are then transformed to a single frame of reference. For a given type of map (core location, arrival direction, profile) the resulting maps from each telescope are interpolated at given locations on a grid with spacing equal to half that of the original parameter step sizes. If the maps from all telescopes which observed the event can be interpolated at the given point, then that point is considered a possible match and the likelihood value assigned to it is the sum of the interpolated values. Summing the log-likelihood values is the same as multiplying the likelihoods, so each point in the summed map is a combined probability for each of the telescopes to observe the shower at that point (lower values being better as the \textit{negative log-likelihoods} are being summed). The final shower parameters are chosen by taking the locations of the minimum values over all bins in each interpolated and summed likelihood map. In other words, the first guess values of (core $x$, core $y$), ($\theta$, $\phi$) and (\Xmax{}, $E$) are obtained separately from each map. An example of the interpolation and summing procedure is shown in Figure \ref{fig:multiTelExamp}. The example shows the interpolated core position likelihood maps of two telescopes from the FAST-MiniV1 layout observing a simulated shower (the top and middle panels). The bottom panel shows the sum of these maps. Compared to each telescope's individual result, the first guess estimate of the core becomes closer to the true value and the uncertainty on the estimate (i.e. the scope of the 1, 2 and 3$\sigma$ contours) decreases. 

\vspace{5mm}

The main issue with this approach is that it doesn't take into account the shape of the likelihood function over the full six-dimensional parameter space. For a single 2-D slice of the likelihood function, say of parameters $a$ and $b$, the method of interpolating and summing the likelihood maps from each telescope \textit{is valid} for obtaining the most probable values of $a$ and $b$. However, doing so for multiple sets of parameters and combing the results fails to account for correlations between parameters in different maps. This can cause a less than optimal set of first guess parameters to be chosen.

\subsection{Possible Extensions}
\label{sec:possExten}
A more advanced smoothing/interpolation technique could in principle be applied over the full six-dimensional likelihood space. For example, it is a reasonable assumption that at sufficiently small scales the region around the minimum likelihood can be approximated by a (six-dimensional) quadratic surface. However, attempts at fitting a surface of the form
\begin{equation}
\label{eqn:quadInterp}
    -2\ln{}\mathcal{L}(\vb{x})=\vb{x}^TQ\vb{x}+b^T\vb{x}+c,
\end{equation}
to the likelihood values from each candidate template, and then estimating the shower parameters from the minimum location, was found to be very challenging. In this equation $\vb{x}$ is the vector of candidate template parameters and $Q$, $b$ and $c$ are matrices/vectors of fitted coefficients. During testing, Equation \ref{eqn:quadInterp} was fit to all points which had a value of $-2\ln{}\mathcal{L}$ less than $5\sigma$ above $(-2\ln{}\mathcal{L})_\textrm{min}$ where $\sigma=7.04$ was the same as when fitting all six parameters in the TDR. In cases where there was a large amount of degeneracy between parameters and hence the $5\sigma$ region very large, the parabolic surface became very flat and the fitted minimum rarely aligned with the location of $(-2\ln{}\mathcal{L})_\textrm{min}$ or the true values. The performance was found not to significantly improve even for more tightly constrained regions. Ultimately, the failure here is due to the comparatively large step size of the templates versus the required density of points around the minimum region to achieve a reasonable quadratic approximation. The necessary step-size is likely too fine for producing templates over the entire parameter space, both in physical memory and computational time requirements. This is precisely the reason for the TDR. Another future extension to the Template Method may be to incorporate information from telescopes which do not observe the shower. This could perhaps be achieved by adding a penalty if a candidate template for one telescope has a geometry which falls in the FOV of another telescope which has no signal. Implementing this robustly is expected to be challenging.

\section{Template Method Performance}
\label{sec:templateMethodPerformance}
\begin{figure}
    \centering
    \includegraphics[width=1\linewidth]{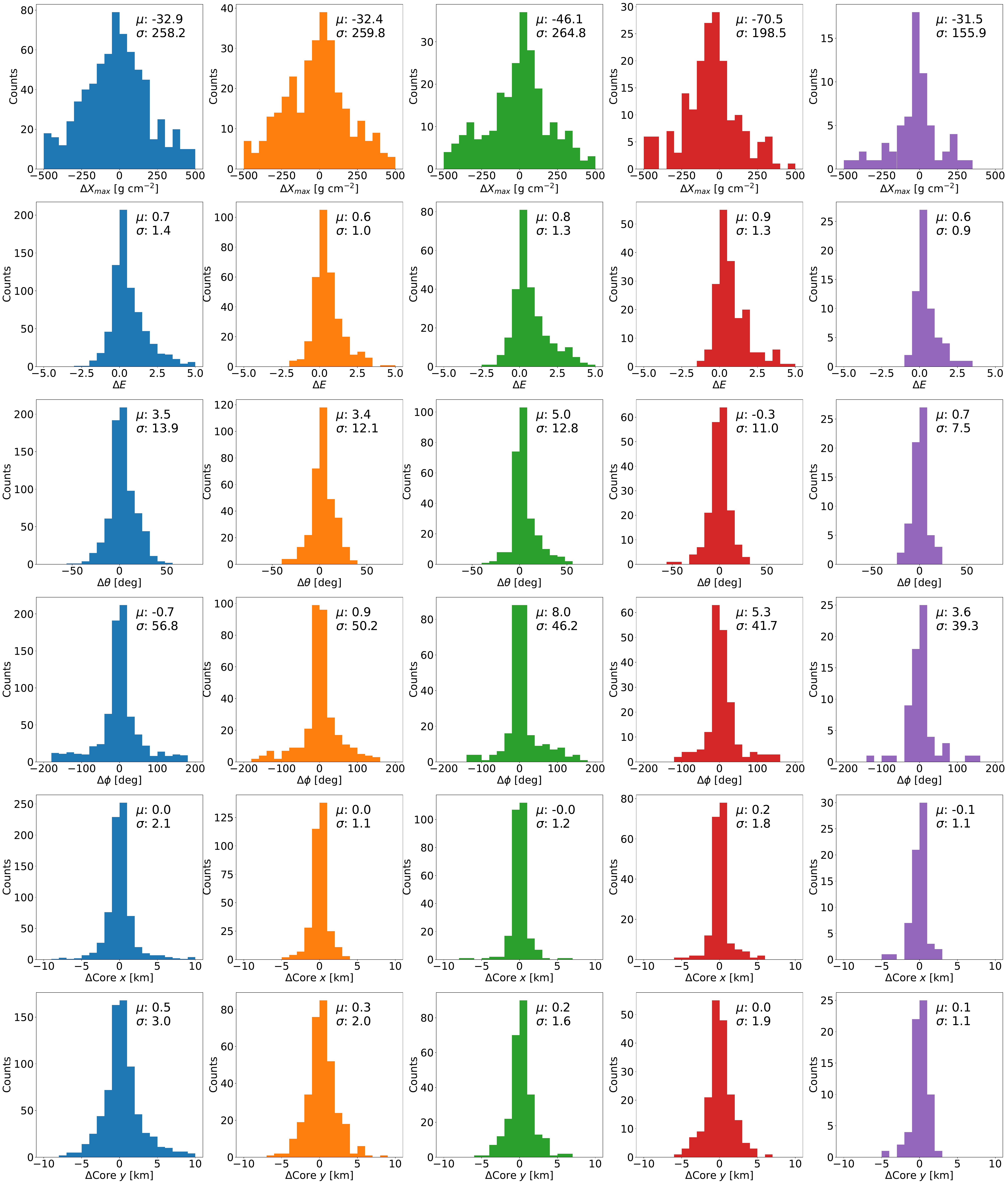}
    \caption{Differences between the true and reconstructed values for each shower parameter, separated by the number of triggered PMTs/eyes. From right to left, the event types are; 1 triggered PMT, 2 triggered PMTs, 3 or more triggered PMTs, 2-Eye stereo and 3-Eye stereo. From top to bottom the reconstructed parameters are \Xmax{}, energy, zenith, azimuth, core $x$ and core $y$. The mean and standard deviation calculated from each distribution (no Gaussian fit) is shown in the top right of each panel.}
    \label{fig:diffHistsTemp}
\end{figure}

\begin{figure}
    \centering
    \includegraphics[width=1\linewidth]{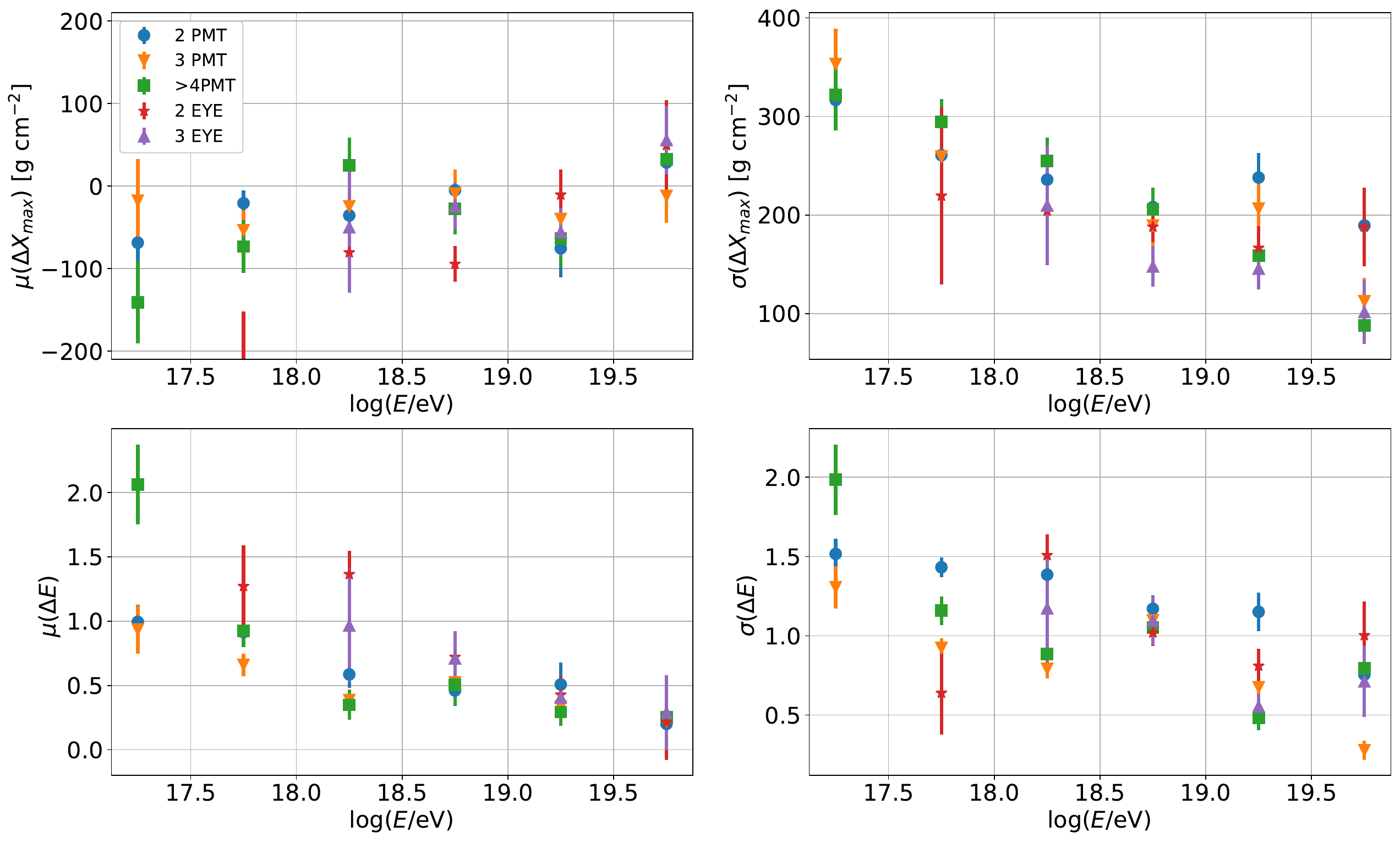}
    \includegraphics[width=1\linewidth]{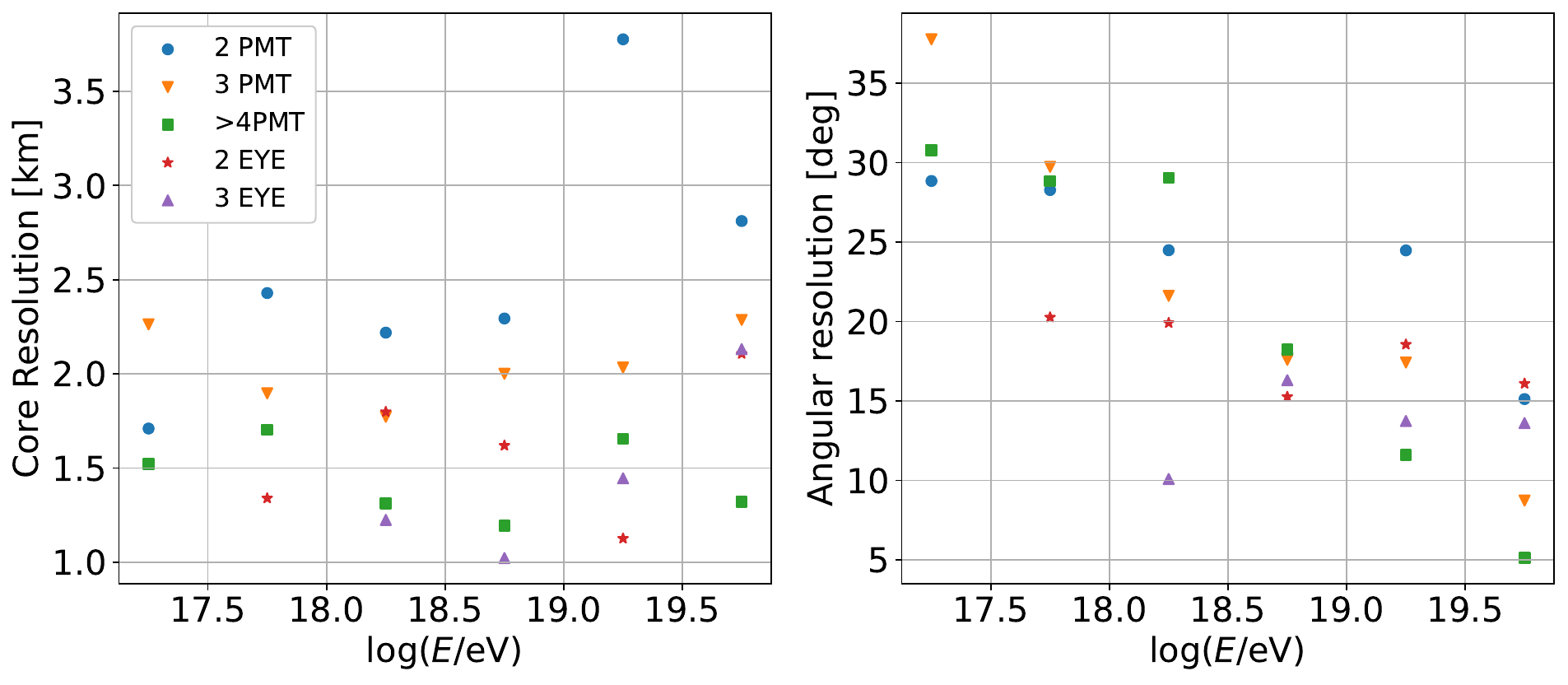}
    \caption{\textit{Top:} Biases and resolutions for \Xmax{} and energy as a function of energy for each different event type when reconstructing using the Template Method. \textit{Bottom:} Core and angular resolutions of the Template Method as a function of energy.}
    \label{fig:tempBiasAndResolutions}
\end{figure}

To test the performance of the Template Method, the method was applied to the same set of showers simulated in Section \ref{sec:combinedMLcheck}, i.e. showers incident on the four different layouts FAST-Single, FAST-TA, FAST-MiniV1 and FAST-MiniV2. Figure \ref{fig:diffHistsTemp} shows histograms of the differences between the reconstructed and true values for each parameter. Figure \ref{fig:tempBiasAndResolutions} breaks these plots down further by showing the biases and resolutions in the \Xmax{} and energy parameters as a function of energy whilst also showing the core and angular resolutions as a function of energy. The distributions in Figure \ref{fig:diffHistsTemp} are separated by how many PMTs/Eyes observed the event. Five classes of event are defined; 1 triggered Eye with either 2 (blue), 3 (orange), or $\geq$4 (green) triggered PMTs, and stereo observation with two Eyes (red) or three Eyes (purple).

\vspace{5mm}

For each parameter, the width of the distributions is largest when just two pixels observe the shower. This matches the findings from Chapter \ref{ch:ML}. For \Xmax{} and energy, the biases and resolutions do not appear to change significantly with the type of event. This is in contrast to the machine learning results where these values were seen to improve with additional pixels and in particular with stereo observation. Thus it seems the Template Method has less sensitivity to these parameters than the machine learning approach. As for the shower geometry, the results for 3/$\geq4$ triggered pixels are mostly similar. The exception to this is the core resolution, where the additional PMTs appear to give better estimates by $\sim500-1000$\,m. This also roughly matches the machine learning results, where the main improvement from having additional triggered pixels came in the core resolution. For stereo observation, the 3-Eye stereo results for core resolution are generally the best out of all event types, again matching the machine learning results. However at the highest energies the best angular resolution is achieved when using a single Eye rather than stereo observation. This goes against both the machine learning results and standard expectations for air shower reconstruction procedures. The reason is thought to be related to the less than ideal method of selecting the best parameters, i.e. the minimum locations of three independently summed and smoothed slices of the likelihood function. Overall, on an absolute scale, the optimal resolutions for the Template Method are worse than those of the machine learning approach. The best \Xmax{} and energy resolutions are roughly 150\gcm{} and 60\% respectively, significantly worse than the best case $75$\gcm{} and 30\% for the TSFEL DNN (with FAST-MiniV2). The geometrical parameters, whilst still not reaching the TSFEL DNN resolutions, are comparatively better, with best case core resolutions around 1 - 1.5\,km and angular resolutions around $10\degree$.

\vspace{5mm}

Naturally the Template Method results could be improved with a denser array of templates. Generating templates with parameter step sizes half that of the current setup would be feasible even if very time intensive. A better approach may be to use a graded spacing such that, in regions of the parameter space where the telescope's response changes quickly, the density of templates is high, whilst in regions where it changes slowly the density is lower. In practice this would correspond to simulating templates with smaller step sizes closer to the telescope and larger step sizes further away. Similarly, sampling values of $\theta$ in even steps of $\sec\theta$ rather than $\theta$ and values of $\phi$ such that showers pointing towards the telescope have a higher density (to account for strongly forward beamed Cherenkov radiation) may be beneficial. An alternative to using showers simulated on a grid could be to use randomly distributed showers. The same signal search criteria developed above could be applied to narrow down the candidate events. Whichever the case, future studies attempting to emulate the method described here should focus on showers within a more confined region, perhaps within 10 - 15\,km of the telescope. This is roughly the maximum distance at which the FAST prototypes have observed air showers thus far (see Section \ref{sec:coincAnal}). This will reduce the space of core locations required to be tested, allowing for finer spacing in the other shower parameters. This finer spacing will be important for implementing more robust interpolation methods.

\section{Shortcomings}
Like the machine learning approach, the Template Method for estimating a first guess comes with its own shortcomings, the most egregious of which is the necessary time required to reconstruct a single event. At worst it may take $\sim5$ minutes for just a \textit{single telescope} to come up with a starting point for the TDR. For a future full sized FAST array detecting potentially hundreds of events per night (depending on the energy threshold), this would clearly be a bottleneck for the analysis. If the goal of the Template Method is to map out the likelihood space and then subsequently give an estimate of the minimum location together with an uncertainty, then the process of template evaluation must be sped up significantly. This could be achieved by making the smallest unit of the templates not a single telescope like done here, but two or three telescopes, as will be arranged for FAST mini-array. Another solution may be random sampling of the candidate templates. Regarding the assignment of a quantitative uncertainty to the first guess, although this was an original goal of the Template Method, the often large/segmented nature of the 1/2/3$\sigma$ contour regions made determining specific values impractical. A proper interpolation/approximation of the six-dimensional likelihood function is likely necessary to obtain reasonable uncertainty estimates. Investigating this is left for future work.

\vspace{5mm}

Another shortcoming is that the true prototype telescopes are unlikely to have a completely symmetric directional efficiency map. If there is significant non-uniformity across the map (for example as seen in Figure \ref{fig:newDirEffMap}), then a single telescope will require twice the amount of simulations to cover the full phase space. Moreover, different telescopes are likely to have slightly different maps, introducing additional systematic uncertainties to the first guess. As with the machine learning results, not accounting for the time-dependent observation conditions and the differences in the properties of the background noise between data and simulations also introduces systematic uncertainties. The quantitative effects of these uncertainties should be investigated in future work. Finally, the method of evaluating the minimum location is not straightforward in the six-dimensional parameter space. The approach adopted here works reasonably well for a single telescope with a large number of triggered PMTs, however combining the results from more than one telescope to achieve the resolutions shown to be possible with the machine learning method will likely require more robust interpolation methods and a denser set of templates.

\section{Summary}
This chapter has developed an alternative method for deriving a set of first guess parameters for the TDR known as the \say{Template Method}. The method involves simulating a library of template traces for a single telescope and comparing these templates to data traces using the likelihood function in Equation \ref{eqn:eventLikelihood}. For events with a single triggered telescope, the first guess is given by the best matching template. For multiple triggered telescopes, likelihood maps in 2-D slices of the parameter space are created for each telescope. These are then interpolated on set grids and summed, with the first guess parameters given by the minimum locations in each summed map. Although imperfect, this is the first method of obtaining a first guess of the shower parameters to be applicable to any number/configuration of FAST telescopes. The primary drawback of the method was found to be the required computation time. This was reduced through a number of different means, such as reducing the trace sizes and selecting only candidate templates which matched the relative signal ratios and timing differences between triggered PMTs. Even with these optimisations however, comparing all candidate templates for a single telescope would occasionally take up to several minutes. Evaluating the performance of the Template Method showed, overall, inferior resolutions to those of the TSFEL DNN. This was thought to be due to the relatively large spacing between the templates and imperfect method for combing results from different telescopes. In particular, the resolutions in reconstructed \Xmax{} and energy failed to reach below 100\gcm{} and 50\%. Despite these shortcomings however, the Template Method does show promise as a versatile first guess approach given its applicability to different telescope layouts and the reasonable core and angular resolutions achieved for stereo events (1 - 1.5\,km and $\sim10$ - $15\degree$ respectively at the highest energies). Future efforts using a denser set of templates and interpolation within the six-dimensional likelihood space are expected to improve these resolutions and provide a more robust estimate of the first-guess uncertainty. 

\chapter{Reconstruction of Data from FAST Prototypes}
\label{ch:REAL}

The final chapter of this thesis utilises the improvements and developments made to the reconstruction procedure in previous chapters to reconstruct data from the FAST prototypes at TA and Auger. The events analysed are \say{coincidence events}, that is events which have been observed by both FAST and its companion experiment i.e. TA/Auger. This allows for the direct event-by-event comparison of the reconstruction results. The ability to perform such an analysis is largely thanks to the development of a new method to search for candidate events developed by a fellow member of the FAST collaboration, Jakub Kmec. This method improves upon previous signal search attempts, identifying a total of $\sim650$ events across both sites. The significant increase in statistics compared to previous work allows for new analyses and comparisons to be performed. As such, this study constitutes the first comprehensive analysis of the FAST reconstruction applied to observed data and is a crucial test for determining how the algorithms perform outside of simulations. The chapter begins by introducing the new signal search method and data set, before comparing expectations from simulations with the data. The coincidence events are then reconstructed and the first measurements of the UHECR energy spectrum and mass composition with FAST made.

\section{New Signal Search Method}
\label{sec:coincEventSearch}
This section provides a basic overview of the new signal search method developed by Kmec. The method is essentially an improved version of the threshold trigger discussed in previous chapters, in that it estimates the SNR of a pixel but accounts for the possibility of a non-zero signal baseline. Since small fluctuations in signal baselines are often observed in the current prototype data, a formulation which accounts for them is critical to accurately determine whether or not a pixel contains signal. Whilst forthcoming updates to the FAST electronics are expected to reduce these fluctuations, the robustness of the algorithm to noise is highly desirable given the large diameter of the FAST PMTs. Note that previous chapters all utilised FAST PMT traces consisting of 1000 bins each of 100\,ns. However PMT traces from the current prototypes actually consist of 5000 bins each of 20\,ns. The algorithm introduced below has been specifically developed for this configuration. Later, during reconstruction, the data traces will be re-binned to the standard 1000 bins at 100\,ns/bin.

\vspace{5mm}

\begin{figure}[t]
    \centering
    \includegraphics[width=1\linewidth]{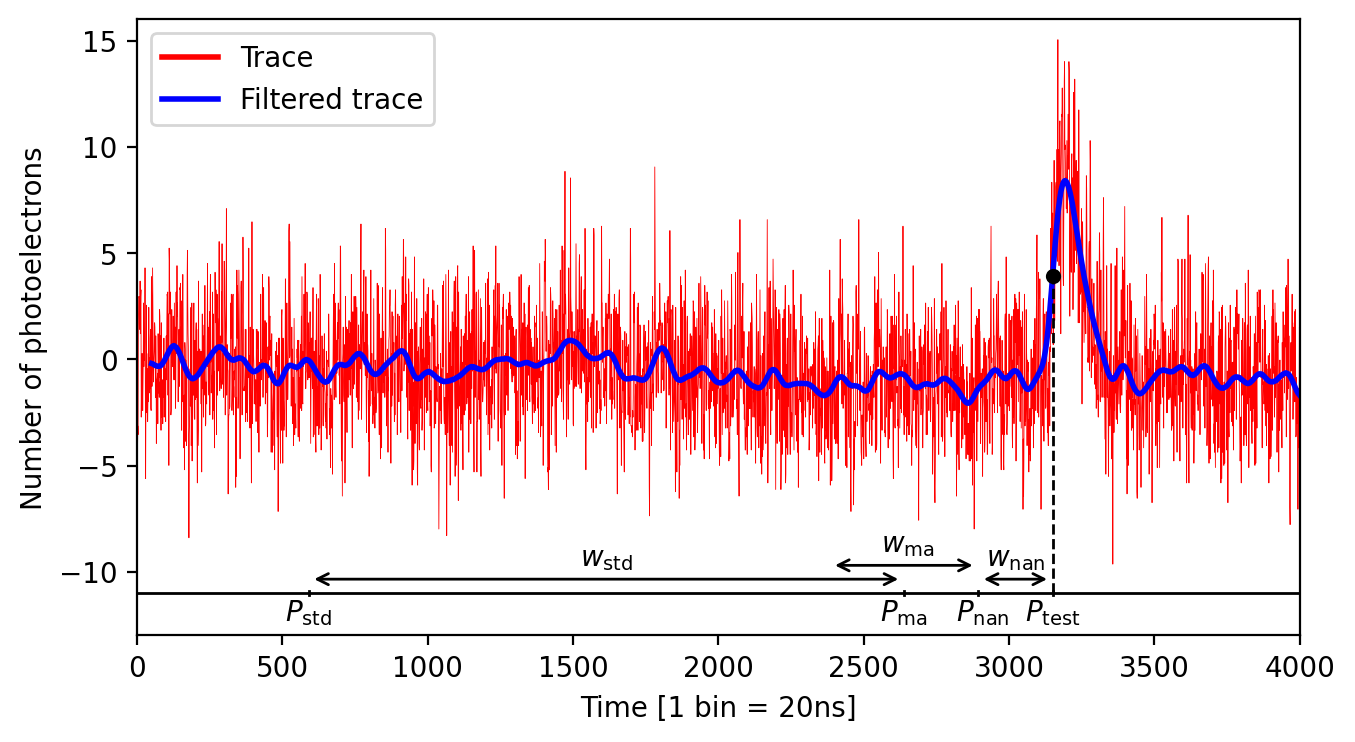}
    \caption{Example of how Equation \ref{eqn:realSignalSearch} is applied to a FAST PMT trace to determine the presence of signal or lack thereof. The original PMT trace is shown by the red line and one of the filtered traces by the blue line. $P_\textrm{test}$ is the bin being tested for signal. Points between $P_\textrm{nan}$ and $P_\textrm{test}$ are ignored as they may be contaminated with signal. The baseline of the original trace is estimated by calculating the average value of the trace over a window of size $w_\textrm{ma}$ centred on $P_\textrm{ma}$. Samples of the FIR filtered trace from $P_\textrm{std}$ to $P_\textrm{ma}$ are used to estimate the standard deviation. From \cite{kmec2023intexttrig}. 
    }
    \label{fig:jakubSignalSearchExample}
\end{figure}

The algorithm is as follows. Given a  PMT trace $T$ with 20\,ns bins, $T$ is smoothed using \gls{fir} filters of differing lengths to give waveforms $F_j$. Specifically, Hamming windows with 5 different window sizes ranging from 25 - 401 bins are used. The different sizes are used so that both long and short signals can be identified. Each bin $i$ in each filtered trace $F_j$ is then tested for signal by calculating the following ratio
\begin{equation}
\label{eqn:realSignalSearch}
    \textrm{SNR}_j(i) = \frac{F_j(i) - T_\mathrm{BG}(i)}{\sigma_{F_j}}.
\end{equation}
Here, $F_j(i)$ is the value of the FIR filtered trace at bin $i$, $T_{\textrm{BG}}(i)$ is the signal baseline (ideally zero) estimated using a section of the original trace between bins $(i-769,i-256)$, and $\sigma_{F_j}$ is the standard deviation of the filtered trace.
If, for at least one filter length $j$, the maximum SNR$_j$ over all bins exceeds the appropriate threshold for that filter length, then the PMT is considered to be triggered. An event is labelled as an EAS candidate if at least one PMT triggers. Note this algorithm is still under development and as such has not been utilised in the previous chapters. However when applied to data from the FAST prototypes, it identifies significantly more events than more standard formulations like Equation \ref{eqn:newsnr}. In particular, it is better at identifying weak signals and signals where the baseline may be significantly shifted from zero \cite{kmec2023intexttrig}.
Figure \ref{fig:jakubSignalSearchExample} shows an example of how the algorithm is applied. 

\section{Coincidence Data Set}
\label{sec:coincAnal}

\begin{table}[]
    \centering
    \small
    \begin{tabular}{c|c|c}
         & \textbf{FAST@TA} & \textbf{FAST@Auger}\\
         \hline\hline
         \textbf{Analysis period} & $2018/03-2018/10$ $(2018/10-2023/02)$ & $2022/07-2022/10$\\
         \textbf{Observation time} & ~65\,hrs (182\,hrs) & 122\,hrs \\
         \textbf{Coincidence events} & 438 & 235
    \end{tabular}
    \caption{Overview of the datasets used in the coincidence analysis. The FAST@TA column contains two values for both the analysis period and observation time. These correspond to the observation periods with two telescopes (left) and three telescopes (right, in brackets).
    }
    \label{tab:coincEvents}
\end{table}

\begin{figure}[t]
    \centering
    \includegraphics[width=0.49\linewidth]{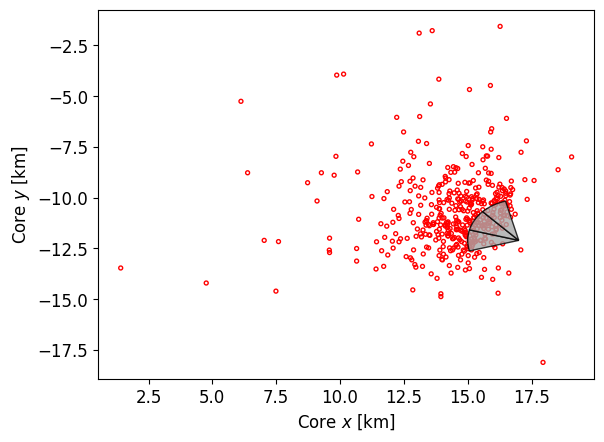}
    \includegraphics[width=0.49\linewidth]{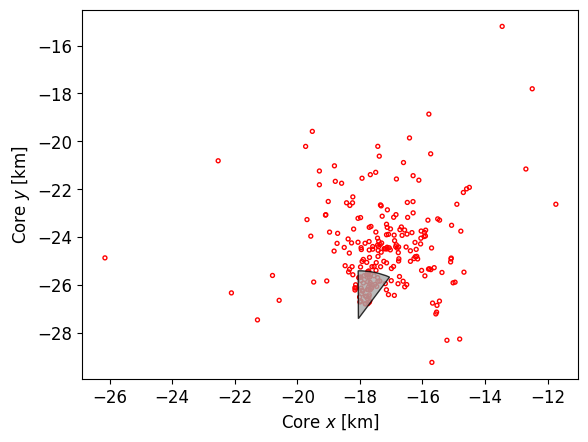}
    \caption{Telescope configurations of FAST@TA (left) and FAST@Auger (right) shown with the core positions, as reconstructed by TA/Auger, of the coincidence events (open red circles).}
    \label{fig:coincDataCores}
\end{figure}

\begin{figure}
    \centering
    \includegraphics[width=0.49\linewidth]{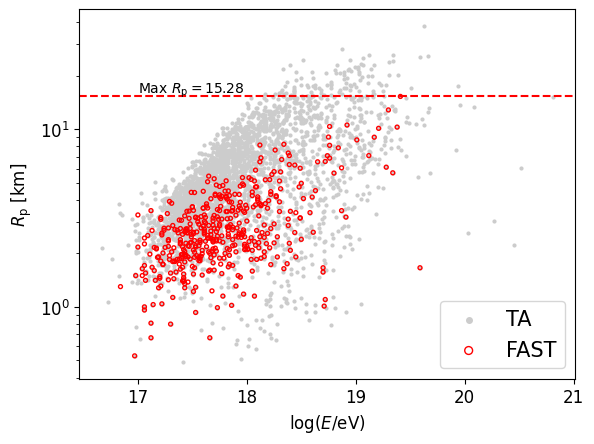}
    \includegraphics[width=0.49\linewidth]{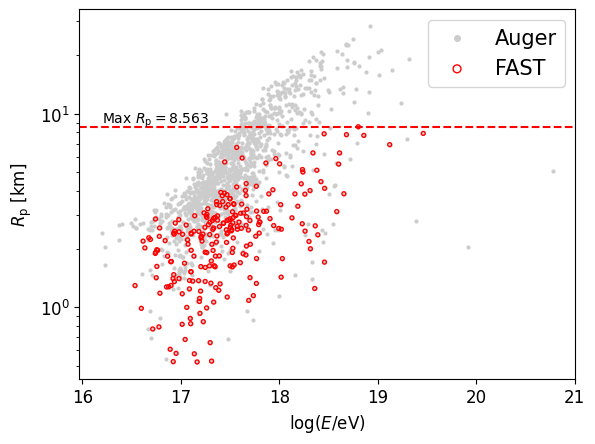}
    \caption{The distance to the shower axis $R_\textrm{p}$ vs. energy for the coincidence events seen by FAST (red open circles) and all candidate events reconstructed by either TA/Auger (grey points). The results from FAST@TA (FAST@Auger) are shown on the left (right).}
    \label{fig:coincRp}
\end{figure}

The data used in the following analysis was taken by FAST@TA and FAST@Auger using external triggers from their respective companion experiment. These triggers were provided by the fluorescence detectors at each experiment which overlook the same field of view as the FAST telescopes, these being the TA fluorescence detector at BRM and the Auger fluorescence detector at LL respectively. To search for coincidence events, rather than checking every single event recorded by FAST, only events which were successfully reconstructed by TA/Auger were checked for signal (see Section \ref{sec:dataMCcomp} for TA/Auger reconstruction details). These events were identified in the FAST data by first calculating the average time offset between the TA/Auger trigger times and the FAST@TA/FAST@Auger trigger times. The event closest in time to the TA/Auger trigger time + offset  was selected. Each of these events (set of PMT traces), were then checked for signal utilising the above algorithm. When applying the algorithm to a PMT trace from prototype data, the last 3,000 bins of the trace (purely noise from the night sky background + electronics) are shifted to the beginning of the trace in order to accurately estimate $\sigma_{F_j}$. With this method 438 coincidence events in the FAST@TA data and 236 coincidence events in the FAST@Auger data were found. This information, together with the analysis periods and corresponding observation times of the coincidence event search, is summarised in Table \ref{tab:coincEvents}. The number of coincidence events and total observation time of FAST@Auger are approximately half the corresponding FAST@TA values, roughly matching expectations.

\vspace{5mm}

The observation times in Table \ref{tab:coincEvents} are calculated as $5$\,minutes$\times$number of FAST event files checked for signal (i.e. the number of data files containing a reconstructed event from either TA/Auger), as each FAST data file covers a 5\,minute interval. The discrepancy between the few hundred hours listed here and the total number of observation hours of the FAST prototypes (see Section \ref{sec:FASTFirstGenPrototypes}), comes from this limited checking and, in the case of FAST@Auger, only considering a short time period where accurate GPS timestamps are available. Ideally, because all 5 minutes from each file are being claimed as observation time, every event in the file should be checked for signal. The assumption being made by only checking the reconstructed events is that all other events would not pass the above signal search. The accuracy of this assumption should be investigated in future work. Note the above does not mean that every TA/Auger reconstructed event will be visible in the FAST data. For now, these observation times will be used in the subsequent data/\gls{mc} comparisons and energy spectrum measurements. 

\vspace{5mm}

The layouts of FAST@TA and FAST@Auger in their respective coordinate systems are shown in Figure \ref{fig:coincDataCores} together with the core locations of the coincidence events as reconstructed by TA/Auger. The majority of events are within $\sim5$\,km of the telescope positions. Figure \ref{fig:coincRp} shows the distributions of $R_\textrm{p}$ vs. shower energy for the coincidence events observed by FAST (red points) and all candidate events reconstructed by TA/Auger (grey points). For a given energy, TA/Auger are able to observe showers at a greater distance. This is due to the smaller pixel size and hence larger SNR since $\textrm{SNR}\propto{}1/\sqrt{\Delta{}\Omega}$ where $\Delta{}\Omega$ is the pixel solid angle (for the same effective collecting area). The maximum $R_\textrm{p}$ of the showers observed by FAST@TA/FAST@Auger is 15.3\,km and 8.6\,km respectively.

\vspace{5mm}

\begin{figure}
    \centering
    \includegraphics[width=0.49\linewidth]{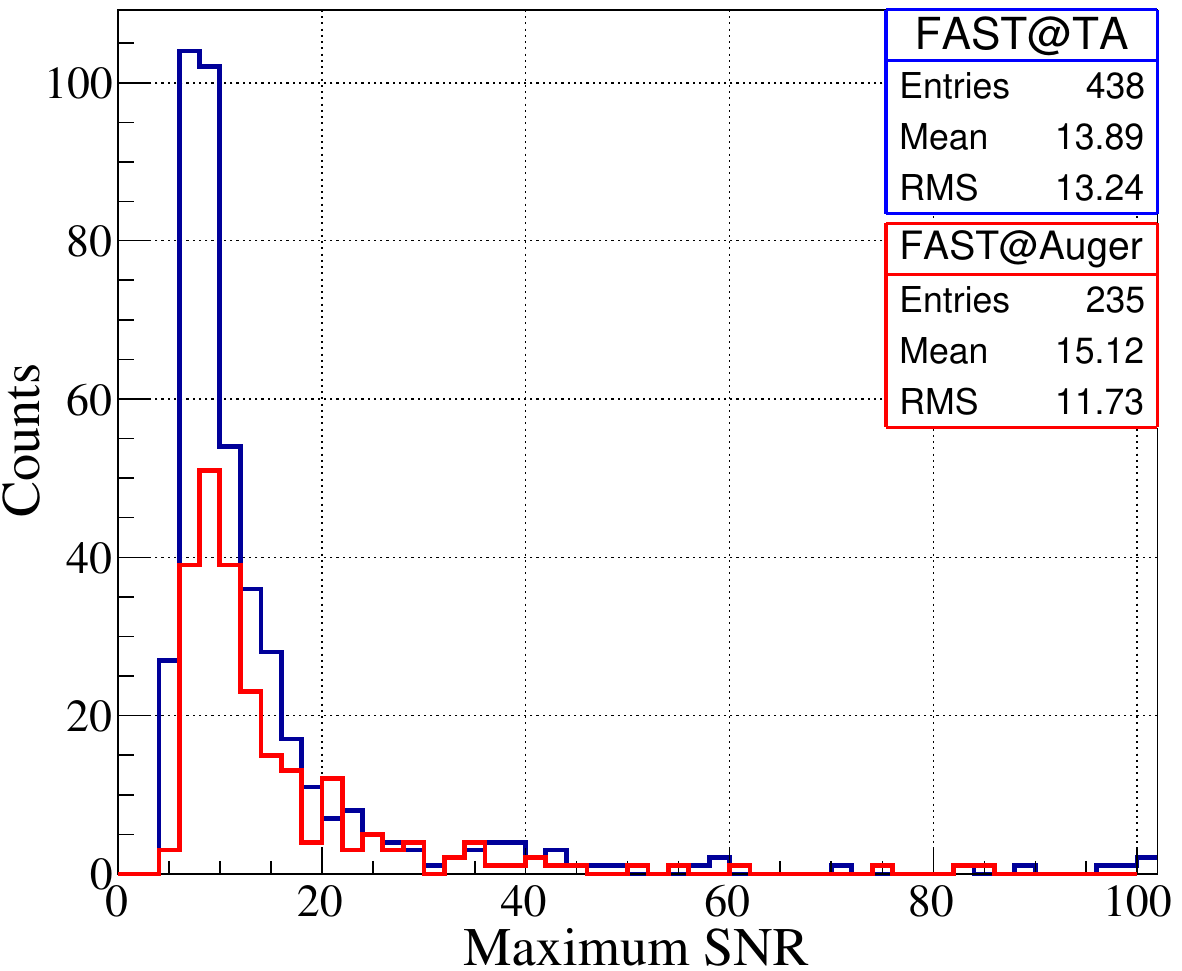}
    \includegraphics[width=0.49\linewidth]{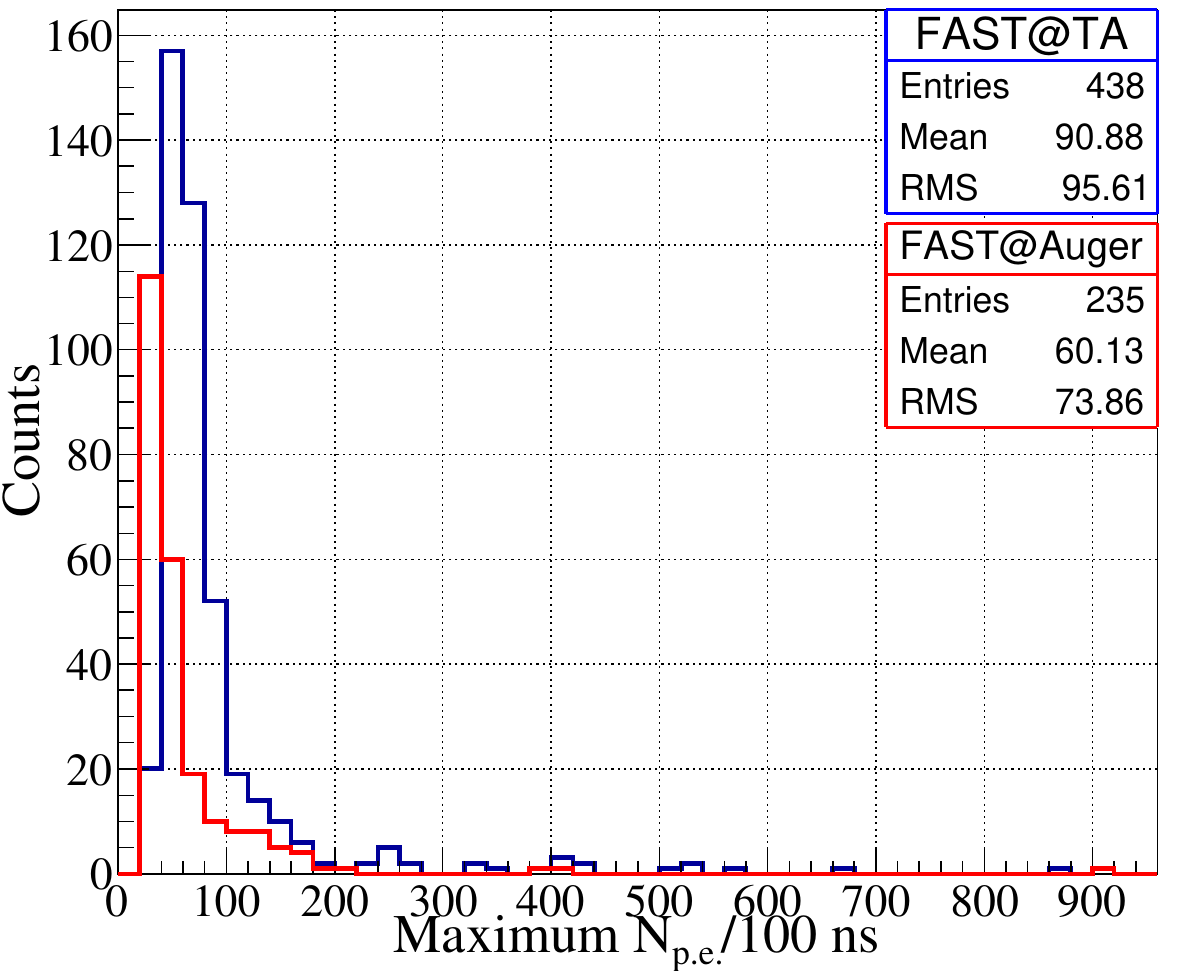}
    \includegraphics[width=0.49\linewidth]{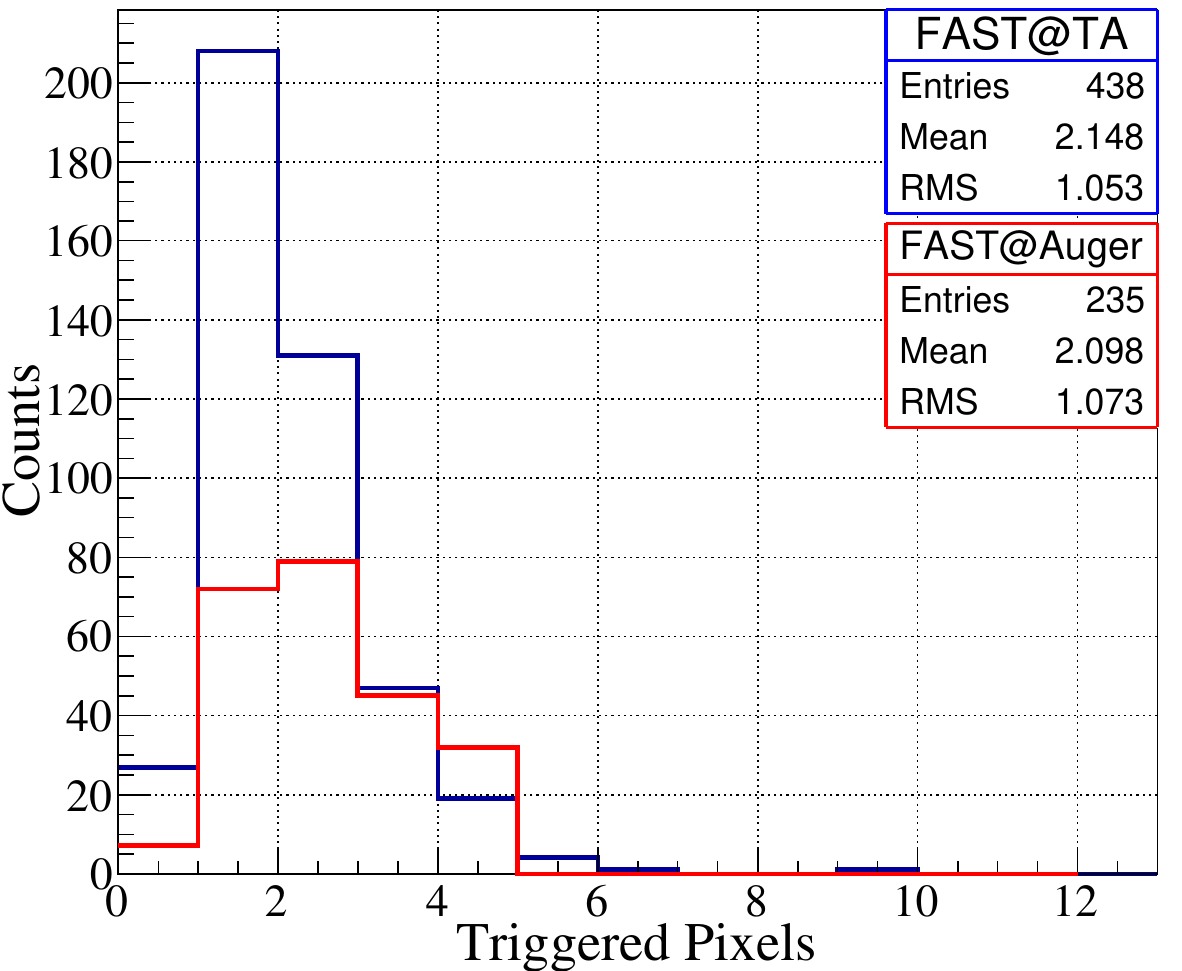}
    \includegraphics[width=0.49\linewidth]{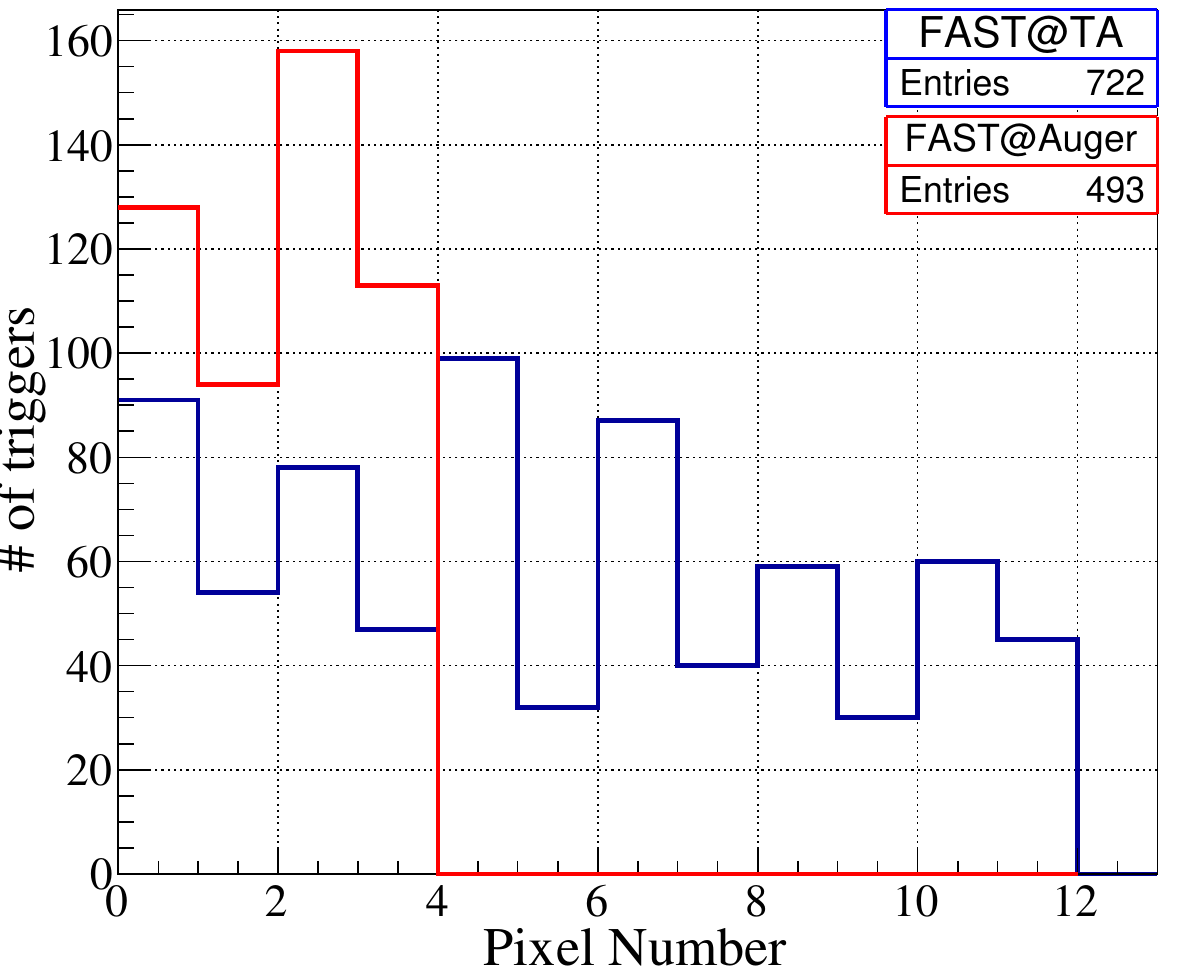}
    \includegraphics[width=0.49\linewidth]{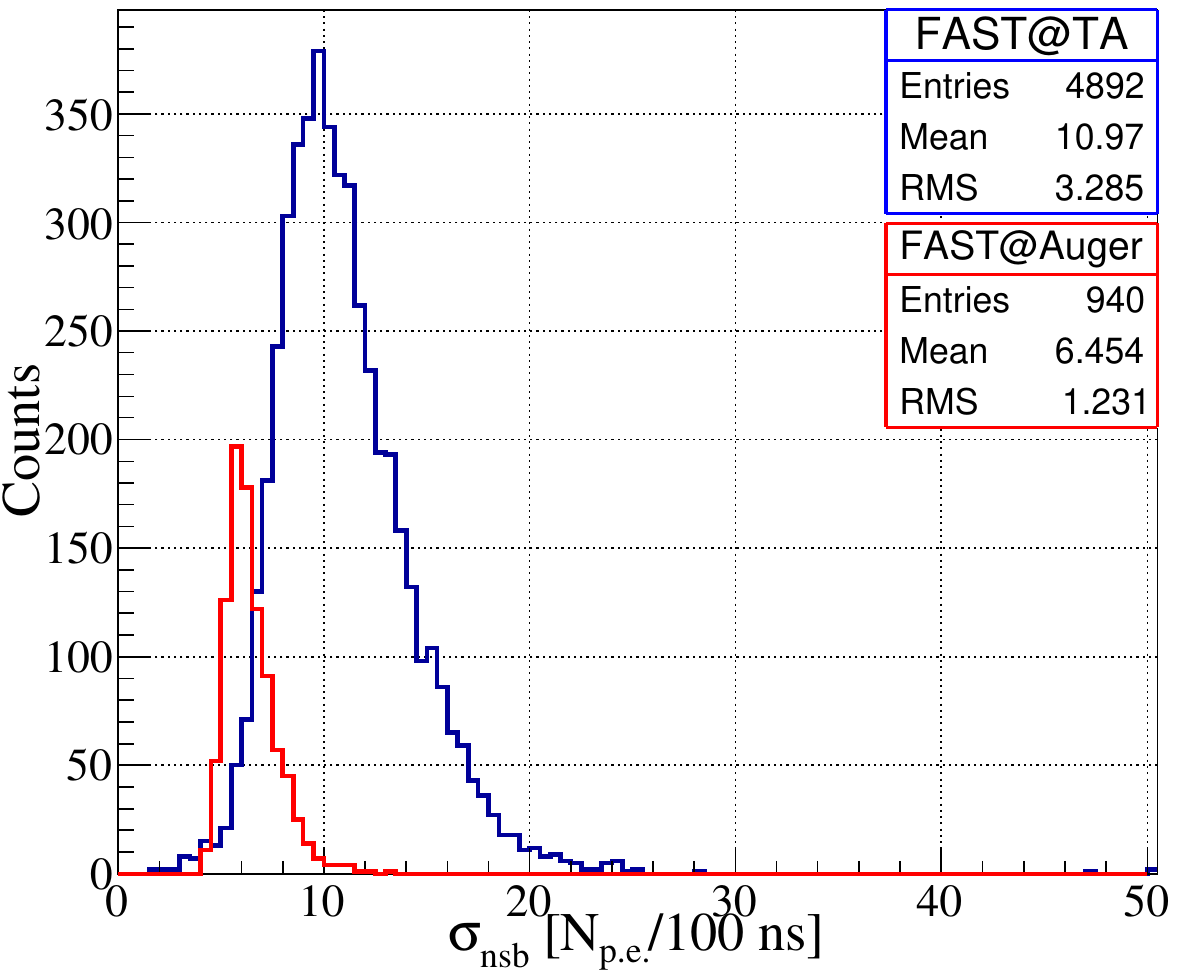}
    \caption{Plots checking low level data for the coincidence events. FAST@TA results are shown in blue, FAST@Auger results in red. \textit{Top left:} Maximum SNRs calculated using Figure \ref{eqn:newsnr}. \textit{Top right:} Maximum number of photo-electrons per event. \textit{Middle left:} Number of triggered pixels. \textit{Middle right:} Which pixels triggered. \textit{Bottom:} Background noise.}
    \label{fig:fundamentalPlots}
\end{figure}

Figure \ref{fig:fundamentalPlots} shows distributions of low level data regarding the coincidence events. These include distributions of the maximum SNR, maximum $N_\textrm{p.e.}$, and number of triggered pixels per event. Also shown are histograms of the distribution of triggered pixels (i.e. which pixels triggered), and the background noise levels estimated from the standard deviation of the first 100 bins of each PMT trace. Note the SNR calculated here comes from Equation \ref{eqn:newsnr} and the trigger condition for a single pixel is as in Section \ref{sec:additionalTriggering} (threshold trigger with threshold = 6, followed by timing check and grouping check). Regarding the distributions of what pixels triggered, the even numbered pixels, which see the upper portion of each telescope's FOV, trigger more often. This is expected since the location of \Xmax{} typically falls within or above the FOV of these pixels. Additionally, FAST 1 and FAST 2 of FAST@TA (pixel numbers 0 - 7) have more triggers than FAST 3 of FAST@TA (pixel numbers 8 - 12) due to extra observation time. As for the other distributions, the background noise level shows a large difference between the installations, with an average value of $\sim11$\,p.e./100\,ns for FAST@TA and $\sim6.5$\,p.e./100\,ns for FAST@Auger. Assuming the average properties of the showers arriving at both installations is the same, a lower background noise level at FAST@Auger would allow signals with a lower maximum $N_\textrm{p.e.}$ to be observed, perhaps accounting for the difference seen in those distributions. This would not change the maximum SNR distribution however, as the overall ratio of background to signal would still be similar to FAST@TA. This is indeed what is observed. Considering the rough agreement in maximum SNR, the differences in the distributions of number of triggered pixels is thought to be due to baseline fluctuations in the FAST@Auger PMTs causing false triggers. The underlying reasons for the difference in background noise level meanwhile are unknown at this time. Possible causes include differences in the magnitude of the NSB at the Auger/TA sites due to location/artificial light sources and/or minor differences between the FAST@TA and FAST@Auger telescopes. The distributions in Figure \ref{fig:fundamentalPlots} may also be affected by differences between the shower selection/reconstruction processes of Auger/TA. Investigation into each of these points is recommended for future work.

\section{Data - MC Comparison}
\label{sec:dataMCcomp}
Before reconstructing the coincidence events with the TDR, the distributions of the shower parameters as estimated by TA/Auger are compared with expected distributions based on calculations with the FAST simulation. Such comparisons are typically called \say{data/MC} comparisons, where MC refers to the results from simulations. Here, the TA values come from the TA monocular reconstruction, whilst the Auger values come from a preliminary hybrid reconstruction. These values will be used throughout the remainder of this chapter as a comparison point for the FAST reconstruction. The data/MC check is performed to gauge whether the FAST simulation is accurately reproducing the observation conditions of the prototype telescopes. The comparison considers only events exceeding 1\,EeV in energy. 

\subsection{Data Set}
\label{sec:realDataAnal}
\begin{figure}[h]
    \centering
    \includegraphics[width=1\linewidth]{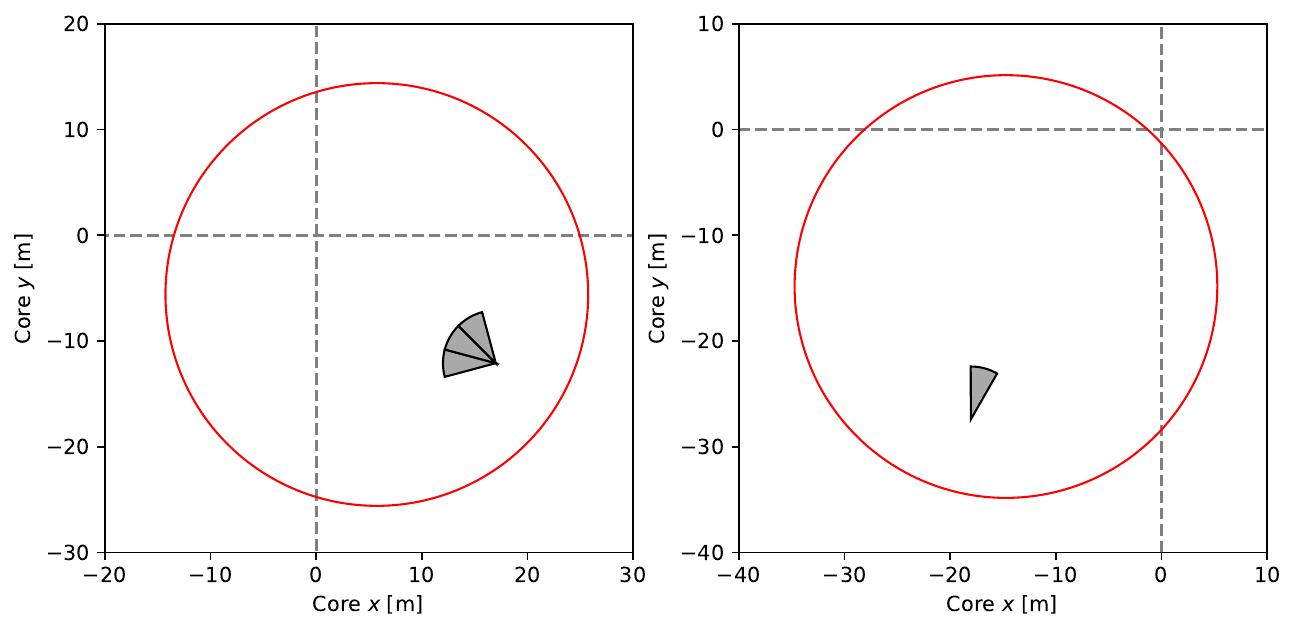}
    \caption{Layouts of the FAST@TA (left) and FAST@Auger (right) telescopes with respect to the TA and Auger coordinate systems respectively. These telescope layouts were used when generating the simulations for comparison with
    coincidence events. The red circles show the range of simulated core positions.}
    \label{fig:expectedPerformanceSimCores}
\end{figure}

For an accurate comparison, the response of the FAST telescopes to all showers which could potentially be observed must be considered. This necessitates simulating showers beyond the observational limits of the telescope i.e. at distances and energies where no significant signal will be observed. Additionally, the distributions of the simulated parameters, with the exception of energy (accounted for later), must be representative of their true distributions. The parameters used to generate the simulations are listed in Table \ref{tab:MCdata}. Mass fractions of (H, He, CNO, Fe) = (0.25, 0.25, 0.25, 0.25) were used for generating \Xmax{}, with the \Xmax{} distributions sampled from being parameterisations of the EPOS-LHC distributions from \cite{blaess2018extracting}. The arrival directions are sampled such that they evenly populate a hemisphere and the core positions such that they extend roughly 7\,km behind and $30$\,km in front of each telescope. The core location distributions for the simulated data sets corresponding to FAST@TA and FAST@Auger are shown in Figure \ref{fig:expectedPerformanceSimCores}. In total, 200,000 events were simulated for each layout. Future work should consider implementing a varying composition with energy which better represents the current knowledge of UHECR mass composition. This could be achieved by using mass fractions estimated by either Auger/TA.

\begin{table}[t!]
    \centering
    \begin{tabular}{|c||c|}
    \hline\hline
        \Xmax & EPOS-LHC ($500$ - 1200\,g\,cm$^{-2}$) \\
        \hline
         Energy & $E^{-1.5}$ ($1$ - 100\,EeV) \\
         \hline
         Zenith & $\sin\theta\cos\theta$ ($0$ - 80$\degree$) \\
         \hline
         Azimuth & Uniform ($0$ - $360\degree$) \\
         \hline
         Core position & \makecell[c]{Uniformly distributed in circle, $r$ = 20\,km, centred at\\$(5741\,\textrm{m},-5060\,\textrm{m})$ (TA) \& $(-14726\,\textrm{m},-14846\,\textrm{m})$ (Auger)} \\
         \hline\hline
    \end{tabular}
    \caption{Parameter distributions used to generate the simulations for comparison with coincidence events. The right column lists the distributions sampled from and range (in brackets) for each parameter shown in the left column. Each of the distributions sampled from were continuous.
    }
    \label{tab:MCdata}
\end{table}

\vspace{5mm}

The simulated responses of the FAST telescopes must also be set to match the responses of the prototypes as closely as possible. This means instead of the ideal directional efficiency map, which has been utilised in all simulations thus far, directional efficiency maps representative of the prototype telescopes (shown in Appendix \ref{apx:direcEffMaps}, Figure \ref{fig:oldDirEffMap}) were used. These maps were constructed using the spatially dependent collection efficiency measurements from \cite{abbasi2010139} and the rotation angles of the PMTs installed on the FAST@TA prototypes. Figure \ref{fig:FAST13Map} is used for FAST 1 and FAST 3 
and the single telescope at FAST@Auger. FAST 2 uses the map in Figure \ref{fig:FAST2Map} due to the positioning and construction of it's camera box being slightly different. 
Lastly, in an ideal comparison, the time dependent state of the detector and surrounding environment would be accounted for and replicated in the simulations. However a lack of precise atmospheric measurements and the setup of the current FAST software framework prevents this from being feasible at the present time. For these simulations then, time-dependent calibration constants and atmospheric states were not considered. The molecular atmospheric profiles for FAST@TA and FAST@Auger were set to the standard US atmosphere and the standard parametric atmosphere used in previous simulations respectively.

\subsection{Calculating the Expected Distributions}
For a single shower parameter $p$, the expected distribution for showers observed by FAST over a time $T$, falling within an area $A$, arriving from solid angles $\Omega$, with energies between $E_1$ and $E_2$, can be written as 
\begin{equation}
\label{eqn:expectedDist}
    D(p;T,A,\Omega) = \int_{E_1}^{E_2}J(E)\times{}\textrm{exposure}(E,A,\Omega,T)\times H(p;E,A,\Omega)\,\mathrm{d}E.
\end{equation}
Here, $J(E)$ is the differential flux of cosmic rays and must be assumed to take some form, for example that of a previously measured spectrum. The \say{exposure} is typically defined as 
\begin{equation}
    \textrm{exposure}(E,A,\Omega,T)=A\times{}\Omega\times{}T\times{}\textrm{efficiency}(E)
\end{equation}
where the efficiency is normally estimated from simulations as
\begin{equation}
    \textrm{efficiency}(E)=\frac{\textrm{\# reconstructed showers}(E)}{\textrm{\# simulated showers}(E)}.
\end{equation}
Note the \say{\#} symbol is used as shorthand for \say{number of}. $H(p;E,A,\Omega)$ represents the normalised distribution of $p$ for showers observed by FAST at a particular energy $E$ and for the given $A$ and $\Omega$. Note that in general, both $H$ and the exposure may vary with time, due to changes in the detector response/detector upgrades and variability in the observational conditions.
These dependencies will not be considered here. However, for the time periods where FAST@TA observed with 2 and 3 telescopes respectively, the efficiencies, exposures and distributions $H$ will be calculated separately and then added to calculate the final distributions.  

\vspace{5mm}

In essence, Equation \ref{eqn:expectedDist} says that the expected distribution of $p$ will be the sum of the number of particles expected to be detected at each energy ($J(E)\times\textrm{exposure($E$)}$) multiplied by the normalised distribution of $p$ for showers detected at that energy, $H(p;E)$ (dropping $A$ and $\Omega$ for brevity). To calculate $H(p;E)$ for each parameter histograms of each parameter were constructed in energy bins of width $\log(E/\textrm{eV})=0.1$ using only those showers \say{seen} by the FAST telescopes. These distributions were then normalised to give the histograms $H^\textrm{norm}_i$ where $i\in$\{[18 - 18.1], [18.1 - 18.2],..., [19.9 - 20]\} represents the energy bin. The expected distribution of $p$ can then be expressed as a weighted sum of each $H_i^\textrm{norm}$ i.e. 
\begin{equation}
    D(p;T,A,\Omega) = A\Omega{}T\sum_{i}J_i\times\Delta{}E_i\times{}\frac{\textrm{\# showers \say{seen} by FAST}_i}{\textrm{\# simulated showers}_i}\times H_i^\textrm{norm}.
\end{equation}
For the setup here, $A=\pi20^2$\,km$^{2}$, $\Omega=2\pi(1-\cos^2(80\degree))$\,sr and $T$ varies depending on the layout (Table \ref{tab:coincEvents}). The values for $J_i$ are taken from the TA energy spectrum as given in \cite{abbasi2016energy} and $\Delta{}E_i$ is the bin width in linear scale. The efficiency term is more challenging to calculate. Note that the number of showers \say{seen} by the FAST telescopes, rather than the number of reconstructed showers, is used. This is because the coincidence events being compared to were not simply detected and then reconstructed by FAST; instead, they have first been selected as a candidate by the TA/Auger triggering system, then successfully reconstructed by TA/Auger, and then passed the FAST signal search (Equation \ref{eqn:realSignalSearch}).

\begin{figure}
    \centering
    \includegraphics[width=0.49\linewidth]{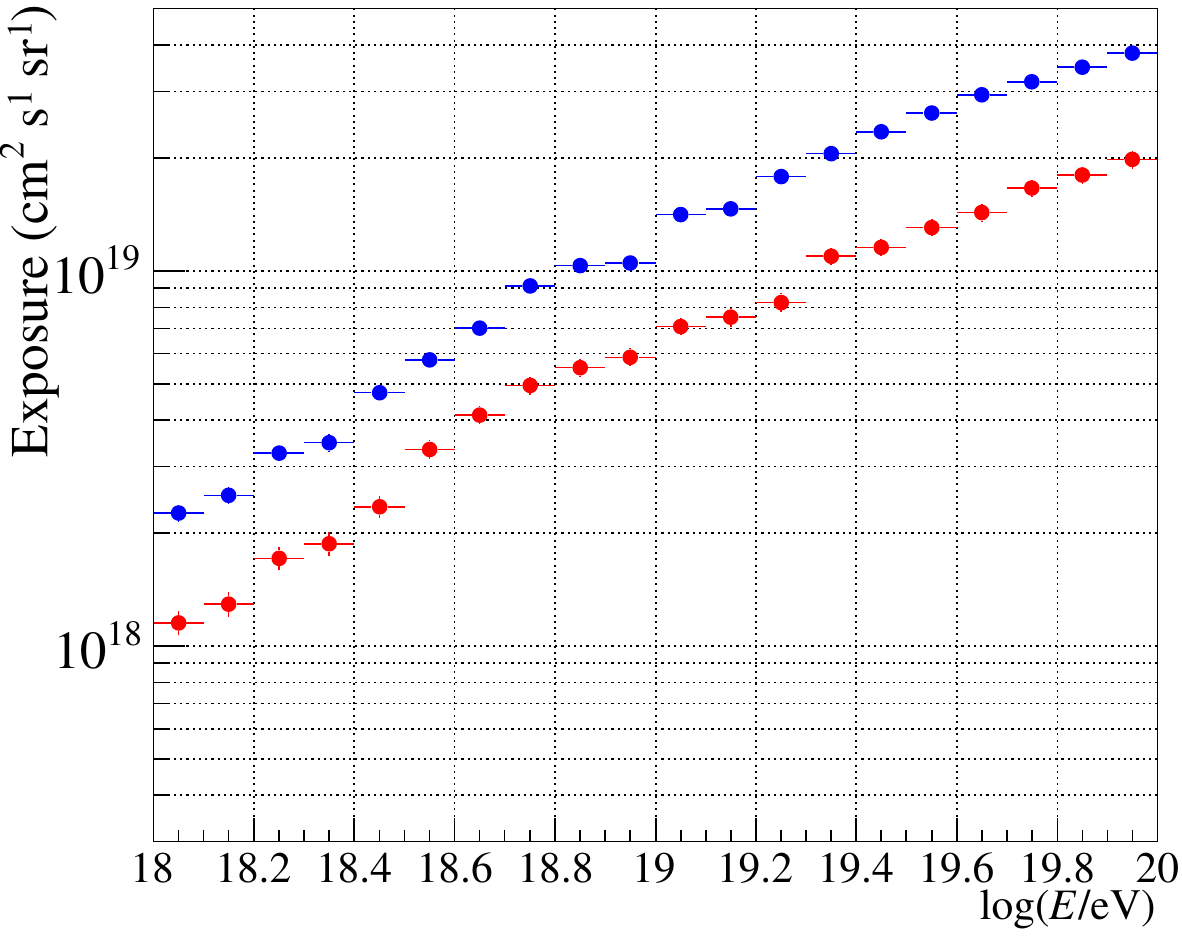}
    \includegraphics[width=0.49\linewidth]{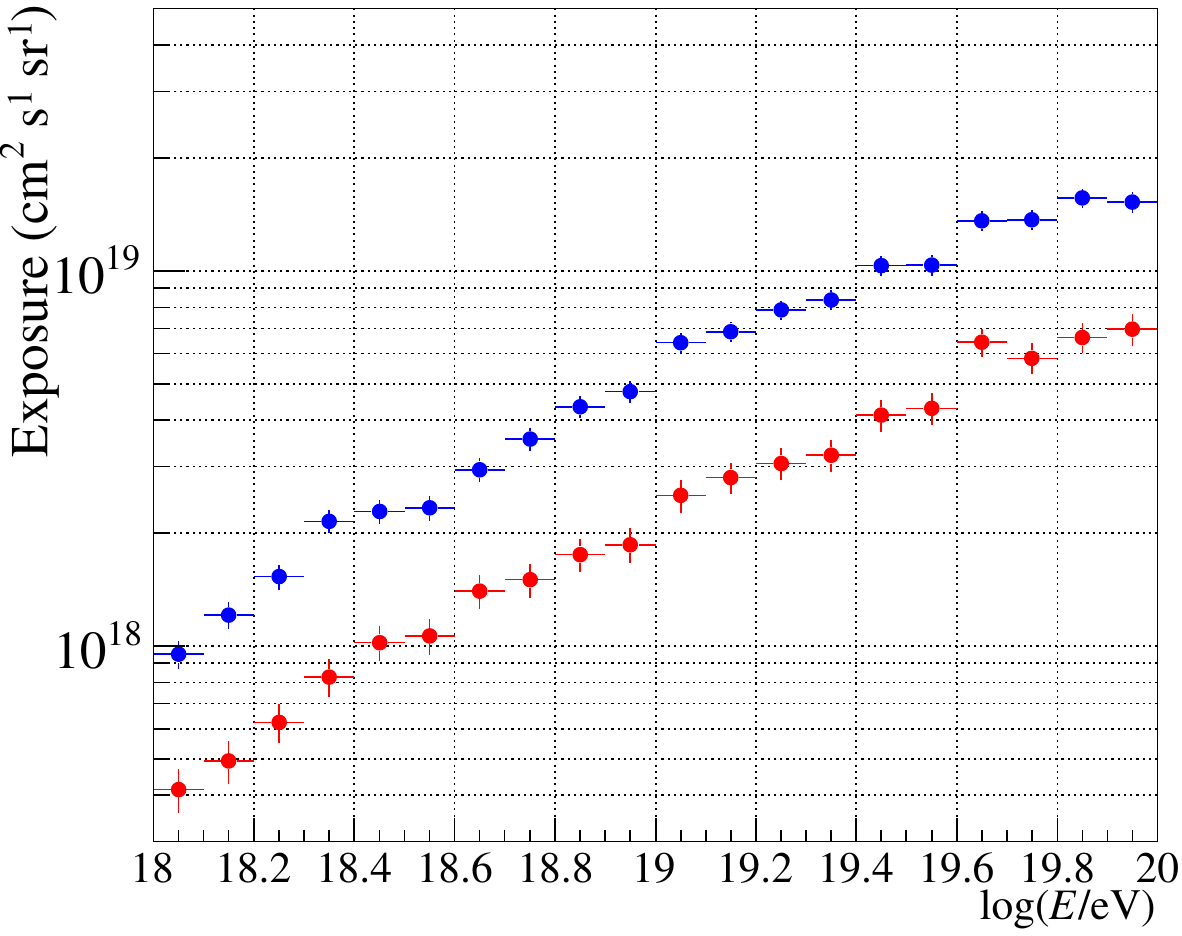}
    \caption{Exposures for FAST@TA (left) and FAST@Auger (right) for 247\,hrs and 122\,hrs of observation respectively.}
    \label{fig:expectedPerformanceExposures}
\end{figure}

\vspace{5mm}

Ideally, this process would be reproduced in the FAST simulations when determining the efficiency term. Unfortunately this is not feasible with the current simulation framework. The best that can be done is to try and imitate these \say{selection criteria} using various trigger conditions. A simulated shower which passes said trigger conditions would be considered equivalent to detecting a coincidence event. In this work two simple trigger conditions are tested, these being if a simulated shower had at least one or at least two PMTs passing the threshold trigger (SNR $>6$) respectively. The definition of SNR here again comes from Equation \ref{eqn:newsnr}. Figure \ref{fig:expectedPerformanceExposures} shows the calculated exposures for FAST@TA and FAST@Auger using both conditions, where the blue (red) points correspond to $\geq1$ PMT ($\geq2$ PMTs). As expected only requiring at least one PMT results in larger exposures. The bumpiness seen across all graphs comes from low statistics, whilst the continuing linear trend at high energies shows that the efficiency does not saturate before 10$^{20}$\,eV for the given simulation setup/triggers. Future data/MC analyses using FAST data taken with external triggers could investigate different triggering conditions which more closely replicate the Auger/TA event selection process. Analyses using data taken with an internal trigger will be able to avoid this challenge entirely and properly incorporate the FAST reconstruction into the efficiency calculation.

\subsection{Results}

\begin{figure}
    \centering
    \includegraphics[width=0.49\linewidth]{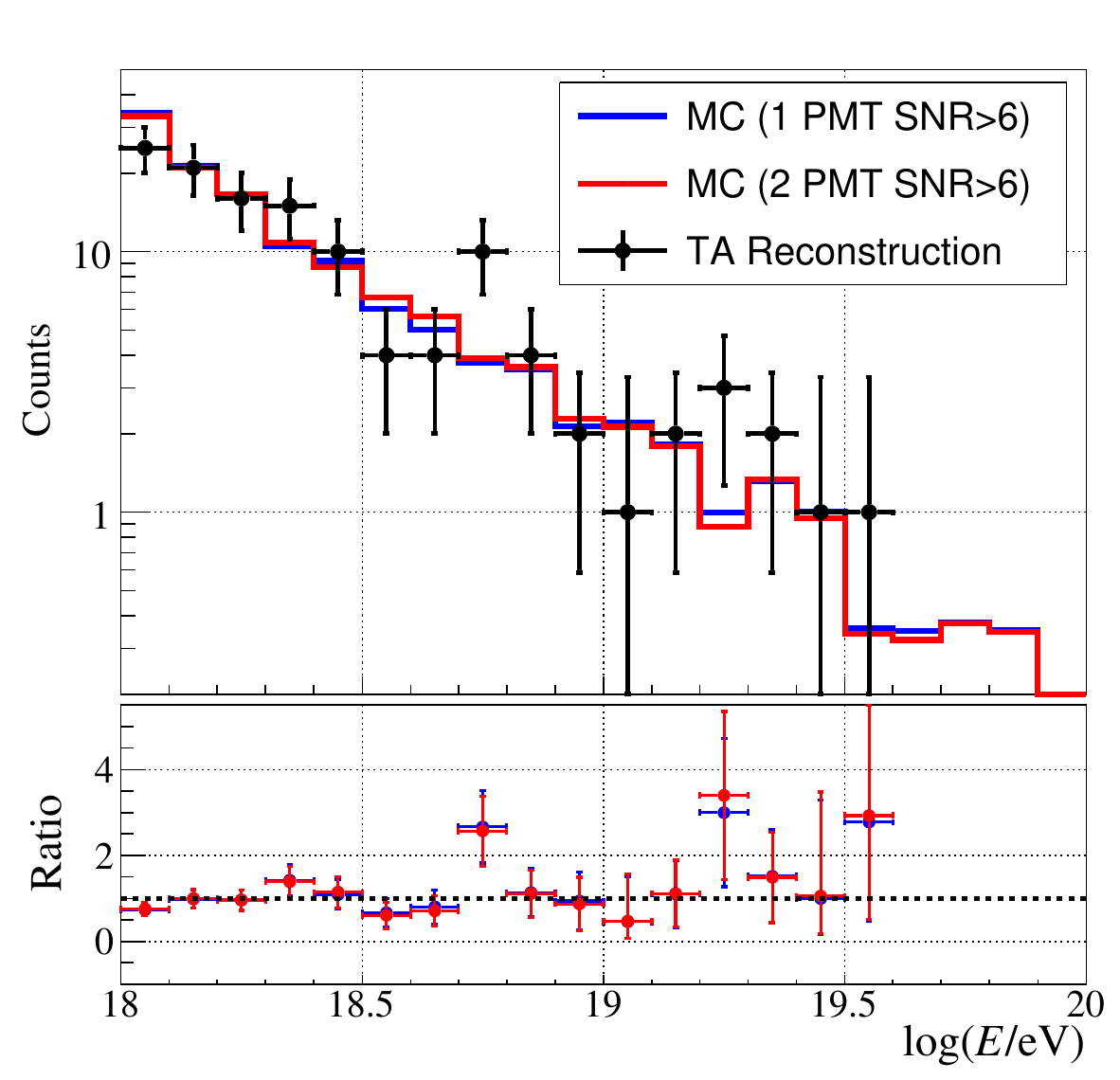}
    \includegraphics[width=0.49\linewidth]{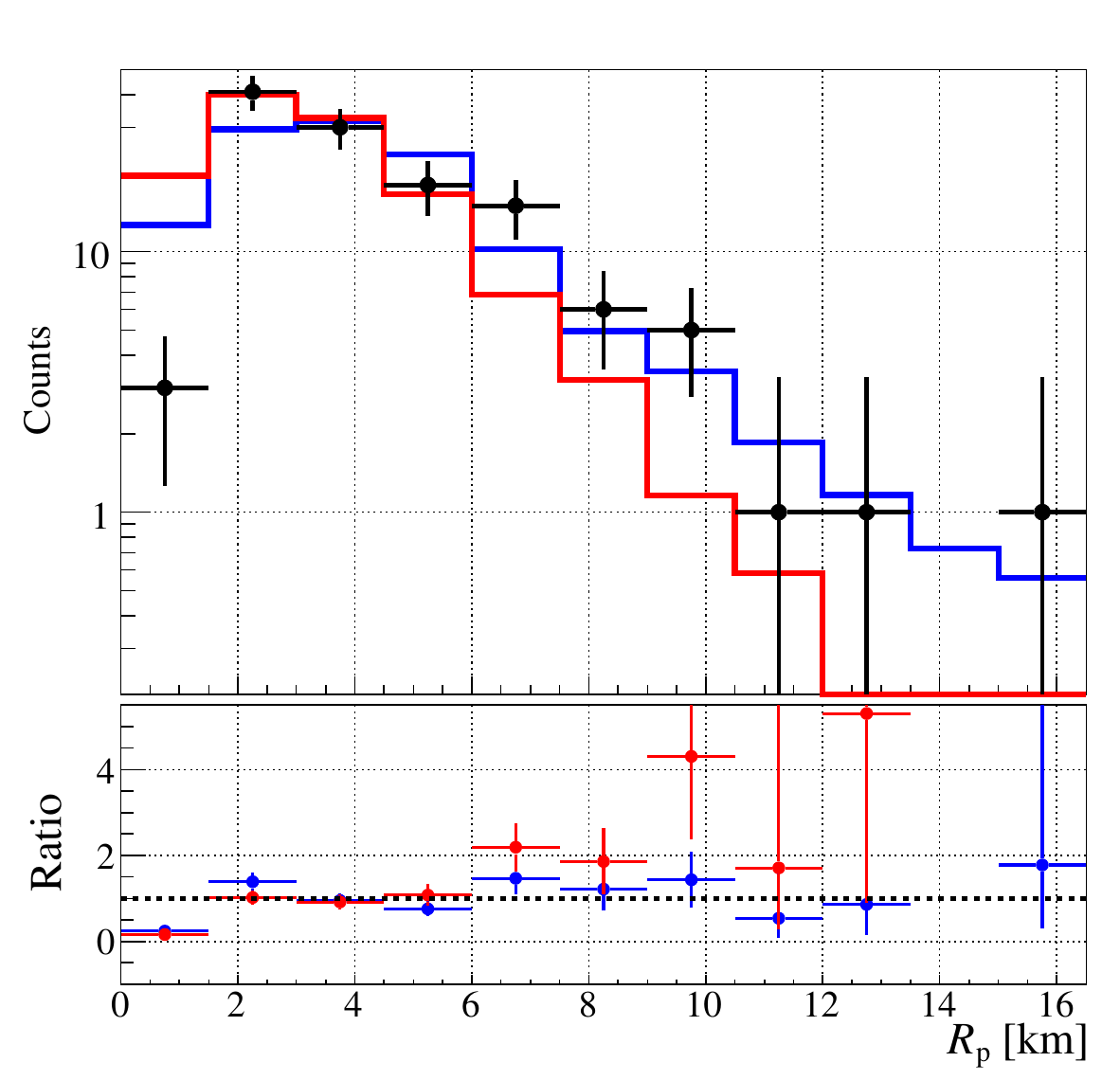}
    \includegraphics[width=0.49\linewidth]{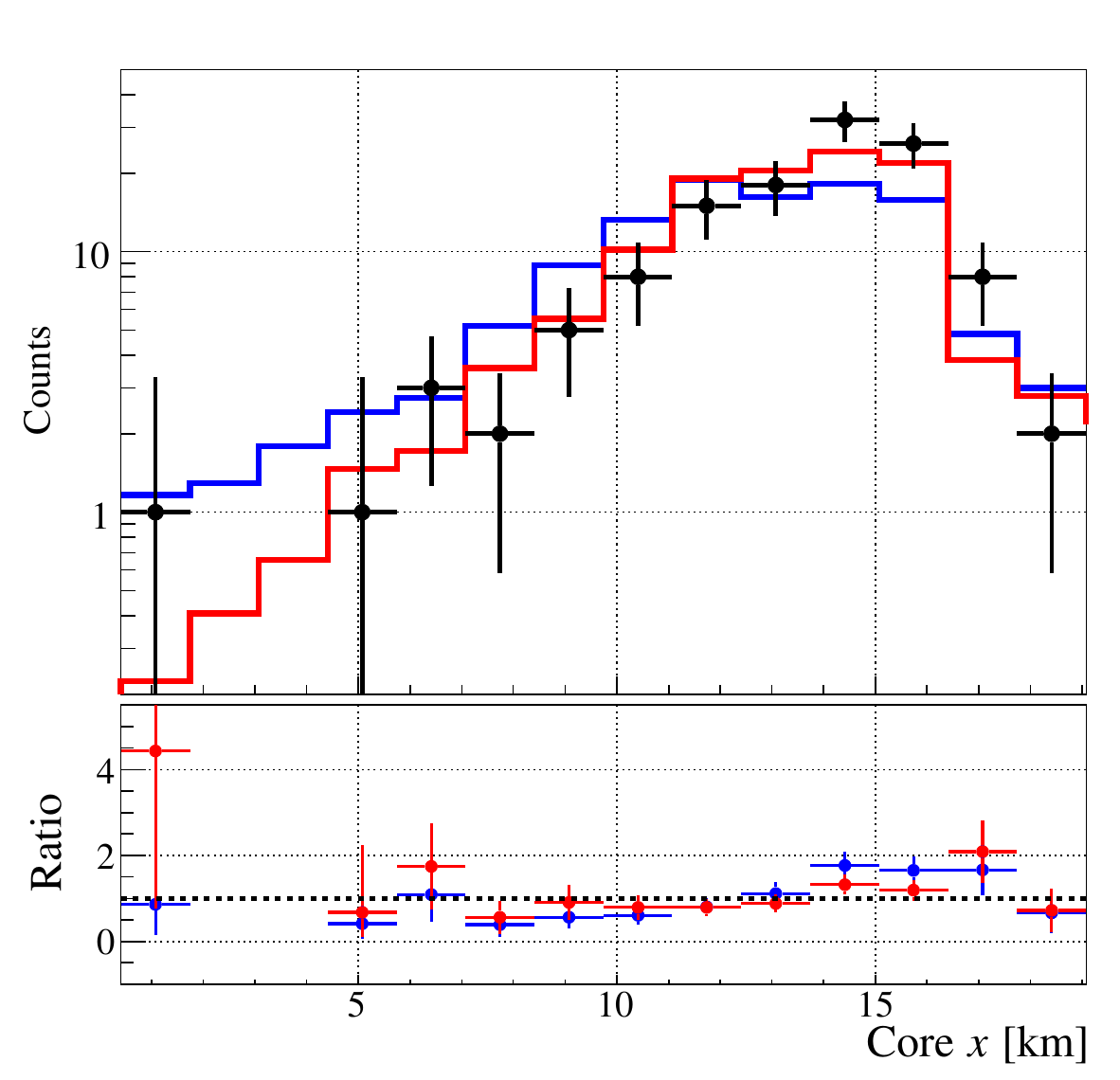}
    \includegraphics[width=0.49\linewidth]{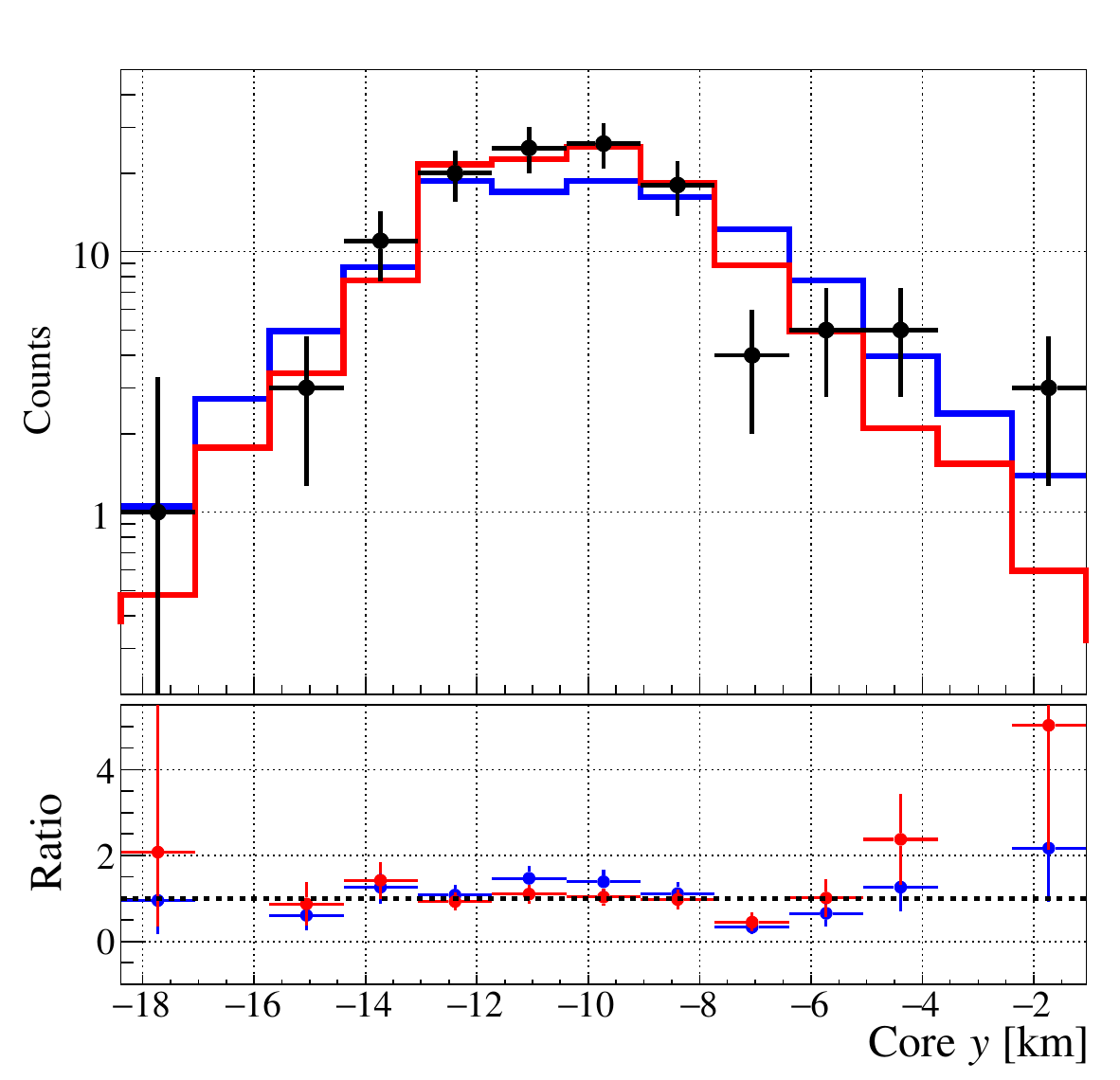}
    \includegraphics[width=0.49\linewidth]{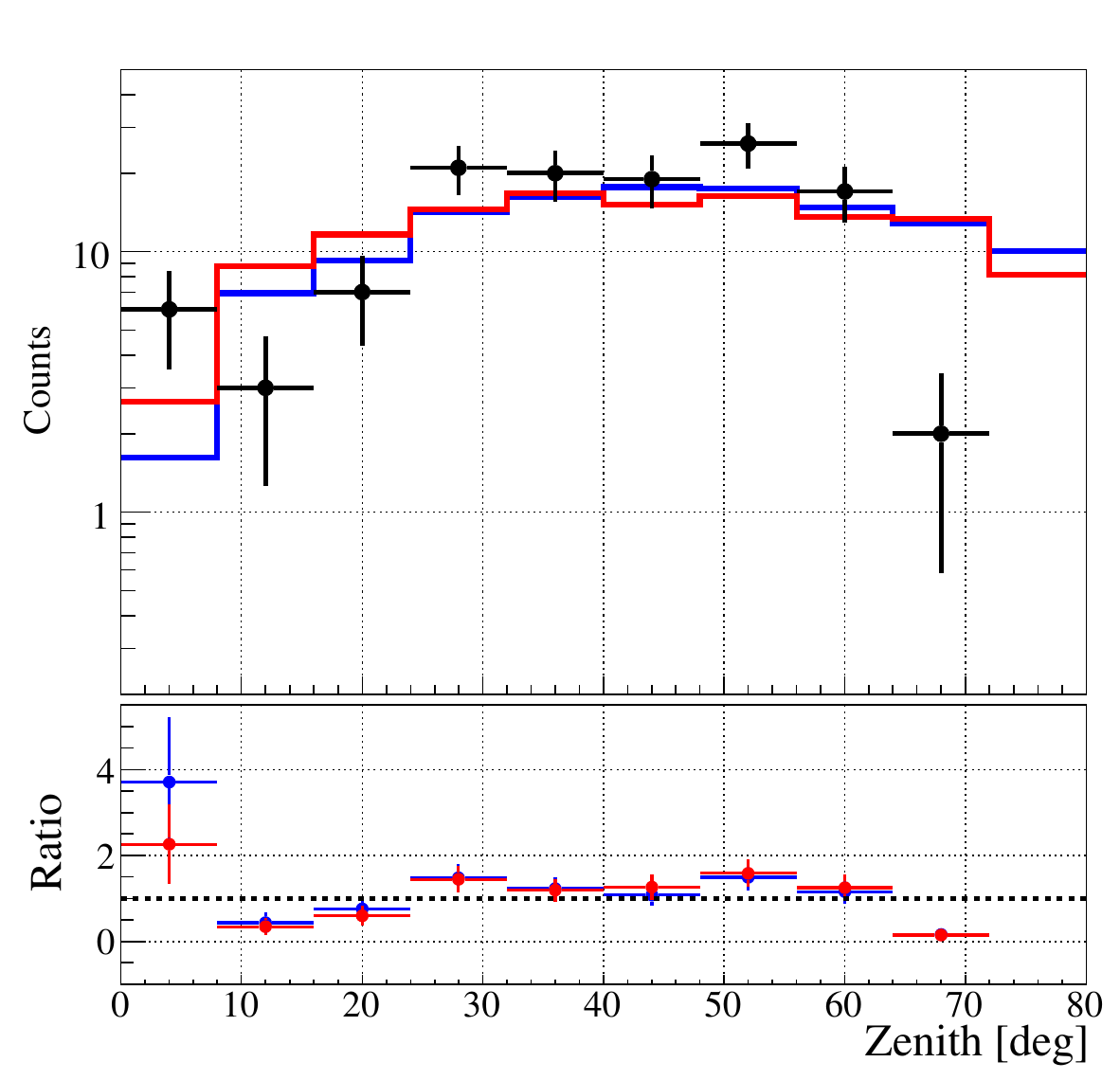}
    \includegraphics[width=0.49\linewidth]{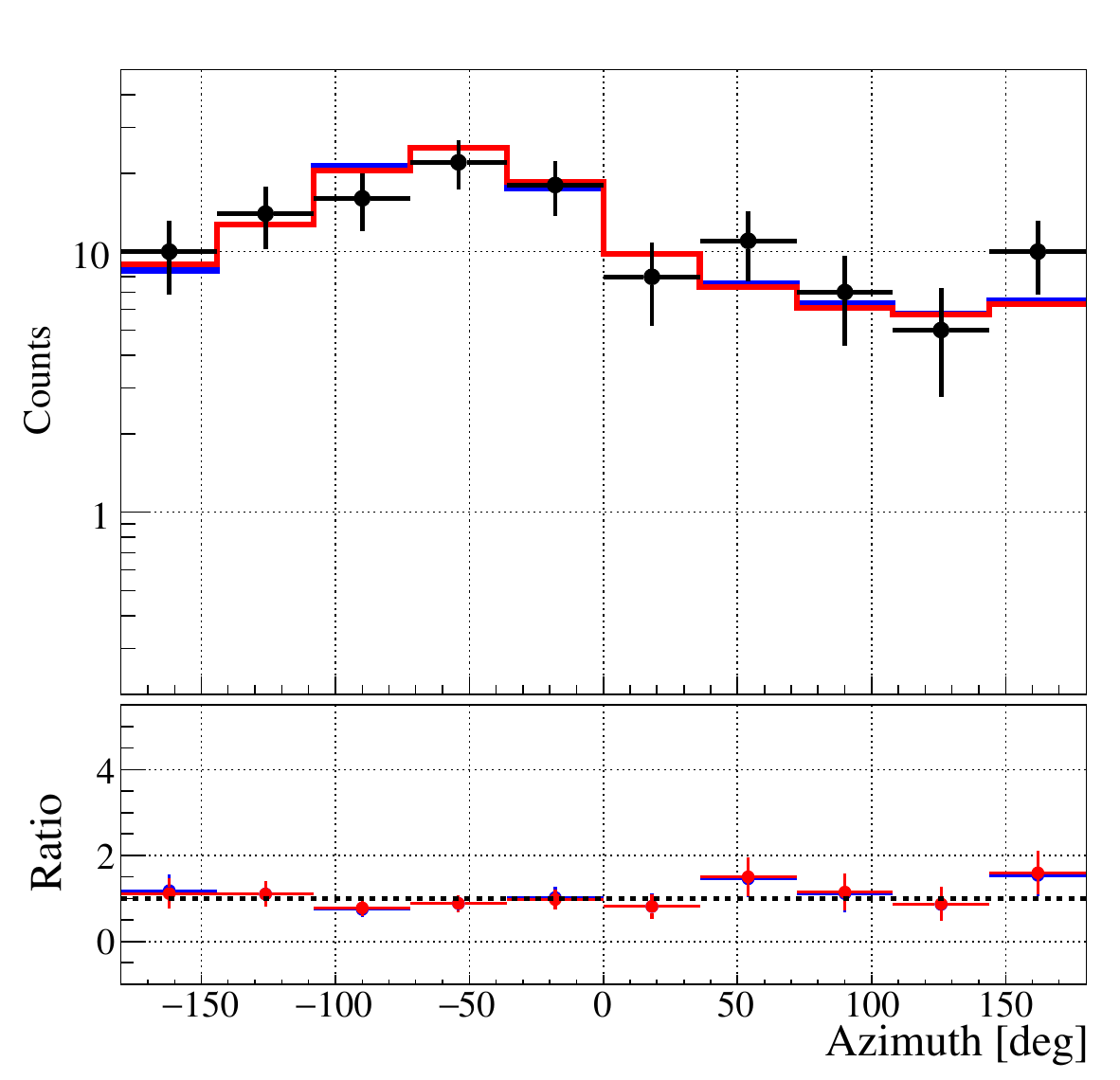}
    \caption{Comparison of coincidence event parameters (black points) as reconstructed by TA with expected distributions from FAST simulations using two different trigger conditions (blue/red histograms). See the text for details.
    }
    \label{fig:originalDataMCcomparisonTA}
\end{figure}

\begin{figure}
    \centering
    \includegraphics[width=0.49\linewidth]{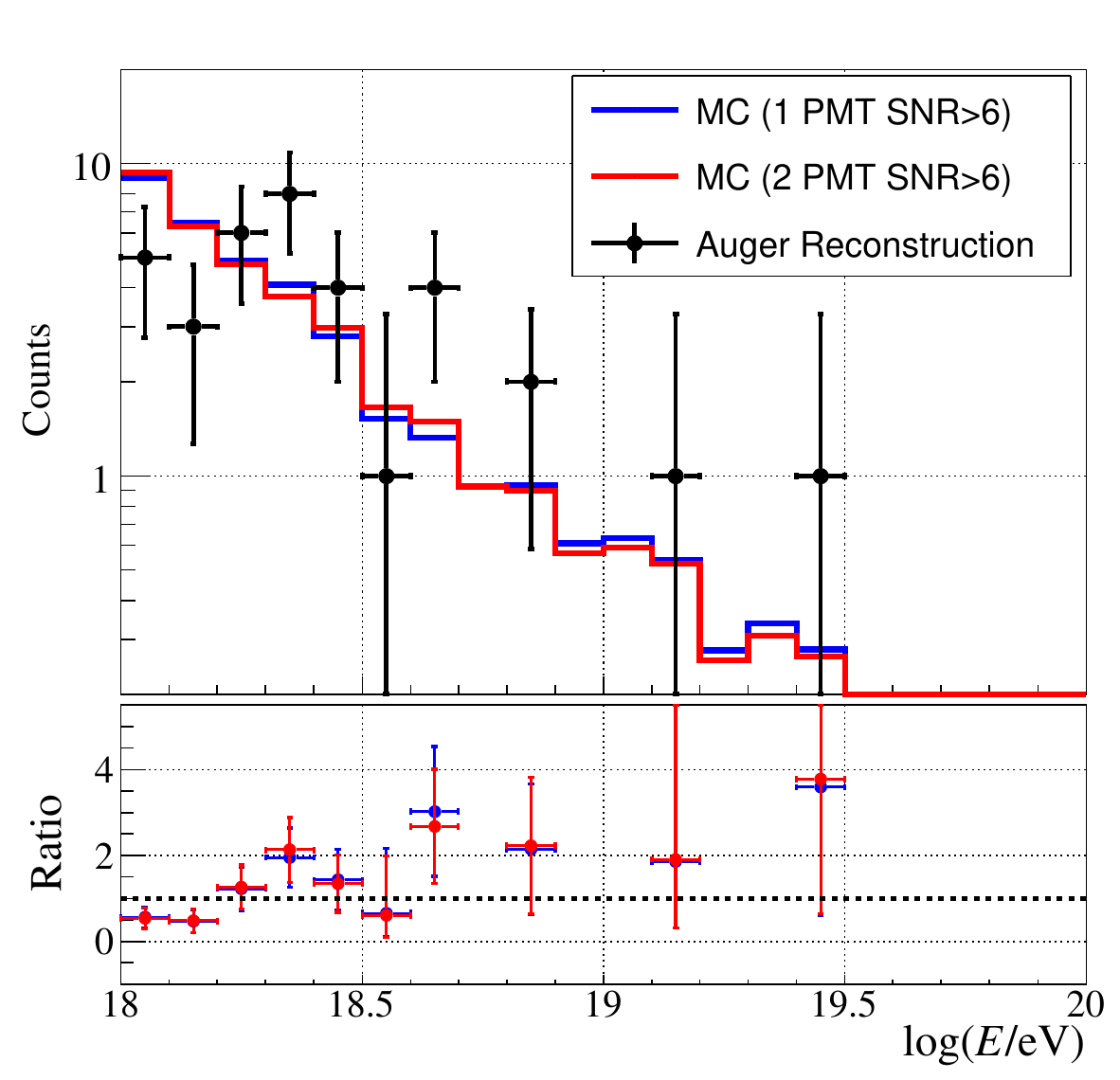}
    \includegraphics[width=0.49\linewidth]{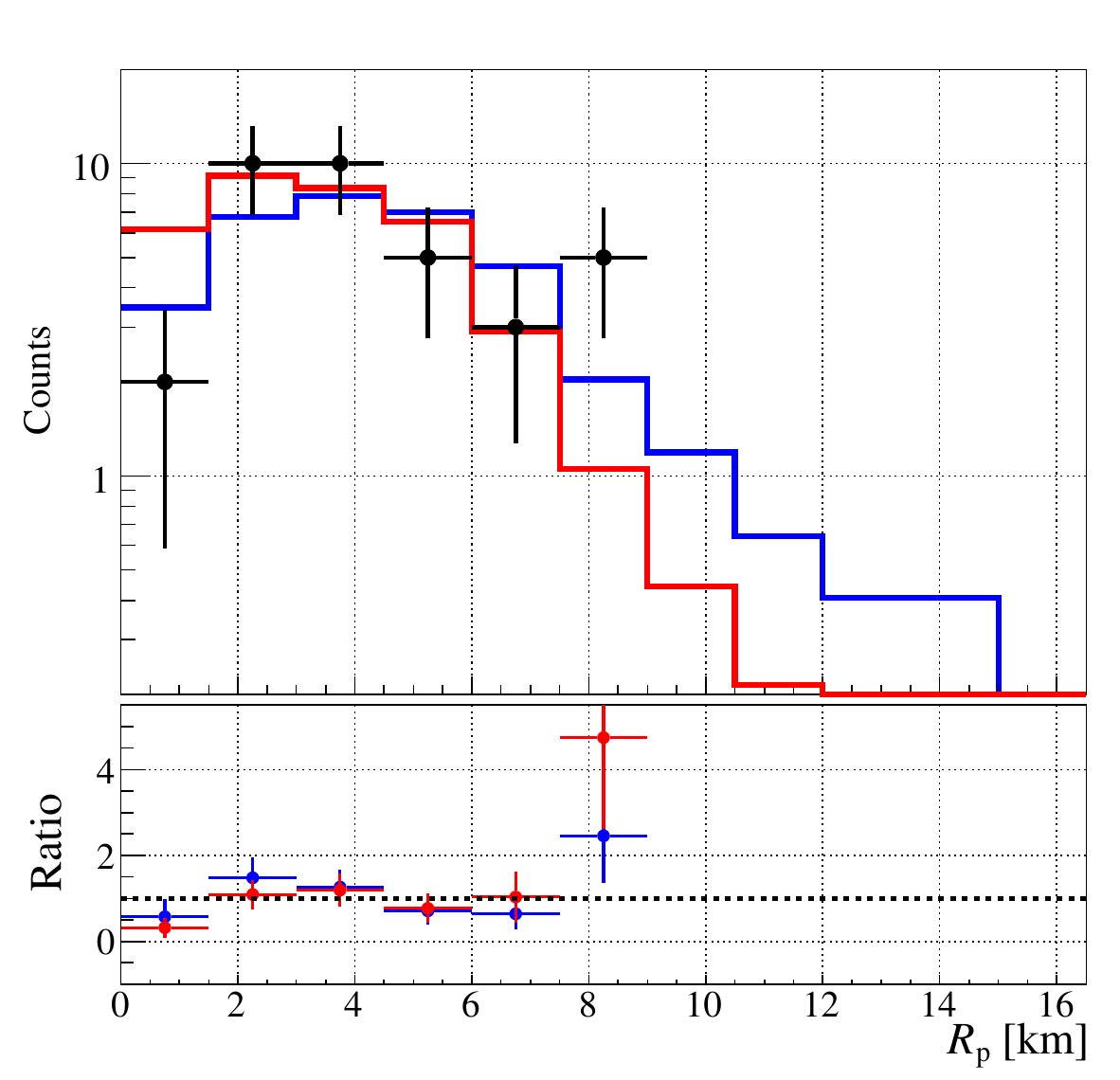}
    \includegraphics[width=0.49\linewidth]{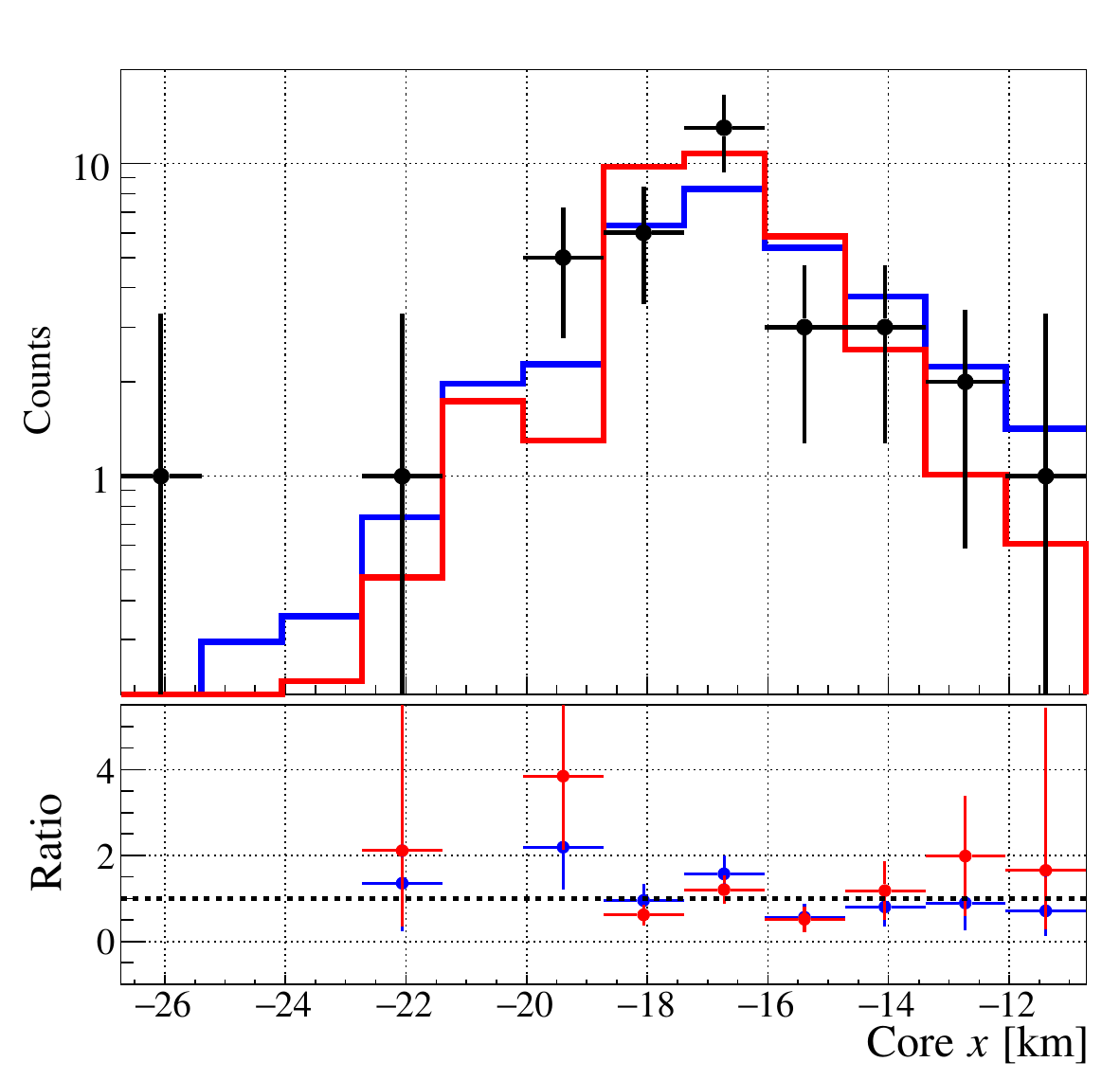}
    \includegraphics[width=0.49\linewidth]{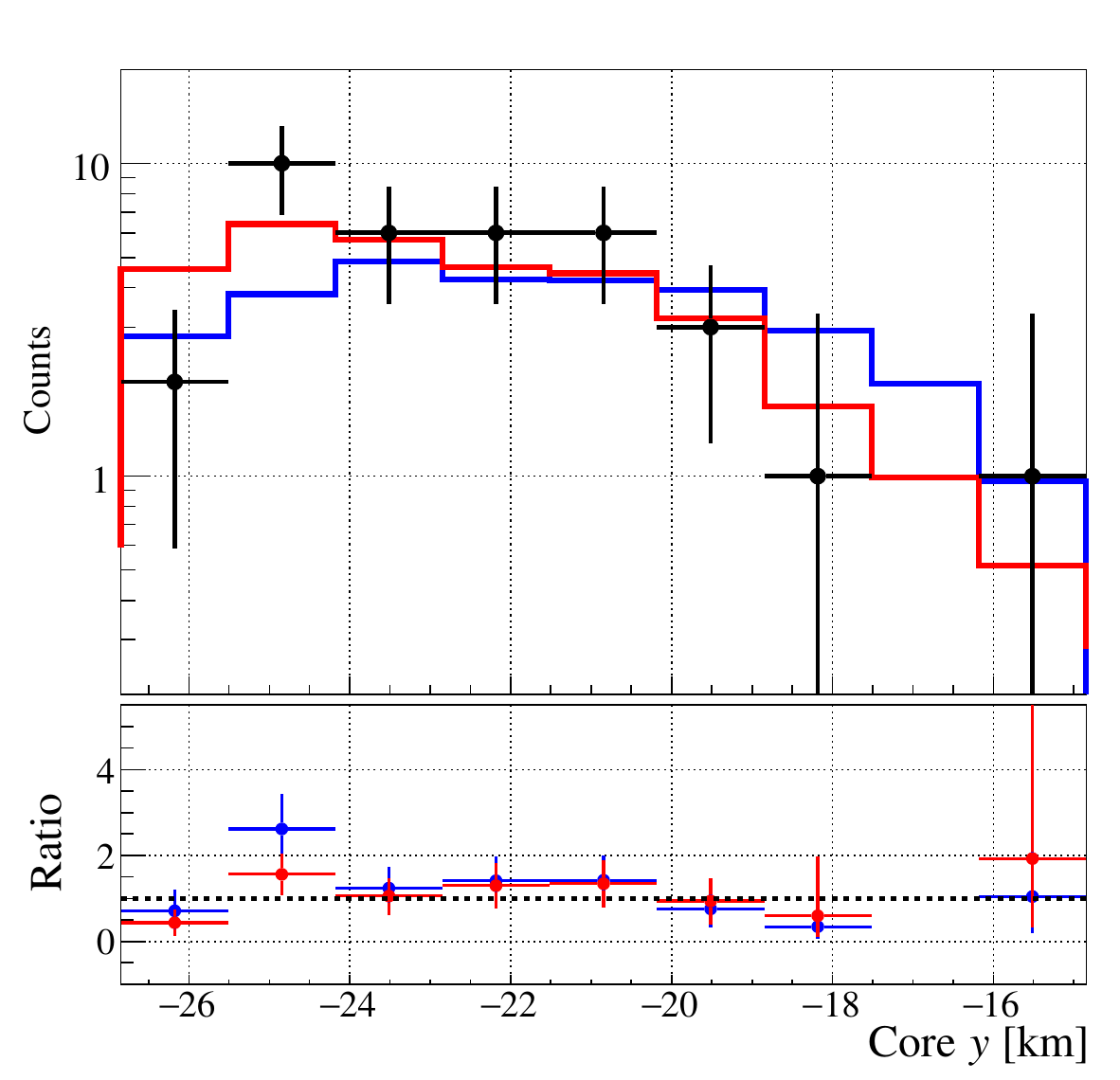}
    \includegraphics[width=0.49\linewidth]{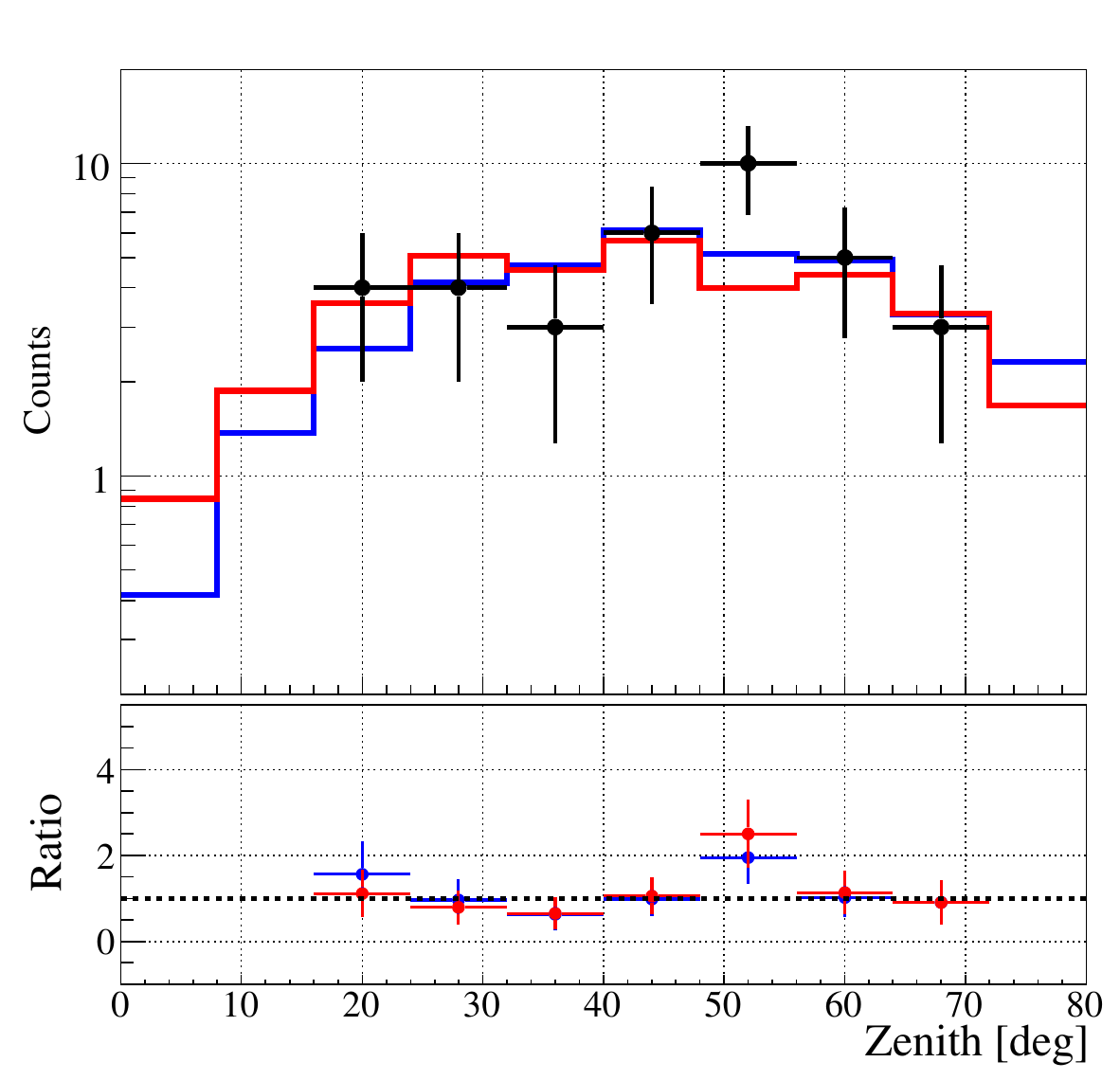}
    \includegraphics[width=0.49\linewidth]{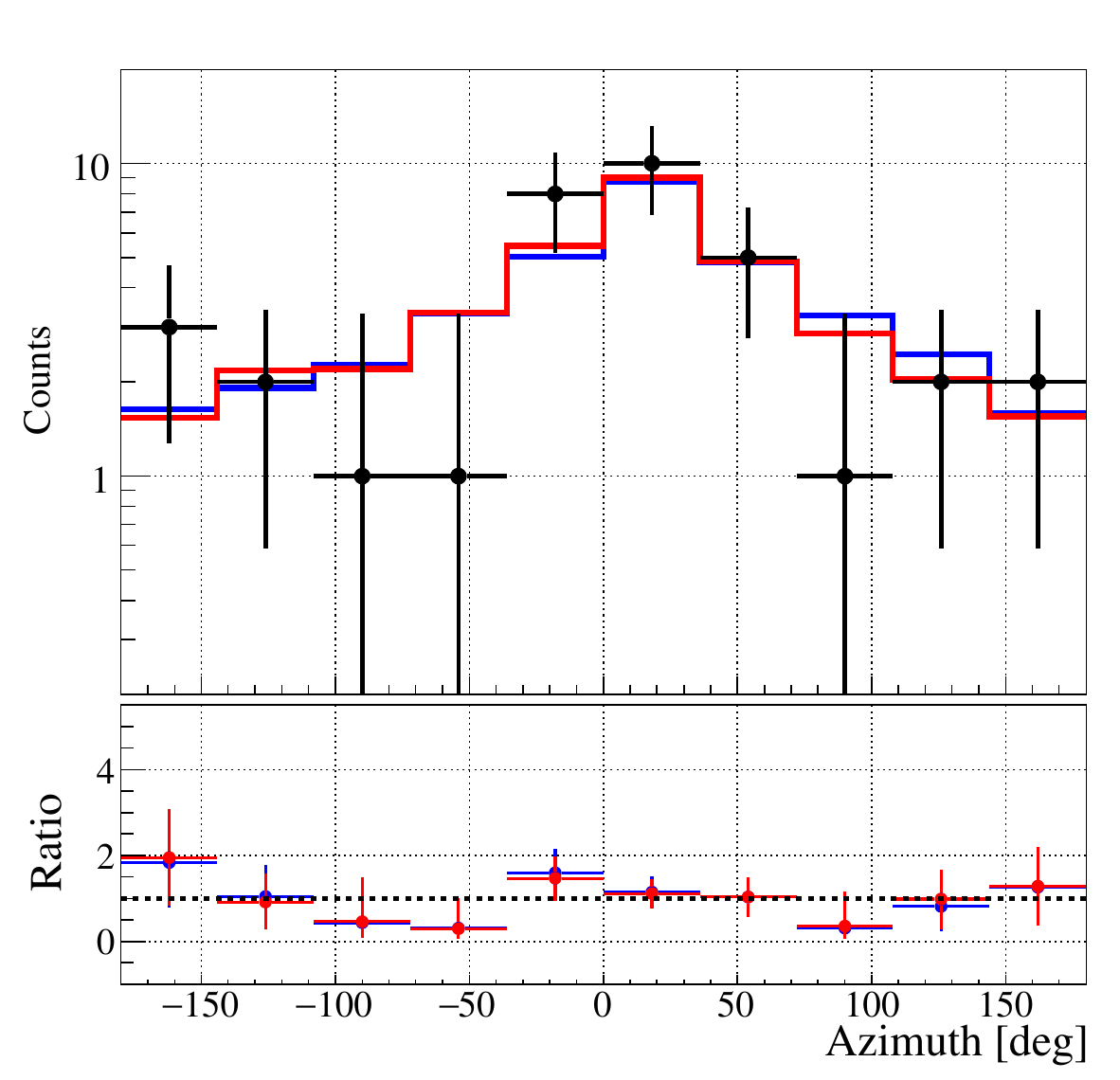}
    \caption{Same as Figure \ref{fig:originalDataMCcomparisonTA} but for Auger/FAST@Auger.}
    \label{fig:originalDataMCcomparisonAuger}
\end{figure}

Figures \ref{fig:originalDataMCcomparisonTA} and \ref{fig:originalDataMCcomparisonAuger} show the data/MC comparisons for FAST@TA and FAST@Auger. The parameters as estimated by the TA/Auger reconstructions are shown as black points, where the error bars are calculated as the square root of the bin content. The exception to this is bins with a single entry. For these bins the error is given by the 68\% Garwood confidence interval of a Poisson distribution with $\lambda=1$ \cite{garwood1936fiducial}. The blue and red histograms show the expected distributions of each parameter when using the $\geq1$ PMT and $\geq2$ PMTs trigger conditions respectively. The ratios between the data points and predicted distributions are shown in the lower panels of each plot. The black dotted line corresponds to a ratio of one. Note that the expected distributions have been scaled to match the area of the data histograms, as at this stage the shape of the distributions is the primary concern. 

\vspace{5mm}

For FAST@TA, the energy and azimuth MC distributions do not differ noticeably between trigger conditions and match the data reasonably well. The same is mostly true for the zenith angle comparison. With the exception of the first and second to last bins, the MC distributions seem consistent with the data, though with slightly more deviation than found in the energy/azimuth plots. For $R_\textrm{p}$, requiring only one triggered PMT appears to match the \say{slope} of the data better than for the $\geq2$ PMTs condition. In both cases though the number of showers expected in the first bin overestimates the data. The core $y$ MC distribution for $\geq1$ PMT is wider than the distribution for $\geq2$ PMTs, however both trigger conditions do seem consistent with the data. Lastly, the TA reconstructed core $x$ parameter appears better explained by the $\geq2$ PMTs condition, somewhat at odds with the $R_\textrm{p}$ result. 

\vspace{5mm}

Similar patterns hold for the FAST@Auger results. Like FAST@TA, the energy, azimuth and zenith MC distributions show only minor differences and seem roughly consistent with the data, whilst the $R_\textrm{p}$ and core position results show a greater difference between MC distributions. The low statistics (36 showers above 10$^{18}$\,eV) make identifying the more appropriate trigger condition challenging. Overall, considering the uncertainty in the trigger condition, the MC distributions for both FAST@Auger and FAST@TA appear broadly consistent with the data, suggesting that the telescope is performing roughly as expected. 
A quantitative analysis on the degree of agreement will not be performed here, though one could be considered in future work, for example by evaluating the pull distributions of each parameter. This decision is made primarily due to the several layers of uncertainty in the analysis, the largest of which is how well the applied trigger condition replicates the coincidence event selection process. One option for future data/MC studies which utilise data from an external trigger could be to search for the optimal trigger condition using a quantitative metric for the comparison. Other uncertainties in the analysis include the simulated telescope response/environmental conditions, quality of the TA/Auger reconstructions and the hadronic interaction model assumed for the \Xmax{} distributions. Additional observations and further efforts into understanding the time-dependent telescope performance + incorporating these measurements into the FAST simulation will be needed before future data/MC analyses may begin to reveal useful discrepancies, if any. Future data/MC checks should also incorporate the low level variables shown in Figure \ref{fig:fundamentalPlots}. This will aid in pin-pointing the source of any significant differences found.

\section{Initial Reconstruction Results}
\label{sec:initRecon}
The initial reconstruction of the coincidence events was performed with the TDR using the reconstructed values from TA/Auger as the first guess. All six shower parameters were reconstructed simultaneously and the absolute time offset fitted. The US Standard Atmosphere and monthly average atmospheres at Malargüe were used for the FAST@TA and FAST@Auger reconstructions respectively.
The cuts applied on the reconstruction results were 
\begin{itemize}
    \item Successful minimisation,
    \item Best fit time offset lay within an expected time window based on the known delay from the external triggers (100$<$$t_\textrm{off}$$<$500),
    \item The reconstructed \Xmax{} lay in the FOV of at least one of the triggered telescopes (using reconstructed shower geometry),
    \item The relative error in both the reconstructed \Xmax{} and reconstructed energy was $<0.5$.
\end{itemize} 
Table \ref{tab:cutsTable} shows the number of events remaining from both datasets after each cut. The \Xmax{} in FOV cut reduces the statistics greatly, by a factor of 2 for FAST@Auger and a factor of 4 for FAST@TA. Why the percentage of showers remaining after the cut differs so significantly between Auger and TA is unknown at this time. It may be related to the cuts each experiment places on their respective events. Poor geometry reconstructions may also impact the effectiveness of this cut, however the difference is not expected to be significant here as both the FAST@TA and FAST@Auger reconstructed geometries show reasonable agreement with the TA/Auger values (see Appendix \ref{apx:NEWSIMrecResults}).

\begin{table}[h]
    \centering
    \begin{tabular}{lcc}
        \toprule
        \textbf{Cut} & \multicolumn{2}{c}{\textbf{Events Remaining}} \\
        \cmidrule(lr){2-3}
                     & \textbf{TA} & \textbf{Auger} \\
        \midrule
        No cuts      &     438      &    235    \\
        Minimisation       &     412      &    215    \\
        Time Offset        &      405     &    211    \\
        \Xmax{} in FOV  &     103      &    116    \\
        Relative Error     &     86      &    86    \\
        \bottomrule
    \end{tabular}
    \caption{Number of events remaining after each cut applied to the reconstructed data from FAST@TA and FAST@Auger.}
    \label{tab:cutsTable}
\end{table}

\begin{figure}
    \centering
    \includegraphics[width=1\linewidth]{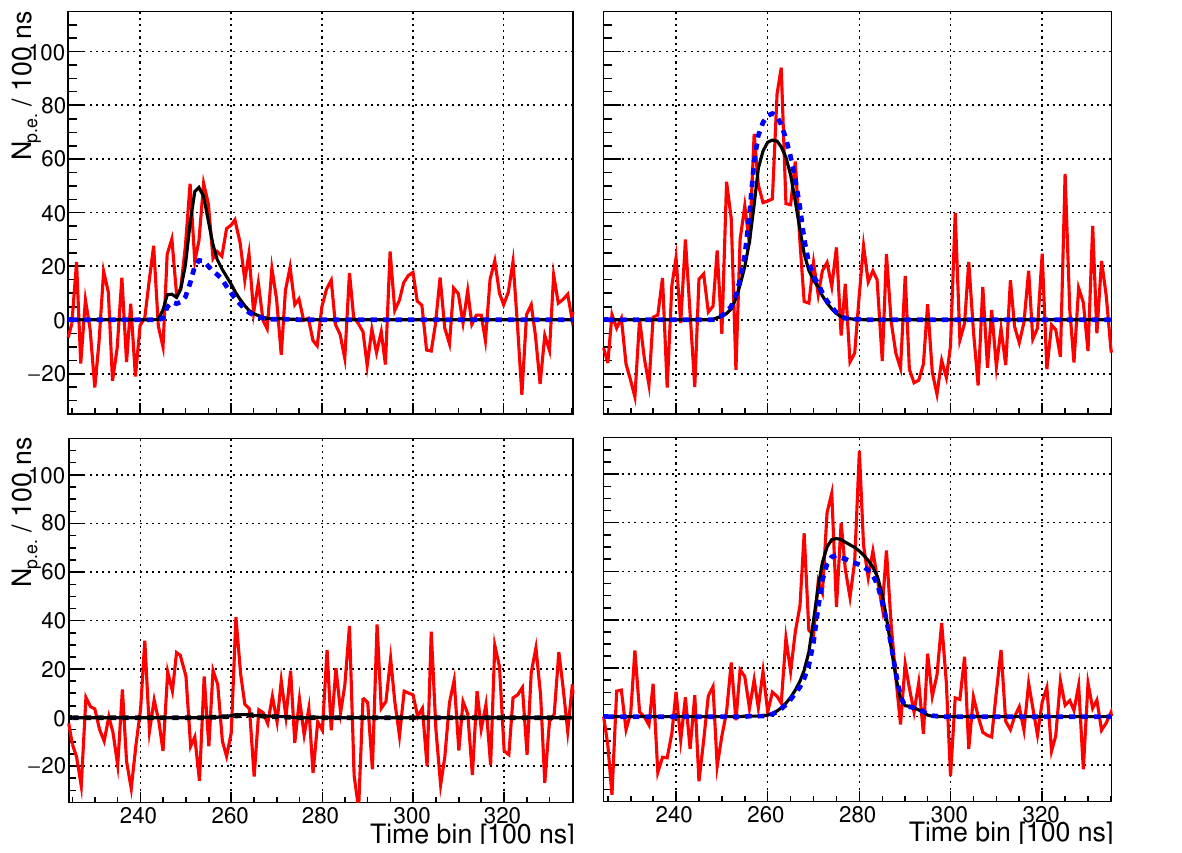}
    \includegraphics[width=1\linewidth]{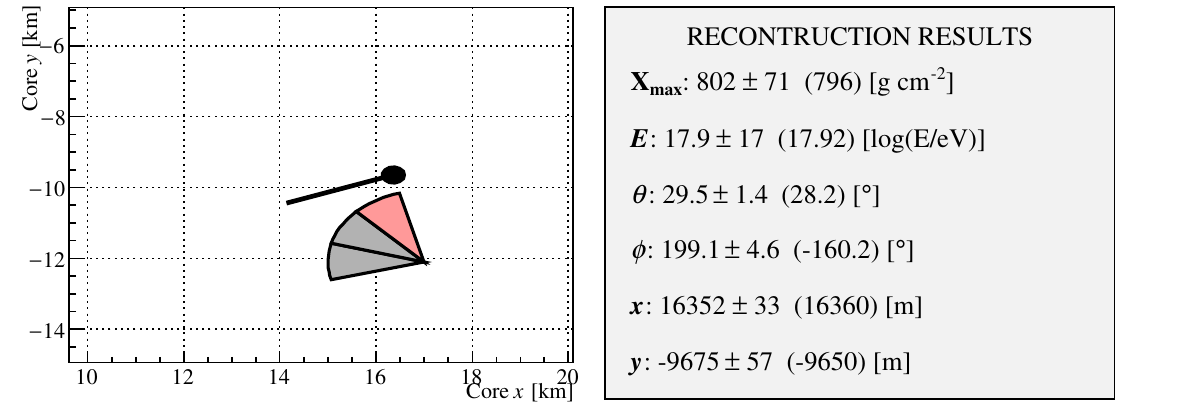}
    \caption{Example of the TDR applied to a coincidence event detected by FAST@TA on 2018/11/05. The blue dotted line shows the simulated trace according to the first guess from TA. The black line is the best fit trace found by the TDR. The geometry of the best fit event in relation to the FAST telescopes is shown in the bottom left panel with the triggered telescope highlighted. The reconstruction results are shown in the bottom right panel. The TA values are indicated in brackets.}
    \label{fig:taExampleRec}
\end{figure}

\begin{figure}
    \centering
    \includegraphics[width=1\linewidth]{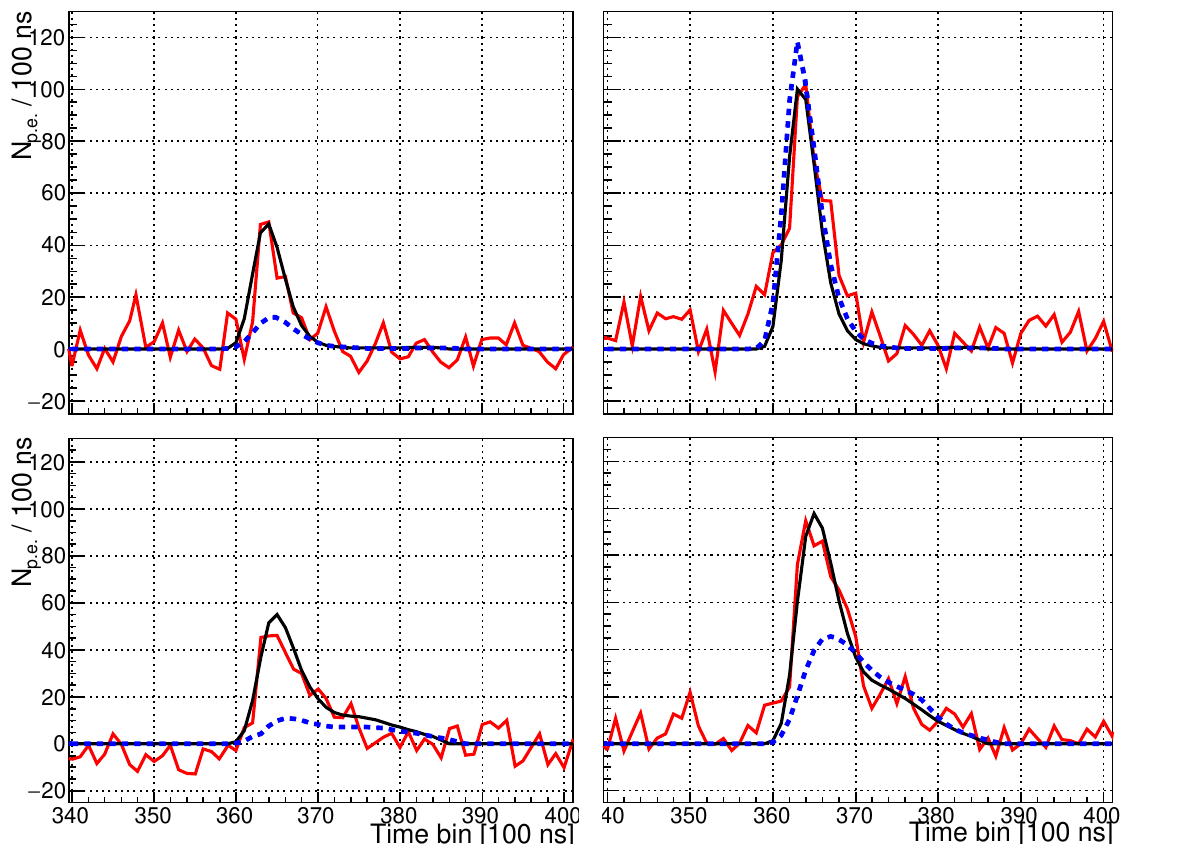}
    \includegraphics[width=1\linewidth]{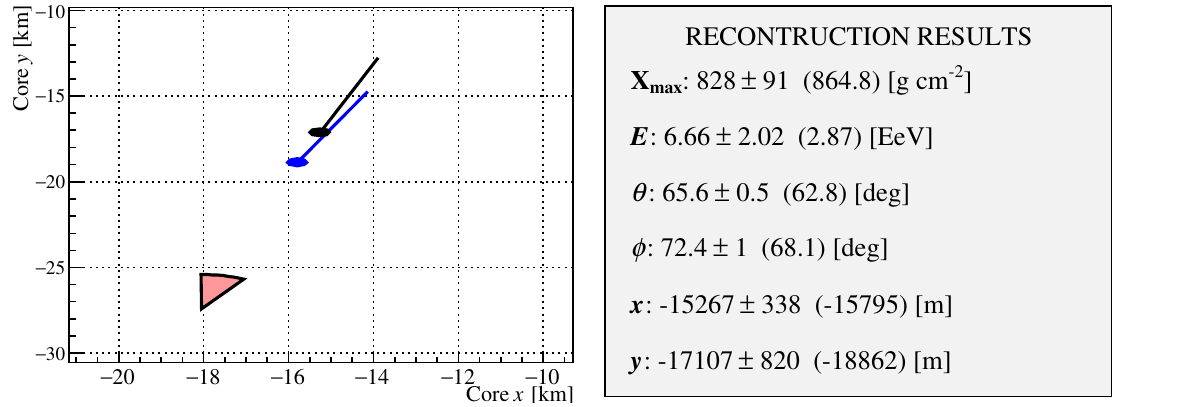}
    \caption{Example of the TDR applied to a coincidence event detected by FAST@Auger on 2022/07/31. The plot layout is the same as in Figure \ref{fig:taExampleRec}. The bottom left panel now shows the geometry of both the first guess (here the Auger reconstructed parameters) and the FAST reconstructed result in blue and black respectively. This matches the colour code used in the trace plots.}
    \label{fig:augerExampleRec}
\end{figure}

Figures \ref{fig:taExampleRec} and \ref{fig:augerExampleRec} show examples of two reconstructions, one for FAST@TA and one for FAST@Auger. In both cases the best-fit trace (black) appears to match the data reasonably well, with only minor changes in the shape and scale from the TA/Auger first guess. The FAST@TA example is an example of a fluorescence dominated event. The TA results lie within the uncertainty bounds on the best fit parameters. The FAST@Auger example showcases a Cherenkov dominated event. Here there is slightly more discrepancy, with only the Auger reconstructed \Xmax{} being within 1$\sigma$ of the best fit values. The difference can likely be attributed to some uncertainty in the Auger measurements and the FAST simulation not reproducing the conditions at FAST@Auger 100\% accurately. The strongly-forward-beamed nature of Cherenkov light means that any small differences between the on-site telescopes and simulations will be amplified. Visually inspecting numerous other fits generally showed good matching between the best-fit traces and data.

\vspace{5mm}

\begin{figure}[t]
    \centering
    \includegraphics[width=0.49\linewidth]{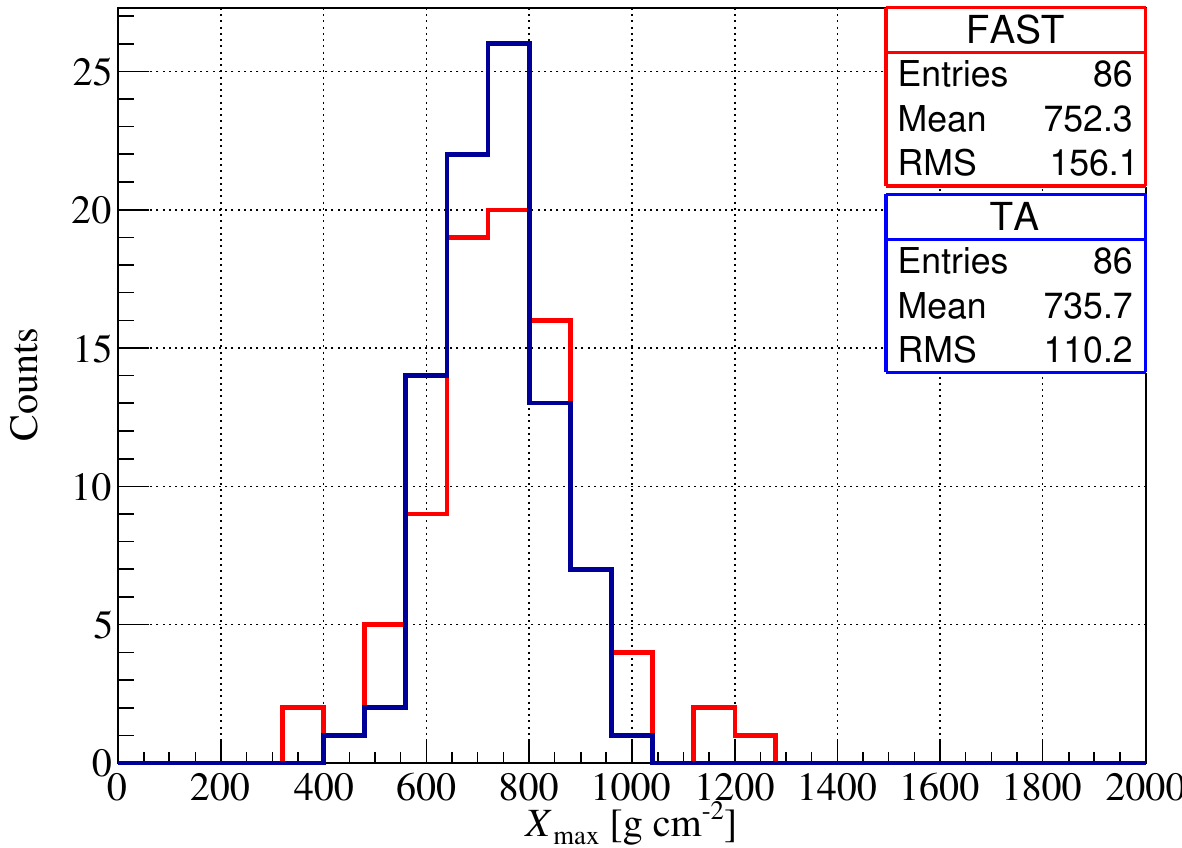}
    \includegraphics[width=0.49\linewidth]{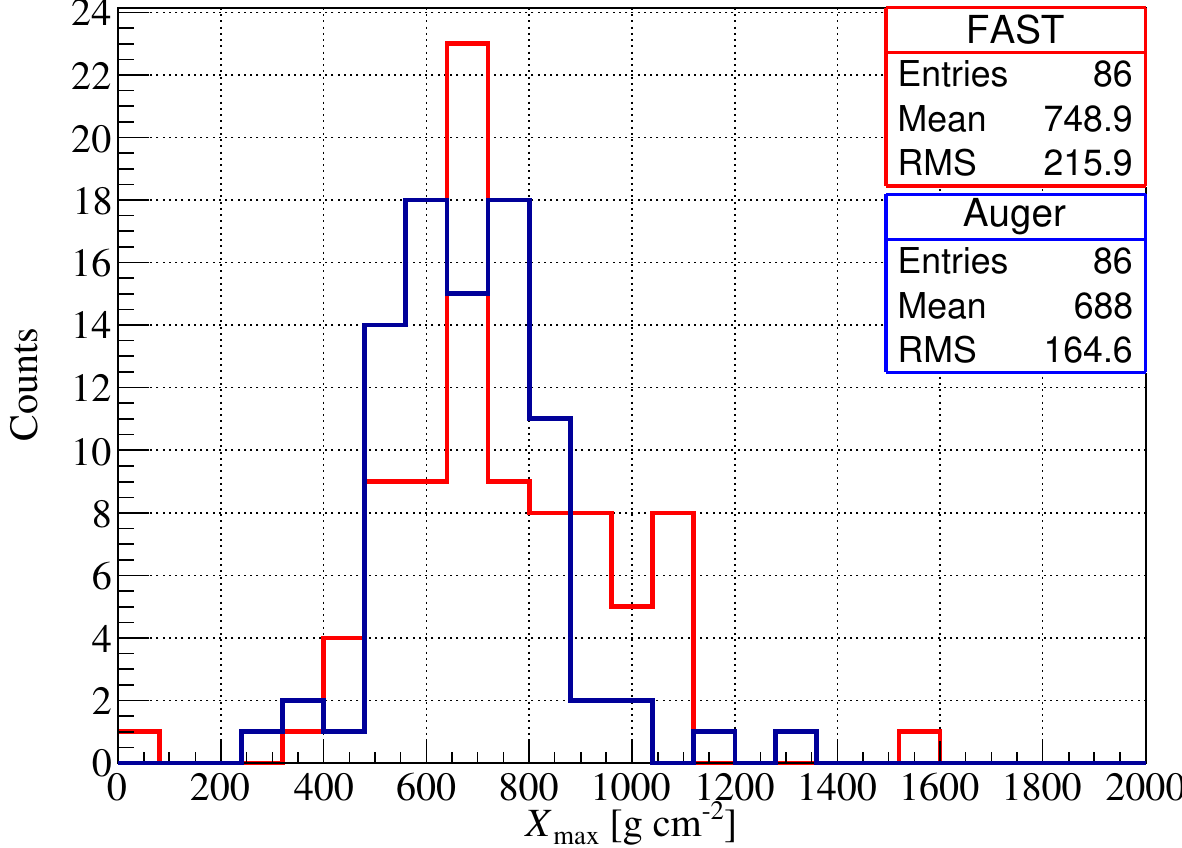}
    \includegraphics[width=0.49\linewidth]{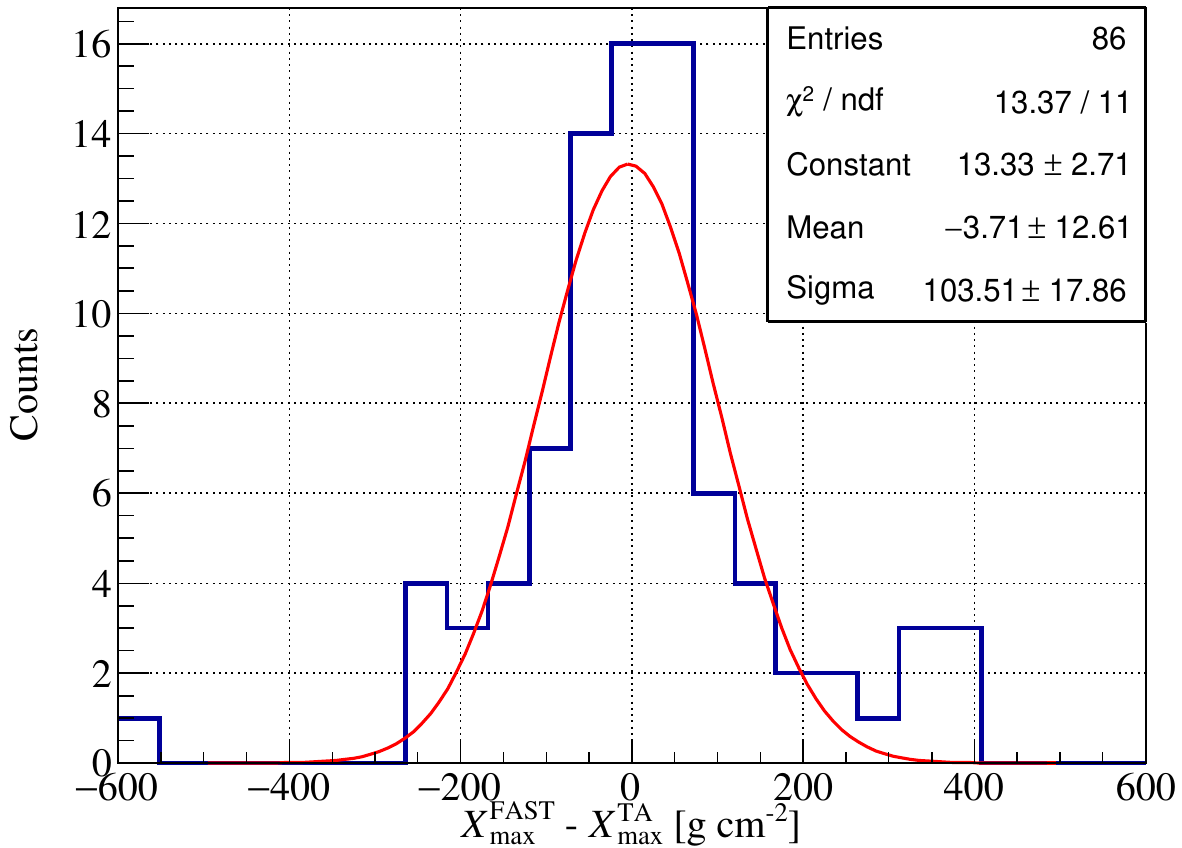}
    \includegraphics[width=0.49\linewidth]{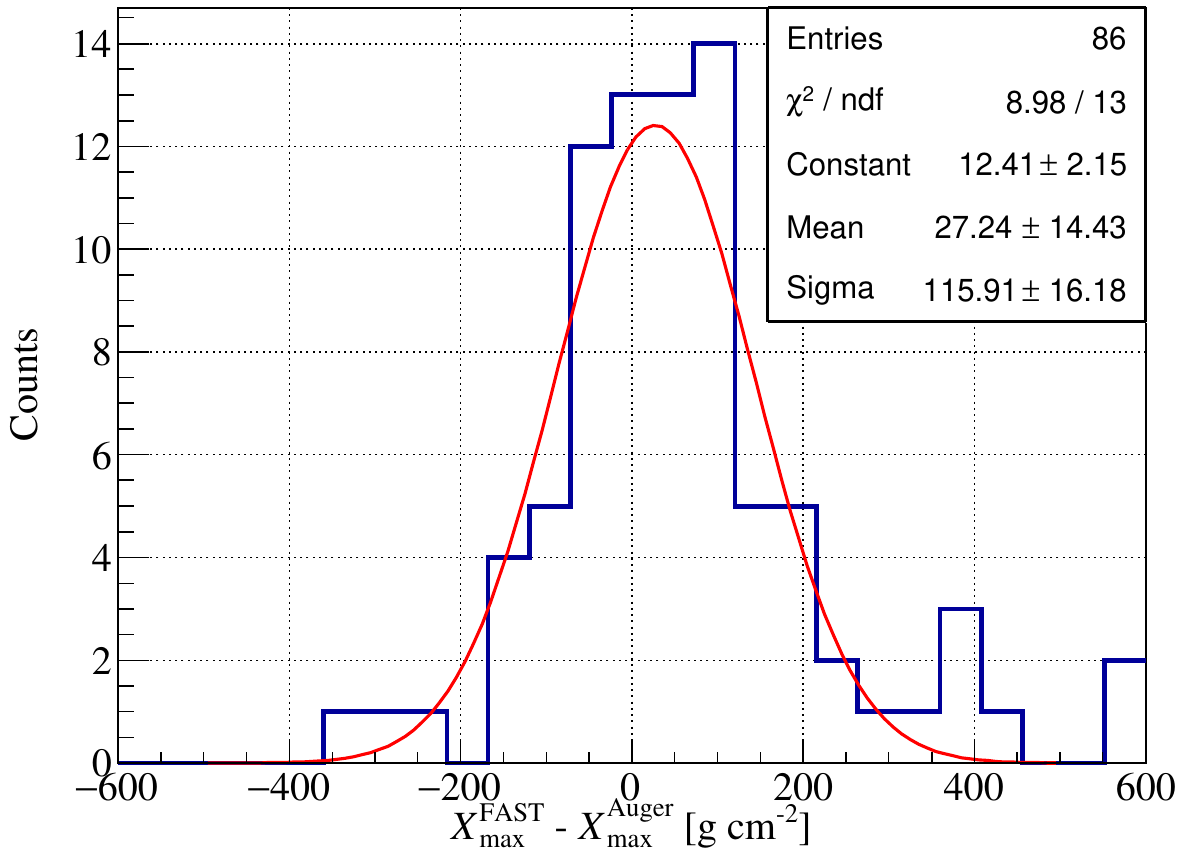}
    \caption{Histograms of the reconstructed \Xmax{} values from FAST@TA (left) and FAST@Auger (right). The FAST results are shown in red. The TA/Auger results are shown in blue. The event by event difference histograms are shown in the bottom panels.}
    \label{fig:recResultCompareXmax}
\end{figure}

\begin{figure}[t]
    \centering
    \includegraphics[width=0.49\linewidth]{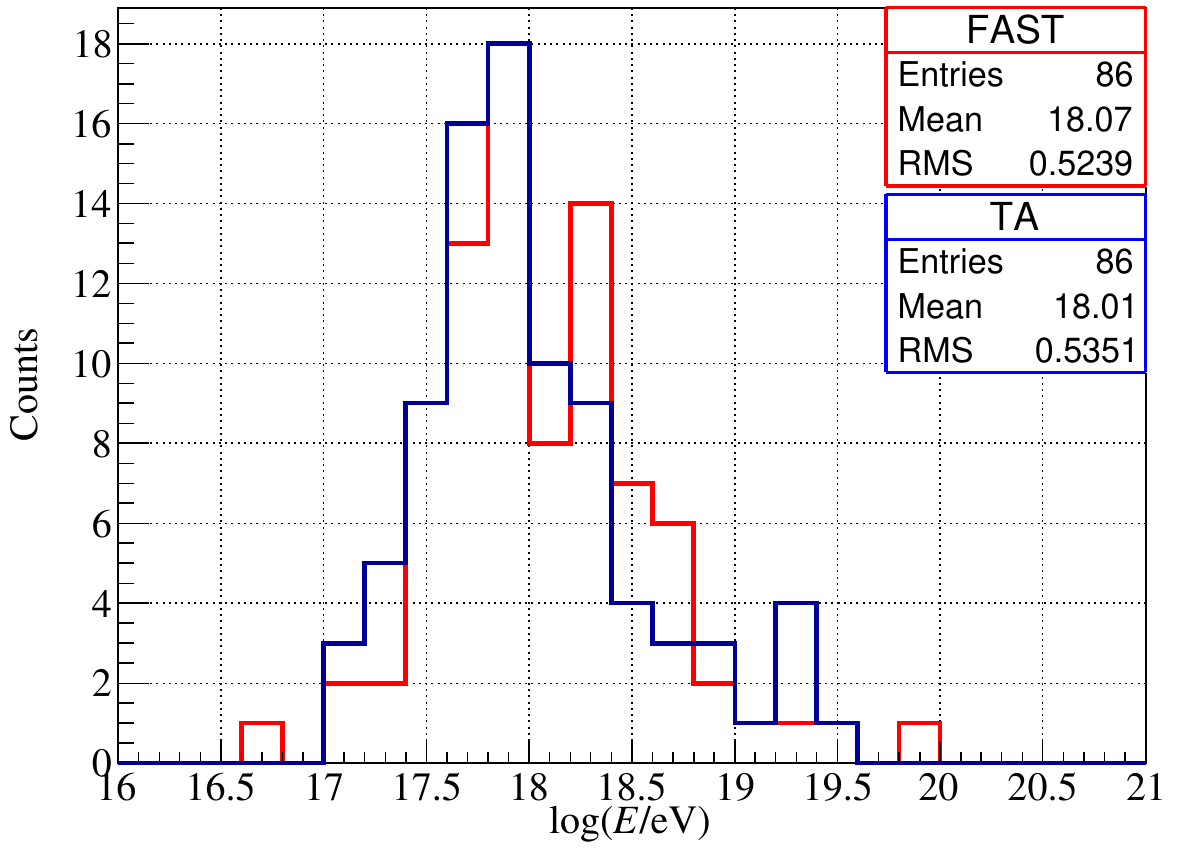}
    \includegraphics[width=0.49\linewidth]{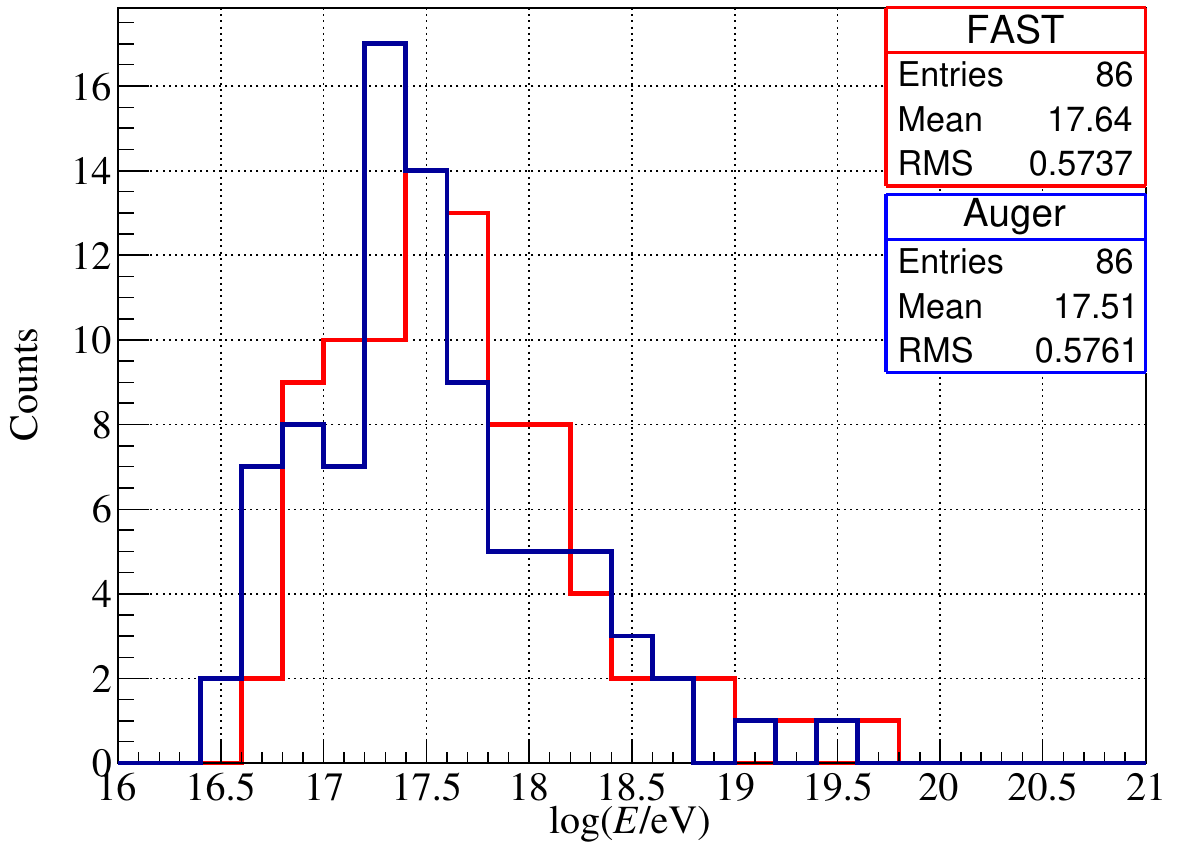}
    \includegraphics[width=0.49\linewidth]{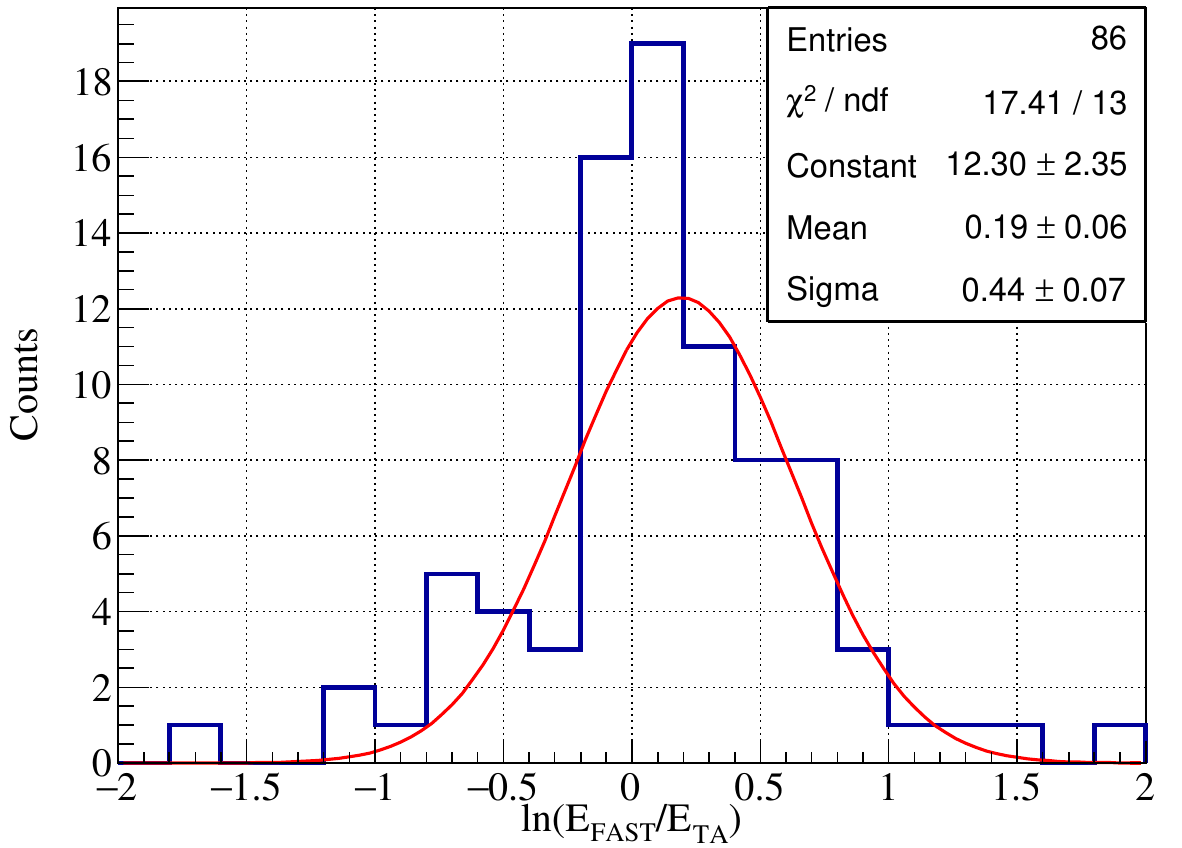}
    \includegraphics[width=0.49\linewidth]{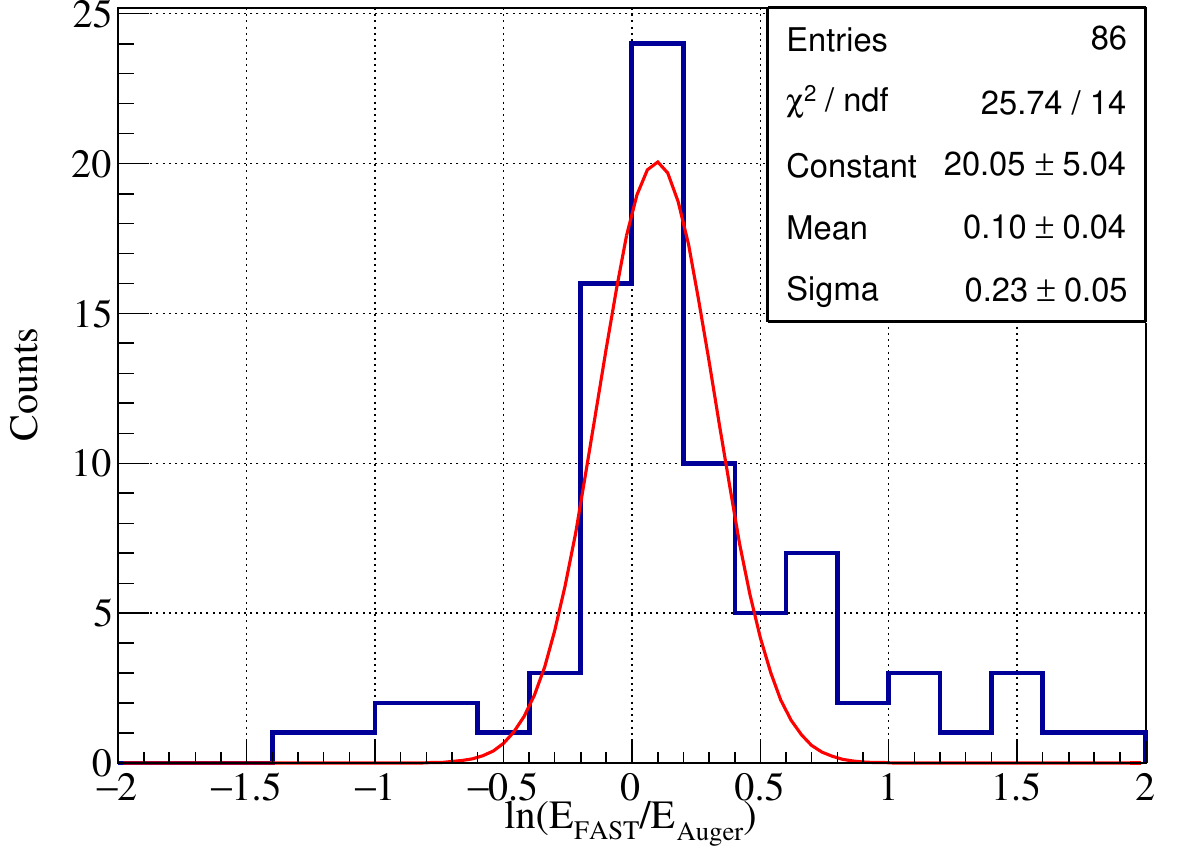}
    \caption{Same as Figure \ref{fig:recResultCompareXmax} but for energy.}
    \label{fig:recResultCompareEnergy}
\end{figure}

Histograms showing the distributions of \Xmax{} and energy for events which passed the above cuts are shown in Figures \ref{fig:recResultCompareXmax} and \ref{fig:recResultCompareEnergy}. The FAST@TA and FAST@Auger results are shown in red and the TA/Auger results in blue. The event-by-event differences are also shown, with Gaussian fits performed to estimate the mean and standard deviation of the distributions. Similar plots for the geometrical parameters can be found in Appendix \ref{apx:NEWSIMrecResults}. 
For FAST@TA, the \Xmax{} reconstruction shows almost no bias compared to the TA result, with an average difference of just $-3$\gcm{}. There is however a somewhat large spread in the event-by-event differences of $\sigma(\Delta{}$\Xmax{}$)\approx100$\gcm{}. The energy reconstruction shows a shift towards larger values for FAST by roughly 20\%, again with a wide spread in the event-by-event differences of $\sigma(\Delta{}E)\approx0.44$. The FAST@Auger results show biases in both the \Xmax{} and energy reconstructions, approximately +25\gcm{} and +10\% respectively. The widths of the event-by-event distributions are roughly 115\gcm{} for \Xmax{} and 0.23 for energy. In both the FAST@TA and FAST@Auger results some reconstructions are very far from the TA/Auger values i.e. $\Delta$\Xmax{}$>400$\gcm{} or $\ln(E_\textrm{rec}/E_\textrm{(TA/Auger)})\approx2$ (seven times the energy).

\vspace{5mm}

Contributions to the width of the distributions include (but are not limited to) the inherent uncertainties in the FAST reconstruction, primarily due to the large background noise, baseline fluctuations and strong dependence on the first guess, and the uncertainties in the TA and Auger monocular reconstructions, which can be conservatively estimated to be on the order of $\sim50$\gcm{} and $\sim15\%$ at these energies \cite{abbasi2016energy}. The degeneracy of the FAST reconstruction for small signals in particular could potentially be causing the outlier events highlighted above. In principle such degeneracy is encoded in the parameter uncertainties estimated by the TDR. Additional quality cuts which incorporate the total amount of signal, number of triggered PMTs, all parameter uncertainties and possibly another method for quantifying the degeneracy in the reconstructed parameters (e.g. something similar to the Template Method) may improve the overall quality of the reconstructed events and remove the observed outliers. As the statistics are already very limited, such cuts are not implemented here, but could be considered in future work.

\vspace{5mm}

The FAST reconstruction of the shower geometry generally agrees well with TA/Auger. The $\Delta{}\theta$ distributions have small positive biases of 1 - 2$\degree$ and $\sigma(\Delta\theta)\approx2.5\sim5\degree$, whilst the difference in reconstructed azimuth shows no major biases. The core position parameters also show no overall bias with the exception of slightly larger core $y$ values for FAST@Auger ($\sim+150$\,m). This is likely related to the energy-core position degeneracy identified in Section \ref{sec:basicDNNperformance} and Appendix \ref{fig:coreyEdegen}. There are also a handful of events with best fit parameters quite different from the Auger/TA values i.e. $\Delta{}x$/$\Delta{}y>1\sim2$\,km or $\Delta\theta\sim20\degree$. The overall level of agreement here is another positive sign that the FAST simulation's model of air-shower physics is valid and that it roughly reproduces the responses of the telescopes. However, even if the FAST simulation were to be reproducing the observational conditions 100\% accurately, the performance of the reconstruction in this case is likely attributable to using the TA/Auger results as a first guess. Reconstructing the coincidence events solely from FAST data is expected to yield a worse overall comparison, particularly considering the majority of events have small signals and only trigger 1 - 2 PMTs. Such events have been shown by both first guess methods investigated to be challenging to reconstruct.

 \vspace{5mm}
 
Lastly, commenting again on the observed \Xmax{} and energy biases, although the FAST@TA \Xmax{} bias is consistent with zero (allowing for a 1$\sigma$ shift based on the estimated parameter errors), the FAST@Auger bias sits just under $2\sigma$ away from zero. The biases in energy are roughly $3\sigma$ and $2.5\sigma$ away from zero for FAST@TA and FAST@Auger respectively. Assuming that the first guess values from TA/Auger are correct, or rather contain no inherent bias, and that the FAST simulation is accurately reproducing the observation conditions, then the FAST simulated traces based on the TA/Auger values should, on average, match the data traces. The FAST reconstructed values should therefore show no bias with respect to the TA/Auger values. Note this does assume that there is limited degeneracy in the TDR. Hence, the observed biases likely point to systematic differences in either the reconstruction processes of FAST and TA/Auger or between the FAST simulation and the true performance of the prototype telescopes. One factor which is not accounted for in the current FAST reconstruction is the missing or \say{invisible} energy, $E_\textrm{inv}$, carried by the non-electromagnetic part of the shower. This fraction has been estimated by Auger and TA to vary as a function of the electromagnetic energy, $E_\textrm{cal}$, from roughly 10 - 20\% at 10$^{17}$\,eV down to 5 - 10\% at 10$^{20}$\,eV \cite{abbasi2016energy, aab2019data}. However this contribution would be added to the FAST reconstructed energies, \textit{increasing} the overall biases. Future work should incorporate the invisible energy into the reconstructed results, either through an energy dependent correction or through the use of CONEX showers for simulations which inherently account for $E_\textrm{inv}$. Another systematic effect not accounted for in the simulations is the reduction in the UV filter transmittance over time. This comes from a build up of contaminants on the filter surface and has been estimated by in-situ measurements to be roughly $-10\%$ after 2 years without cleaning (see Section \ref{sec:cleaning}). Once again though, accounting for this effect would actually increase the reconstructed energies. Aside from possible fitting degeneracies, one effect which may lower the reconstructed energies is the directional efficiency maps used in simulations vs. those of the true prototypes. Significant differences between such maps can yield large differences in the shape and scale of the traces and thus could potentially introduce a bias/additional uncertainty on both \Xmax{} and energy. There may also be errors in the absolute/relative calibration of the PMTs, and/or possibly a reduction in the PMT gain over time which is contributing to the energy biases.

\vspace{5mm}

Investigating all the sources of uncertainty and possible systematic differences mentioned above is beyond the scope of the current work. As such, attention is focused only on differences between directional efficiency maps. Specifically, measurements taken of the FAST@TA PMTs' uniformity in early 2024 revealed unexpectedly high variation across the PMT surfaces. This non-uniformity is not seen in the ray trace maps used for the initial reconstruction (Figure \ref{fig:oldDirEffMap}) and could relate to the large event by event differences and/or biases observed in the reconstructed vs. TA parameters. The following section provides a brief overview of these measurements and there incorporation into a new set of directional efficiency maps for FAST@TA. The maps are then used to re-reconstruct the FAST@TA coincidence events.

\section{PMT Gain Measurements at FAST@TA}
\label{sec:recBias}
\begin{figure}[t]
    \centering
    \includegraphics[width=0.4\linewidth]{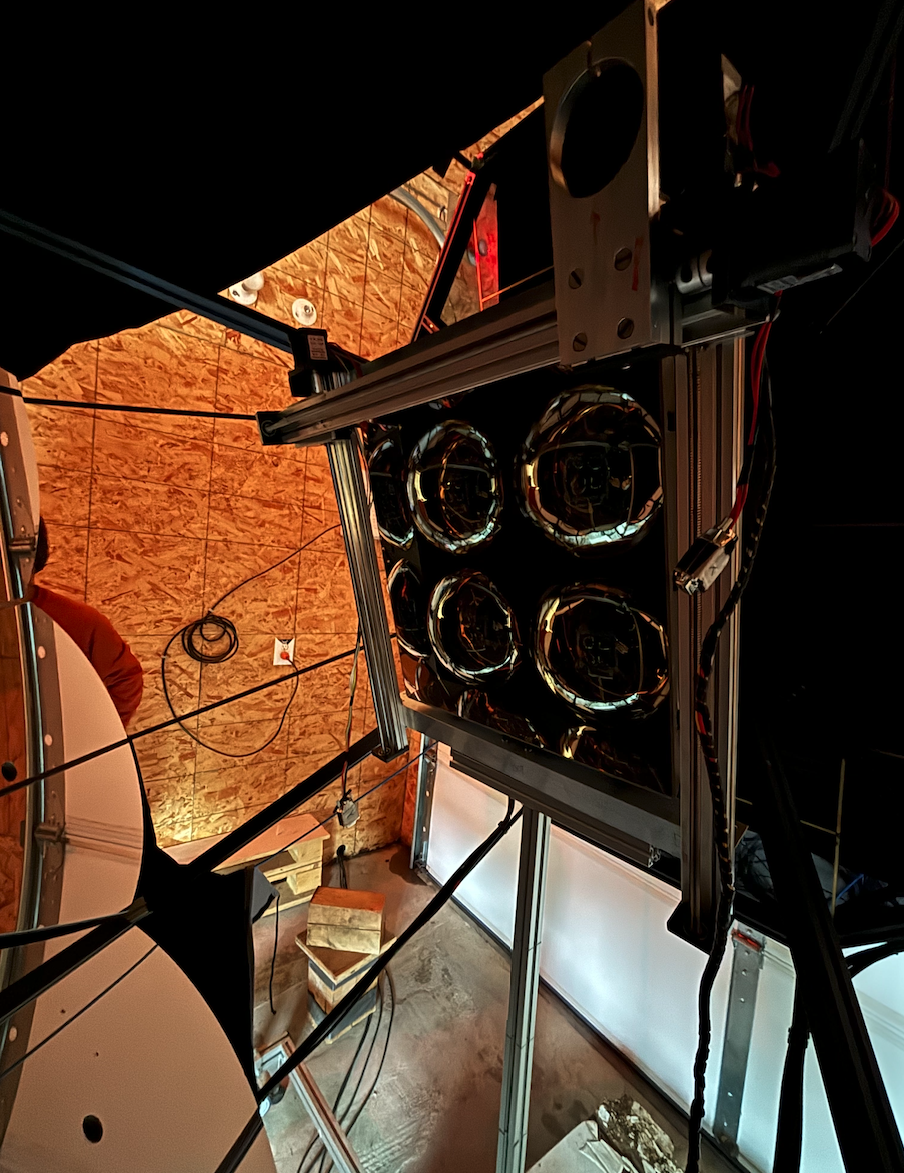}
    \includegraphics[width=0.59\linewidth]{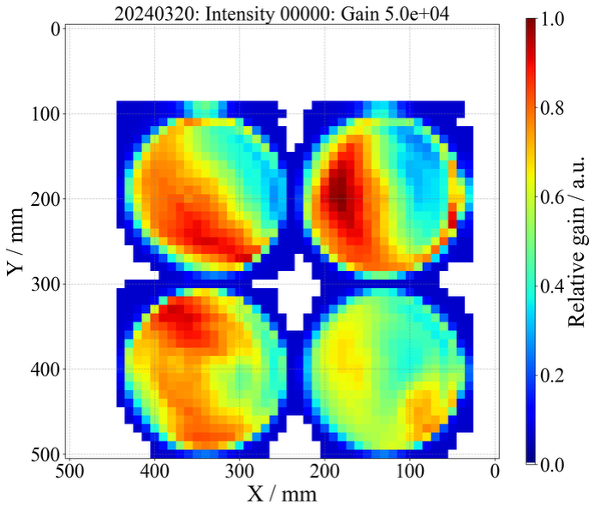}
    \caption{\textit{Left:} Photo of the PMT scanner installed on the camera box of FAST 2. \textit{Right:} Measurements of the non-uniformity in the PMT gain response for FAST 3 after scaling and smoothing.}
    \label{fig:pmtscannerSetup}
\end{figure}

As the TDR relies on matching simulations to data, it is critical that the simulations accurately reflect the conditions/performance of each telescope. In particular, it is necessary to account for the non-uniformity of each PMT's gain response across its surface. Although these measurements had been performed in a controlled lab setting, they had not been done in situ for the FAST prototypes, where the geomagnetic field, amongst other factors, may alter the response. To this end, in March 2024, the Author travelled to the FAST@TA installation in Utah, USA, to assist in measurements of the in situ PMT gain response. The measurements were taken over a four day period using a device called the \say{PMT scanner}, a photo of which is shown on the left side of Figure \ref{fig:pmtscannerSetup}.

\vspace{5mm}

The PMT scanner is essentially a lightweight, aluminium, square frame which can be easily mounted to the FAST telescope camera boxes. The vertical bars support a horizontal bar which is free to move up and down via motorised wheels. Attached to this bar is a holder for a light source which can be moved horizontally. The position of the holder (and thus light source) in front of the camera box can be set via computer software to within sub-millimetre accuracy to any position within a 40\,cm$\times$40\,cm region. The light source used was a flasher with wavelength 350\,nm, with measurements taken via a portable oscilloscope. The direction of the light source was kept fixed, in this case perpendicular to the motor axes. This meant, due to the curvature of the FAST PMTs, that the light was incident perpendicular to each PMT surface \textit{only} at the PMT apex. At all other points the light hit the PMT surfaces at an angle. This is expected to decrease the efficiency compared to normal incidence. Determining this difference quantitatively is planned for future work. 

\vspace{5mm}

The gain response for all three sets of four PMTs were measured using both 10\,mm and 20\,mm steps. The measurements were performed at night inside the FAST huts with doors/shutter closed to prevent excess light reaching the PMTs. The result for FAST-3 (post processing) is shown on the right side of Figure \ref{fig:pmtscannerSetup}. Similar maps for FAST 1 and 2 can be found in Appendix \ref{fig:PMTscannerMeasurements}. The measurements show a large degree of non-uniformity, with up to a $\sim0.6$\% difference in the relative gain between the maximum and minium values in each map. 
\begin{figure}[t!]
    \centering
    \includegraphics[width=0.49\linewidth]{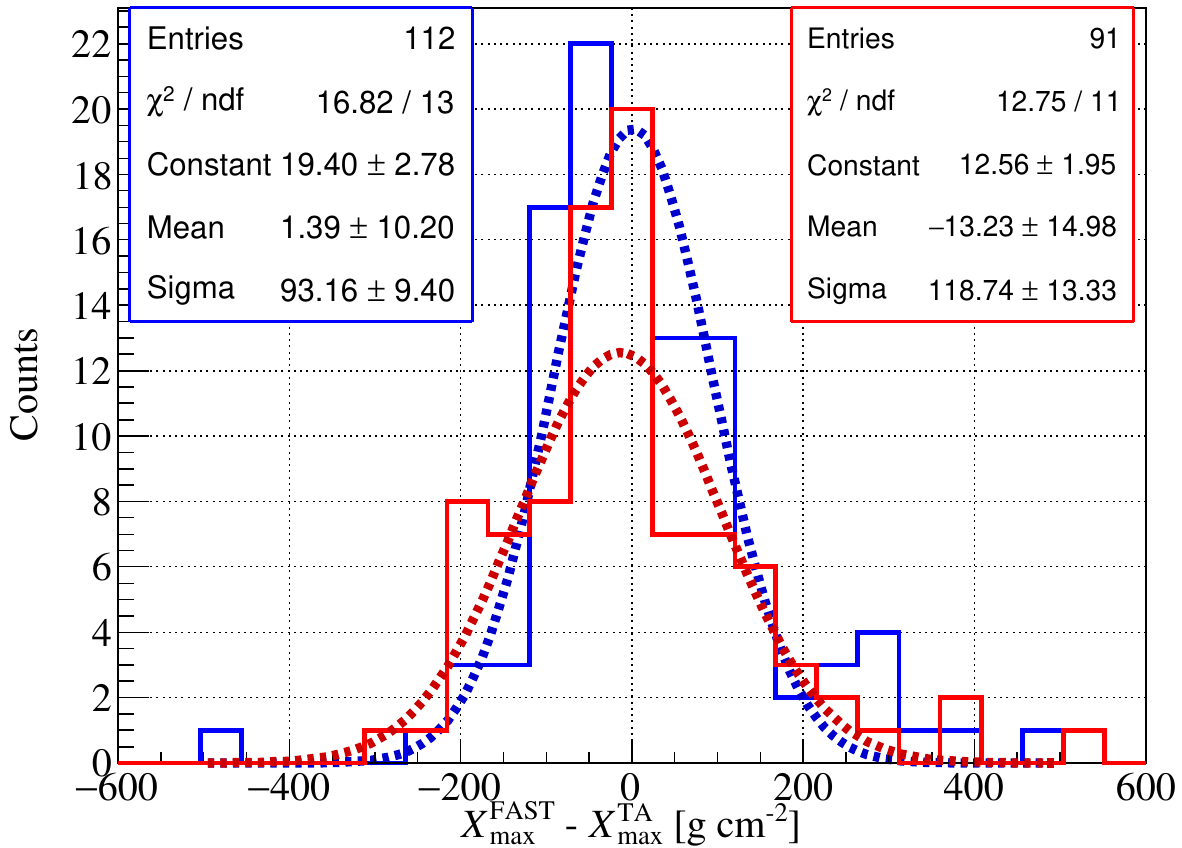}
    \includegraphics[width=0.49\linewidth]{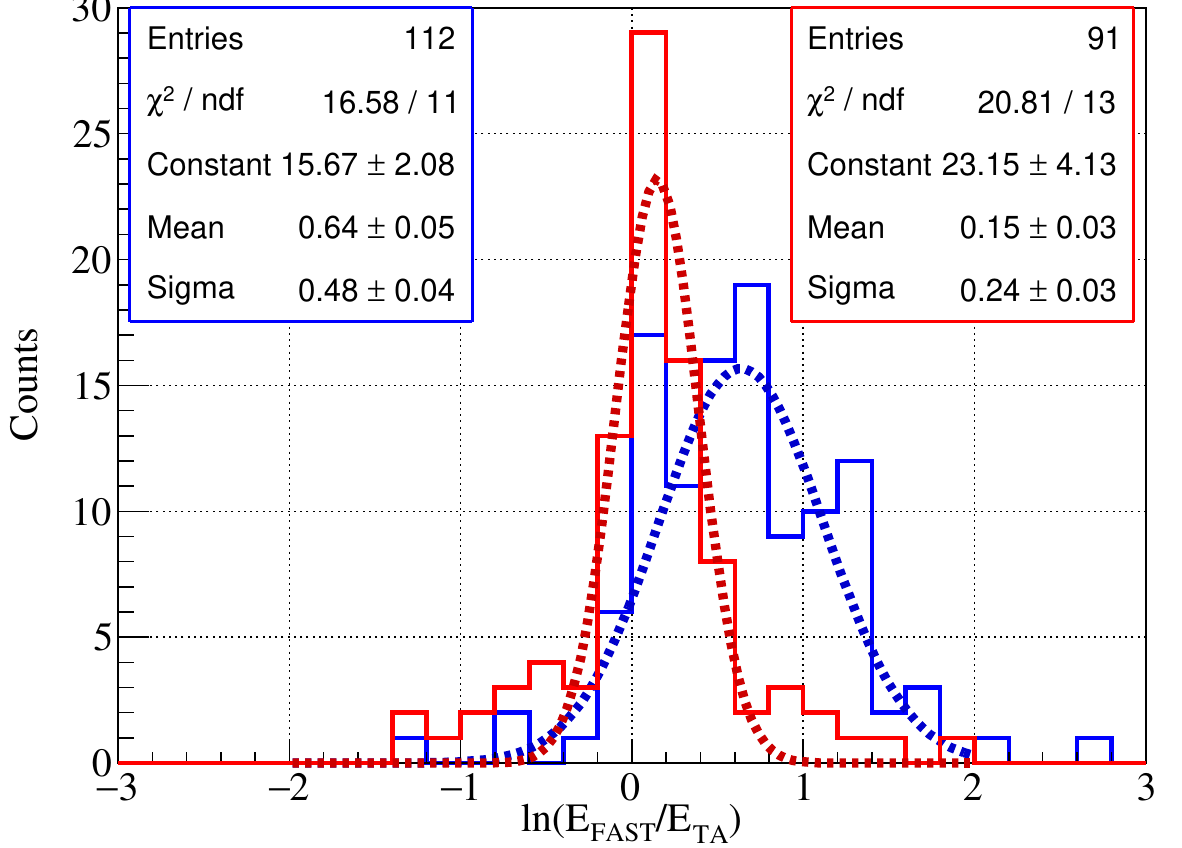}
    \includegraphics[width=0.49\linewidth]{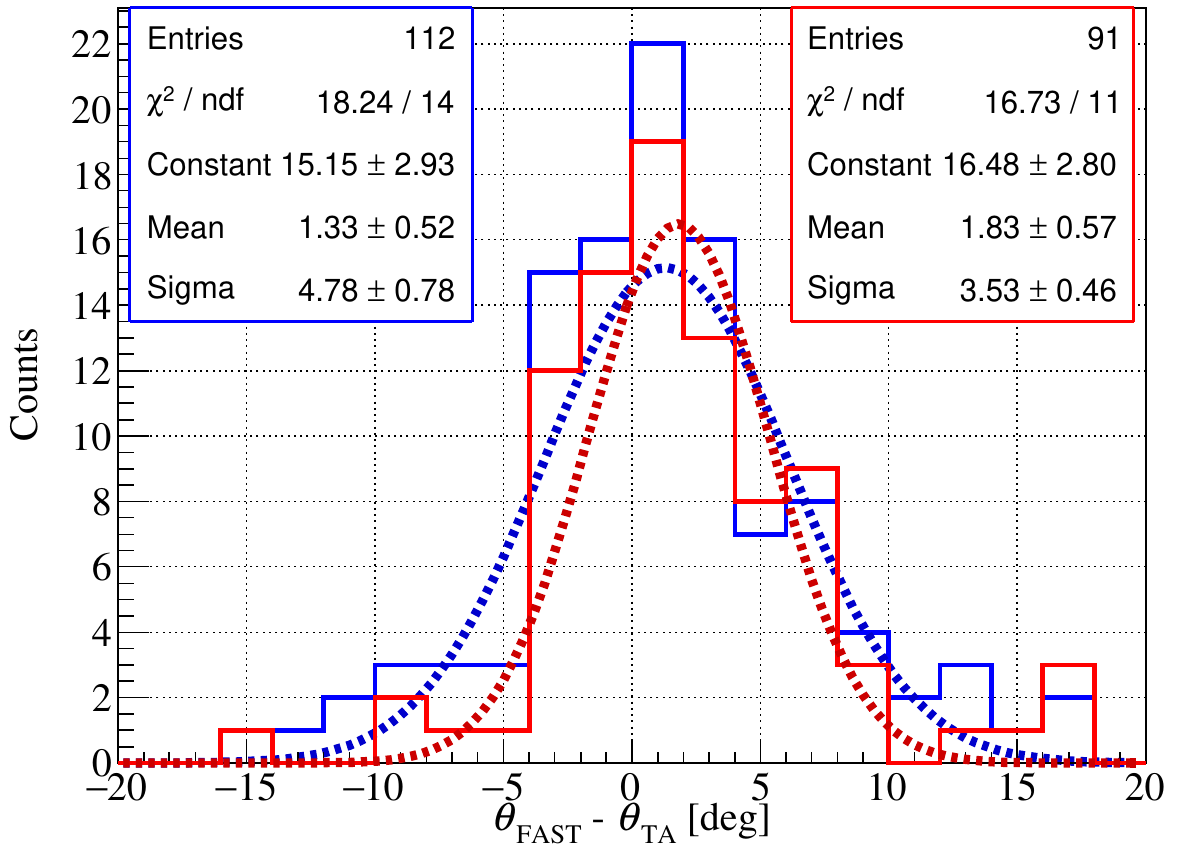}
    \includegraphics[width=0.49\linewidth]{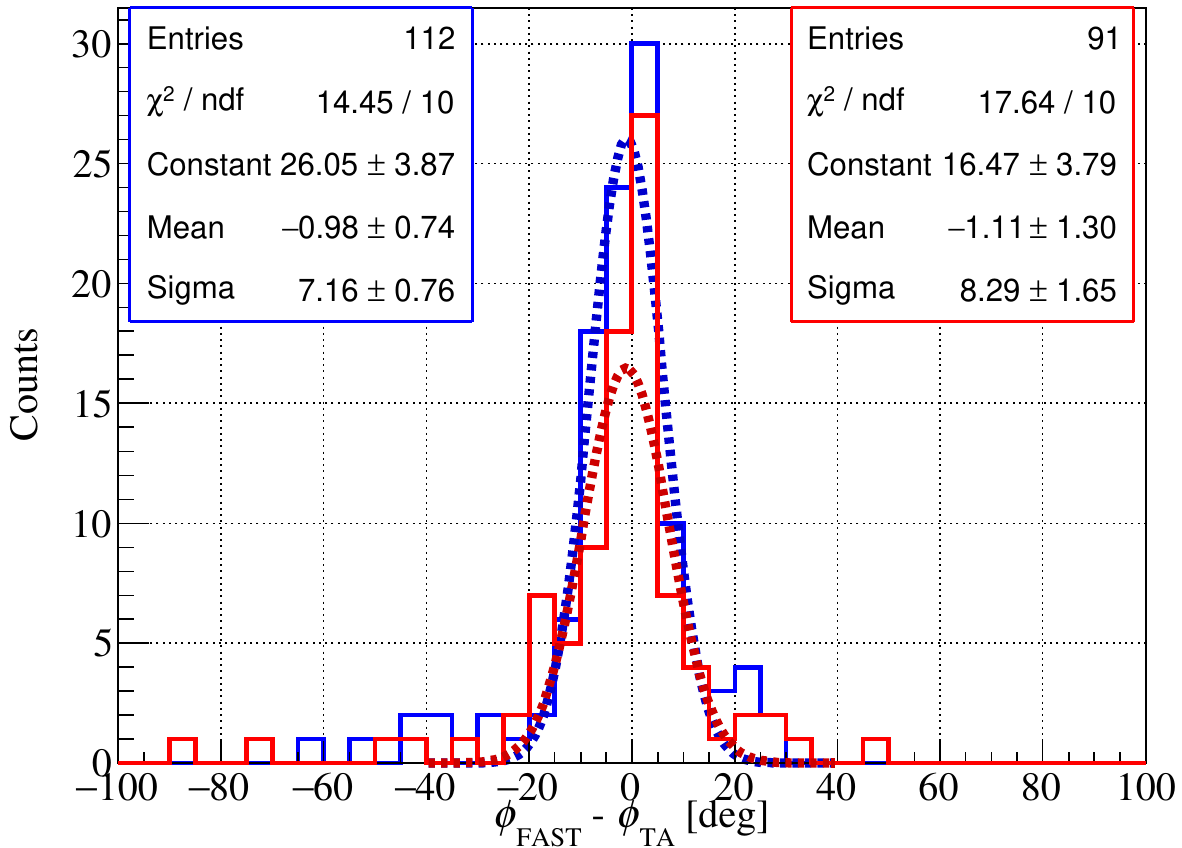}
    \includegraphics[width=0.49\linewidth]{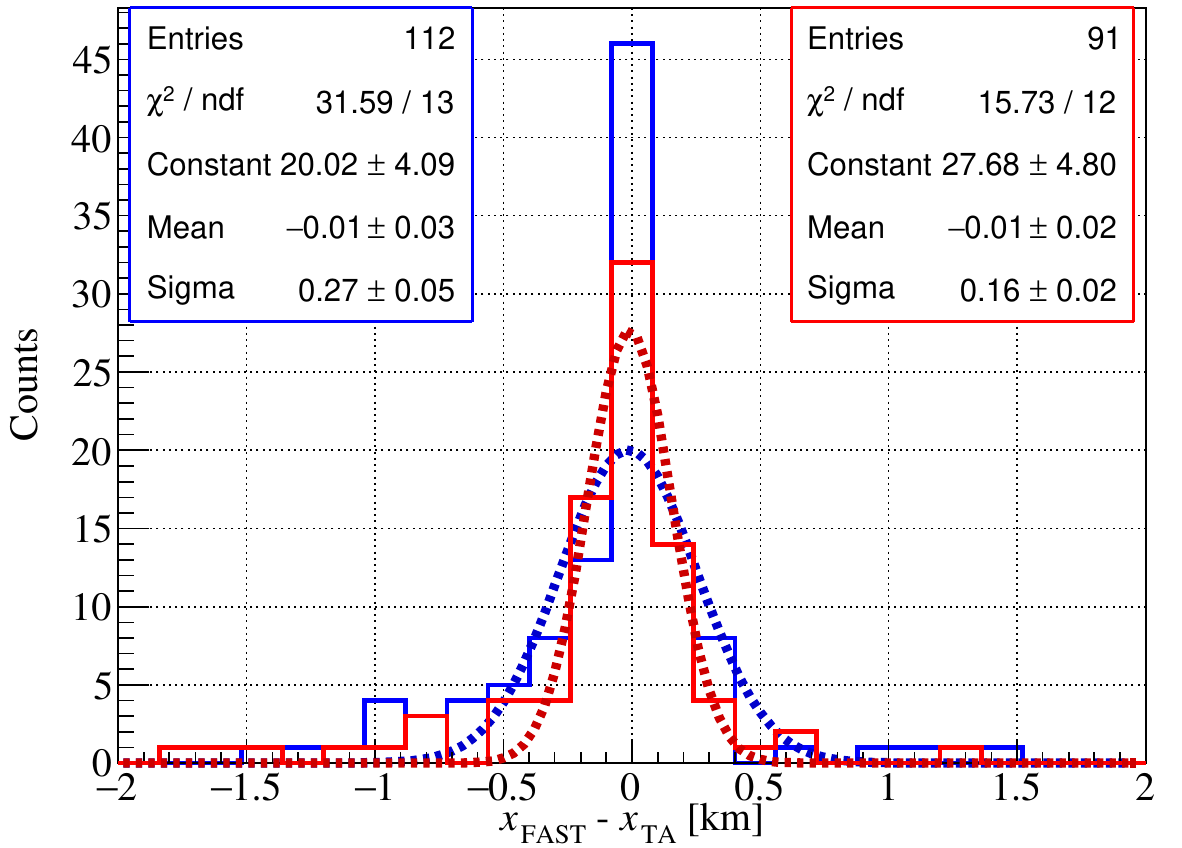}
    \includegraphics[width=0.49\linewidth]{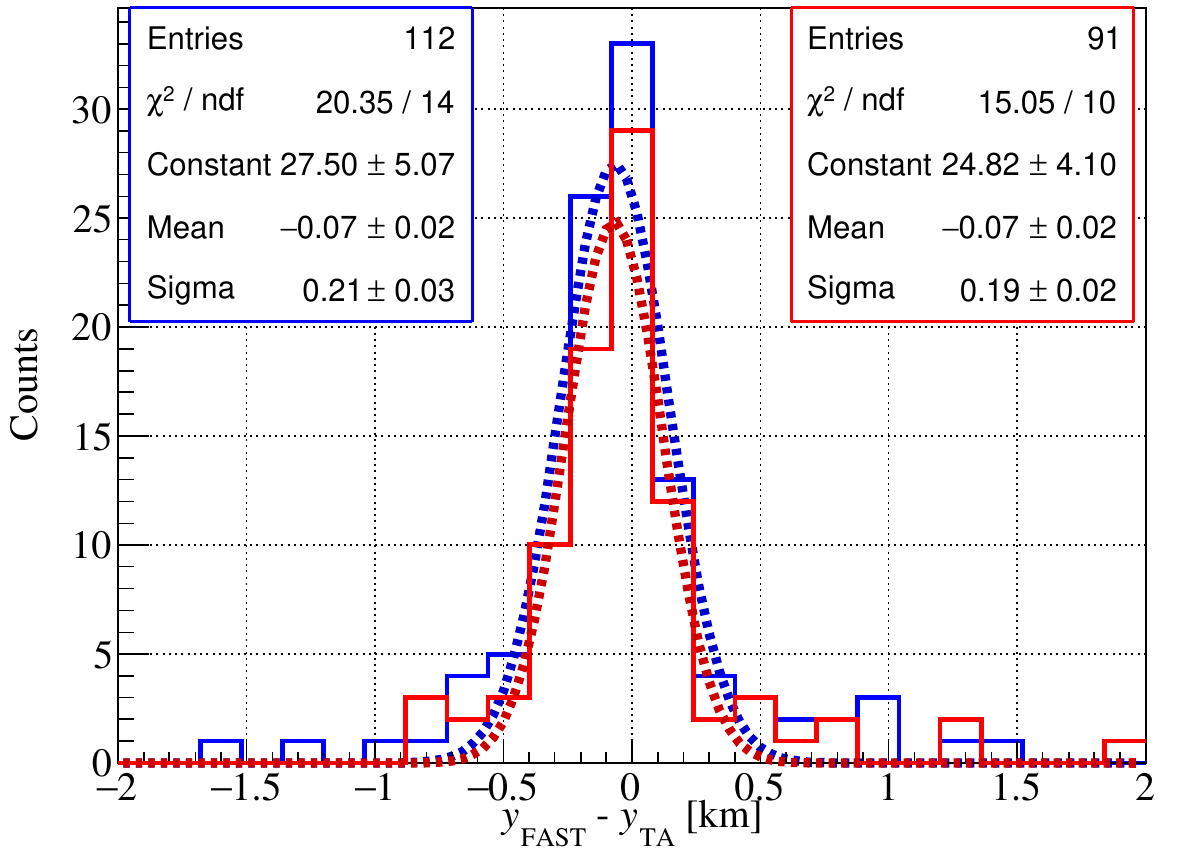}
    \caption{Updated reconstruction results using the new directional efficiency maps. Results for the unscaled (scaled) maps are shown in blue (red).}
    \label{fig:updatedResultDifferences}
\end{figure}
\begin{figure}[t]
    \centering
    \includegraphics[width=1\linewidth]{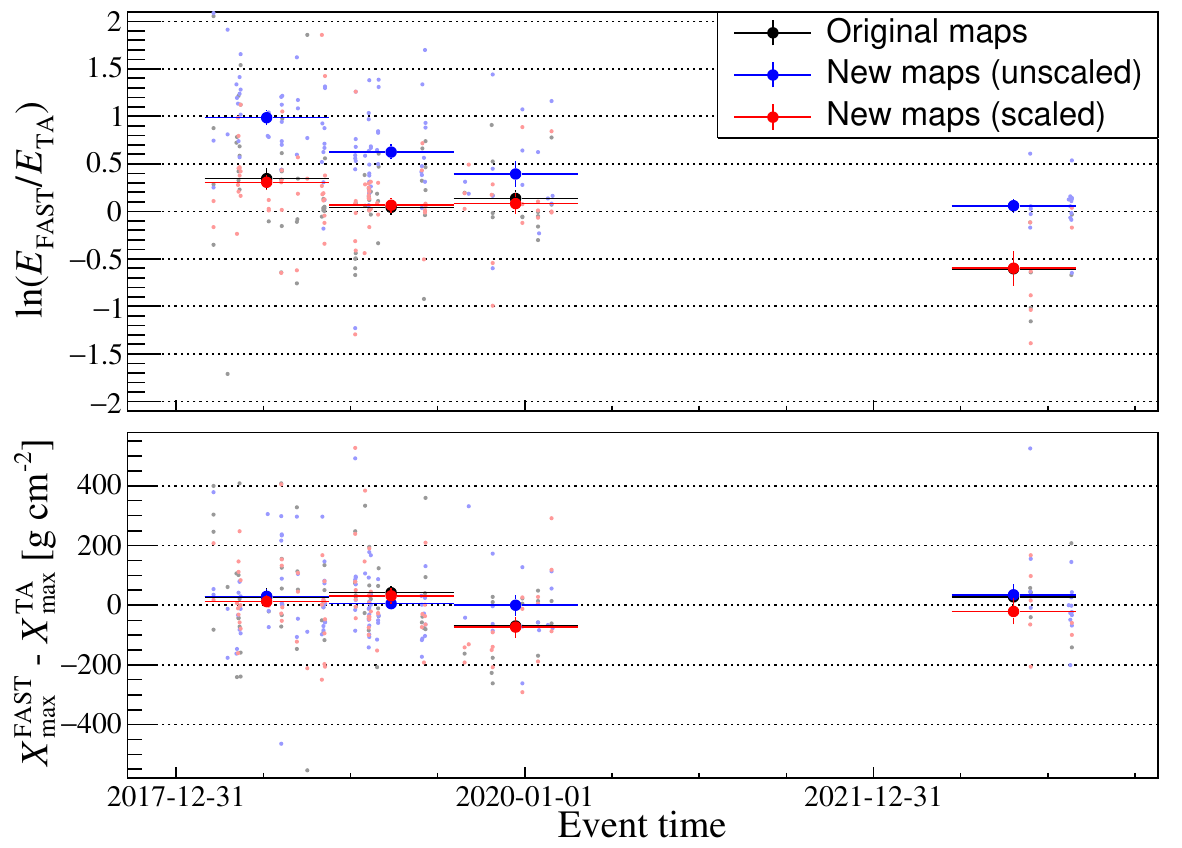}
    \caption{Time dependence of $\ln(E_\textrm{rec}/E_\textrm{TA})$ and $\Delta{}$\Xmax{} for the reconstruction results using the original maps (black points), the new maps unscaled (blue points) and the new maps scaled (red). The light dots represent individual events. The large dots show the average values of the individual points in four different time bins. The error bars are the standard error on the mean.}
    \label{fig:timeDependence}
\end{figure}
The measurements were then used in a full ray-tracing simulation of the telescope (including all optical effects) using the ZEMAX software to produce new directional efficiency maps for each telescope. Note when used inside the ZEMAX simulation the relative scale \textit{between} the maps was ignored. This relative scale was re-introduced after the directional efficiency maps had been calculated. The new directional efficiency maps are shown in Appendix \ref{fig:newDirEffMap}. 
Compared to the previous maps there is a large amount of non-uniformity. This will significantly change the shape of the simulated pulses. Additionally, the new and old maps differ in absolute scale by roughly a factor of 1.5. This is expected to cause the reconstructed energy to be significantly larger than before, as it will take a higher energy shower to produce the same simulated signal which matches the data traces. The change in absolute scale arises because the areas of high intensity in the ray tracing simulation and high intensity in the PMT gain response align with each other \textit{less} than what they did in the previous measurements \cite{martin}. Whilst the bias in energy may worsen, it is still valuable to assess the reconstruction performance when using the new maps as the updated structure could in principle reduce the event-by-event differences. Future work may be able to validate the overall structure of different maps by simulating the TA/Auger first guess for well (TA/Auger) reconstructed events which have geometries crossing multiple FAST pixels with high SNRs. Equation \ref{eqn:eventLikelihood} could then be used to quantitatively assess which maps reproduce the shape of the detected pulses better.

\subsection{Reconstruction Results with the New Maps}
Re-reconstructing the FAST@TA coincidence events with the new maps gives the event by event difference histograms and fits (blue) shown in Figure \ref{fig:updatedResultDifferences}.  An additional 26 events pass the reconstruction cuts. The geometry reconstruction results remain largely unchanged, with only slightly larger widths in the azimuth and core $x/y$ parameters. The $\Delta{}$\Xmax{} bias and width remain the same (within fitted parameter errors), whilst the difference in energy has increased to $\mu(\ln(E_\textrm{FAST}/E_\textrm{TA}))\approx0.64$. One explanation for the observed behaviour is that, although the structure of the new maps may be correct, the absolute difference in scale is causing the minimiser to prioritise fitting the energy and therefore the reconstruction of the other parameters is not significantly improved with respect to the TA values. 

\vspace{5mm}

To test this hypothesis and rectify the energy bias, the new maps were scaled such that the average integral of the new maps matched the average integral of the old maps. \say{Integral} here refers to the sum of efficiency values over all bins in a single directional efficiency map. This was done by solving the following equation for $C$,
\begin{equation}
    C(\textrm{FAST1}_\textrm{new} + \textrm{FAST2}_\textrm{new} + \textrm{FAST3}_\textrm{new}) = 2\times\textrm{FAST1}_\textrm{old} + \textrm{FAST2}_\textrm{old}
\end{equation}
where FAST\#$_\textrm{new/old}$ are the integrals of the corresponding maps, \say{\#} representing 1, 2 or 3 in this case. This gave a value of $C\approx1.7$. Scaling each new map by this value and reconstructing the FAST@TA coincidences once again gave the red histograms in Figure \ref{fig:updatedResultDifferences}. The bias in energy is slightly reduced compared to the original results, though still within the $1\sigma$ parameter errors. The width however has been almost halved, from $0.44\pm0.07$ to $0.24\pm0.03$. The $\Delta$\Xmax{} distribution remains consistent with the original results, whilst the width of the $\Delta{}\theta$ distribution is slightly reduced. The differences between the FAST and TA values for the other geometrical parameters do not appreciably differ from the original results. Overall, assuming the above scaling is valid, introducing the updated directional efficiency maps has improved the event-by-event agreement between the FAST and TA reconstructed energies. This scaling is somewhat arbitrary however and a proper end-to-end absolute calibration of the FAST prototypes is needed to verify the energy scale. The discrepancies between the FAST and TA reconstructions of the other parameters seem reasonable considering the quality of the observed coincidences (number of triggered PMTs, average SNR etc.) and the sources of uncertainty discussed in Section \ref{sec:initRecon}.  

\vspace{5mm}

Despite the improved agreement between the FAST and TA energy reconstructions, the bias is still $+15\%$, and this is without considering the aforementioned invisible energy and filter transmittance corrections. It is likely then that there are still other systematic differences yet to be accounted for. One possibility is a change in the detector response over time. Plotting $\Delta$\Xmax{} and $\ln(E_\textrm{FAST}/E_\textrm{TA})$ as a function of time when using the original (black) and newly calculated unscaled (blue)/scaled (red) directional efficiency maps gives the results in Figure \ref{fig:timeDependence}. Irrespective of the maps used, there appears to be a decreasing linear trend in the energy difference. Additionally, the absolute scale of the maps seems to be the most important factor for determining the energy as indicated by the black and red points aligning in the top plot. It is also curious that, for the unscaled maps, the most recent time bin yields an energy difference near zero. This time bin is the closest to when the measurements were taken, however agreement with the TA energy is not necessarily expected since the measurements taken were purely relative. As for $\Delta{}$\Xmax{}, no clear trend is present. The unscaled new map appears the most stable over time, however this could again be related to the minimiser prioritising the energy fit. 

\vspace{5mm}

There are two possibilities to explain the decreasing trend in energy. If the FAST calibration is assumed to be perfectly correct throughout time, then it would imply that the structure of the PMT response has changed over time. This would go partway to explaining why the old maps match towards the beginning of the FAST@TA observations and the new maps to the most recent data. However a change in structure with time is generally not expected, save for physical alterations to the PMT/telescope orientations. A more likely scenario is that there is an unaccounted for, time-dependent calibration factor, perhaps a decrease in the absolute PMT gain from deterioration. A decrease in gain over time is expected though the rate at which this occurs is currently uncertain. Characterising this time-dependence will be crucial for understanding the long-term potential of the FAST design. In the short term, knowing the dependence will allow both past and future measurements to be better calibrated, ultimately leading to more accurate reconstruction results. Further investigation into this topic is recommended for future work.

\section{FAST Only Reconstruction}
The previous section has demonstrated that when using a first guess of the shower parameters from either TA/Auger, the FAST reconstruction results show reasonable agreement with the TA/Auger reconstructions (ignoring the bias in energy). Whilst this is promising, a future FAST array must be able to produce a first guess of the shower parameters independently. In this section, the first guess methods developed in Chapters \ref{ch:ML} and \ref{ch:TEMP} are applied to the coincidence events. The reconstructed values from the first guesses and the full reconstruction (first guess + TDR) are then compared to the TA/Auger results. Note that both the TSFEL DNN and Template Method assume that the input data has the same properties as the ideal simulations used to develop them. This assumption clearly does not hold for either the FAST@TA or FAST@Auger data in general, given the different directional efficiency maps, atmospheric conditions, baseline fluctuations in traces etc. These systematic differences are kept in mind when interpreting the results below. Two additional notes; the TDR here uses the \textit{original} directional efficiency maps (Figure \ref{fig:oldDirEffMap}) and \say{triggered PMTs} will refer to PMTs passing the event level trigger condition (threshold trigger of SNR $>6$ + timing check + grouping).

\subsection{TSFEL DNN First Guess + TDR}

\begin{figure}[]
    \centering
    \begin{subfigure}[b]{1\textwidth}
        \centering
        \includegraphics[width=1\linewidth]{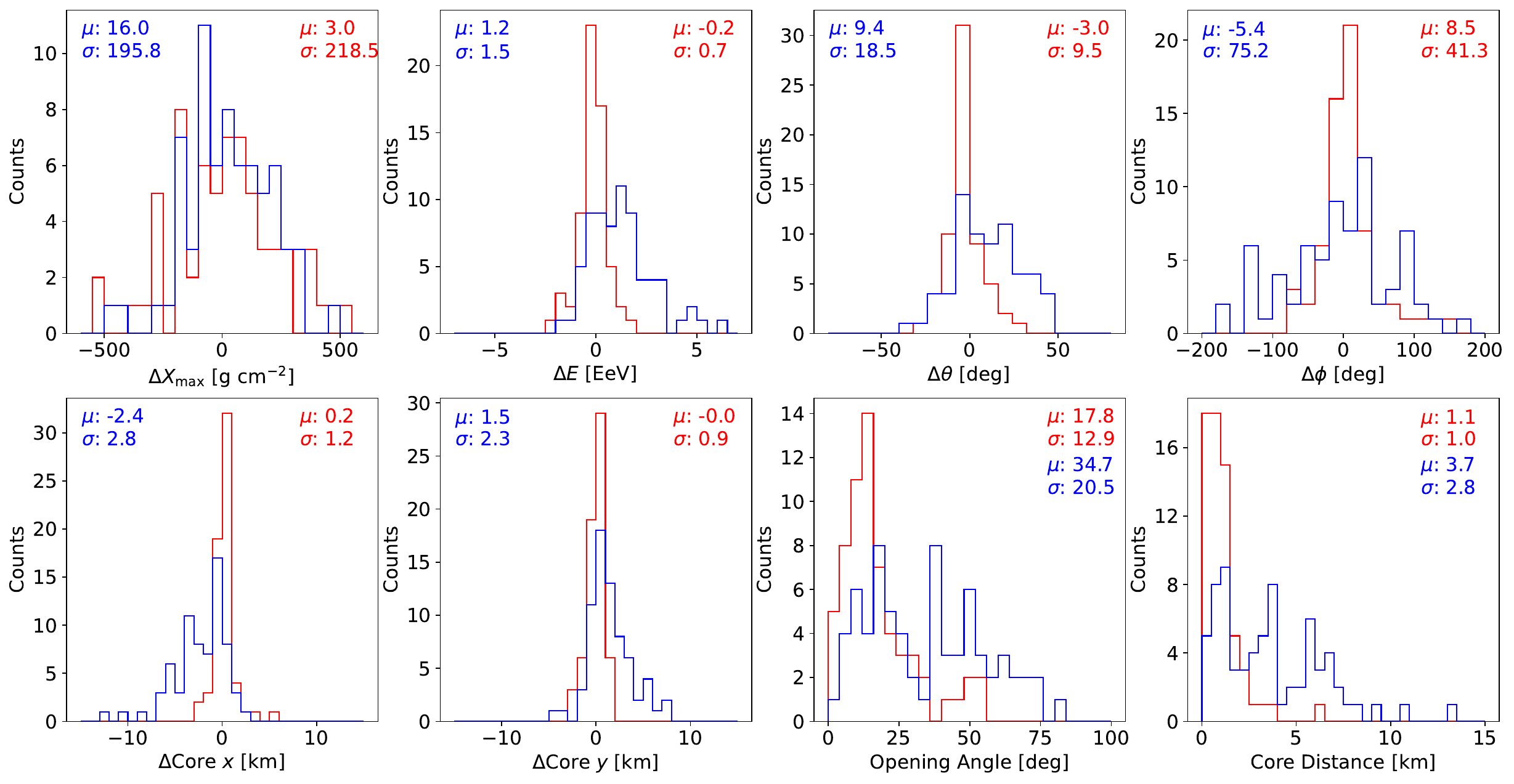}
        \caption{FAST@TA}
        \label{fig:mlFirstGuessTA}
    \end{subfigure}
    \begin{subfigure}[b]{1\textwidth}
        \centering
        \includegraphics[width=1\linewidth]{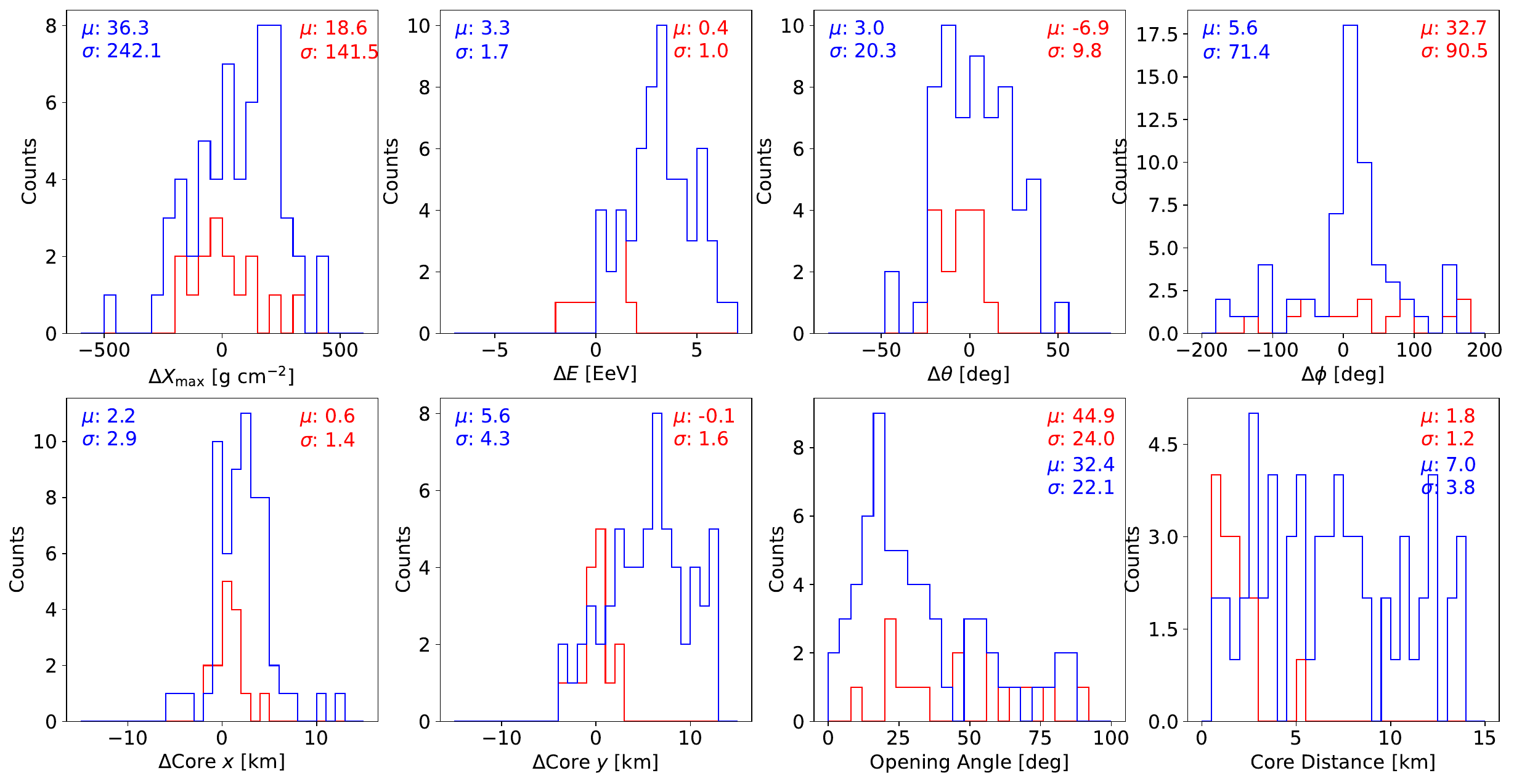}
        \caption{FAST@Auger}
        \label{fig:mlFirstGuessAuger}
    \end{subfigure}
    \caption{Event by event differences between the TA \textbf{(a)} / Auger \textbf{(b)} reconstructed parameters for the coincidence events and; \textit{Red}: TSFEL DNN reconstructed parameters for FAST simulations using the TA/Auger values. \textit{Blue}: TSFEL DNN reconstructed parameters for the FAST coincidence data. The means and standards deviations of each distribution (no Gaussian fits) are shown in the top corners of the plots. The opening angle and core distance distributions are shown for completeness. Note that for parameter $x$, $\Delta{}x=x_\textrm{FAST} - x_\textrm{TA/Auger}$, except for energy which is $\ln(x_\textrm{FAST}/x_\textrm{TA/Auger})$.}
    \label{fig:mlFirstGuessRealData}
\end{figure}

\begin{figure}[]
    \centering
    \includegraphics[width=1\linewidth]{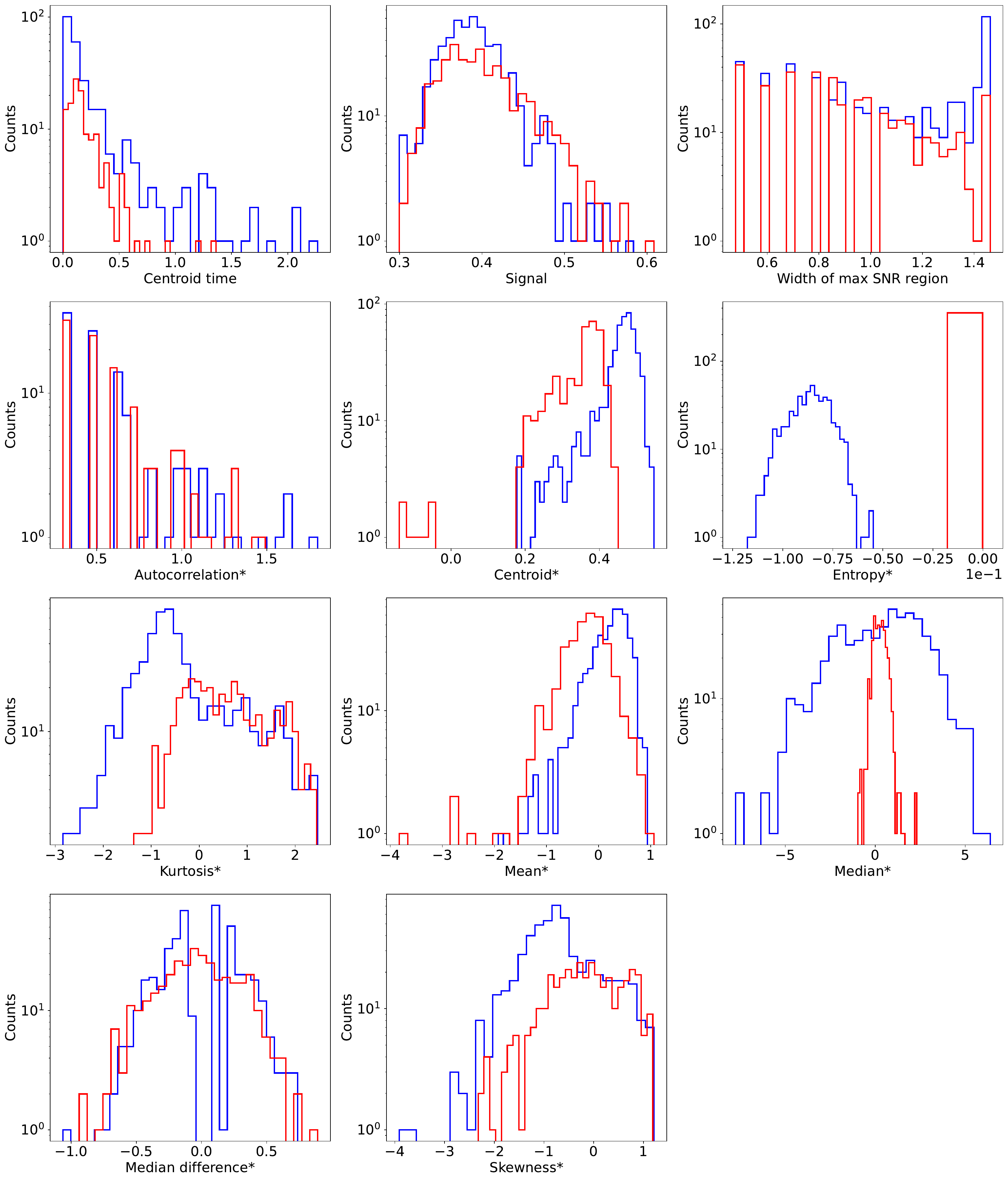}
    \caption{Distributions of the input parameters to the TSFEL DNN model for simulations using the Auger reconstructed values of the coincidence events (red) and the FAST@Auger coincidence event data (blue).}
    \label{fig:mlFirstGuessInputsAuger}
\end{figure}

\begin{figure}[]
    \centering
    \includegraphics[width=1\linewidth]{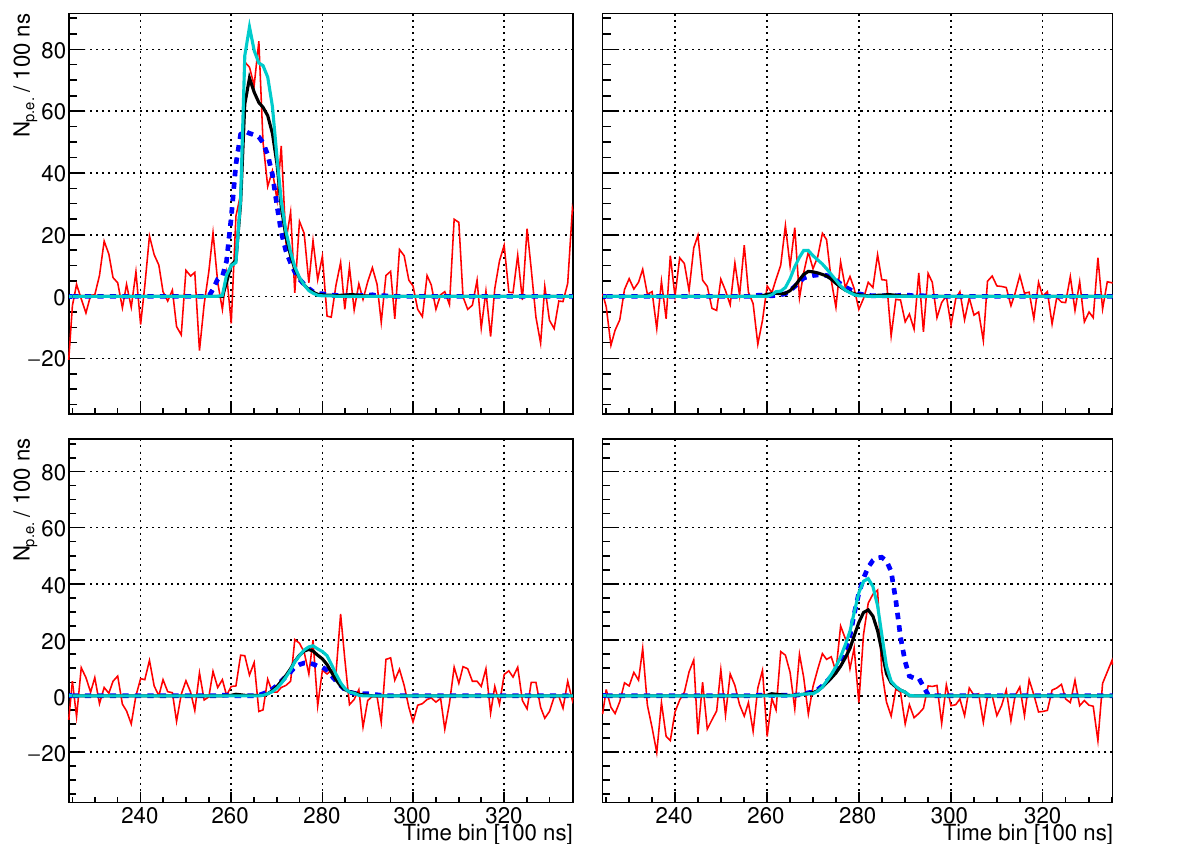}
    \includegraphics[width=1\linewidth]{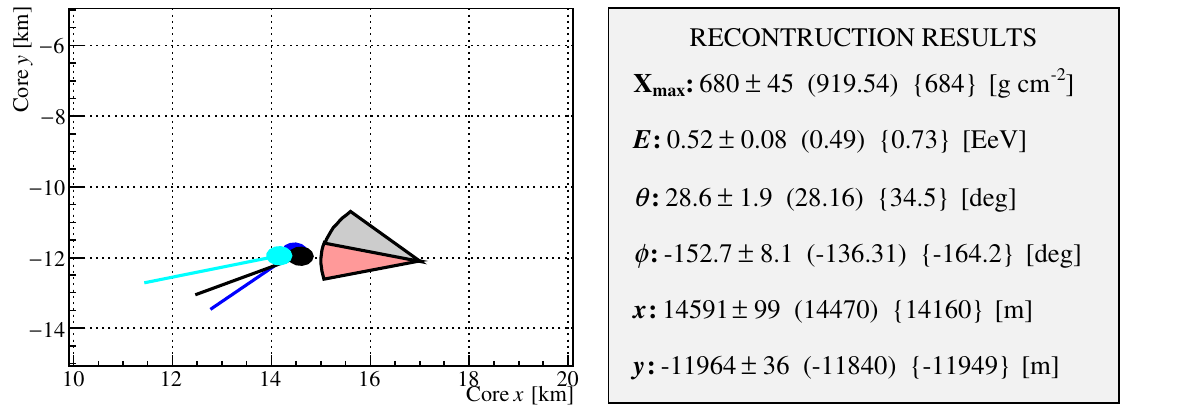}
    \caption{Reconstruction of an event observed by FAST@TA on 2018/05/11. The event has been reconstructed using the TSFEL DNN to provide a first guess to the TDR (TSFEL DNN + TDR). The data traces are shown in red, the best fit TSFEL DNN + TDR traces as black solid lines and the first guess traces from the TSFEL DNN as dotted blue lines. The cyan traces show the best fit found when using the TA values as the first guess (TA + TDR). The bottom right panel shows the reconstructed results. From left to right the values correspond to the TSFEL DNN + TDR reconstructed values with errors, the TSFEL DNN first guess values (round brackets) and the TA + TDR reconstructed values (curly braces). The bottom left panel shows the reconstructed geometries, with the black, blue, and cyan lines corresponding to the TSFEL DNN + TDR reconstruction, TSFEL DNN first guess, and the TA + TDR reconstruction respectively (as in the trace plots).}
    \label{fig:taMLexamp1}
\end{figure}

\begin{figure}[]
    \centering
    \includegraphics[width=1\linewidth]{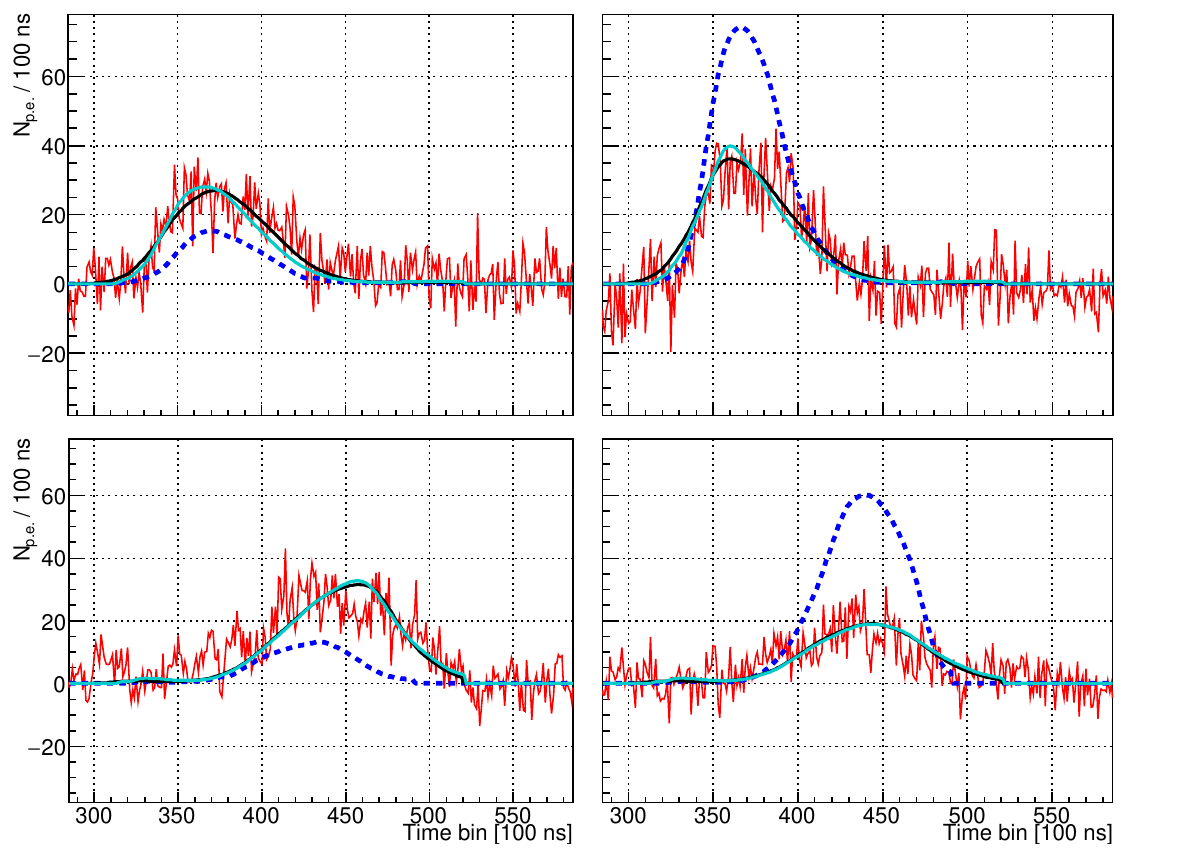}
    \includegraphics[width=1\linewidth]{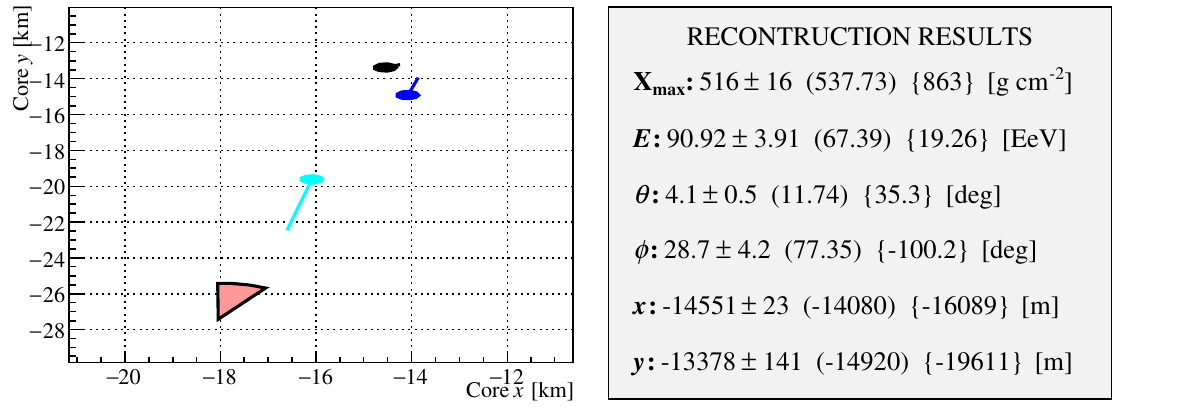}
    \caption{Reconstruction of an event observed by FAST@Auger on 2022/10/21 using the TSFEL DNN + TDR. The format of the figure is the same as Figure \ref{fig:taMLexamp1}. Notice how both the TSFEL DNN + TDR reconstructed traces and the Auger + TDR reconstructed traces both match the data well and yet the corresponding shower parameters are very different.}
    \label{fig:paoMLexamp1}
\end{figure}

\begin{figure}
    \centering
    \begin{subfigure}[b]{1\textwidth}
        \centering
        \includegraphics[width=1\linewidth]{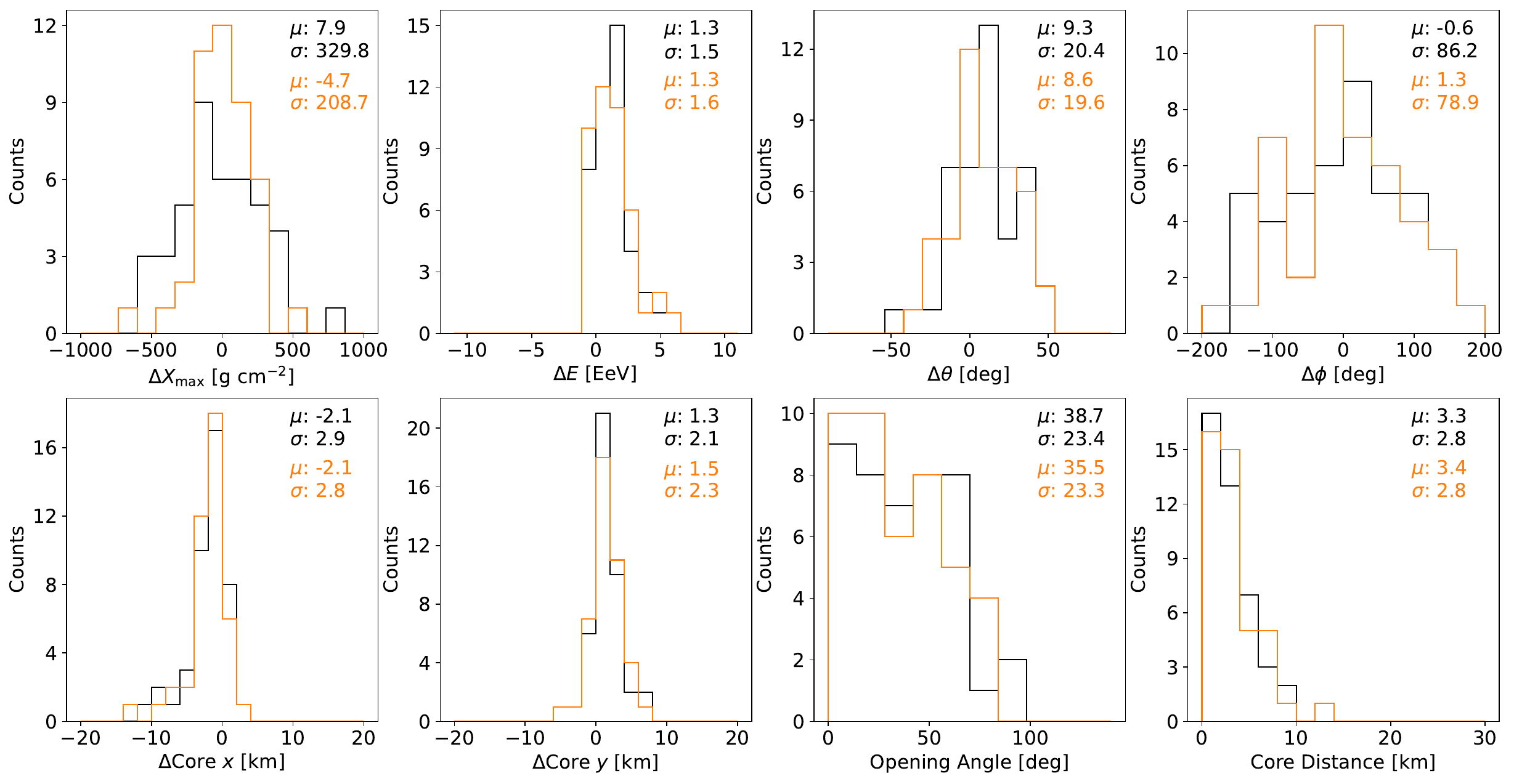}
        \caption{}
        \label{fig:mlFirstGuessTA_vTDR}
    \end{subfigure}
    \begin{subfigure}[b]{1\textwidth}
        \centering
        \includegraphics[width=1\linewidth]{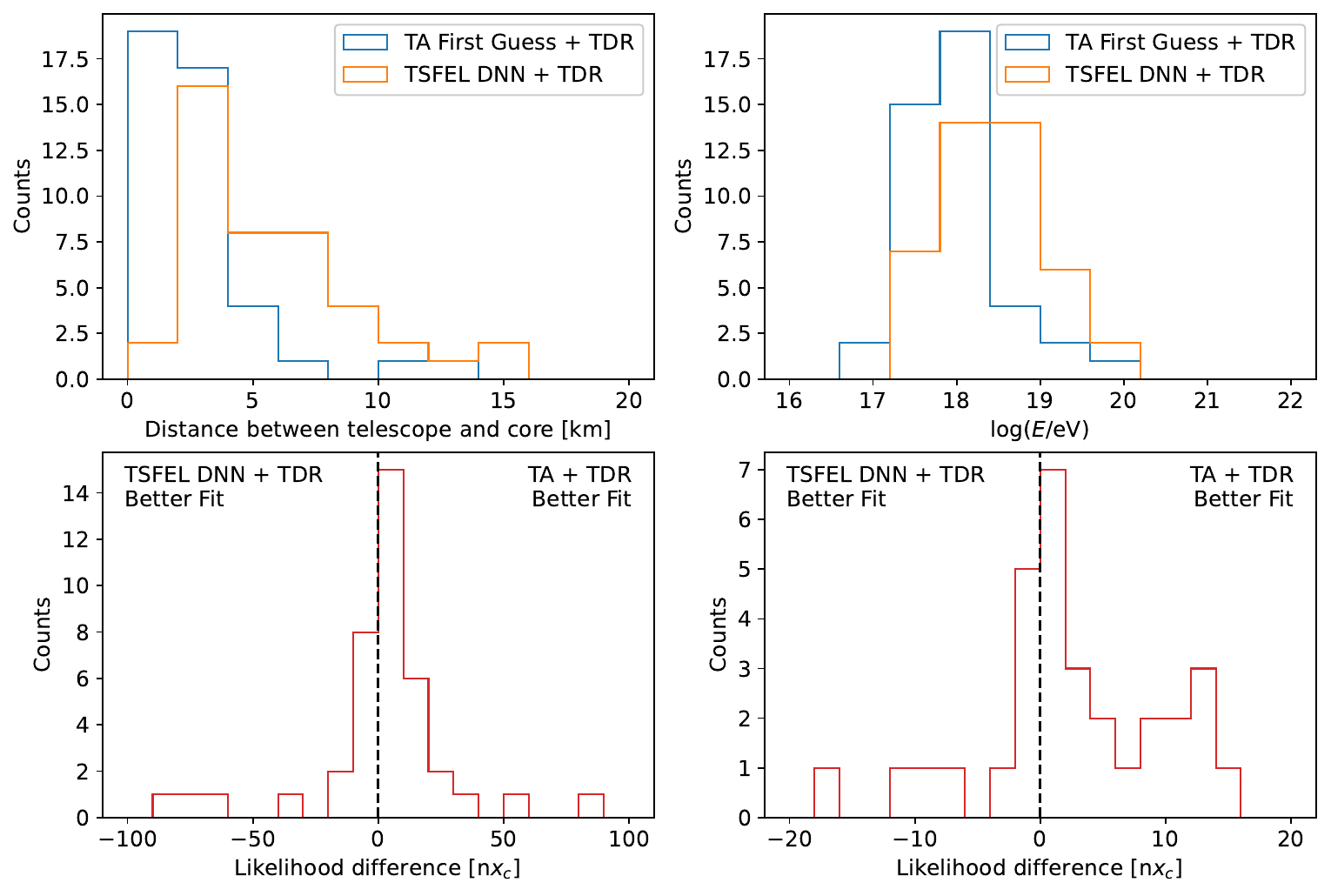}
        \caption{}
        \label{fig:mlFirstGuessTAdegen}
    \end{subfigure}
    \caption{\textbf{(a)} Event-by-event differences between the TSFEL DNN first guess and TA values (black) and between the TSFEL DNN + TDR and TA values (orange). \textbf{(b)} \textit{Top panels:} Distributions of the distance between the telescope location and core position (left) and distributions of reconstructed shower energies (right) obtained from the TA first guess + TDR (blue) and TSFEL DNN + TDR (orange). \textit{Bottom panels:} Differences in event-by-event likelihoods of the two fits, with the left (right) plot showing a zoomed out (zoomed in) view of the distribution.}
    \label{fig:TAMLdiff}
\end{figure}

\begin{figure}
    \centering
    \begin{subfigure}[b]{1\textwidth}
        \centering
        \includegraphics[width=1\linewidth]{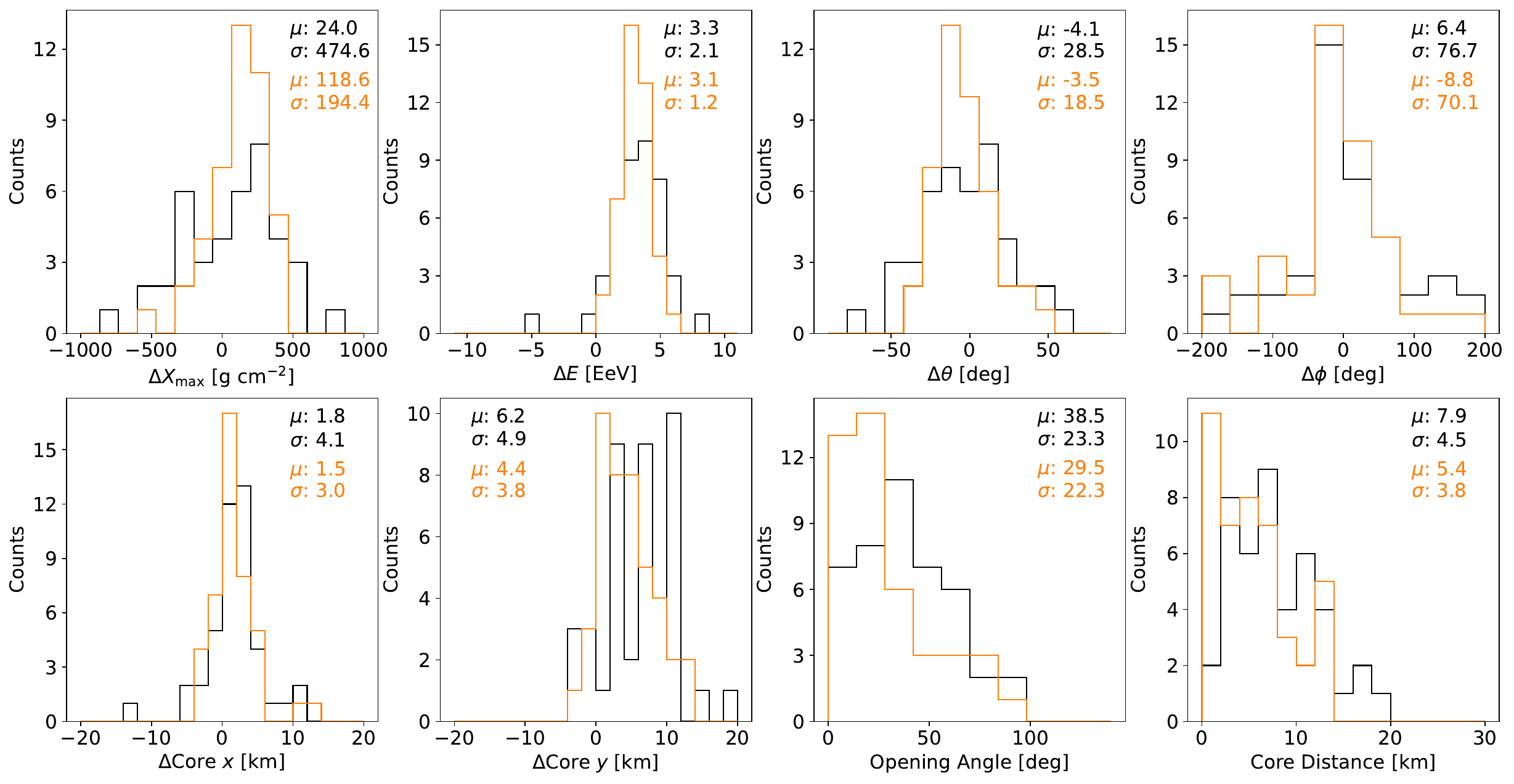}
        \caption{}
        \label{fig:mlFirstGuessAuger_vTDR}
    \end{subfigure}
    \begin{subfigure}[b]{1\textwidth}
        \centering
        \includegraphics[width=1\linewidth]{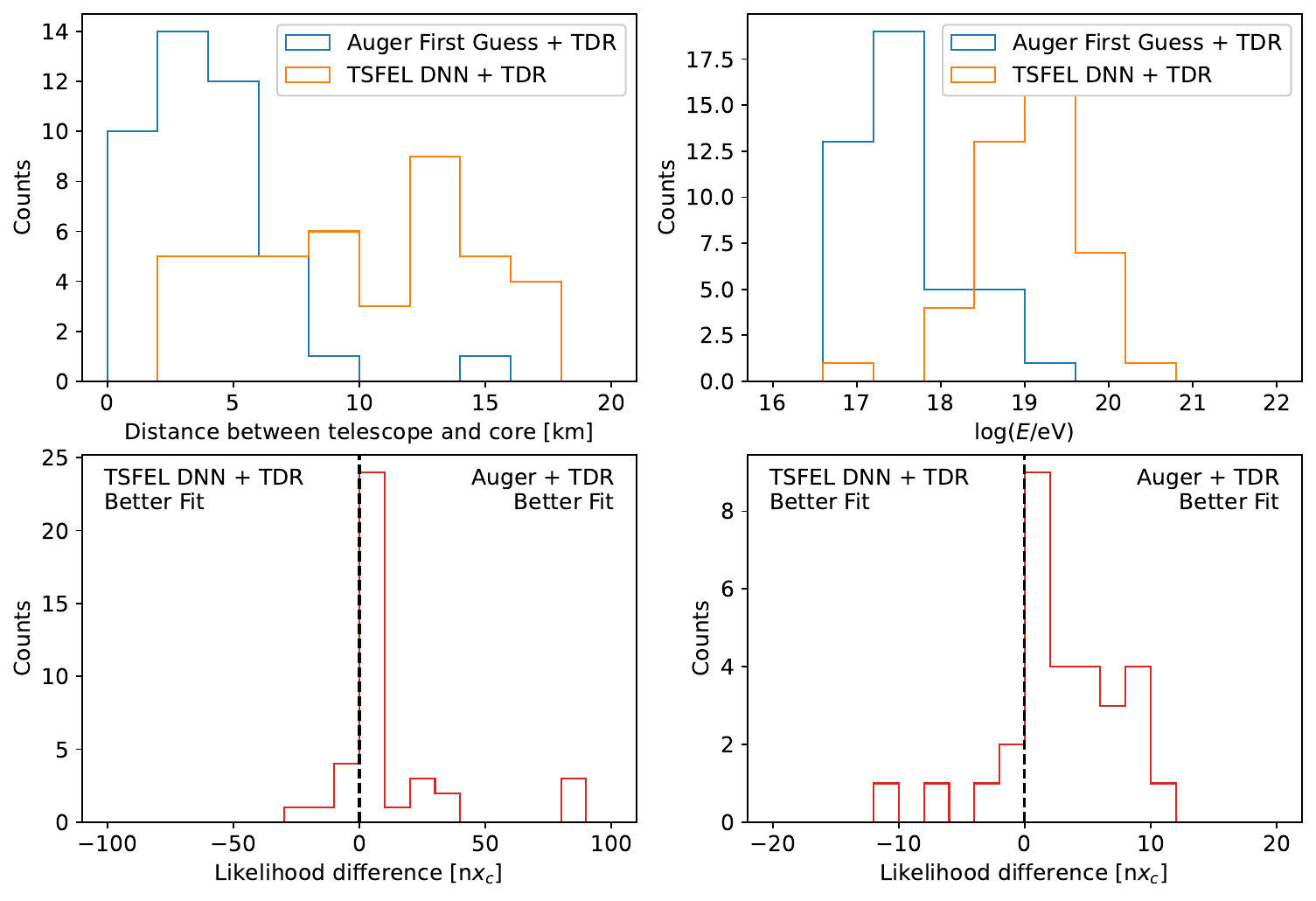}
        \caption{}
        \label{fig:mlFirstGuessAugerdegen}
    \end{subfigure}
    \caption{Same as Figure \ref{fig:TAMLdiff} but for FAST@Auger.}
    \label{fig:AugerMLdiff}
\end{figure}

Before applying the TSFEL DNN to the FAST data, FAST simulations of the coincidence events using the TA/Auger reconstructed values were created. Background noise was added to the simulated traces using $\sigma_\textrm{nsb}=10$\,p.e./100\,ns for FAST@TA (identical to the simulations used to train the TSFEL DNN) and $\sigma_\textrm{nsb}=7$\,p.e./100\,ns for FAST@Auger (slightly above the mean on-site background noise level). These simulated showers were then processed and reconstructed using the corresponding TSFEL DNN model (FAST-Single or FAST-TA) trained in Section \ref{sec:TSFELDNNperformance}. The event-by-event differences between the model estimates and the simulation parameters were then plotted as histograms. The results are shown in red in Figure \ref{fig:mlFirstGuessRealData}. This was done to check the inherent uncertainty in the TSFEL DNN estimation. In other words, if the FAST data were to \say{exactly match} simulated traces generated using TA/Auger reconstructed values, how well would the TSFEL DNN reproduce these values? For FAST@TA, events before 2018/10/06 where only two telescopes were present had all inputs for the 3rd telescope set to 0. 
In theory this may introduce a bias from showers which would normally be seen by the third telescope. However accounting for such cases is beyond the scope of the current work.
Only the results from events with three or more triggered pixels are shown. This choice was made based on the performance evaluation in Section \ref{sec:TSFELDNNperformance}.
Considering the low statistics, the biases (generally close to zero) and widths of these distributions match the performance estimates in Section \ref{sec:TSFELDNNperformance} as expected.
The one large discrepancy is in the FAST@Auger estimate of $\phi$ and in turn opening angle. The cause of this is not clear, though it may simply be an unlucky coincidence considering the relatively few number of events.

\vspace{5mm}

With the best case scenario understood, the same processing and reconstruction steps were applied directly to the FAST coincidence event data. Again only events with three or more triggered pixels were reconstructed. The results are shown as the blue histograms in Figure \ref{fig:mlFirstGuessRealData}. Large biases are now present in the energy and core position. The direction of the energy/core position shift matches the expected degeneracy i.e. a low energy shower close to the telescope/s looks similar to a high energy shower far away. Differences in the size and shape of the data traces vs. traces from simulations with the TA/Auger values obviously contribute to the increased widths of the distributions. Moreover, the energy biases found in the previous section imply that the FAST traces are systematically larger than the reconstructed values of TA/Auger would expect. This will manifest as an energy bias in the TSFEL DNN reconstruction. However the previous energy biases were on the order of 10 - 20\%. This is far less than the factor of $\sim3$ for FAST@TA and $\sim10$ for FAST@Auger seen here.
 In addition, there is a large disparity in the reconstructed opening angles/core distances and the number of events with $>2$ triggered pixels for FAST@Auger. These observations may be explainable by the various systematic differences alluded to at the beginning of this section, in particular the differences in the \textit{properties} of the PMT traces between data and simulations (e.g. fluctuating baselines). These can be partly checked by looking at the input parameters to the network. 

\vspace{5mm}

Figure \ref{fig:mlFirstGuessInputsAuger} shows the inputs to the TSFEL DNN network for the simulations using Auger reconstructed values (blue) and the FAST@Auger coincidence event data (red). A similar plot for FAST@TA can be found in Appendix \ref{apx:mlFirstGuessInputsTA}. For FAST@Auger there are five parameters which appear to have systematically different distributions for the simulated and coincidence event data; centroid*, entropy*, kurtosis*, mean* and median*. For FAST@TA only the entropy* and median* parameters are appreciably different. This likely explains why the TSFEL DNN reconstruction matches the TA results better than it does the Auger results. The systematically larger centroid* and mean* values could be an indicator of the signal baselines being, on average, positively offset from zero in the FAST@Auger coincidence data. These discrepancies, together with the differences in absolute scale of the entropy* distributions and width of the median* distributions for both FAST@TA and FAST@Auger, reiterate the importance of ensuring that the trace features which are utilised are consistent between data and the simulations used for model training.
The more features which are used, the more checking needs to be performed to ensure consistency - this is one reason to favour simpler models. 

\vspace{5mm}

Putting aside these differences for now, the TSFEL DNN first guess values were used as input to the TDR and the coincidence events with $>2$ triggered PMTs reconstructed. As the number of events with $>2$ triggered PMTs is only 70 for FAST@TA and 75 for FAST@Auger, the \Xmax{} in FOV cut was relaxed for these reconstructions. The number of showers passing the remaining quality cuts were 43 for both FAST@TA and FAST@Auger. Visually inspecting the fits, roughly 1/2 of the reconstructions for FAST@Auger and 2/3 of reconstructions for FAST@TA found a set of traces which appeared to reasonably match the data. Reconstructions which failed to find a good match generally had a first guess which did not show signal in one or more of the triggered PMTs. Figure \ref{fig:taMLexamp1} shows the reconstruction of an event which passed all cuts for FAST@TA. Here, the core position, arrival direction and energy of the TSFEL DNN + TDR are roughly similar to the result when using the TA values as the first guess. Figure \ref{fig:paoMLexamp1} shows another example reconstruction, this time for FAST@Auger. For this event, both the TSFEL DNN + TDR and Auger first guess + TDR give traces which appear to match the data well. However the reconstructed shower parameters are very different. This is another demonstration of the degeneracy which makes reconstructing events with only a single Eye challenging. Additional reconstruction examples can be found in Appendix \ref{apx:tsfelTDRexamples}.

\vspace{5mm}

The event-by-event differences between the TSFEL DNN + TDR reconstructed values and the TA/Auger reconstructed values are shown in Figures \ref{fig:mlFirstGuessTA_vTDR} and \ref{fig:mlFirstGuessAuger_vTDR} by the orange histograms. The event-by-event differences using just the TSFEL DNN are shown for reference in black. The results for the TSFEL DNN + TDR are largely unchanged compared to using only the TSFEL DNN. This shows that the best fit shower parameters found by the TDR were generally in the vicinity of the TSFEL DNN first guess. On the one hand, for the events which showed reasonable matching between the best fit traces and data, this is promising since it demonstrates that the general reconstruction principle of having a rough first guess optimised by the TDR can work well on data. However, the large differences between the \say{true} (TA/Auger) parameters and the best fit parameters highlights the degeneracy issues which have been observed throughout this thesis. 

\vspace{5mm}

Figures \ref{fig:mlFirstGuessTAdegen} and \ref{fig:mlFirstGuessAugerdegen} provide perhaps the clearest demonstration of the energy/core position degeneracy. The top two panels show distributions of the distance between the reconstructed core position and telescope (left) and shower energy (right) for the TSFEL DNN + TDR (orange) and TA/Auger first guess + TDR (blue). The TA/Auger first guess + TDR is used for comparison to ensure both reconstructions represent a \say{best match} to the data found with the TDR. The bottom two panels show two versions of the same plot - the differences in the negative-log likelihoods between the TSFEL DNN + TDR and the TA first guess + TDR. The differences are shown as a multiple of $x_c=7.04$, which, recall from Section \ref{sec:FASTreconstruction}, is the amount by which the negative-log likelihood must be increased to obtain $1\sigma$ uncertainties in the parameters. The left (right) plot shows a zoomed out (in) view of the distribution. Events to the left of the zero line have lower likelihoods (and thus \say{better fits} when using the TSFEL DNN + TDR), whilst events to the right of the zero line have lower likelihoods (and thus \say{better fits} for the TA/Auger first guess + TDR). The extreme differences beyond $nx_c=\pm50$ seen in the zoomed out plots show events where one of the reconstructions failed to find a good match to the data whilst the other method did. For FAST@TA, the TSFEL DNN + TDR shows better fits for roughly 1/3 of events, whilst for FAST@Auger only 5/43 events were better reconstructed using the TSFEL DNN first guess. Although the TA/Auger first guess + TDR has, on average, found better fits here, the majority of differences are within a few multiples of $x_c$, indicating that the reconstructed traces are reasonably close - and yet the reconstructed energy and core positions are very different, particularly for FAST@Auger. Overall, these results emphasise the importance of stereo observation to reduce the observed degeneracies (see the parameter biases in Figures \ref{fig:TSFELex} - \ref{fig:TSFELzRes}).

\subsection{Template Method First Guess + TDR}

\begin{figure}
    \centering
    \begin{subfigure}[b]{1\textwidth}
        \centering
        \includegraphics[width=1\linewidth]{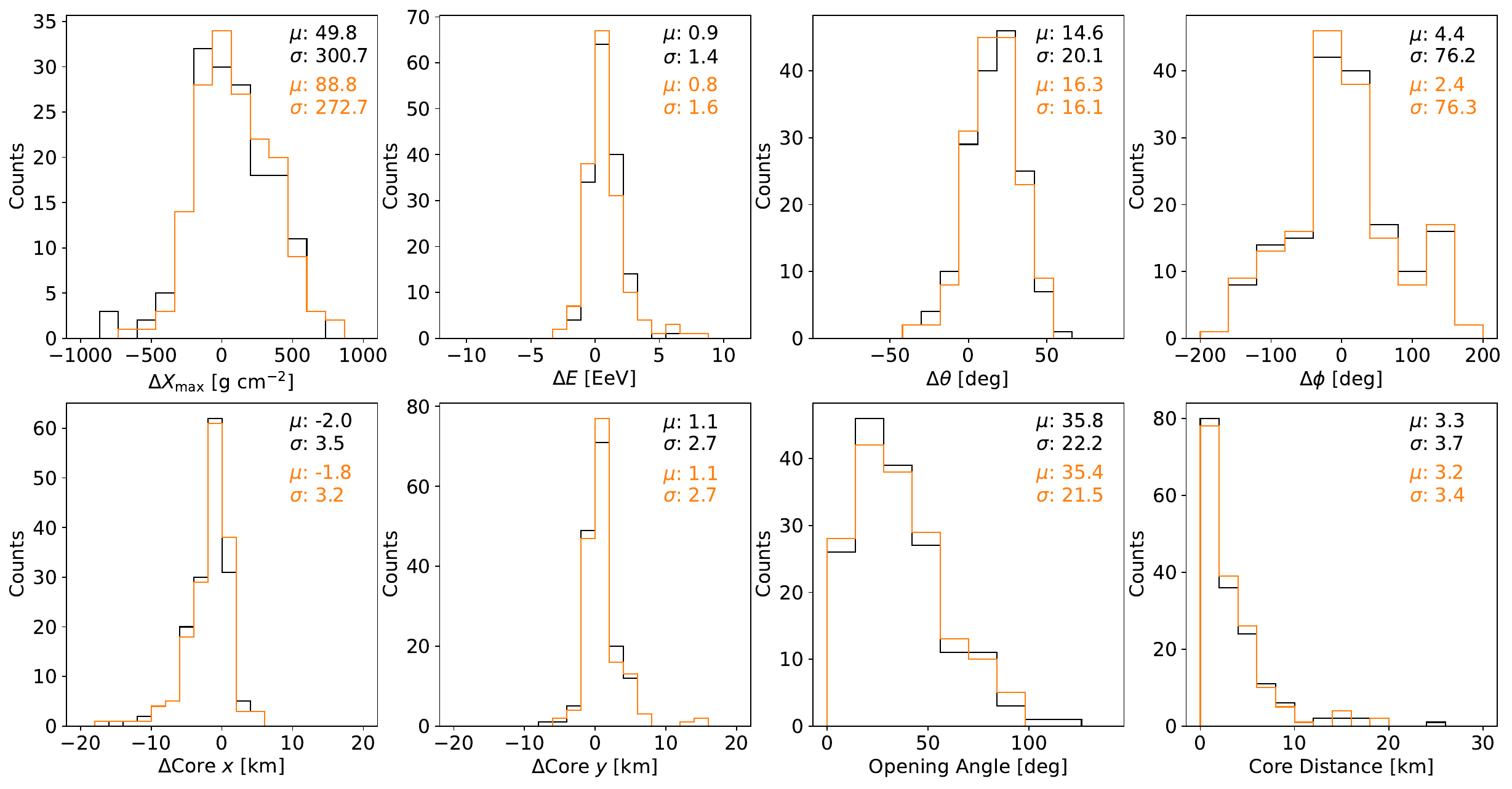}
        \caption{}
        \label{fig:realtempdiffta}
    \end{subfigure}
    \begin{subfigure}[b]{1\textwidth}
        \includegraphics[width=1\linewidth]{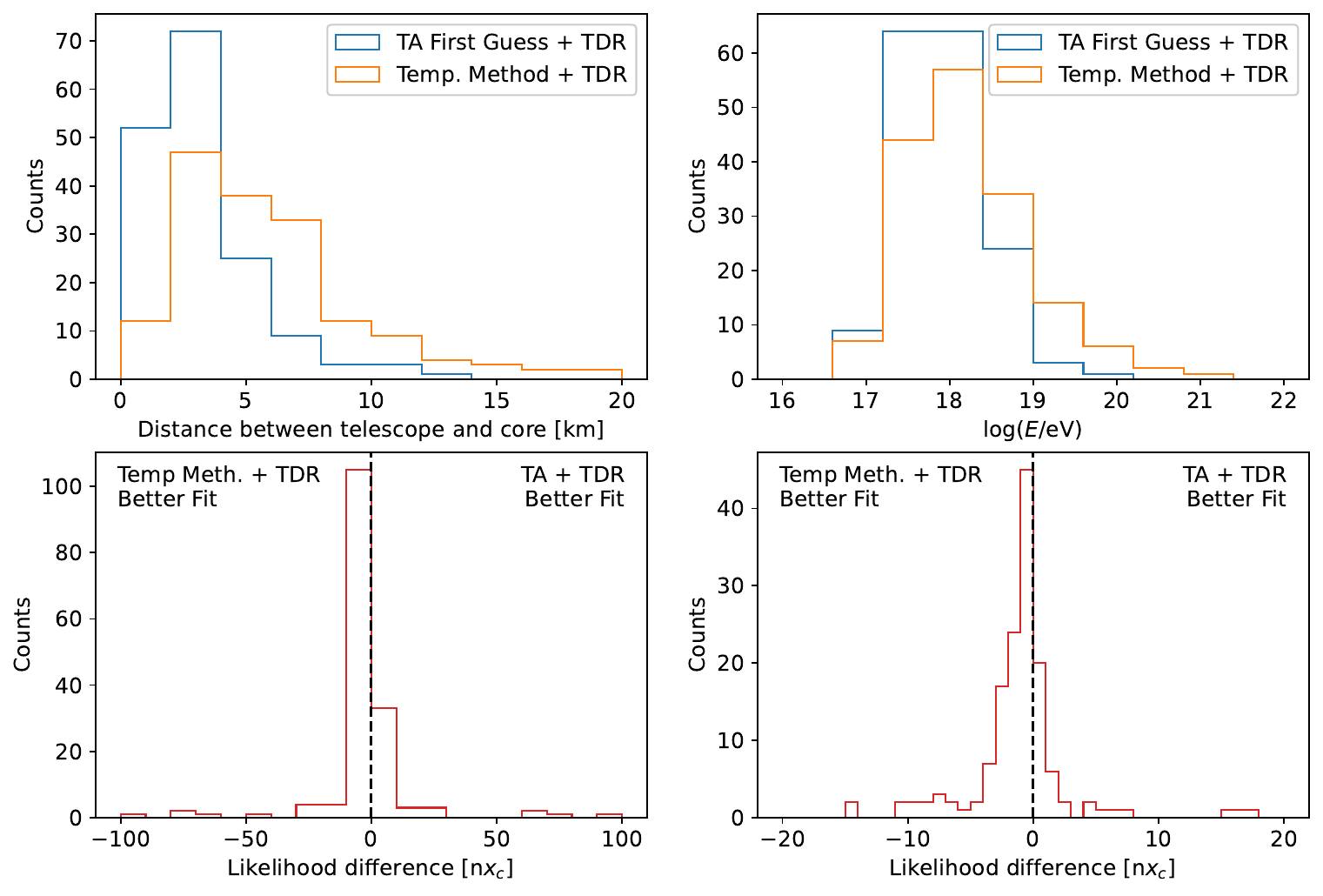}
        \caption{}
        \label{fig:tempdegenta}
    \end{subfigure}
    \caption{\textbf{(a)} Event-by-event differences between the Template Method first guess and TA values (black) and between the Template Method + TDR and TA values (orange). \textbf{(b)} \textit{Top panels:} Distributions of the distance between the telescope location and core position (left) and distributions of reconstructed shower energies (right) obtained from the TA first guess + TDR (blue) and Template Method + TDR (orange). \textit{Bottom panels:} Differences in event-by-event likelihoods of the two fits, with the left (right) plot showing a zoomed out (zoomed in) view of the distribution.}
    \label{fig:tempMethodDiffTA}
\end{figure}

\begin{figure}
    \centering
    \begin{subfigure}[b]{1\textwidth}
        \centering
        \includegraphics[width=1\linewidth]{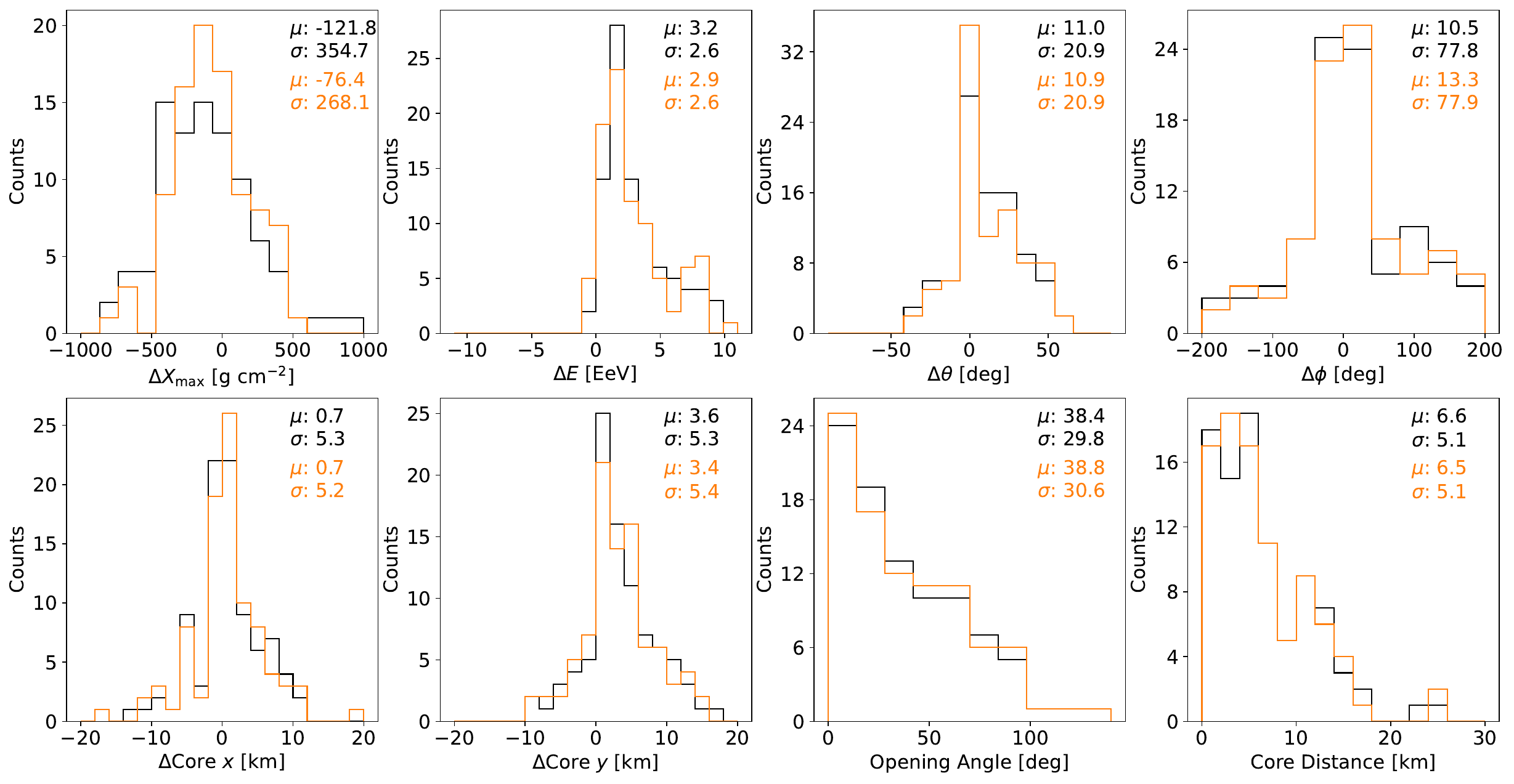}
        \caption{}
        \label{fig:realtempdiffauger}
    \end{subfigure}
    \begin{subfigure}[b]{1\textwidth}
        \includegraphics[width=1\linewidth]{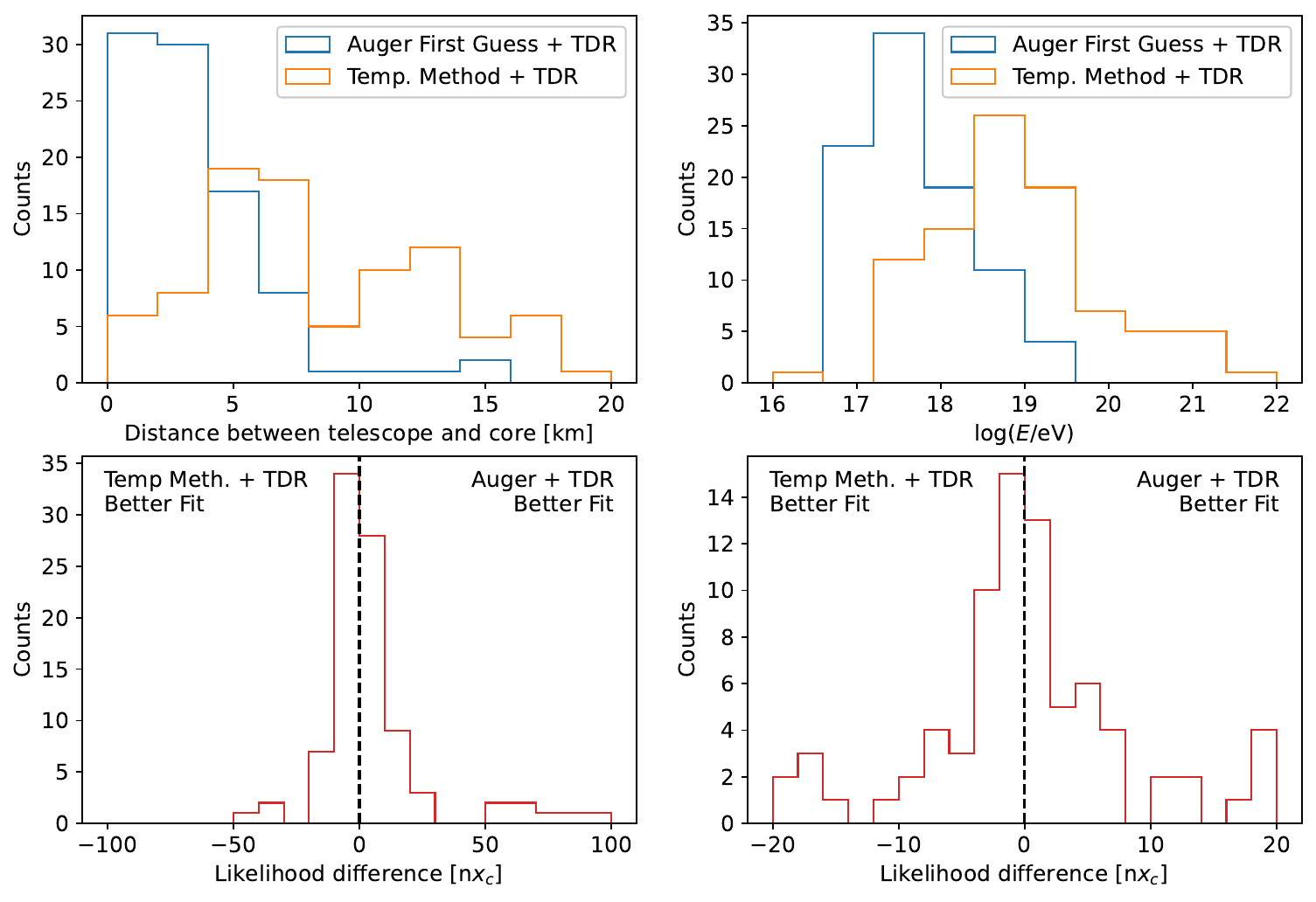}
        \caption{}
        \label{fig:tempdegenauger}
    \end{subfigure}
    \caption{Same as Figure \ref{fig:tempMethodDiffTA} but for FAST@Auger.}
    \label{fig:tempMethodDiffAuger}
\end{figure}

\begin{figure}
    \centering
    \includegraphics[width=0.65\linewidth]{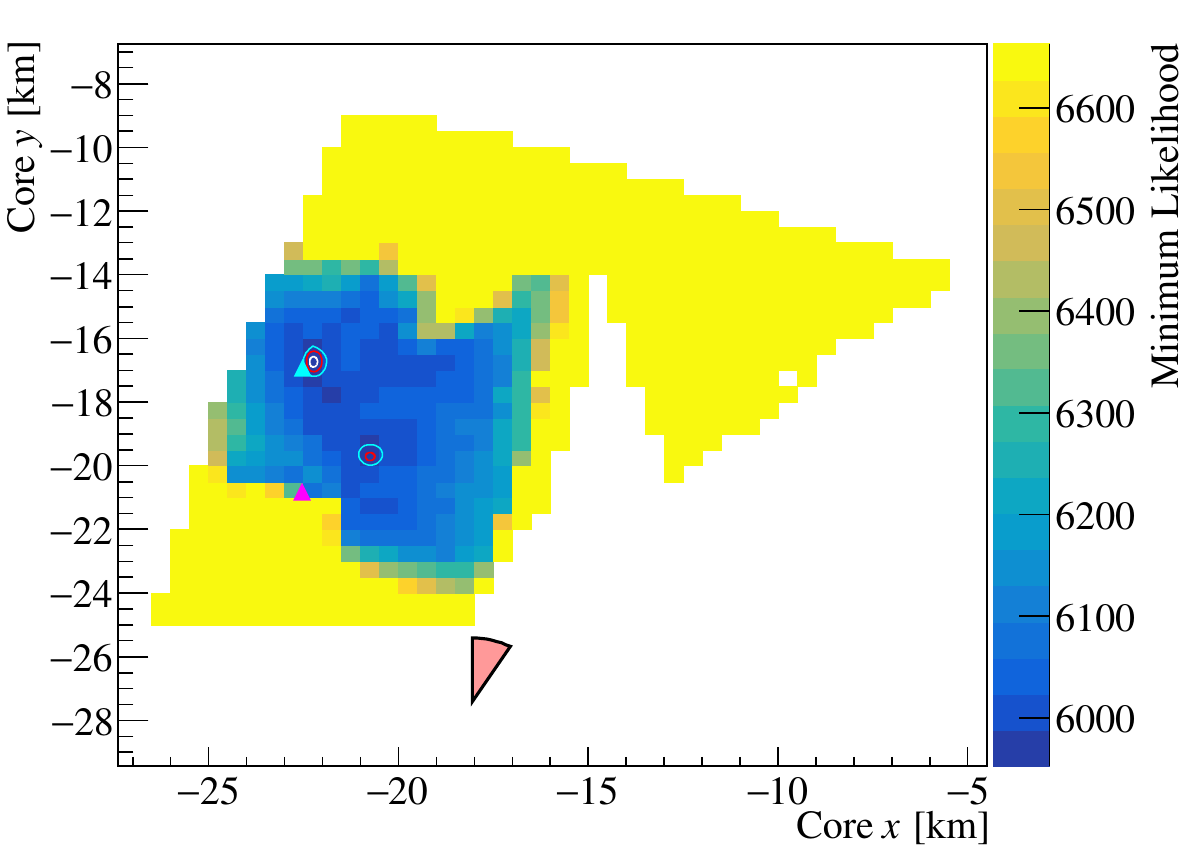}
    \includegraphics[width=0.65\linewidth]{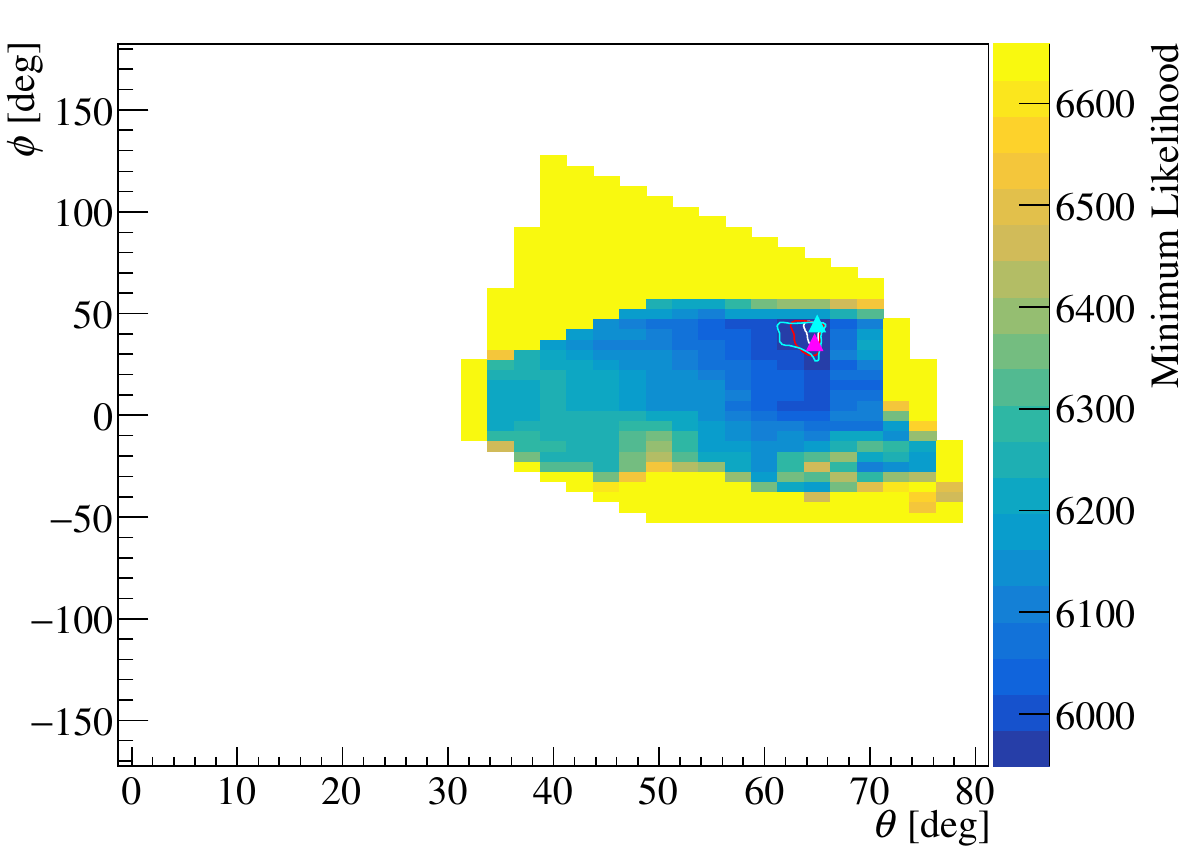}
    \includegraphics[width=0.65\linewidth]{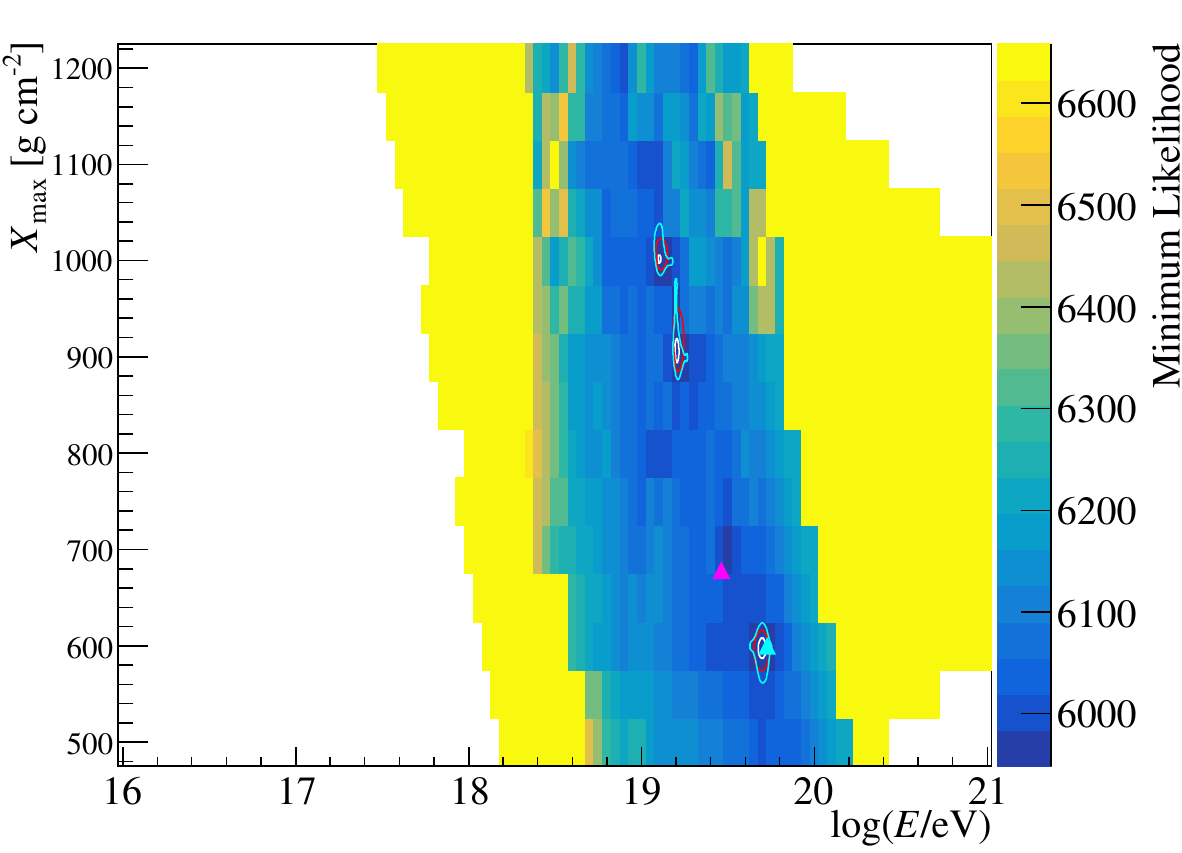}
    \caption{
    Likelihood maps from the Template Method applied to a coincidence event observed by FAST@Auger on 2022/08/30. The white/red/cyan lines correspond to the 1$\sigma$, 2$\sigma$ and 3$\sigma$ contours above the minimum likelihood. The cyan triangles show the positions of the found minima, and the pink triangles the TA reconstructed values.}
    \label{fig:paoTMexampMaps}
\end{figure}

\begin{figure}
    \centering
    \includegraphics[width=1\linewidth]{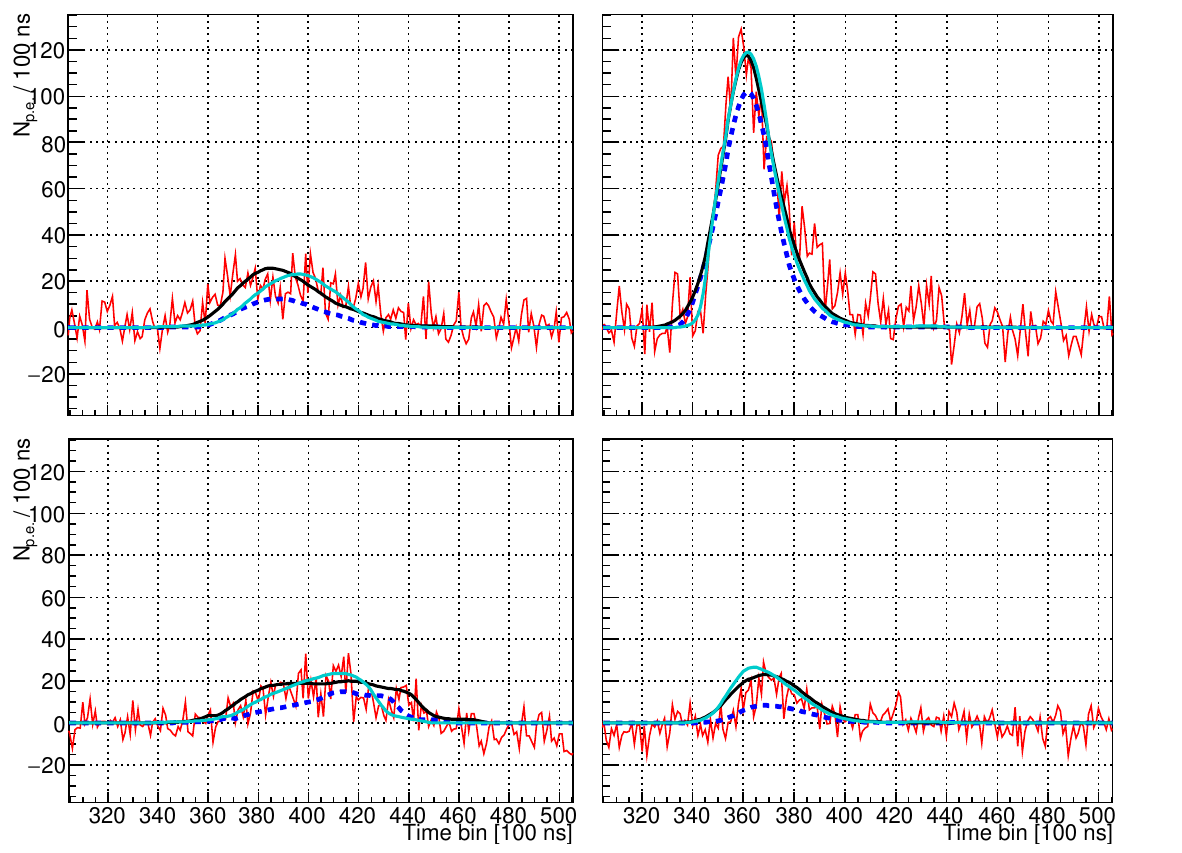}
    \includegraphics[width=1\linewidth]{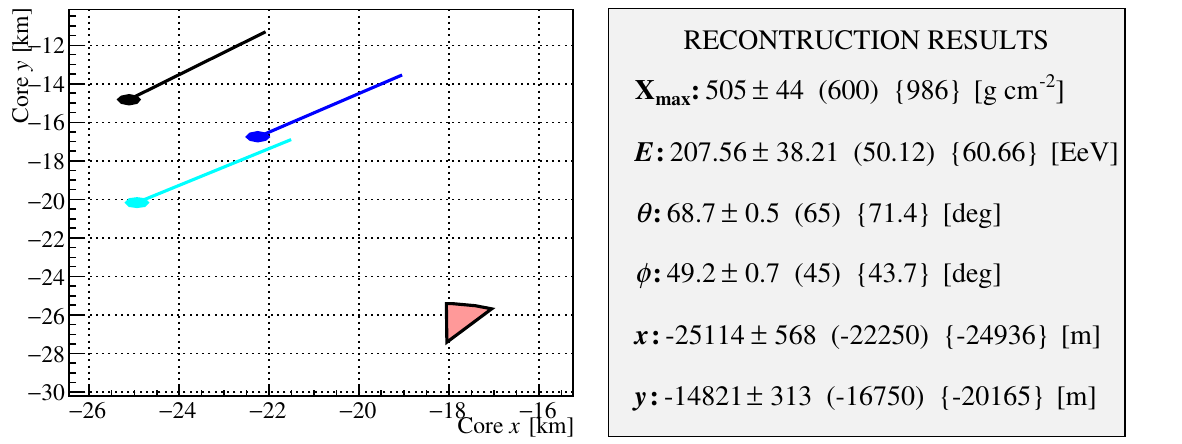}
    \caption{Template Method + TDR result for the event in Figure \ref{fig:paoTMexampMaps}. The format of the figure is the same as Figure \ref{fig:taMLexamp1}. As a reminder, the data traces are shown in red, whilst the blue, black and cyan traces / shower axes in the bottom left plot show the results for the Template Method first guess, Template Method + TDR best fit, and the Auger first guess + TDR best fit respectively. The reconstruction results are shown from left to right in the bottom right panel as Template Method + TDR best fit, Template Method first guess and Auger first guess + TDR best fit.}
    \label{fig:paoTMexamp}
\end{figure}

\begin{figure}
    \centering
    \includegraphics[width=0.65\linewidth]{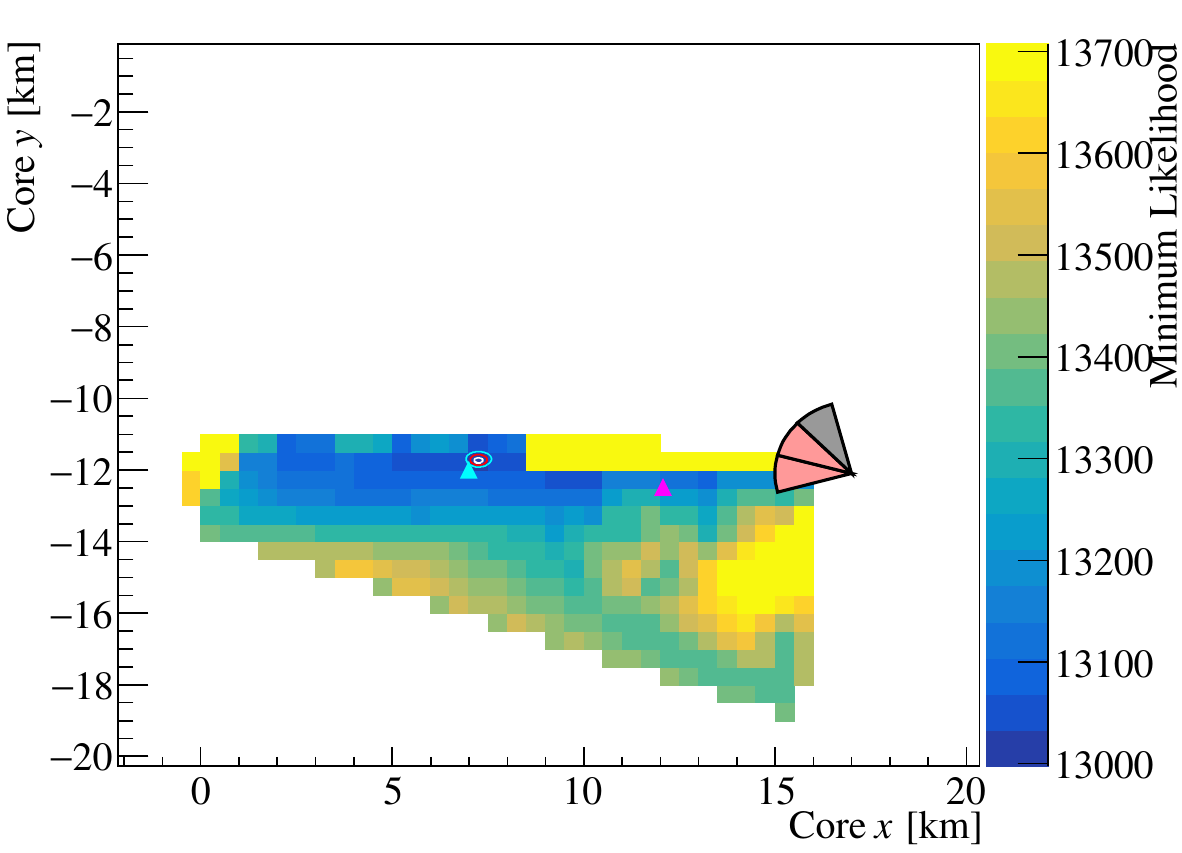}
    \includegraphics[width=0.65\linewidth]{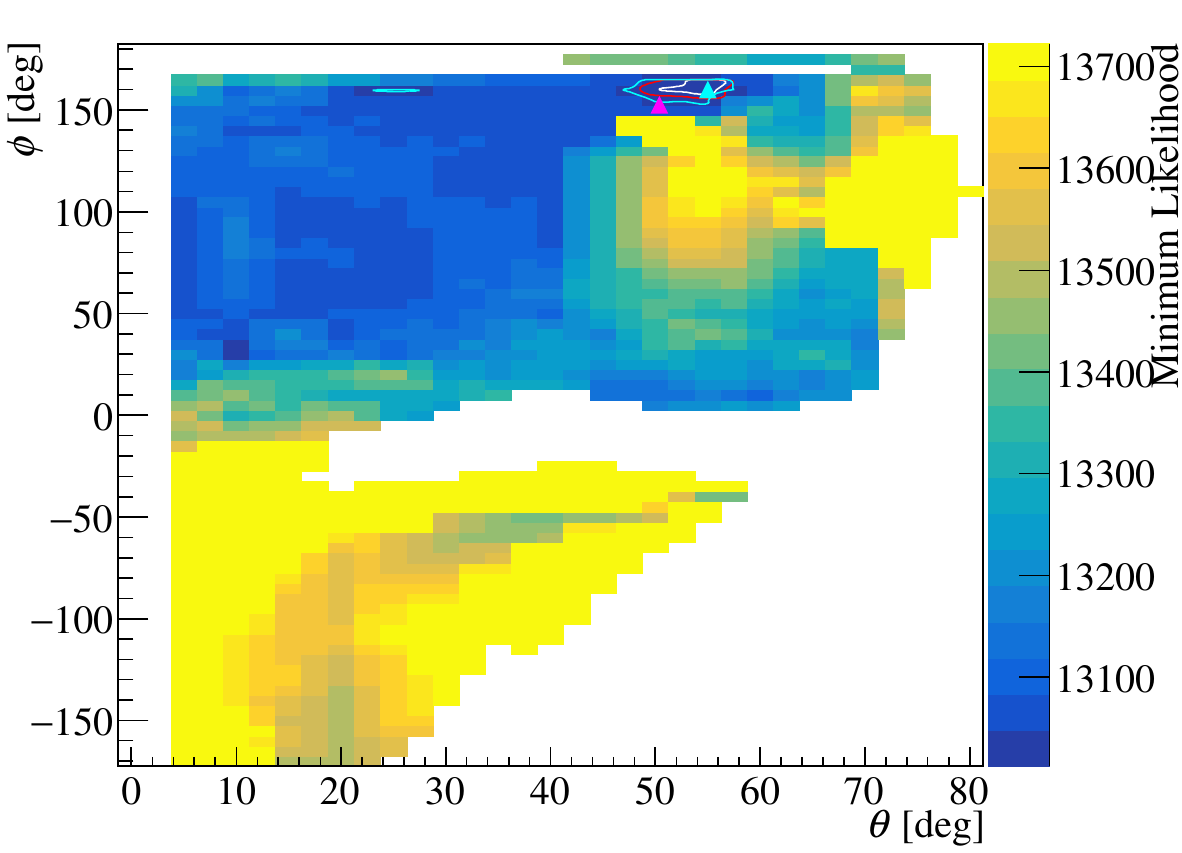}
    \includegraphics[width=0.65\linewidth]{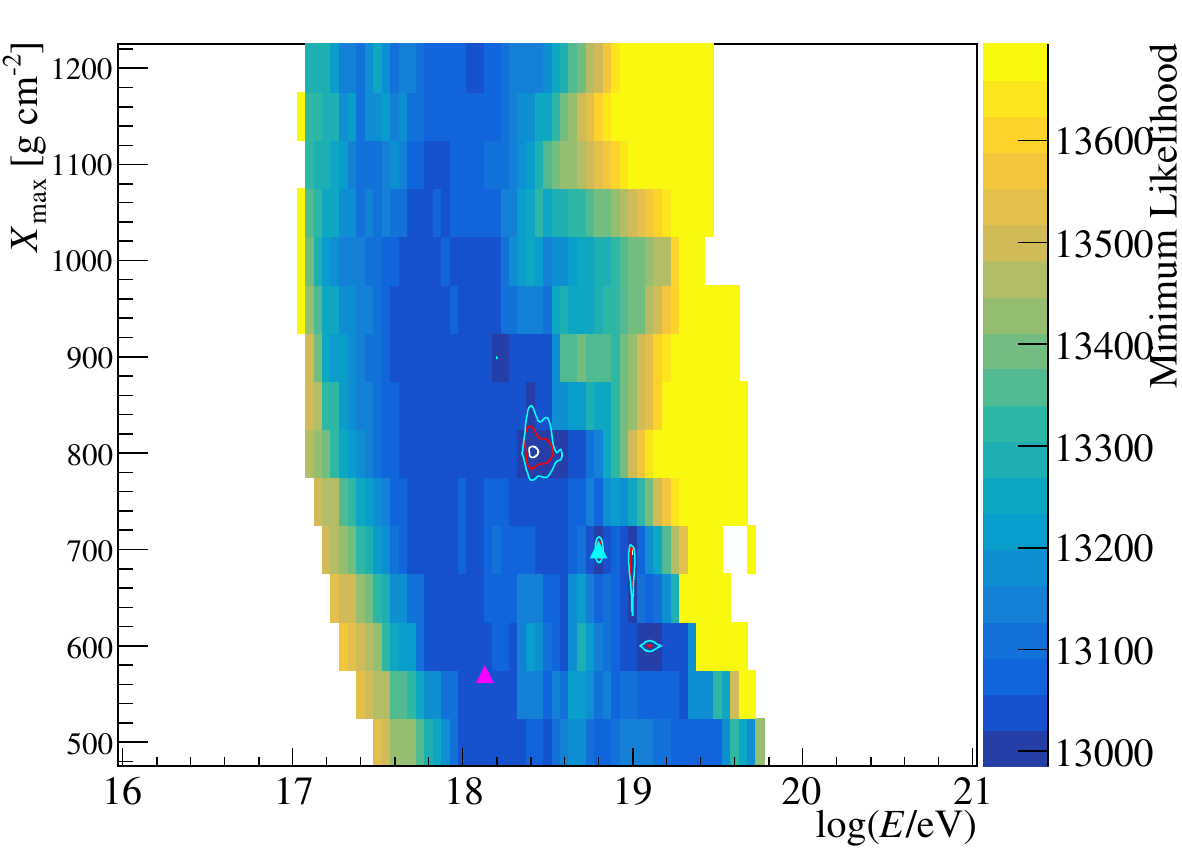}
    \caption{Likelihood maps from the Template Method applied to a coincidence event observed by FAST@TA on 2019/06/07. The layout of the figure is the same as Figure \ref{fig:paoTMexampMaps}.}
    \label{fig:taTMexampMaps}
\end{figure}

\begin{figure}
    \centering
    \includegraphics[width=0.75\linewidth]{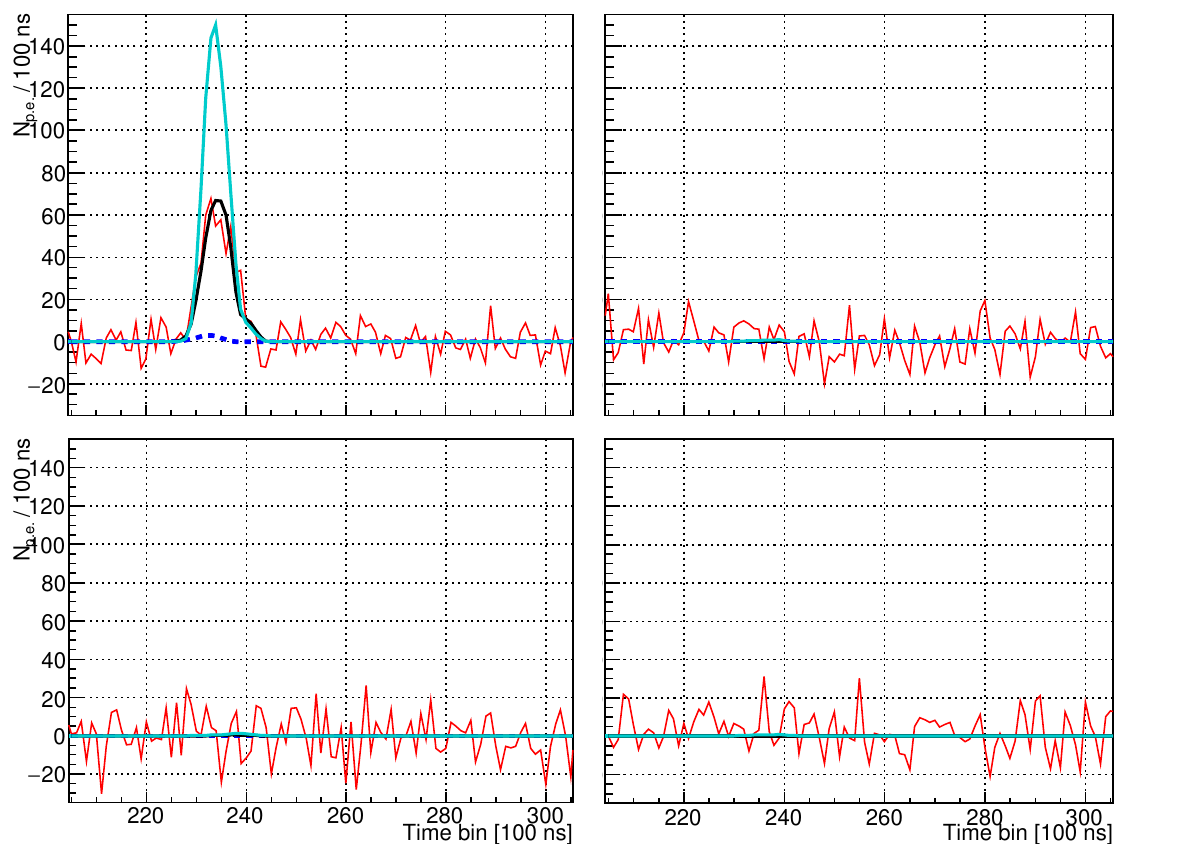}
    \includegraphics[width=0.75\linewidth]{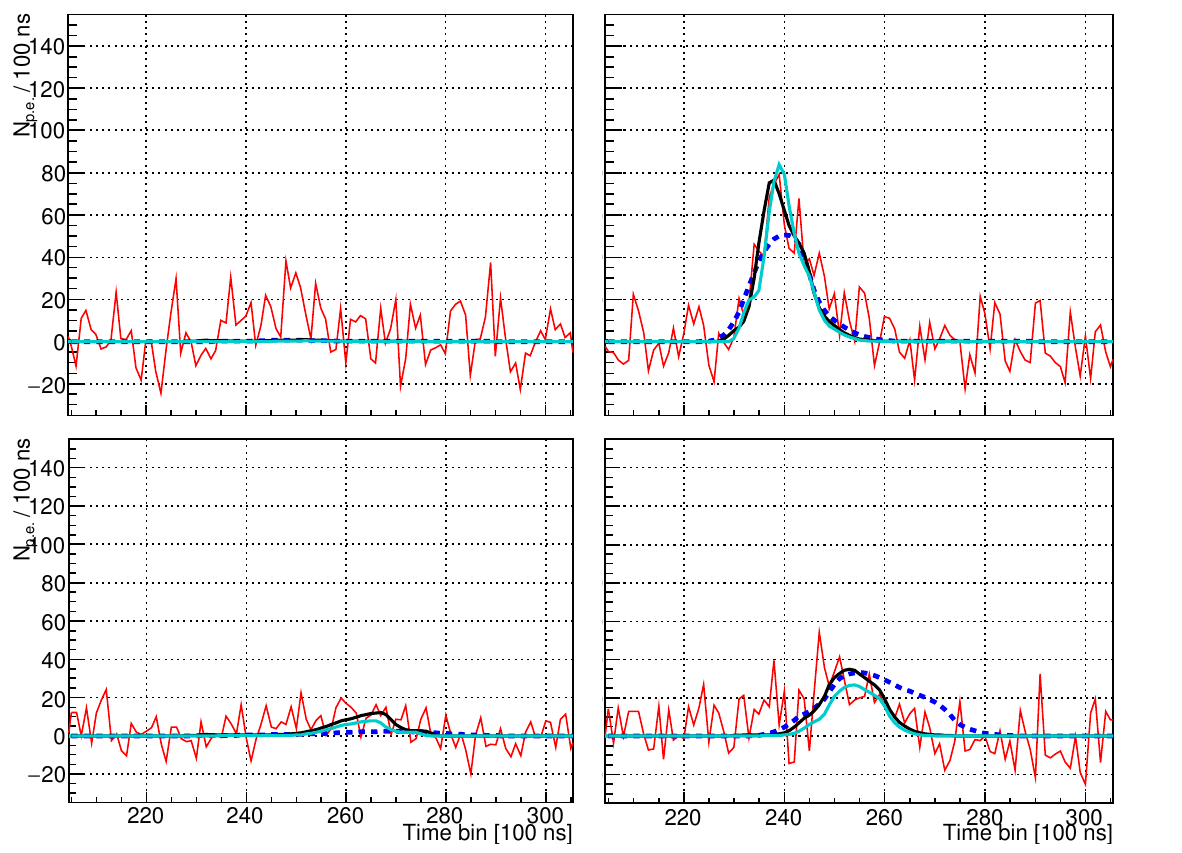}
    \includegraphics[width=1\linewidth]{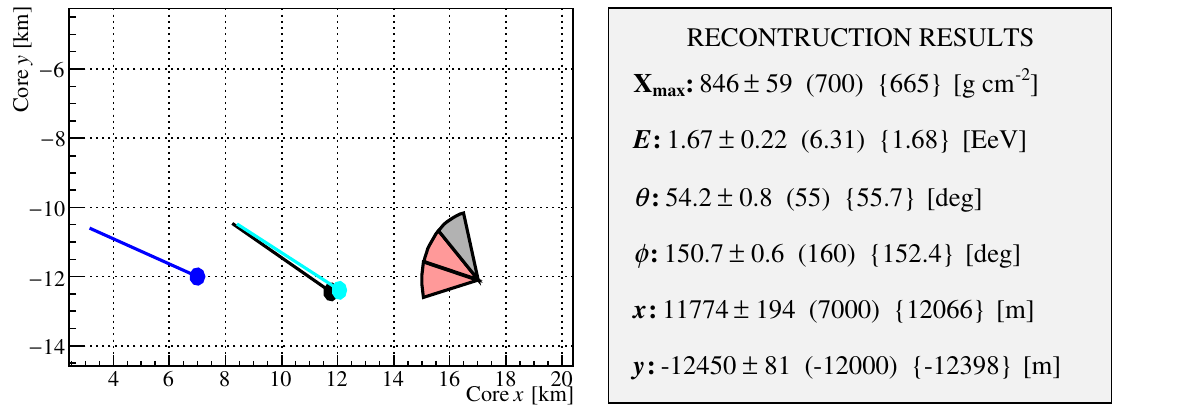}
    \caption{Template Method + TDR result for the event in Figure \ref{fig:taTMexampMaps}. The top (middle) set of traces correspond to the middle (bottom) telescope. The format of the figure is the same as Figure \ref{fig:paoTMexamp}.}
    \label{fig:taTMexamp}
\end{figure}

No special considerations were made for applying the Template Method to coincidence data. The method as presented in Chapter \ref{ch:TEMP} was applied to all events with \textit{at least} two triggered PMTs. This was 203 events for FAST@TA and 160 events for FAST@Auger. Using the Template Method first guess as inputs, the TDR was then applied. The number of events remaining after quality cuts was 165 for FAST@TA and 91 for FAST@Auger (again no \Xmax{} in FOV cut). The larger percentage of showers cut from FAST@Auger comes from more failed/poor minimisations. These occur because of the larger baseline fluctuations for FAST@Auger, causing the number of PMTs passing the event level trigger which do not contain any significant signal to be greater. When the template method is applied to an event with such pixels, the calculation of candidate templates based on the signal ratio and centroid time differences is affected and can cause templates not representative of the data to be selected, ultimately leading to a poor first guess/final reconstruction. The event-by-event differences between the Template Method first guess / Template Method + TDR and TA values are shown in Figure \ref{fig:realtempdiffta} as the black and orange histograms respectively (for showers which passed the TDR cuts). A similar plot for FAST@Auger is shown in \ref{fig:realtempdiffauger}. 

\vspace{5mm}

For the Template Method first guess, the same degeneracy trends as found with the machine learning method are present here, with large biases in the energy and core parameters. There is also a bias in zenith angle where the template method guesses larger values. The cause of this is not understood. Perhaps most surprising is that the FAST first-guess energy is occasionally $\sim10^{4}$ times greater than that of it's companion experiment. This is caused by the energy calculation method in Equation \ref{eqn:templateEnergy} - when testing templates which have extremely small signals (e.g. far from the telescope) but shapes which match the data, the energy of the shower may be estimated to be on the order of 10$^{20}$ - 10$^{21}$\,eV. Combined with a low reconstructed TA/Auger energy, such differences can arise. In principle there is no issue with a large reconstructed energy, after all these are the showers FAST aims to measure. However the frequency with which the current first guess methods seem to predict them does not match existing measurements of the energy spectrum. Once again stereo observation is critical here to reduce the energy/core position degeneracy.

\vspace{5mm}

Looking at the Template Method + TDR results, there is, as was the case with the TSFEL DNN results, minimal change in the overall distribution compared to just the first guess. A visual inspection of the Template Method + TDR fits showed reasonable matching with the data for $\sim75\%$ of events for FAST@Auger and $\sim90\%$ of events for FAST@TA. The increase in the percentage of \say{good} fits over the TSFEL DNN + TDR is because the Template Method actively searches for a set of traces which \say{match} the data, whilst the TSFEL DNN simply predicts the shower parameters. Figures \ref{fig:paoTMexampMaps} - \ref{fig:taTMexamp} show two examples of Template Method + TDR reconstructions. Both the likelihood maps for the first guess determination and final reconstructed traces/parameters are shown. The FAST@TA example shows an instance where the Template Method + TDR reconstruction matches the TA + TDR result, whilst the Auger example is another case where both sets of traces match the data but the final parameters are very different. Figures \ref{fig:tempdegenta} and \ref{fig:tempdegenauger} show the same 
telescope/core distance, reconstructed energy, and likelihood difference distributions as in Figure \ref{fig:mlFirstGuessTAdegen} but now when using the Template Method instead of the TSFEL DNN. For FAST@TA, the Template Method + TDR shows, on average, slightly better fits than the TA first guess + TDR, whilst for FAST@Auger the results are split roughly 50/50. These results further solidify the degeneracy issue as, even though the quality of the fits is similar, the reconstructed energy and core positions are very different, particularly for FAST@Auger. 

\vspace{5mm}

Overall, the above analysis of the FAST-Only reconstruction has shown that the principle of a first guess method + TDR can work well when applied to data, in so far as a set of traces which (reasonably) match the data can be found. However, when using only a single Eye, significant degeneracies between the fitted energy and core position arise with the current first guess methods. Although biases towards larger energies/further away core positions could be treated in the first guess methods (with different distributions of templates/training data for example), the fact that such degeneracies are even possible stresses the importance of stereo observation for FAST to achieve a reliable, accurate reconstruction of the shower parameters.

\section{UHECR Energy Spectrum with FAST}

The following two sections show the first cosmic-ray energy spectra and cosmic ray mass composition measurements from FAST. In light of the large biases observed in energy when using the TSFEL DNN and Template Method first guess estimates above, the TDR results when using the TA/Auger reconstructed values as a first guess will be used instead. This section looks at calculating the UHECR energy spectrum. Note that this calculation is simply a check of whether the current prototype measurements and analysis procedures give rough agreement with TA/Auger. There are several factors that remain to be properly considered on both the analysis and calibration fronts, let alone the low number of events, before serious statistical interpretation/comparison of the results can me made.

\vspace{5mm}

\begin{figure}[t!]
    \centering
    \includegraphics[width=1\linewidth]{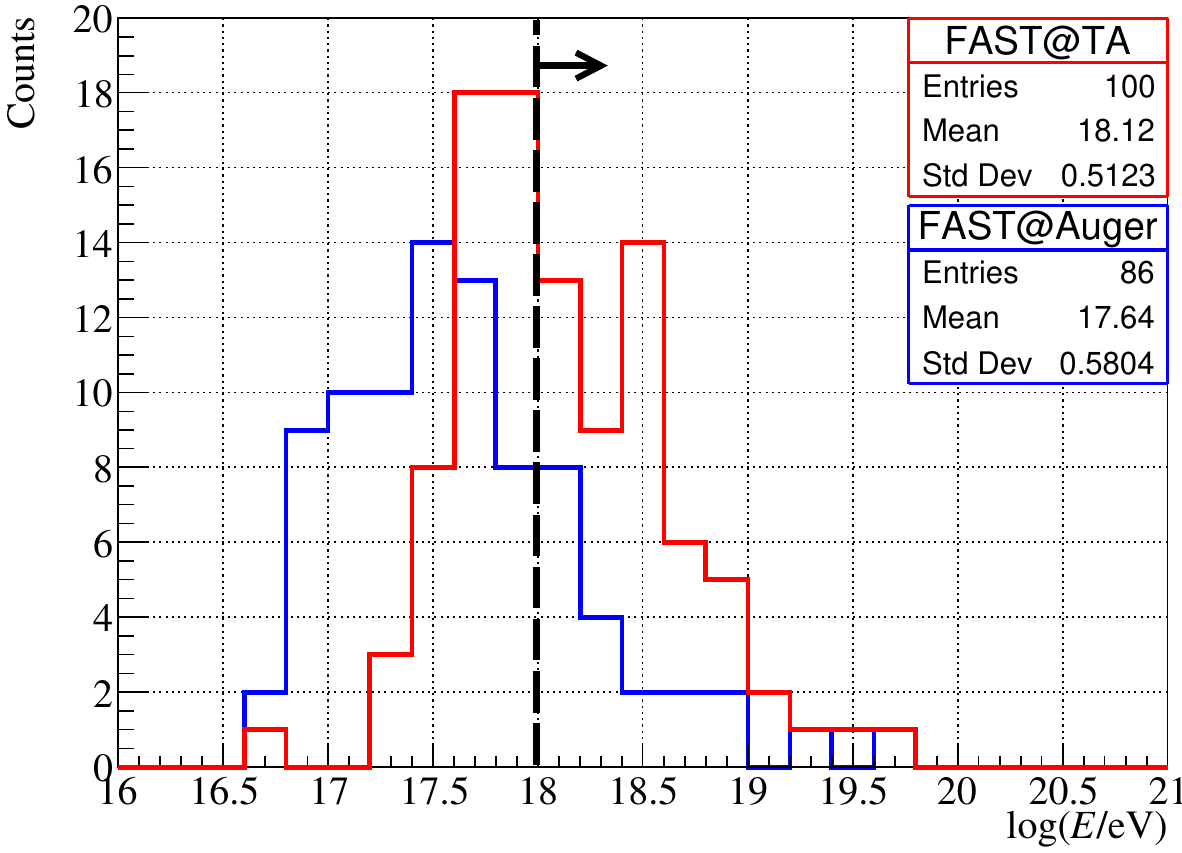}
    \caption{Distribution of reconstructed energies from FAST@TA (red) and FAST@Auger (blue). Events to the right of the black dotted line are used to calculate the energy spectra in Figure \ref{fig:energySpectrumComparison}. See the text for details.}
    \label{fig:finalEnergyDist}
\end{figure}

For FAST@Auger, the energy spectrum was calculated using the reconstructed energies from the initial reconstruction. For FAST@TA, the results in Figure \ref{fig:updatedResultDifferences} clearly indicate that calculating the energy spectrum using only the results from the unscaled directional efficiency maps would result in a heavily biased spectrum with respect to TA. However, relying entirely on TA for the correct energy scale, for example by performing a fit to the linear trend in Figure \ref{fig:timeDependence}, would remove all but the exposure calculation from the comparison. Therefore the reconstructed energies using the new, unscaled directional efficiency maps for the last time bin and the scaled versions for the first three time bins in Figure \ref{fig:timeDependence} were used. This approach still heavily relies on the TA measurements as a reference point, but assumes that the structure of the new directional efficiency maps is correct with simply two different time periods each corresponding to a different absolute calibration. Future analysis using internal triggers and purely FAST data for the entire reconstruction will (in general) not have the luxury of coincident events to check the energy scale. As such understanding the trend in Figure \ref{fig:timeDependence} is vital for the long-term success of FAST. The distributions of the reconstructed energies from FAST@TA and FAST@Auger are shown in Figure \ref{fig:finalEnergyDist}. The significant difference in the means of the distributions can be primarily attributed to the differences in the triggering/reconstruction of Auger and TA (as the FAST reconstruction using the TA/Auger reconstructed values as a first guess roughly agrees with the TA/Auger results). Only events with $E>10^{18}$\,eV were used to calculate the spectrum.

\vspace{5mm}

\begin{figure}[t!]
    \centering
    \includegraphics[width=1\linewidth]{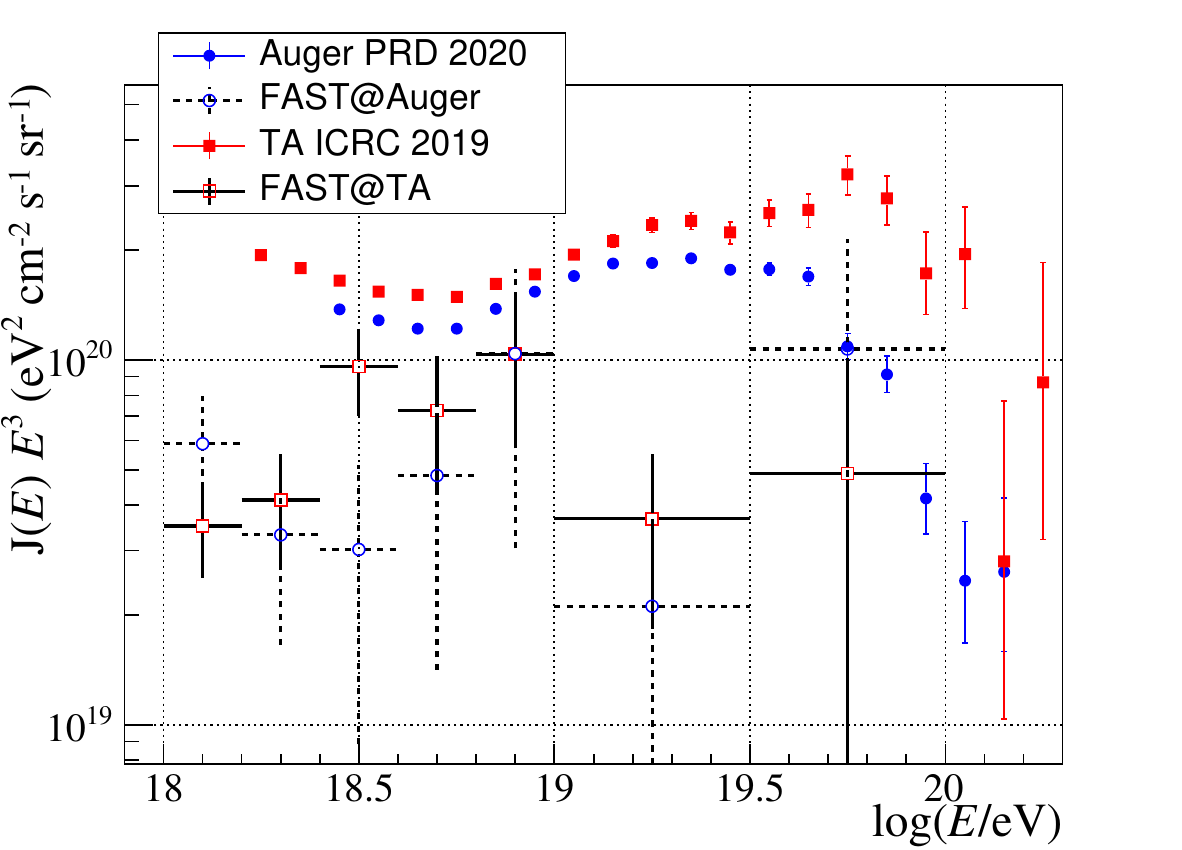}
    \caption{Energy spectra from FAST@TA and FAST@Auger estimated using the regular FAST@Auger reconstruction and the updated maps scaled and unscaled for FAST@TA. Spectra from TA \cite{ivanov2019energy} and Auger \cite{aab2020measurement} are shown for comparison.}
    \label{fig:energySpectrumComparison}
\end{figure}

The energy spectra are calculated using
\begin{equation}
\label{eqn:energySpectrum}
    J(E)=\frac{N(E)}{\textrm{exposure}(E)\times{}\textrm{d}E}
\end{equation}
where $N(E)$ is the number of events in each energy bin and d$E$ is the bin width. The exposures here are slightly different to those calculated in Figure \ref{fig:expectedPerformanceExposures}. Firstly, when calculating the efficiency, the condition requiring at least two triggered PMTs was used. This was done as, in the data/MC comparison, the ratio between the total number of coincidence events and the integral of the expected distributions was closer to one for the 2 triggered PMT histograms. Additionally, an \Xmax{} in FOV cut was incorporated into the efficiency calculation for both FAST@Auger and FAST@TA. This was done to roughly match the conditions for a successfully reconstructed event. For FAST@TA, three separate exposures were calculated; one for the time period with 2 telescopes, one for the time period with 3 telescopes before 2021 and one for the time period after 2021 also with three telescopes. These exposures were calculated using the same setup as described in Section \ref{sec:dataMCcomp} but with the new maps (unscaled after 2021, scaled before 2021). The three exposures were added to give the final exposure for FAST@TA. Using Equation \ref{eqn:energySpectrum} $J(E)$ was then calculated for FAST@TA and FAST@Auger and plotted against TA/Auger spectra for reference. The result is shown in Figure \ref{fig:energySpectrumComparison}.

\vspace{5mm}

The energy spectra for FAST@TA and FAST@Auger generally agree within statistical uncertainty. They are both lower than the TA/Auger spectra by a factor of $2\sim10$ depending on the energy. Considering the relatively small biases present in \textit{these} FAST reconstructed energies (10 - 20\%), the primary source of discrepancy is believed to be the exposure calculation. Further improvements to the exposure estimation, for example the inclusion of realistic background noise in the simulations along with the other suggestions made at the end of Section \ref{sec:dataMCcomp}, together with greater statistics and systematic uncertainty estimates will be needed to better the comparison.

\section{UHECR Composition with FAST}

\begin{figure}[t!]
    \centering
    \includegraphics[width=1\linewidth]{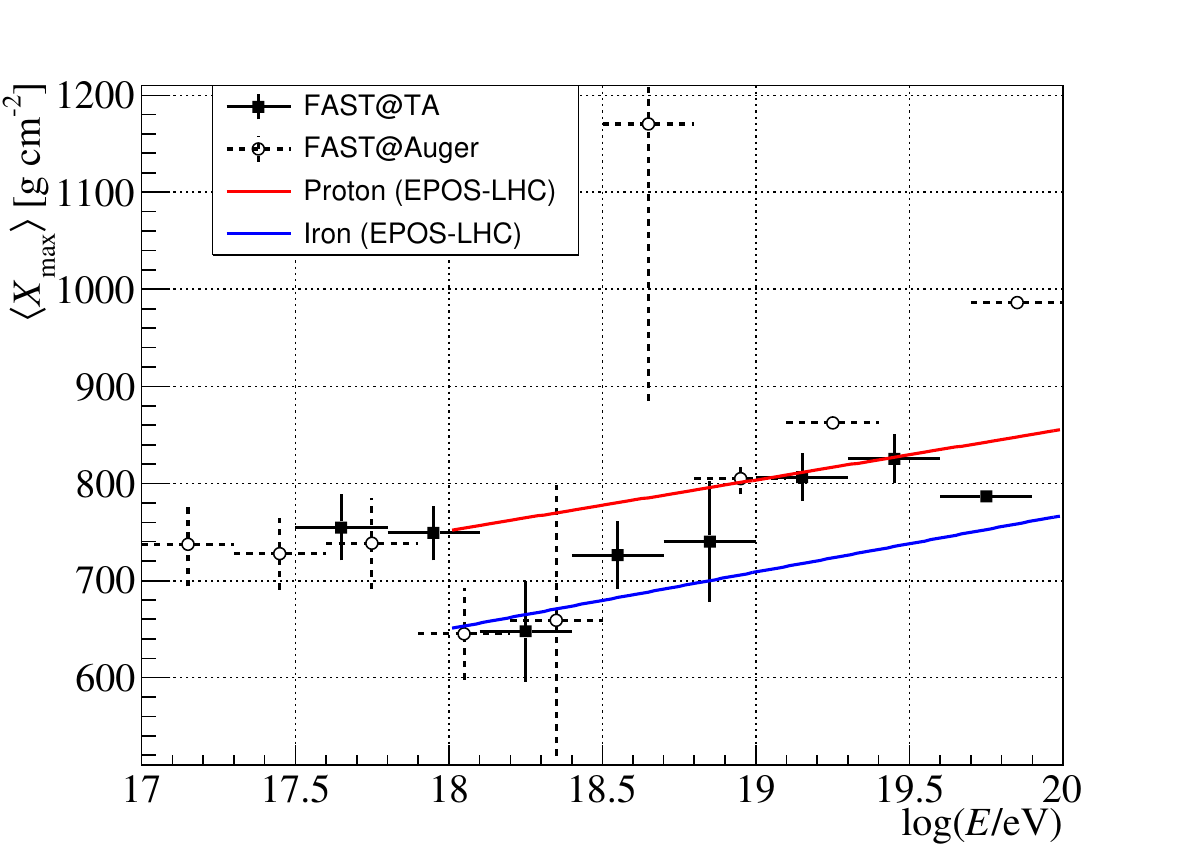}
    \caption{Average \Xmax{} as a function of energy (elongation rate) as estimated by FAST@TA (black squares) and FAST@Auger (open circles). The results are compared to the average reconstructed values of purely proton (red) and iron (blue) showers from simulations. 
    }
    \label{fig:elongationRate}
\end{figure}

To measure the cosmic ray mass composition using fluorescence detectors, typical studies analyse the elongation rate i.e. the change in average \Xmax{} with energy. Using the same set of reconstructed data as outlined in the previous section, the elongation rate for both FAST@Auger and FAST@TA was plotted. The results are shown in Figure \ref{fig:elongationRate}. Different binning is used between FAST@TA and FAST@Auger due to the offset in reconstructed energy distributions. Once again note that the goal of this section is not to shed new insights into cosmic ray mass composition, but simply to check whether the current results roughly agree with those of other experiments. To check whether the measured data (black) is consistent with a lighter/heavier composition, predictions from purely proton/iron showers are plotted for comparison. These predictions are commonly referred to as \say{rails} and, in this case, are constructed using data from FAST simulations. This is because the measured data includes detector effects/quality cuts and therefore so too should the rails for a valid comparison. 

\begin{figure}[t!]
    \centering
    \includegraphics[width=1\linewidth]{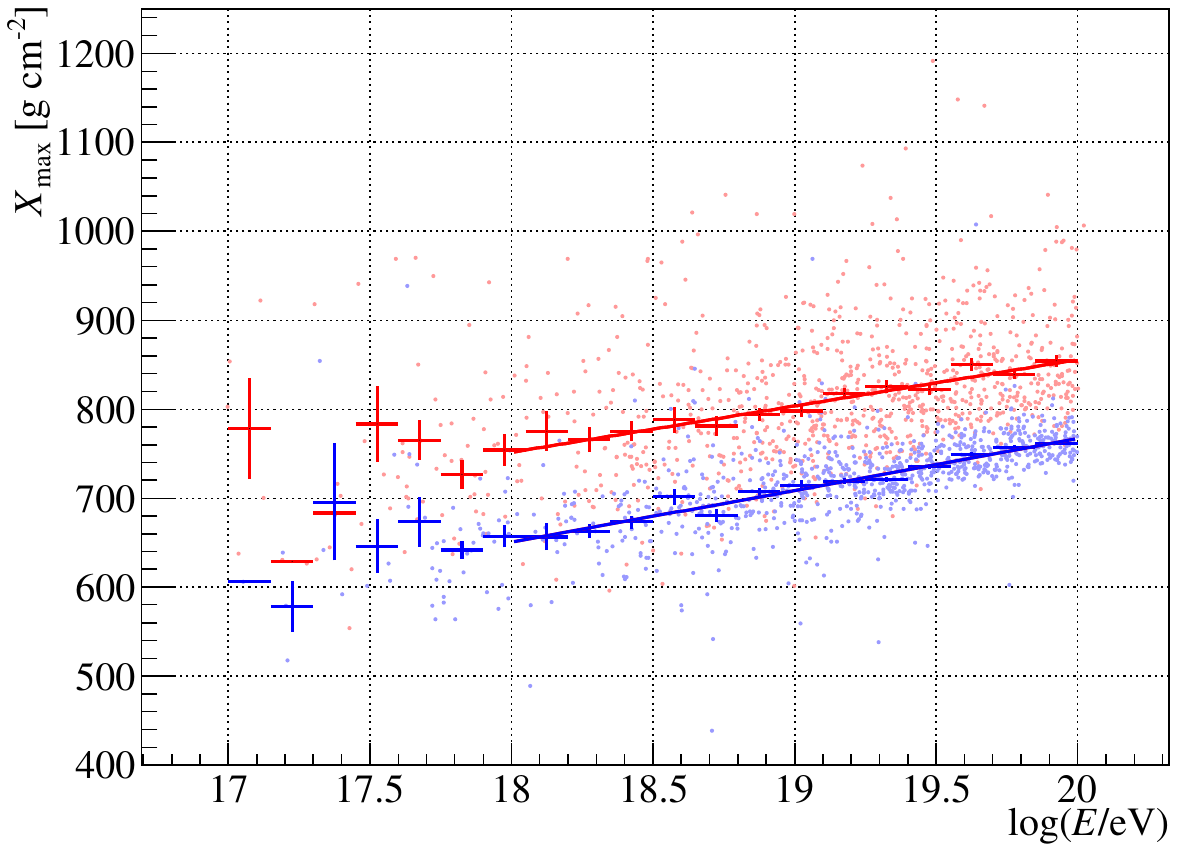}
    \caption{Construction of the \Xmax{} rails used in Figure \ref{fig:elongationRate}. The red (blue) dots represent individual \Xmax{}/energy reconstructions for proton (iron) events. The overlaid markers represent the average \Xmax{} in each bin (bin widths indicated by the horizontal bars). The fitted rails are shown by the red/blue lines.}
    \label{fig:railFits}
\end{figure}

\vspace{5mm}

The simulated showers were generated in two energy bands, $17<\log(E\mathrm{/eV})<18$ and $18<\log(E\mathrm{/eV})<20$. Each energy band contained 10,000 showers. Unlike advanced air-shower simulation tools such as CORSIKA, the FAST simulation has no method of simulating air showers from different primary nuclei. The only factor which separates proton and iron showers in the FAST simulation is their distribution of \Xmax{} values. These distributions come from parameterisations of the \Xmax{} distributions of different nuclei estimated with different hadronic interaction models (EPOS-LHC, Sibyll2.3 and QGSJetII-04) \cite{blaess2018extracting}. For these simulations the EPOS-LHC parameterisations were used. In each energy band, half of all showers had simulated \Xmax{} values sampled from the proton distribution parameterisation, and half from the iron distribution parameterisation. The showers were set incident on a single FAST telescope located at (0,0) pointing along the $y$-axis, with a realistic distribution of arrival directions and core positions uniformly distributed in a circle centred at (0\,km,8\,km) with $r=10$\,km. 

\vspace{5mm}

The simulated showers were then reconstructed using a first guess with geometry fixed to the simulated values and \Xmax{}/energy randomly smeared from their true values by 30\gcm{} and 10\% respectively. Only showers with at least one PMT passing the threshold trigger and passing the reconstructoin cuts outlined in Section \ref{sec:initRecon} (with the exception of the timing cut), were used to estimate the rails. The fits were performed over the energy range $18<\log(E\mathrm{/eV})<20$ and are shown in Figure \ref{fig:railFits}. Even though many coincidence events passing the reconstruction cuts have an energy lower than 10$^{18}$\,eV, the rails are not extended below this energy for two reasons. Firstly, the number of simulated showers with a reconstructed energy below 10$^{18}$\,eV is very low. Secondly, lower energy showers, which develop higher up in the atmosphere and have smaller \Xmax{} values, will only pass the \Xmax{} in FOV cut if they possess a relatively \textit{large} \Xmax{} for their energy. This causes the \say{detected} average \Xmax{} at lower energies to be artificially high and creates the flattening below $\sim10^{18}$\,eV observed in Figure \ref{fig:railFits}. The appropriate functional form to model the detected average \Xmax{} in this region is uncertain. When combined with the low statistics, it was decided to restrict the fits to data above 10$^{18}$\,eV.

\vspace{5mm}

Regarding the results in Figure \ref{fig:elongationRate}, the FAST@TA and FAST@Auger measurements appear roughly consistent, ignoring the bins which contain only a single data point (those with no error bars). Taking the data together as a whole, between approximately $10^{18}$ - $10^{19.5}$\,eV, the average \Xmax{} appears to trend upwards, from a heavier composition to lighter composition. However, considering the highly limited statistics in this range ($\sim$20 showers) no definite conclusions can be drawn at this time. The elongation rate as measured by other experiments shows a trend from heavier composition to lighter composition between $10^{17}\sim10^{18.5}$ - $10^{19}$\,eV depending on the experiment. As observation with FAST continues and improvements to the analysis made, it will be telling to see whether FAST observes this key feature.

\section{Summary}
This chapter has presented the first results of applying the entire FAST reconstruction process to a large coincidence data set from the prototype telescopes at FAST@TA and FAST@Auger. Thanks to a new triggering algorithm, roughly 650 coincidence events were identified between the FAST installations and TA/Auger. Comparisons between the TA/Auger reconstructed values for these showers and the expectations from FAST simulations showed reasonable agreement. Reconstructing the coincidence events using the TA/Auger reconstructed values as the first guess yielded 86 showers from each installation passing the quality cuts. The FAST reconstructed energies for these showers were found to be 10 - 20\% larger than those of TA/Auger. The FAST@TA reconstructions were performed again after accounting for the non-uniformity of the PMT gain response measured in situ at FAST@TA. The reduced absolute scale of the new directional efficiency maps calculated from these measurements introduced an even larger bias in the energy. This indicated an unaccounted for calibration factor. Scaling the new maps to match the overall integral of the original maps reverted the energy bias to its previous size and improved the event-by-event agreement in energy between FAST and TA. A key finding was that the difference in FAST and TA reconstructed energies seems to be decreasing with time. Understanding this discrepancy is paramount for the long-term deployment of FAST. 

\vspace{5mm}

The first guess methods developed in Chapters \ref{ch:ML} and \ref{ch:TEMP} were then applied to the coincidence data. Results showed that, at least on the level of the PMT traces themselves, there are some fundamental differences between data and simulations which are unaccounted for. Moreover both methods showed a tendency to estimate the coincidence event parameters  with higher energies and core distances further away from the telescope than reconstructed by TA/Auger. The subsequent application of the TDR was found not to significantly improve the agreement, however the majority of showers passing all reconstruction cuts were found to have best fitting traces which reasonably matched the data. This is a promising sign for FAST and verifies that the first guess + TDR method can be \say{successfully} applied to data. Introducing stereo observation is expected to greatly reduce the degeneracies in the shower parameters and improve agreement between the TA/Auger results.

\vspace{5mm}

Finally, the first measurements of the UHECR energy spectrum and composition with FAST were made. The energy spectra from FAST@TA and FAST@Auger were both found to be lower than the TA/Auger spectra. The primary source of discrepancy is believed to be the exposure calculation. Future analyses will require improvements to the exposure calculation and quality cuts, in addition to more statistics, for quantitative comparisons to be made.

\chapter{Conclusions}

This thesis has focused on the development of the reconstruction procedure of a next-generation cosmic ray experiment; The Fluorescence detector Array of Single-pixel Telescopes. The primary aim of FAST - to observe UHECRs with unprecedented statistics and thus uncover the nature of their sources - cannot be achieved without a precise and accurate reconstruction of the extensive air showers initiated by these particles. 
Prior to this work, studies into the \say{top-down reconstruction (TDR)} method used by FAST were mainly limited to those investigating the performance of the FAST-3 Eye layout in simulations. Although promising, these studies provided limited insights into the expected performance of the current or near future prototype installations. Furthermore, the reconstruction had not been applied to a large sample of data from the prototype telescopes with which to test its efficacy. Previous studies also found two main difficulties with the TDR. The first of these was a strong reliance on an accurate first guess for successful minimisation, and the second was a significant issue regarding the reconstruction efficiency at high energies. This thesis has attempted to address these points. In doing so, a greater understanding of the potential and limitations of the FAST design has been obtained.

\vspace{5mm}

In Chapter \ref{ch:TDR}, the primary issue with the TDR, a decrease in the reconstruction efficiency with energy, was fixed. This was done by amending several sections of the FAST simulation code which were causing discontinuities in the likelihood function. Whilst the cause of some of these discontinuities were simple oversights, the discontinuities caused by the previous shower-axis binning and trace re-binning methods in particular were found to arise uniquely because of the likelihood function used. Specifically, calculating the likelihood as a sum of contributions from each individual bin means that, for the area around the minimum to be smooth (and thus achieve a successful minimisation using gradient based approaches), the \textit{value of every single bin} must change smoothly as a function of the shower parameters at small scales. This was not the case for the original shower-axis binning and trace-re-binning methods. These discontinuities were enough to cause many failed minimisations, leading to a decrease in reconstruction efficiency. By tweaking the methods to ensure smoothness in the likelihood function, the reconstruction efficiency increased to a roughly constant 95\% across all energies. This improvement will allow FAST to more accurately and reliably reconstruct detected showers.
The second issue addressed with the TDR was a correlation between \Xmax{} and energy when fitted simultaneously. The correlation was found to arise when attempting to reconstruct showers with \Xmax{} outside the FOV of the triggered telescopes. For such showers, only the tail of the generated Gaisser-Hillas profile is visible, causing a large degeneracy in the fitted energy and \Xmax{}. The problem was resolved by simply removing such showers from the analysis, similar to the FOV cuts performed by Auger and TA. Using simulations, including the cut was shown to improve the resolutions in the reconstructed \Xmax{} and energy.
Future work should focus on investigating the remaining fail states of the TDR, estimating the systematic errors in the reconstruction (i.e. dependence of final result on the first guess) and on speeding up the simulation procedure to allow for reconstructions, and hence analyses, to be performed much more efficiently.


\vspace{5mm}

Chapter \ref{ch:ML} was the first of two chapters to investigate possible \say{first guess methods} for use by the TDR. It presented the first results of using neural networks with the current and near future FAST-prototype layouts to obtain a first guess of the shower parameters. Three different models were tested. The first of these, the Basic DNN, had the same architecture and inputs as previous machine learning studies with FAST. The second model, the TSFEL DNN, extended the Basic DNN by adding eight additional inputs per PMT. The third model, the LSTM network, utilised the Long-Short Term memory architecture. The TSFEL DNN was found to be the best overall choice for the simulated data set, giving similar validation losses as the LSTM network for significantly less computational effort. The model was found to perform best when reconstructing showers observed in stereo or triggering $\gtrsim4$ pixels in a single eye. The best resolutions were obtained with the FAST-MiniV2 model above a true energy of $10^{19}$\,eV for 3-Eye stereo observation. The \Xmax{}, energy, core and angular resolutions were 75\gcm{}, 30\%, 750\,m and $6.5\degree$ respectively. 
When using the output of the TSFEL DNN as the input to the TDR, the performance was found to improve across all layouts. For FAST-MiniV2 specifically, the resolutions obtained fell within the requirements of a future GCOS candidate, namely $\sigma$(\Xmax)<30\gcm{}, $\sigma{}(E)<10\%$ and angular resolution $<1\degree$. This is a significant achievement for FAST. The large caveat here is that these results were obtained with perfectly ideal simulations and no considerations for systematic uncertainties/differences between simulations/data. Future work should look to include such uncertainties and re-evaluate the expected resolutions. The other primary take away from this chapter was that, for just one or two triggered PMTs in a single FAST Eye, accurate reconstruction appears extremely difficult. Future studies should investigate the impact on the efficiency and potential biases introduced by ignoring such showers. Further hyper-parameter optimisation of the current model and/or testing more sophisticated architectures which could account for slight variations in telescope position/pointing (such as transformers or graph neural networks) are additional avenues for future work. 

\vspace{5mm}

In Chapter \ref{ch:TEMP} an alternative method for deriving the first guess parameters was studied. This method, dubbed the \say{Template Method}, utilised a library of templates which consisted of sets of traces from a single telescope. Given a set of data traces from a single telescope, Equation \ref{eqn:TDLikelihood} was used to evaluate the likelihood of each template vs. the data. By sampling the six-dimensional likelihood function for each telescope which observed the shower separately and then combining the results, a first guess of the shower parameters was able to be obtained regardless of the number/layout of telescopes. This is the first method for estimating the shower parameters (outside of the TDR) where this is possible. This flexibility 
is a major advantage over the TSFEL DNN developed in Chapter \ref{ch:ML}, which would require new models to be trained every time there is an additional telescope added to the current prototype installations or changes to the positioning/pointing direction of current telescopes are made. Such changes are likely to be frequent as FAST begins to deploy more and more telescopes. Another benefit of the approach is the ability to visualise the uncertainty on the first guess parameters. The heavy computational requirements of the Template Method were reduced by only comparing a set of data traces to templates which had similar relative signal ratios and timing differences between triggered PMTs. This reduced the number of templates required to be checked from over one million to typically $\sim5,000$ - $50,000$, with the necessary amount depending strongly on the number of triggered PMTs. Evaluating the performance of the method showed some similar trends to the machine learning results, though overall the resolutions were inferior. This was thought to be due to the relatively large spacing between the templates and not accounting for all correlations between showers parameters when combing results from different telescopes. Both of these points can be addressed in future work. In particular, an interpolation of the likelihood function in the six-dimensional parameter space using a denser set of templates is expected to improve the resolutions in the reconstructed parameters. Overall, provided some additional improvements to the speed and accuracy of the method can be made, the Template Method is a promising approach to estimate the first guess parameters. This is because it directly searches for a good match to the data traces and doesn't rely on complicated relationships between specific trace features and outputs present in machine learning approaches, which may be more vulnerable to systematic differences between simulations (used for training) and data.

\vspace{5mm}

The final chapter of this thesis applied the improved reconstruction procedure to data from the FAST prototypes. A total of approximately 650 coincidence events, identified by way of an improved signal search algorithm, were analysed. This analysis presented the first application of the entire FAST reconstruction chain to a large sample of prototype data. Initial comparisons between the TA/Auger reconstructed values for these showers and expectations from FAST simulations showed reasonable agreement. However the difficulty in accurately reproducing the efficiency for detecting such events, which requires reproducing the Auger/TA triggering and reconstruction processes, together with poor statistics above the minimum energy for the analysis (10$^{18}$\,eV), meant no quantitative comparisons were performed. Reconstructing the coincidence events yielded 86 showers from each installation passing the quality cuts. The applied cuts are the bare minimum for what should be considered an acceptable event. It is likely that future cuts, either on the number of PMTs or on the minimum/maximum SNR of the detected signals, will improve results. Showers passing the cuts showed slight biases in the reconstructed energy of 10 - 20\%. 
After accounting for non-uniformity across the PMT surfaces obtained via in-situ measurements, the FAST@TA events were reconstructed once again. This revealed a time-dependent decrease in the FAST reconstructed energy with respect to TA. The cause of this trend is unknown, though it could feasibly be connected to deterioration of the PMTs. If this is the case then accurate measurements will need to taken over an extended period of time to properly correct for this effect. No other significant biases were found, and there appeared reasonable agreement between the FAST reconstructed values and those of TA/Auger. Although this is likely due to using the TA/Auger reconstructed values as the first guess, it was still encouraging to verify that the TDR can (in principle, given a high quality first guess) accurately reconstruct the shower parameters.
The first guess methods developed in Chapters \ref{ch:ML} and \ref{ch:TEMP} were then applied to the coincidence data. Results showed promising signs that the first guess + TDR approach can be reliably applied to data. The majority of events passing all reconstruction cuts had a set of best fit traces which matched the data at least as well as the best fit traces from the TA/Auger first guess + TDR. However, there were significant differences in the reconstructed core positions and energies between these reconstructions, again highlighting the difficulty in reconstructing the shower parameters with a single FAST Eye and underscoring the need for stereo observation. 
Finally, measurements of the UHECR energy spectrum and composition with FAST were made. These were basic checks to confirm whether the analysis procedures yielded results consistent with those of other experiments. Differences were found in the measured energy spectrum and composition, however the low statistics and uncertainty in the exposure calculation mean no quantitative conclusions were drawn. 

\vspace{5mm}

In conclusion, this thesis has achieved it's primary goals of addressing issues with the TDR, developing first guess methods applicable to the current and near-future FAST prototype layouts, and applying these developments to data. The primary takeaway is that stereo observation is essential for a reliable reconstruction using the current FAST design. The upcoming FAST mini-array plans to verify this. Immediate focus should be directed towards improving the first guess of the shower parameters for the FAST mini-array layouts, as this will be pivotal in achieving accurate reconstructions. If this can be accomplished, it will make a strong case for the large-scale deployment of FAST, which will in turn help to elucidate the origins of UHECRs.

\appendix
\chapter{Top Down Reconstruction Improvements: Supplemental Plots}

\section{Improved Reconstruction: Parameter Resolutions}
\label{apx:improvedReconFits}
\begin{figure}[h!]
    \centering
    \includegraphics[width=1\linewidth]{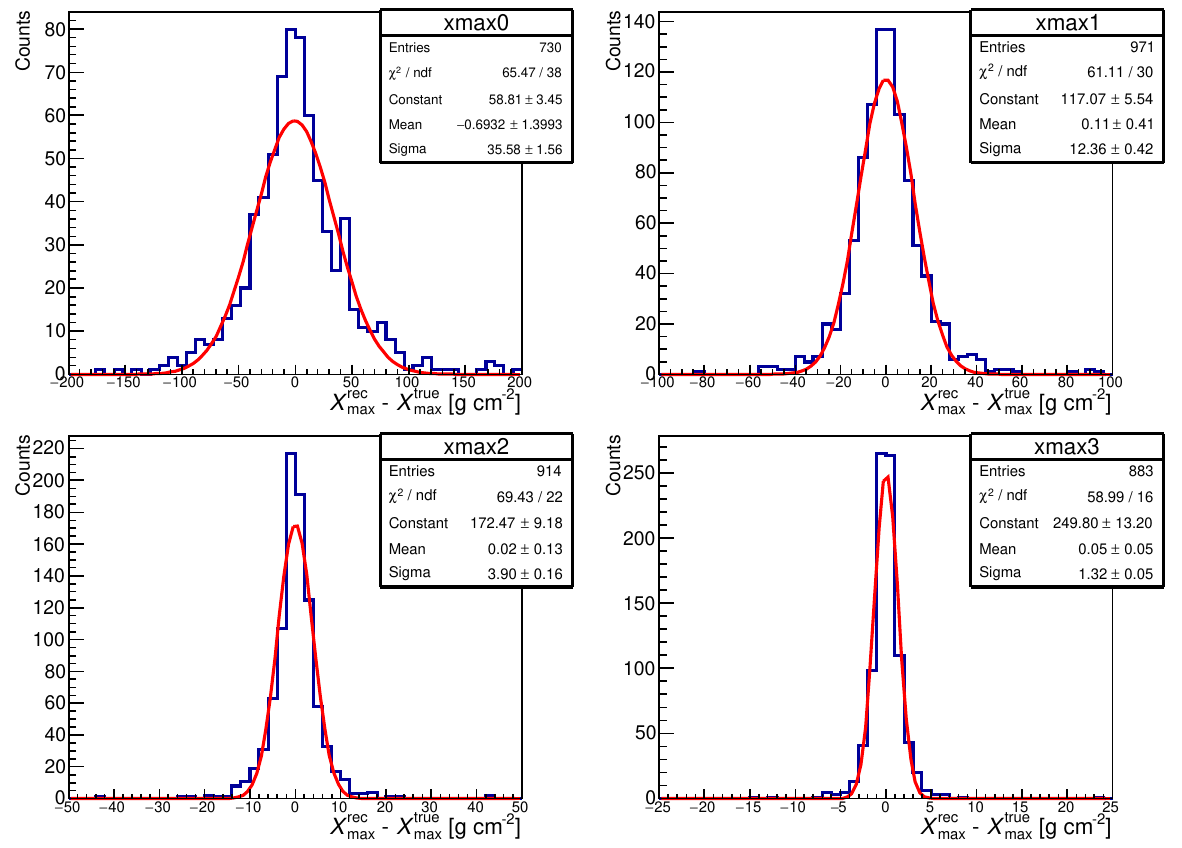}
    \caption{Gaussian fits to the distributions of the differences between true and reconstructed \Xmax{} values from the reconstructions performed in Section \ref{sec:CheckingRecon}. \textit{Top left:} $\textrm{log}(E/\textrm{eV})=18.5$. \textit{Top right:} $\textrm{log}(E/\textrm{eV})=19$. \textit{Bottom left:} $\textrm{log}(E/\textrm{eV})=18.5$. \textit{Bottom right:} $\textrm{log}(E/\textrm{eV})=20$.}
    \label{fig:backupfitsXmax}
\end{figure}

\begin{figure}
    \centering
    \includegraphics[width=1\linewidth]{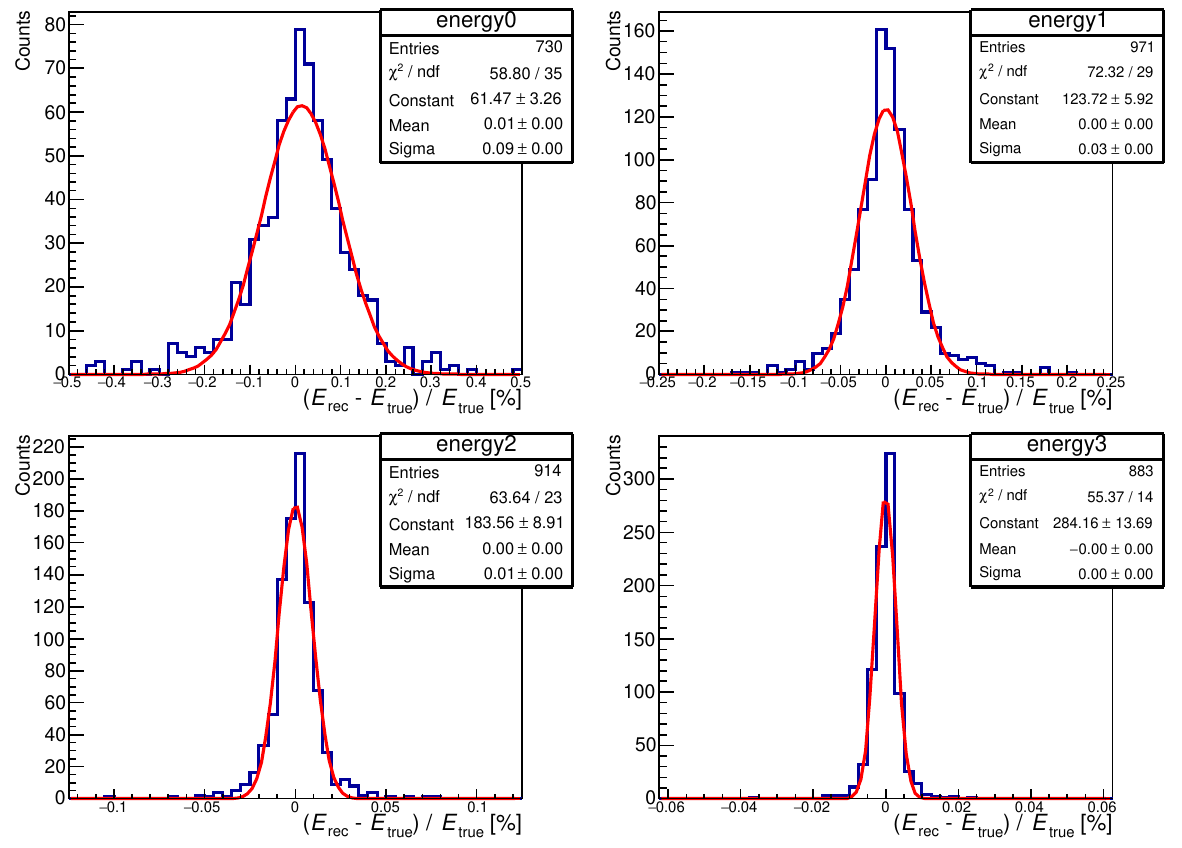}
    \caption{Same as Figure \ref{fig:backupfitsXmax} but for energy.}
    \label{fig:backupFitsEnergy}
\end{figure}

\begin{figure}
    \centering
    \includegraphics[width=1\linewidth]{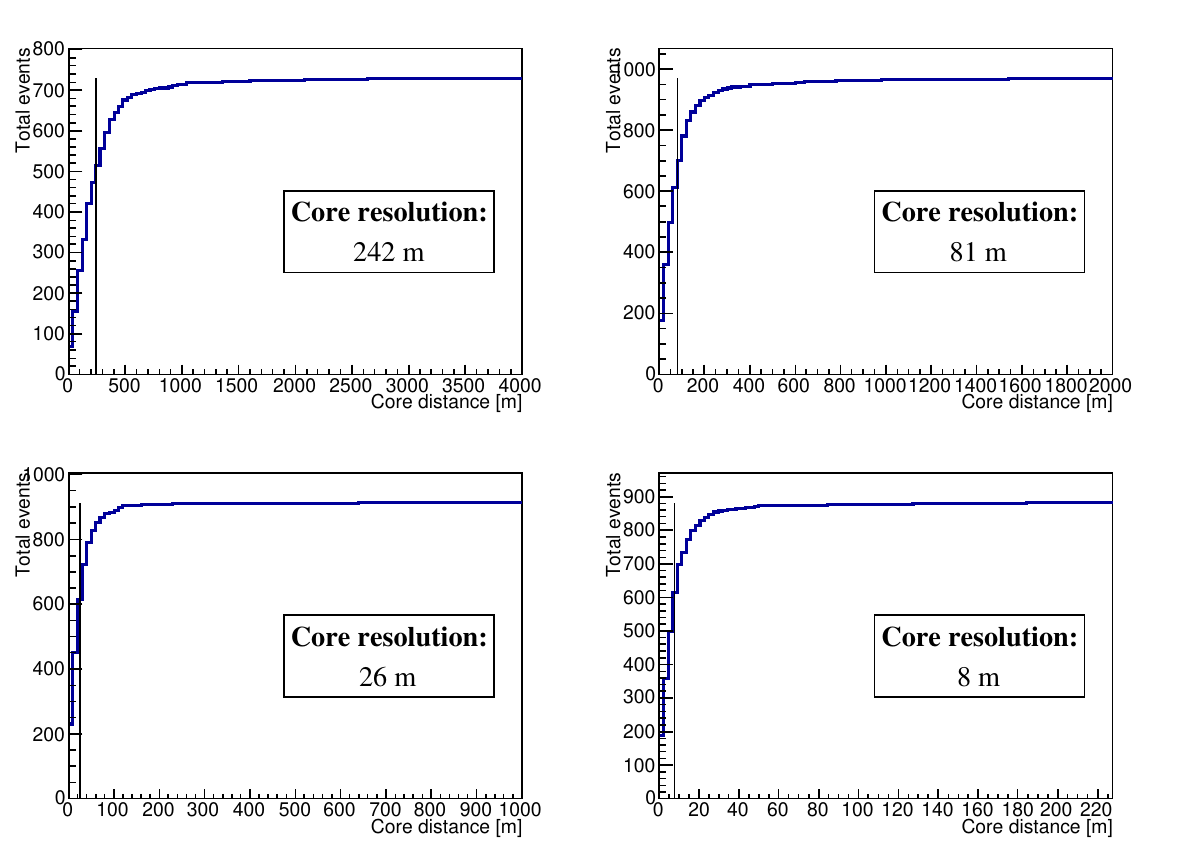}
    \caption{Core resolution determination for the reconstructions in Section \ref{sec:CheckingRecon}. Layout of the plot is the same as in Figure \ref{fig:backupfitsXmax}.}
    \label{fig:corefits}
\end{figure}

\begin{figure}
    \centering
    \includegraphics[width=1\linewidth]{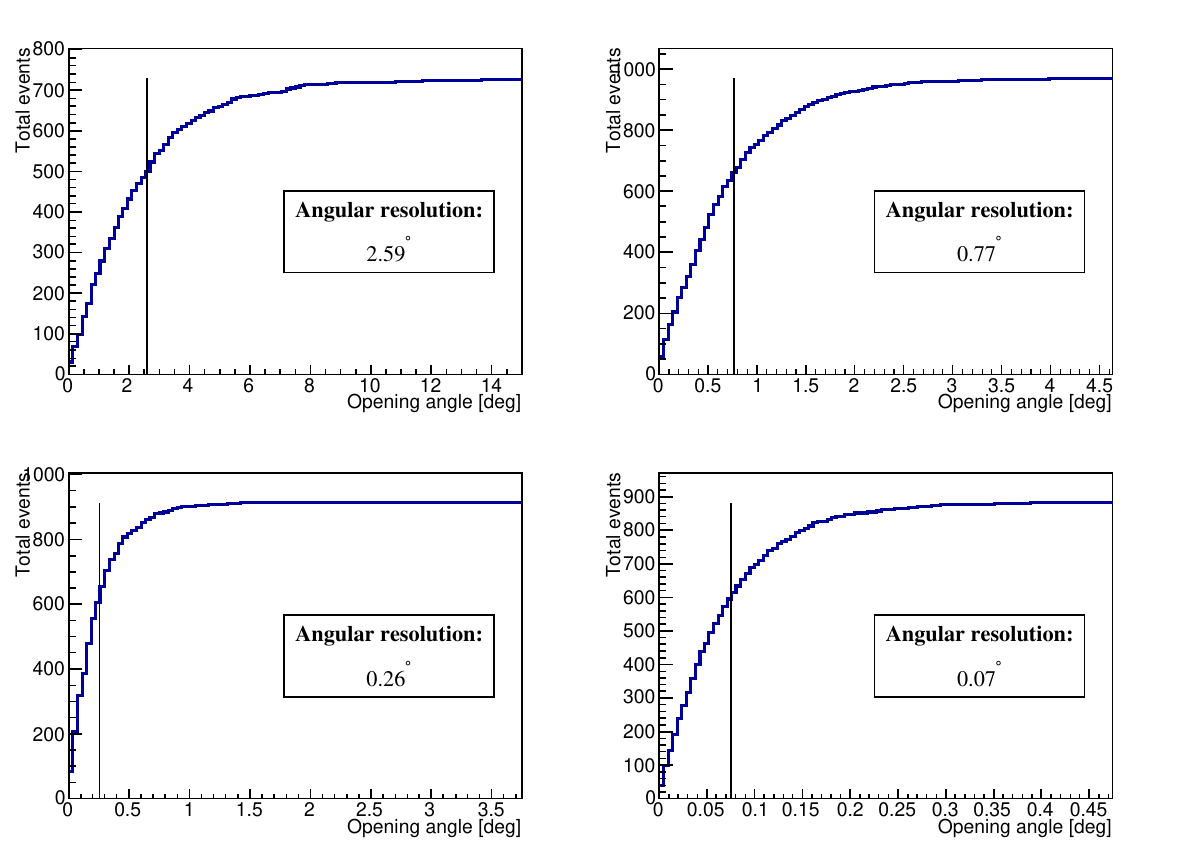}
    \caption{Same as Figure \ref{fig:corefits} but for the angular resolution.}
    \label{fig:angularFits}
\end{figure}
\chapter{First Guess I - Neural Networks: Supplementary Plots}
\section{Basic DNN Plots} 
\label{apx:basicDNNplots}
\begin{figure}[h!]
    \centering
    \includegraphics[width=1\linewidth]{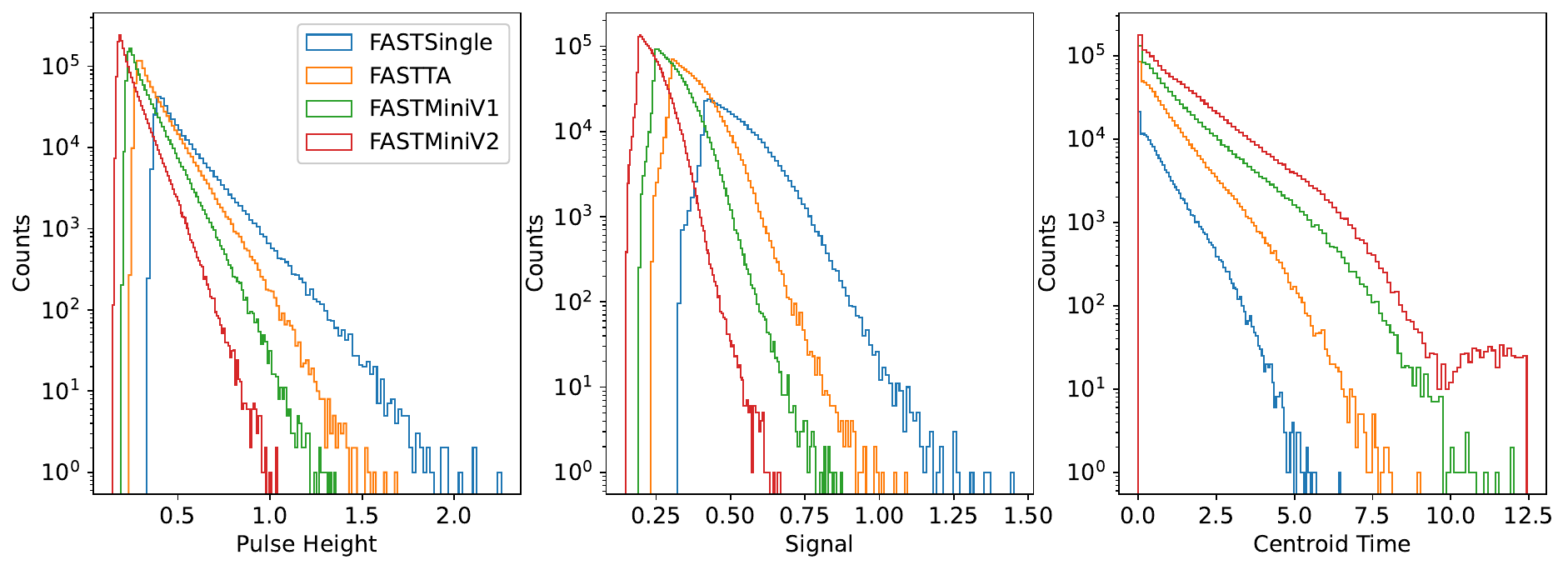}
    \caption{Normalised input parameters used for the Basic DNN model for each layout.}
    \label{fig:justinModelInputs}
\end{figure}

\clearpage

\begin{figure}[]
    \centering
    \includegraphics[width=0.97\linewidth]{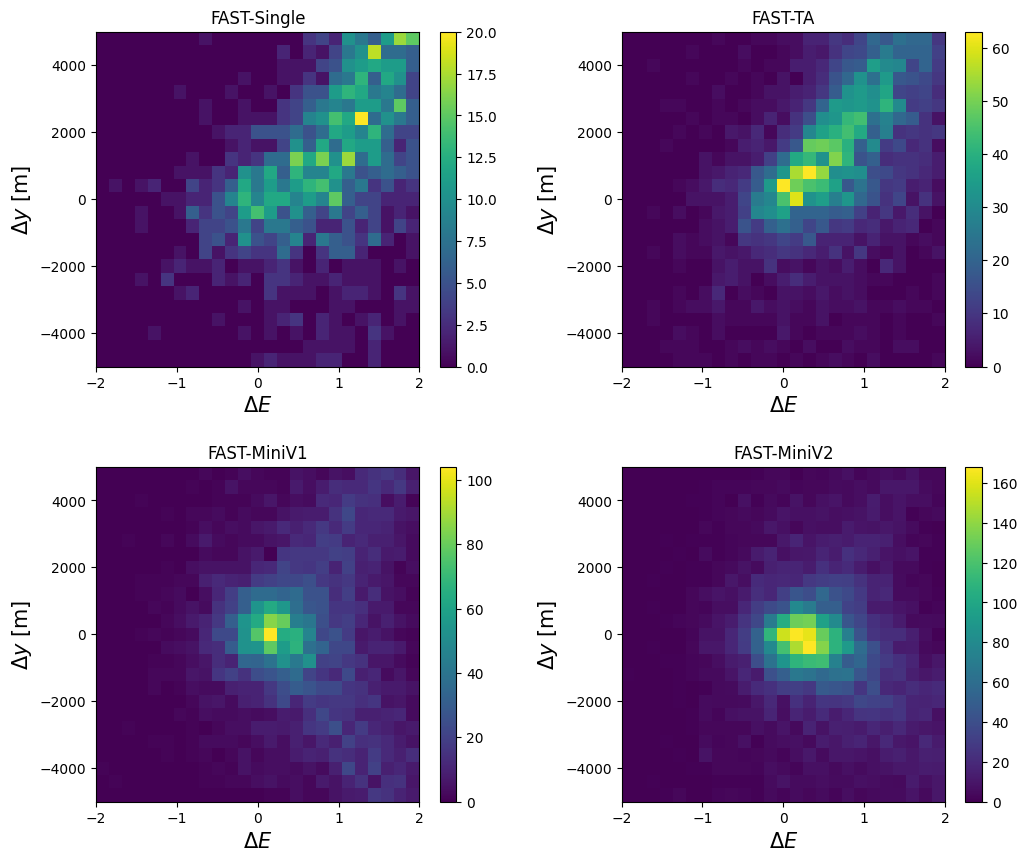}
    \caption{2-D histograms of the differences between the reconstructed and true values of energy and core $y$ for each layout. Only the results from showers with a true energy between 17.5<log($E$/eV)<18.5 are used. Notice the positive correlation in the FAST-Single and FAST-TA histograms, indicative of a degeneracy between the two parameters.}
    \label{fig:coreyEdegen}
\end{figure}

\clearpage

\section{TSFEL DNN and LSTM Network Trace Selection}
\label{apx:binCheck}
\begin{figure}[h!]
    \centering
    \includegraphics[width=1\linewidth]{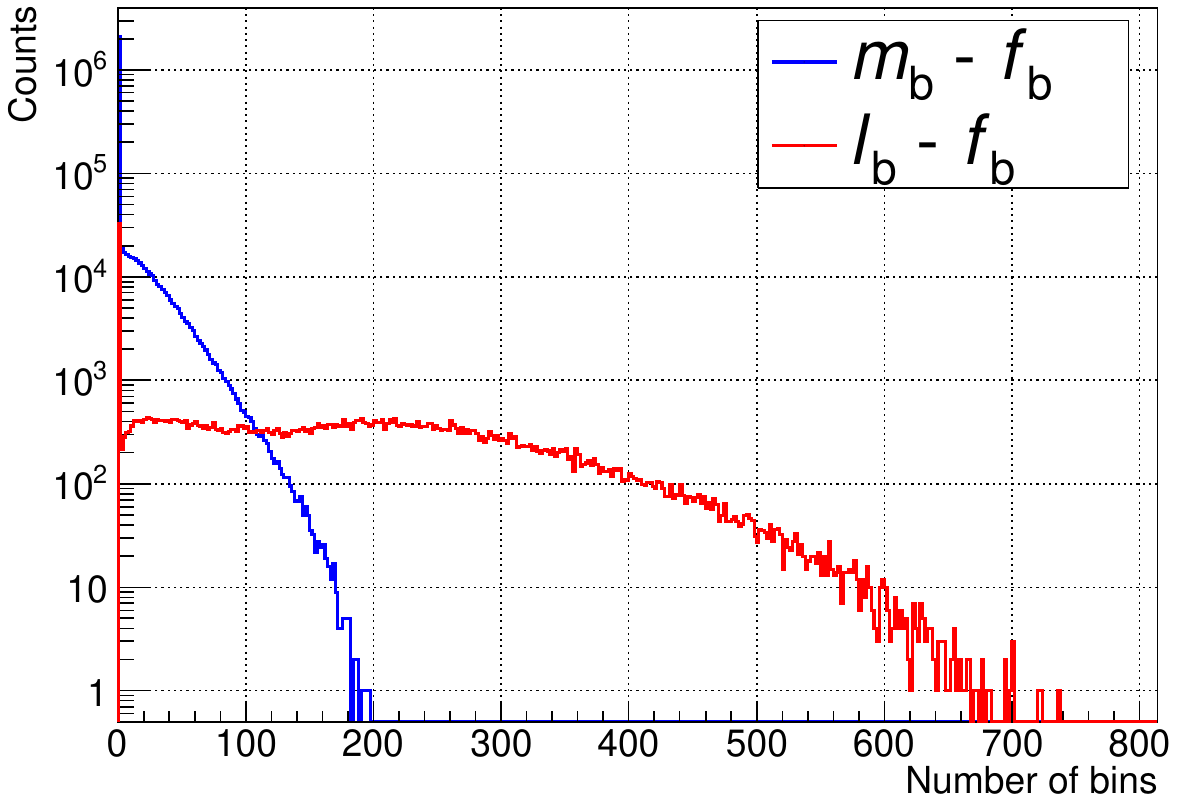}
    \caption{Histograms of the difference between $f_\textrm{b}$ and $l_\textrm{b}$ (red), and $m_\textrm{b}$ and $f_\textrm{b}$ (blue). See the text for details.}
    \label{fig:checkingAppropriateTraceSection}
\end{figure}

To determine the appropriate section of the PMT traces to extract for both the LSTM and TSFEL networks, two histograms were constructed. The first was the difference between the first and last bin above a threshold of 5\,p.e./100\,ns in traces containing zero additional noise. The first and last bins are labelled $f_\textrm{b}$ and $l_\textrm{b}$. Any usable/relevant signal to a FAST event should be located between these two bins. The other histogram was the difference between the maximum bin $m_\textrm{b}$ (bin containing the maximum number of photoelectrons over all bins) in the first triggered PMT and $f_\textrm{b}$. Figure \ref{fig:checkingAppropriateTraceSection} shows the $\left(f_\textrm{b}-l_\textrm{b}\right)$ histogram in red and the $\left(m_\textrm{b}-f_\textrm{b}\right)$ histogram in blue. Since $\left(f_\textrm{b}-l_\textrm{b}\right)$ is almost always < 600, and $\left(m_\textrm{b}-f_\textrm{b}\right)$ almost always < 150, these values were chosen for the length of each trace to extract and the starting bin i.e. $f_\textrm{b}-150$.

\newpage

\section{TSFEL Features}
\label{apx:TSFELfeatList}
\begin{figure}[p]
    \centering
    \includegraphics[width=1\linewidth]{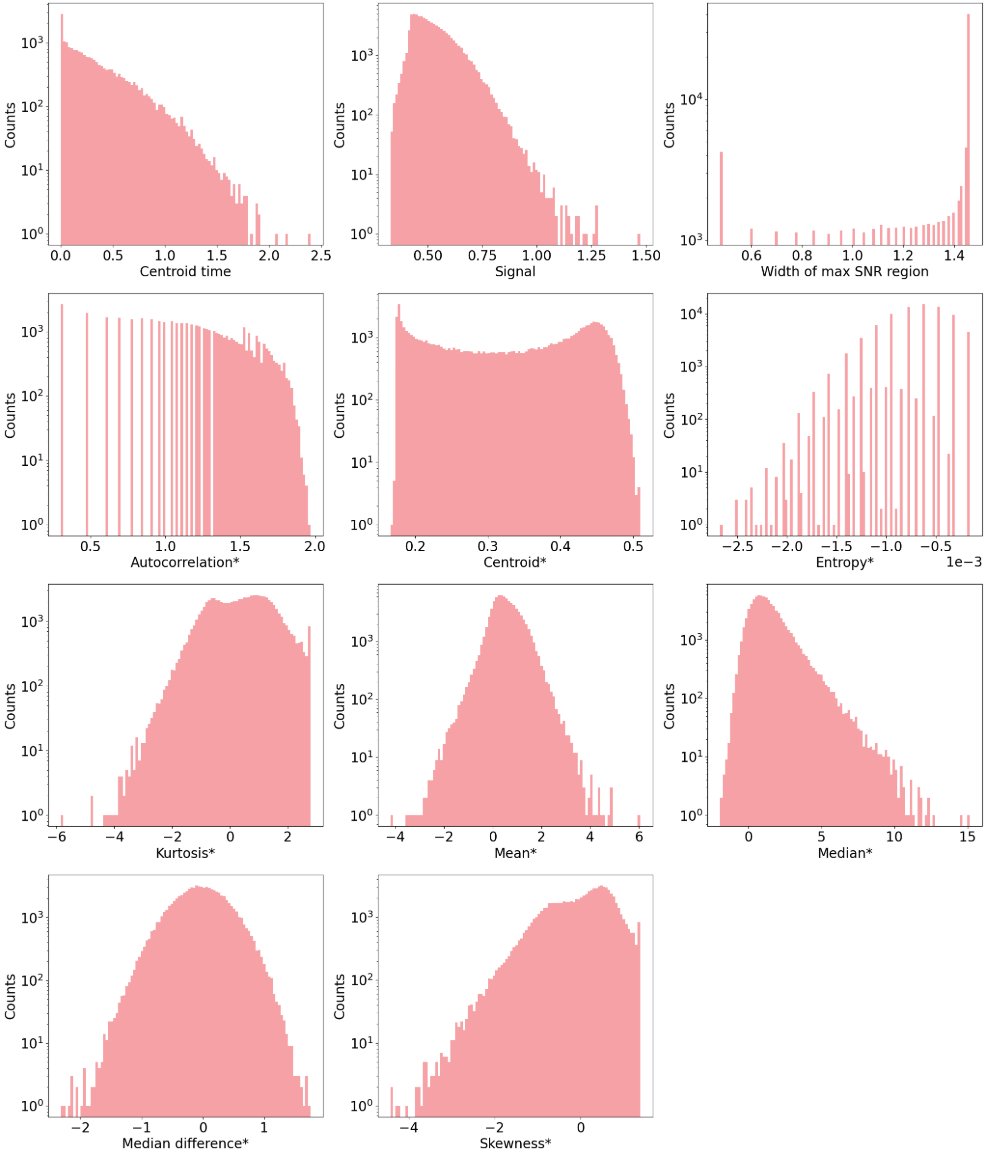}
    \caption{Distributions of the input parameters for the TSFEL DNN from a single PMT.}
    \label{fig:tsfelInputs}
\end{figure}

The features extracted from each PMT trace and used as input for the TSFEL DNN model are listed below. Features marked with a \say{*} are from the TSFEL library. The descriptions of these features are taken from \url{https://tsfel.readthedocs.io/en/latest/index.html}. Additional information on each feature can be found here. 
\begin{itemize}
    \item \textbf{Centroid time} - $\hat{\hat{\bar{t}}}$ as calculated in Section \ref{sec:basicDNN}.
    \item \textbf{Signal} - $\hat{\hat{\bar{S}}}$ as calculated in Section \ref{sec:basicDNN}.
    \item \textbf{Width of max SNR region} - $k_\textrm{stop}-k_\textrm{start}$
    \item \textbf{Autocorrelation*} - The first time lag at which the autocorrelation function drops below $1/e$.
    \item \textbf{Centroid*} - Temporal centroid 
    \item \textbf{Entropy*} - Normalised entropy of the signal calculated using Shannon Entropy.
    \item \textbf{Kurtosis*} - Kurtosis of the signal
    \item \textbf{Mean*} - Mean value of the signal
    \item \textbf{Median*} - Median value of the signal
    \item \textbf{Median diff*} - Median value of the differences between subsequent values of the signal 
    \item \textbf{Skewness*} - Skewness of the signal
\end{itemize}
Figure \ref{fig:tsfelInputs} shows the typical input parameter distributions for a single PMT.

\newpage

\section{LSTM Explanation}
\label{apx:LSTMexplanation}
\begin{figure}[h!]
    \centering
    \includegraphics[width=0.8\linewidth]{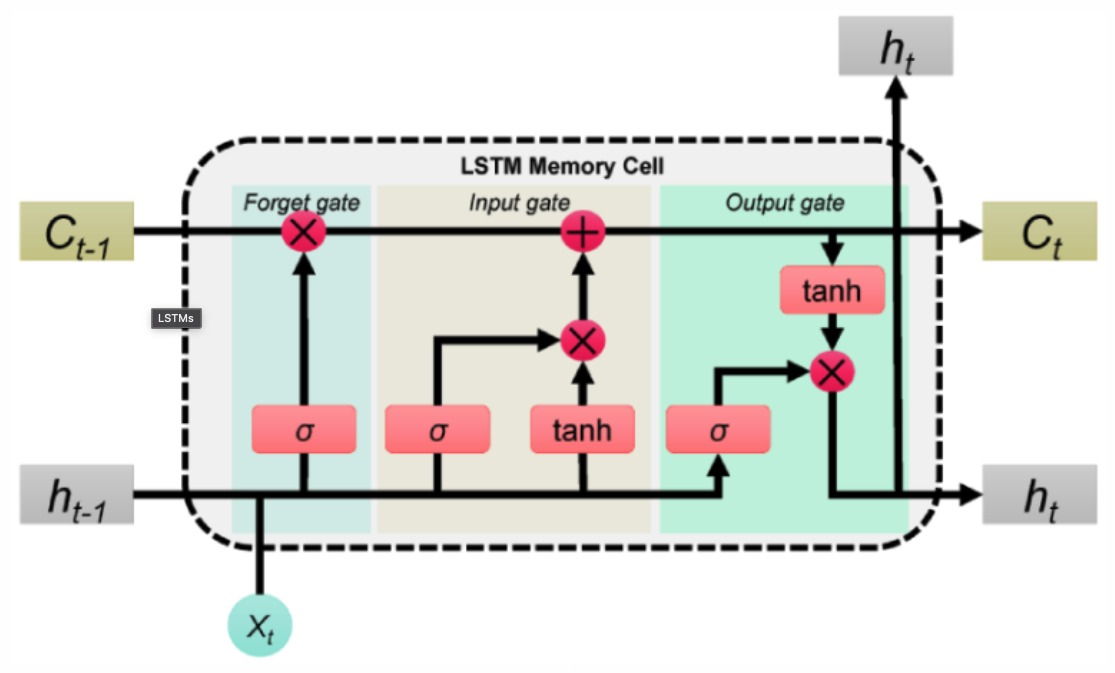}
    \caption{Diagram of the memory cell used by an LSTM network. From \cite{lstmdiagram}.}
    \label{fig:LSTMcell}
\end{figure}

To understand how an LSTM is able to learn the salient features from a time series, a simple example is provided below. 
Consider a PMT trace with values $\{x_1,x_2,...,x_n\}$ which is passed to an LSTM layer for processing. The output from the LSTM layer will be a vector of $m$ features which (abstractly) characterise the trace. The LSTM layer processes each $x_t$ sequentially using a \textit{memory cell}, a diagram of which is shown in Figure \ref{fig:LSTMcell}. This cell possesses an \say{internal state}, $C_t$, which represents the cell's long-term memory, and a \say{hidden state}, $h_t$, which represents the cell's short term memory. At each time step $t$, the input $x_t$, together with the previous internal and hidden states $C_{t-1}$ and $h_{t-1}$, are combined in the memory cell to update the internal and hidden states to $C_t$ and $h_t$. The interior logic of the memory cell is not explained in detail here, but essentially consists of several gates which use either sigmoid or tanh activation functions to alter the long/short-term memory. Importantly, the input to a gate $g$, $I_g$, is
\begin{equation}
    \vec{I_g} = U_g\vec{h_{t-1}}+W_g\vec{x_{t}}+\vec{b_g}
\end{equation}
where $U_g$ and $W_g$ are matrices of \textit{learned} weights, and $b_g$ is a \textit{learned} bias term. The final output of the LSTM layer is the final state of the short term memory $h_n$. This vector can then be used as the input to a series of dense layers which ultimately predict the final parameters of interest. In this example, training the LSTM can be thought of as updating the internal weights and biases ($U, W, b$) such that the final output $h_n$ is essentially a vector of features characterising the trace, similar to the feature vectors used in the feed-forward models \cite{hochreiter1997long}. Of course, what each value in $h_n$ represents is very abstract and not easily deciphered.

\newpage

\section{LSTM Network inputs}
\label{apx:LSTMInputs}

\begin{figure}[h!]
    \centering
    \includegraphics[width=0.9\linewidth]{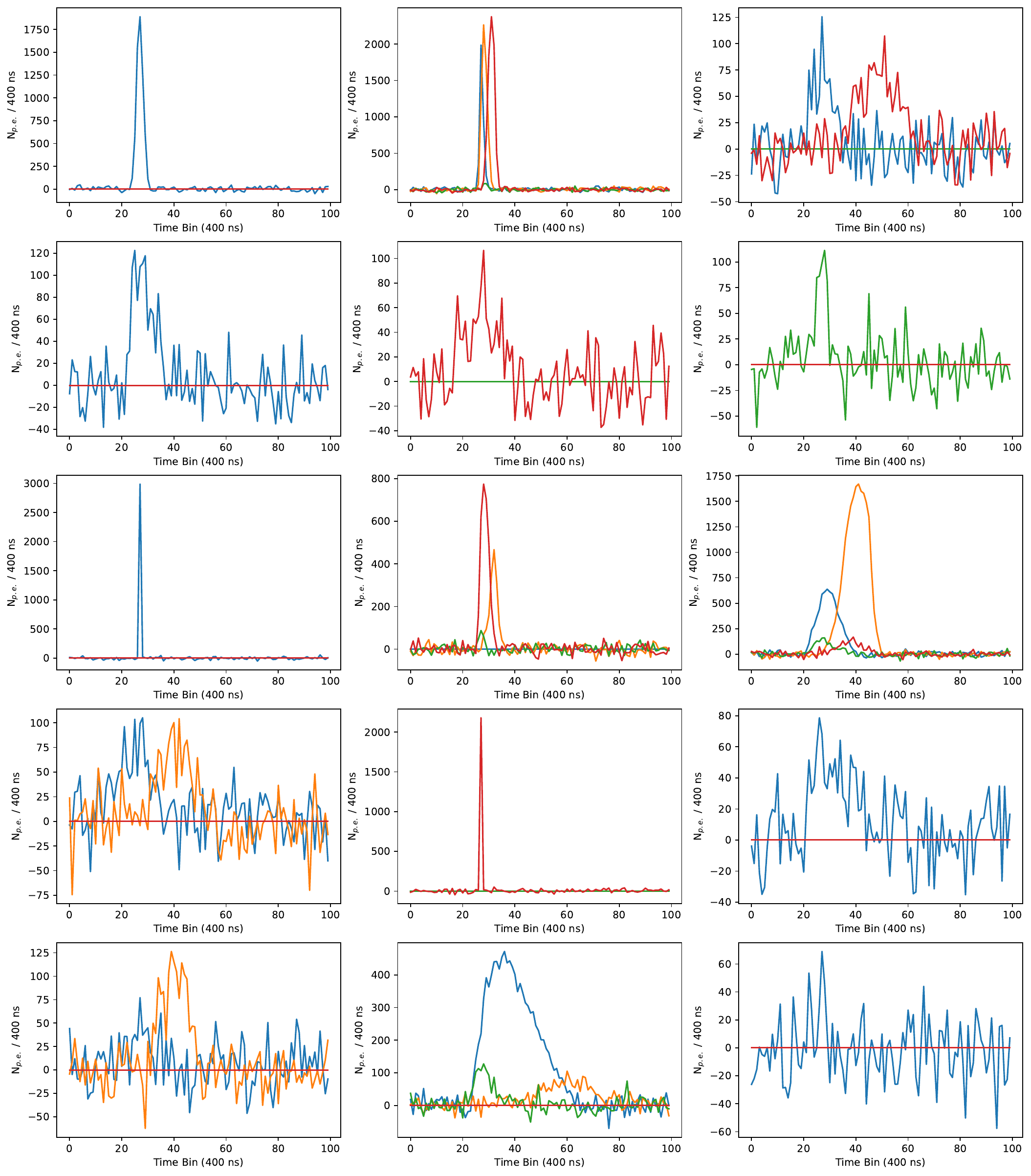}
    \caption{Examples of the traces passed to the LSTM network as inputs. These traces are from the FAST-Single setup. The blue, orange, green and red lines correspond to the PMT viewing the top-left, bottom-left, top-right and bottom-right sections of the sky respectively.}
    \label{fig:LSTMinputs}
\end{figure}

\newpage

\section{TSFEL DNN + TDR Results}
\label{sec:additionalMLTDRplots}
\begin{figure}[h!]
    \centering
    \begin{subfigure}[b]{1\textwidth}
        \centering
        \includegraphics[width=1\linewidth]{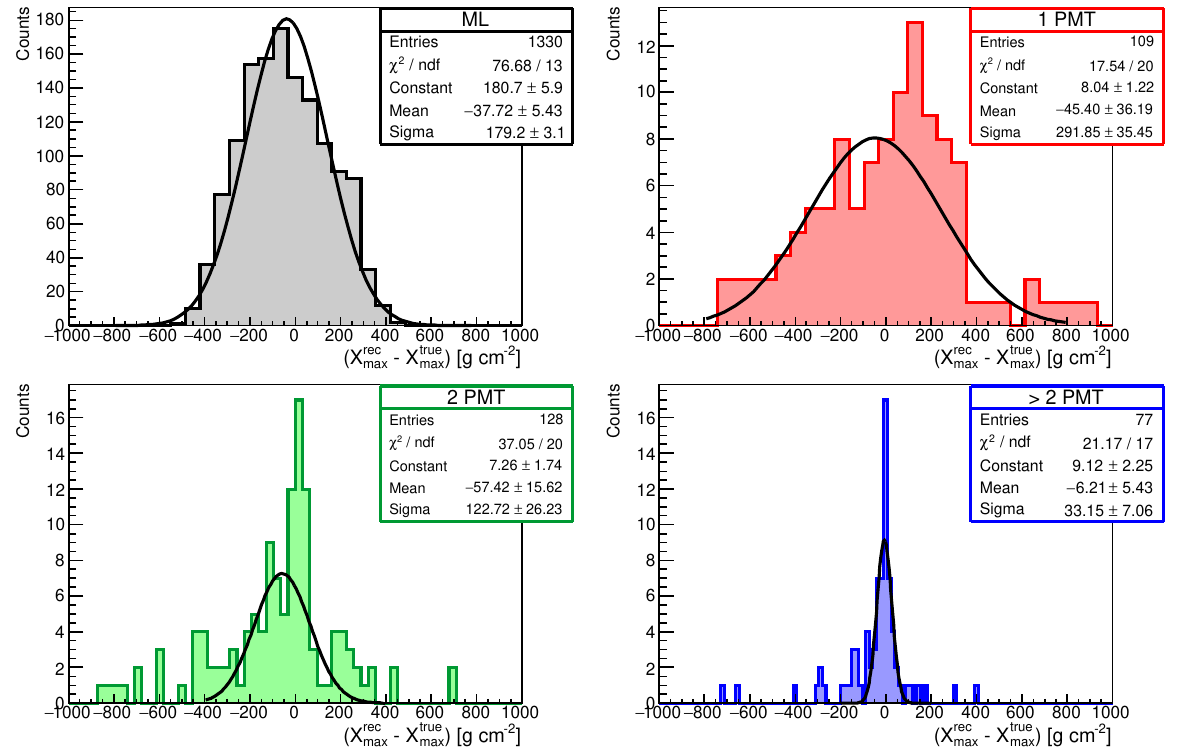}
        \caption{FAST-Single: \Xmax{}}
        \label{fig:xmaxsingfits}
    \end{subfigure}
    \begin{subfigure}[b]{1\textwidth}
        \centering
        \includegraphics[width=1\linewidth]{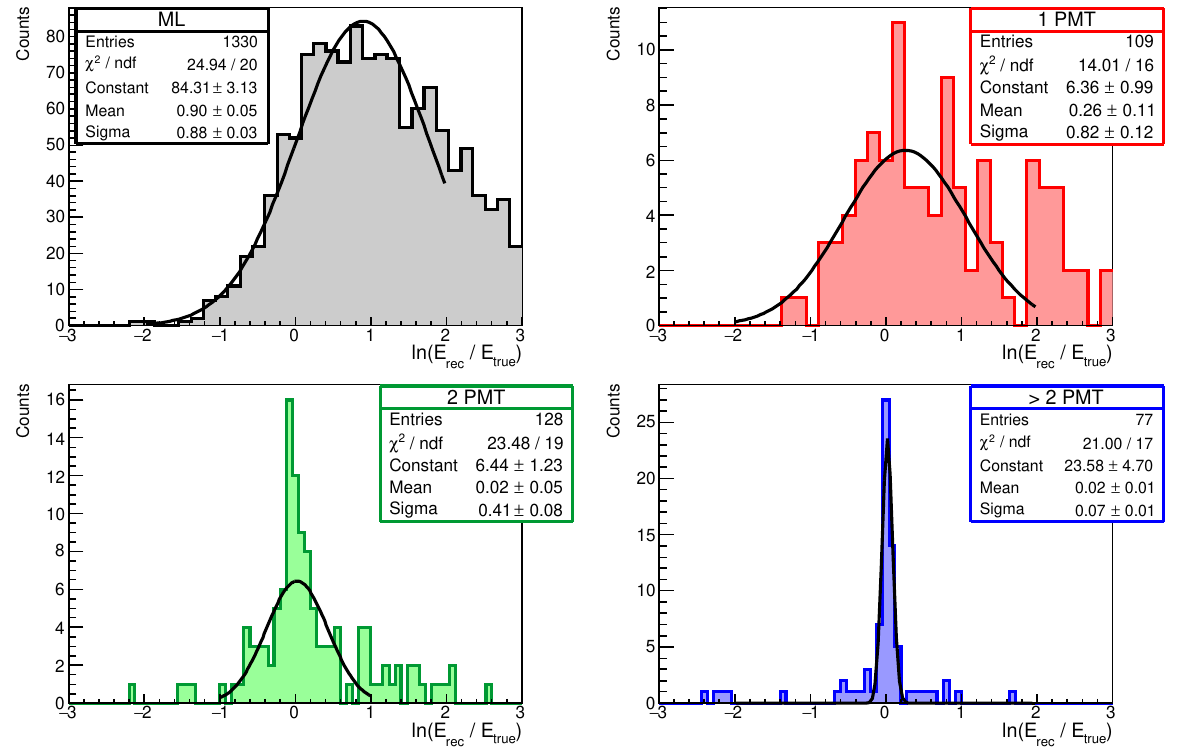}
        \caption{FAST-Single: Energy}
        \label{fig:enesingfits}
    \end{subfigure}
    \caption{}
    \label{fig:xmax_energy_diffmltdrSingle}
\end{figure}

\begin{figure}
    \centering
    \begin{subfigure}[b]{1\textwidth}
        \centering
        \includegraphics[width=1\linewidth]{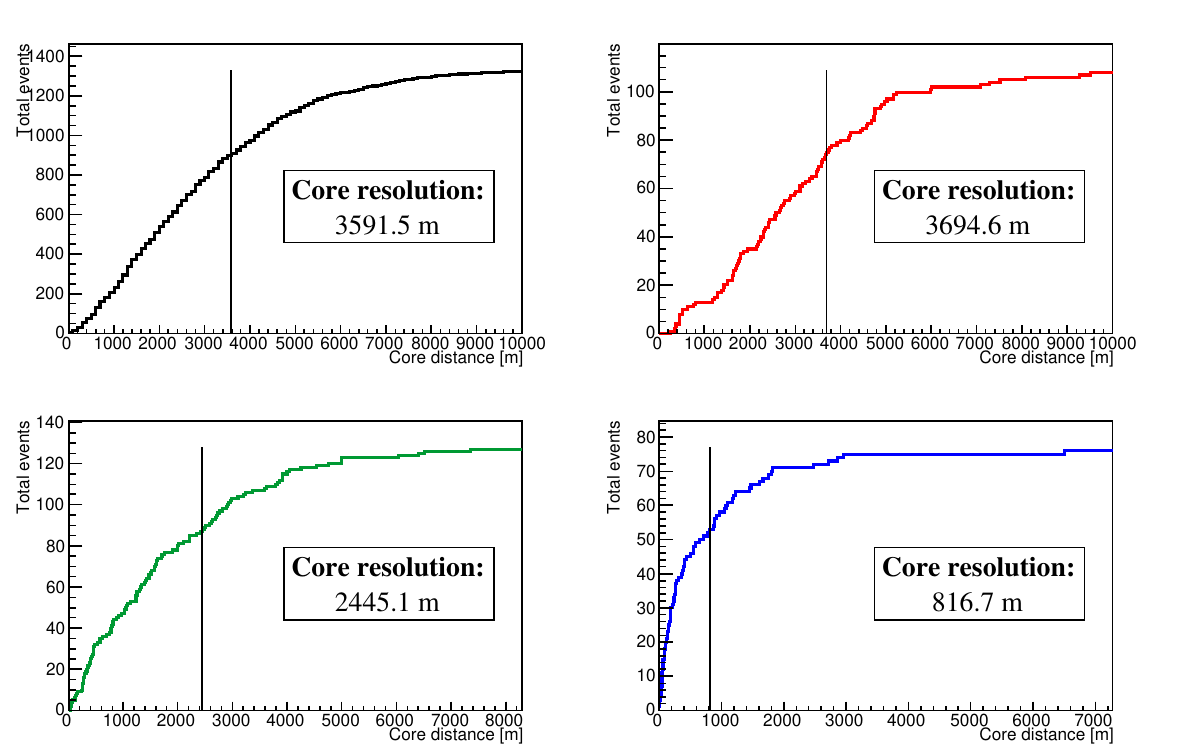}
        \caption{FAST-Single: Core resolution}
        \label{fig:coresingfits}
    \end{subfigure}
    \begin{subfigure}[b]{1\textwidth}
        \centering
        \includegraphics[width=1\linewidth]{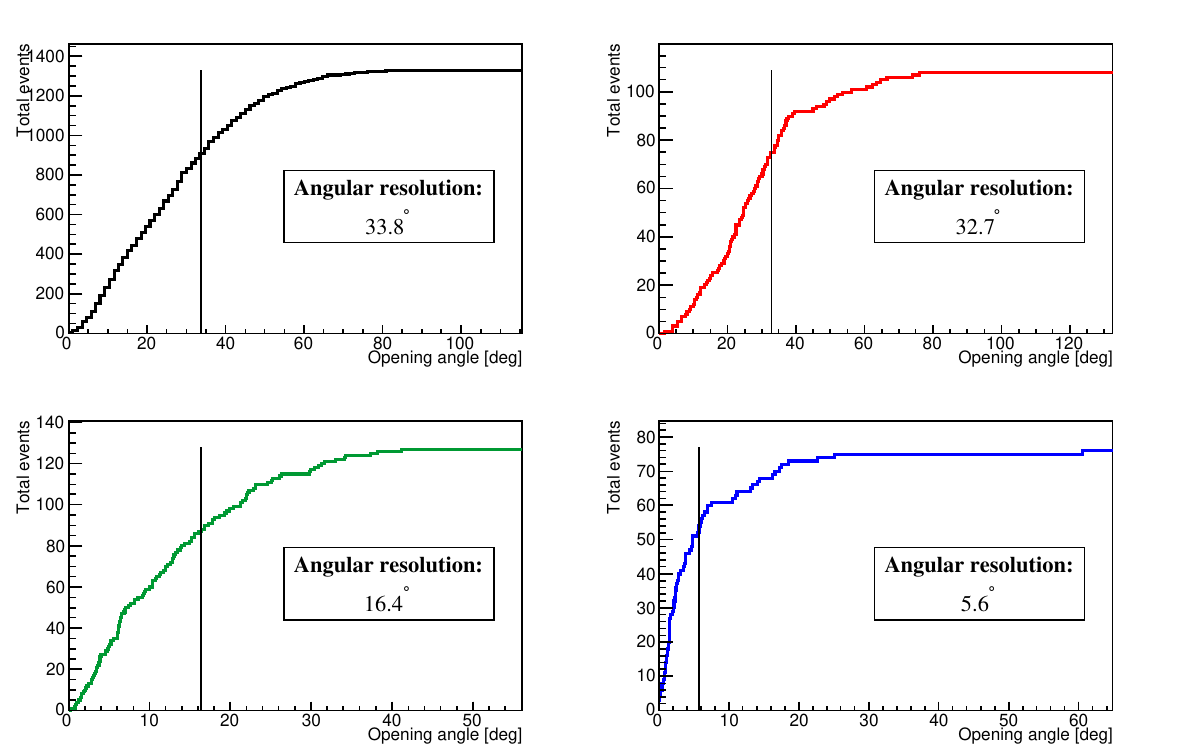}
        \caption{FAST-Single: Angular resolution}
        \label{fig:angsingfits}
    \end{subfigure}
    \caption{}
\end{figure}

\begin{figure}
    \centering
    \begin{subfigure}[b]{1\textwidth}
        \centering
        \includegraphics[width=1\linewidth]{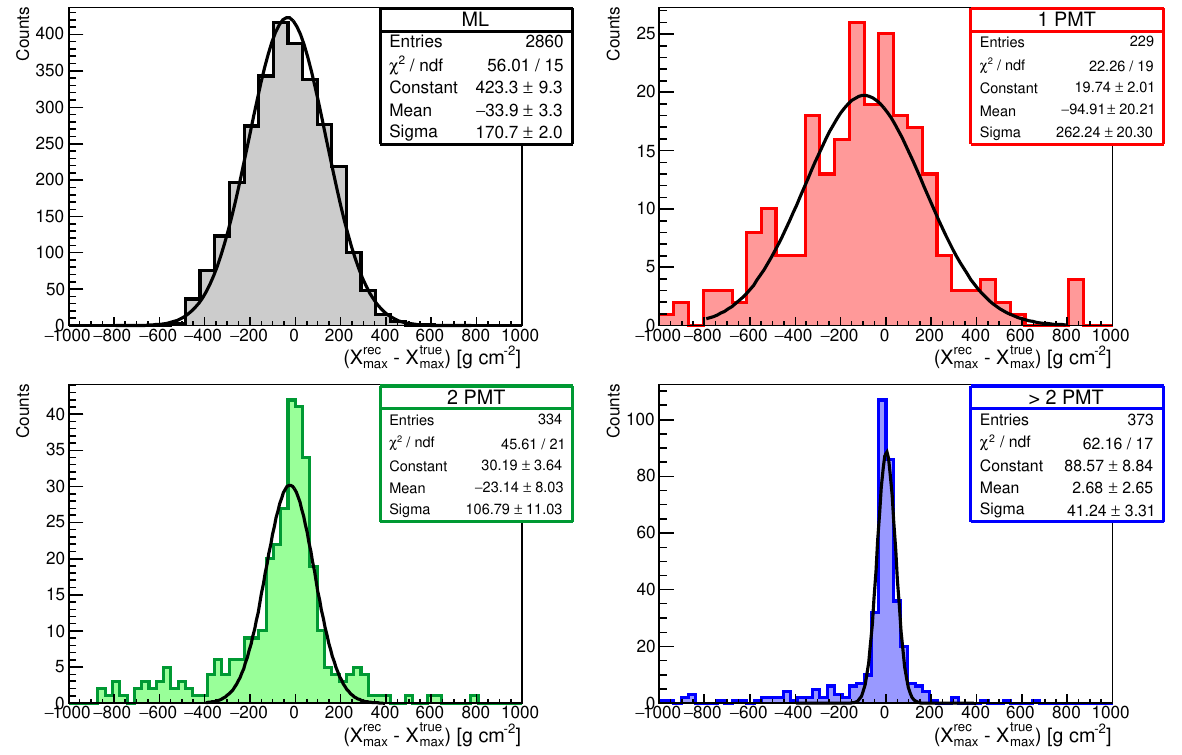}
        \caption{FAST-TA: \Xmax{}}
        \label{fig:xmaxtafits}
    \end{subfigure}
    \begin{subfigure}[b]{1\textwidth}
        \centering
        \includegraphics[width=1\linewidth]{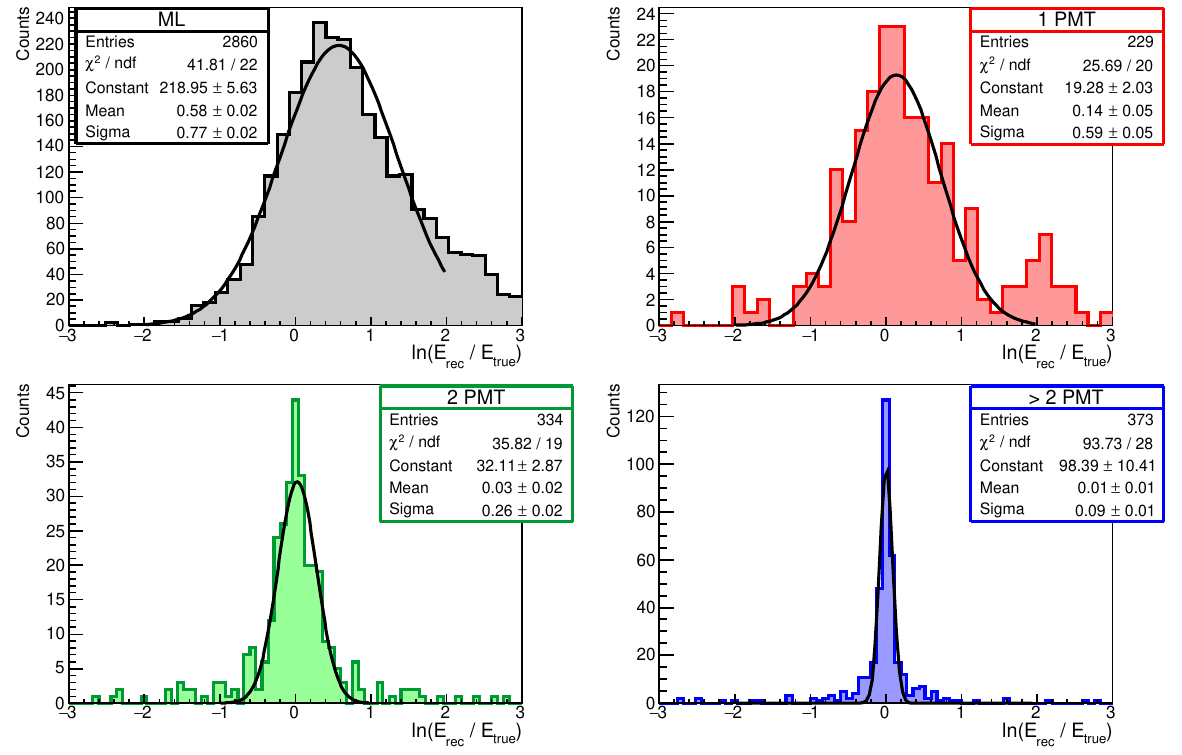}
        \caption{FAST-TA: Energy}
        \label{fig:enetafits}
    \end{subfigure}
    \caption{}
    \label{fig:xmax_energy_diffmltdrTA}
\end{figure}

\begin{figure}
    \centering
    \begin{subfigure}[b]{1\textwidth}
        \centering
        \includegraphics[width=1\linewidth]{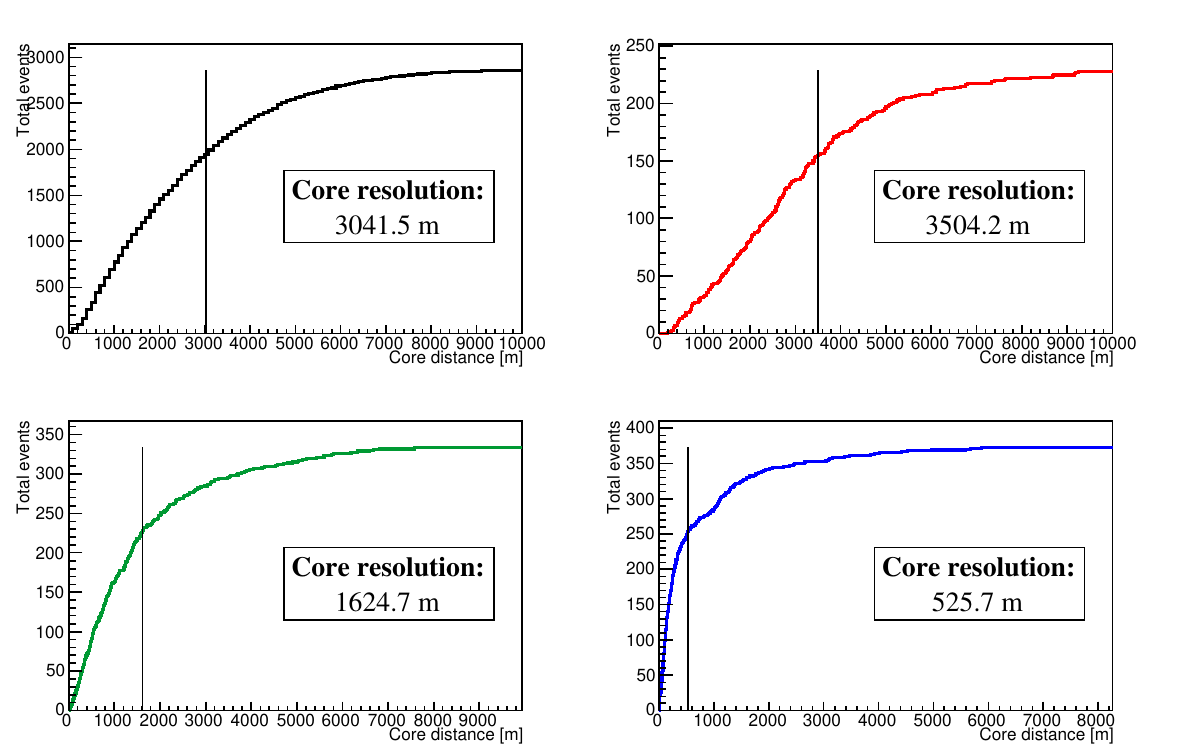}
        \caption{FAST-TA: Core resolution}
        \label{fig:coretafits}
    \end{subfigure}
    \begin{subfigure}[b]{1\textwidth}
        \centering
        \includegraphics[width=1\linewidth]{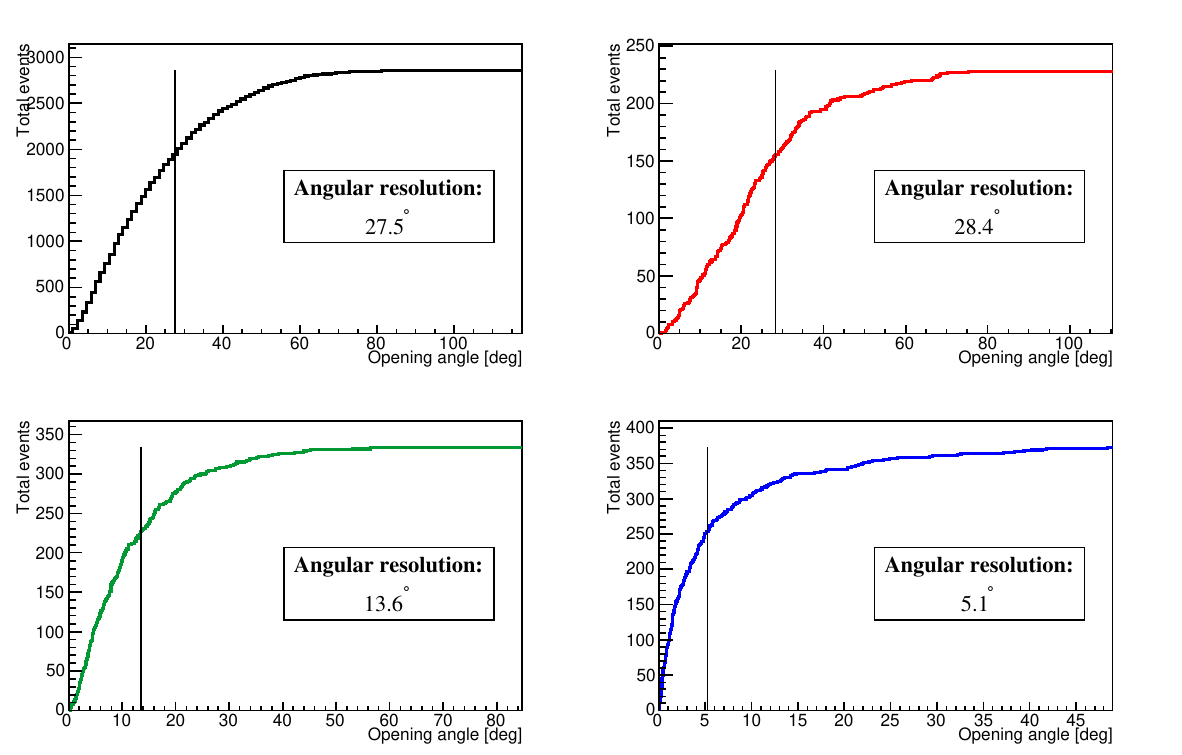}
        \caption{FAST-TA: Angular resolution}
        \label{fig:angtafits}
    \end{subfigure}
    \caption{}
    \label{fig:core_ang_diffmltdrTA}
\end{figure}

\begin{figure}
    \centering
    \includegraphics[width=1\linewidth]{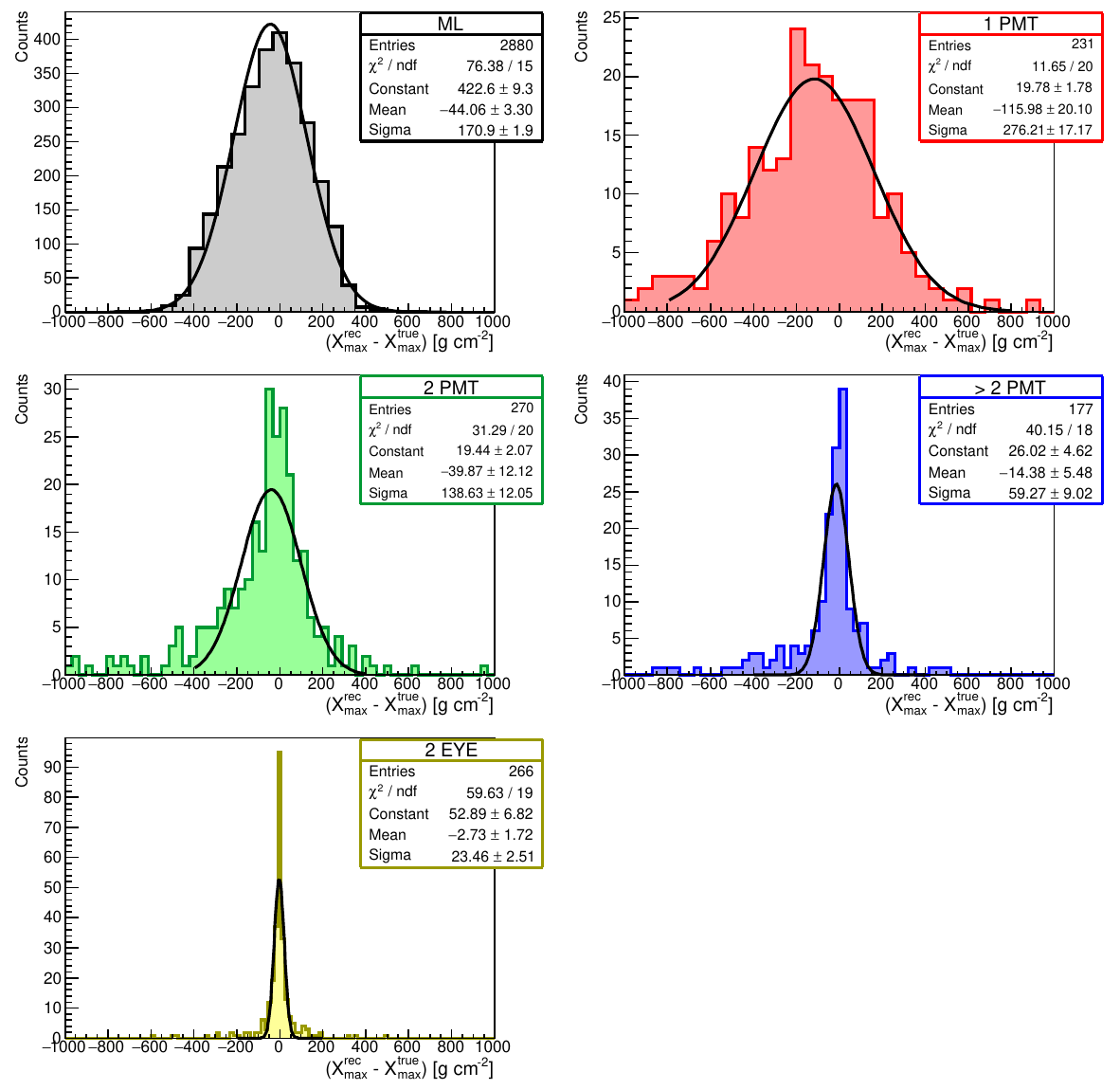}
    \caption{FAST-MiniV1: \Xmax{}.}
    \label{fig:xmaxmini1fits}
\end{figure}

\begin{figure}
    \centering
    \includegraphics[width=1\linewidth]{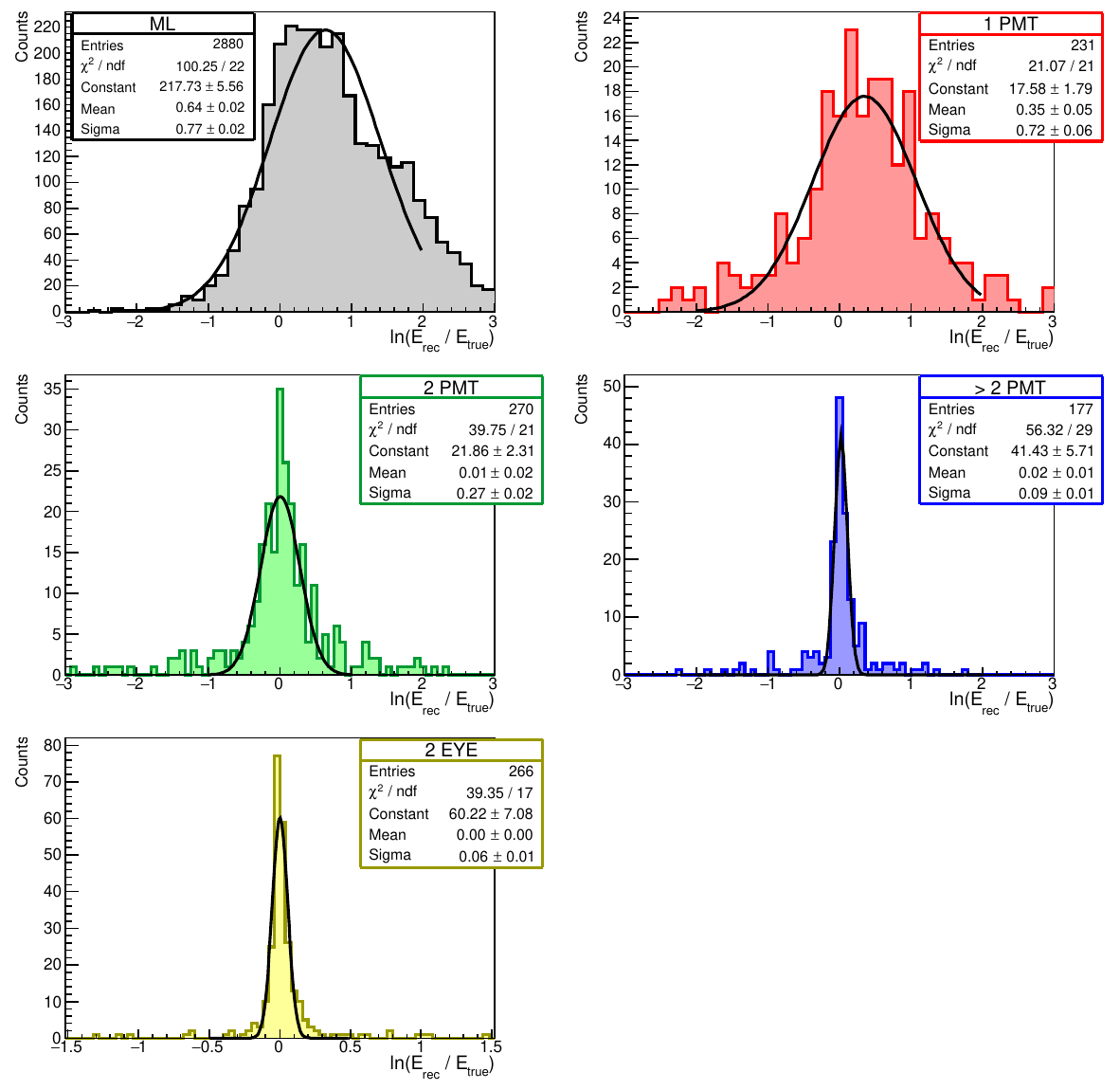}
    \caption{FAST-MiniV1: Energy. The 2-Eye histogram has been zoomed in to better view the central part of the distribution. This causes 3 events below $-1.5$ to not be visible.}
    \label{fig:enemini1fits}
\end{figure}

\begin{figure}
    \centering
    \includegraphics[width=1\linewidth]{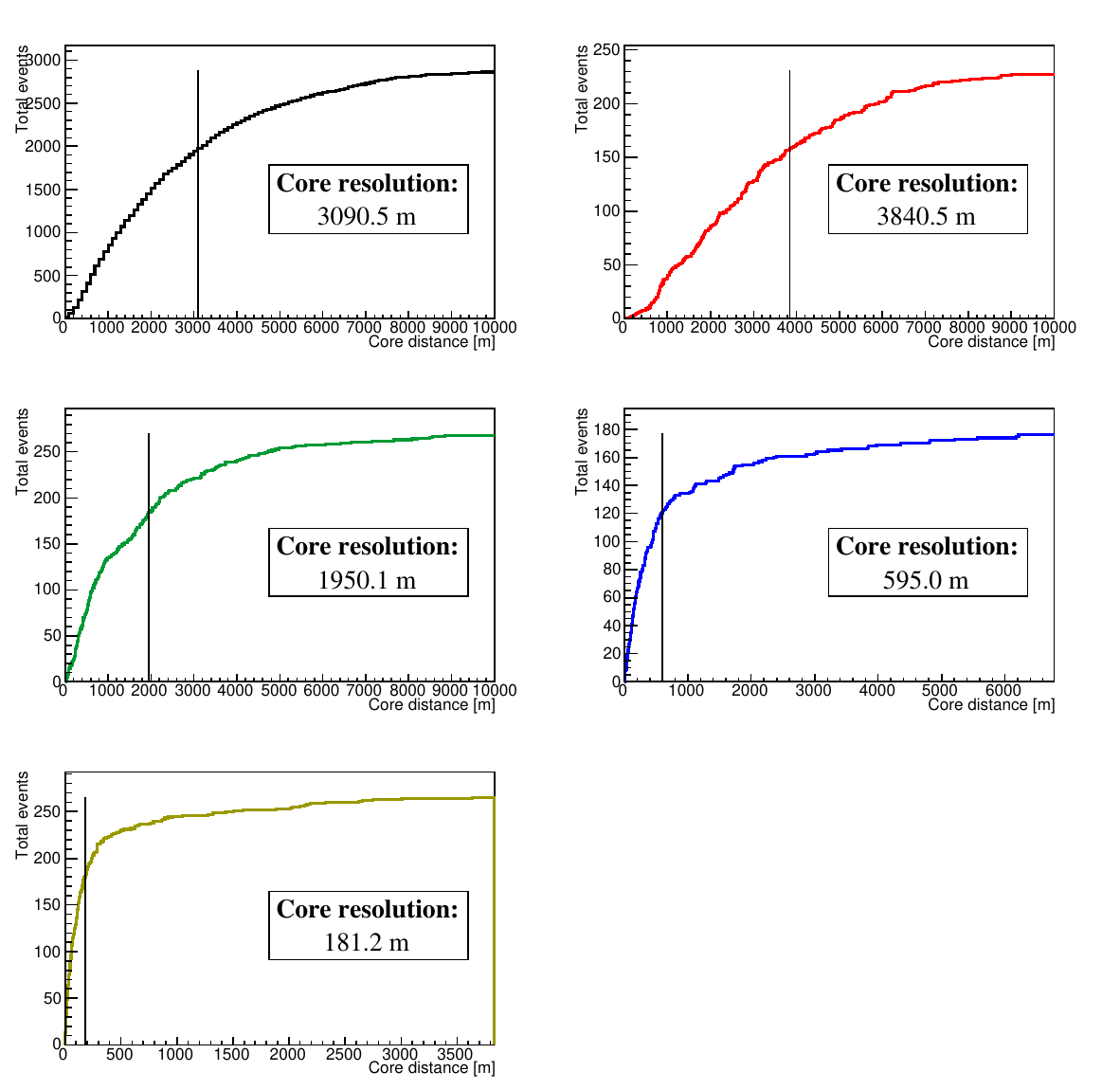}
    \caption{FAST-MiniV1: Core resolutions}
    \label{fig:coremini1fits}
\end{figure}

\begin{figure}
    \centering
    \includegraphics[width=1\linewidth]{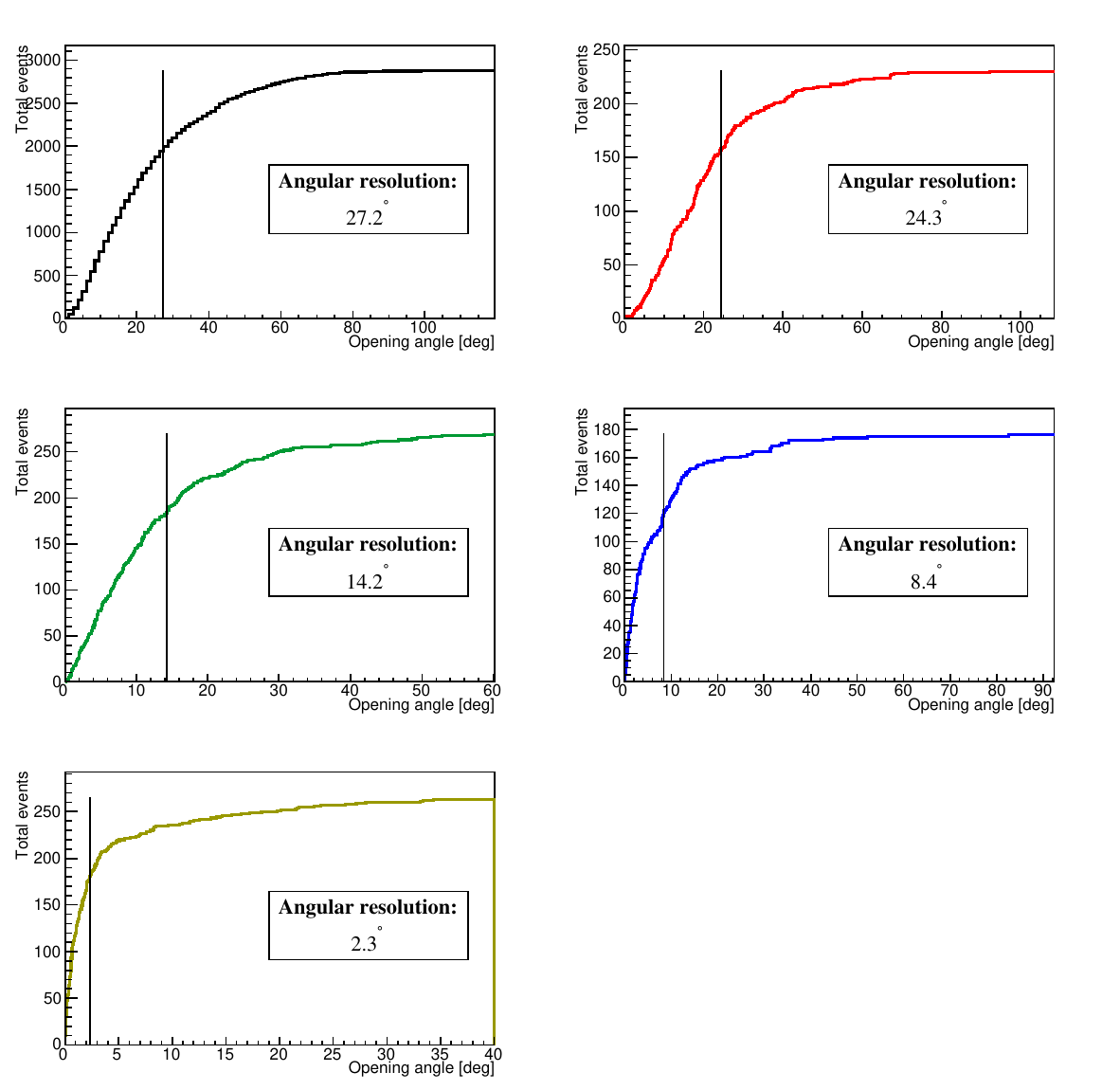}
    \caption{FAST-MiniV1: Angular resolutions}
    \label{fig:angmini1fits}
\end{figure}

\chapter{First Guess II - Template Method: Supplementary Plots}

\section{Non-Zero Baseline Example}
\label{apx:baseline}
\begin{figure}[h!]
    \centering
    \includegraphics[width=1\linewidth]{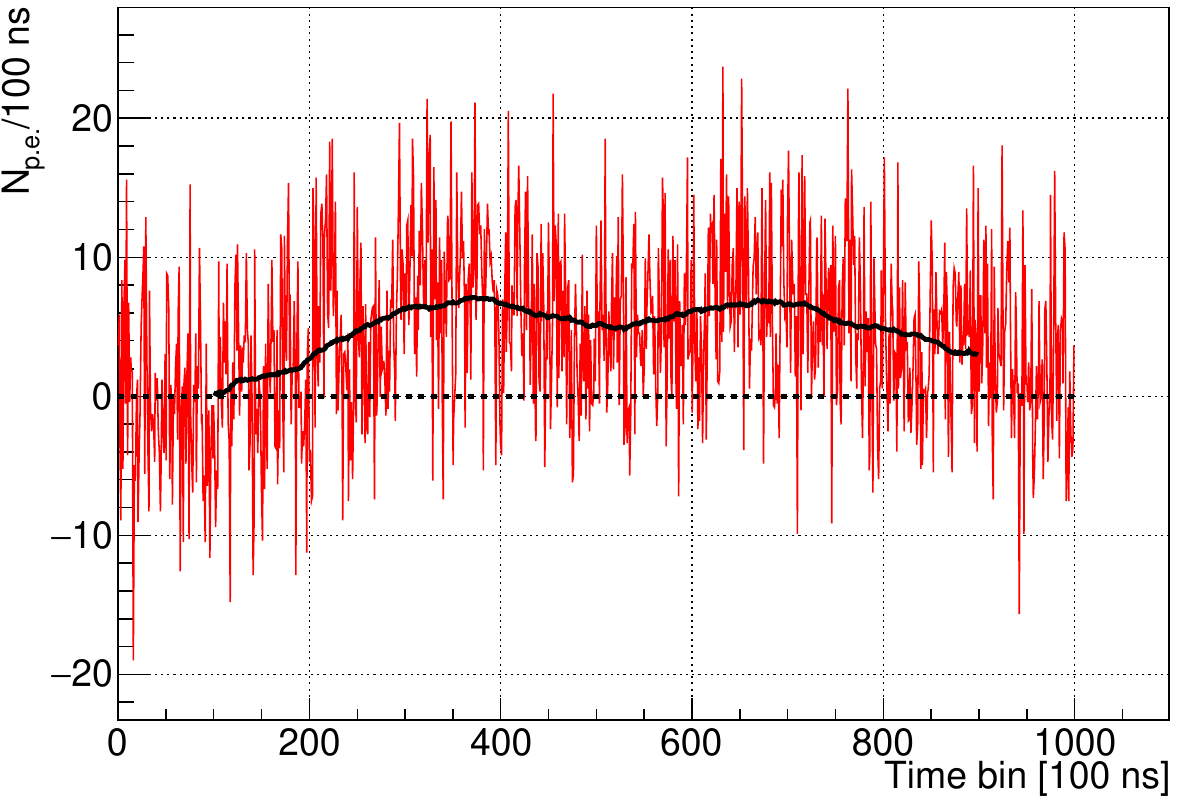}
    \caption{Example of a PMT trace from data exhibiting a fluctuating, non-zero baseline. This example comes from a coincidence event observed by FAST@Auger. The red line is the data. The black line is the result after filtering the data using a moving average filter with window length 201 bins (applied between bins 101 and 899). }
    \label{fig:baselineExample}
\end{figure}

\begin{figure}
    \centering
    \includegraphics[width=1\linewidth]{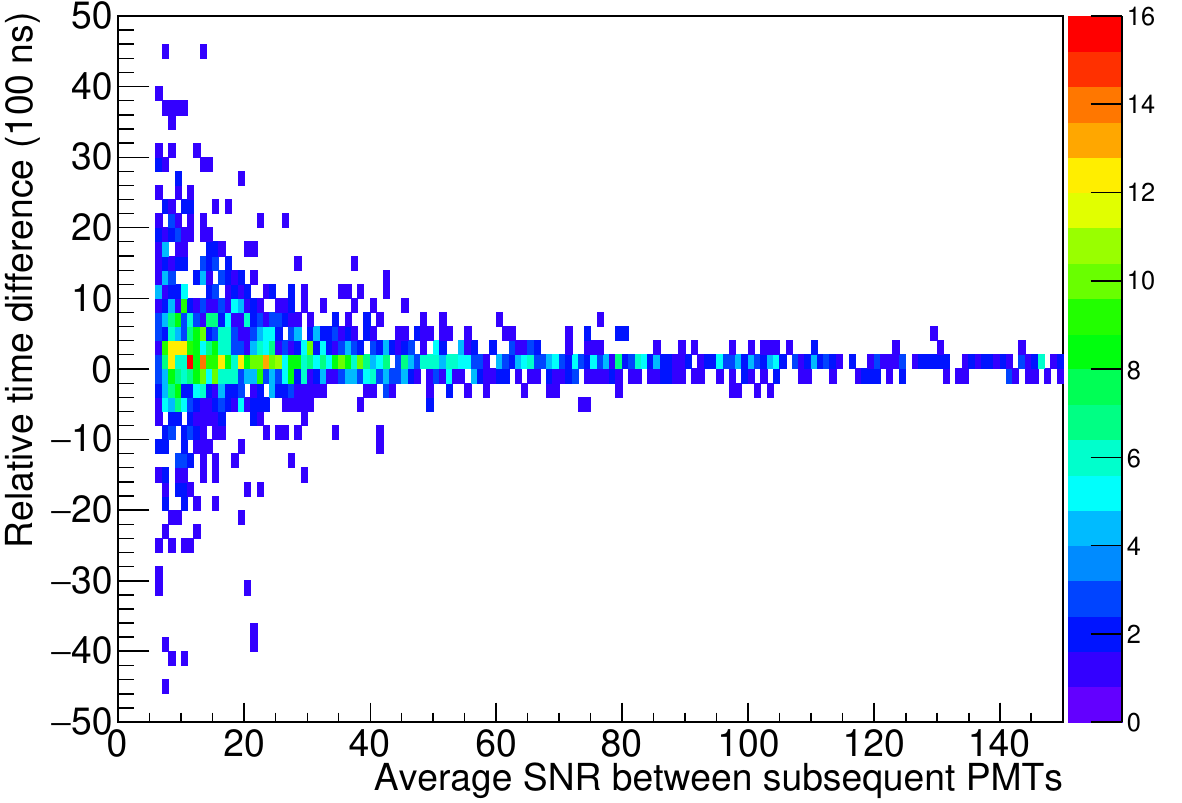}
    \includegraphics[width=1\linewidth]{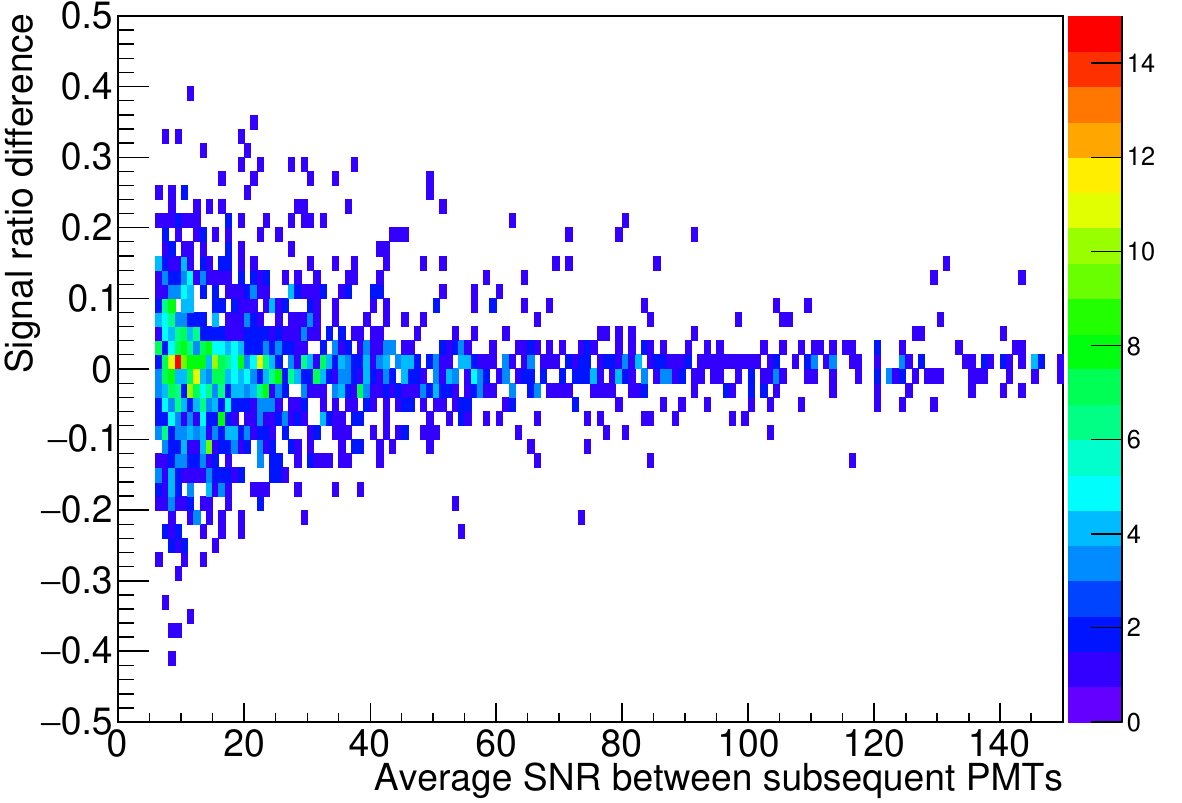}
    \caption{2D-Histograms of the relative time/signal ratio differences and the average SNR between subsequent PMTs.}
    \label{fig:templateTimeSigSNR}
\end{figure}
\chapter{PMT Directional Efficiency Maps}
\label{apx:direcEffMaps}

Find here the different PMT directional efficiency maps used throughout this thesis.

\begin{figure}[h!]
\label{fig:idealRayTraceMap}
    \centering
    \includegraphics[width=1\linewidth]{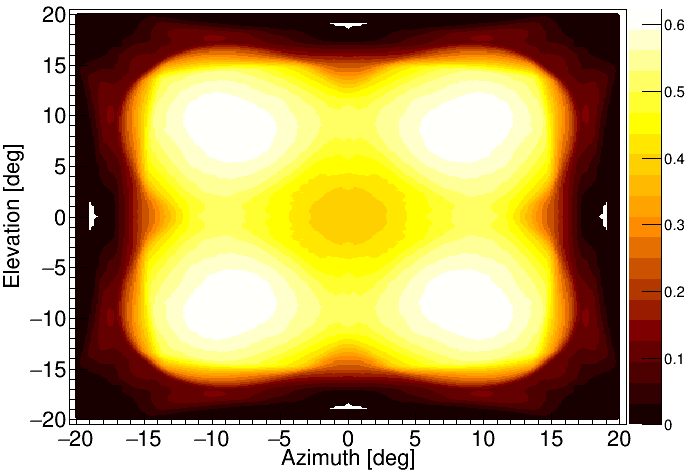}
    \caption{Ideal ray trace map used in simulations for machine learning (Section \ref{sec:MLdataset}) and template creation Section \ref{sec:tempDataset}.}
    \label{fig:idealDirEffMap}
\end{figure}

\begin{figure}
\label{fig:orginialTraceMaps}
    \centering
     \begin{subfigure}[b]{1\textwidth}
        \centering
        \includegraphics[width=0.8\linewidth]{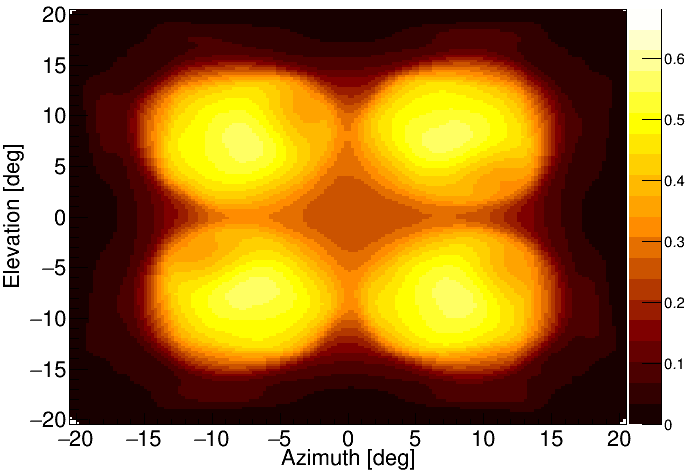}
        \caption{Original directional efficiency map for FAST@TA telescopes 1 and 2, and FAST@Auger.}
        \label{fig:FAST13Map}
    \end{subfigure}
    \begin{subfigure}[b]{1\textwidth}
        \centering
        \includegraphics[width=0.8\linewidth]{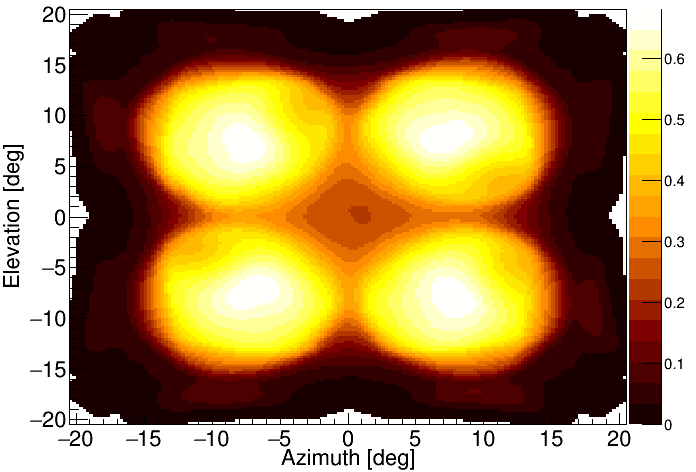}
        \caption{Original directional efficiency map for FAST@TA telescope 2.}
        \label{fig:FAST2Map}
    \end{subfigure}
    \caption{Ray trace maps using the measured response of a single FAST PMT and rotating based on PMT orientations at telescopes 1/3 (top) and 2 of FAST@TA. The absolute scale difference is due to the slightly different optical structures of the telescopes.}
    \label{fig:oldDirEffMap}
\end{figure}

\begin{figure}
\label{fig:TA2024maps}
    \centering
    \includegraphics[width=0.7\linewidth]{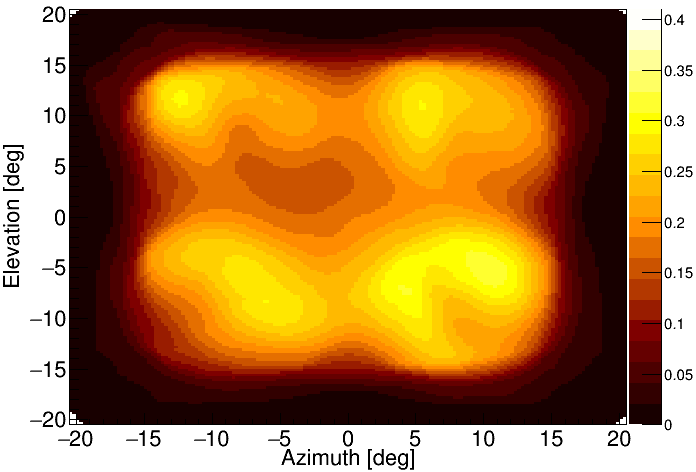}
    \includegraphics[width=0.7\linewidth]{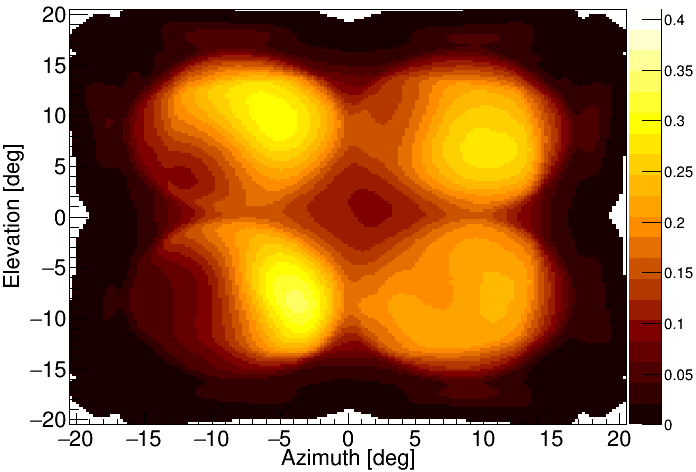}
    \includegraphics[width=0.7\linewidth]{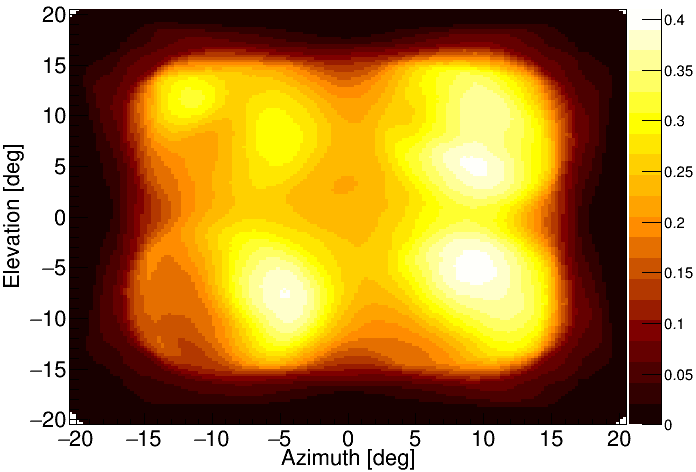}
    \caption{The ray trace maps of the FAST@TA telescopes derived from gain measurements made in 2024. From top to bottom the maps are for FAST1, FAST2 \& FAST3.}
    \label{fig:newDirEffMap}
\end{figure}

\chapter{Reconstruction of Prototype Data: Supplementary Plots}

\section{Initial Reconstruction Results: Geometry}
\label{apx:NEWSIMrecResults}
\begin{figure}[h!]
    \centering
    \includegraphics[width=0.49\linewidth]{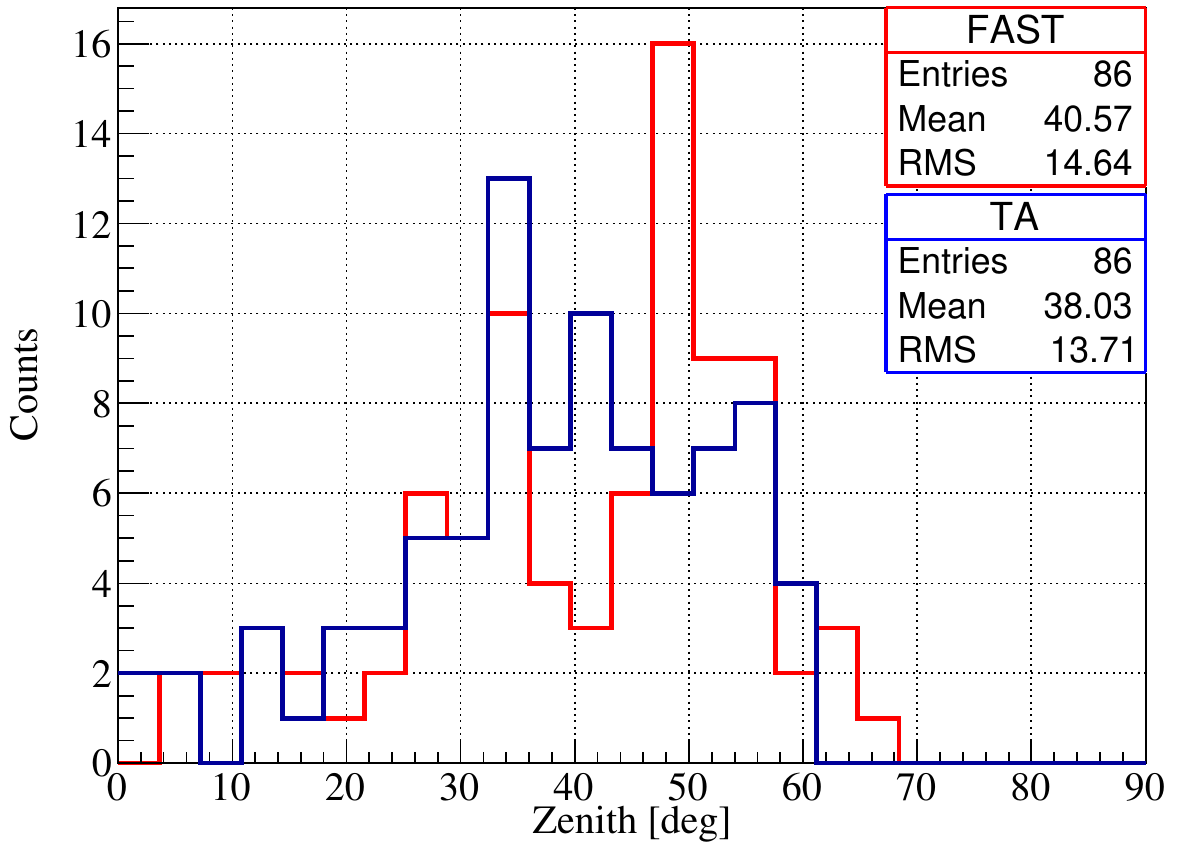}
    \includegraphics[width=0.49\linewidth]{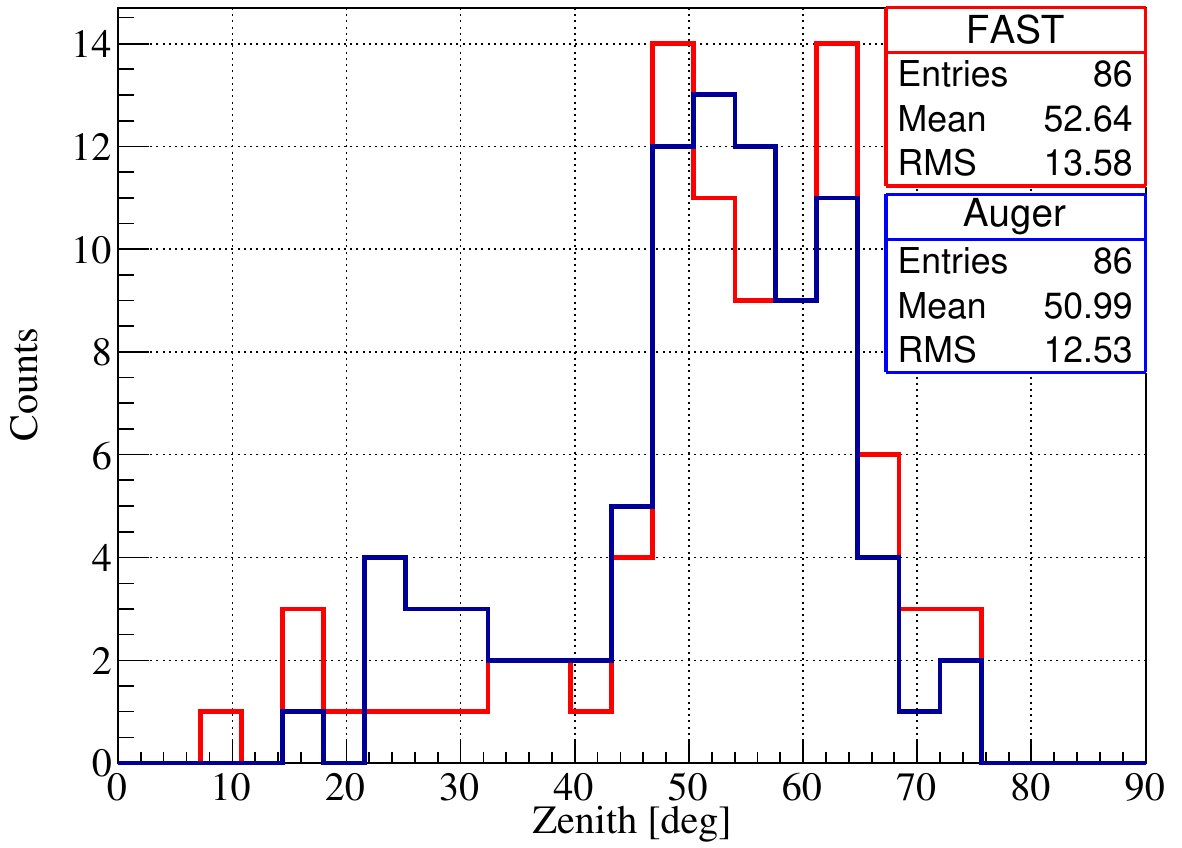}
    \includegraphics[width=0.49\linewidth]{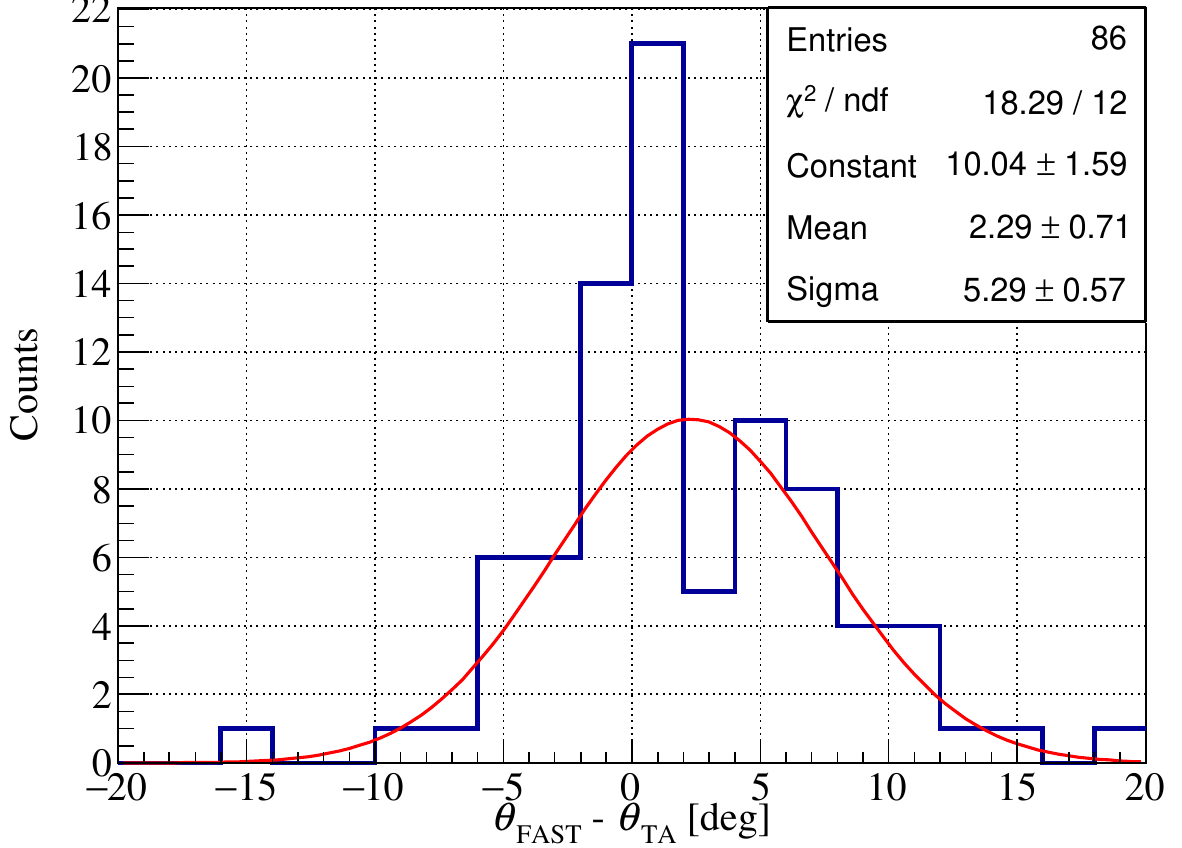}
    \includegraphics[width=0.49\linewidth]{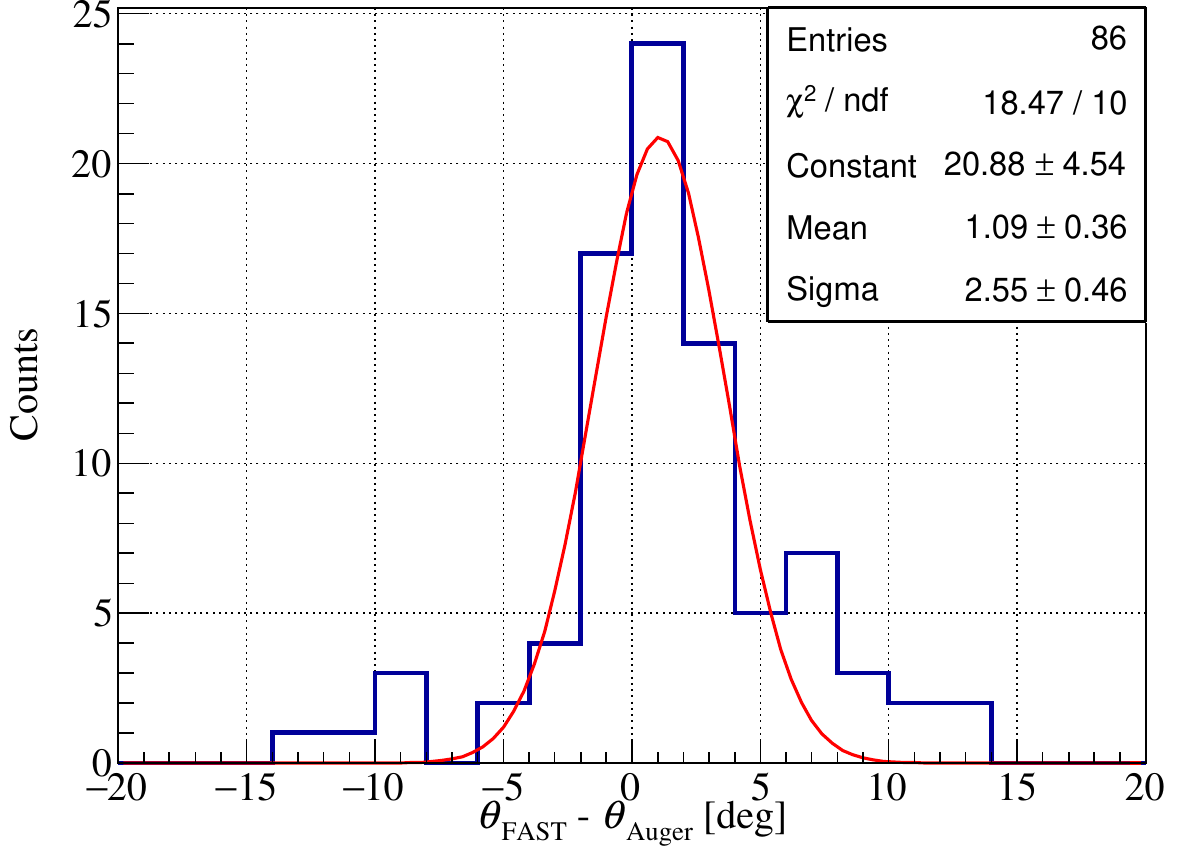}
    \caption{Histograms of the reconstructed zenith values from FAST@TA (left) and FAST@Auger (right) in the initial reconstruction (see Section \ref{sec:initRecon}). The FAST results are shown in red. The TA/Auger results are shown in blue. The event by event difference histograms are shown in the bottom panels.}
    \label{fig:firstreczen}
\end{figure}

\clearpage
\begin{figure}
    \centering
    \includegraphics[width=0.49\linewidth]{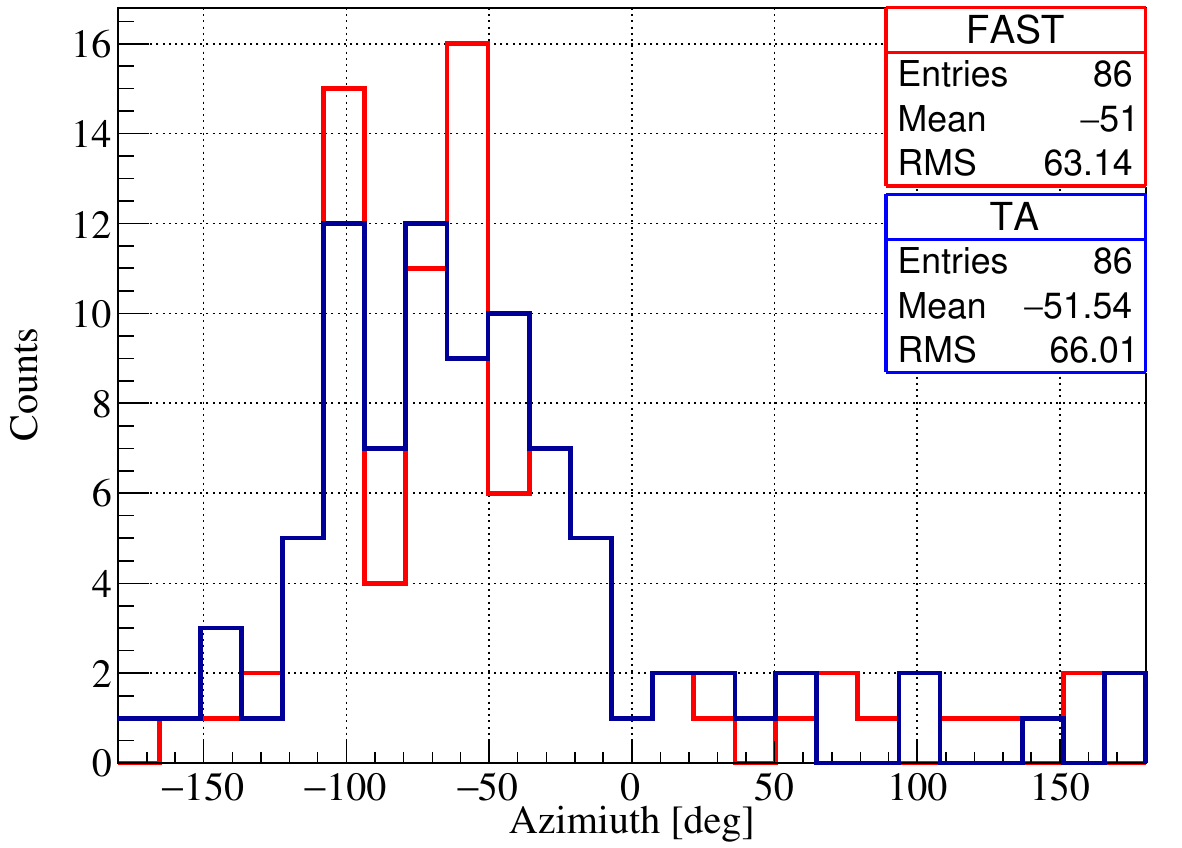}
    \includegraphics[width=0.49\linewidth]{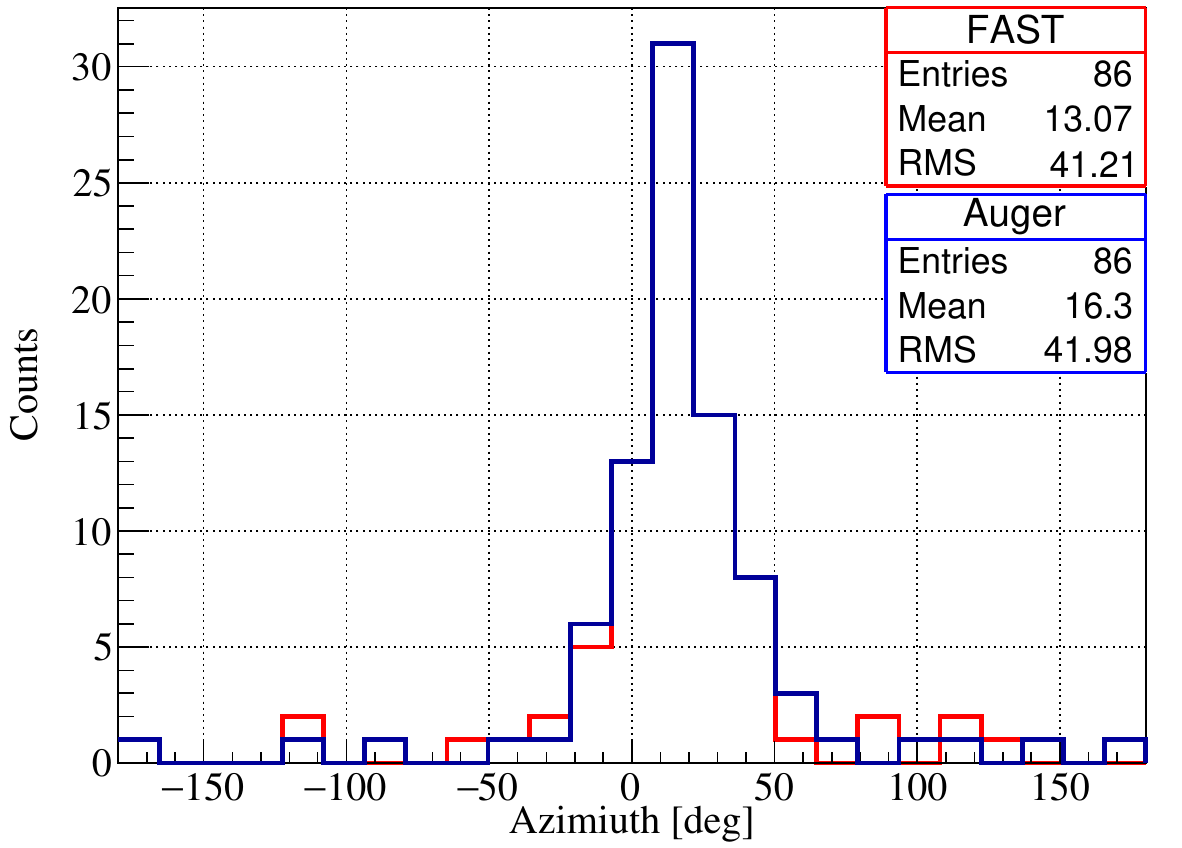}
    \includegraphics[width=0.49\linewidth]{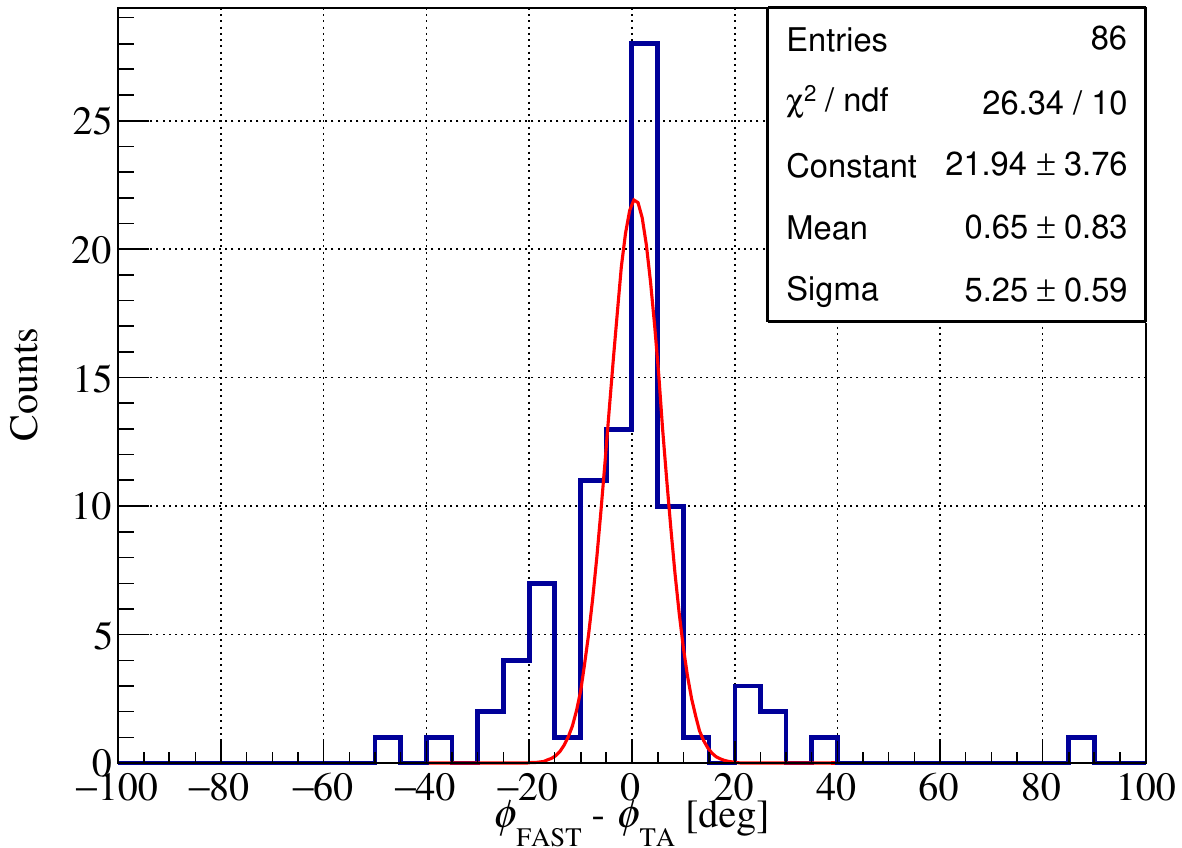}
    \includegraphics[width=0.49\linewidth]{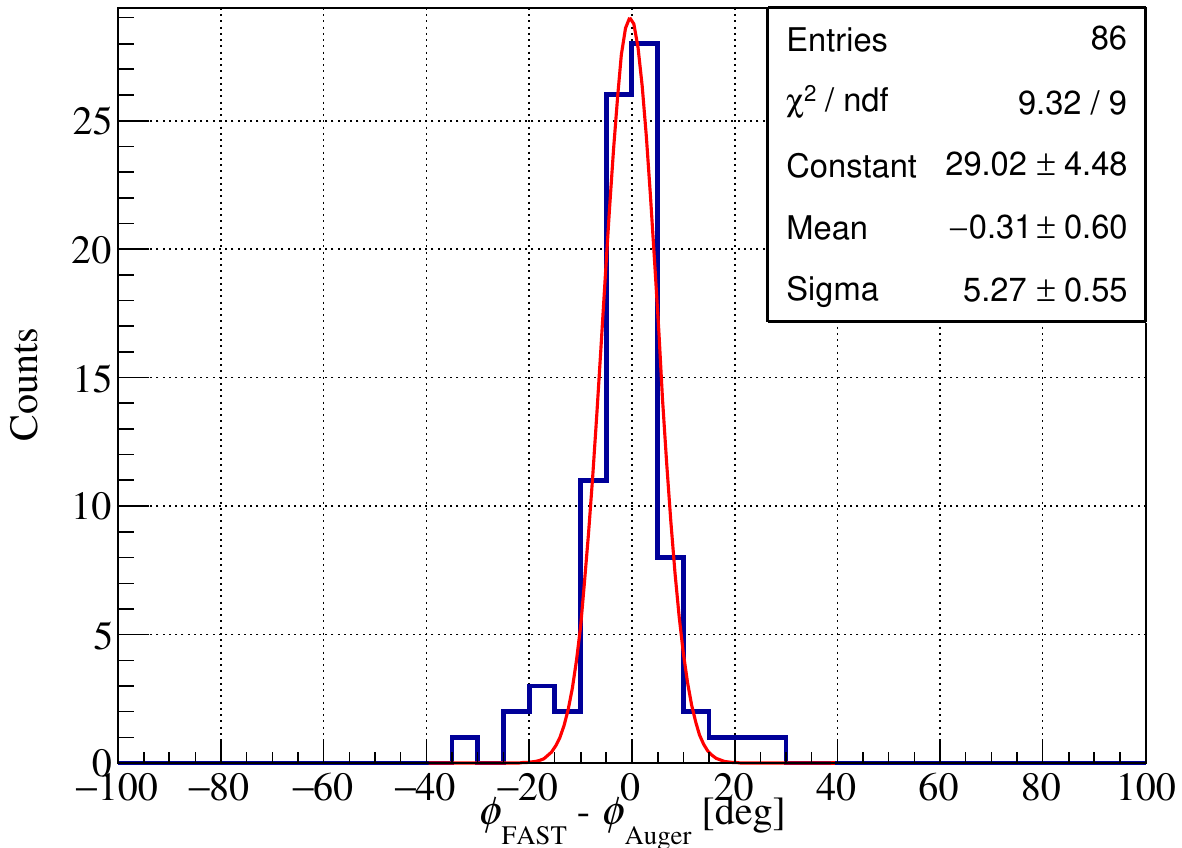}
    \caption{Same as Figure \ref{fig:firstreczen} but for azimuth.}
    \label{fig:firstrecazi}
\end{figure}

\clearpage
\begin{figure}
    \centering
    \includegraphics[width=0.49\linewidth]{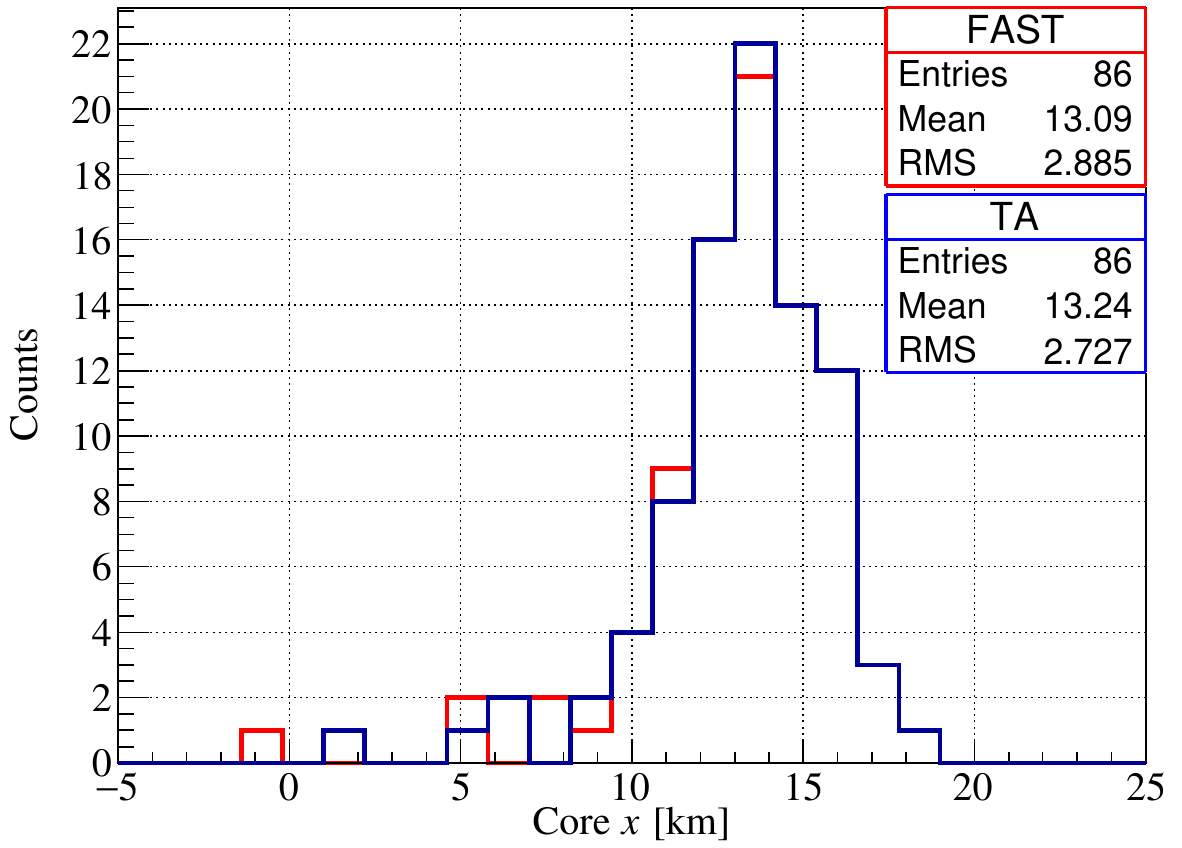}
    \includegraphics[width=0.49\linewidth]{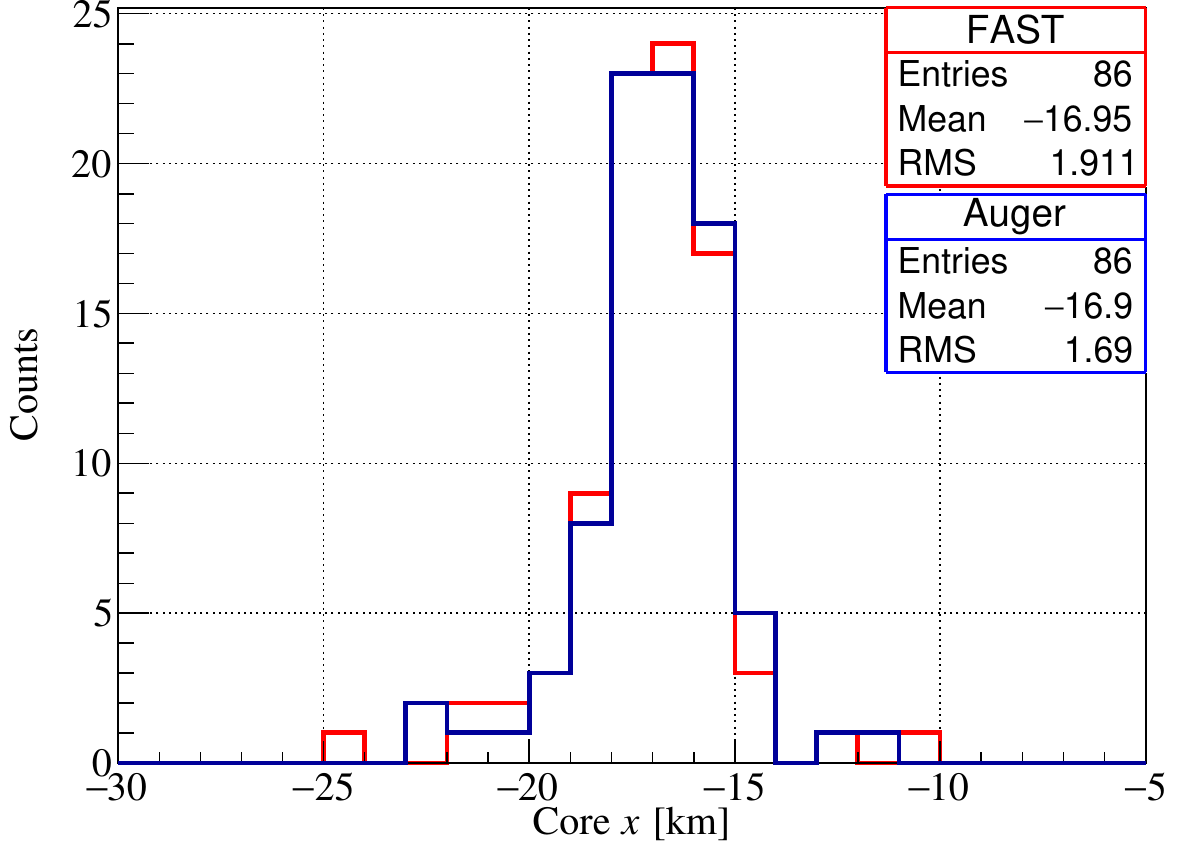}
    \includegraphics[width=0.49\linewidth]{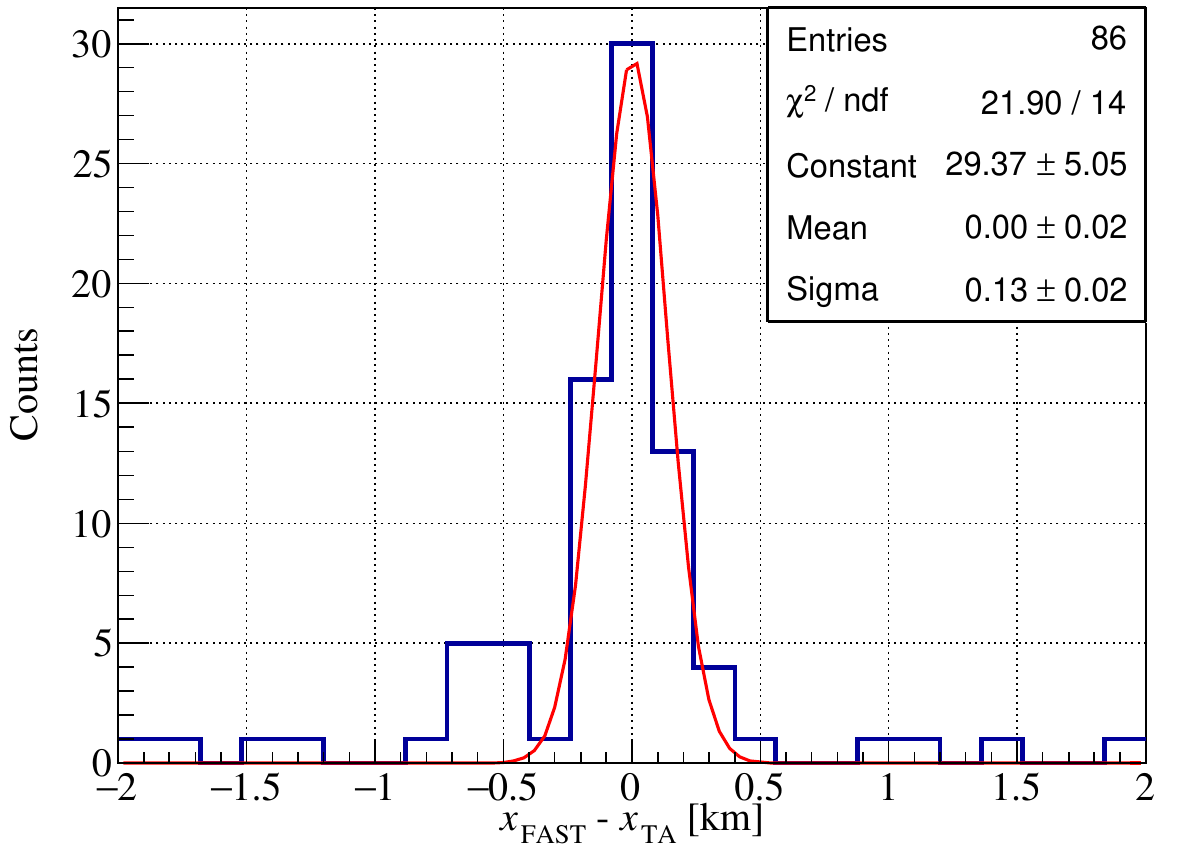}
    \includegraphics[width=0.49\linewidth]{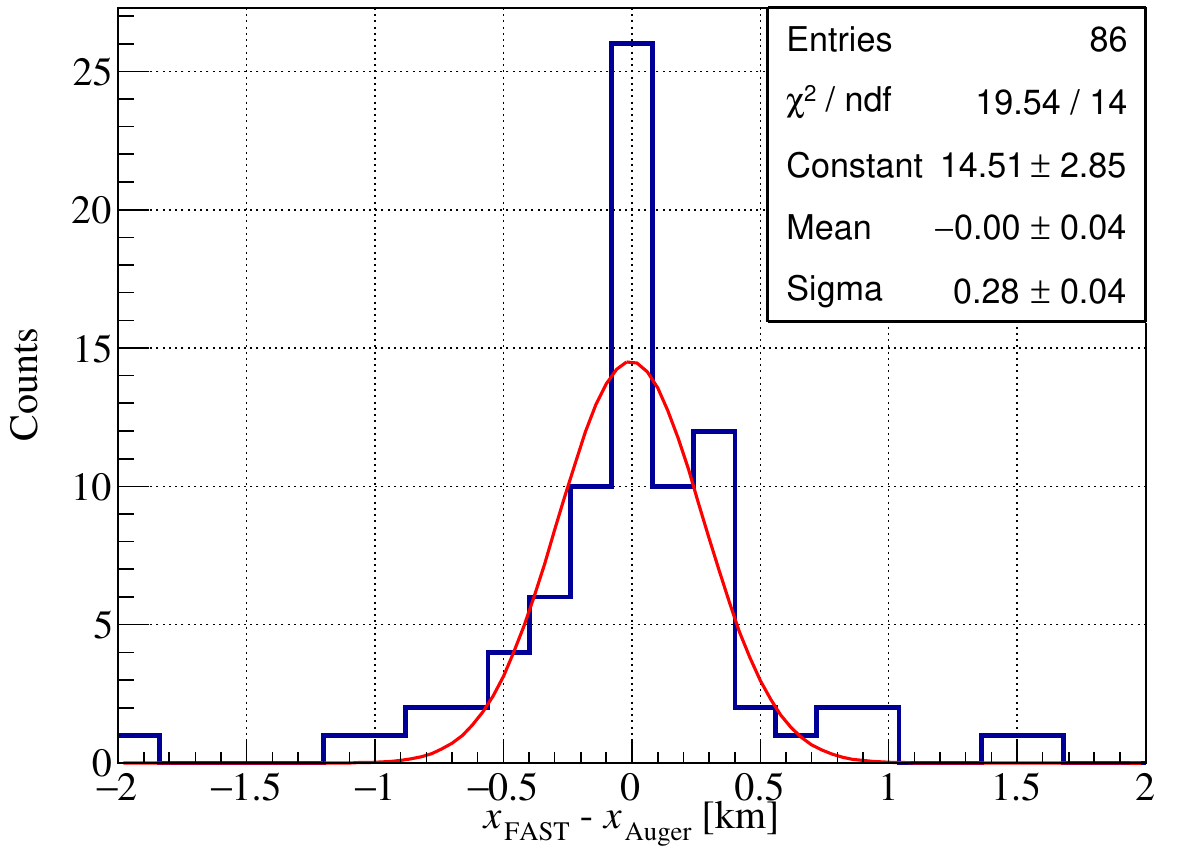}
    \caption{Same as Figure \ref{fig:firstreczen} but for core $x$.}
    \label{fig:firstreccx}
\end{figure}

\clearpage
\begin{figure}
    \centering
    \includegraphics[width=0.49\linewidth]{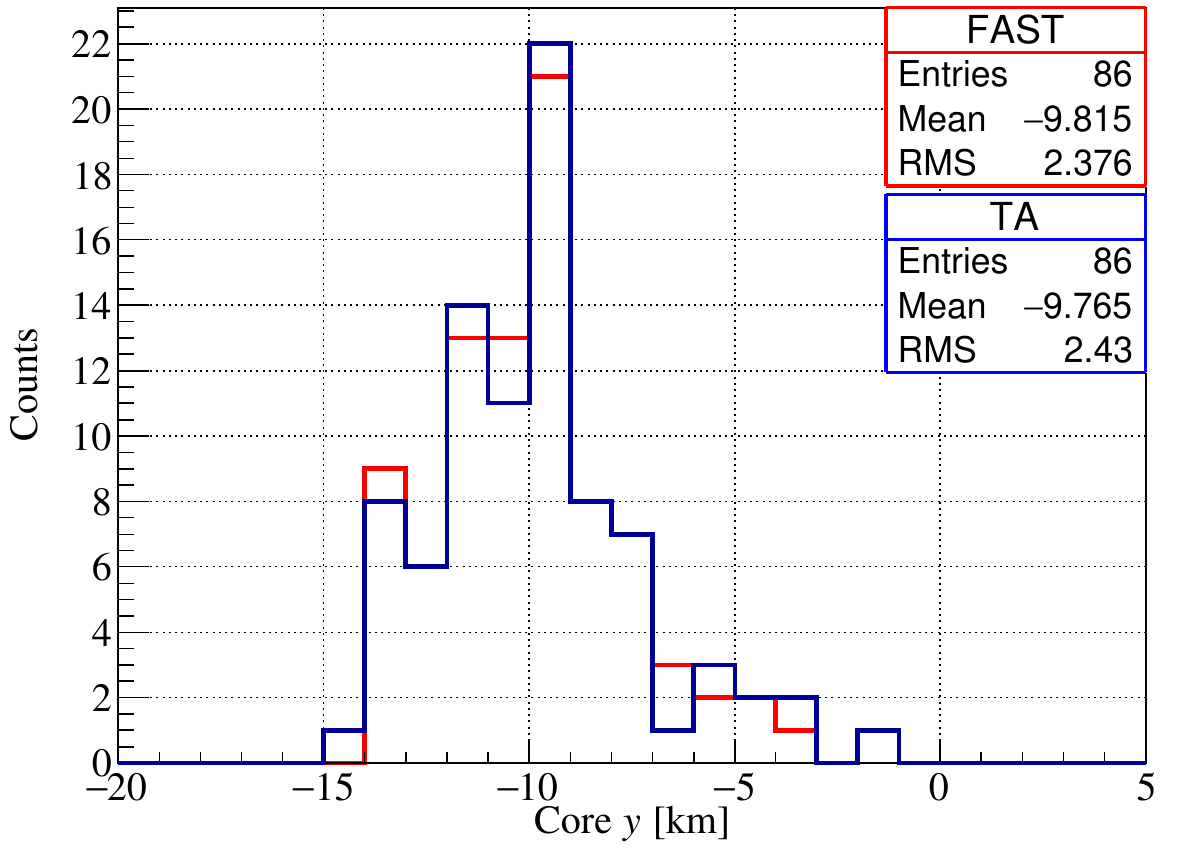}
    \includegraphics[width=0.49\linewidth]{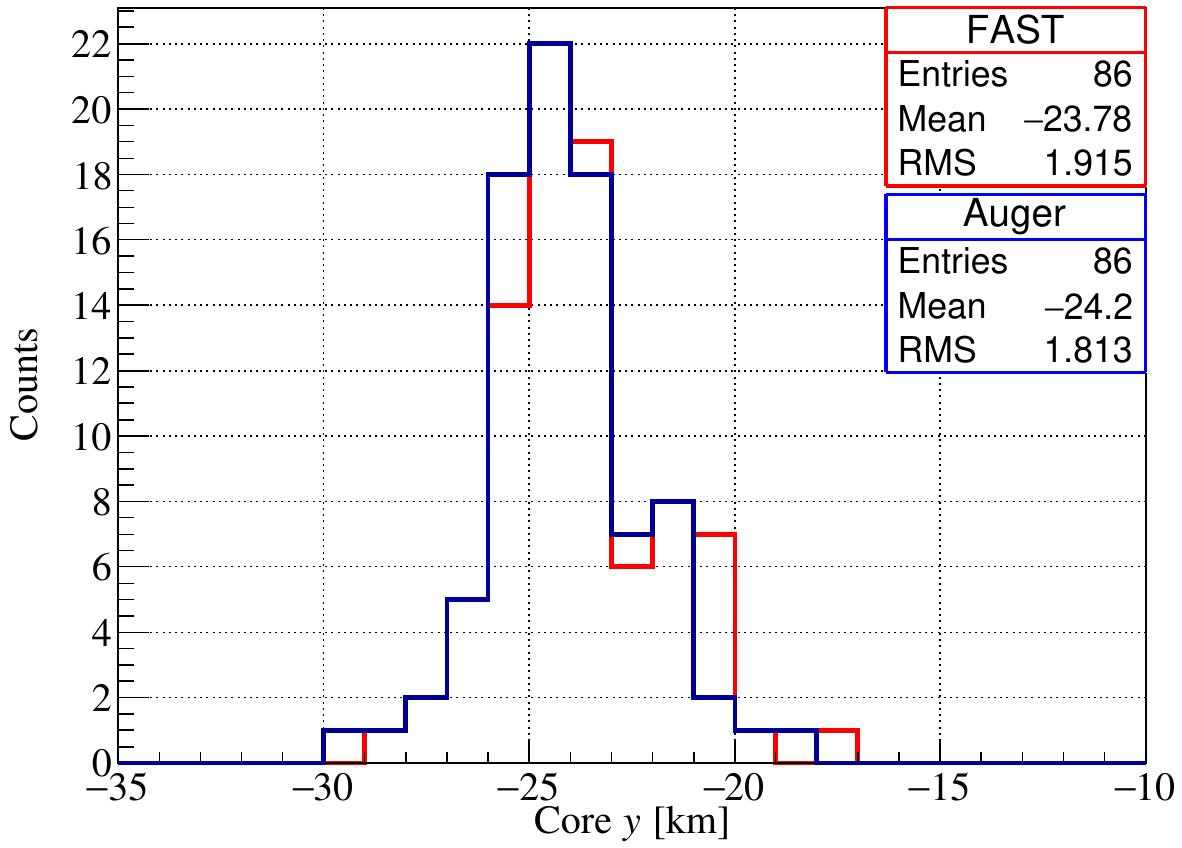}
    \includegraphics[width=0.49\linewidth]{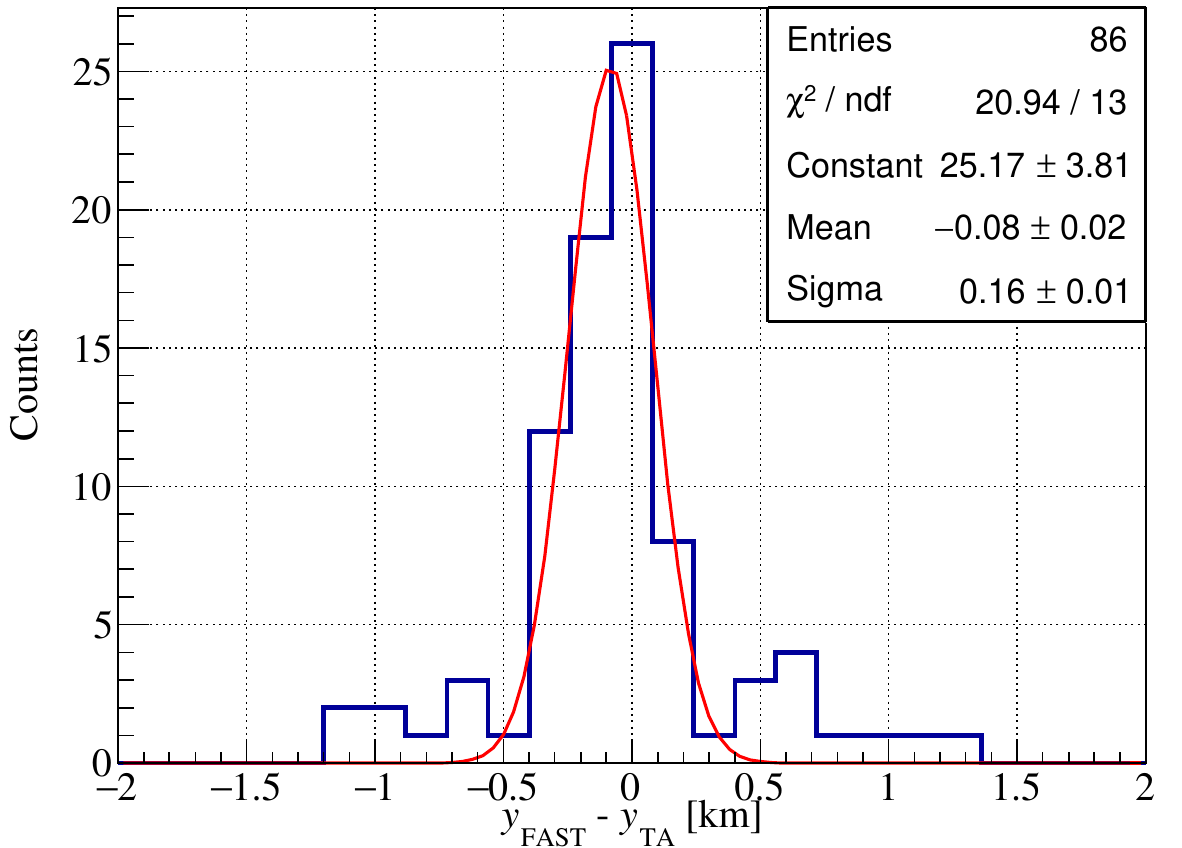}
    \includegraphics[width=0.49\linewidth]{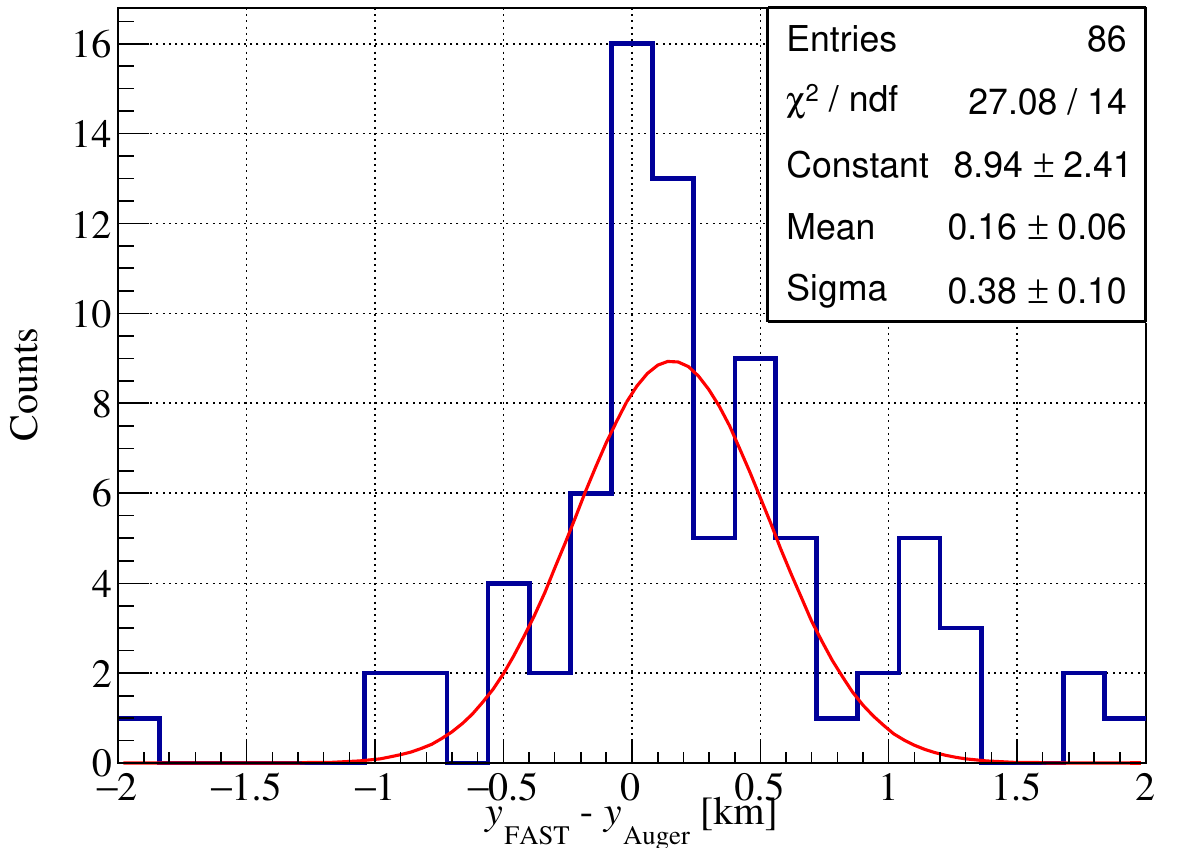}
    \caption{Same as Figure \ref{fig:firstreczen} but for core $y$.}
    \label{fig:firstreccy}
\end{figure}

\clearpage
\section{PMT Scanner Measurements: FAST 1 and 2}
\label{sec:PMTscannerMeasurements}
\begin{figure}[h!]
    \centering
    \includegraphics[width=0.7\linewidth]{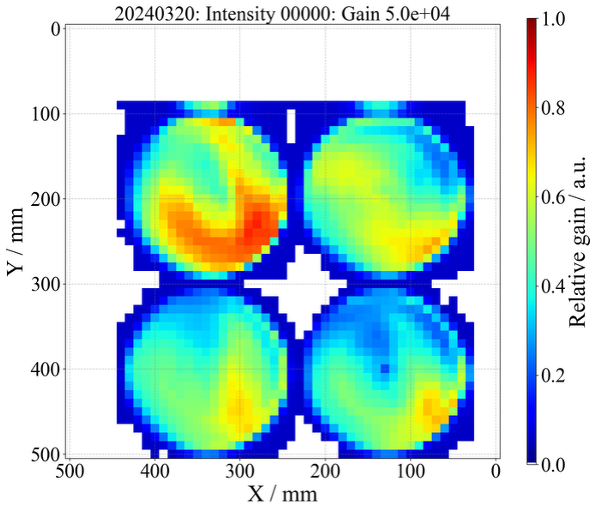}
    \includegraphics[width=0.7\linewidth]{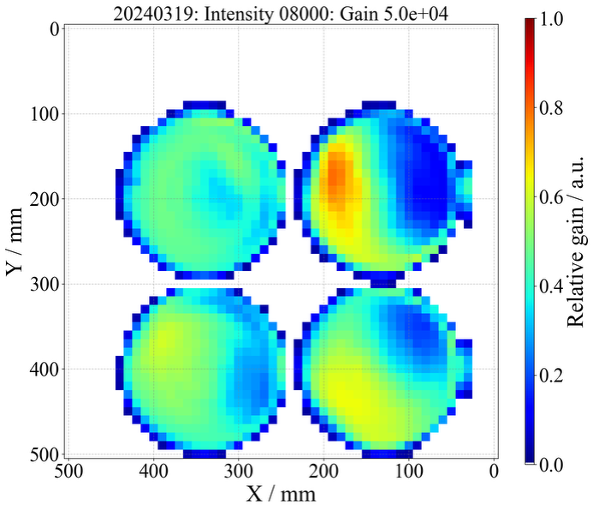}
    \caption{Measurements of the non-uniformity in the PMT gain response for FAST 1 (top) and FAST 2 (bottom) after scaling and smoothing.}
    \label{fig:PMTscannerMeasurements}
\end{figure}

\clearpage

\section{TSFEL DNN Inputs: FAST@TA Coincidences}
\label{apx:mlFirstGuessInputsTA}
\begin{figure}[h!]
    \centering
    \includegraphics[width=1\linewidth]{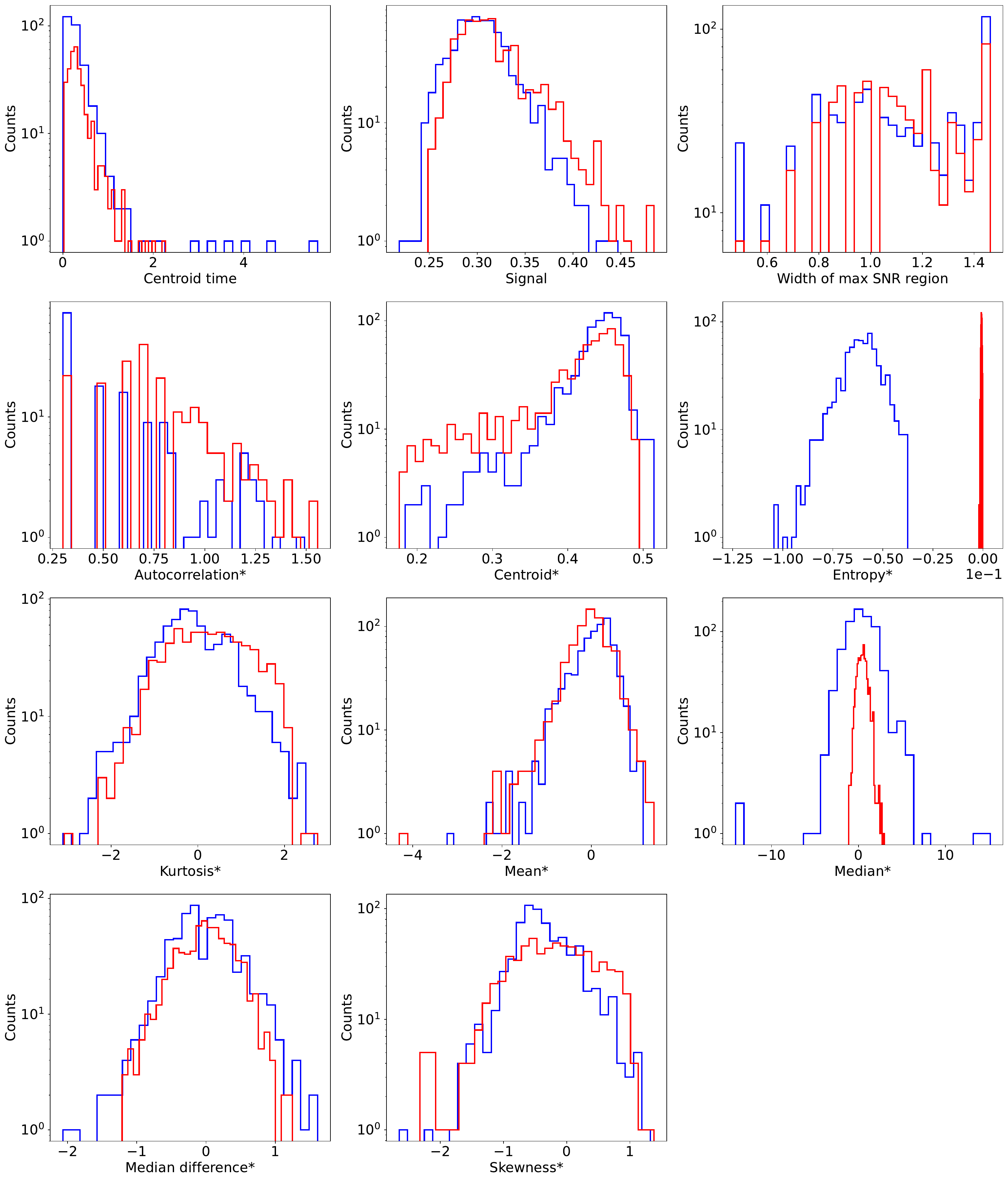}
    \caption{Inputs into the TSFEL DNN model for the FAST@Auger data. The coincidence data results are shown in blue and those from the FAST simulation of the Auger values in red.}
    \label{fig:mlFirstGuessInputsTA}
\end{figure}

\clearpage

\section{First guess + TDR Examples}
\label{apx:tsfelTDRexamples}

\begin{figure}[h!]
    \centering
    \includegraphics[width=1\linewidth]{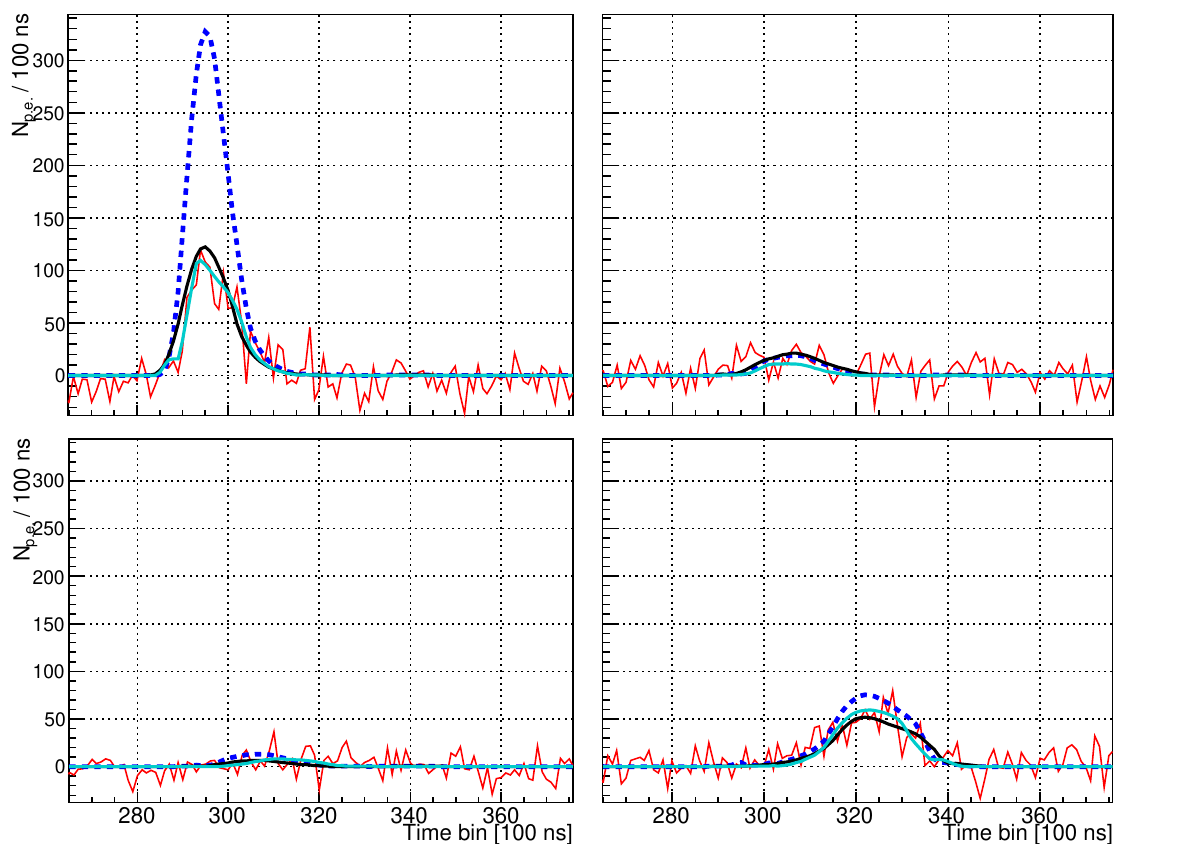}
    \includegraphics[width=1\linewidth]{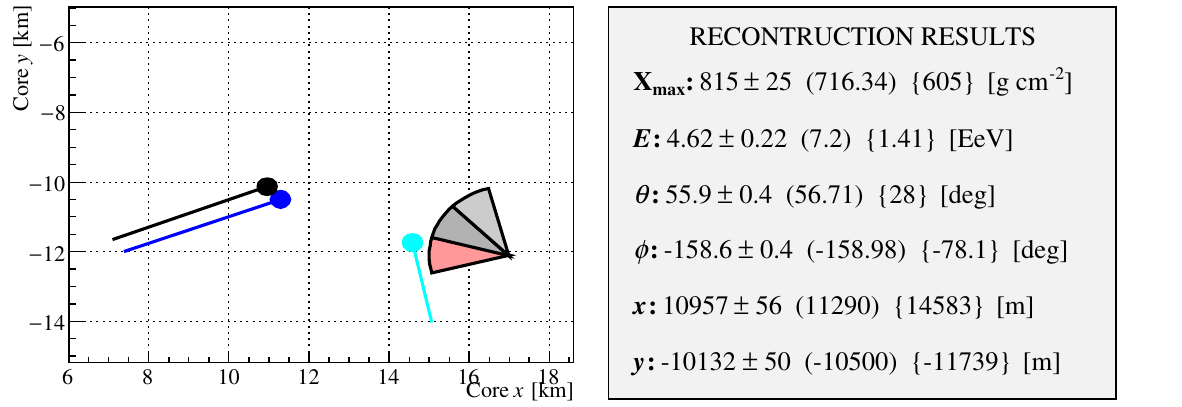}
    \caption{Reconstruction of an event observed by FAST@TA on 2018/11/07 using the TSFEL DNN and TDR. The format of the figure is explained in Figure \ref{fig:taMLexamp1}. As a reminder, the data traces are shown in red, whilst the blue, black and cyan traces / shower axes in the bottom left plot show the results for the TSFEL DNN first guess, TSFEL DNN + TDR best fit, and the TA first guess + TDR best fit respectively. The reconstruction results are shown from left to right in the bottom right panel as TSFEL DNN + TDR best fit, TSFEL DNN first guess and Auger first guess + TDR best fit.}
    \label{fig:taMLexamp2}
\end{figure}

\begin{figure}[]
    \centering
    \includegraphics[width=1\linewidth]{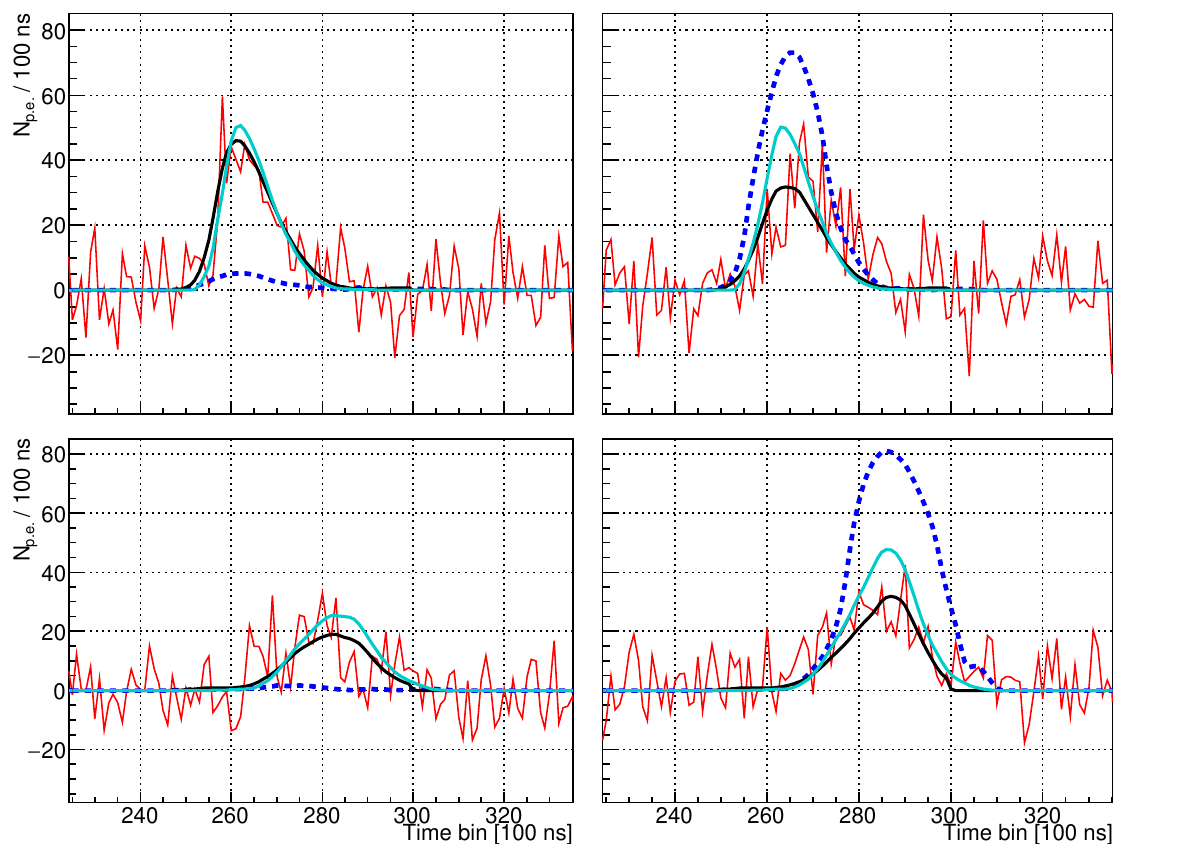}
    \includegraphics[width=1\linewidth]{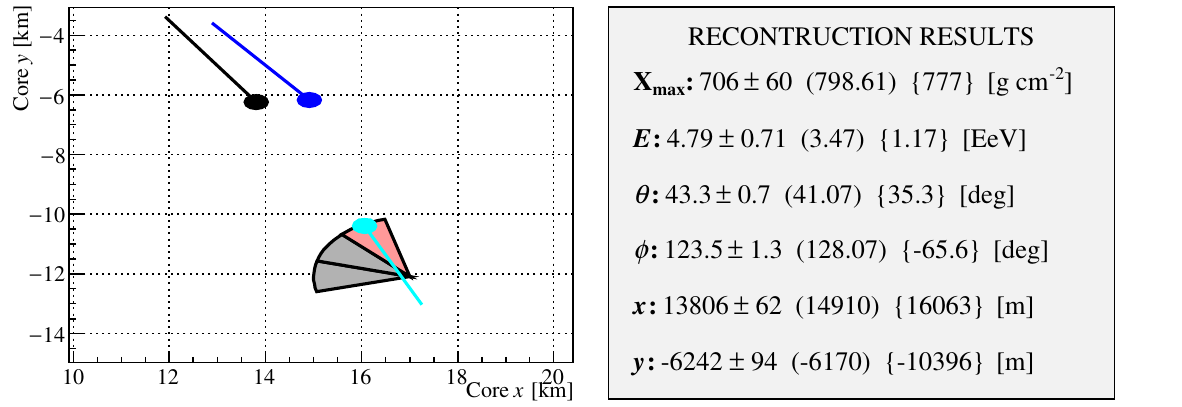}
    \caption{Reconstruction of an event observed by FAST@TA on 2019/02/09 using the TSFEL DNN and TDR. See Figure \ref{fig:taMLexamp2} or \ref{fig:taMLexamp1} for an explanation of the figure layout.}
    \label{fig:taMLexamp3}
\end{figure}

\begin{figure}[]
    \centering
    \includegraphics[width=1\linewidth]{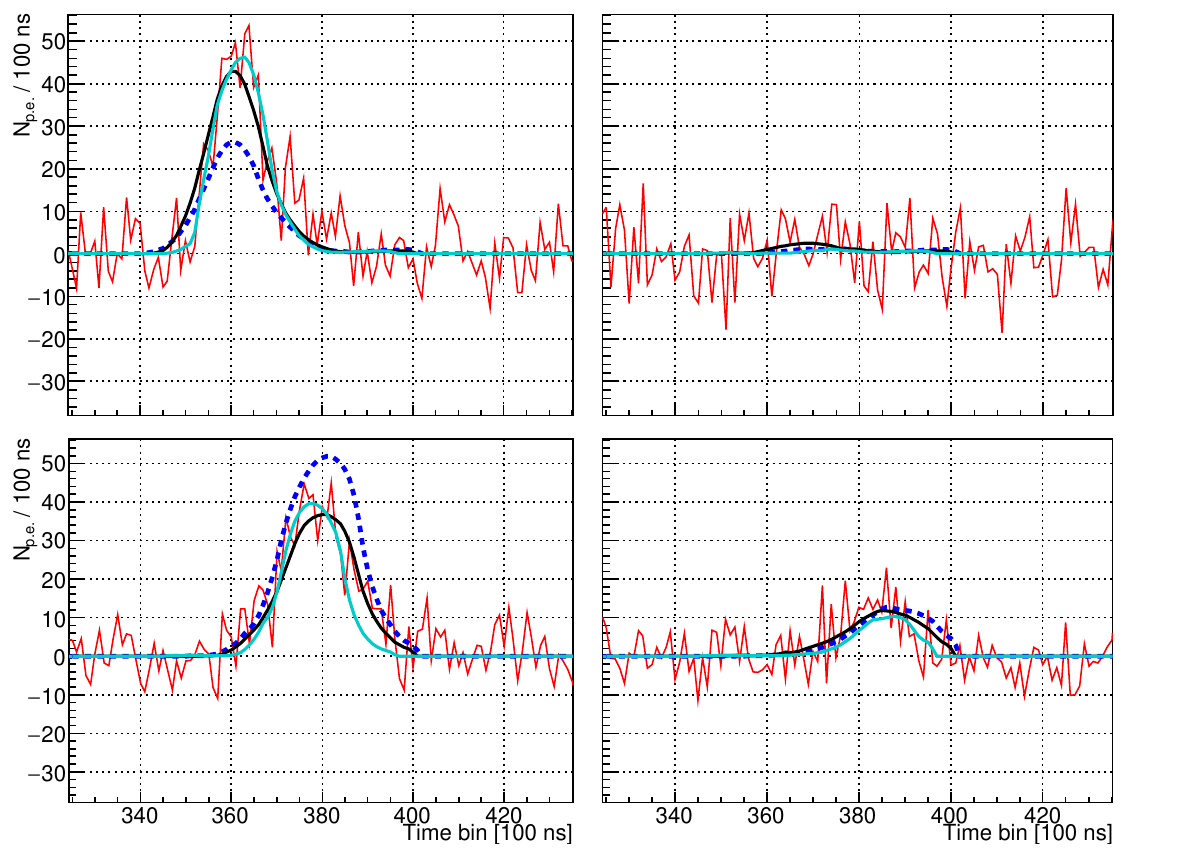}
    \includegraphics[width=1\linewidth]{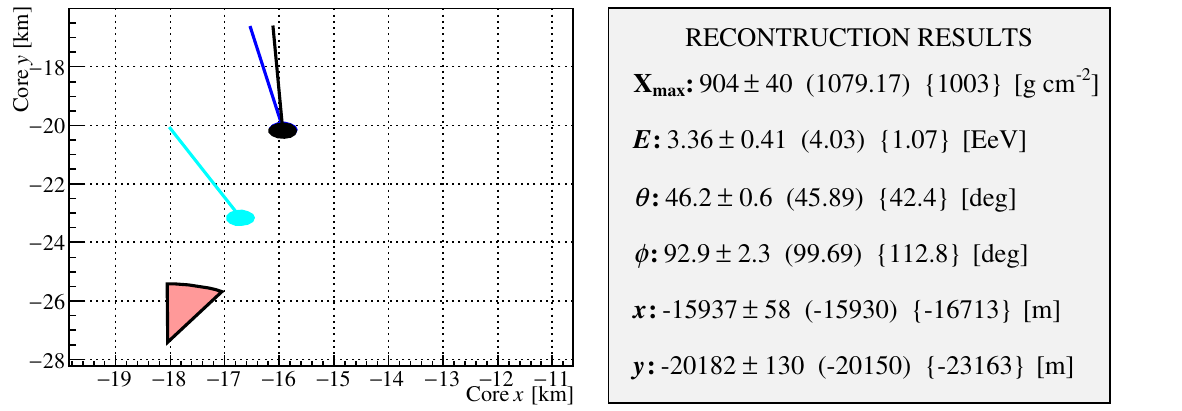}
    \caption{Reconstruction of an event observed by FAST@Auger on 2022/09/17 using the TSFEL DNN and TDR. See Figure \ref{fig:taMLexamp2} or \ref{fig:taMLexamp1} for an explanation of the figure layout.}
    \label{fig:paoMLexamp2}
\end{figure}

\clearpage

\begin{figure}
    \centering
    \includegraphics[width=0.65\linewidth]{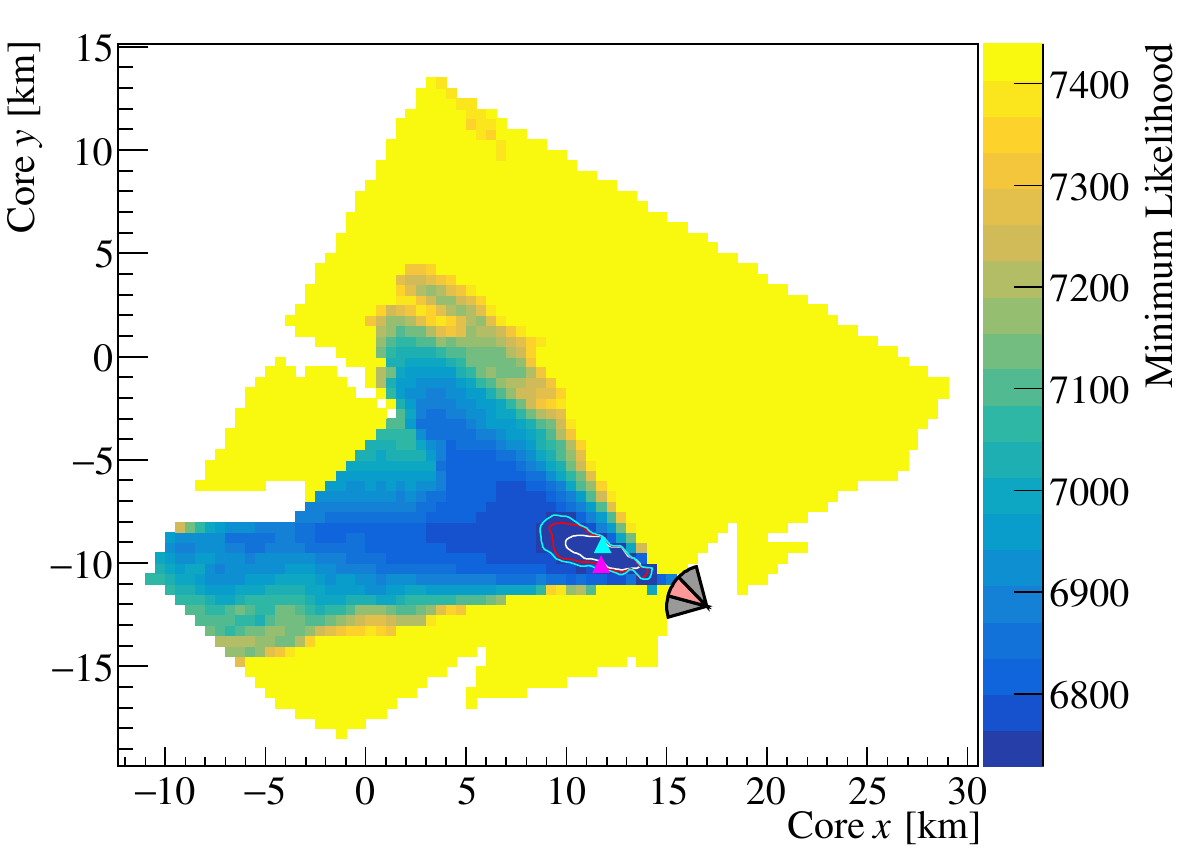}
    \includegraphics[width=0.65\linewidth]{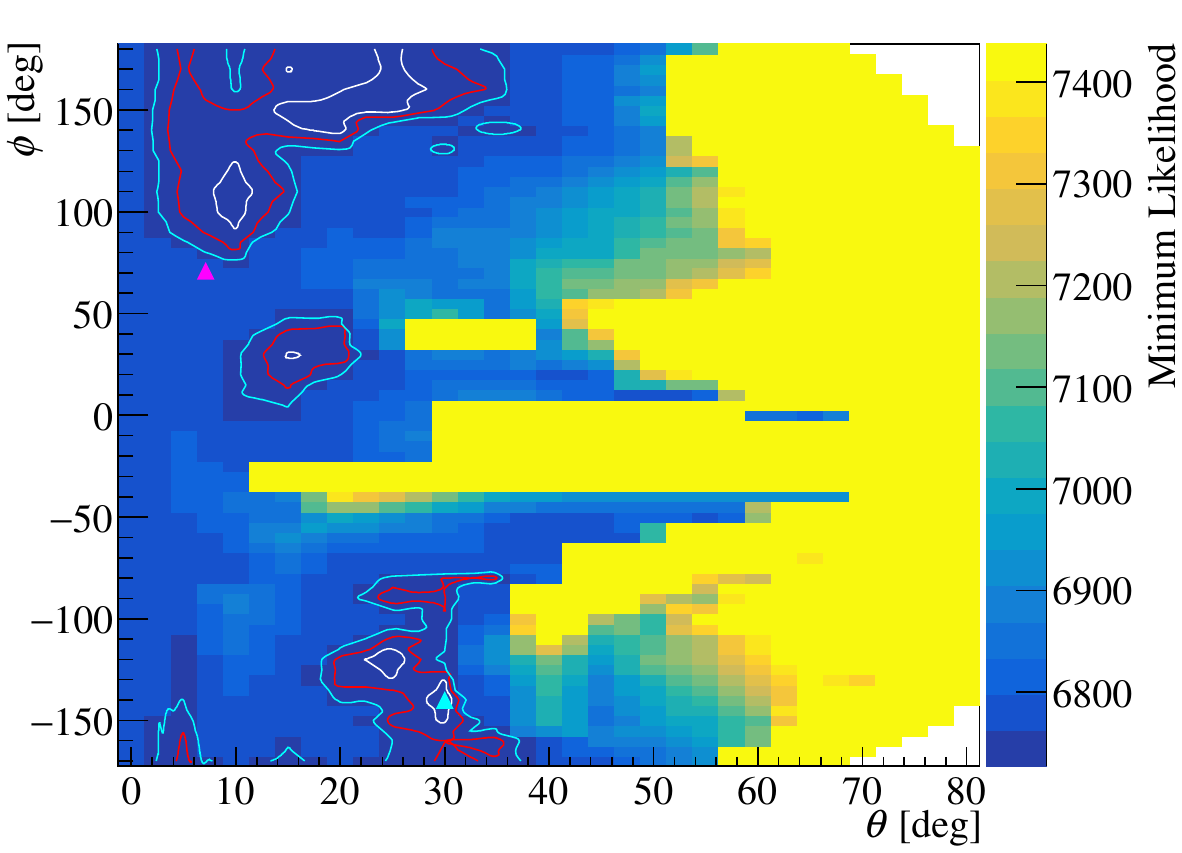}
    \includegraphics[width=0.65\linewidth]{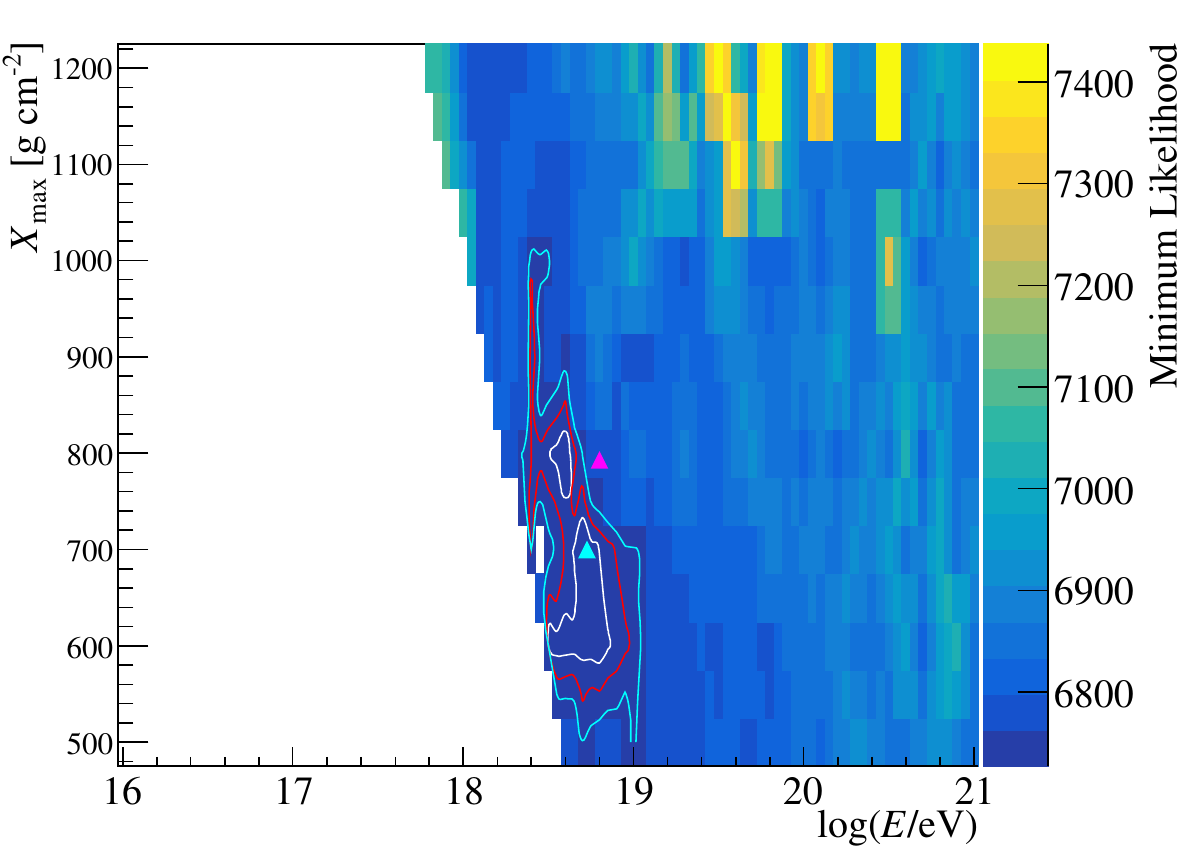}
    \caption{Likelihood maps from the Template Method applied to a coincidence event observed by FAST@TA on 2019/10/25. The layout of the figure is the same as Figure \ref{fig:taTMexampMaps}.}
    \label{fig:taTMexampMaps2}
\end{figure}

\begin{figure}
    \centering
    \includegraphics[width=1\linewidth]{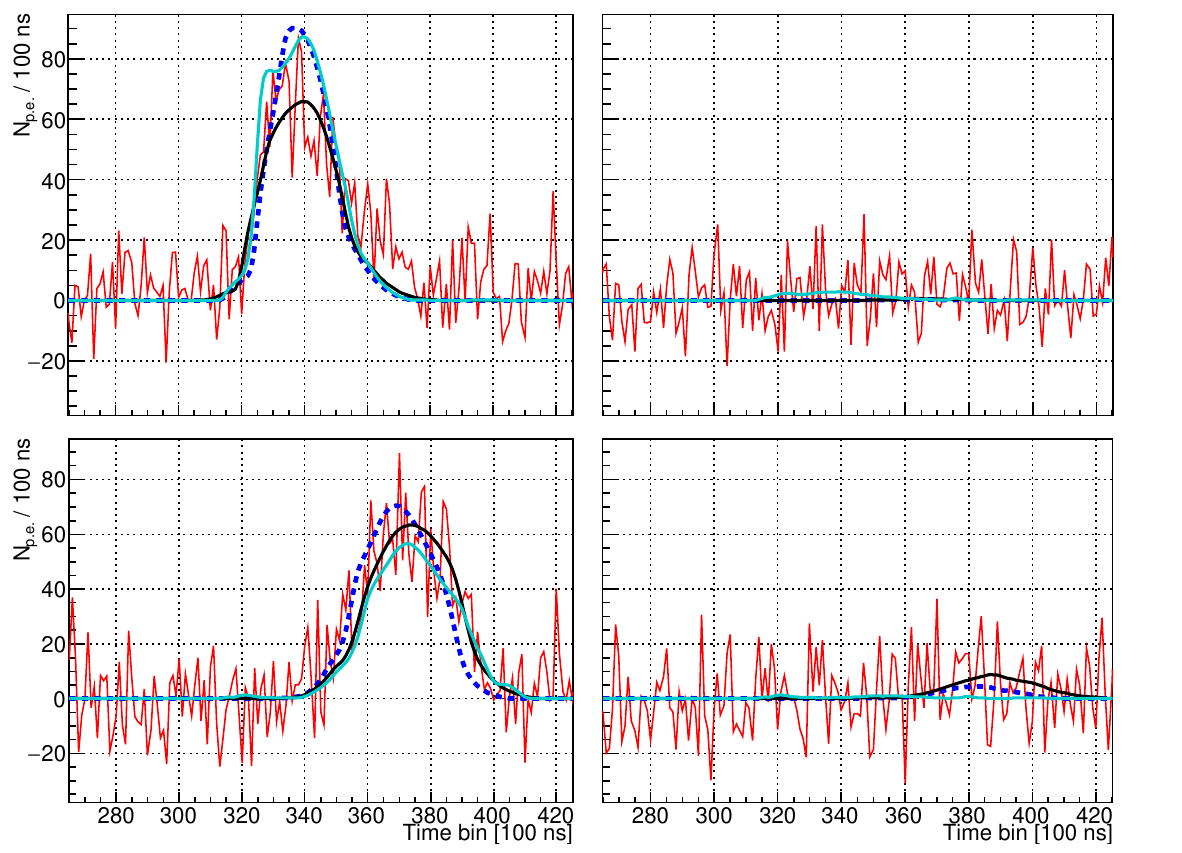}
    \includegraphics[width=1\linewidth]{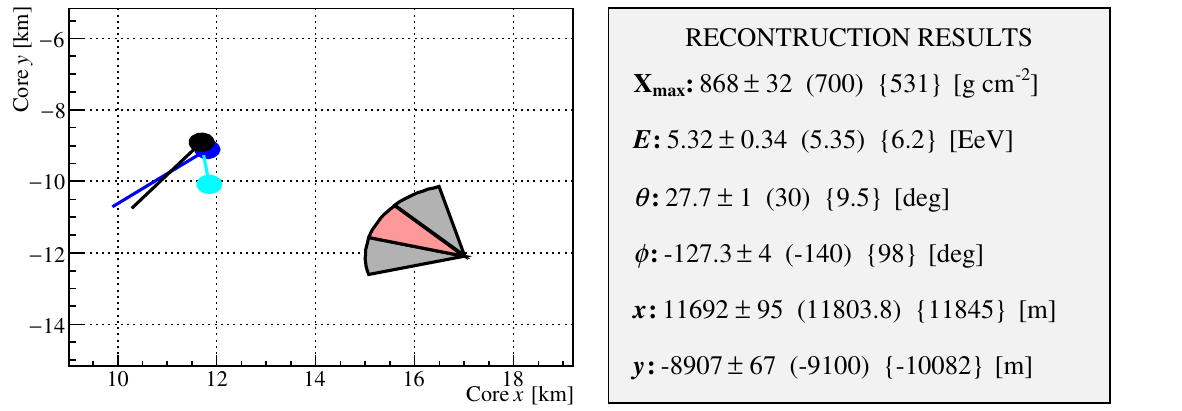}
    \caption{Template Method + TDR result for the event in Figure \ref{fig:taTMexampMaps2}. See Figure \ref{fig:taMLexamp2} or \ref{fig:taMLexamp1} for an explanation of the figure layout.}
    \label{fig:tempTDRresExamp_ta}
\end{figure}

\begin{figure}
    \centering
    \includegraphics[width=0.65\linewidth]{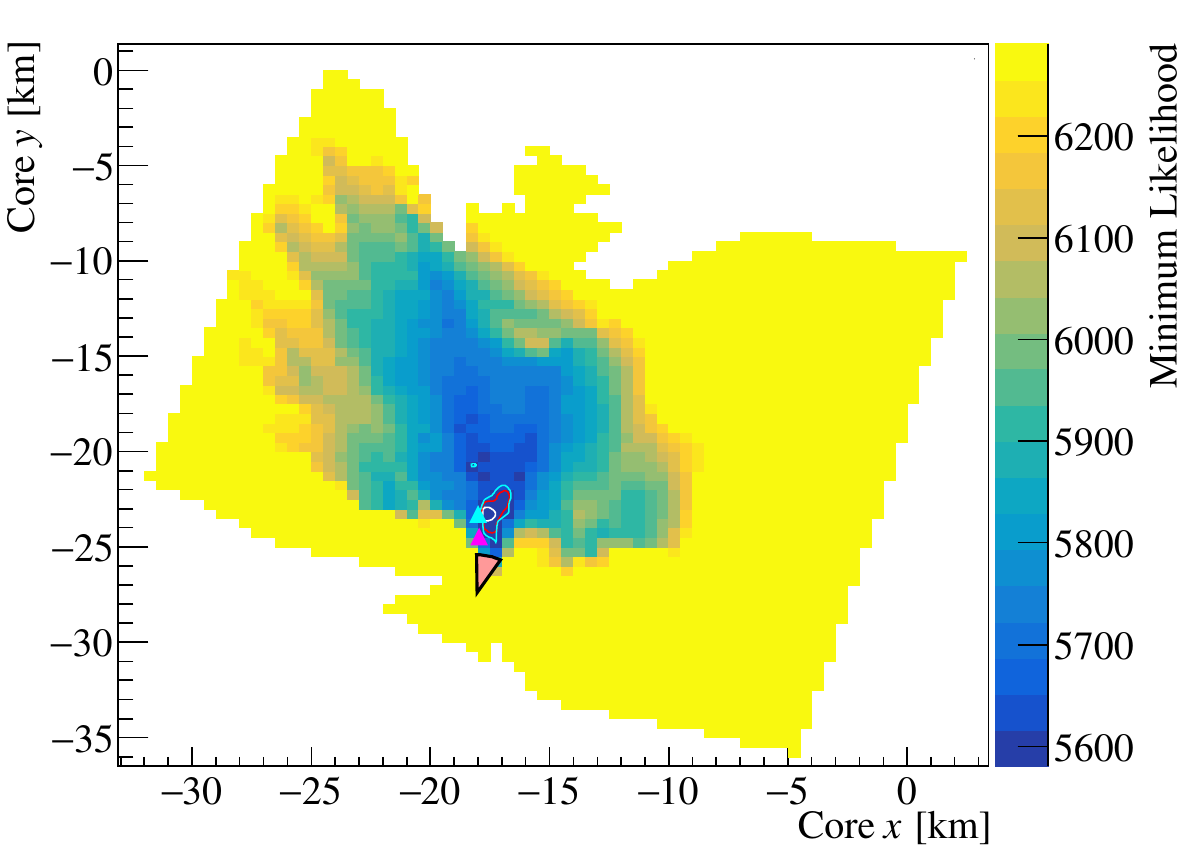}
    \includegraphics[width=0.65\linewidth]{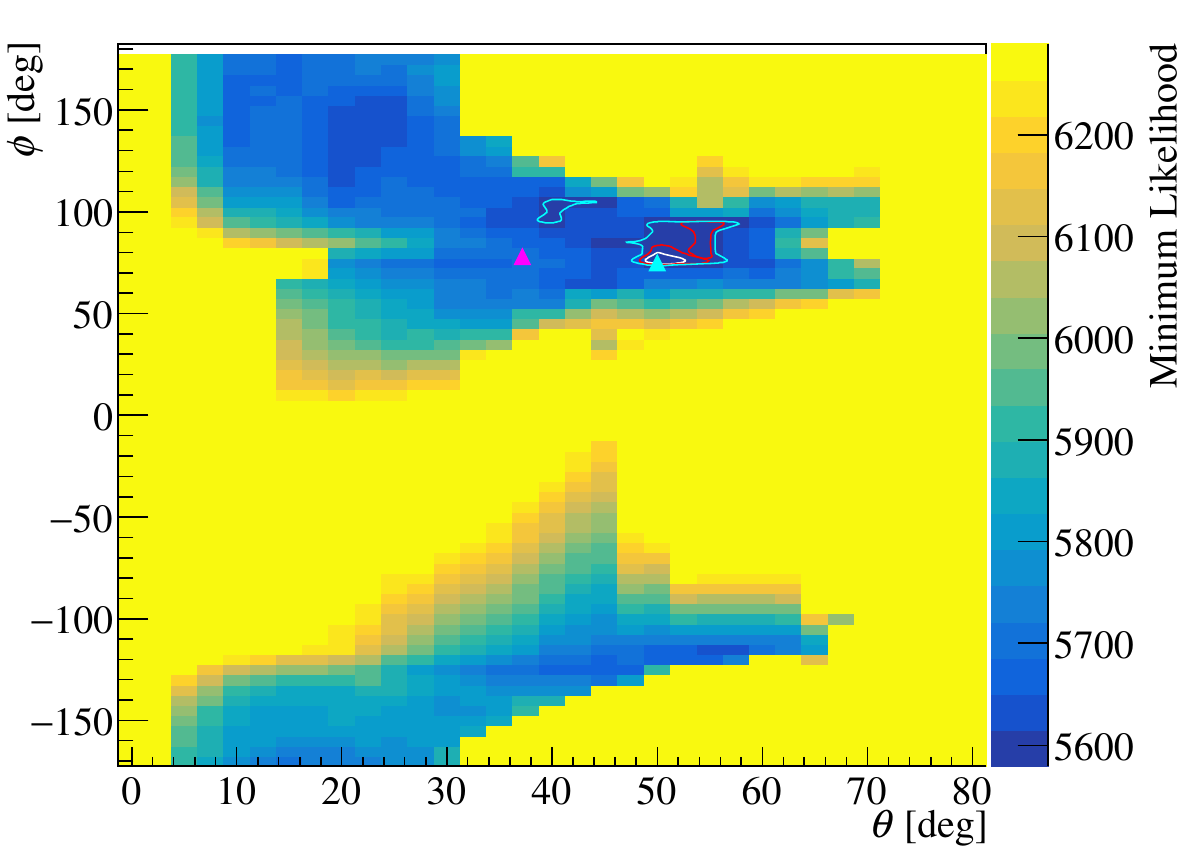}
    \includegraphics[width=0.65\linewidth]{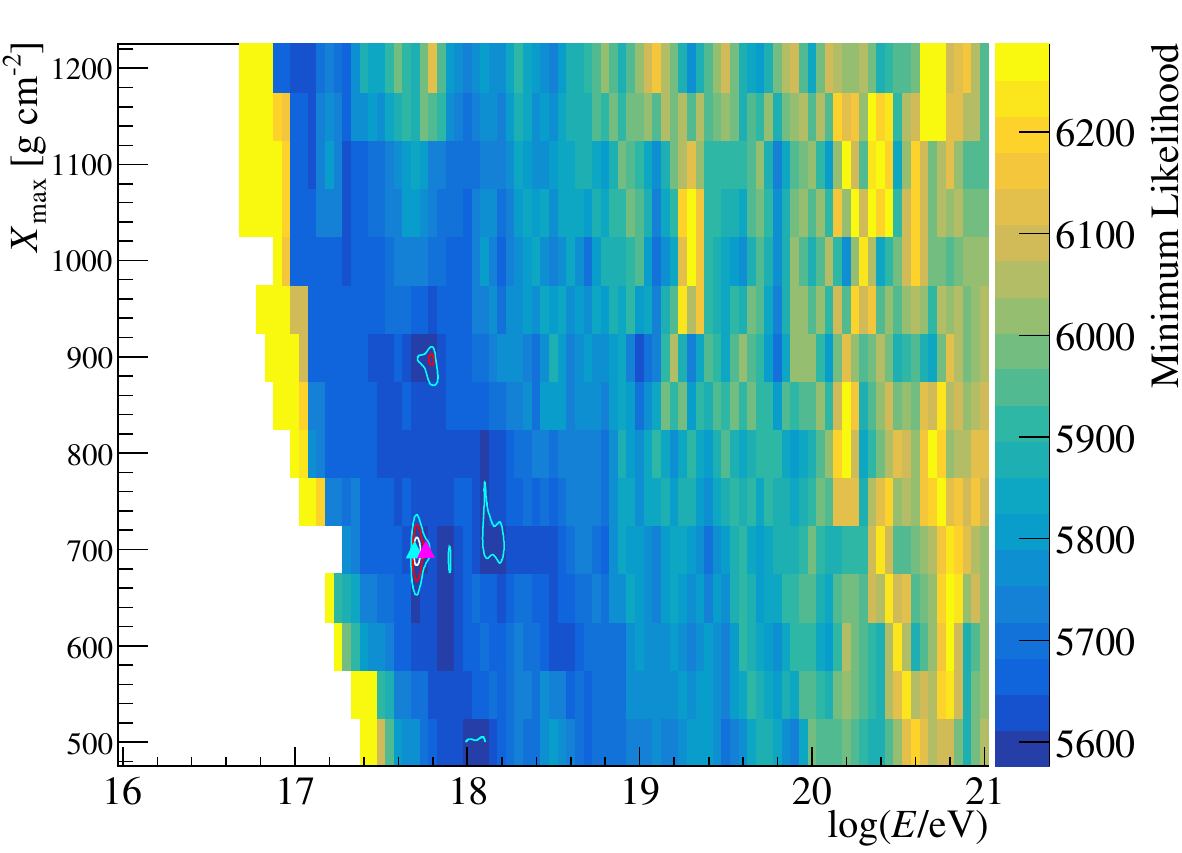}
    \caption{Likelihood maps from the Template Method applied to a coincidence event observed by FAST@Auger on 2022/07/06. The layout of the figure is the same as Figure \ref{fig:taTMexampMaps}.}
    \label{fig:paoTMexampMaps2}
\end{figure}

\begin{figure}
    \centering
    \includegraphics[width=1\linewidth]{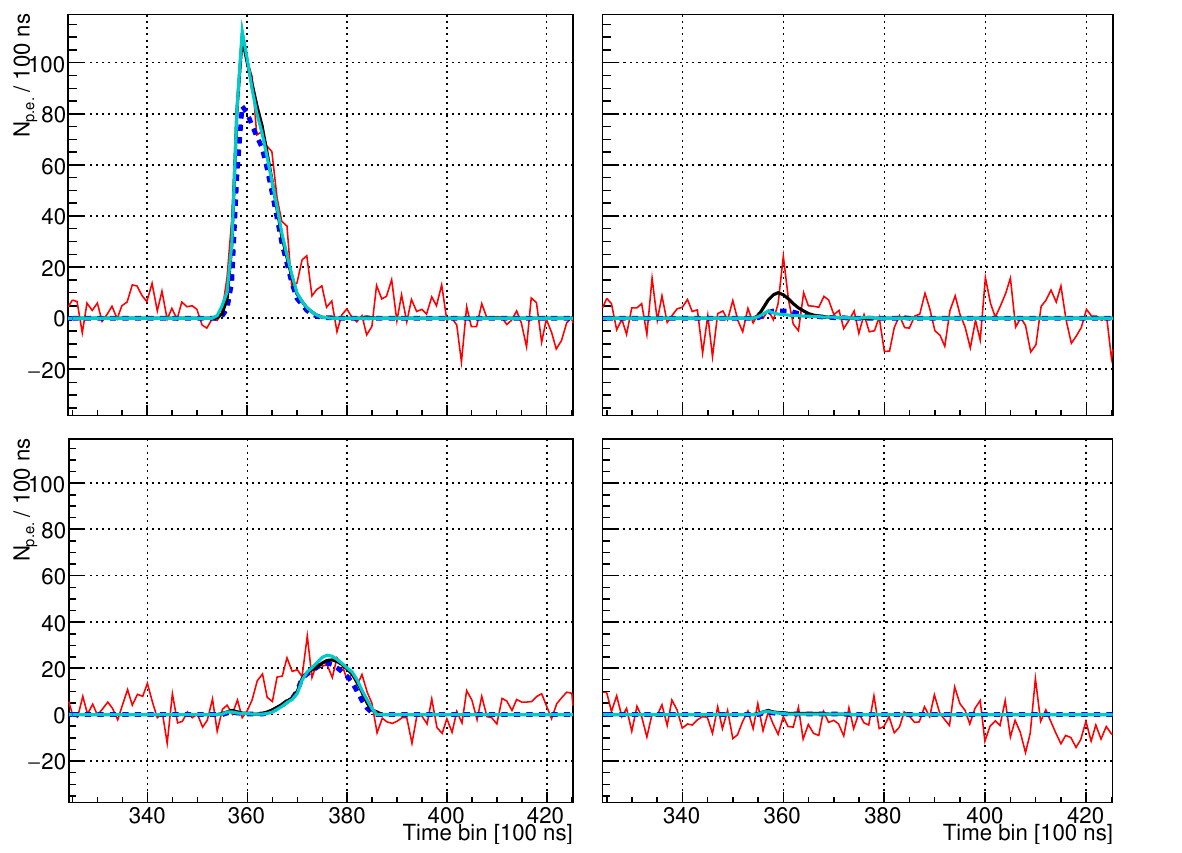}
    \includegraphics[width=1\linewidth]{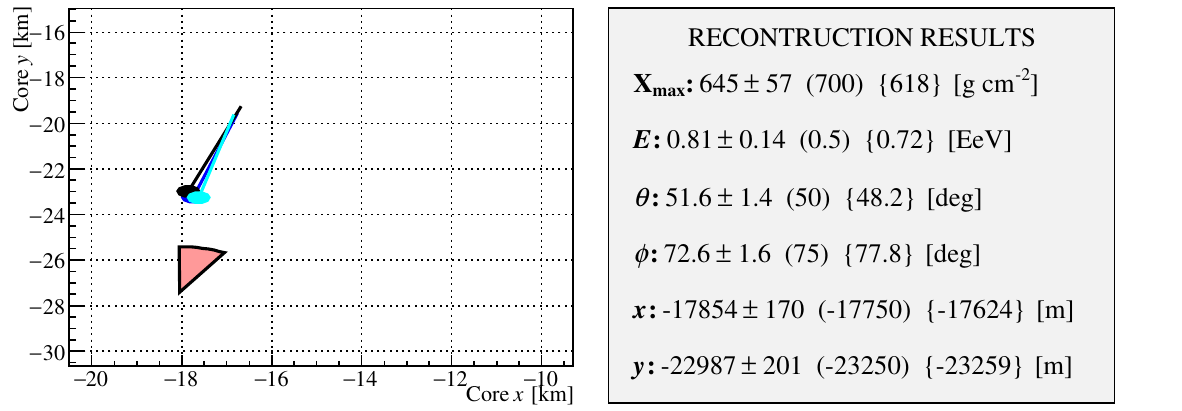}
    \caption{Template Method + TDR result for the event in Figure \ref{fig:paoTMexampMaps2}. See Figure \ref{fig:taMLexamp2} or \ref{fig:taMLexamp1} for an explanation of the figure layout.}
    \label{fig:tempTDRresExamp_pao}
\end{figure}

\cleardoublepage
\chapter*{Funding and Support}

This work has been funded by the former University Fellowship Founding Project (Sep. 2022 - Apr. 2024) and the current Support for Pioneering Research Initiated by the Next Generation / SPRING (Rhizome-Type Researcher Development for Interdisciplinary Knowledge Blooming) (Apr. 2024 - Sep. 2025) programs from Osaka Metropolitan University. The Author wishes to express their gratitude for the support provided by these programs over the course of this PhD.

\vspace{5mm}

The author would like to thank colleagues at the Joint Laboratory of Optics of Palacký University and the Institute of
Physics of the Czech Academy of Sciences for their assistance and fruitful discussions throughout this research. In particular, thank you to Professor Petr Travnicek and Dr. Jakub Kmec for proofreading the thesis and proving valuable comments.

\vspace{5mm}

Lastly, the author would like to thank the Pierre Auger and Telescope Array Collaborations for providing logistic support, part of the instrumentation to operate the FAST telescopes, and data for analysis purposes.

\begin{figure}[h!]
    \centering
    \includegraphics[width=\linewidth]{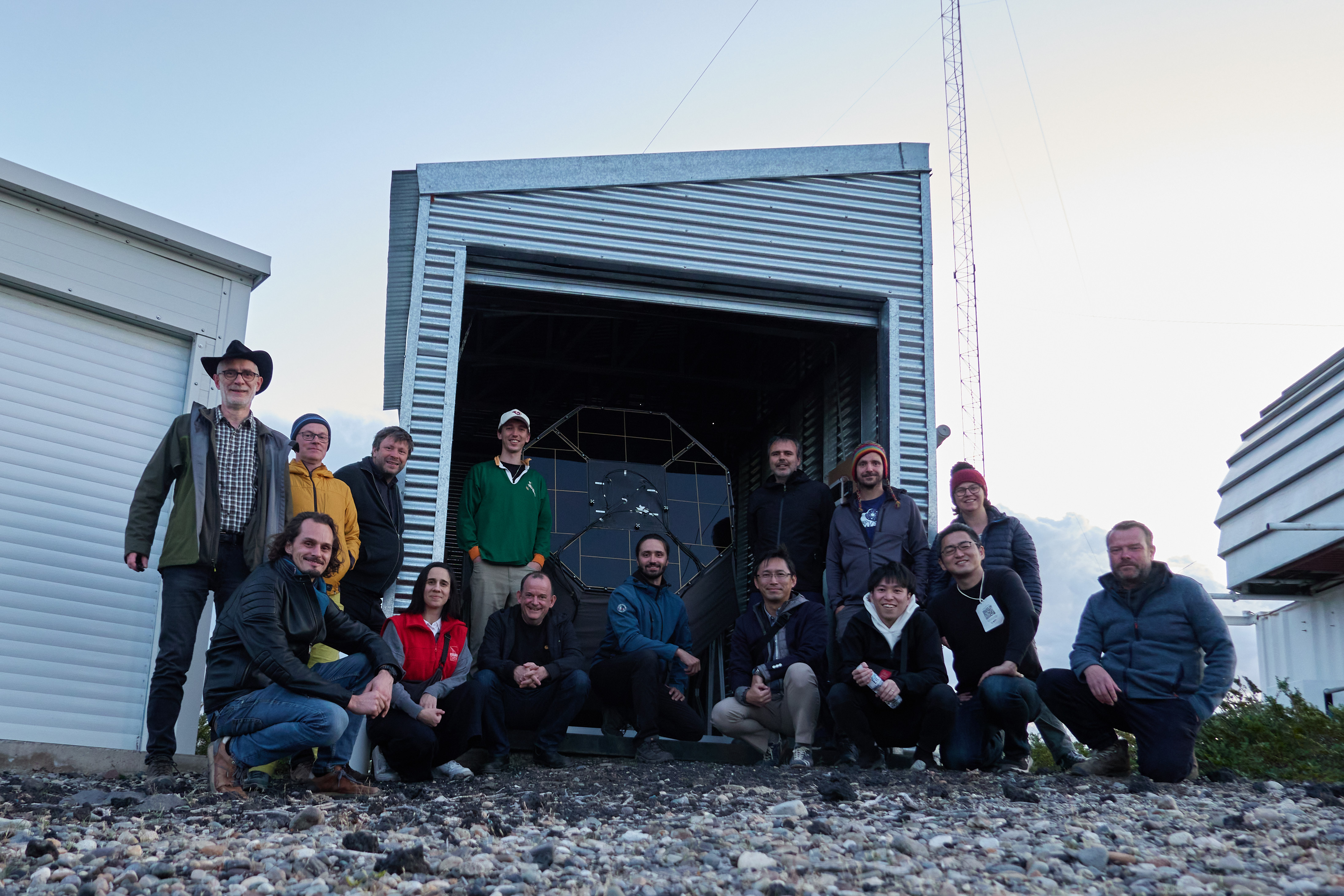}
\end{figure}
\addcontentsline{toc}{chapter}{Funding and Support}

\setglossarystyle{list}
\printglossary[title=Acronyms, type=\acronymtype]

\printbibliography[heading=bibintoc]

\end{document}